\documentclass[a4paper]{cmmp}
\pdfoutput=1

\usepackage{hyperref}

\usepackage{makeidx}
\usepackage[noadjust]{cite}
\makeindex

\usepackage{subdepth}
\usepackage{layouts}
  \usepackage{framed}

\usepackage{graphicx}
\usepackage{amsmath,amssymb,amsbsy,amstext, simplewick}
 \usepackage{exscale,relsize}
  \usepackage{dsfont}

\allowdisplaybreaks[1]

\usepackage{colortbl}
\definecolor{lightgreen}{cmyk}{0.2, 0, 0.2, 0.2}
\definecolor{lightgray}{cmyk}{0.1,0.2,0,0.1}
\definecolor{lightgray2}{cmyk}{0.1,0.1,0,0.1}
\usepackage{booktabs}

\def\beq{\begin{equation}}
\def\eeq{\end{equation}}
\def\bea{\begin{eqnarray}}
\def\eea{\end{eqnarray}}

\def\d{{\rm d}}
\def\x{\boldsymbol{x}}

\def\k{\boldsymbol{k}}

\def\n{\boldsymbol{n}}
\def\e{\boldsymbol{e}}
\def\R{{\cal R}}

\def\fnl{f_{\mathsmaller{\rm NL}}}
\def\Mp{M_{\rm pl}}
\def\Mpl{M_{\rm pl}}
\def\MKK{M_{\rm KK}}
\def\MSUSY{M_{\rm SUSY}}

\makeatletter

\newlength{\apb@width}
\newcommand{\autoparbox}[2][c]{\settowidth{\apb@width}{#2}\parbox[#1]{\apb@width}{#2}}
\newcommand{\includegraphicsbox}[2][]{\autoparbox{\includegraphics[#1]{#2}}}
\makeatother

\RequirePackage{color}

\begin{document}

\begin{titlepage}

\setcounter{page}{1} \baselineskip=15.5pt \thispagestyle{empty}

\bigskip\

\begin{center}
{\fontsize{22}{28} \selectfont  \sffamily \bfseries  Inflation and String Theory}
\end{center}

\vspace{0.2cm}

\begin{center}
{\fontsize{13}{30}\selectfont  Daniel Baumann$^{\bigstar}$ and Liam McAllister$^{ \blacklozenge}$}
\end{center}

\begin{center}

\vskip 8pt
\textsl{$^\bigstar$ D.A.M.T.P., Cambridge University, Cambridge, CB3 0WA, UK}
\vskip 7pt
\textsl{$^\blacklozenge$ Department of Physics, Cornell University, Ithaca, NY 14853, USA}

\end{center}

\vspace{1.2cm}
\hrule \vspace{0.3cm}
{ \noindent {\sffamily \bfseries Abstract} \\[0.1cm]
We review cosmological inflation and its realization in quantum field theory and in string theory. This material is a portion of a book, also entitled \emph{Inflation and String Theory}, to be published by Cambridge University Press.

\noindent
}
 \hrule

\end{titlepage}

\chapter*{Contents}
\input bookArXiv.toc

\chapter*{Preface}
\addcontentsline{toc}{chapter}{Preface}

The past two decades of advances in observational cosmology have brought about a revolution in our understanding of the universe.
Observations of type Ia supernovae~\cite{Perlmutter:1998np, Riess:1998cb}, measurements of temperature fluctuations in the cosmic microwave background~(CMB)---particularly by the {\it Wilkinson Microwave Anisotropy Probe} (WMAP)~\cite{Spergel:2003cb, Spergel:2006hy, Komatsu:2008hk, Komatsu:2010fb, Hinshaw:2012uq} and the {\it Planck} satellite~\cite{PlanckParameters, PlanckInflation, PlanckNG}---and maps of the distribution of large-scale structure~(LSS)~\cite{Abazajian:2008wr} have established a standard model of cosmology, the $\Lambda$CDM model. This is a universe filled with
68\% dark energy, 27\% dark matter, and only 5\% ordinary atoms~\cite{PlanckParameters}.
There is now decisive evidence that large-scale structures formed via gravitational instability of primordial density fluctuations, and
that these initial perturbations originated from quantum fluctuations~\cite{Mukhanov:1981xt, Chibisov:1982nx,Guth:1982ec,Hawking:1982cz, Starobinsky:1982ee, Bardeen:1983qw}, stretched to cosmic scales during a period of inflationary expansion~\cite{Guth:1980zm, Linde:1981mu, Albrecht:1982wi}.
However, the microphysical origin of inflation remains a mystery, and it will require a synergy of theory and observations to unlock it.

\vskip 4pt
In the standard cosmology without inflation, causal signals travel a finite distance between the time of the initial singularity and the time of formation of the first neutral atoms.
However, the CMB anisotropies display vivid correlations on scales larger than this distance.  This  causality puzzle  is known as the {\it horizon problem}.
The horizon problem is resolved if the early universe went through an extended period of {\it inflationary expansion}, i.e.~expansion at a nearly constant rate, with
\beq
|\dot H| \ll H^2\ , \nonumber
\eeq
where $H \equiv \dot a/a$ is the Hubble parameter associated with a Friedmann-Robertson-Walker spacetime,
\beq
\d s^2 = - \d t^2 + a^2(t) \d \x^2\ . \nonumber
\eeq
Because space expands quasi-exponentially during inflation, $a(t) \propto e^{Ht}$, homogeneous initial conditions on subhorizon scales are stretched to apparently acausal superhorizon scales. Besides
explaining the overall homogeneity of the universe, inflation also creates
small
primordial inhomogeneities, which eventually provide the seeds for large-scale structures. These perturbations are inevitable in a quantum-mechanical treatment of inflation:
viewed
as a quantum field,  the expansion rate $H$ experiences local zero-point fluctuations, $\delta H(t,\x)$, which lead to spatial variations in the density after inflation, $\delta \rho(t,\x)$.  If inflation is correct, then 
CMB observations are probing the quantum origin of structure in the universe.
By measuring the statistical properties of the CMB anisotropies we 
learn about the physics of inflation and about the precise mechanism that created the primordial seed fluctuations.

\vskip 4pt
In this book we will describe two intertwined approaches to the physics of inflation: from the bottom up in effective field theory (EFT), and from the top down in string theory.

\vskip 4pt
We speak of an {\it effective theory} when we do not resolve
the small-scale (or high-energy) details of a more fundamental theory.
Often this coarse-graining is done so automatically that it is not emphasized explicitly: for instance, we describe fluid dynamics and thermodynamics without reference to atoms, and computations  of atomic spectra are in turn insensitive to the quark substructure of nucleons.
Reasoning in terms of theories valid up to a critical energy scale is also how the history of particle physics developed, long before Wilson formalized the concept of effective theories.
Effective theories are
particularly useful when the full theory is unknown, or is specified but not computable.  In that case, one parameterizes the unknown physics  associated with degrees of freedom at a high energy scale $\Lambda$  by a collection of non-renormalizable interactions in the EFT, known as irrelevant interactions.
At low energies, irrelevant interactions are suppressed by powers of $E/\Lambda$. In the limit $E/\Lambda \to 0$,
the high-scale  degrees of freedom {\it decouple}.  However,  in some contexts a low-energy observable is  strongly affected by irrelevant interactions:  such an observable is termed {\it{ultraviolet (UV) sensitive}}.
As we shall explain, inflation is an ultraviolet-sensitive phenomenon.

The ultraviolet behavior of gravity is a foundational question for cosmology.  To understand the nature of  general relativity at high energies,  we recall that the interactions  dictated by the Einstein-Hilbert action can be encoded in Feynman rules,  just as in ordinary quantum field theory (see \cite{Bern:2002kj} for a modern perspective).  The coupling strength  is set by Newton's constant $G$, which has negative mass dimension,  so the interaction becomes stronger at higher energies.  Moreover, when divergences do arise,  they cannot be absorbed by renormalization of the terms  in the classical  Einstein-Hilbert Lagrangian: on dimensional grounds, the factors of the gravitational coupling from  graviton loops must be offset by additional derivatives  compared to the  classical terms.   General relativity is therefore
non-renormalizable,\footnote{Pure Einstein gravity is
free of one-loop divergences,  but diverges at two loops.    Gravitational theories  including matter fields  typically diverge at one loop,  except in  supersymmetric  cases \cite{Bern:2002kj}.}
and for energies above the Planck scale,
\beq
M_{P} \equiv \sqrt{\frac{\hbar c}{G}} = 1.2 \times 10^{19}\, {\rm GeV}/c^2\ , \nonumber \label{equ:MP}
\eeq
the theory stops making sense as a quantum theory: it violates unitarity.
The conservative interpretation of this finding is that new physics has to come into play at some energy below the Planck scale, and any quantum field theory that is coupled to gravity should then be interpreted as an effective theory valid at energies below the Planck scale.
This is precisely what happens in string theory: strings of characteristic size $\ell_{\rm s}$
cut off the divergences in graviton scattering  at  energies of order $1/\ell_{\rm s}$,
where the extended nature of the string becomes important. The result is a finite quantum  theory of gravity, whose long-wavelength description, at energies $E \ll 1/\ell_{\rm s}$, is an effective quantum field theory  that includes gravity, and whose  non-renormalizable interactions include terms suppressed by the Planck scale (or the string scale).
String theory  therefore  provides an  internally consistent framework for  studying  quantum fields  coupled to general relativity.

A striking feature of effective theories that support inflation is that they are sensitive to  Planck-suppressed interactions:
an otherwise successful model of inflation can be ruined by altering the spectrum and interactions of Planck-scale degrees of freedom.  In {\it{every}} model of inflation,  the  duration of the inflationary expansion  is affected by at least a small number of  non-renormalizable interactions suppressed by the Planck scale.    In a special class of scenarios called {\it large-field} models, an infinite series of  interactions,  of arbitrarily high dimension, affect the dynamics:  this  corresponds to  extreme sensitivity to Planck-scale physics.
The universal sensitivity of inflation to Planck-scale physics
implies
that a treatment in a theory of quantum gravity is required in order to address critical questions about the inflationary dynamics.
This is the cardinal motivation for pursuing an understanding of inflation in string theory.

\vskip 4pt
A primary subject of this book is the challenge of realizing inflationary dynamics
in string theory (recommended reviews on the subject include~\cite{Baumann:2009ni, McAllister:2007bg, Burgess:2007pz, Kallosh:2007ig, Baumann:2008aq, Baumann:2009ds, Linde:2005dd,Quevedo:2002xw, Cicoli:2011zz, Burgess:2013sla, Silverstein:2013wua}).
Let us set the stage for our discussion by outlining the range of gains that can be expected from this undertaking.

\vskip 4pt
The most conservative goal of studies of inflation in string theory is to place field-theoretic models of inflation on firmer logical footing, giving controlled computations of quantum gravity corrections to these models.
In particular, ultraviolet completion can clarify and justify
symmetry assumptions made in the EFT approach.
For example, realizing chaotic inflation~\cite{Linde:1983gd} through axion monodromy in string theory
\cite{McAllister:2008hb} gives a microphysical understanding of the shift symmetry, $\phi \mapsto \phi + const.$, that
ensures radiative stability of the low-energy~EFT.  Inflationary models relying on the shift symmetries of axions in string theory --- variants of `natural inflation'~\cite{Freese:1990rb} --- have provided one of the best-controlled paths to ultraviolet-complete scenarios yielding significant gravitational waves.
In favorable cases, the embedding into quantum gravity can also entail small modifications of the theory that lead to additional observational signatures.
For example, in axion monodromy inflation nonperturbative corrections introduce modulations of the power spectrum \cite{Flauger:2009ab} and the bispectrum \cite{Flauger:2010ja}.
This is an example where the structure of the ultraviolet completion could potentially be inferred from correlated signatures.

String theory is a far more constrained framework than effective field theory, and some effective theories that appear consistent at low energies do not admit ultraviolet completions in quantum gravity.
Enforcing the restrictions imposed by ultraviolet completion winnows the possible models, leading to improved predictivity.
For example, the DBI scenario~\cite{Silverstein:2003hf} may be viewed as a special case of k-inflation~\cite{ArmendarizPicon:1999rj}.
While most versions of k-inflation are radiatively unstable, string theory makes it possible to control an infinite series of higher-derivative terms.  In this case, a higher-dimensional symmetry significantly restricts the form of the four-dimensional effective action.  Unlike its field theory counterpart, the observational signatures of DBI inflation are correspondingly specific~\cite{Alishahiha:2004eh}.

String theory has also been an important source of inspiration for the development of novel effective field theories.  The geometric perspective afforded by compactifications, and by D-branes moving inside them, complements the more algebraic tools used to construct effective theories in particle physics.
The effective theories in D-brane inflation \cite{Kachru:2003sx,Baumann:2010sx}, DBI inflation~\cite{Silverstein:2003hf}, fibre inflation \cite{Cicoli:2008gp}, and axion monodromy inflation \cite{McAllister:2008hb}, for example, all exist in their own right as low-energy theories, but would likely have gone undiscovered without the approach provided by string theory.
Generating effective theories from the top down in string theory also leads to modified notions of what constitutes a natural inflationary model, or a minimal one.  Although we are very far from a final understanding of naturalness in string theory, one broad characteristic of existing geometric constructions is the presence of many light scalar fields, the {\it moduli} of the compactification.  Moduli play a central role in inflation, and can affect both the background evolution and the perturbations.  While theories with many `unnecessary' fields might be considered non-minimal in field-theoretic model-building, they are extremely common in string theory.

\vskip 4pt
The boldest hope for the use of string theory in cosmology is that string theory will open entirely new dynamical realms that cannot be described in any effective quantum field theory with a finite number of fields, and the resulting cosmic histories will avoid or overcome the limitations of contemporary models. While this enticing prospect has inspired work in string cosmology for more than two decades, in our opinion string theory is
not yet understood at the level required for such a dramatic step.  Even the low-energy effective actions governing the interactions of massless string states in non-supersymmetric vacua are not adequately characterized at present, while computing dynamics driven by the full tower of massive strings is a distant dream.  Fundamental advances in understanding time-dependent solutions of string theory with string scale curvatures will be required if we are to move outside the aegis of the effective theory for the massless modes.
In this book
we will restrict our attention to conservative applications of string theory to the study of inflation: we will survey the substantial literature in which string theory underpins or informs inflationary effective theories, but does not replace them outright.

\vskip 4pt
The task of making predictions in string theory is overshadowed by the problem of the {\it{landscape}}, i.e.~the fact that string theory has an astronomical number of vacua (see \cite{Douglas:2006es} for a review).
Although the dynamics that populates the landscape is poorly understood, false vacuum eternal inflation seems to be an unavoidable consequence.  The cosmological constant problem, the question of pre-inflationary initial conditions, and the challenge of defining a probability measure for eternal inflation are all facets of the fundamental problem of understanding the landscape and making predictions within it.   The number of vacua  is too large for
enumeration to be a realistic possibility \cite{Denef:2006ad},
but it does not follow that  in the landscape, `everything goes'.  Instead, there seem to exist strong
structural constraints on the properties of the vacua in the landscape.  For example, axion decay constants
appear to be smaller than the Planck mass in all computationally controllable vacua \cite{Banks:2003sx,Svrcek:2006yi}.  As we will see, this has important consequences for inflationary model-building in the context of string theory.
Moreover, all four-dimensional de Sitter vacua in supersymmetric string theories are metastable,
essentially because ten-dimensional Minkowski space is supersymmetric and therefore has zero energy, while a de Sitter solution has positive vacuum energy.
Constructing  a metastable de Sitter solution is much more difficult than finding a supersymmetric vacuum, and correspondingly, determining the prevalence of de Sitter vacua is far more subtle than counting supersymmetric solutions.
In fact, de Sitter solutions appear to be exponentially sparse in comparison to unstable saddle points \cite{Marsh:2011aa}.
The  formidable challenges of constructing and surveying the landscape compel us to understand dynamical selection effects in the early universe, but we have yet to see the first glimmering of a solution.

\vskip 10pt
The organization of this book is as follows:
in Chapter~\ref{sec:dS}, we define inflation as an extended period of quasi-de Sitter evolution,
and show how quantum fluctuations during this era lead to primordial density fluctuations and anisotropies in the CMB.  We review the current observational evidence in favor of the inflationary hypothesis.
In Chapter~\ref{sec:EFT},  we discuss the effective field theory approach to the physics of inflation.  We explain why the effective theories supporting inflation are unusually sensitive to UV physics, and highlight the importance of symmetries for the radiative
stability of inflationary models.
In Chapter~\ref{sec:StringTheory}, we  provide the groundwork for a discussion of inflation in string theory.
We first give a brief overview of string theory, emphasizing those aspects that are particularly relevant for research in string cosmology.  We examine string compactifications, discuss some leading mechanisms for moduli stabilization, and critically analyze proposals for metastable de Sitter vacua.
In Chapter~\ref{sec:StringInflation}, we then outline how inflation can arise in this context.
In Chapter~\ref{sec:Examples}, we provide a more detailed discussion of several classes of inflationary models in string theory.
We end, in Chapter~\ref{sec:Outlook}, by 
describing some challenges and opportunities for the field.

In an effort to make this book self-contained, and accessible for a reader who is entering the field,
we have included extensive background material in the appendices.\footnote{The appendices will appear in the final version of the book \cite{Book}, but are omitted from the arXiv version.}
In Appendix~A,
we collect mathematical concepts, definitions, and results that will be helpful for following the discussion in Chapters~\ref{sec:StringTheory}--\ref{sec:Examples}.
In Appendix~B,
we present the effective theory of adiabatic fluctuations during inflation~\cite{Creminelli:2006xe, Cheung:2007st}.
In Appendix~C, we introduce cosmological perturbation theory and derive the primordial perturbations from inflation.

\vskip 10pt
We are indebted to our colleagues and collaborators for sharing their insights
on the material presented in this book. Special thanks go to Peter Adshead, Nima Arkani-Hamed, Valentin Assassi, Thomas Bachlechner, Neil Barnaby, Cliff Burgess, Anthony Challinor, Xingang Chen, Miranda Cheng, David Chernoff, Michele Cicoli, Joseph Conlon, Paolo Creminelli, Sera Cremonini, Csaba Cs\'aki, Anne Davies, Anatoly Dymarsky,  Richard Easther, Raphael Flau\-ger, Daniel Green, Michael Green, Arthur Hebecker, Shamit Kachru, Renata Kallosh, Marc Kamion\-kowski, Igor Klebanov, Eiichiro Komatsu, Hayden Lee, Andrei Linde, Connor Long, Juan Maldacena, David Marsh, Paul Mc\-Guirk, Alberto Nicolis, Enrico Pajer, Hiranya Peiris, Maxim Perelstein, Rafael Porto, Fernando Quevedo, S\'ebastien Renaux-Petel, Raquel Ribeiro, Leonardo Senatore, David Seery, Paul Shellard, Eva Silverstein, Marko Simonovi\'c, David Sper\-gel, Paul Steinhardt, Andrew Tolley, David Tong, Sandip Trivedi, Henry Tye, Erik Verlinde, Herman Verlinde, Filippo Vernizzi, Yi Wang, Scott Watson, Alexander Westphal, Timm Wrase, Gang Xu, and Matias Zaldarriaga.

We are grateful to Valentin Assassi, Marcus Berg, Michele Cicoli, Joseph Conlon, Daniel Green, Emil Martinec, Enrico Pajer, and Fernando Quevedo for comments on the draft, and we are particularly indebted to John Stout and Alexander Westphal  for extensive  corrections.

Finally, we thank our editor, Vince Higgs of Cambridge University Press, for his guidance and support.

\vskip 4pt
D.B.~gratefully acknowledges support from the European Research Council (ERC STG grant 279617), the Science and Technology Facilities Council (STFC) and the Centre for Theoretical Cosmology in Cambridge.  L.M.~is grateful for support provided by the
National Science Foundation under grant PHY-0757868, by an NSF CAREER~award, and by a Simons Fellowship.

\vskip 20pt
\hfill Daniel Baumann and Liam McAllister

\vskip 2pt
\hfill Cambridge and Ithaca, 2014. 
\chapter*{Notation and Conventions}
\addcontentsline{toc}{chapter}{Notation and Conventions}

Throughout this book, we will employ natural units with $\hbar = c \equiv 1$.
Moreover, the reduced Planck mass,
$$
M_{\rm pl}^{-2} \equiv 8\pi G = \left(2.4 \times 10^{18}\, {\rm GeV} \right)^{-2}\ ,
$$
is often set equal to one.

\vskip 4pt
Our metric signature is mostly plus, $(-+++ \cdots)$.
We use $t$ for physical time and $\tau$ for conformal time.
We denote ten-dimensional spacetime coordinates by $X^M$, four-dimensional spacetime coordinates by $x^\mu$, three-dimensional spatial coordinates by~$x^i$, and three-dimensional vectors by $\x$. The coordinates of extra dimensions are $y^m$.  Worldsheet coordinates of strings and branes are $\sigma^a$.  The spacetime metric in ten dimensions is $G_{MN}$, while for the four-dimensional counterpart we use $g_{\mu \nu}$. The spatial 3-metric of the extended spacetime is $g_{ij}$, while the spatial 6-metric of the compact space is $g_{mn}$.  The worldsheet metric is $h_{ab}$. The notation $(\partial \phi)^2$ means $g^{\mu \nu} \partial_\mu \phi \partial_\nu \phi$ or $G^{MN} \partial_M \phi \partial_N\phi$, depending on the context.

\vskip 4pt
The letter $\pi$ stands both for $3.14159\cdots$ and for the Goldstone boson of spontaneously broken time translations.
We use $\R$ (not $\zeta$) for the curvature perturbation in comoving gauge.
Our Fourier convention is
$$
\R_{\k} = \int \d^3 x \, \R(\x) \,e^{i \k\cdot \x} \ .
$$
The power spectrum for a statistically homogeneous field is defined by
$$
\langle \R_{\k\vphantom{'}} \R_{\k'} \rangle = (2\pi)^3 P_\R(k) \delta(\k + \k') \ .
$$
We also use the dimensionless power spectrum
$$
\Delta_\R^2(k) \equiv \frac{k^3}{2\pi^2} P_\R(k)\ .
$$

\vskip 4pt
The Hubble slow-roll parameters are
$$
\varepsilon  \equiv -  \frac{\dot H}{H^2}\ , \qquad \tilde \eta  \equiv \frac{\dot \varepsilon}{H \varepsilon}\ , 
$$
where overdots stand for derivatives with respect to physical time $t$.
The potential slow-roll parameters are
$$
\epsilon \equiv \frac{\Mp^2}{2} \left( \frac{V'}{V}\right)^2 \ , \qquad \eta \equiv \Mp^2 \frac{V''}{V}\ ,
$$
where primes are derivatives with respect to the inflaton $\phi$, and $V(\phi)$ is the potential energy density.

\vskip 4pt
We define the string length and the string mass, respectively, as
$$
\ell_{\rm s}^2 \equiv \alpha' \ , \qquad M_{\rm s}^2 \equiv \frac{1}{\alpha'}\ ,
$$
where $\alpha'$ is the Regge slope.
Beware of factors of $2\pi$ in alternative definitions of these quantities  in the literature.
The ten-dimensional gravitational coupling  is
$$
 2 \kappa^2 = (2\pi)^7 (\alpha')^4\ .
$$

\chapter{Inflation: Theory and Observations}
\label{sec:dS}

A fundamental observational fact about our universe is that on large scales it is well-described by the spatially flat
Friedmann-Robertson-Walker (FRW) metric\index{FRW metric}
\beq
\d s^2 = - \d t^2 + a^2(t) \hskip 1pt \d \x^2 \ . \label{equ:FRW}
\eeq
In \S\ref{sec:horizon}, we first explain why the homogeneity, isotropy, and flatness of the universe encoded in (\ref{equ:FRW}) are puzzling in the standard cosmology. We then show how an early phase of quasi-de Sitter evolution drives the primordial universe towards these conditions, even if it started in an inhomogeneous,  anisotropic,  and curved initial state.
In~\S\ref{sec:zeta}, we argue that quantum fluctuations during inflation are the origin of all structure in the universe, and we derive the power spectra of scalar and tensor fluctuations.
In \S\ref{sec:Obs}, we describe the main cosmological observables, which are used, in \S\ref{sec:obs}, to obtain constraints on the inflationary parameters. We then review recent experimental results.
Finally, in \S\ref{sec:future}, we discuss future prospects for testing the physics of inflation with cosmological observations.

 \section{Horizon Problem}
\label{sec:horizon}

\subsection{Radiation-Dominated Universe}

\begin{figure}[h!]
   \centering
      \includegraphics[width=0.9\textwidth]{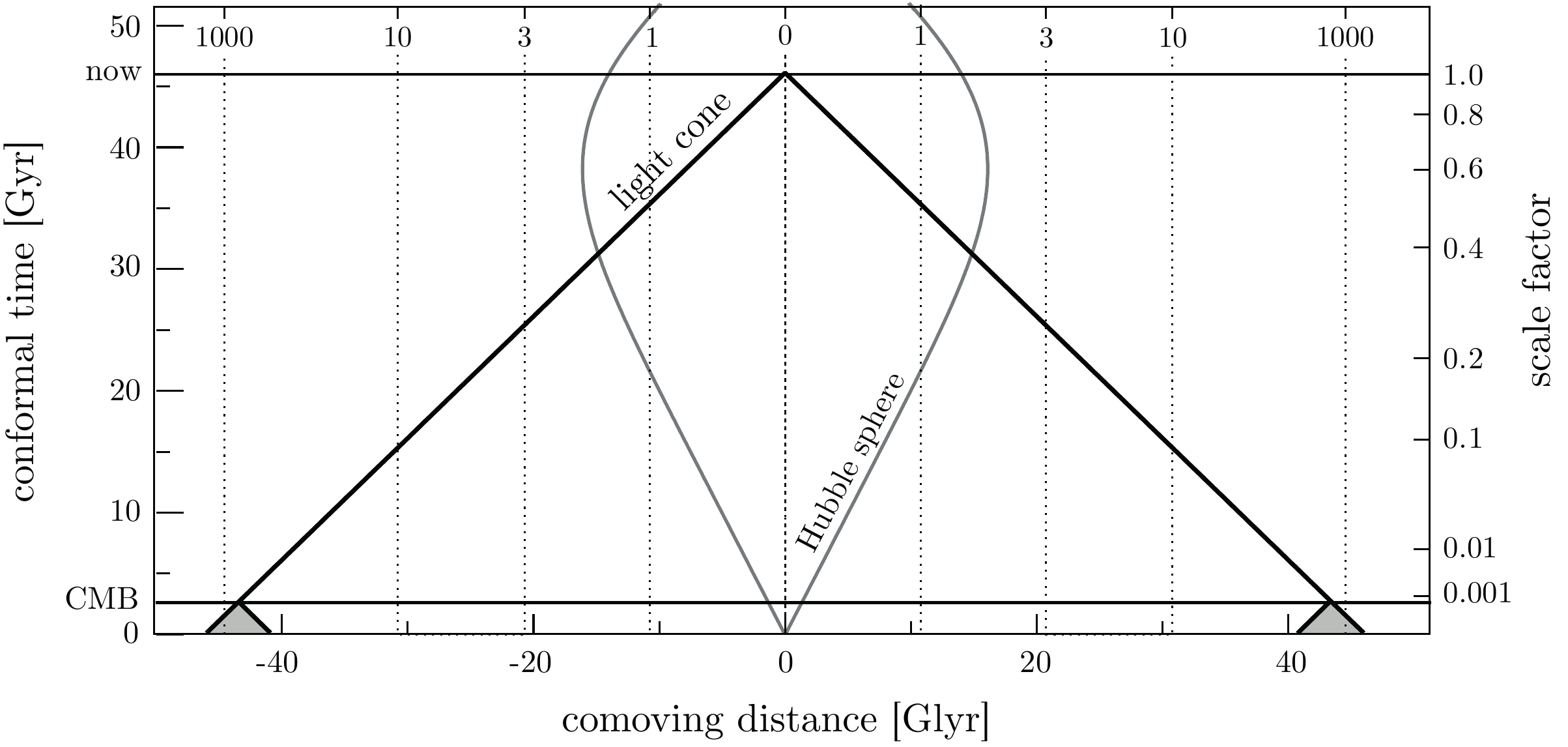}
   \caption{Spacetime diagram illustrating the horizon problem
   in comoving coordinates (figure adapted from \cite{Lineweaver:2003ie}). The dotted vertical lines correspond to the worldlines of comoving objects.  We are the central worldline.  The current redshifts of the comoving galaxies are labeled on each worldline.  All events that we currently observe are on our past light cone. The intersection of our past light cone with the spacelike slice labeled CMB corresponds to two opposite points on the CMB surface of
   last-scattering.
   The past light cones
   of these points, shaded gray, do not overlap, so the points appear never to have been in causal contact.}
  \label{fig:FRW-horizon}
\end{figure}

To discuss the causal structure of the FRW spacetime, we write the metric (\ref{equ:FRW}) in terms of conformal time:\index{conformal time}
\beq
\d s^2 = a^2(\tau) \left[ - \d \tau^2 +  \hskip 1pt \d \x^2 \right] \ , \label{equ:FRW2}
\eeq so that the distance
$|\Delta \x|$ (the {\it comoving} distance) that a particle can travel between times $\tau_1$ and $\tau_2 = \tau_1 + \Delta\tau$  is simply
$|\Delta \x| = \Delta \tau$, for any $a(\tau)$.  In the standard Big Bang cosmology, the expansion at early times is driven by the energy density of radiation, and by tracing the evolution backward one finds that $a \to 0$ at sufficiently early times, and the metric becomes singular at this point. We choose coordinates so that the initial singularity is  at $t=0$.
At some time $t>0$, the maximum comoving distance a particle can  have traversed since the initial singularity (a.k.a the {\it{particle horizon}}) is given by
\beq
\Delta \tau = \int_0^{\hskip 1pt t} \frac{\d t'}{a(t')} = \int_0^{\hskip 1pt a} \frac{\d \ln a}{aH}\ , \qquad {\rm where} \quad H \equiv \frac{1}{a} \frac{da}{dt}\ . \label{equ:tau}
\eeq
During the standard Big Bang evolution, $\ddot{a}<0$ and the comoving Hubble radius $(aH)^{-1} = (\dot a)^{-1}$ grows with time.
The integral in~(\ref{equ:tau}) is therefore dominated by the contributions from late times.
This leads to the so-called {\it horizon problem}.\index{horizon problem}
The amount of conformal time that
elapses between the singularity and the formation of the cosmic microwave background (an event known as recombination)
is much smaller than the conformal time between recombination and today (see fig.~\ref{fig:FRW-horizon}).
Quantitatively, one finds that points in the CMB that are separated by more than one degree were never in causal contact, according to the standard cosmology:
their past light cones do not overlap before the spacetime is terminated by the initial singularity. Yet their temperatures are observed to be the same to one part in $10^4$.
Moreover, the observed temperature fluctuations are {\it correlated} on what seem to be superhorizon scales at recombination.  Not only
must we explain why the CMB is so uniform,
we must also explain
why its small fluctuations are correlated on apparently acausal scales.

\begin{figure}[h!]
   \centering
      \includegraphics[width=0.9\textwidth]{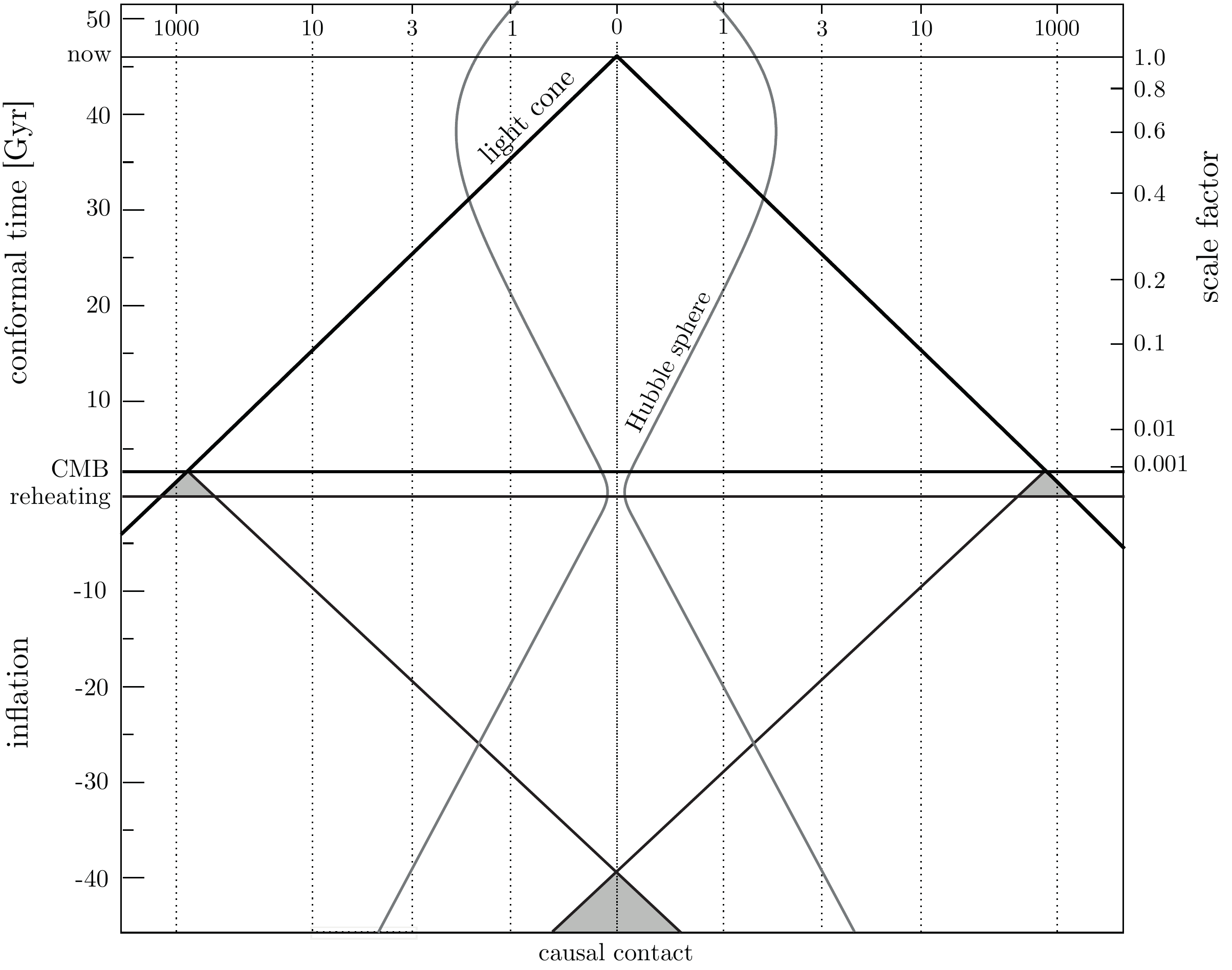}
   \caption{Inflationary solution to the horizon problem. The comoving Hubble sphere shrinks during inflation and expands during the conventional Big Bang evolution (at least until dark energy takes over). Conformal time during inflation is negative. The spacelike singularity of the standard Big Bang is replaced by the reheating surface: rather than marking the beginning of time, $\tau=0$ now corresponds to the transition from inflation to the standard Big Bang evolution. All points in the CMB have overlapping past light cones and therefore originated from a causally connected region of space.}
  \label{fig:Inf-horizon}
\end{figure}

 \subsection{Cosmic Inflation}
\label{sec:CosmicInflation}

To address the horizon problem, we may postulate that the comoving Hubble radius was decreasing in the early universe, so that the integral in~(\ref{equ:tau}) is dominated by the contributions from early times. This introduces an additional span of conformal time between the singularity and recombination (see fig.~\ref{fig:Inf-horizon}): in fact, conformal time now extends to negative values.
If the period of decreasing comoving Hubble radius is sufficiently prolonged, all points in the CMB originate from a causally connected region of space.
The observed correlations can therefore result from ordinary causal processes at early times.

In an expanding universe, a shrinking comoving Hubble sphere implies
\beq
\frac{d}{dt} (aH)^{-1} =  - \frac{1}{a} \left[ \frac{\dot H}{H^2} + 1 \right] < 0 \qquad \Rightarrow \qquad \varepsilon \equiv - \frac{\dot H}{H^2} < 1\ . \label{equ:eps}
\eeq
We will take the slow evolution  of the Hubble parameter, $\varepsilon < 1$,  as our definition of inflation.
This definition includes, but is not limited to, the dynamics of a slowly rolling scalar field (see \S\ref{sec:SR}).
In the de Sitter limit, $\varepsilon \to 0$, the space grows exponentially,
\beq
a(t) \propto e^{Ht}\ ,
\eeq
with $H \approx const$.

Inflationary expansion requires a somewhat unconventional matter content.
 In a spatially-flat FRW universe supported by a perfect fluid, the Einstein equations lead to the Friedmann equations
 \begin{align}
 3\Mp^2 H^2 &\, =\, \rho\ , \label{equ:Fried1}  \\
 6\Mp^2(\dot{H}+H^2) &\,=\,-(\rho+3P) \ ,\label{equ:Fried2}
 \end{align}
 where $\rho$ and $P$ are the energy density and pressure of the fluid. Combining (\ref{equ:Fried1}) and (\ref{equ:Fried2}), we find
 \begin{equation}
2\Mp^2 \dot{H} = - (\rho + P)\ , \label{equ:Hdot}
\end{equation}
 and hence
 \beq
 \varepsilon = \frac{3}{2}\left(1 + \frac{P}{\rho} \right)\ .
 \eeq
Inflation therefore occurs when $P < - \frac{1}{3}\rho$, corresponding to a violation of the
strong energy condition (SEC).\footnote{For a perfect fluid, the SEC states that $\rho + P \ge 0$ and $\rho + 3P \ge 0$.}  One simple energy source that can drive inflation is
a positive potential energy density of a scalar field with negligible kinetic energy, but we will encounter a range of alternative mechanisms.

\section{Primordial Perturbations}
\label{sec:zeta}

\vspace{0.3cm}
\begin{quote}
{\footnotesize With the new cosmology the universe must have been started off in some very simple way. What, then, becomes of the initial conditions required by dynamical theory? Plainly there cannot be any, or they must be trivial. We are left in a situation which would be untenable with the old mechanics. If the universe were simply the motion which follows from a given scheme of equations of motion with trivial initial conditions, it could not contain the complexity we observe. Quantum mechanics provides an escape from the difficulty. It enables us to ascribe the complexity to the quantum jumps, lying outside the scheme of equations of motion. The quantum jumps now form the uncalculable part of natural phenomena, to replace the initial conditions of the old mechanistic view.} \vskip 0.1pt \hfill
{\footnotesize P.~A.~M.~Dirac~\cite{Dirac}.}
\end{quote}
\vspace{0.2cm}

Inflation not only explains the homogeneity of the universe, but also provides a mechanism to create the primordial inhomogeneities required for structure formation~\cite{Mukhanov:1981xt, Chibisov:1982nx,Guth:1982ec,Hawking:1982cz, Starobinsky:1982ee, Bardeen:1983qw}. This process happens automatically when we treat the inflationary de Sitter phase quantum mechanically.
Here, we briefly sketch the quantum generation of primordial fluctuations.  We also present the modern view of inflation as a symmetry breaking phenomenon~\cite{Creminelli:2006xe,Cheung:2007st}.
For more details, see Appendices~B and C.

\subsection{Goldstone Action}
\label{sec:Gold}

By definition, inflation is a transient phase  of accelerated expansion,  corresponding approximately, but not exactly, to a de Sitter solution.
In order for inflation to end, the time-translation invariance  present in an eternal de Sitter  spacetime must be broken.
The slow evolution of the Hubble parameter $H(t)$  serves as a clock that measures the progress of inflation,
breaking time translation invariance and defining a preferred time slicing of the  spacetime.
The isometries of de Sitter space, ${\rm SO}(4,1)$, are spontaneously broken down to just spatial rotations and translations.
It is often useful to think of the time slicing  as being defined by the time-dependent expectation value $\psi_m(t)$ of one or more  bosonic fields $\psi_m$.

 \begin{figure}[h!]
   \centering
      \includegraphics[scale=0.37]{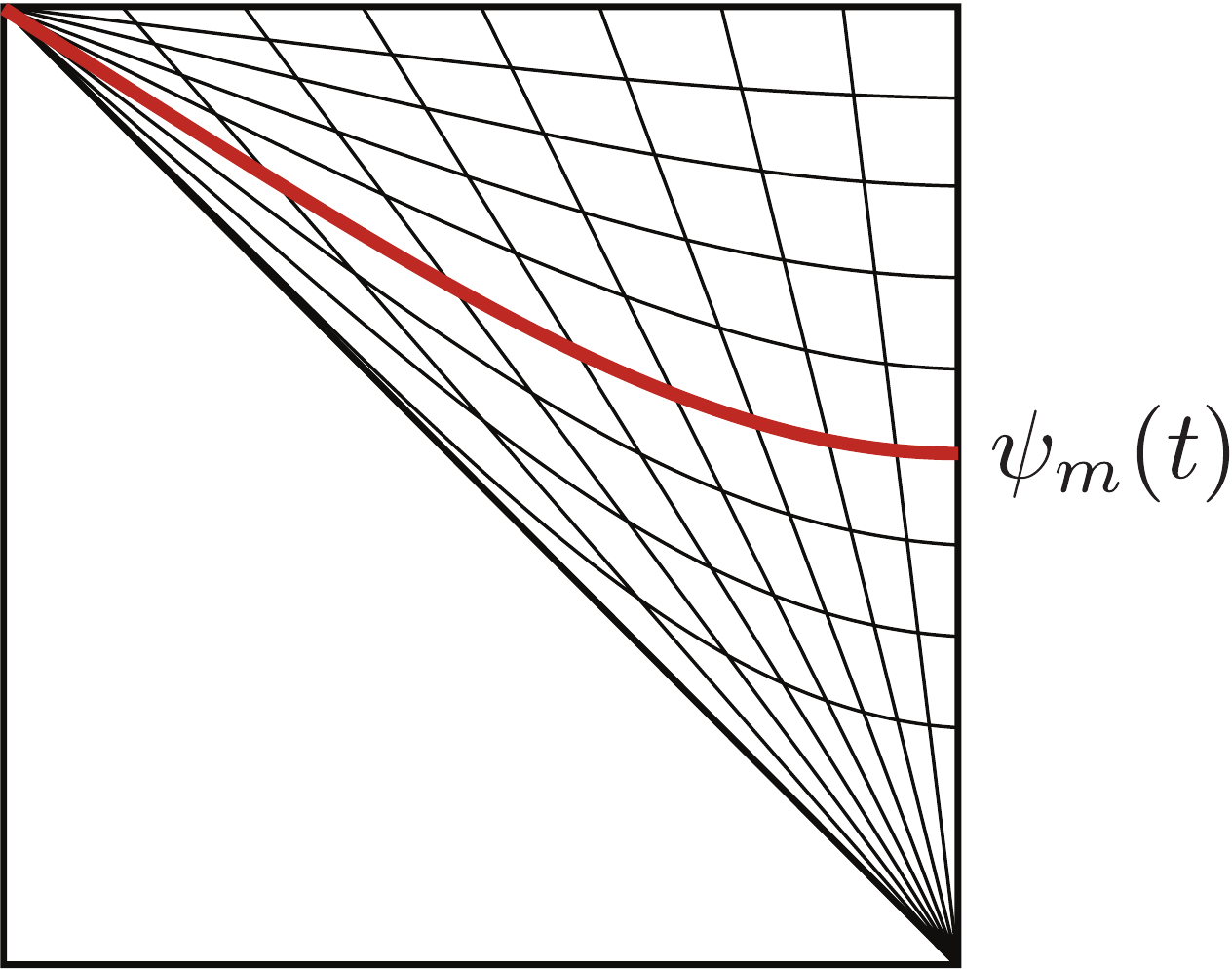}
   \caption{Time-dependent background fields $\psi_m(t)$ introduce a preferred time slicing of de Sitter space.}
  \label{fig:dS-Penrose}
\end{figure}

As with spontaneously broken symmetries  in  flat-space quantum field theory (see e.g.~\cite{Weinberg:1996kr}), the broken symmetry is {\it nonlinearly realized} by a Goldstone boson. Focusing on symmetry breaking and on the physics of the Goldstone boson allows a model-insensitive description of fluctuations during inflation~\cite{Cheung:2007st}.
In particular, we can defer  consideration of the dynamics that created the background evolution $H(t)$,
though ultimately we will return to explaining the background.

The Goldstone boson associated with the spontaneous  breaking of time translation invariance
is introduced as a spacetime-dependent transformation along the direction of the broken symmetry, i.e.~as a spacetime-dependent shift of the time coordinate~\cite{Creminelli:2006xe}
\beq
U(t,\x) \equiv
t + \pi(t,\x) \ .
\eeq
The Goldstone mode $\pi$ parameterizes {\it adiabatic fluctuations} of the fields $\psi_m$, i.e.~perturbations corresponding to a common, local shift in time of the homogeneous fields\index{adiabatic perturbations}
 \beq
\delta \psi_m(t,\x) \equiv \psi_m\bigl(t+\pi(t,\x )\bigr) - \psi_m(t)\ .  \label{defofpi1}
\eeq
The Einstein equations couple the Goldstone boson $\pi$ to metric fluctuations~$\delta g_{\mu \nu}$.
A
convenient gauge for describing these fluctuations is the  {\it spatially flat gauge}, where the spatial part of the metric is unperturbed
\beq
g_{ij} = a^2(t) \hskip 1pt  \delta_{ij} \ .
\eeq
The remaining metric fluctuations $\delta g_{00}$ and $\delta g_{0i}$ are related to $\pi$ by the Einstein constraint equations.
The dynamics of the coupled  Goldstone-metric system can therefore be described by $\pi$ alone.

A second description of the same physics is sometimes convenient, especially in the cosmological context.
First, we note that, for purely adiabatic fluctuations, we can perform a time reparameterization that removes all matter fluctuations, $\delta \psi_m \mapsto 0$. This takes us to {\it comoving gauge}, where the field $\pi$ has been `eaten' by the metric $g_{\mu \nu}$.  The spatial part of the metric can now be written as
\beq
g_{ij} = a^2(t) \hskip 1pt e^{2\R(t,\x)} \hskip 1pt \delta_{ij}\ ,
\eeq
where $\R$ is called the {\it comoving curvature perturbation}.\index{curvature perturbation} The other components of the metric are related to $\R$ by the Einstein constraint equations 
(see Appendix~C).
The relationship between $\pi$ (in spatially flat gauge) and $\R$ (in comoving gauge) is
\beq
\R = - H \pi + \cdots\ , \label{equ:ZetaPi}
\eeq
where the ellipses denotes terms that are higher order in $\pi$.
This links the comoving curvature perturbation $\R$ with the Goldstone boson $\pi$ of spontaneous symmetry breaking during inflation~\cite{Weinberg:2003sw, Hinterbichler:2012nm}.

The Goldstone mode $\pi$ exists in every model of inflation.  In single-field inflation, $\pi$ is the unique fluctuation mode~\cite{Cheung:2007st}, while in multi-field inflation, additional light fields can contribute to $\R$: see Appendix~B.
As we will see in Chapter~\ref{sec:StringInflation}, string theory strongly motivates  considering  scenarios  in which multiple fields are light during inflation.
However, from a purely bottom-up perspective, extra light fields during inflation are not required by present observations,  and in this section we will focus
on the minimal case  of a single light field.

One can learn a great deal about the CMB  perturbations
by studying the Goldstone boson fluctuations alone.  The physics of the Goldstone boson is described by the low-energy effective action for $\pi$, which can be obtained by writing down  the most general Lorentz-invariant action for the field $U \equiv t + \pi$:
\beq
S = \int \d^4 x \sqrt{-g} \, {\cal L}[U,(\partial_\mu U)^2, \Box U, \cdots] \ . \label{equ:SU}
\eeq
The action (\ref{equ:SU}) is manifestly invariant under spatial diffeomorphisms, but because $\pi$  transforms nonlinearly under time translations, one says that time translation symmetry is  nonlinearly realized in (\ref{equ:SU}).
Expanding (\ref{equ:SU}) in powers of $\pi$ and derivatives gives the effective action for the Goldstone mode.
We derive the Goldstone action in detail in Appendix~B,
via an alternative geometric approach~\cite{Creminelli:2006xe,Cheung:2007st}, and present only the main results here.
At quadratic order in $\pi$, and to leading order in derivatives, one finds
\beq
S^{(2)}_\pi =   \int \d^4 x\,\sqrt{-g}\,\,   \frac{M_{\rm pl}^2 |\dot H| }{c_s^2} \left[ \dot \pi^2 - \frac{c_s^2}{a^2} (\partial_i \pi)^2 + 3 \varepsilon H^2 \pi^2 \right]\ , \label{equ:piAction}
\eeq
where $(\partial_i \pi)^2 \equiv \delta^{ij} \partial_i \pi \partial_j \pi$.
Since Lorentz symmetry is broken by the time-dependence of the background, we have the possibility of a nontrivial speed of sound $c_s$;\index{speed of sound}
standard slow-roll inflation (see \S\ref{sec:SR}) is recovered for $c_s =1$.
The field $\pi$ has a small mass term, which arises from the mixing between $\pi$ and the metric fluctuations.
Using (\ref{equ:ZetaPi}), we can write (\ref{equ:piAction}) in terms of the curvature perturbation $\R$,
\beq
S^{(2)}_{\R} = \frac{1}{2} \int \d^4 x\,\, a^3 \,  y^2(t)\, \left[ \dot \R^2 - \frac{c_s^2}{a^2} (\partial_i \R)^2 \right]\ ,  \label{equ:zetaAction}
\eeq
where
\beq
y^2\equiv 2 M_{\rm pl}^2 \frac{ \varepsilon}{c_s^2}\ .
\eeq
The field $\R$ is therefore massless, implying ---  as we shall see ---  that it is conserved on superhorizon scales~\cite{Weinberg:2003sw}.

\begin{figure}[h!]
    \centering
        \includegraphics[width=0.95\textwidth]{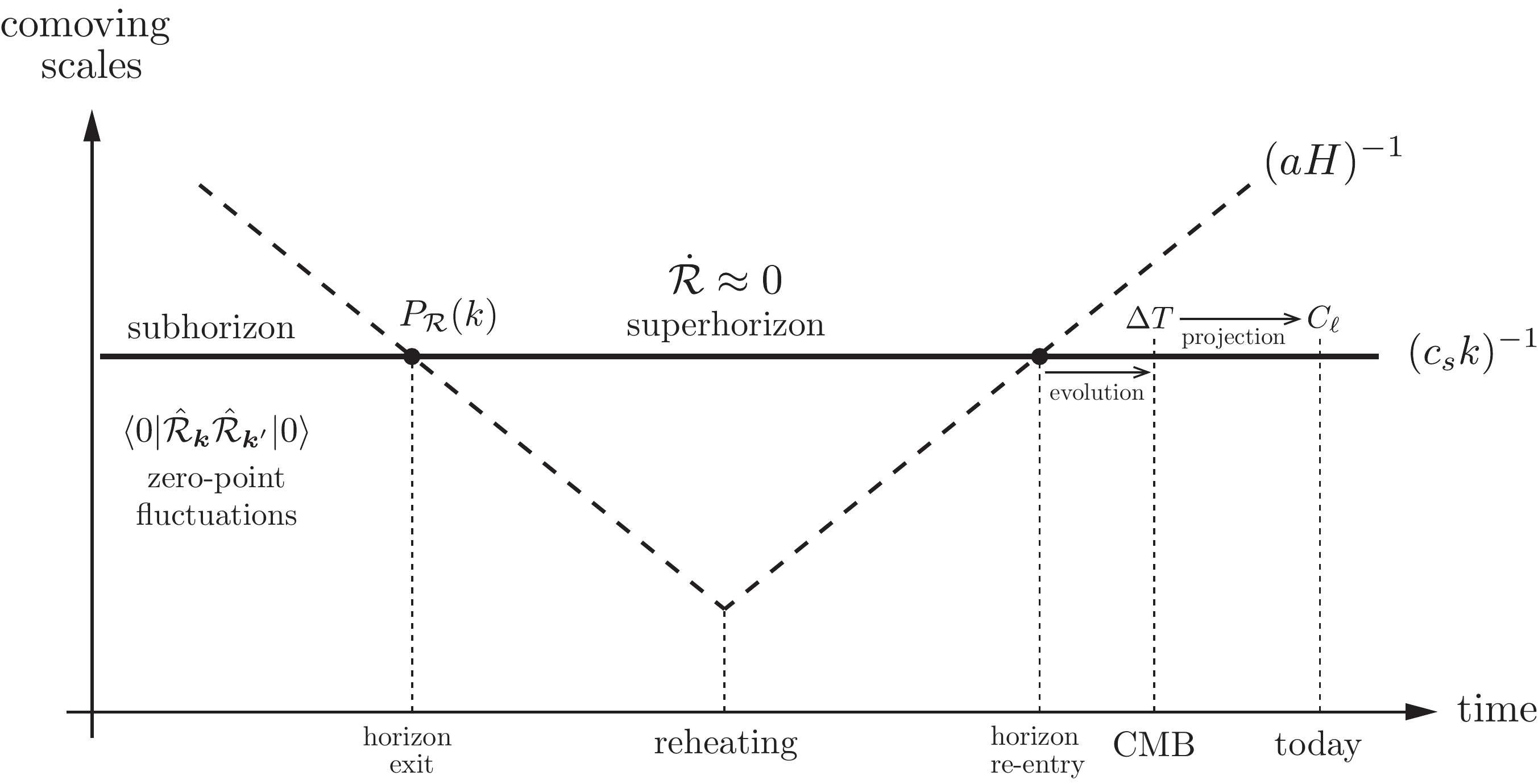}
    \caption{The evolution of curvature perturbations during and after inflation: the comoving horizon $(aH)^{-1}$ shrinks during inflation and grows in the subsequent FRW evolution. This implies that comoving scales $(c_s k)^{-1}$ exit the horizon at early times and re-enter the horizon at late times. In physical coordinates, the Hubble radius $H^{-1}$ is constant and the physical wavelength grows exponentially, $\lambda \propto a(t) \propto e^{Ht}$.  For adiabatic fluctuations, the curvature perturbations $\R$  do not  evolve outside of the horizon, so the power spectrum $P_\R(k)$ at horizon exit during inflation can be related directly to CMB observables at late times. }
    \label{fig:summary}
\end{figure}

For simplicity, we will assume that $\varepsilon$ and $c_s$ are nearly constant, so that the overall normalization of the action can be absorbed into the definition of a new, canonically-normalized, field
\beq
v \equiv y\, \R = \int \d^3 k\,\left[ v_{k}(t) \, a_{\k} \, e^{i\k \cdot \x} + c.c. \right] \ . \label{equ:v}
\eeq
We have written $v$
in terms of time-independent stochastic parameters $a_{\k}$ and time-dependent mode functions~$v_k(t)$.
The mode functions satisfy the {\it Mukhanov-Sasaki equation}\index{Mukhanov-Sasaki equation}
\beq
\ddot v_k + 3 H \hskip 1pt \dot v_k + \frac{c_s^2 k^2}{a^2} v_k = 0 \ . \label{equ:MS}
\eeq
This is the equation of a simple harmonic oscillator with a friction term provided by the expanding background. The oscillation frequency depends on the physical momentum and is therefore time-dependent:
\beq
\omega_k(t) \equiv \frac{c_s k}{a(t)}\ .
\eeq
At early times (small $a$), $\omega_k \gg H$ for all modes of interest. In this limit, the friction is irrelevant and the modes oscillate.
However, the frequency of each mode drops exponentially during inflation.
At late times (large $a$), the dynamics is dominated by friction and the mode has a constant amplitude. We say that the mode `freezes' at {\it{horizon crossing}}, i.e.~when $\omega_k(t_\star) = H$ or $c_s k = aH(t_\star)$.
It is these constant superhorizon fluctuations that eventually become the density fluctuations that we  observe in the CMB or in LSS (see fig.~\ref{fig:summary}).\footnote{Recall that we are assuming adiabatic initial conditions.
The presence of entropy perturbations,  as in  multi-field models, can complicate the relation between the curvature perturbations at horizon crossing and the late-time observables.}

\subsection{Vacuum Fluctuations}
 \label{sec:vacuum}

 The initial conditions for $v$ (or $\R$) are computed by treating it as a quantum field in a classical inflationary background spacetime.
This calculation has become textbook material~\cite{Dodelson:2003ft, Mukhanov:2005sc} and can
also be found in many reviews (e.g.~\cite{Jacobson:2003vx, Baumann:2009ds}). We present the details in Appendix~C.
Here, we will restrict ourselves to a simplified, but intuitive, computation~\cite{Hollands:2002yb}.

\vskip 4pt
The Fourier modes of the classical field $v$ are promoted to quantum operators
\beq
\hat v_{\k} = v_k(t) \hat a_{\k} + h.c.
\eeq
We have seen that at sufficiently early times
all modes of cosmological interest were deep inside the Hubble radius. In this limit, each mode behaves as an ordinary harmonic oscillator.
The operators $\hat a_{\k}$ play the role of the annihilation operators of the quantum oscillators. The vacuum state is defined by $\hat a_{\k} | 0 \rangle = 0$. The oscillation amplitude will experience the same zero-point fluctuations as an oscillator in flat space,  $\langle 0| \hat v_{\k\vphantom{'}} \hat v_{\k'} | 0 \rangle = (2\pi)^3 |v_k|^2 \delta(\k + \k')$,
where
\begin{align}
|v_k|^2 &= \frac{1}{a^3}  \frac{1}{2 \omega_k}\ . \label{equ:zero}
 \end{align}
The factor of $a^{-3}$ arises from the physical volume element in the Lagrangian (\ref{equ:zetaAction})---note
that the Fourier mode~$v_k$ was defined using the comoving coordinates rather than the physical coordinates. The second factor, $1/(2\omega_k)$, is the standard result for the variance of the amplitude of a harmonic oscillator in
its ground state. (In inflation, this state is the {\it Bunch-Davies vacuum}.)\index{Bunch-Davies vacuum}
As long as the physical wavelength of the mode is smaller than the Hubble radius, the ground state will evolve adiabatically.  Eq.~(\ref{equ:zero}) then continues to hold and the precise time at which we define the initial condition is not important.
Once
a given
mode gets stretched outside the Hubble radius, the adiabatic approximation breaks down and the fluctuation amplitude freezes at
\beq
| v_k|^2 = \frac{1}{2}\frac{1}{a_\star^3} \frac{1}{c_s k/a_\star}\ , \label{equ:zero1}
\eeq
where $a_\star$ is the value of the scale factor at horizon crossing,
\beq
\frac{c_s k}{a_\star} = H\ . \label{equ:cross}
\eeq
Combining (\ref{equ:cross}) and (\ref{equ:zero1}), we get
\beq
| v_k|^2 = \frac{1}{2} \frac{H^2}{(c_sk)^3}\ , \label{equ:vResult}
\eeq
where from now on it is understood implicitly that the right-hand side is evaluated at horizon crossing.

 \subsection{Curvature Perturbations}

 Using (\ref{equ:v}), we obtain the {\it power spectrum} of primordial curvature perturbations
\beq
P_\R(k) \equiv | \R_k|^2 =  \frac{1}{4} \frac{H^4}{M_{\rm pl}^2 |\dot H| c_s} \frac{1}{k^3}\ .
\eeq
The variance in real space is $\langle \R^2 \rangle = \int \d \ln k \,\, \Delta_\R^2(k)$,
where we have defined the dimensionless power spectrum
\beq
\Delta_\R^2(k) \equiv \frac{k^3}{2\pi^2} P_\R(k)  =\frac{1}{8 \pi^2} \frac{H^4}{M_{\rm pl}^2 |\dot H| c_s} \ . \label{equ:DZ}
\eeq
Since the right-hand side is supposed to be evaluated at horizon crossing, $c_s k = a H$,
 any time dependence of $H$ and $c_s$ translates into a scale dependence of the power spectrum. Scale-invariant fluctuations correspond to $\Delta_\R^2(k) = const.$, and
deviations from scale invariance are quantified by the
{\it spectral tilt}\index{scalar spectral index}
\beq
n_s - 1 \equiv \frac{d \ln \Delta_\R^2}{d \ln k}= -2\varepsilon - \tilde \eta - \kappa\ ,
\eeq
where we have defined two additional expansion parameters,
\beq
\tilde \eta \equiv \frac{\dot \varepsilon}{H \varepsilon} \quad\ {\rm and} \quad\ \kappa \equiv \frac{\dot c_s}{H c_s} \ . \label{equ:Hubble2}
\eeq
Inflationary backgrounds typically satisfy
$\{\hskip 1pt \varepsilon, |\tilde \eta|, |\kappa| \hskip 1pt \} \ll 1$ and hence predict $n_s \approx 1$.
Inflation would not end if the slow-roll parameters vanished, so importantly we also expect
a finite deviation from perfect scale-invariance, $n_s \ne 1$.

 \subsection{Gravitational Waves}

 Arguably the cleanest prediction of inflation is a  spectrum of primordial gravitational waves. These are tensor perturbations to the spatial metric,
\beq
g_{ij} =a^2(t)(\delta_{ij} + 2 h_{ij})\ ,
\eeq
where $h_{ij}$ is transverse and traceless.
Expanding the Einstein-Hilbert action leads to the quadratic action for the tensor fluctuations:
\beq
S^{(2)}_h = \frac{1}{2} \int  \d^4 x\,\, a^3  \,  y^2\, \left[ ( \dot h_{ij})^2 -  \frac{1}{a^2}(\partial_k h_{ij})^2 \right] \ , \label{equ:hAction}
\eeq
where
\beq
y^2\equiv \frac{1}{4} M_{\rm pl}^2\ .
\eeq
The structure of the action is identical to that of the scalar fluctuations, eq.~(\ref{equ:zetaAction}), except that tensors do not have a nontrivial sound speed and the relation to the canonically-normalized field does not include $\varepsilon$,
because at linear order tensors do not feel the symmetry breaking due to the background evolution.
The quantization of tensor fluctuations is therefore the same as for the scalar fluctuations. In particular, eq.~(\ref{equ:vResult}) applies to each polarization mode of the gravitational field. Adding the power spectra of the two polarization modes, one finds~\cite{Starobinsky:1979ty}
\beq
\Delta_h^2(k) \equiv  \frac{k^3}{2\pi^2}P_h(k) \, =\, \frac{2}{\pi^2} \frac{H^2}{M_{\rm pl}^2} \ , \label{equ:Dh}
\eeq
where the right-hand side is evaluated at horizon crossing, $k=aH$.
While the power spectrum of scalar fluctuations, eq.~(\ref{equ:DZ}), depends on $H$, $\dot H$, and $c_s$, the power spectrum of tensor fluctuations is only a function of the de Sitter expansion rate $H$.
Tensor  fluctuations are therefore a direct probe of the
energy scale at which inflation took place.
The scale-dependence of the tensor modes is determined by the time-dependence of~$H$,
\beq
n_t \equiv \frac{d \ln \Delta_h^2}{d \ln k} = - 2 \varepsilon \ .
\eeq

Observational constraints on tensor modes are usually expressed in terms of the tensor-to-scalar ratio\index{tensor-to-scalar ratio}
\beq
r \equiv \frac{\Delta_h^2}{\Delta_\R^2}  \ . \label{equ:r}
\eeq
Since the amplitude of scalar fluctuations has been measured,
the tensor-to-scalar ratio quantifies the size of the tensor fluctuations.
Using (\ref{equ:Dh}), we can write
\beq
 \frac{H}{M_{\rm pl}} = \pi\, \Delta_\R(k_\star) \sqrt{\frac{r}{2}}\ , \label{HMp}
\eeq
which on
substituting $\Delta_\R(k_\star) = 4.7 \times 10^{-5}$ becomes
\beq
H= 3 \times 10^{-5} \left( \frac{r}{0.1}\right)^{1/2} \, M_{\rm pl}\ . \label{equ:HMp}
\eeq
Detecting inflationary tensor  perturbations at the level $r \gtrsim 0.1$ would imply that the expansion rate during inflation was about $10^{-5} M_{\rm pl}$.
This is sometimes expressed in terms of the {\it energy scale of inflation}
\beq
E_{\rm inf}  \equiv (3H^2\Mp^2)^{1/4}  = 8 \times 10^{-3} \left( \frac{r}{0.1}\right)^{1/4} \, M_{\rm pl}\ . \label{equ:Einf}
\eeq
Note that reducing $r$ by four orders of magnitude reduces $E_{\rm inf}$ by only one order of magnitude.  Gravitational waves from inflation are only observable if inflation occurred near the GUT scale, $E_{\rm inf} \sim 10^{-2}\Mp \sim 10^{16}$ GeV.

\section{Cosmological Observables}
\label{sec:Obs}

When the curvature perturbation $\R$ re-enters the horizon it
sources fluctuations in the primordial plasma.
These matter perturbations evolve into anisotropies in the cosmic microwave background\index{cosmic microwave background} (CMB)~\cite{Hu:2008hd, Dodelson:2003ft} and inhomogeneities in the large-scale structure (LSS).
In this section, we describe these key cosmological observables.
In the next section, we will show how these observables are used to constrain both the composition of the universe and its initial conditions.

\subsection{CMB Anisotropies}

\begin{figure}[h!]
    \centering
        \includegraphics[width=0.9\textwidth]{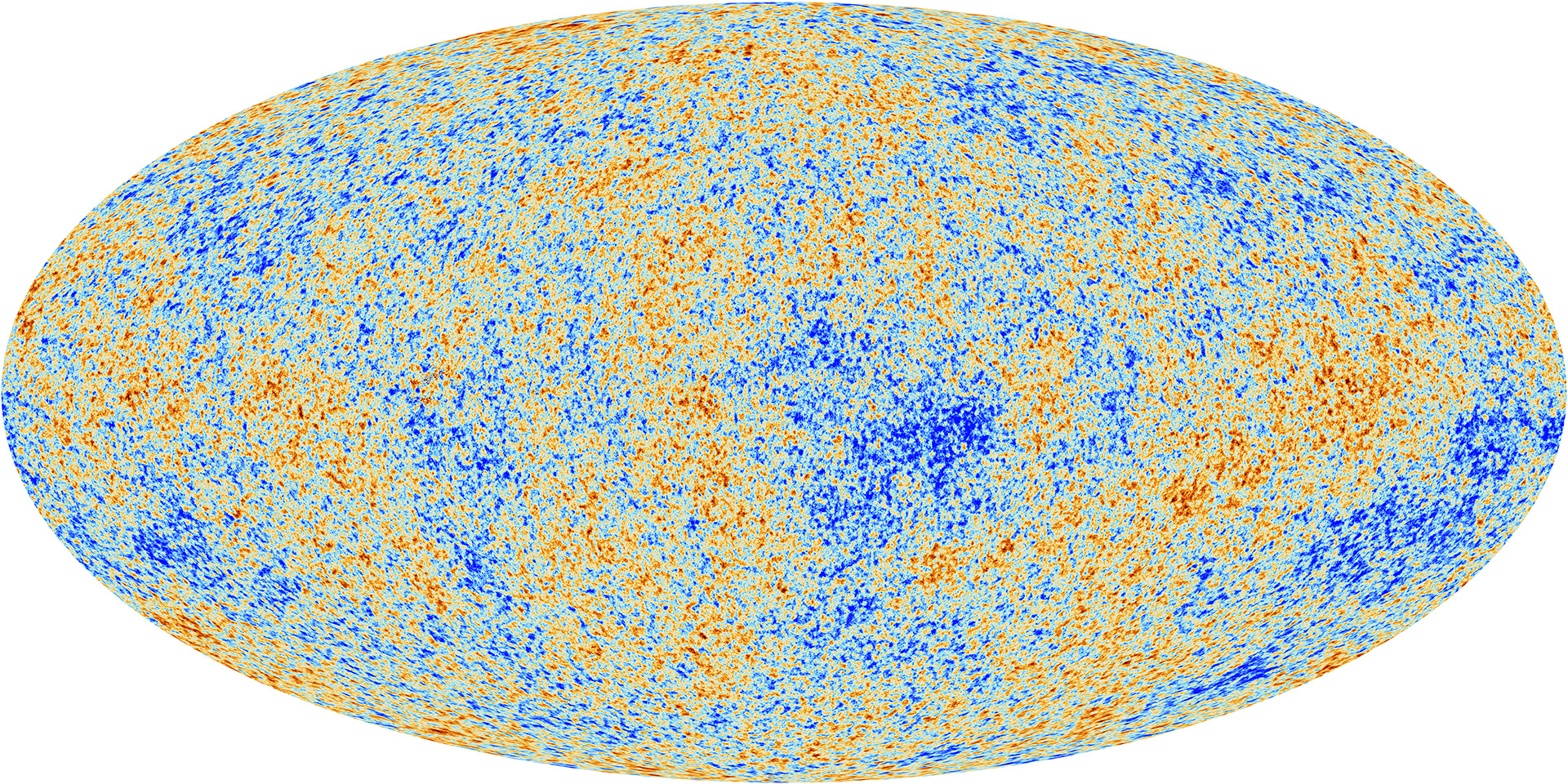}
    \caption{CMB anisotropies as observed by the Planck satellite. Red (blue) spots are hotter (colder) than the average temperature, reflecting density variations at recombination.}
    \label{fig:planck-cmb}
\end{figure}

In the very early universe, photons had a small mean free path due to the high density of charged particles.
At a temperature of about 0.3 eV, the formation of neutral hydrogen,
\begin{equation}
e+p \rightarrow H + \gamma\ ,
\end{equation}
termed {\it recombination}, became entropically favored.  The free electron density dropped rapidly and Thomson scattering between electrons and photons, $e + \gamma \leftrightarrow e+ \gamma$, became inefficient: the photons {\it decoupled}.
Since the moment of {\it last scattering} at $t \approx 380,000$ yrs, these primordial photons have been streaming freely through the universe, reaching our detectors 13.7 billion years later~\cite{Penzias:1965wn}.
The observed frequency spectrum is that of an almost perfect black body with a mean temperature $\bar T = 2.72548 \pm 0.00057$ K~\cite{Fixsen:2009ug}.  Fig.~\ref{fig:planck-cmb} shows the variation of the CMB temperature as a function of direction $ \n$ on the sky,
\beq
\Delta T( \n) \equiv T(\n) - \bar T\ .
\eeq
These anisotropies reflect inhomogeneities in the density of the primordial plasma, which can be traced back to the curvature perturbations calculated in the previous section.

\vskip 4pt
For Gaussian
initial conditions, complete information about the temperature map is contained
in the correlations between the temperatures at pairs of distinct
points $ \n$ and $ \n'$,
\beq
C(\theta) \equiv \left\langle \frac{\Delta T}{\bar T}( \n) \frac{\Delta T}{\bar T}( \n') \right\rangle \ , \label{equ:Ct}
\eeq
where $\cos \theta \equiv  \n \cdot \n'$, and the angle brackets denote an ensemble average.\footnote{Recall that in \S\ref{sec:vacuum} we computed a quantum average. This is related to the ensemble average after decoherence turns the quantum state into a single classical state of the ensemble:
see~e.g.~\cite{Guth:1985ya, Polarski:1995jg, Perez:2005gh, Burgess:2006jn, Kiefer:2008ku}.}
It is convenient to describe the same information in {harmonic space}, by expanding the temperature field in
spherical harmonics,
\beq
\label{equ:Tharm}
\frac{\Delta T(\n)}{\bar T}= \sum_{\ell =0}^{\infty} \sum_{m=-\ell}^{+\ell} a_{\ell m} Y_{\ell m}(\n)\ .
\eeq
where $\ell$ and $m$ are eigenvalues of differential operators on the sphere, with $\nabla^2 Y_{\ell m} = - \ell(\ell+1) Y_{\ell m}$ and $\partial_\phi Y_{\ell m}=im Y_{\ell m}$. Reality of the temperature field imposes $a_{\ell m}^* = (-1)^m a_{\ell - m}$. Statistical isotropy constrains the two-point correlation function of the multipole moments $a_{\ell m}$ to be of the form
\beq
\label{equ:Cl}
\langle a_{\ell\vphantom{'} m\vphantom{'}} a_{\ell' m'}^* \rangle = C_\ell \, \delta_{\ell \ell'} \delta_{m m'}\ .
\eeq
The {\it angular power spectrum}, $C_\ell$, is the Legendre transform of the two-point function (\ref{equ:Ct}):\index{CMB power spectrum}
\beq
C_\ell = 2\pi \int_{-1}^1 \d \cos \theta \, \, C(\theta)\, P_{\ell}(\cos\theta)\ .
\eeq

\begin{figure}[h!]
   \centering
  \includegraphics[width=0.9\textwidth]{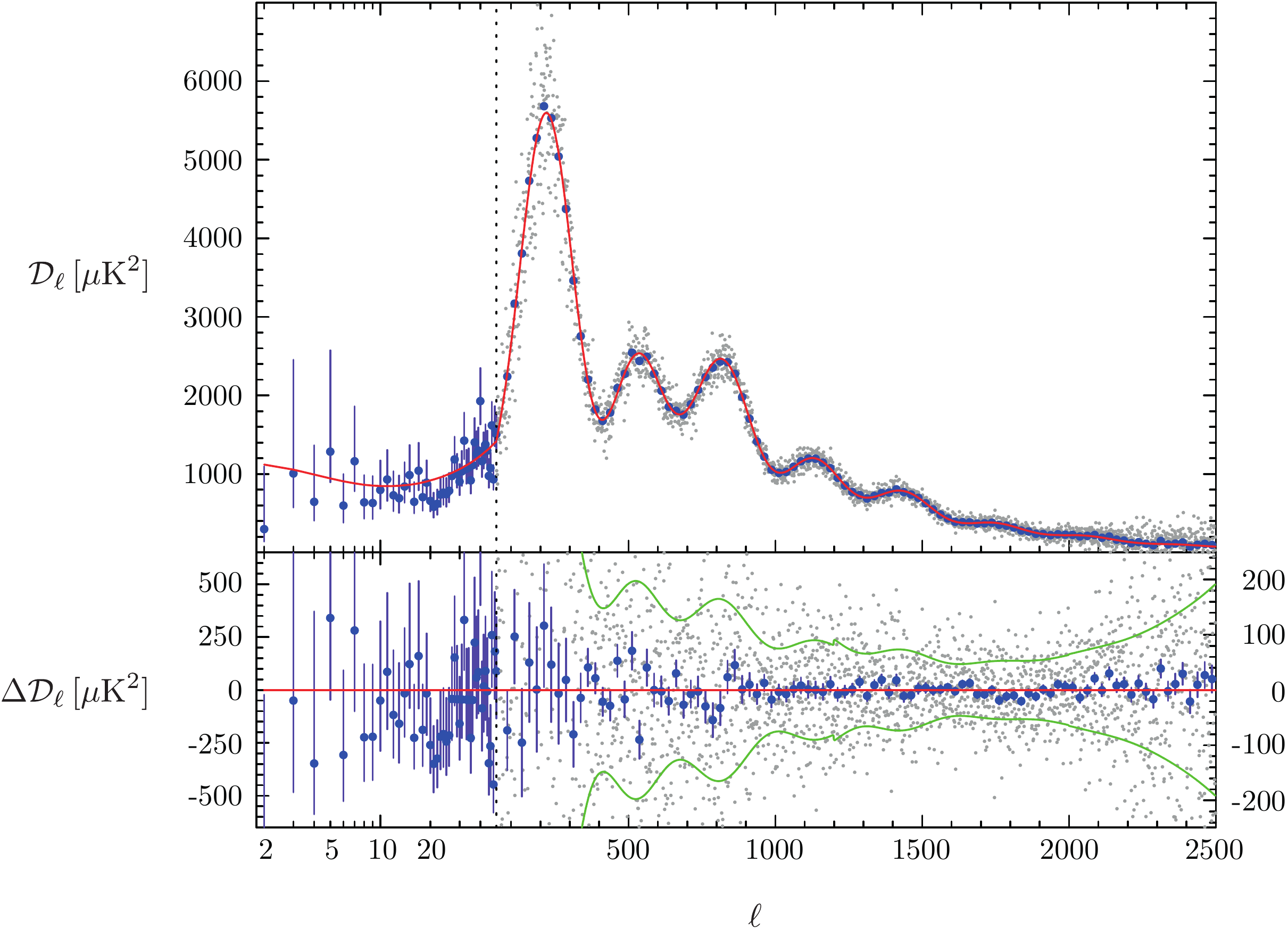}
   \caption{The power spectrum of CMB temperature anisotropies measured by the Planck satellite (figure adapted from~\cite{PlanckParameters}). Plotted is the combination ${\cal D}_\ell \equiv \ell (\ell + 1) C_\ell/2\pi$. Shown are both the data for individual multipoles (gray points), as well as binned averages (blue points with error bars). The lower plot shows the residuals with respect to the best-fit $\Lambda$CDM model.}
  \label{fig:Planck-TT}
\end{figure}

Although the theory predicts ensemble-averaged quantities, we only observe a single realization of the ensemble. After extracting the multipole moments of the measured temperature map, we can construct an {\it estimator} for the angular power spectrum,
\beq
\hat C_\ell = \frac{1}{2\ell +1} \sum_m |a_{\ell m}|^2 \ .
\eeq
This estimator is unbiased, in that $\langle \hat C_\ell \rangle = C_\ell$.
The variance of the estimator is called {\it cosmic variance}:
\beq
{\rm var}(\hat C_\ell) \equiv \langle \hat C_\ell \hat C_\ell \rangle - \langle \hat C_\ell \rangle^2 = \frac{2}{2\ell+1} C_\ell^2\ .
\eeq
This irreducible error arises from having only $2\ell+1$ modes at each multipole moment $\ell$ to estimate the variance of their distribution.  Fig.~\ref{fig:Planck-TT} shows the CMB power spectrum as measured by the Planck satellite. The error bars include both cosmic variance and measurement noise, but the former dominates up to $\ell \sim 2000$.

\vskip 4pt
The shape of the CMB power spectrum is well-understood theoretically.
Before neutral hydrogen formed, photons and baryons were strongly coupled and acted as a single fluid in which the photon pressure sustained acoustic oscillations (i.e.~sound waves) driven by the gravitational force induced by the curvature perturbations.
The observed CMB fluctuations are a snapshot of these density waves.
For adiabatic initial conditions, the angular power spectrum is predicted to be
\beq
\label{equ:Cl2}
C_\ell = \int  \d \ln k \, \Delta^2_\R(k) \, T_\ell^2(k) \ ,
\eeq
where the transfer function $T_\ell(k)$ describes both the {evolution} of the initial fluctuations from the moment of horizon entry to the time of recombination, as well as the {projection} from recombination to today~\cite{Hu:2008hd, Dodelson:2003ft}.
Since the transfer function depends only on known physics it is computable using a set of coupled Einstein-Boltzmann equations for the primordial plasma~\cite{Seljak:1996is, Lewis:1999bs}.
The knowledge of $T_\ell(k)$ allows us to use the observed $C_\ell$ as a probe of the initial conditions $\Delta_\R(k)$.  The theoretical curve in fig.~\ref{fig:Planck-TT} assumes a nearly scale-invariant spectrum as predicted by inflation.

\subsection{CMB Polarization}
\label{sec:CMBpol}

Recombination was not an instantaneous process.
In the time it took protons and electrons to combine into neutral hydrogen,
the photons developed a quadrupole anisotropy in the local electron rest frame.
Thomson scattering converted this
 into an anisotropy  of the CMB  polarization~\cite{Rees, Polnarev, Hu:1997hv}.

\begin{figure}[h!]
   \centering
      \includegraphics[width=0.4\textwidth]{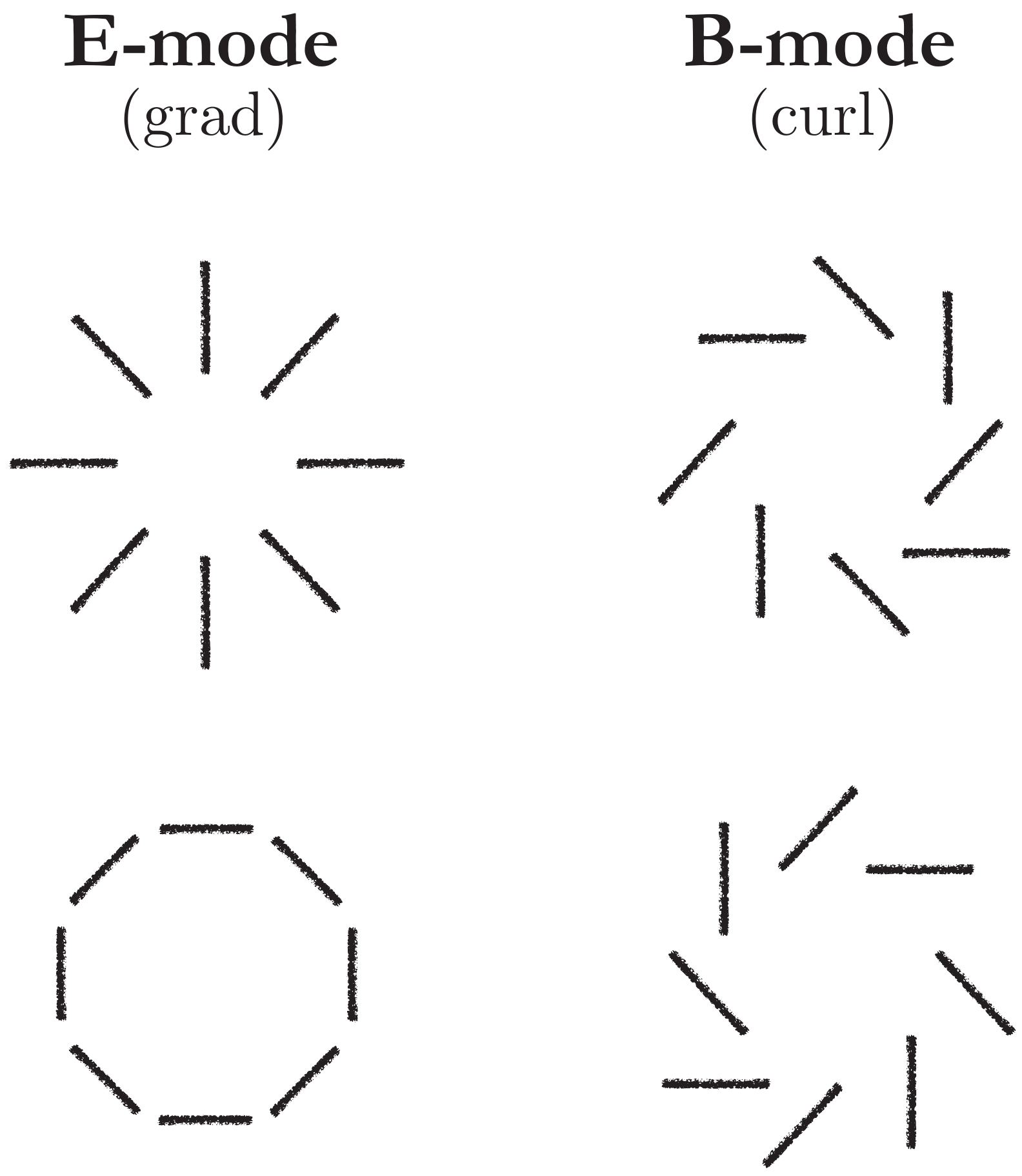}
   \caption{Examples of E-mode and B-mode patterns of CMB polarization. While the E-mode patterns are mirror-symmetric, the B-mode patterns are interchanged under reflection about a line going through the center.}
  \label{fig:EB-modes}
\end{figure}

\vskip 4pt
 Linear polarization can be measured in terms of the Stokes parameters $Q$ and $U$~\cite{chandrasekhar2013radiative}. Let ${\n}$ be the direction of observation and ($\e_1,\e_2$) be a basis of orthogonal unit vectors. The Stokes parameters are not invariant under a change of these coordinates, rotating the basis ($\e_1,\e_2$) by an angle $\psi$ leads to
 \beq
 (Q\pm iU)'(\n) = e^{\mp 2i\psi} (Q\pm i U)(\n) \ .
 \eeq
 This identifies  $Q\pm i U$
 as a spin-$2$
 field to be expanded in terms of spin-weighted spherical harmonics~\cite{Newman:1966ub}
 \beq
 (Q\pm i U)(\n) = \sum_{\ell m} a_{\pm 2, \ell m}\, {}_{\pm 2} Y_{\ell m}(\n)\ .
 \eeq
 Acting twice with a spin-lowering operator on $Q + i U$ and twice with a spin-raising operator on $Q-iU$ produces scalar (spin-0) quantities. These scalars can be collected according to their transformations under parity (the operation which takes $\n$ into $-\n$):
 \begin{align}
 E(\n) \equiv a_{E, \ell m} Y_{\ell m}(\n) \ , \qquad  a_{E, \ell m} \equiv - \frac{a_{2,\ell m} + a_{-2, \ell m}}{2} \ , \\
  B(\n) \equiv a_{B, \ell m} Y_{\ell m}(\n)\ , \qquad a_{B, \ell m} \equiv - \frac{a_{2,\ell m} - a_{-2, \ell m}}{2i}\ .
 \end{align}
 The E-modes are parity-even, while the B-modes are parity-odd.
 Roughly, we can think of the E-mode as the gradient of a scalar and the B-mode as the curl of a vector. Typical E- and B-patterns are shown in fig.~\ref{fig:EB-modes}.
 Given $T$, $E$ and $B$, we can form several types of correlation functions
 \beq
\langle a_{X\vphantom{'} ,\ell\vphantom{'}  m\vphantom{'} } a_{Y, \ell' m'}^* \rangle = C_\ell^{XY} \delta_{\ell \ell'} \delta_{m m'} \ , \quad X,Y \equiv \{T,E,B\}\ .
 \eeq
 Since B is parity-odd, while T and E are parity-even, we expect $C_\ell^{TB} = C_\ell^{EB} = 0$.

Discussing polarization in terms of E-modes and B-modes has several distinct advantages.
First of all, unlike the Stokes parameters, the parameters E and B  are independent of the choice of coordinates. More importantly, symmetry forbids the generation of  B-modes by scalar fluctuations~\cite{Kamionkowski:1996ks, Zaldarriaga:1996xe}. B-modes are therefore a crucial signature of the presence of tensor (or vector)
fluctuations.

\subsection{Large-Scale Structure}

The density perturbations are small at recombination, but under the influence of gravity they grow~\cite{Lemaitre:1931zz}, eventually forming
the large-scale structure of the universe.
A linear order, the initial conditions from inflation are related to the dark matter density contrast $\delta \equiv \delta \rho/\rho$ at redshift~$z$  via a transfer function $T_\delta(z,k)$:
\beq
P_\delta(z,k) = T_\delta^2(z,k) P_\R(k) \ .
\eeq
On large scales, the transfer function is relatively easy to calculate in perturbation theory~\cite{Bernardeau:2001qr}, while on small scales numerical N-body simulations~\cite{Springel:2005mi} are required.

\begin{figure}[h!]
   \centering
  \includegraphics[width=0.8\textwidth]{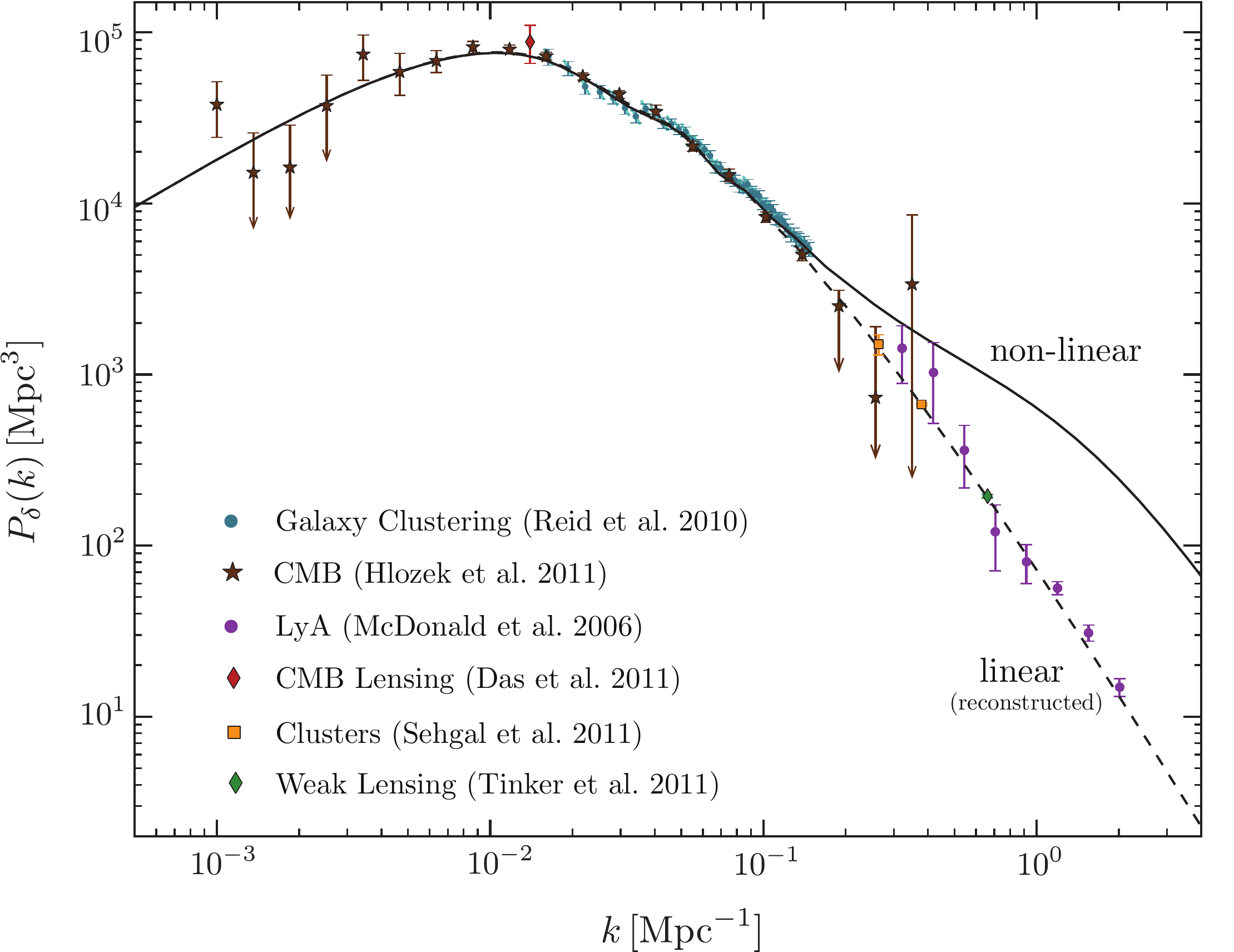}
   \caption{Compilation of measurements of the matter power spectrum (figure adapted from~\cite{Hlozek:2011pc}).}
  \label{fig:Pk}
\end{figure}

\vskip 4pt
With the exception of gravitational lensing~\cite{Bartelmann:1999yn, Refregier:2003ct, Lewis:2006fu}, we do not measure the dark matter  density $\delta$
directly. Instead, we observe biased baryonic tracers of the dark matter field, such as galaxies, clusters,  and Ly$\alpha$ fluctuations:
see the compilation of recent measurements in fig.~\ref{fig:Pk}.
On large scales, the density contrast of these tracers, $\delta_g$, has a linear and deterministic relationship to the underlying dark matter field,\index{galaxy bias}
\begin{equation}
\delta_g(z,\x) = b(z)\hskip 1pt \delta(z,\x) \ .
\end{equation}  where $b(z)$  parameterizes the biasing.
On small scales, however, the biasing can become non-linear, non-local and stochastic. 
This makes it challenging to relate large-scale observations to the initial conditions.

\vskip 4pt
The Einstein equations couple the oscillations in the photon-baryon fluid to the dark matter density. The same oscillations that we observe in the CMB power spectrum are therefore also imprinted in the matter power spectrum.
These oscillations are barely visible in fig.~\ref{fig:Pk} between $k=0.01$ ${\rm Mpc}^{-1}$ and  $0.1$ ${\rm Mpc}^{-1}$.
\begin{figure}[h!]
   \centering
  \includegraphics[scale=0.4]{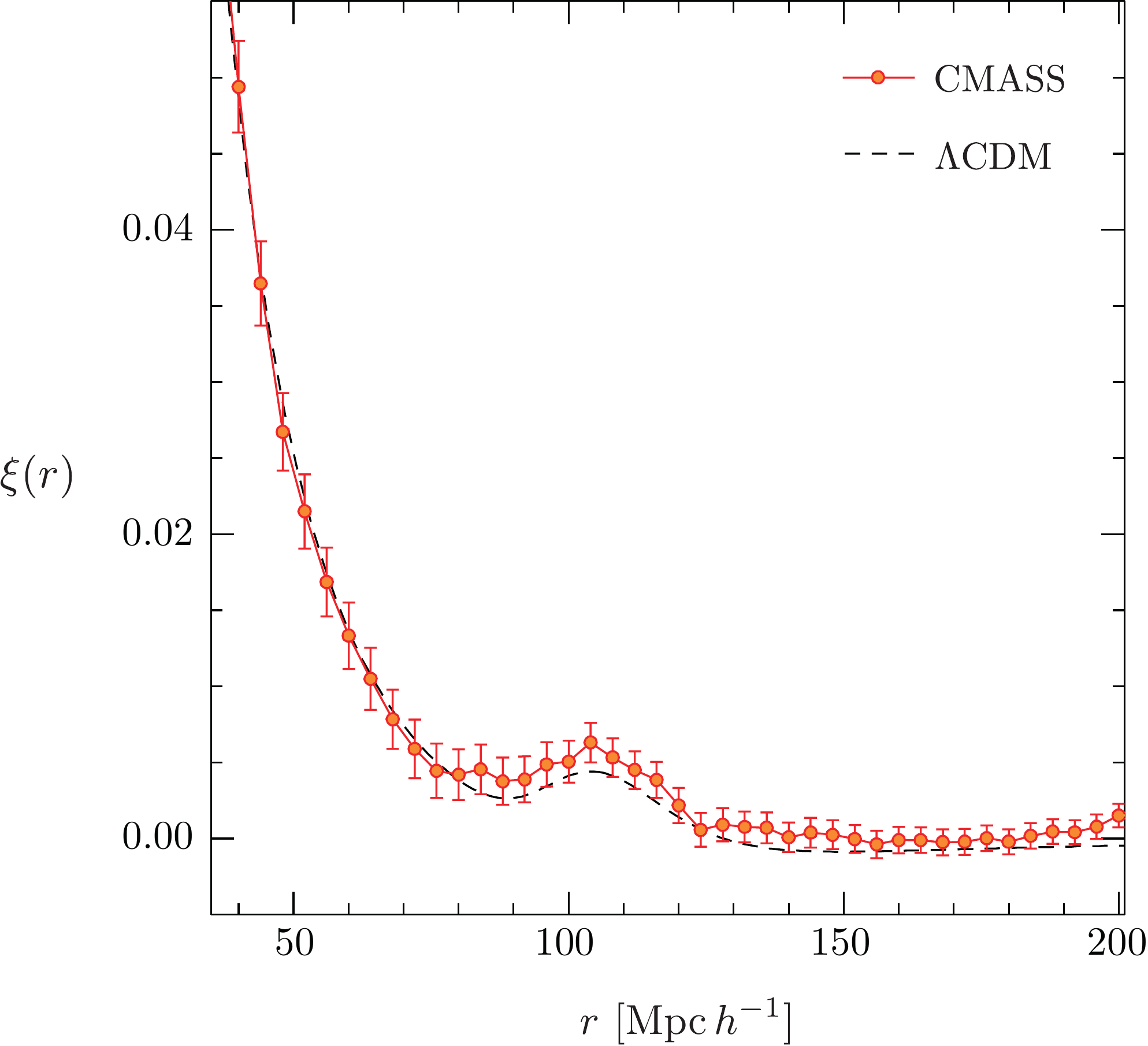}
   \caption{Spherically averaged redshift-space correlation function of the CMASS Data Release 9 (DR9) sample of the
Baryonic Oscillation Spectroscopic Survey (BOSS). The dashed line corresponds to the best-fitting $\Lambda$CDM model.
(Figure adapted from~\cite{Sanchez:2012sg}).}
  \label{fig:BAO}
\end{figure}
Detections of these {\it baryon acoustic oscillations} (BAO) were first reported in~\cite{Cole:2005sx, Eisenstein:2005su}  and more fully characterized in~\cite{Percival:2009xn, Padmanabhan:2012hf, Anderson:2012sa, Blake:2011en, Beutler:2011hx} (for a review of BAO see~\cite{Weinberg:2012es}).\index{baryon acoustic oscillations (BAO)}
Fig.~\ref{fig:BAO} shows the measured matter two-point function in real space, $\xi(r)$. The BAO feature is clearly visible at about 100 Mpc.

Both the CMB observations and the BAO observations measure the sound horizon of the photon-baryon plasma.
The observed scale in the CMB measurements depends on the angular diameter distance to recombination, $D_A(z_{\rm rec})$.
For BAO, the observed scale depends on the (spherically-averaged) distance to the effective survey redshift $\bar z$,
which is a combination of the angular diameter distance and the Hubble parameter:
\beq
D_V(\bar z) \equiv \left[ (1+\bar z)^2 D_A^2(\bar z) \frac{c\bar z}{H(\bar z)} \right]^{1/3}\ .
\eeq
Comparing CMB and LSS measurements provides important information about the evolution of the universe after recombination and helps to break an important geometric degeneracy~\cite{Bond:1997wr, Zaldarriaga:1997ch} that exists in the CMB-only analysis.
Alternatively, the degeneracy can be broken by using the gravitational lensing of the CMB anisotropies~\cite{Lewis:2006fu}.

\section{Current Tests of Inflation}
\label{sec:obs}

In March 2013, the {\it Planck} collaboration released its first cosmological analysis~\cite{PlanckParameters}. Together with the measurements of the CMB damping tail by the {\it Atacama Cosmology Telescope} (ACT)~\cite{Das:2010ga, Sievers:2013wk} and the {\it South Pole Telescope} (SPT)~\cite{Keisler:2011aw, Story:2012wx} this provides a beautiful picture of the first seven acoustic peaks of the CMB power spectrum.
In this section, we summarize how the CMB results have tested the physics of inflation~\cite{PlanckInflation, PlanckNG}.
Errors quoted in this section are $1\sigma$ errors (68\% limits) unless otherwise specified.

\subsection{$\Lambda$CDM Model}

The standard model of cosmology has six free parameters: the physical baryon density, $\omega_b \equiv \Omega_b h^2$; the physical density of cold dark matter (CDM), $\omega_c \equiv \Omega_c h^2$; the dark energy density, $\Omega_\Lambda$; the optical depth~$\tau$; and the amplitude $A_s$ and the spectral index $n_s$ in the power law ansatz for the initial conditions\hskip 2pt\footnote{In the Planck analysis, $A_s$ and $n_s$ are defined at the pivot scale $k_\star = 0.05$ Mpc${}^{-1}$.  }
\beq
\Delta_\R^2(k) = A_s \left(\frac{k}{k_\star} \right)^{n_s-1} \ . \label{equ:PL}
\eeq
This simple model provides a superb fit
to a wide range of cosmological data, from CMB to LSS. Fig.~\ref{fig:Planck-TT}  shows the power spectrum of CMB temperature fluctuations measured by Planck, as well as the best-fit curve of the $\Lambda$CDM model.
Table~\ref{Table:LCDM} summarizes the best-fit parameters.
The Planck data is precise enough to determine all six parameters at the percent level without recourse to external datasets.
A small degeneracy between $\tau$ and $A_s$ (and/or $n_s$) is broken by the addition of WMAP low-$\ell$ polarization data~\cite{Hinshaw:2012uq} (or Planck lensing data~\cite{PlanckLensing}).

\begin{figure}[h!]
   \centering
     \includegraphics[width=0.7\textwidth]{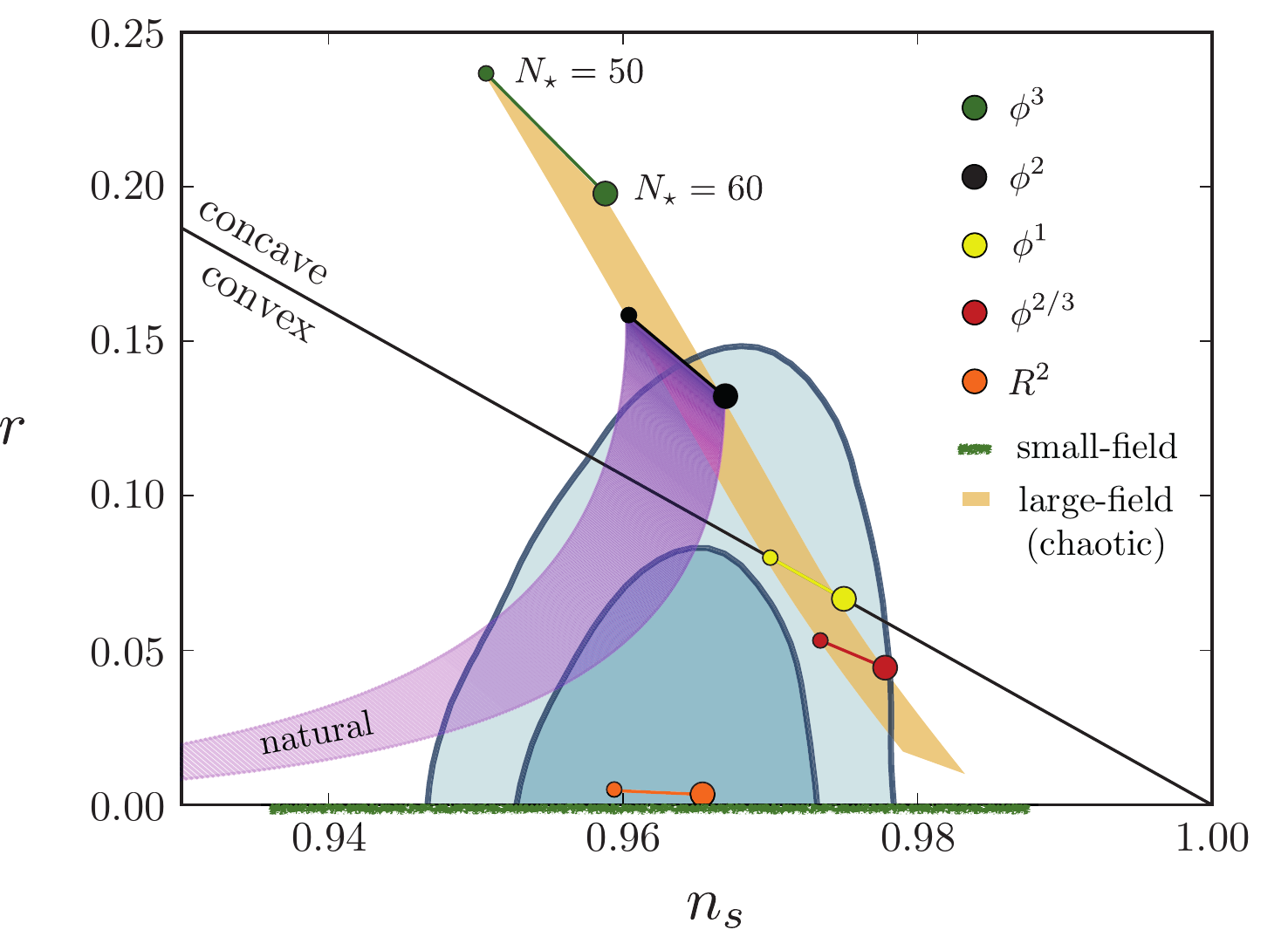}
   \caption{Planck+WMAP+BAO constraints on $n_s$ and $r$ (figure adapted from~\cite{PlanckInflation}). }
  \label{fig:nsr}
\end{figure}

 \begin{table}[h!]

	\heavyrulewidth=.08em
	\lightrulewidth=.05em
	\cmidrulewidth=.03em
	\belowrulesep=.65ex
	\belowbottomsep=0pt
	\aboverulesep=.4ex
	\abovetopsep=0pt
	\cmidrulesep=\doublerulesep
	\cmidrulekern=.5em
	\defaultaddspace=.5em
	\renewcommand{\arraystretch}{1.6}

	\begin{center}
		\small
		\begin{tabular}{lllll}

			\toprule
		Parameter  & Planck & $\cdots$ + WMAP + ACT &  CMB + BAO \\
			\midrule
		  $\Omega_b h^2$  & $0.02207\pm0.00067$  & $0.02207 \pm 0.00054$ & $0.02214 \pm 0.00048$ \\[-1mm]
		   $\Omega_c h^2$  & $0.1196\pm0.0061$ & $0.1198 \pm 0.0052$ & $0.1187 \pm 0.0034$ \\[-1mm]
		  $\Omega_\Lambda$  & $0.683 \pm 0.040$  & $0.685 \pm 0.033$ & $0.692 \pm 0.021$ \\[-1mm]
		  $\tau$  & $0.097 \pm 0.080$  & $0.091 \pm 0.027$ & $0.092 \pm 0.026$ \\[1mm]
		\midrule
		 $10^{9}A_s$  & $2.23 \pm 0.32$  & $2.20 \pm 0.11$ & $2.20 \pm 0.11$ \\[-1mm]
		 $n_s$  & $0.962 \pm0.019$ & $0.959 \pm 0.014$ & $0.961 \pm 0.011$ \\[1mm]
 			\bottomrule
		\end{tabular}
	\end{center}
	\caption{Parameters of the $\Lambda$CDM baseline model (with $2\sigma$ errors). The first four parameters describe the composition of the universe, the last two its initial conditions.
The BAO data improves the constraint on $\Omega_\Lambda$.  The small-scale CMB data hardly affect the constraints but help with a characterization of foregrounds, which becomes essential when going beyond the $\Lambda$CDM model.
	\label{Table:LCDM}}
	\end{table}
The best-fit value for the scalar amplitude is
\beq
A_s = \left(\, 2.196^{+0.051}_{-0.060}\, \right) \times 10^{-9}\ . \label{equ:DZ0}
\eeq
A scale-invariant primordial power spectrum is now excluded at almost 6$\sigma$ significance,
\beq  \label{nslimit}
n_s = 0.9603 \pm 0.0073  \ .
\eeq
This result assumes that tensor fluctuations make a negligible contribution to the temperature fluctuations. Allowing for tensors introduces a new parameter, the tensor-to-scalar ratio~$r$, cf.~(\ref{equ:r}).
With earlier datasets, including $r$ in the fit weakened the evidence for $n_s < 1$, but with Planck this result is now robust: see figure~\ref{fig:nsr} and table~\ref{Table:BLCDM}.

\subsection{Inflation after Planck}
\label{sec:tests}

The Planck collaboration has tested for deviations from the standard assumptions for the initial conditions, such as deviations from Gaussianity, adiabaticity, power law scaling, and flatness.  Here, we summarize their findings.

\subsection*{Geometry}

Inflation very effectively solves the flatness problem~\cite{Guth:1980zm}.
The baseline analysis of Planck has therefore fixed the curvature parameter to be vanishing, $\Omega_K = 0$.
On the other hand, including $\Omega_K$ in the fit allows a test of this key prediction of inflation.
Table~\ref{Table:OmegaK} shows the
constraints on the parameter $\Omega_K$, after marginalizing over the other parameters of the $\Lambda$CDM model. Here, the BAO data plays a crucial role in breaking the
geometric degeneracy between $\Omega_m$ and $H_0$ and reducing the error on $\Omega_K$ by an order of magnitude. Even at this new level of precision the observable patch of the universe is consistent with spatial flatness. Planck has also tested the isotropy assumption~\cite{PlanckIsotropy}. Except perhaps on the largest scales, the universe indeed seems to be statistically isotropic.	

\vspace{0.3cm}
 \begin{table}[h!]

	\heavyrulewidth=.08em
	\lightrulewidth=.05em
	\cmidrulewidth=.03em
	\belowrulesep=.65ex
	\belowbottomsep=0pt
	\aboverulesep=.4ex
	\abovetopsep=0pt
	\cmidrulesep=\doublerulesep
	\cmidrulekern=.5em
	\defaultaddspace=.5em
	\renewcommand{\arraystretch}{1.6}

	\begin{center}
		\small
		\begin{tabular}{lllll}

			\toprule
		Parameter  & Planck & $\cdots$+ WMAP + ACT & CMB + BAO \\
			\midrule
		 $\Omega_K$  & $-0.072\pm 0.081$ & $-0.037 \pm 0.049$   & $-0.0005 \pm 0.0066$  \\[1mm]
 			\bottomrule
		\end{tabular}
	\end{center}
	\caption{Constraints on the geometry of the universe (with $2\sigma$ errors). The inclusion of BAO data plays an important role.
	\label{Table:OmegaK}}
	\end{table}

\vspace{-0.5cm}
\subsection*{Scalar Fluctuations}

The observations of the primordial scalar fluctuations are in striking agreement with the predictions of inflation, both qualitatively and quantitatively:

\begin{itemize}
\item[$\triangleright$] {\it Coherent phases.}---A
telling feature of the CMB anisotropies is that they span {\it superhorizon} scales at recombination (corresponding to $\ell < 200$) and have {\it coherent phases}. This fact is observed unambiguously through the low-$\ell$ peak in the cross-correlation between temperature fluctuations and $E$-mode polarization (see fig. \ref{fig:TE}).
In the absence of phase coherence, this peak would disappear~\cite{Spergel:1997vq, Dodelson:2003ip}.
It is easy to see why the inflationary mechanism for generating fluctuations leads to phase coherence.
Modes freeze when their physical wavelengths become larger than the Hubble radius and only start evolving again when they re-enter the horizon.
All modes with the same wavenumber $k$, but possibly distinct wavevectors $\k$, therefore start their evolution at the same time.  This phase coherence allows for constructive interference of the modes and yields acoustic oscillations in the CMB.  Alternative mechanisms for structure formation
involving topological defects (e.g.~cosmic strings,  see \S\ref{cosmicstrings}) source perturbations with  incoherent phases,  smearing out the peaks \cite{Albrecht:1995bg},
and are therefore  ruled out by the CMB observations.
Isocurvature fluctuations also destroy some of the phase coherence\footnote{In contrived scenarios, causal evolution  inside the horizon yields isocurvature perturbations  that lead to acoustic peaks \cite{Turok:1996wa} --- see the review \cite{Durrer:2001cg}.} and are hence significantly constrained by the data (see below).
\index{phase coherence}

\begin{figure}[h!]
   \centering
  \includegraphics[width=0.8\textwidth]{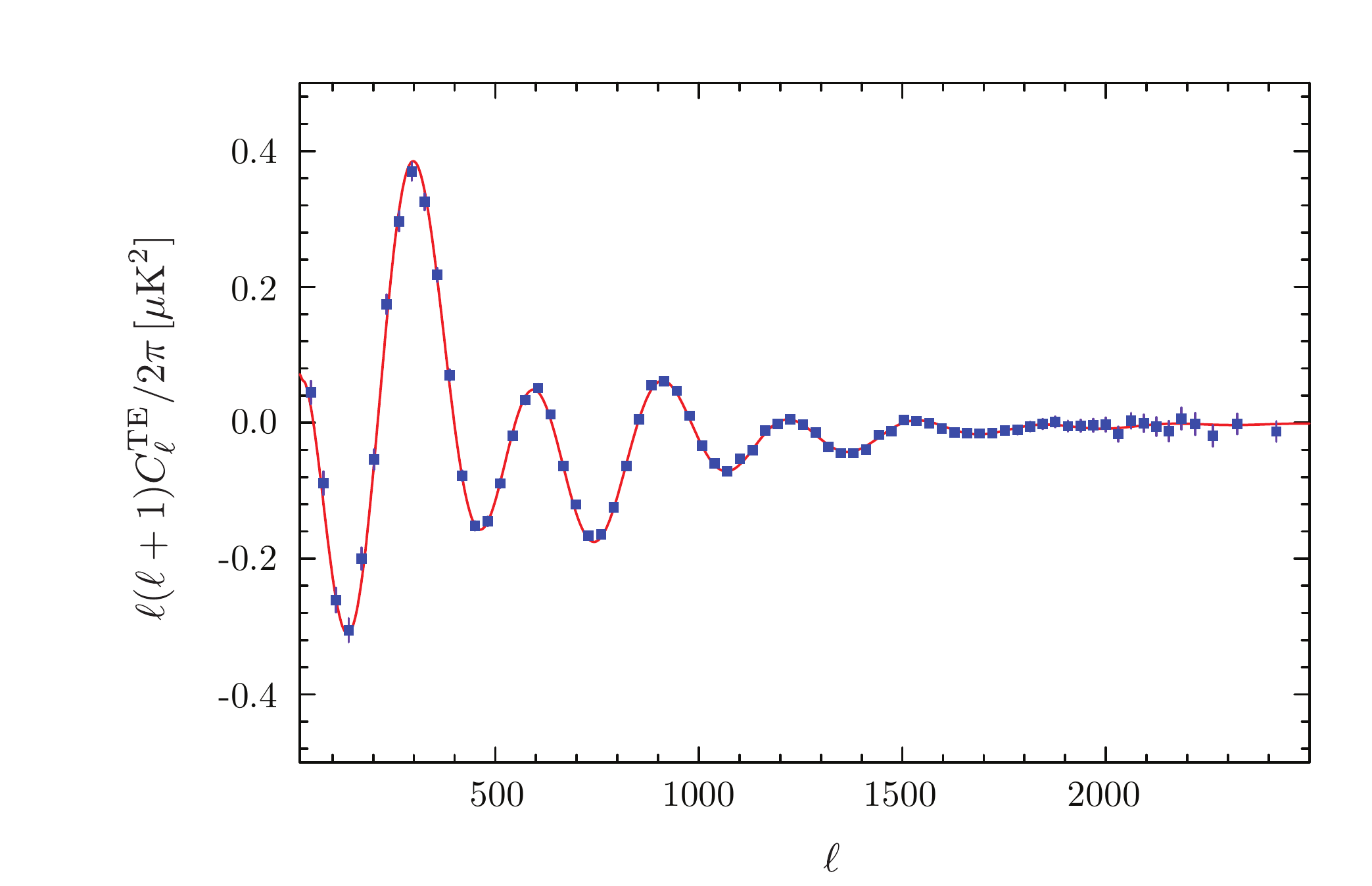}
   \caption{The cross-correlation of CMB temperature anisotropies and E-mode polarization  (figure adapted from~\cite{PlanckParameters}). The curve is not a fit, but a prediction! The low-$\ell$ peak is a signature of phase coherence of the initial conditions.}
  \label{fig:TE}
\end{figure}

\item[$\triangleright$]  {\it Power law spectrum.}---We have seen above
that slow-roll inflation predicts a power law spectrum with a percent-level deviation from perfect scale-invariance,
which Planck has detected at high significance.
At second order in the slow-roll expansion, inflation predicts a small correction to the power law spectrum
\beq
\Delta_\R^2(k) = A_s \left(\frac{k}{k_\star} \right)^{n_s-1 + \frac{1}{2}\alpha_s \ln(k/k_\star)} \ .
\eeq
The data is not yet precise enough to detect the expected {\it running} of the spectrum, $\alpha_s \sim (n_s-1)^2$, and a detection of running at a level accessible to Planck would in fact be in conflict with the inflationary expectation.  It is nevertheless interesting to include $\alpha_s$ as a free parameter in the fit.
Table~\ref{Table:BLCDM} summarizes the latest constraints on $\alpha_s$, which depend on whether the
tensor-to-scalar
ratio~$r$ is included as a parameter or is set to zero.
At present, there are no clear indications for a departure from the inflationary power law spectrum.
\end{itemize}

 \begin{table}[h!]

	\heavyrulewidth=.08em
	\lightrulewidth=.05em
	\cmidrulewidth=.03em
	\belowrulesep=.65ex
	\belowbottomsep=0pt
	\aboverulesep=.4ex
	\abovetopsep=0pt
	\cmidrulesep=\doublerulesep
	\cmidrulekern=.5em
	\defaultaddspace=.5em
	\renewcommand{\arraystretch}{1.6}

	\begin{center}
		\small
		\begin{tabular}{lllll}

			\toprule
		Parameter  & Planck & $\cdots$ + WMAP + ACT & CMB + BAO \\
			\midrule
	 $n_s$  & $0.963 \pm 0.019$   & $0.960 \pm 0.014$ & $0.962 \pm 0.011$  \\[-1mm]
		 $r$  &  $< 0.115 $   & $< 0.117$ & $< 0.119$ \\[1mm]
			\midrule
			 $n_s$  & $0.974 \pm 0.030$  & $0.955 \pm 0.015$ & $0.960 \pm 0.012$ \\[-1mm]
		 $\alpha_s$  & $-0.034 \pm 0.035$   & $-0.015 \pm 0.017$  & $-0.013 \pm 0.018$ \\[1mm]
			\midrule
			 $n_s$  & $0.976 \pm 0.030$  & $0.957 \pm 0.015$ &  $0.959 \pm 0.011$ \\[-1mm]
			 $r$  &  $<0.228$  & $<0.230$ & $<0.235$ \\[-1mm]
		 $\alpha_s$  & $-0.041 \pm 0.037$  & $-0.022 \pm 0.021$  & $-0.022  \pm 0.022$ \\[1mm]
 			\bottomrule
		\end{tabular}
	\end{center}
	\caption{Constraints on tensor modes and on deviations from the power law spectrum (with $2\sigma$ errors).
	\label{Table:BLCDM}}
	\end{table}

	\vspace{-0.5cm}
\subsection*{Tensor Fluctuations}

Tensor modes contribute to the CMB temperature power spectrum in a specific way and are therefore constrained by the Planck analysis.
Fig.~\ref{fig:nsr} shows the current constraints on the parameters $n_s$ and~$r$. Marginalizing over $n_s$ gives an upper limit on the tensor-to-scalar ratio~\cite{PlanckInflation}
\begin{align}
r &< 0.12  \quad  \mbox{(95\% limit)} \ . \label{equ:PlanckR}
 \end{align}
This constraint is at the limit of what can be achieved with CMB temperature data alone~\cite{Knox:1995dq}.  To probe smaller values of $r$ requires measurements of CMB polarization:  as we explained in \S\ref{sec:CMBpol}, B-modes are a unique signature of inflationary tensor modes. The BICEP2 collaboration has recently reported a detection of primordial B-modes.  We discuss this result in \S\ref{sec:BICEP}.

\subsection*{Non-Gaussianity}

The CMB power spectrum in fig.~\ref{fig:Planck-TT} reduces the Planck data from about 50 million pixels to $10^3$~multipole moments. This enormous data compression is justified if the primordial perturbations are isotropic and Gaussian. On the other hand, a wealth of information may be contained in deviations from a perfectly Gaussian distribution~\cite{Komatsu:2009kd, Chen:2010xka, Wang:2013zva}.   Among the primary accomplishments of the Planck mission are the significant upper bounds placed on higher-order CMB correlations, or {\it non-Gaussianity}~\cite{PlanckNG}.\index{non-Gaussianity} (For previous results from WMAP see~\cite{Bennett:2012fk, Senatore:2009gt}.)
This has allowed the study of primordial quantum fields to move beyond the free field limit and start to place meaningful constraints on  interactions.

 \vskip 4pt
In \S\ref{sec:zeta}, we computed the two-point function (or power spectrum) of primordial curvature perturbations,
\beq
\langle 0 | \hskip 1pt \hat \R_{{\k}_1} \hat \R_{{\k}_2} \hskip 1pt  | 0 \rangle =(2\pi)^3 \, P_\R(k_1)\,  \delta({\k}_1 + {\k}_2)\ , \label{equ:2ptX}
 \eeq
 where $ | 0 \rangle$ denotes the vacuum state and $\hat \R$ is the quantum operator associated with the field $\R$.
In principle, there is more information in the vacuum expectation
 values of higher-order $n$-point functions.
Schematically, we can write these as the following path integral
\beq
\langle \Omega | \hskip 1pt \hat \R_{{\k}_1} \cdots \hat \R_{{\k}_n}\hskip 1pt| \Omega \rangle \propto \int [{\cal{D}} \R] \, \R_{{\k}_1} \cdots \R_{{\k}_n}\, e^{i S[\R]} \ ,
\eeq
where $S$ is the inflationary action and $ | \Omega \rangle $ is the vacuum of the interacting theory. For a {\it free} field theory, the action is a quadratic functional $S_{(2)}$, cf.~eq.~(\ref{equ:zetaAction}), and the $e^{iS}$ weighting of the path integral is a Gaussian (after Wick rotating to Euclidean time).
All correlation functions with $n$ odd then vanish, while those with $n$ even are completely determined by the two-point function~(\ref{equ:2ptX}).
However, including nontrivial interactions in the action, $S_{\rm int} = S_{(3)} + S_{(4)} + \cdots$, makes the $e^{iS}$ weighting of the path integral {\it non-Gaussian}.
This allows non-zero $n$-point functions for all $n$.

The primary diagnostic for primordial non-Gaussianity is the three-point function (or bispectrum),
\beq
\langle \Omega | \hskip 1pt \hat \R_{{\k}_1} \hat \R_{{\k}_2}\hat \R_{{\k}_3} \hskip 1pt| \Omega \rangle =(2\pi)^3 \, B_\R(k_1,k_2,k_3)\,  \delta({\k}_1 + {\k}_2 + {\k}_3)\ .
 \eeq
 The delta-function is a consequence of statistical homogeneity:
it enforces that the three momentum vectors form a closed triangle.
The momentum dependence of the bispectrum determines the amount of non-Gaussianity associated with
triangles of different shapes.
A useful measure of the size of the non-Gaussianity is the parameter
 \beq
f_{\mathsmaller{\rm NL}}  \equiv \frac{5}{18} \frac{ B_\R(k,k,k)}{P_\R^2(k)}\ ,
\eeq
i.e.~the normalized amplitude of the bispectrum in the equilateral configuration, $k_1=k_2=k_3 \equiv k$.
The momentum dependence of the bispectrum $B_\R(k_1,k_2,k_3)$ potentially contains substantial information about the physics that generated the primordial perturbations.
The Planck analysis~\cite{PlanckNG} has tested for shapes of non-Gaussianity parameterized by the following templates:
\begin{align}
B_{\rm local} &\equiv \frac{6}{5} \Big( P_1 P_2 + \mbox{\rm perms.} \Big) \ , \label{equ:Bloc} \\
B_{\rm equil} &\equiv \frac{3}{5} \Big( 6\, (P_1^3 P_2^{2} P_3)^{1/3} - 3 P_1 P_2 - 2\, (P_1 P_2 P_3)^{2/3} + \mbox{\rm perms.} \Big) \ , \label{equ:Bequil}\\
B_{\rm ortho} &\equiv  \frac{3}{5} \Big( 18\, (P_1^3 P_2^{2} P_3)^{1/3} - 9 P_1 P_2 - 8\, (P_1 P_2 P_3)^{2/3} + \mbox{\rm perms.} \Big) \, ,\label{equ:Bortho}
\end{align}
where $P_i \equiv P_\R(k_i)$. We comment briefly on the physical motivations for these choices of bispectrum shapes:

\begin{itemize}
\item[$\triangleright$]  {\it Local non-Gaussianity.}---The shape (\ref{equ:Bloc}) arises from the following ansatz in real space~\cite{Gangui:1993tt, Komatsu:2001rj}:
\beq
\R(\x) \equiv \R_g(\x) + \frac{3}{5} f^{\rm local}_{\mathsmaller{\rm NL}}  \Big[ \R_g^2(\x) - \langle \R^2_g \rangle \Big]\ ,
\eeq
where $\R_g(\x)$ is a Gaussian random field. In momentum space, the signal peaks for
{\it squeezed} triangles, e.g.~$k_1 \ll k_2 \sim k_3$ (see fig.~\ref{fig:LocalShape}).
This shape of non-Gaussianity arises in models of {\it multi-field inflation}--- see Appendix C. 
On the other hand, in single-field inflation (i.e.~in models in which only the adiabatic mode $\pi$ is excited) the signal vanishes in the squeezed limit.
This important theorem is known as
 the {\it single-field consistency relation}~\cite{Maldacena:2002vr, Creminelli:2004yq}. Under mild assumptions about the
inflationary action and the initial state, it is possible to show that the bispectrum in single-field inflation satisfies\footnote{This theorem can be interpreted as a Ward identity associated with the non-linearly realized dilatation symmetry of the background~\cite{Assassi:2012zq, Hinterbichler:2012nm, Hinterbichler:2013dpa, Goldberger:2013rsa}, and is the analog of the Adler zero in pion physics.}
\beq
\lim_{k_1 \to 0} \frac{ B_\R(k_1,k_2,k_3)}{P_\R(k_1)P_\R(k_2) } = (1-n_s) \ll 1\ , \label{equ:CONSIST}
\eeq
In terms of the shapes~(\ref{equ:Bloc})--(\ref{equ:Bortho}), this implies that
$f^{\rm local}_{\mathsmaller{\rm NL}} \ll 1$, as only the local shape peaks in the squeezed limit.
Observing a signal in the squeezed limit ($f^{\rm local}_{\mathsmaller{\rm NL}} \gtrsim 1$)
would rule out {\it all} models of single-field inflation, not just slow-roll models.  Planck has now severely constrained this possibility (see below).

\begin{figure}[h!]
   \centering
      \includegraphics[width=0.95\textwidth]{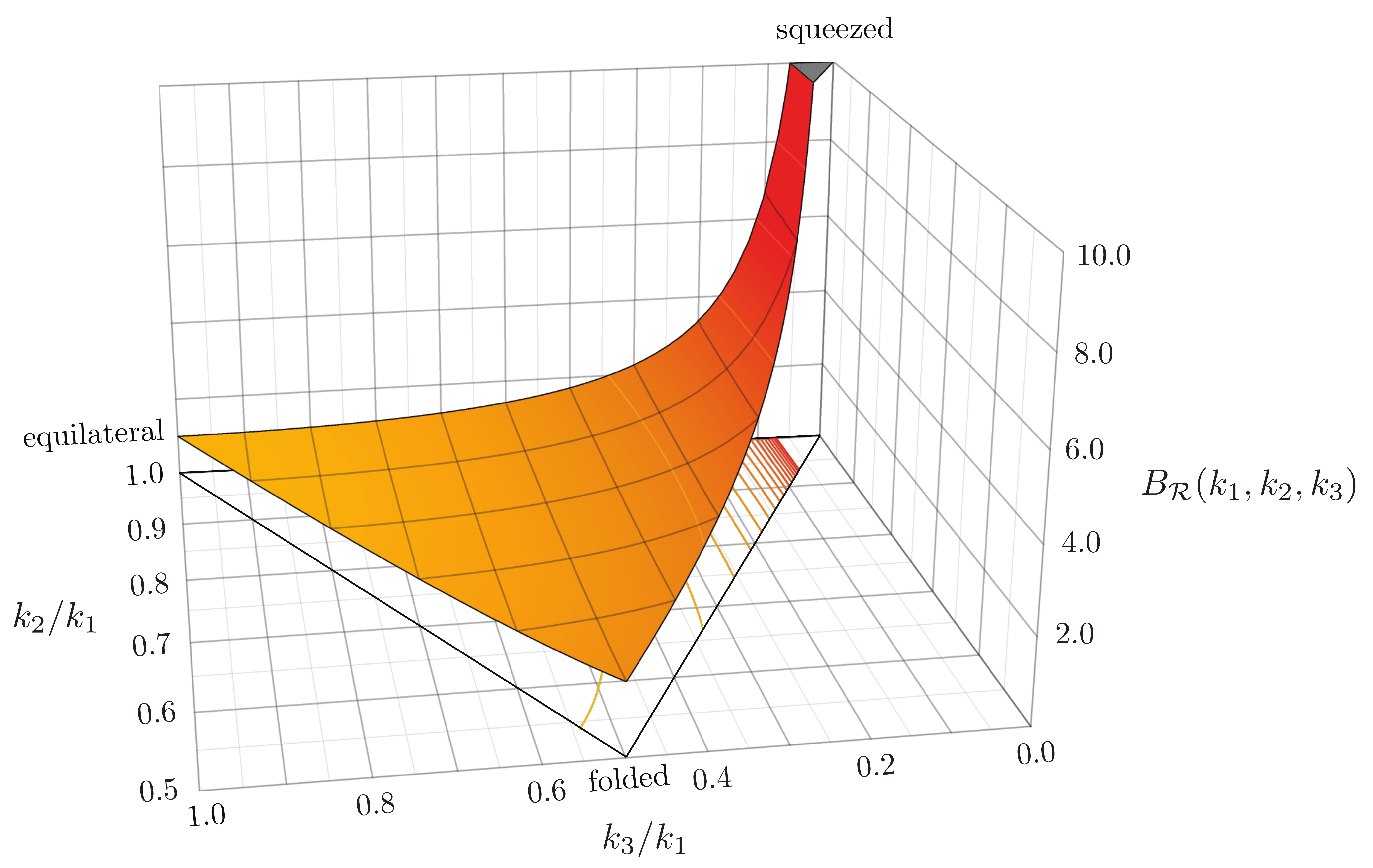}
      \vspace{-0.1cm}
   \caption{Bispectrum of the local ansatz. The signal is peaked for squeezed triangles.}
  \label{fig:LocalShape}
\end{figure}

\item[$\triangleright$]  {\it Equilateral non-Gaussianity.}---Large non-Gaussianity in single-field inflation can nevertheless arise from higher-derivative interactions \cite{Chen:2006nt, Seery:2005wm}. This leads to signals that peak in equilateral triangle configurations, i.e.~$k_1 \sim k_2 \sim k_3$.
To characterize this type of non-Gaussianity, we return to the Goldstone action.
At cubic order and to lowest order in derivatives, we get~\cite{Cheung:2007st} (see 
Appendix~B
 for the derivation)
\beq
S^{(3)}_\pi = \int \d^4 x\, \sqrt{-g} \,\, \frac{M_{\rm pl}^2 \dot H}{c_s^2} (1 - c_s^2) \left( \frac{\dot \pi (\partial_i \pi)^2}{a^2} + \frac{A}{c_s^2} \dot \pi^3\right)\ . \label{equ:S3}
\eeq
We have two cubic operators, $\dot \pi (\partial_i \pi)^2$ and $\dot \pi^3$, but only one new parameter, $A$. This is a consequence of the
nonlinearly-realized
time translation symmetry, which relates the amplitude of the operator $\dot \pi (\partial_i \pi)^2$ to the sound speed. In DBI inflation  (see \S\ref{ssec:DBI}) one has $A= -1$~\cite{Alishahiha:2004eh}, while more generally
naturalness arguments suggest $A \sim {\cal O}(1)$~\cite{Senatore:2009gt}.
 Both $\dot \pi (\partial_i \pi)^2$ and $\dot \pi^3$ produce bispectra that are well approximated by the equilateral template (\ref{equ:Bequil}) (see fig.~\ref{fig:EquilShape}).

 \begin{figure}[h!]
   \centering
      \includegraphics[width=0.95\textwidth]{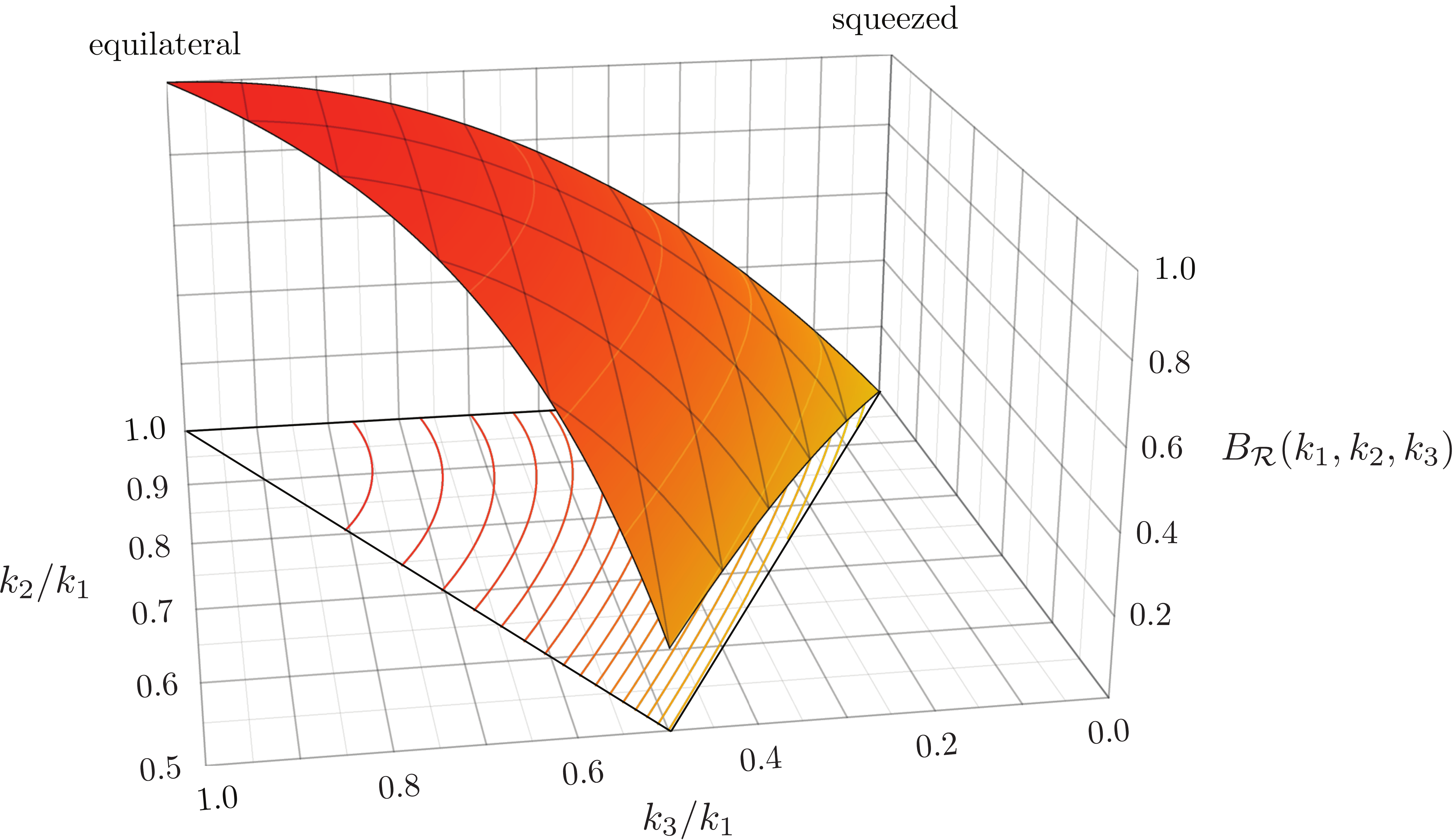}       \vspace{-0.1cm}
   \caption{Bispectrum of the interaction $\dot \pi (\partial_i \pi)^2$.  The signal is peaked for equilateral triangles.}
  \label{fig:EquilShape}
\end{figure}

\item[$\triangleright$]   {\it Orthogonal non-Gaussianity.}---The two equilateral bispectra are not identical, so one can find a linear combination of the two operators  $\dot \pi (\partial_i \pi)^2$ and $\dot \pi^3$ that is orthogonal in a well-defined sense~\cite{Babich:2004gb} to the shape (\ref{equ:Bequil}), and also to the local shape (\ref{equ:Bloc}).  This is the {\it{orthogonal}} template (\ref{equ:Bortho})~\cite{Senatore:2009gt}.
In terms of the parameters of the Lagrangian~(\ref{equ:S3}), the signal is mostly of the orthogonal shape --- specifically,
with greater than 70\% correlation with the orthogonal template --- for $3.1 \lesssim A \lesssim 4.2$.

\end{itemize}

\begin{figure}[h!]
    \centering
        \includegraphics[width=.6\textwidth]{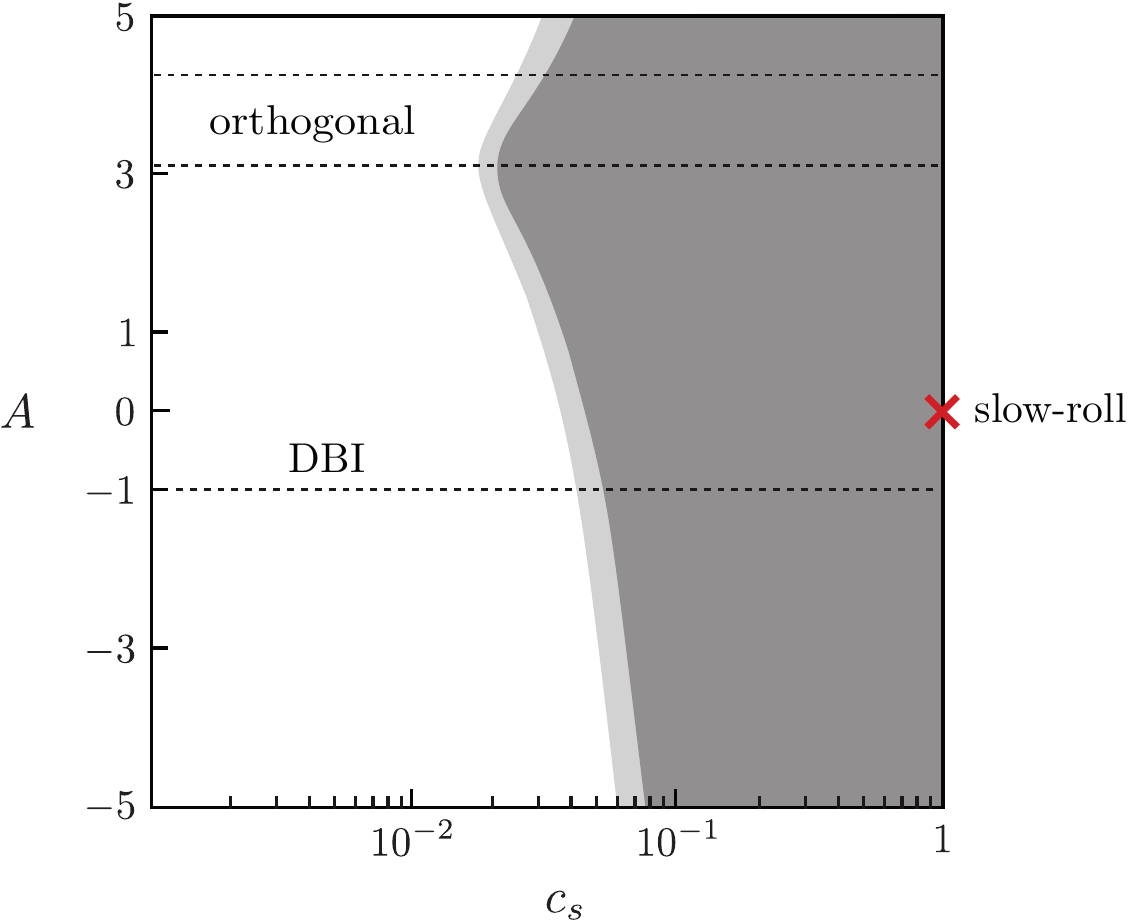}
    \caption{Planck constraints on non-Gaussianity in single-field inflation (figure adapted from~\cite{Creminelli:2013xfa}).
    Shown are the 68\% and 95\% constraints on the sound speed $c_s$ and interaction coefficient $A$; cf.~eq.~(\ref{equ:S3}).}
    \label{fig:CMBprecise}
\end{figure}

\noindent
The Planck collaboration has reported the following constraints on the amplitudes of the templates (\ref{equ:Bloc}), (\ref{equ:Bequil}) and (\ref{equ:Bortho})~\cite{PlanckNG}:
\begin{align}
f^{\rm local}_{\mathsmaller{\rm NL}}  &= 2.7 \pm 5.8\ , \label{equ:fl}\\
f^{\rm equil}_{\mathsmaller{\rm NL}} &= -42\pm 75 \ , \label{equ:fe}\\
f^{\rm ortho}_{\mathsmaller{\rm NL}}  & = - 25 \pm 39\ . \label{equ:fo}
\end{align}
Eq.~(\ref{equ:fl}) is a very strong constraint on multi-field inflation.
The limits~(\ref{equ:fe}) and (\ref{equ:fo}) are strong, but they do not make a future detection inconceivable (see \S\ref{sec:future}).
   Observational constraints on the parameters in the Goldstone action (\ref{equ:S3}) are shown in fig.~\ref{fig:CMBprecise}.

\subsection*{Non-Adiabaticity}

As we have seen, single-field inflation predicts initial fluctuations that are {\it adiabatic}.
Adiabatic perturbations have the property that the local state of matter (determined, for example, by the energy density $\rho$) at some spacetime point ($t,\x$) of the perturbed universe is the same as in the background universe at some slightly different time $t +\pi(t,\x)$.  That is, some parts of the universe are `ahead' and others `behind' in the evolution.
At recombination, the universe consists of a mixture of several fluids: photons ($\gamma$), baryons ($b$), dark matter ($c$) and neutrinos ($\nu$). For adiabatic initial conditions, the density perturbations in each species `$I$' are proportional to the Goldstone boson of broken time translations
\beq
\delta_I(t,\x) \equiv \frac{\bar \rho_I(t+\pi(t,\x)) - \bar \rho_I(t)}{\bar \rho_I(t)} \approx \frac{\dot{\bar \rho}_I}{\bar \rho_I} \, \pi(t,\x)\ .
\eeq
All matter perturbations therefore have the same density contrast (e.g.~$\delta_b=\delta_c$) and are proportional to the radiation perturbations (e.g.~$\delta_c = \frac{3}{4}\delta_\gamma$).
For adiabatic initial conditions, all species fluctuate synchronously and lead to the curvature perturbation $\R$.

In multi-field inflation, it is possible to generate so-called {\it isocurvature} perturbations, where an overdensity in one species compensates for an underdensity in another, resulting in no net curvature perturbation. For example, we can define the following isocurvature perturbation for dark matter and photons,
\beq
{\cal S} \equiv \delta_c- \tfrac{3}{4} \delta_\gamma \ .
\eeq
If this field were significantly different from zero it would lead to a measurable effect in the CMB power spectrum.

We digress briefly to describe two classic distinctions between isocurvature perturbations and curvature perturbations,
at the level of the acoustic peaks \cite{Hu:1996vq,Hu:1996yt}.
The first distinction involves the angular positions of successive peaks.   Adiabatic  perturbations from  single-field inflation  have fixed amplitude outside the horizon,  and begin to evolve upon entering the horizon.   The resulting evolution may be thought of as a cosine mode.   The curvature perturbations  sourced by  cosmic defects, in contrast, have  negligible amplitude as they enter the horizon, and grow subsequently through causal processes.   The result is typically a sine mode.   These two cases  make different predictions for the angular
positions of subsequent peaks,  which are in the ratio $1:2:3$  in the cosine case, and $1:3:5$  in the sine case. The  relative heights of even and odd peaks  provide another  means  of testing adiabaticity.
Acoustic peaks corresponding to compression waves ---  namely, the odd peaks ---  are enhanced  compared to even peaks  in the adiabatic case,  but suppressed compared to even peaks  in the isocurvature case.

Definitive evidence  against isocurvature models involving  causal evolution inside the horizon,  without an inflationary phase,  comes from measurements of CMB  polarization. A characteristic signature of these models  is that  the temperature and E-mode  polarization perturbations are positively correlated  on large angular scales \cite{Hu:1996vq},  while in inflation these perturbations are anti-correlated.  The measurement of TE anti-correlation on  superhorizon scales \cite{Peiris:2003ff} shows that superhorizon adiabatic perturbations were present when the CMB decoupled.

Although  purely  isocurvature perturbations are  now ruled out, it is possible that the observed  anisotropies originate from a combination of adiabatic and isocurvature perturbations.
To quantify the isocurvature contribution, it is conventional to define
the relative amplitude of the power spectra of the isocurvature field and the curvature perturbation
\beq
\alpha \equiv \frac{P_{\cal S}}{P_\R}\ .
\eeq
Assuming that ${\cal S}$ and $\R$ are uncorrelated (motivated by axion isocurvature models\cite{Seckel:1985tj, 1991PhLB..259...38L, Turner:1990uz}), Planck has  constrained this ratio~\cite{PlanckInflation},
 \beq
\alpha_0 < 0.036  \ .
\eeq
The constraint strengthens if ${\cal S}$ and $\R$ are perfectly correlated (as in curvaton isocurvature models~\cite{Lyth:2001nq, Moroi:2001ct}),
\beq
\alpha_{+1} < 0.0025\ .
\eeq

Observing an isocurvature contribution to the primordial fluctuations is another way to rule out single-field inflation, since only the presence of additional light fields can give rise to non-adiabaticity. Unfortunately, the amplitude of the signal depends on the post-inflationary evolution:
the  primordial perturbations become adiabatic if the particles produced after inflation reach a suitable thermal equilibrium~\cite{Weinberg:2004kf}.
Correspondingly, observable isocurvature is possible only  when one or more particle species has an abundance determined by physics beyond thermal equilibrium.

\subsection{Inflation after BICEP2}
\label{sec:BICEP}

In March 2014, the {\it BICEP2} collaboration announced the first detection of primordial B-modes~\cite{Ade:2014xna}.  The most straightforward interpretation\footnote{Even if the BICEP2 measurement turns out to be correct in every detail, further experimental and theoretical work will be required to exclude alternative explanations for a spectrum of primordial gravitational waves, although no alternative is nearly as compelling as the inflationary prediction.
One way to confirm the inflationary origin of the signal would be by establishing the superhorizon nature of the B-modes at recombination~\cite{Baumann:2009mq}.}
of the
signal seen by BICEP2
is as the imprint of primordial gravitational waves from quantum fluctuations of the gravitational field during inflation, as in (\ref{equ:Dh}).
An unambiguous detection of {inflationary}
gravitational waves would pinpoint the energy scale of inflation at the GUT scale,
$E_{\rm inf}  \sim 10^{16}$ GeV,
and also provide 
experimental evidence that gravity is quantized.
If the BICEP2 result is confirmed, it will stand as one of the pivotal discoveries in the history of cosmology.

\begin{figure}[h!]
   \centering
      \includegraphics[width=0.95\textwidth]{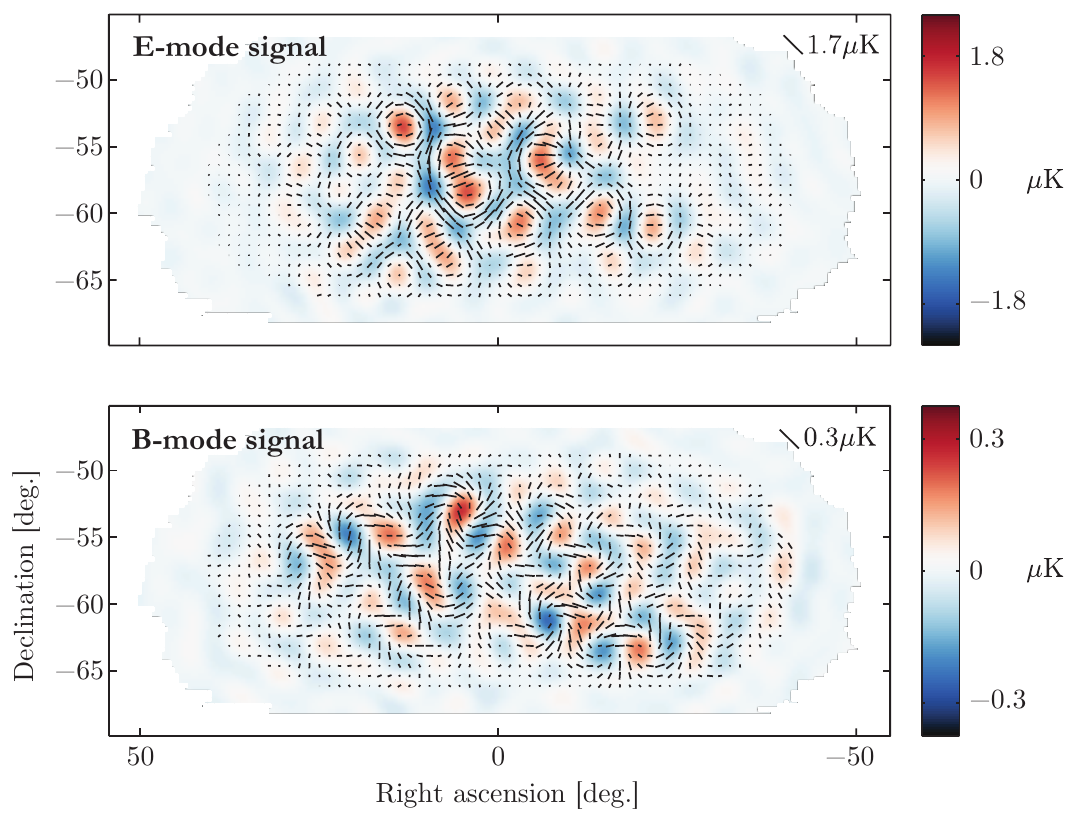}
   \caption{E-mode and B-mode maps measured by the BICEP2 experiment (figure adapted from~\cite{Ade:2014xna}). An excess over the lensing B-mode is detected with high signal-to-noise. }
  \label{fig:BICEP-Maps}
\end{figure}

\begin{figure}[h!]
   \centering
      \includegraphics[width=0.8\textwidth]{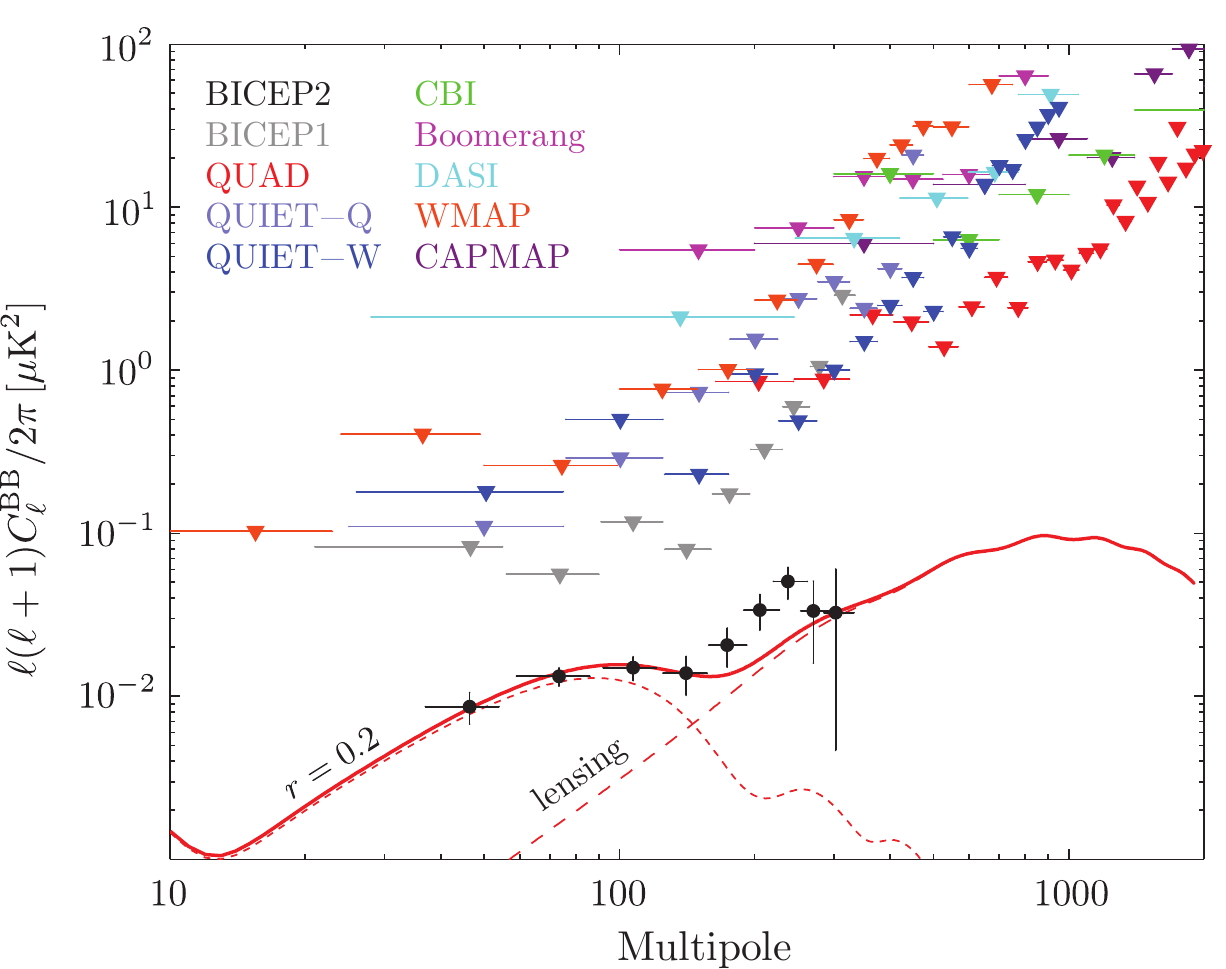}
   \caption{B-mode power spectrum measured by BICEP2, as well as $95\%$ upper limits from several previous experiments (figure adapted from~\cite{Ade:2014xna}). Also shown is the best-fit theoretical curve for $r=0.2$. This has two components: one from primordial tensors that peaks around $\ell \sim 80$, and one from the lensing conversion of E-modes that peaks around $\ell \sim 1000$.  }
  \label{fig:BBspectrum}
\end{figure}

\vskip 4pt
The BICEP experiment was designed specifically to search for the primordial B-mode signal on degree angular scales. Located at the South Pole, it observed a small and
exceptionally clean patch of the sky, the `Southern Hole'.
The observation frequencies were chosen to avoid contamination from synchrotron radiation and from emission by dust.
The first version of the experiment, BICEP1, observed at two frequencies: 100 GHz and 150 GHz. Collecting data from 2006 to 2008, it obtained the first significant upper limit on $r$ from polarization measurements alone~\cite{Chiang:2009xsa}
\beq
r < 0.73 \quad \mbox{(95\% limit)}\ .
\eeq
BICEP2 observed at only one frequency, 150 GHz, but with ten times as many detectors as BICEP1.
Data was taken over three seasons from 2010
to
2012.
The final polarization maps are shown in fig.~\ref{fig:BICEP-Maps}.
Even by eye, the B-mode pattern is clearly visible!
The derived B-mode power spectrum is shown in fig.~\ref{fig:BBspectrum}.
The best-fit value for the tensor-to-scalar ratio is\footnote{The central value of $r$ claimed by BICEP2 seems somewhat in tension with the Planck upper bound (\ref{equ:PlanckR}).  This issue is currently under active investigation~\cite{Smith:2014kka, Dvorkin:2014lea}, so we will limit ourselves to a few remarks on this issue.
Most importantly, the measured $r$ has not yet stabilized to changes in the analysis.
For example, there is still a large spread in the maximum likelihood values of $r$ for different models of foregrounds, roughly $0.12 < r < 0.21$.  We therefore caution against a premature judgement of the issue.
Even if the apparent tension survives further scrutiny, it has to be recognized that Planck and BICEP are sensitive to tensors in different ways. While BICEP measures tensors quite directly via their imprint on B-mode polarization, Planck constrains the combined effect of tensors and scalars on the temperature power spectrum. The Planck constraint on $r$ is therefore model-dependent and weakens if the scalar power is suppressed on large scales.}
\beq
r = 0.2^{+0.07}_{-0.05}\ . \label{equ:rBICEP}
\eeq
The null hypothesis $r=0$ is rejected at almost 7$\sigma$.

The result in (\ref{equ:rBICEP}) was obtained without any foreground subtraction.
The signal is large enough to dominate over available estimates of the polarized foregrounds (synchrotron and dust), but a more direct and convincing exclusion of a foreground explanation would be a detection of B-modes at a second frequency that confirms the expected thermal spectrum of the cosmological signal.  Using foreground models to correct for any residual foreground contamination tends to reduce the maximum likelihood value of $r$, but not by enough to seriously weaken the significance of the claimed detection.

The BICEP team took exceptional care to test for systematic errors.
They performed a large suite of so-called jackknife tests. Here, the data is split according to various criteria
and then the difference of the two sets is taken. The signal will cancel but any systematic effects that vary between the two sets may remain and can therefore be identified.
No failures of any jackknife tests have been reported.

Despite the strong evidence already provided by BICEP2, a detection of primordial B-modes is such a significant result that one should insist on confirmation by a second, independent experiment looking at a different part of the sky.  There are many experiments looking for B-modes on degree scales, including KeckArray~\cite{Ade:2014gua}, {EBEX}~\cite{ReichbornKjennerud:2010ja}, {SPIDER}~\cite{Fraisse:2011xz}, {ABS}~\cite{EssingerHileman:2010hh}, and {CLASS}~\cite{Eimer:2012ny}); and on arcminute scales, including {POLARBEAR}~\cite{Kermish:2012eh}, {SPTpol}~\cite{Austermann:2012ga}, and {ACTpol}~\cite{Niemack:2010wz}).
Finally,  given the  size of the signal seen by
the
BICEP2 collaboration, the Planck satellite has a chance to see the reionization peak in the B-mode spectrum at low $\ell$.

\section{Future Tests of Inflation}
\label{sec:future}

Cosmological observations show no signs of slowing down.
CMB observations continue to provide important measurements of the primordial fluctuations, especially on small angular scales~\cite{Das:2010ga, Sievers:2013wk, Keisler:2011aw, Story:2012wx}.
A large number of ground-based and balloon-borne experiments are targeting high-precision measurements of CMB polarization.
Current and future large-scale structure surveys will provide additional information (see table~\ref{tab:lss}).  In this section, we discuss what one can hope to learn from measuring the primordial perturbations with increased precision and over a wider range of scales.

	
\subsection{Tensor Tilt}	

In inflation, the tilt of the tensor spectrum is related to the time-evolution of the Hubble parameter:
\beq
n_t = 2 \frac{\dot H}{H^2}\ .  \label{nteqn}
\eeq
Finding a nearly scale-invariant spectrum, $|n_t| \ll 1$, would confirm that $|\dot H| \ll H^2$ in the early universe.  Since $|\dot H| \ll H^2$ was our definition of inflation in \S\ref{sec:CosmicInflation}, this would be as much of a `proof' of inflation as we can ever hope to get.

The sign of $n_t$ is also informative.
Consider a spatially-flat FRW universe filled with a perfect fluid with pressure $P$ and energy density $\rho$.
From (\ref{equ:Hdot}) we see
that
$\dot{H}>0$ is only possible if $\rho + P > 0$,
corresponding to a violation of the null energy condition (NEC).\footnote{The NEC states that the stress tensor satisfies $T_{\mu \nu} n^\mu n^\nu \ge 0$, for all null vectors $n^\mu$.  For a perfect fluid, the NEC reduces to $\rho + P \ge 0$.}
Thus, in all theories for which $n_t$  is given by (\ref{nteqn}) and the NEC holds, we predict $n_t < 0$.
 
Finally, in single-field slow-roll inflation, a consistency relation links the tensor tilt to the tensor-to-scalar-ratio:
\beq
n_t = - \frac{r}{8}\ . \label{sec:}
\eeq	
If the BICEP2 finding of $r \sim {\cal O}(0.1)$ is confirmed, then we expect $n_t \sim {\cal O}(0.0125)$. Measuring the tensor tilt at this level will be very challenging, but does not seem impossible. Testing the consistency relation will be one of the main targets of future CMB polarization experiments. Forecasts of experimental sensitivities can be found in~\cite{Knox:1995dq, Katayama:2011eh, Caligiuri:2014sla, Dodelson:2014exa}.

\subsection{Scalar Tilt and Running}	

Models of inflation make specific predictions for the parameters $n_s$ and $r$.
Improving the measurements of either of these parameters will therefore play a vital role in narrowing down the number of viable models~\cite{Creminelli:2014oaa}. Future galaxy surveys~\cite{Levi:2013gra} may reduce the error on $n_s$ by a factor of 5.
At the same time, future CMB polarization experiments~\cite{Baumann:2008aq, Andre:2013afa, Bouchet:2011ck, Kogut:2011xw, Matsumura:2013aja} have the potential to reduce the error on $r$ to the percent level.

A test of the slow-roll paradigm may come from measurements of the running of the scalar spectrum.
At second order in the Hubble slow-roll parameters, the running of the scalar spectrum is~\cite{Easther:2006tv}
\beq
\alpha_s = 16\hskip 1pt \varepsilon^2 - 6 \hskip 1pt \varepsilon \tilde \eta + \tilde \eta \chi \ ,
\eeq
where $\chi \equiv \dot{\tilde \eta}/(H\tilde \eta)$.
Measuring $\alpha_s$ would test the consistency of the slow-roll expansion.
However,  because the running is second order in slow-roll, we expect it to be small, $\alpha_s \sim (n_s-1)^2$.
Current bounds on $\alpha_s$ are still two orders of magnitude larger than this  target, but future galaxy surveys may allow such a measurement~\cite{Takada:2005si} (see also~\cite{Adshead:2010mc}).  Any detection of a larger level of running would be a challenge for slow-roll inflation and would require additional physics to explain.

\subsection{Non-Gaussianity}


The constraints on primordial non-Gaussianity from the CMB have almost reached their limit.
Silk damping of the small-scale anisotropies prohibits using multipoles larger than $\ell_{\rm max} \sim 2000$ to extract information about initial conditions.
This limits the number of modes available in the CMB to
\beq
{\cal N}^{\rm CMB} \sim \left(\frac{\ell_{\rm max}}{\ell_{\rm min}}\right)^2 \sim  10^6\ , \label{equ:Ncmb}
\eeq
which is nearly saturated by the recent Planck measurements.

More modes are in principle accessible through large-scale structure measurements.
This is because galaxy surveys probe the three-dimensional cosmic density field, while the CMB is only a  two-dimen\-sional projection. Hence,
while ${\cal N}^{\rm CMB} \propto \ell_{\rm max}^2$ for the CMB, we have ${\cal N}^{\rm LSS} \propto k_{\rm max}^3$ for LSS (but see~\cite{Rimes:2005xs, Crocce:2005xz}), where $k_{\rm max}$ is associated with the smallest scale that is both measurable and  under theoretical control. Pushing to smaller scales (larger $k_{\rm max}$) therefore increases
rather dramatically
the amount of information that can be extracted from the data.
The total number of (quasi-)linear modes in LSS is estimated to be
\beq
{\cal N}^{\rm LSS}_{\rm linear} \sim  \left(\frac{k_{\rm max}}{k_{\rm min}}\right)^3 \sim  10^9\ ,
\eeq
where we have taken $k_{\rm max} \sim 0.1\, {\rm Mpc}^{-1}$ and $k_{\rm min} \sim 10^{-4}\, {\rm Mpc}^{-1}$.  Although this shows the great potential of LSS observations, it assumes that we measure the entire volume at low redshift.
More realistically, 
we have $k_{\rm min} \sim 10^{-3}\, {\rm Mpc}^{-1}$ (e.g.~for the Euclid mission), and hence~\cite{Amendola:2012ys}
\beq
{\cal N}^{\rm Euclid}_{\rm linear} \sim  \left(\frac{k_{\rm max}}{k_{\rm min}}\right)^3 \sim  10^6\ ,
\eeq
which is comparable to the result (\ref{equ:Ncmb}) for the CMB.
However, while $\ell_{\rm max}$ for the CMB cannot be extended, for LSS
 $k_{\rm max}$
might be
pushed
to larger values through a better understanding of
non-linearities in the dark matter evolution, the biasing, and the redshift space distortions.
This is one of the objectives of the recently developed `effective theory of large-scale structure'~\cite{Baumann:2010tm, Carrasco:2012cv}~(see also~\cite{Carrasco:2013mua, Porto:2013qua, Carroll:2013oxa, Mercolli:2013bsa, Pajer:2013jj, Hertzberg:2012qn}).

\begin{table}[t]

	\heavyrulewidth=.08em
	\lightrulewidth=.05em
	\cmidrulewidth=.03em
	\belowrulesep=.65ex
	\belowbottomsep=0pt
	\aboverulesep=.4ex
	\abovetopsep=0pt
	\cmidrulesep=\doublerulesep
	\cmidrulekern=.5em
	\defaultaddspace=.5em
	\renewcommand{\arraystretch}{1.6}

	\begin{center}
		\small
		\begin{tabular}{llcccc}

			\toprule
		Name & $z_{\rm max}$ & ${V}\hskip 1pt\big[({\rm Gpc}/h)^3\big]$ & $n_g\hskip 1pt \big[ ({\rm Mpc}/h)^{-3} \big]$ & $k_{\rm max}\hskip 1pt \big[h/{\rm Mpc} \big]$ & $\Delta \fnl^{\rm local}$ \\
			\midrule
		{\rm SDSS LRG}  \ \ &  0.315 & 1.48   & $1.36 \times 10^{-3}$ & 0.1 & $5.62$
				\\[1mm]
				{\rm BOSS} &  0.35  & 5.66 &  $0.27 \times 10^{-3}$ & 0.1 & 3.34\\[1mm]
		{\rm Big-BOSS }  &  0.5  & 13.1 &  $0.30  \times 10^{-3}$ & 0.1 & 2.27\\[1mm]

				{\rm HETDEX} &  2.7  & 2.96 & $0.27  \times 10^{-3}$ & 0.2 & 3.65 \\[1mm]
				{\rm CIP} &  2.25  & 6.54  & $0.50  \times 10^{-3}$ & 0.2  & 1.03 \\[1mm]
			{\rm EUCLID} &  1.0  & 102.9 & $0.16  \times 10^{-3}$ & 0.1  & 0.92 \\[1mm]
				{\rm WFIRST} &  1.5  &  107.3  & $0.94  \times 10^{-3}$ & 0.1 & 1.11  \\[1mm]
 			\bottomrule
		\end{tabular}
	\end{center}
	\caption{Compilation of current and future LSS surveys.  Here, $z_{\rm max}$ refers to the maximal redshift of the survey, $V$ is the survey volume and $n_g$ is the mean comoving number density of objects. The projected errors on $\fnl^{\rm local}$ are from measurements of the galaxy bispectrum. (Data collected by Donghui Jeong.)
	\label{tab:lss}}
	\end{table}


\vskip 4pt
Even from the CMB alone, 
the limit on the amplitude of local non-Gaussianity, cf.~(\ref{equ:fl}), is getting close to an interesting threshold for multi-field inflation.
The conversion of hidden sector non-Gaussianity during reheating~\cite{Dvali:2003em, Zaldarriaga:2003my, Kofman:2003nx} or after inflation~\cite{Linde:1996gt, Lyth:2002my} typically leads to
\beq
|\fnl^{\rm local}| \gtrsim {\cal O}(1)\ .
\eeq
This possibility is now highly constrained, and further data from Planck and LSS surveys (see table~\ref{tab:lss}) has the potential to rule out the natural parameter space of a large class of multi-field models.

\vskip 4pt
As we explain in 
Appendix~B,
a similar threshold exists for equilateral non-Gaussianity~\cite{Abazajian:2013mma}:
\beq
\fnl^{\rm equil} \sim {\cal O}(1)\ . \label{equ:EqThresh}
\eeq
 Not seeing a signal at the level of (\ref{equ:EqThresh})
 would allow us to conclude that the UV-completion of the effective theory is slow-roll inflation, up to perturbative higher-derivative corrections.
Conversely, a detection of equilateral non-Gaussianity with $\fnl^{\rm equil} > {\cal O}(1)$ would imply that the theory has to be UV-completed by something other than slow-roll inflation, such as DBI inflation.  The threshold (\ref{equ:EqThresh}) therefore provides an important observational distinction between UV-completions of inflation corresponding to {weakly-coupled} backgrounds and those that involve {strongly-coupled} backgrounds.
Unfortunately, $\fnl^{\rm equil} \sim {\cal O}(1)$ is almost two orders of magnitude smaller than the CMB bound~(\ref{equ:fe}).  Future CMB data may improve the bound by a factor of a few, but not by enough to reach
the threshold (\ref{equ:EqThresh}).
However, optimistic estimates of galaxy lensing tomography suggest that this may not be completely out of reach for future LSS observations~\cite{Giannantonio:2011ya}.

\chapter{Inflation in Effective Field Theory}
\label{sec:EFT}

Inflation is a well-understood phenomenon in quantum field theory coupled to gravity,
and many field theories that support inflationary phases have been proposed.
Nevertheless,  deriving the inflationary action from a more fundamental principle, or in the context of a well-motivated parent theory, remains a central problem.

There are two approaches or  perspectives that can be used to obtain a quantum field theory suitable for inflation.  In the `top-down' approach,  one begins with a complete theory in the ultraviolet (UV),  such as string theory,  and tries to derive inflation as one of its low-energy consequences.
This undertaking is  discussed at length in Chapters~\ref{sec:StringInflation} and~\ref{sec:Examples}.
The more conservative `bottom-up' approach starts from the low-energy (IR) degrees of freedom and parameterizes our ignorance about the UV theory.
Both approaches arrive at an {\it effective field theory} (EFT) that  is valid at inflationary energies, but they do so from opposite directions.  The two approaches are  complementary and can inform each other.

\vskip 4pt
The outline of this chapter is as follows: in \S\ref{sec:EFT2}, we provide a general overview of the essential principles of effective field theory.\footnote{More details on effective field theory can be found in~\cite{Burgess:2003jk, Kaplan:2005es, Skiba:2010xn, Burgess:2007pt}.}
We apply these concepts to inflation in \S\ref{sec:EFT-Inflation}, highlighting the sensitivity of inflation to Planck-scale physics in \S\ref{ssec:UV}.

\begin{figure}[h!]
   \centering
      \includegraphics[width=0.8\textwidth]{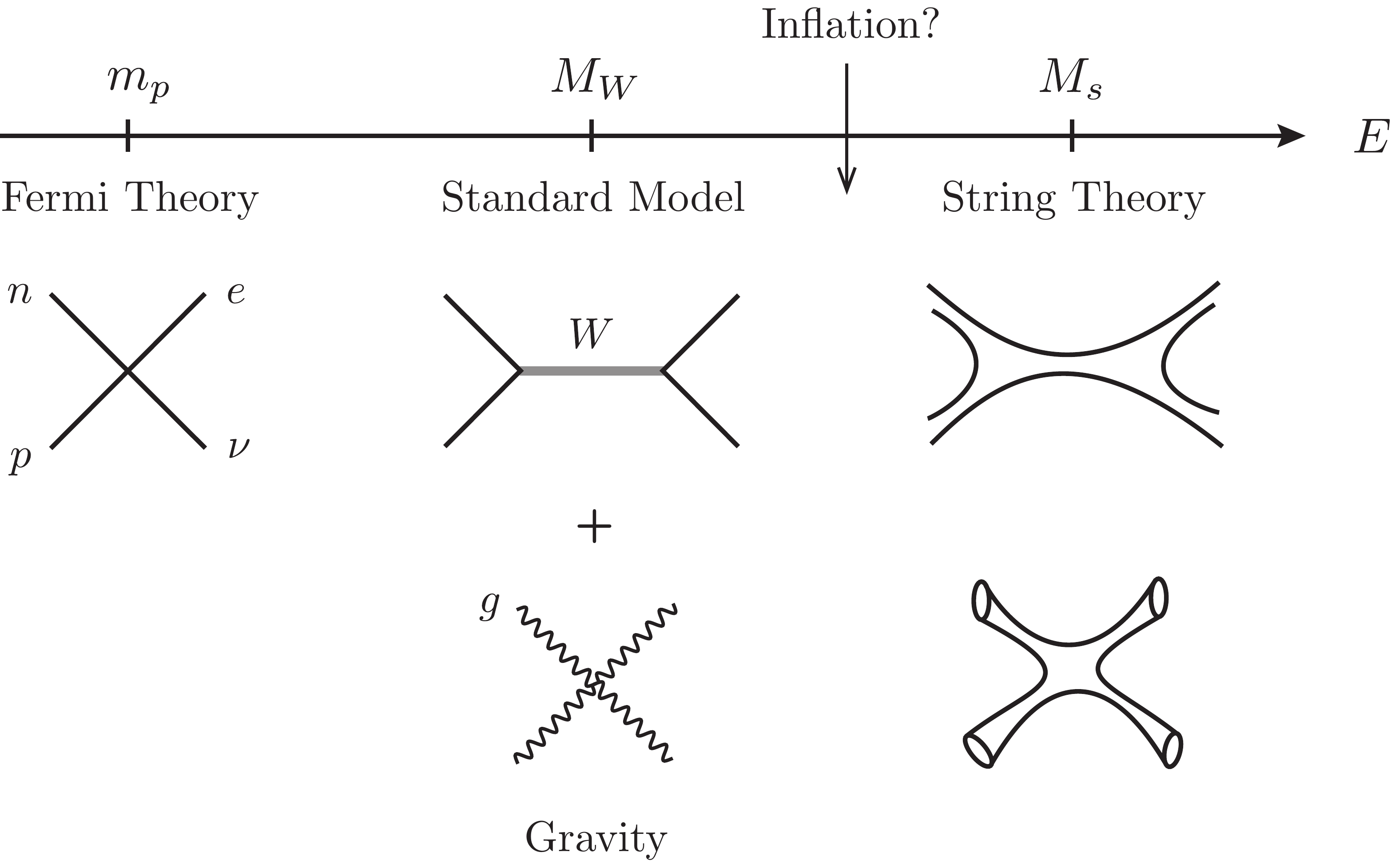}
   \caption{
   Effective field theories in particle physics. Both Fermi theory and general relativity are non-renormalizable and are interpreted as effective theories.}
  \label{fig:EFT}
\end{figure}

\section{Principles of Effective Field Theory}
\label{sec:EFT2}

Natural phenomena occur across a vast range of length scales.  Fortunately, in many cases one can analyze physical processes involving distinct scales by examining one relevant scale at a time. Fig.~\ref{fig:EFT} illustrates this logic using a few famous examples from the history of particle physics.
For instance, at low energies, Fermi theory describes neutron-proton interactions by a four-fermion contact interaction with coupling constant $G_F = (293.6\, {\rm GeV})^{-2}$. This theory is incomplete and breaks down (violates perturbative unitarity) at about 100 GeV.  What actually happens close to 100~GeV is that we start to resolve the $W$-boson exchange interaction and Fermi theory is replaced by the electroweak theory of the Standard Model.
Similarly, interactions of the Standard Model fields with gravitational degrees of freedom are determined by Newton's constant $G_N = (1.2 \times 10^{19}\, {\rm GeV})^{-2}$.  Just like the Fermi theory, this theory breaks down at high energies, this time at the Planck scale
\beq
M_{\rm pl}  \equiv \frac{1}{\sqrt{8\pi G_N}} = 2.4 \times 10^{18}\, {\rm GeV}\ .
\eeq
The Standard Model plus general relativity should therefore also be viewed as an effective theory to be replaced by a more fundamental theory at some energy
at or
below the Planck scale.  In much of this work we will assume that this more fundamental theory is string theory.

\subsection{Effective Action}
\label{ssec:Principles}

The first step in constructing effective field theories is identifying the degrees of freedom that are relevant for the measurements of interest.  For instance, in particle physics we distinguish light and heavy degrees of freedom on the basis of whether the corresponding particles can be produced on-shell at the energies available to the experiment.
Formally, we introduce a cutoff scale $\Lambda$ to define the regime of validity of the EFT.
{\it Light} particles~$\phi$, with masses $m < \Lambda$, are included in the effective theory, while {\it heavy} particles $\Psi$, with masses $M > \Lambda$, are `integrated out', in a sense that we will make precise.

\vspace{0.5cm}
\begin{figure}[h!]
   \centering
      \includegraphics[scale=0.36]{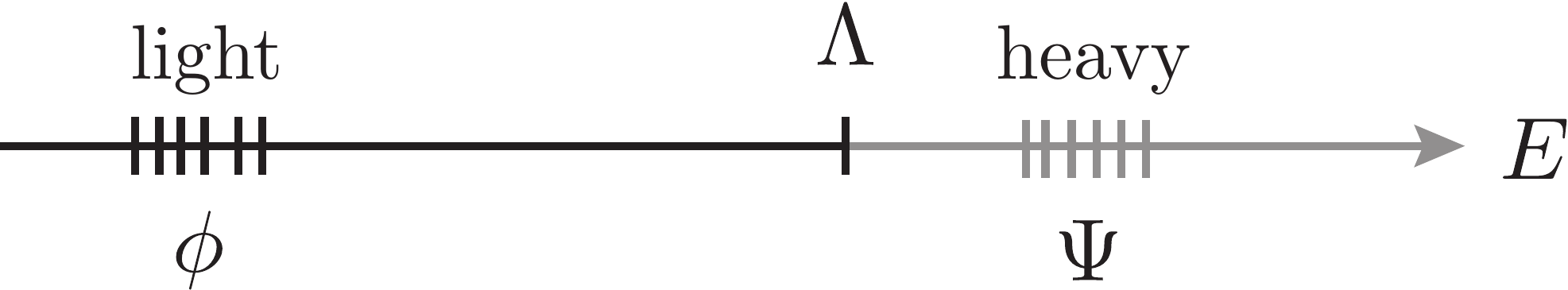}
   \caption{
   Effective field theories describe the physics of light degrees of freedom below a cutoff scale~$\Lambda$. We arrive at these theories either by integrating out the heavy fields (if the complete UV theory is known) or by parameterizing their effects (if the UV theory is not known or is not computable). In the latter case, symmetries inform the choice of allowed interactions.}
\end{figure}

\vspace{-0.3cm}
\subsection*{Top down: \hskip 1pt Integrating out\hskip 1pt}

Imagine that we knew the full Lagrangian of the UV theory,
\beq
{\cal L}[\phi,\Psi] = {\cal L}_l[\phi] + {\cal L}_h[\Psi] + {\cal L}_{lh}[\phi,\Psi] \ ,
\eeq
where ${\cal L}_l$ (${\cal L}_h$) describes the part of ${\cal L}$ involving only the light (heavy) fields, and ${\cal L}_{lh}$ includes all interactions involving both sets of fields.
The {\it Wilsonian effective action} $S_{\rm eff}$
is defined
via
a path integral over the heavy modes (and over the high-frequency contributions of the light fields):
\beq
e^{i S_{\rm eff}[\phi]} = \int [{\cal{D}} \Psi] \, e^{i S[\phi,\Psi]} \ . \label{equ:Seff}
\eeq
In practice, the effective action is rarely found by performing the path integral. Instead a so-called matching calculation order by order in perturbation theory is usually more practical~\cite{Burgess:2003jk, Kaplan:2005es, Skiba:2010xn, Burgess:2007pt}.

In the classical approximation,  performing the path integral over the heavy modes corresponds to using the
equations of motion to eliminate the heavy field $\Psi$. In the language of Feynman diagrams, this is the tree-level approximation. The complete path integral, however, also includes loops of the heavy fields. These loops describe how the heavy degrees of freedom are eliminated from the quantum theory.
The result is usually
non-local, meaning that it contains terms such as $\phi \hskip 1pt (-\Box + M^2)^{-1} \hskip 1pt \phi$.
However, at low energies, $E \ll \Lambda \le M$, these terms can be expanded in derivatives---for example,
\begin{equation}
\phi \hskip 1pt (-\Box + M^2)^{-1} \hskip 1pt \phi \, =\,  \frac{\phi}{M^2} \left(1 + \frac{\Box}{M^2} + \cdots \right) \phi \ ,
\end{equation}
and the EFT becomes approximately local.
In other words, the effective action admits a systematic expansion in powers of the ratio $(E/M)$,
\beq
{\cal L}_{\rm eff}[\phi] = {\cal L}_l[\phi] + \sum_i c_i(g) \hskip 1pt \frac{{\cal O}_i[\phi]}{M^{\delta_i -4}} \ , \label{equ:Leff}
\eeq
where $c_i$ are dimensionless constants that depend on the couplings $g$ of the UV theory,
and ${\cal O}_i$ are {local} operators of dimension $\delta_i$.
This procedure typically generates all operators ${\cal O}_i$ consistent with the symmetries of the UV~theory.
The  absence  from the effective theory of an operator allowed by the symmetries of the UV theory, or an anomalously small coefficient for such an operator, is described as a fine-tuning.\footnote{An important exception is an {\it accidental  symmetry}: if all  operators in the UV theory  violating a symmetry ${\cal S}$  are irrelevant in the sense of the renormalization group (RG),  then ${\cal S}$  is an approximate symmetry of the  low-energy theory.   However, in this case the  smallness of the ${\cal S}$-violating terms is  not mysterious:  it is simply a consequence of RG flow.}

In (\ref{equ:Leff}) we have split the effective action into a renormalizable part ${\cal L}_l$ and a sum of non-renormalizable corrections.  Note that non-renormalizable terms arise in the EFT even if the UV theory is renormalizable.
Operators of dimensions less than four (in four spacetime dimensions) are called {\it relevant operators}. They dominate in the IR and become small in the UV. Unsurprisingly, operators of dimensions greater than four are called {\it irrelevant operators}.\footnote{Operators of dimension equal to four are called {\it marginal operators}. Quantum corrections decide if a marginal operator is relevant or irrelevant in the IR.} These operators dominate in the UV but become small in the IR: the contribution  of an operator ${\cal O}_i$ of dimension $\delta_i$ to low-energy observables is proportional to $(E/M)^{\delta_i - 4}$.  As a result, although the sum in (\ref{equ:Leff}) includes operators of arbitrarily large dimension $\delta_i$, only a finite number of terms are required to predict the results of experiments to a given accuracy.
On the other hand, by studying the low-energy effects of irrelevant operators, we can learn about the structure of the UV theory. Measuring or constraining irrelevant operators can therefore be very informative.

\vskip 4pt
\noindent
{\it A toy model.}---Let us illustrate this procedure with a simple toy example: we take the Lagrangian of the UV theory to be\footnote{This is a simplified version of the example studied in \cite{Burgess:2007pt}. }
\beq
{\cal L}[\phi,\Psi] = -\frac{1}{2}(\partial \phi)^2 - \frac{1}{2}m^2 \phi^2  - \frac{1}{4!}\lambda \phi^4  -\frac{1}{2}(\partial \Psi)^2 - \frac{1}{2}M^2 \Psi^2 - \frac{1}{4} g \phi^2 \Psi^2\ . \label{equ:toy}
\eeq
Note that this Lagrangian respects
the $\mathbb{Z}_2$ symmetry $\phi \to -\phi$.
The effective Lagrangian for the light field $\phi$ takes the   form
\begin{align}
{\cal L}_{\rm eff}[\phi] &= - \frac{1}{2}(\partial \phi)^2 -  \frac{1}{2} m^2_{\rm R} \phi^2 - \frac{1}{4!} \lambda_{\rm R} \phi^4 \nonumber \\
&\ \ \ -\, \sum_{i=1}^\infty \left( \frac{c_i(g)}{M^{2i}}\, \phi^{4+2i} + \frac{d_i(g)}{M^{2i}}\, (\partial \phi)^2\phi^{2i} + \cdots \right)\ . \label{equ:Leff2}
\end{align}
The parameters in (\ref{equ:Leff2})
can be
determined in an expansion in the couplings of the UV theory (here, $\lambda$ and $g$).
For example, the bare values of the mass $m$ and the quartic coupling $\lambda$ receive loop corrections with the heavy particle~$\Psi$ running in the loop,\footnote{Of course, there are similar diagrams with the light particle $\phi$ running in the loop. For simplicity, we hide those terms in the ellipses of eqs.~(\ref{equ:mR}) and (\ref{equ:lR}).}
\begin{align}
m_{\rm R}^2 &\ =\ \includegraphicsbox[scale=.65]{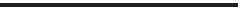} \ + \ \includegraphicsbox[scale=.65]{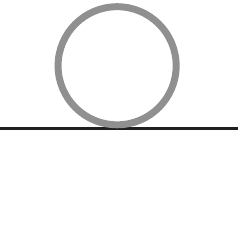} \ +\cdots \ , \label{equ:mR}\\
\lambda_{\rm R} &\ =\  \includegraphicsbox[scale=.65]{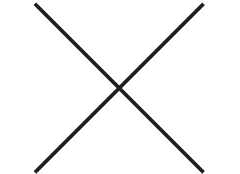} \ + \ \includegraphicsbox[scale=.65]{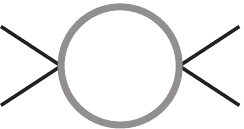} \ +\cdots\ , \label{equ:lR}
\end{align}
The loop contributions diverge in the UV and have to be regularized. Cutting off the (Euclidean) momentum integrals at the scale $\Lambda$, we find~\cite{Burgess:2003jk}
\begin{align}
m_{\rm R}^2 &\ =\ m^2 + \frac{g}{32\pi^2}  (\Lambda^2 -M^2 L) +\cdots \ , \label{equ:mRX}\\
\lambda_{\rm R} &\ =\ \lambda - \frac{3 g^2 }{32 \pi^2}  L + \cdots\ , \label{equ:lRX}
\end{align}
where $L \equiv \ln(\Lambda^2/\mu^2)$, with $\mu$ being an arbitrary renormalization scale.
Dimensional regularization would give the same result, except that we would not find the $\Lambda^2$
term in (\ref{equ:mRX}),
and we would replace $L$ by \beq L \to \frac{1}{\epsilon} + \gamma -\ln(4\pi) \ ,\eeq
where $\epsilon \equiv 4-d$ and $\gamma \equiv 0.577 \cdots$.   We see that the quadratic divergence in (\ref{equ:mRX}) is scheme-dependent and hence not physical.  However, notice that the unphysical term comes with the same coupling~$g$ as the physical contribution to the renormalized mass proportional to $M^2$.  It is therefore common to use the dependence on the cutoff $\Lambda$ as a proxy for the physical dependence on the mass of the heavy particles, $M \ge \Lambda$. However, see \cite{Burgess:1992gx} for examples where this logic fails.
Finally, the Wilson coefficients in (\ref{equ:Leff2})
can also be computed in a loop expansion;
e.g.~the coupling of the operator $\phi^6$ is
\beq
c_1 \ =\  \includegraphicsbox[scale=.65]{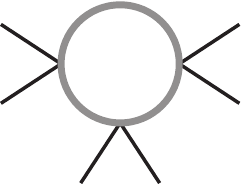} \ + \cdots\ \sim\  {\cal O}(g^3) + \cdots\ .
\eeq

\vskip 4pt
\noindent
{\it The trouble with light scalars.}---From (\ref{equ:mRX}) we see that the effective mass $m_{\rm R}$ gets a large contribution from the mass of the heavy field $M$.
Having a light scalar field in the EFT is therefore {\it unnatural}~\cite{PhysRevD.20.2619} in the sense that large quantum corrections, ${\cal O}(M)$, must be canceled by a large bare mass $m$ with opposite sign to achieve $m_{\rm R} \ll M$ (see~\S\ref{ssec:natural} for a detailed discussion).
This is a real problem, since for natural values of $m_{\rm R}$ the fields $\phi$ are not even part of the low-energy EFT!
The apparent need to fine-tune the Higgs boson mass is the famous electroweak hierarchy problem of the Standard Model.  As we will see in \S\ref{ssec:UV}, a qualitatively similar (but quantitatively less dramatic) hierarchy problem exists in inflationary models driven by scalar fields.
Notice that because the loop correction is proportional to the coupling $g$, a hierarchy $m_{\rm R} \ll M$ can be natural if we have reason to believe that $g\ll 1$ (and $m \ll M$).  Thus, the strength of the coupling between the light and heavy fields is a critical parameter, and symmetry structures in the UV theory that achieve $g \ll 1$ play an important role in discussions of light scalar fields (see~\S\ref{sss:symmetry}).

\vskip 4pt
\noindent
{\it Decoupling.}---All the divergences in (\ref{equ:mRX}) and (\ref{equ:lRX}) can be absorbed into a renormalization of the parameters of the Lagrangian.
Moreover, in the limit $M \to \infty$ the effects of the heavy particles disappear completely.
This {\it decoupling} of UV physics~\cite{Appelquist:1974tg} ensures that the physical effects of massive particles are suppressed for large~$M$.
Decoupling is one reason that the Standard Model of particle physics was constructed by focusing only on renormalizable theories, although today we view it as an effective theory.

\subsection*{Bottom up: \hskip 1pt Parameterizing ignorance\hskip 1pt}

Often we do not know the complete UV theory, so that we cannot construct the EFT explicitly by integrating out the heavy modes. Instead we parameterize our ignorance about the UV physics, by making assumptions about the symmetries of the UV theory, and writing down the most general effective action consistent with these symmetries:
\beq
{\cal L}_{\rm eff}[\phi] = {\cal L}_l[\phi] + \sum_i c_i\hskip 1pt \frac{{\cal O}_i[\phi]}{\Lambda^{\delta_i -4}} \ , \label{equ:Leff3}
\eeq where the sum runs over all operators ${\cal O}_i[\phi]$, of dimension $\delta_i$, allowed by the symmetries of the UV theory. The size of the higher-dimension operators is estimated in terms of the cutoff scale $\Lambda$, while the prefactors $c_i$ are dimensionless {\it{Wilson coefficients}}. Eq.~(\ref{equ:Leff3}) will be our starting point for discussing inflation in EFT.

Several comments are important at this stage.  First, making assumptions about the symmetries of the UV theory can be non-trivial: not all low-energy symmetries have to admit ultraviolet completions.  In \S\ref{ssec:gravity}, we discuss this issue in the context of string theory.  Second, in writing (\ref{equ:Leff3}) we have introduced an energy scale $\Lambda$ and a collection of dimensionless
coefficients $c_i$.\footnote{At the level of (\ref{equ:Leff3}), the joint rescaling $\Lambda \to \kappa \Lambda$ and $c_i \to \kappa^{\delta_i-4} c_i$ leaves the theory unchanged.  Formalizing this leads to the renormalization group and to the running of the renormalized couplings with energy.}
It is clearly important to  understand how to assign values for the scale $\Lambda$ and the coefficients $c_i$.  The guiding principle for this undertaking, and more generally for the construction and interpretation of effective field theories, is {\it{naturalness}}.

\subsection{Naturalness}
\label{ssec:natural}

Naturalness arguments work in two directions, from the top down and from the bottom up.
We will discuss these two aspects of naturalness in turn.

\subsection*{Top-down naturalness}

The top-down version of naturalness asserts that the Wilson coefficients in an
EFT will be
of order unity if the cutoff $\Lambda$ is chosen to be
of order the characteristic mass scale $M$ of the UV theory.\footnote{The  dimensionless couplings $g$ of the light fields to the heavy fields  are assumed to be of order unity for this purpose. Systematically weaker couplings should be incorporated by defining a higher effective mass scale, $\tilde{M} \sim M/\sqrt{g}$.}

We emphasize that the coefficients in question are those of operators ${\cal O}_i$ {\it{allowed}} by the symmetries of the UV theory: approximate or exact symmetries of the ultraviolet theory can lead to small or vanishing coefficients for the corresponding operators in the effective theory.  This top-down version of naturalness is merely a formalization of the expectation of genericity.
When the ultraviolet theory is computable, naturalness is hardly needed, as the effective theory can be constructed directly by integrating out the heavy modes, and the Wilson coefficients can be obtained explicitly in terms of the parameters of the UV~theory.  However, this favorable circumstance is very rare, and in particular it does not arise in presently-studied compactifications of string theory. We often only have partial information about the UV~theory. For instance, we might know the relevant scales, but not the couplings to all light fields.
 Top-down naturalness is then widely used as a systematic framework for guessing how the calculation of the effective theory would turn out if we were strong enough to perform it.

\subsection*{Bottom-up naturalness}

The bottom-up version of naturalness allows one to
infer, based on properties of a given low-energy effective theory, the plausible scale $M$ of new physics.
Here, new physics refers to the scale of the lightest degrees of freedom that are part of the ultraviolet theory but not of the effective theory.
Because we generally learn about nature beginning at low energies and proceeding to higher energies, the bottom-up version of naturalness can be a predictive tool for the results of experiments.

\vskip 4pt
The logic of bottom-up naturalness arguments is the following:  suppose we are presented with partial information about an EFT like (\ref{equ:Leff3}). Imagine that we know all the renormalizable couplings of the light fields, but the UV cutoff~$\Lambda$ and the higher-dimension contributions are unknown ---   this is, for example, the situation for the Standard Model of particle physics.  We would like to make an educated guess about the size of $\Lambda$.
In the low-energy theory, we can calculate loop corrections to the parameters of the renormalizable Lagrangian as functions of an unknown cutoff scale $\Lambda$.  The parameters are said to be bottom-up natural\footnote{We should stress that what we call `bottom-up naturalness' here is universally referred to as `naturalness', but we find the distinction useful for the present exposition.} as long as their measured values are larger than the loop corrections.  As we extrapolate the EFT to higher energies and increase $\Lambda$, some parameter may become unnatural, and insisting on natural parameters therefore defines a maximal scale for the effective theory,
$\Lambda = \Lambda_{\rm{max}}$.
Bottom-up naturalness asserts that `new physics' should appear at some scale $\Lambda \le \Lambda_{\rm{max}}$ and modify
the effective theory, thereby explaining the smallness of the measured  parameter values.
The unnatural alternative would be that multiple loop and/or bare contributions cancel against each other for  reasons beyond the purview  of the effective theory.  Bottom-up naturalness  is a formalization of the  expectation --- or more properly, the hope ---  that  this is not the case.
In the quintessential example of the Higgs boson, bottom-up naturalness predicts that new physics should appear at ${\cal O}(10^2-10^3)$ GeV  to cut off the quadratic divergence of the Higgs mass $m_H$ and explain why $m_H=125$ GeV.

\vskip 4pt
The naturalness criterion has been profoundly influential in motivating physics beyond the Standard Model and it often plays an important role in inflationary model-building.  However, there are reasons to use it with care.
We digress briefly to discuss a few instructive examples illustrating both successes and failures of the naturalness principle.\footnote{These comments are based on \cite{Murayama:2000dw, Giudice:2008bi} and on private communications with Nima Arkani-Hamed. }

\vskip 4pt
\noindent
{\it Successes of naturalness.}---Let us first look at examples where insisting on natural parameter values has led (or could have led) to the correct physics.

\begin{itemize}
\item[$\triangleright$]  {\it Positron.}---In classical
electromagnetism
the mass of the electron is unnatural..
The electric field around an electron carries energy $\Delta E = \alpha/r_e$, where $\alpha \approx 1/137$ is the fine structure constant and $r_e$ is the size of the electron, which is introduced to regulate the divergence.
This Coulomb self-energy of the electron contributes to its mass:
\beq
\Delta m_e = \alpha\, \Lambda\ , \label{equ:me}
\eeq
where $\Lambda \equiv r_e^{-1}$.
In order for the observed mass of the electron ($m_e \approx 0.511$ MeV) to be natural we require $\Lambda < 70$~MeV.
Indeed, in quantum field theory, new physics in the form of the positron comes to the rescue.
In \cite{Weisskopf:1939zz}, Weisskopf showed that virtual positrons surrounding the electron precisely cancel the linear divergence in (\ref{equ:me}), leaving only a logarithmic dependence on the cutoff,
\beq
\Delta m_e = \alpha\, m_e \ln \left({\Lambda/m_e} \right)\ .
\eeq
In the new effective theory, containing both the electron and the positron, the small electron mass is natural even for large $\Lambda$.

\item[$\triangleright$]  {\it Rho meson.}---The mass difference between the charged pions and the neutral pion receives a quantum correction from photon loops
\beq
m_{\pi^+}^2 - m_{\pi^0}^2 = \frac{3 \alpha}{4\pi} \, \Lambda^2\ , \label{equ:ex2}
\eeq
where $\Lambda$ is the UV cutoff of the effective theory of pions. In order for~(\ref{equ:ex2}) not to exceed the measured mass splitting, $m_{\pi^+}^2 - m_{\pi^0}^2 = (33.5\, {\rm MeV})^2$, we require $\Lambda < 850$~MeV.  New physics in the form of the rho meson with $m_\rho = 770$ MeV comes in at exactly the scale suggested by naturalness. The charged pion mass is natural in the new EFT that includes the rho meson.

\item[$\triangleright$]  {\it Charm quark.}---Historically, one of the most interesting applications of the naturalness principle is $K^0$-$\bar K^0$ mixing.
In an effective theory valid below the kaon mass scale, the mass splitting between the $K_L^0$ and $K_S^0$ states takes the form
\beq
\frac{m_{K_L^0} - m_{K_S^0}}{m_{K_L^0}} = \frac{G_F^2 f_K^2}{6\pi^2} \sin^2 \theta_c\, \Lambda^2\ , \label{equ:ex3}
\eeq
in terms of the cutoff $\Lambda$, the Cabibbo angle $\sin \theta_c \approx 0.22$, and the kaon decay constant $f_K = 114$ MeV.
For~(\ref{equ:ex3}) to be compatible with the measured splitting $(m_{K_L^0} - m_{K_S^0})/{m_{K_L^0}} = 7 \times 10^{-15}$,
we require $\Lambda < 2$ GeV.
In fact, new physics in the form of the charm quark, with $m_c \approx 1.3$ GeV, modifies the UV behavior of the theory.
Gaillard and Lee used this naturalness argument in a successful prediction of the charm quark mass~\cite{Gaillard:1974hs}.
\end{itemize}

\vskip 4pt
\noindent
{\it Failures of naturalness?}---Naturalness arguments are not always applicable, and need to be used with care.

\begin{itemize}

\item[$\triangleright$]  {\it Phase transitions.}---Condensed matter systems near critical points are described by effective theories with fine-tuned parameters, and correspondingly large correlation lengths.  This `unnatural' situation is simply a consequence of explicit fine-tunings performed by the experimentalist.

\item[$\triangleright$]  {\it Nuclear physics.}---An example of fine-tuning in nature occurs in nuclear physics~\cite{Kaplan:1998we} (for reviews see \cite{Bedaque:2002mn, Kaplan:2005es}).
In~\cite{Weinberg:1990rz, Weinberg:1991um, Weinberg:1992yk}, Weinberg observed  that  the scattering lengths  measured in low-energy nucleon-nucleon scattering  are larger than  would be expected from chiral perturbation theory.
 The fundamental scale of the EFT is the Compton wavelength of the pion, $m_\pi^{-1} = (140\hskip 2pt {\rm MeV})^{-1}$, but the
     scattering lengths in the spin singlet state,
     $a_s \approx (8\hskip 2pt {\rm MeV})^{-1}$,
     and in the spin triplet state,
     $a_t \approx (36\hskip 2pt {\rm MeV})^{-1}$, are much larger than  $m_\pi^{-1}$.
     Correspondingly, two neutrons fail to form a bound state by only $60$ keV, and the deuteron binding energy is just $2$ MeV, even though  the natural expectation would involve energies of order $m_\pi =  140$ MeV.  These results can be attributed to approximate cancellations of the kinetic and potential energies of the nucleons, but the underlying reason for these cancellations is poorly understood.

\item[$\triangleright$]  {\it Electroweak scale.}---At the time of writing, experiments at the LHC have discovered the Higgs boson, but have not yet revealed whether the physics determining the hierarchy between the electroweak scale and the Planck scale is natural.
In the Standard Model, the dominant quantum correction to the Higgs mass comes from the coupling to the top quark,
\beq
\Delta m_H^2 \sim \frac{y_t^2}{(4\pi)^2} \hskip 1pt \Lambda^2\ ,
\eeq
where $y_t \sim 1$ is the top Yukawa coupling. For the observed value of the Higgs mass ($m_H = 125$ GeV) to be natural, we require $\Lambda < 1.5$ TeV.
This argument, which suggests that physics beyond the Standard Model should appear at or below the TeV scale,  has had far-reaching impact on decades of work in theoretical and experimental particle physics.
Many physicists have anticipated detecting evidence for a natural mechanism stabilizing the electroweak hierarchy, e.g.~supersymmetric partners of known particles, both in earlier experiments and in the first stages of the LHC.  No such direct evidence has yet materialized, and we must continue to wait for guidance from experiment.
We should emphasize that properly defining and characterizing the predictions of natural mechanisms, such as supersymmetry, is a major area of research, and it is far too soon to conclude that all such mechanisms are excluded.
Due to the close analogy between the electroweak hierarchy problem and the problem of naturalness of the inflaton mass (see \S\ref{ssec:UV}), we may hope that the ultimate resolution of the former will suggest a particular approach to the latter.

\item[$\triangleright$]  {\it Dark energy.}---No discussion of naturalness is complete without addressing the cosmological constant problem.  The vacuum of a
    quantum field theory with local Lorentz invariance corresponds to a stress-energy tensor of the form
\beq
\langle T_{\mu \nu} \rangle = - \rho_{\rm vac} \, g_{\mu \nu}\ .
\eeq
Quantum contributions to the vacuum energy scale as
\beq
\Delta \rho_{\rm vac} \sim \Lambda^4 \ .
\eeq
Naturalness of the observed vacuum energy, $\rho_{\rm vac} \sim (10^{-3} \,{\rm eV})^4$, therefore suggests new physics at $\Lambda \lesssim 10^{-3}$ eV.
Indeed, if the world were supersymmetric down to $10^{-3}$ eV, the small value of the cosmological constant would be natural. But the world is not supersymmetric at low energies, and we have also not seen any other new physics at $10^{-3}$ eV that could account for the smallness of the vacuum energy.\footnote{One or two neutrino masses may have the correct scale, but this has not led to a solution to the cosmological constant problem.}
In the absence of a mechanism explaining the small value of the cosmological constant, we have to entertain the possibility that it is simply a fine-tuned parameter.  Moreover, in the string theory landscape it is conceivable that the observed value is environmentally selected~\cite{Bousso:2000xa}, consistent with anthropic arguments~\cite{Weinberg:1987dv}.
\end{itemize}

It seems clear from these examples that naturalness can at best serve as a tentative guide towards new physics, rather than as a law of nature.

\subsection{Symmetries}
\label{sss:symmetry}

The interplay between symmetry structures in ultraviolet theories and light scalars in effective theories is crucial for understanding inflation in effective field theory and string theory, so we now discuss these issues in more depth.  A pivotal question in inflationary model-building in effective field theory is whether light scalars with $m \ll H$ can be natural (see \S\ref{ssec:UV}).
As we have just explained, whether a given effective theory can be considered natural depends on the properties of the ultraviolet theory, whether known or assumed. Symmetries often play a
central role in the radiative stability of the low-energy theory. In this section, we explain this fact for theories
 in flat space. In the next section, we will discuss some subtle aspects that arise in the generalization to ultraviolet completions that include gravity.

\vskip 4pt
\noindent
{\it SUSY in flat space.}---We have seen that, in the absence of symmetries, scalar masses receive loop corrections of the form
\beq
\Delta m^2 \propto \Lambda^2\ .
\eeq
There are only a few known ways to protect scalars from these effects. One elegant possibility is unbroken supersymmetry (SUSY), which  obliges boson and fermion loops to cancel, so that the scalar mass is not renormalized.   However, as we will see, SUSY is necessarily broken during inflation, generating a mass of the order of the Hubble scale $H$.
Although $m \sim H$ can be significantly smaller than $\Lambda$, it still inhibits successful inflation.
Below we will have more to say about this.

\vskip 4pt
\noindent
{\it Global symmetries in flat space.}---Another possibility to stabilize light scalars is a global  internal symmetry.
As a concrete example,
suppose that the renormalizable part of the EFT, ${\cal L}_{l}[\phi]$, respects the
{\it shift symmetry}
\beq
\phi \mapsto \phi + const. \label{equ:shift}
\eeq
This may arise, for example, if $\phi$ is the Goldstone boson of a spontaneously broken $U(1)$ symmetry (corresponding to the angular flat direction in the  familiar Mexican hat potential). If (\ref{equ:shift}) is exact, it forbids the mass term, or any potential terms for that matter. To get nontrivial dynamics, we are usually interested in the case where the shift symmetry (\ref{equ:shift}) is only {\it approximate}. Concretely, we imagine that the symmetry is broken by a small mass term,  $\Delta V = \frac{1}{2} m^2 \phi^2$, with $m \ll \Lambda$.
Loop corrections to the tree-level mass must then scale with the symmetry breaking parameter (which is $m$), so that
\beq
\Delta m^2 \propto m^2 \ .
\eeq
At most, the correction can now scale logarithmically with the cutoff $\Lambda$.
Moreover, in the limit $m \to 0$, the exact symmetry (\ref{equ:shift}) is restored and $\phi$ becomes massless.  A small mass for $\phi$ is therefore {\it technically natural}~\cite{tHooft:1979bh}: the smallness of the symmetry-breaking parameter controls the renormalization.
At the level of model building, one is free to set the mass at any desired level  without risking destabilization through quantum effects.
On the other hand, it still makes sense to ask  whether the fact that the symmetry is weakly broken in the first place  is  dictated by some mechanism
and is natural in the top-down sense.  In principle, a symmetry can be broken explicitly by an operator whose  coefficient  is small purely because of fine-tuning,  and the resulting small parameters are technically natural but not top-down natural.

\vskip 4pt
\noindent
{\it Ultraviolet completion.}---Exact or approximate symmetries of the UV theory can control the sizes of the Wilson coefficients in the non-renormalizable part of the effective Lagrangian.
If the symmetry is weakly broken by the heavy degrees of freedom, or if the light fields  couple only weakly  to the symmetry-breaking terms, then the EFT  enjoys an approximate symmetry, and
the Wilson coefficients of all symmetry-breaking operators will be naturally small.
This can be seen in our toy model (\ref{equ:toy}): the coupling $g \hskip 1pt \phi^2 \Psi^2$ breaks the shift symmetry in the UV.
In the EFT, this breaking shows up through symmetry-breaking irrelevant operators.
Since the symmetry is restored in the limit $g \to 0$, the Wilson coefficients of all symmetry-breaking operators in (\ref{equ:Leff2}) must satisfy
\beq
\lim_{g \to 0} c_i(g) = 0 \ .
\eeq
For finite $g$, the $c_i$ are proportional to
positive powers of
the symmetry-breaking parameter $g$.
An approximate symmetry in the UV would explain $g \ll 1$ and hence $c_i \ll 1$ in the EFT.
We emphasize that assuming $c_i \ll 1$ in the effective theory amounts to assuming something about the couplings to the degrees of freedom at the cutoff scale $\Lambda$.
Whether a given low-energy theory is deemed natural can hinge on which symmetries are thought to be permissible in the UV completion. Consulting a UV-complete theory like string theory can be valuable when general reasoning about what ought to be typical does not give a sharp answer.

\subsection{Gravity}
\label{ssec:gravity}

Gravity plays a fundamental role in any description of cosmology, so our effective theory must include gravitational degrees of freedom.
Moreover, semi-classical gravity itself has a limited range of validity. At or below the Planck scale, graviton-graviton scattering violates perturbative unitarity, and we expect new degrees of freedom to become relevant. In this section, we discuss to what extent the UV completion of gravity can affect the matter interactions in the low-energy effective theory.

\vskip 4pt
\noindent
{\it Gravity as an effective theory.}---The low-energy degree of freedom of gravity is the spacetime metric $g_{\mu \nu}$, whose leading interactions are determined by the Einstein-Hilbert action,
\beq
S_{\rm EH} =  \frac{M_{\rm pl}^2}{2}  \int \d^4 x \sqrt{-g} \,R  \ .
\eeq
This theory is non-renormalizable and should be understood as an effective theory~\cite{Donoghue:1995cz, Burgess:2003jk}.
To see this, let us expand the metric in terms of small perturbations around flat space,
$g_{\mu \nu} \equiv \eta_{\mu \nu} + \frac{1}{M_{\rm pl}} h_{\mu \nu}$. Schematically, the Einstein-Hilbert action then becomes
\beq
S_{\rm EH} = \int \d^4 x \left[ (\partial h)^2 + \frac{1}{M_{\rm pl}} h (\partial h)^2 + \frac{1}{M_{\rm pl}^2} h^2 (\partial h)^2  + \cdots\right] \ .
\eeq
This weak-field expansion of the Einstein-Hilbert action looks similar to the action of Yang-Mills theory,
\beq
S_{\rm YM} = \int \d^4 x  \left[ (\partial A )^2 + g A^2 \partial A + g^2 A^4\right]\ .
\eeq
However, while the Yang-Mills action terminates at a finite order, the expansion of the Einstein-Hilbert action contains an infinite number of terms, coming from the expansion of $\sqrt{-g} $ and $g^{\mu \nu}$.
Gravity is therefore interpreted as an effective quantum field theory with cutoff $\Lambda = M_{\rm pl}$.
The quantum perturbation theory of gravitons is organized in terms of the dimensionless ratio $(E/M_{\rm pl})^2$, where $E$ is the energy of the process, and
this theory
breaks down at the Planck scale. At this point either new degrees of freedom become important (like the massive excitations in string theory; see \S\ref{ssec:StringTheory}) or a nonperturbative miracle happens (as in asymptotic safety~\cite{Zichichi:1978gb}).
In the absence of detailed information about the UV completion, the simplest assumption is that the low-energy effective action contains all terms that are consistent with general coordinate invariance. We can organize this as a derivative expansion,\footnote{There is no $R_{\mu \nu \sigma \rho} R^{\mu \nu \sigma \rho}$ term, since by the generalized Gauss-Bonnet theorem $R^2 - 4 R_{\mu \nu} R^{\mu \nu} + R_{\mu \nu \sigma \rho} R^{\mu \nu \sigma \rho}$
is topological, and equals the Euler characteristic  of the spacetime.}
\begin{align}
\label{gravityexpansion}
S_g &=   \int \d^4 x \sqrt{-g}  \left[M_\Lambda^4 + \frac{M_{\rm pl}^2}{2}  R + c_1 {R^2} + c_2 {R_{\mu \nu} R^{\mu \nu}}\right. \nonumber \\& \left. \hspace{4.7cm}+ \frac{1}{M^2} (d_1 R^3 + \cdots) + \cdots \right]\ ,
\end{align}
where $c_i$ and $d_i$ are dimensionless numbers that may be expected to be of order unity.
For pure gravity, the scale $M$ should also be the Planck scale~$M_{\rm pl}$. However, couplings to matter fields might lead to a hierarchy between $M$ and $M_{\rm pl}$. Note also that the renormalized value of the cosmological constant $M_\Lambda$
deduced from cosmological experiments is  extremely
far from its natural value~$M_{\rm pl}$.  This is, of course, the famous cosmological constant problem.

In many string theories, the higher-curvature terms in (\ref{gravityexpansion}) can be computed order by order in the $\alpha^{\prime}$ and $g_{\rm s}$ expansions detailed in \S\ref{ssec:StringTheory}.  An important example is type IIB string theory in ten dimensions, where
one finds\footnote{The ten-dimensional cosmological constant is set to zero in (\ref{IIBgravityexpansion}) because we are displaying the supersymmetric effective action, but the four-dimensional cosmological constant  arising upon compactification and supersymmetry breaking  is subject to the usual cosmological constant problem.} \cite{Gross:1986iv}
\beq \label{IIBgravityexpansion}
S_g =   \int \d^{10} X \sqrt{-G}  \left[\frac{M_{10}^2}{2}  R +  \frac{\zeta(3)}{3\cdot 2^{5}} \frac{1}{M^6} {\cal{R}}^4 + \cdots \right]\ ,
\eeq
where $M_{10}$ is the ten-dimensional Planck mass, $\zeta$ denotes the Riemann zeta function, with $\zeta(3) \approx 1.202$; $ {\cal{R}}^4$ is a particular quartic invariant constructed from the Riemann tensor; and notably the omitted terms include additional contributions at ${\cal O}(1/M^6)$, and at higher orders, as well as terms that are subleading in the string coupling $g_{\rm s}$.
The mass $M$  appearing in (\ref{IIBgravityexpansion})  corresponds to the mass of the first excited level of the type II  superstring, given by
\begin{equation}
M^2 = \frac{4}{\alpha^{\prime}}\ ,
\end{equation}  where  $\alpha^{\prime}$ is
the inverse string tension defined in \S\ref{ssec:StringTheory}.
This is the proper physical cutoff scale because the higher-derivative  term in (\ref{IIBgravityexpansion})  arises upon integrating out the massive excitations of the string, which have  the mass spectrum $m^2=4N/\alpha^\prime$, $N \in \mathbb{Z}$.   In the  regime of weak coupling and weak curvature where the  corrections to the  Einstein-Hilbert action are small,  the string scale $M$  is small compared to the  Planck scale $M_{\rm pl}$, so the  higher-derivative  contributions shown above are more significant in string theory than general reasoning about quantum gravity would suggest.  On the other hand, the coefficient $\zeta(3)/(3\cdot 2^{5})\approx   10^{-2}$ is rather small, illustrating that  the general expectation of order-unity Wilson coefficients  should not be viewed as a precise and immutable law.

\vskip 4pt
\noindent
{\it Global symmetries in quantum gravity.}---A number of `folk theorems'  state that exact continuous global symmetries are impossible in a theory of quantum gravity. Instead, any continuous global symmetry must be merely an accidental symmetry of the low-energy effective theory, broken by irrelevant operators at a scale no larger than $M_{\rm pl}$.
We will briefly recall some of the arguments; see \cite{Banks:2010zn} for a modern discussion of related  issues.

The simplest argument against global symmetries in quantum gravity is that Hawking evaporation of a black hole can destroy global charges.
For example, imagine tossing a substantial clump of baryons into a  macroscopic black hole,  assumed to be large enough so that  protons and neutrons make up an arbitrarily small fraction of the Hawking quanta.   The black hole  will lose  most of its mass to light  quanta  that carry zero baryon number,  and by the time that the Hawking temperature  is high enough for baryons to be  emitted,  the black hole mass  will be smaller than the mass of the initial clump of baryons.   Unless the theory contains states  with an  arbitrarily high ratio of baryon number to mass,  a sufficiently large black hole will be unable to radiate away all of its initial baryon number in Hawking quanta, or to  deposit this baryon number in a highly charged remnant.
Baryon number is therefore violated in the black hole evaporation process and cannot be an exact symmetry.
The same applies to any continuous global internal symmetry with a well-defined conserved charge.\footnote{Shift symmetries are an important example where the absence of a conserved charge requires a refinement of the black hole arguments.}
Another class of arguments  appeals to the destruction of global charges by wormholes~\cite{Abbott:1989jw,Coleman:1989zu,Kallosh:1995hi}.

We should be clear that although these arguments show that global symmetries are broken by Planck-scale effects, they do not show that the breaking is necessarily of order unity.
Indeed, it is conceivable the Wilson coefficients for the symmetry-breaking higher-dimension operators might be  suppressed for some reason.  We know too little about the degrees of freedom at the Planck scale to make definitive statements about the strength of the symmetry breaking,
though in concrete examples in string theory it is often possible to compute the symmetry-breaking effects.\footnote{For example, in \S\ref{sec:AxionInflation}, we will study axions in string theory.
An axion $\phi$ with infinite periodicity ($f/M_{\rm pl} \to \infty$)  enjoys the exact  shift symmetry $\phi \mapsto \phi + const.$, so the above arguments  suggest that such axions are not possible in quantum gravity. And indeed, direct searches for axions with $f \gtrsim M_{\rm pl}$ in  parametrically controlled string compactifications  have been unsuccessful \cite{Banks:2003sx, Svrcek:2006yi}.
A different argument against axions with $f \gg M_{\rm pl}$ appears in \cite{Conlon:2012tz}.}

\vskip 4pt
\noindent
{\it Global symmetries in string theory.}---The absence of exact continuous global internal\footnote{Exact Lorentz symmetry is possible in string theory, cf.~\cite{Banks:1988yz}.} symmetries is also a theorem in perturbative string theory~\cite{Banks:1988yz}.
Suppose that there is an exactly conserved global symmetry of the conformal field theory on the string worldsheet, so that by Noether's theorem there is a corresponding conserved current on the worldsheet.
This current can be used to construct a vertex operator that corresponds to the emission of a massless excitation of the string, which turns out to be nothing other than a gauge boson associated to the symmetry \cite{Banks:1988yz}.  Thus, the postulated symmetry must be a gauge symmetry in the target spacetime.

\vskip 4pt
We conclude that general arguments in quantum gravity, and specific findings in string theory, limit the sorts of global symmetries that are allowed in an ultraviolet completion.
Asserting a symmetry structure for the UV theory and taking natural coefficients for the operators in the resulting effective action may lead to an effective theory that is consistent at low energies but cannot be embedded in a theory of quantum gravity.  Because constraints from quantum gravity can play a critical role in determining the effective action, we view it as prudent to examine any postulated symmetry structure in a theory of quantum gravity.

\vskip 4pt
\noindent
{\it Coupling quantum field theory to gravity.}---Thus far we have  discussed  flat-space quantum field theories,  as well as purely gravitational theories, but  the  theories of interest in cosmology are quantum field theories coupled to gravity.
Let us illustrate this by coupling the  toy model of (\ref{equ:Leff2})  to a gravitational theory with higher-curvature corrections.  The resulting effective theory  takes the form
\beq
S_{\rm eff}[\phi,g] = S_g + S_{\rm eff}[\phi] + S_{g,\phi}\ ,
\eeq
where $S_g$ is given in (\ref{gravityexpansion}), $S_{\rm eff}[\phi]$  is the action corresponding to  the Lagrangian density (\ref{equ:Leff2}), and
\beq
S_{g,\phi} = \int \d^4 x \sqrt{-g} \left[ \sum_{i} c_{i} \frac{{\cal O}_i[g,\phi]}{\Lambda^{\delta_i -4}} \, \right]\ . \label{equ:LeffphiG}
\eeq Here, ${\cal O}_i[g,\phi]$ are operators constructed from curvature invariants and from $\phi$ and  its derivatives.
In spacetimes  where the curvature  is small in units of the cutoff $\Lambda$,  the only important coupling in $S_{g,\phi}$ is
\beq
S^{(4)}_{g,\phi} = \int \d^4 x \sqrt{-g} \, \xi \, \phi^2 R \ , \label{equ:LeffphiR}
\eeq  where $\xi$ is a dimensionless coefficient.   One can perform a Weyl  rescaling of the metric,
\beq \label{equ:WeylR}
g_{\mu\nu} \mapsto \bar{g}_{\mu\nu} \equiv e^{2\omega(\phi)}g_{\mu\nu}\ ,
\eeq
so that by a suitable choice of the function $2\omega(\phi)$, one arrives at $\xi=0$,  known as {\it{minimal coupling}}.  However, the rescaling (\ref{equ:WeylR}) also changes any other terms in the full action that are not conformally invariant.

\subsection{Time-Dependence}
\label{ssec:Time}

To complete our survey of the basic principles of effective field theory, we need to discuss if and how effective field theory applies to time-dependent settings such as those arising in cosmology.
(For further discussion, see~\cite{Burgess:2003jk, Collins:2012nq, Burgess:2014lwa}.)

\vskip 4pt
An immediate concern might be that we have classified heavy and light states relative to a cutoff {\it energy}, but energy conservation is inapplicable in time-dependent backgrounds, and when the background evolution is rapid,
high-energy modes can be produced out of low-energy modes.
Fortunately, in many cosmological applications the time-dependence of the background is sufficiently slow to be treated {\it adiabatically}.  We can then define an adiabatic notion of energy at a given time,
and define the split into light and heavy fields relative to a slowly evolving cutoff $\Lambda(t)$.
When the background evolution is sufficiently  rapid to allow the production of heavy states,
$|\dot \Lambda|/\Lambda^2 \gg 1$, then the system may not admit a description in terms of an effective theory  containing only the light fields: the solutions to the equations of motion of the EFT will contain only the adiabatic solutions of the full theory.
In inflation, the adiabatic approximation is justified as long as {\it i}) we start in the {\it Bunch-Davies vacuum} (also called the adiabatic vacuum) and {\it ii}) the subsequent evolution is adiabatic.
This situation applies to a very broad range of inflationary models, but there are interesting exceptions: for a recent discussion of this issue see \cite{Avgoustidis:2012yc, Burgess:2012dz}.

Even for slowly-evolving backgrounds, there may be {\it level crossings}: the slow evolution of the cutoff, $\Lambda(t)$, may cause some light fields to leave the EFT, and/or may draw in light fields that were previously heavy enough to integrate out. Thus, one effective theory evolves into another over time. We will encounter this possibility in 
large-field inflation (see \S\ref{ssec:UV}).

\vskip 4pt
In summary, the methods of effective field theory are applicable in backgrounds whose
time evolution is sufficiently adiabatic, provided also that the initial state is the Bunch-Davies vacuum.   In this setting we can  focus on the evolution of low-energy states, without having to worry about the production of high-energy states.
In the rest of this work, we will mostly consider adiabatic evolution, and any violations of adiabaticity will be noted.

\section{Effective Theories of Inflation}
\label{sec:EFT-Inflation}

In Chapter~\ref{sec:dS}, we defined inflation as an extended period of quasi-de Sitter evolution, with $-\dot H \ll H^2$,
but we did not specify the physical origin of the inflationary background $H(t)$.  In this section, we will
show that the dynamics of a slowly rolling scalar field leads to inflation.
However, at this level, models of slow-roll inflation are toy models that lack a clear connection to the rest of physics.  To make the models more realistic, we will embed them into low-energy effective theories, allowing
us to discuss high-scale corrections to the slow-roll actions.  A striking feature is that the inflationary dynamics is sensitive even to Planck-suppressed corrections, as
we will explain in \S\ref{ssec:UV}.

\subsection{Slow-Roll: Dynamics and Perturbations}
\label{sec:SR}

\begin{figure}[htbp]
    \centering
        \includegraphics[width=0.6 \textwidth]{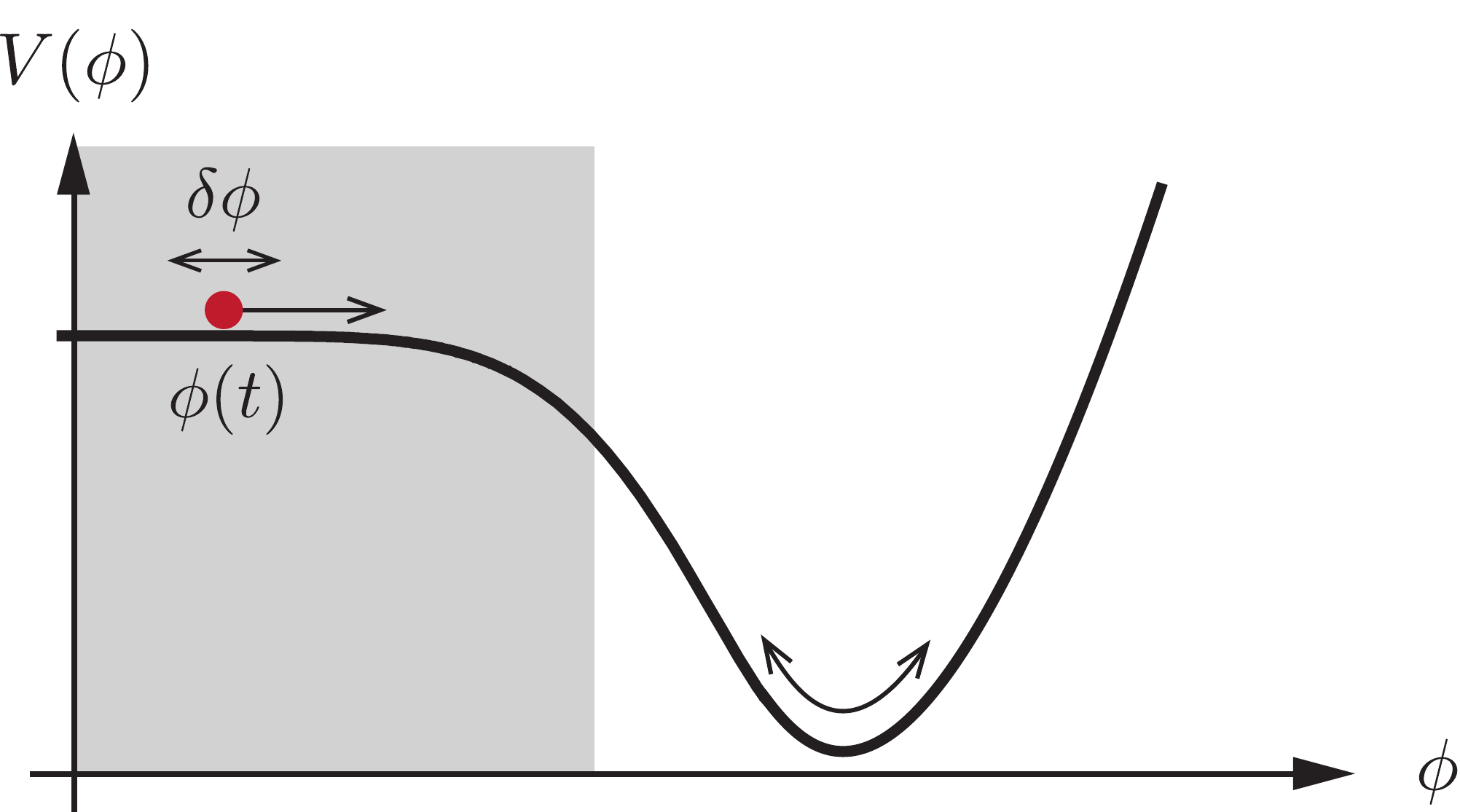}
\caption{Example of a slow-roll potential.  Inflation occurs in the shaded part of the potential.
In addition to the homogeneous evolution $\phi(t)$, the inflaton experiences spatially-varying quantum fluctuations $\delta \phi(t,\x)$.
}   \label{fig:SR}
\end{figure}

One of the earliest  and most influential
models of inflation uses a single scalar field, the {\it inflaton}~$\phi$, minimally coupled to
 gravity~\cite{Linde:1981mu, Albrecht:1982wi},
\beq
S =  \int \d^4 x \sqrt{-g} \, \left[  \frac{M_{\rm pl}^2}{2}R - \frac{1}{2} (\partial \phi)^2 - V(\phi) \right]\ , \label{equ:Vminimal}
\eeq
where we have allowed for an arbitrary inflaton potential $V(\phi)$ (see fig.~\ref{fig:SR}).

\vskip 4pt
\noindent
{\it Classical dynamics.}---The Friedmann equation and the Klein-Gordon equation for the homogeneous background field $\phi(t)$ are
\beq
3M_{\rm pl}^2 H^2 = \frac{1}{2} \dot{\phi}^2 + V \qquad {\rm and} \qquad \ddot{\phi} + 3 H \dot{\phi} = - V'\ , \label{equ:Friedmann}
\eeq
where $V' \equiv \partial_\phi V$.
These equations can be combined into
\beq
\varepsilon = - \frac{\dot H}{H^2} = \frac{\frac{1}{2} \dot{\phi}^2}{M_{\rm pl}^2 H^2}\ . \label{equ:varE}
\eeq
Inflation ($\varepsilon < 1$) therefore occurs when the potential energy of the field dominates over the kinetic energy, $V \gg \frac{1}{2} \dot{\phi}^2 $.
The kinetic energy stays small and slow-roll persists if
the acceleration of the field is small, $|\ddot{\phi}| \ll 3 H |\dot{\phi}|$.

The conditions for
prolonged
slow-roll inflation can
 be expressed as conditions on the shape of the potential~\cite{Steinhardt:1984jj}:
\beq
\epsilon \equiv \frac{M_{\rm pl}^2}{2} \left( \frac{V'}{V}\right)^2 \ll 1 \quad , \quad |\eta| \equiv M_{\rm pl}^2 \frac{|V''|}{V} \ll 1\ . \label{equ:eta}
\eeq
During a slow-roll period, the `potential slow-roll parameters' $\epsilon$ and $\eta$ are related to the `Hubble slow-roll parameters' $\varepsilon$ and $\tilde \eta$ (see \S\ref{sec:zeta}) via $\epsilon \approx \varepsilon$ and $\eta  \approx \tilde \eta + \varepsilon$.
We will see that realizing the slow-roll conditions
(\ref{equ:eta}) in a theory of fundamental physics is a nontrivial task.
In particular,
$|\eta| \ll 1$ requires a small hierarchy between the inflaton mass and the Hubble scale, $m^2 = V'' \ll 3 H^2 \approx V/\Mp^2$.  Explaining the small inflaton mass is one of the key challenges for any microscopic theory of inflation.

\vskip 4pt
\noindent
{\it Quantum fluctuations.}---As we explained in Chapter~\ref{sec:dS}, light fields experience quantum fluctuations during inflation.
As a result of fluctuations in the inflaton, $\delta \phi(t, {\x})$,
some regions
of space
remain potential-dominated longer than others, and different parts of the universe undergo slightly different evolution. After inflation, 
these differences
 in the evolution 
induce 
curvature perturbations $\R(t, {\x})$, 
which lead to density perturbations $\delta \rho(t,\x)$.
 
We note that the inflaton fluctuation $\delta \phi$ plays the role of the Goldstone boson $\pi$ of broken time translations (see \S\ref{sec:zeta}). In spatially flat gauge the two are simply proportional,
 \beq
\pi = \frac{\delta \phi}{\dot \phi}\ .
\eeq
Using (\ref{equ:ZetaPi}), we then
find
\beq
\R(t,\x) =  - H \pi(t,\x) = - \frac{H}{\dot{\phi}} \hskip 1pt \delta \phi(t,\x)\ .
\eeq
During the slow-roll period, the sound speed of the Goldstone boson is equal to the speed of light, $c_s = 1$ (see Appendix~B).
The analysis of \S\ref{sec:zeta} then implies the following results for the
spectra of scalar and tensor fluctuations:
\beq
\Delta_\R^2 =  \frac{1}{24 \pi^2} \frac{1}{\epsilon} \frac{V}{M_{\rm pl}^4} \quad ,  \quad \Delta_h^2 = \frac{2}{3\pi^2} \frac{V}{M_{\rm pl}^4}\ .  \label{naivespectrum}
\eeq
The scalar spectral index and the tensor-to-scalar ratio are
\begin{align}
n_s - 1 &\, =\, 2 \hskip 1pt \eta - 6 \hskip 1pt \epsilon \\
r &\,=\, 16 \hskip 1pt \epsilon \ .     \label{naivetilt}
\end{align}
These observables should be evaluated at the time when the {\it pivot scale}\footnote{In the WMAP analysis the pivot scale was chosen to be $k_\star =0.002$ Mpc${}^{-1}$, while for Planck $k_\star =0.05$ Mpc${}^{-1}$.}---a representative scale among the scales probed by the CMB---exited the horizon.
This moment corresponds to a specific point in field space, $\phi_\star$, at which the number of $e$-folds of inflation remaining is
(for $\phi_\star > \phi_{\rm end}$)
\beq
 N_\star \, =\, \int_{\phi_{\rm end}}^{\phi_\star} \frac{\d \phi}{M_{\rm pl}} \frac{1}{\sqrt{2 \epsilon}} \ .  \label{nstaris}
\eeq
The value of $N_\star$ depends on the inflationary model and
on the details of reheating.
Typically, one finds $40 \lesssim N_\star \lesssim 60$.
In the next subsection, we give a few examples of specific slow-roll models and their observational predictions.

\subsection{Slow-Roll: Selected Models}
\label{sec:SRM}

\vspace{0.3cm}
\begin{quote}
{\footnotesize You know how sometimes you meet somebody and they're really nice, so you invite them over to your house and you keep talking with them and they keep telling you more and more cool stuff? But then at some point you're like, maybe we should call it a day, but they just won't leave and they keep talking and as more stuff comes up it becomes more and more disturbing and you're like, just stop already? That's kind of what happened with inflation.} \vskip 0.1pt \hfill
{\footnotesize Max Tegmark~\cite{Tegmark}.}
\end{quote}
\vspace{0.3cm}

We  will not provide a comprehensive
account
of the vast landscape of slow-roll models, but instead give a brief sketch of some of the most important classes of models.  For more details on slow-roll model-building we refer the reader to~\cite{Lyth:1998xn, Lyth:2009zz, Lyth:2007qh, Martin:2013tda}.

\begin{figure}[h!]
    \centering
        \includegraphics[width=0.9 \textwidth]{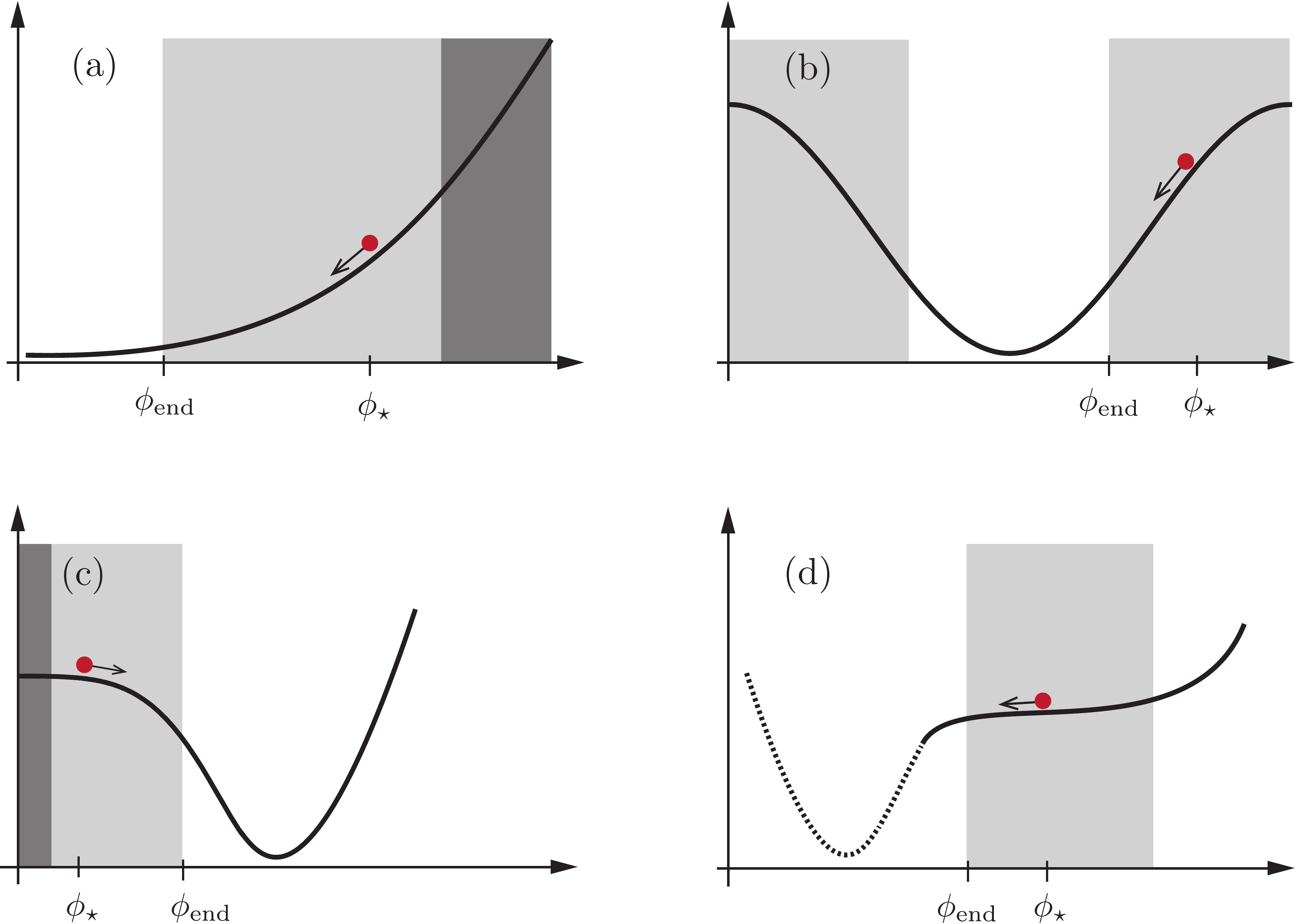}
\caption{Examples of different classes of slow-roll potentials:  (a) chaotic inflation,  (b) natural inflation, (c) hilltop inflation  and  (d) inflection point inflation. The light grey regions indication the parts of the potential where slow-roll inflation occurs. The dark grey regions denote regions of {\it eternal inflation}. The figures are {\it not} drawn to scale: (a)+(b) correspond to large-field models ($\Delta \phi > M_{\rm pl}$), while (c)+(d) are small-field models ($\Delta \phi < M_{\rm pl}$).}   \label{fig:SR-examples}
\end{figure}

\vskip 4pt
\noindent
{\it Chaotic inflation.}---An important  class of inflationary models arises when the potential is a simple monomial,
\beq
V(\phi) = \mu^{4-p} \phi^p\ ,
\eeq where $p>0$, and $\mu$ is a parameter with the dimensions of mass.  For historical reasons, such models are called {\it chaotic inflation}~\cite{Linde:1983gd}.

The slow-roll parameters in chaotic inflation are
\begin{align}
\epsilon = \frac{p^2}{2} \left(\frac{M_{\rm pl}}{\phi}\right)^2 \quad , \quad \eta = p(p-1)\left(\frac{M_{\rm pl}}{\phi}\right)^2 \ . \label{equ:chaoticepseta}
\end{align}
Notice that $\epsilon$ and $\eta$ do not depend on the scale $\mu$.
Using (\ref{nstaris}), the number of $e$-folds occurring  in the region $\phi \le \phi_{\star}$  is  found to be
\begin{equation}
N_\star \approx \frac{1}{2p}\left(\frac{\phi_{\star}}{M_{\rm pl}}\right)^2 \ , \label{nstarchaotic}
\end{equation} implying that in a model of chaotic inflation, prolonged inflationary  expansion requires a super-Planckian displacement, $\phi_{\star} \gg M_{\rm pl}$.
At the pivot scale,
the spectral index and the tensor-to-scalar ratio are
\begin{align}
n_s - 1 &= - \frac{(2+p)}{2N_\star} \quad , \quad r = \frac{4p}{N_\star} \ . \label{equ:chaotic}
\end{align}
Let us illustrate these results in a few simple cases, setting $N_\star=60$ for definiteness:
\begin{align}
p=1: \qquad n_s &\approx 0.975\ , \qquad  r \approx 0.07\ , \qquad \phi_{\star} \approx 11 M_{\rm pl} \ , \\[4pt]
p=2: \qquad n_s &\approx 0.967\ , \qquad  r \approx 0.13\ , \qquad \phi_{\star} \approx 15 M_{\rm pl} \ , \\[4pt]
p=3: \qquad n_s &\approx 0.958\ , \qquad  r \approx 0.20\ , \qquad \phi_{\star} \approx 19 M_{\rm pl} \ , \\[4pt]
p=4: \qquad n_s &\approx 0.950\ , \qquad  r \approx 0.27\ , \qquad \phi_{\star} \approx 22 M_{\rm pl} \  .
\end{align}
An approximate shift symmetry can make a  model of chaotic inflation bottom-up natural 
(see \S\ref{ssec:gwaves}), but  to establish top-down naturalness a realization in string theory is  necessary.  
Attempts to embed chaotic inflation in string theory are described in \S\ref{sec:AxionInflation}.

\vskip 4pt
\noindent
{\it Natural inflation.}---An  influential idea in  inflationary model-building  is that the inflaton could be a pseudoscalar {\it{axion}}.
At the perturbative level, an axion enjoys a continuous shift symmetry, but this is broken nonperturbatively  to a discrete symmetry,  leading to 
a potential  of the form
\beq
V(\phi) = \frac{V_0}{2}  \left[ 1 - \cos\left( \frac{\phi}{f}\right) \right] \ ,  \label{naturalqft}
\eeq
where $f$ is the axion decay constant.
For $f \gtrsim 4 M_{\rm pl}$, the potential (\ref{naturalqft})  supports {\it natural inflation}~\cite{Freese:1990rb}. Due to the shift symmetry, the model is bottom-up natural. Establishing top-down naturalness requires finding axions in string theory whose effective decay constant can be larger than the Planck scale --- see \S\ref{sec:AxionInflation}.
At the pivot scale, one finds the following expressions for the scalar tilt and the tensor-to-scalar ratio~\cite{Lyth:2009zz}:
\begin{align}
n_s -1 &\,=\, - \alpha\, \frac{e^{N_\star \alpha} +1}{e^{N_\star \alpha} - 1}  \ \xrightarrow{\ \alpha \ll 1\ }\ - \frac{2}{N_\star} \ ,\\
 r &\,=\, 8\alpha\, \frac{1}{ e^{N_\star \alpha} -1} \ \,\, \xrightarrow{\ \alpha \ll 1\ }\ + \frac{8}{N_\star} \ ,
\end{align}
where we have defined $\alpha \equiv M_{\rm pl}^2/f^2$. As expected, the predictions for natural inflation reduce to those of $m^2 \phi^2$ chaotic inflation for $f \gg M_{\rm pl}$: 
cf.~eq.~(\ref{equ:chaotic}) with $p=2$.

\vskip 4pt
\noindent
{\it Hilltop inflation.}---Consider the situation where inflation occurs near the fixed point of a symmetry, that is at a point in field space with $V'_0=0$.  Expanding the potential around this point gives \begin{align}
V(\phi) &= V_0 + \frac{1}{2} m^2 \phi^2 + \cdots \ . \end{align}
For positive $m^2$, the symmetry is intact, while for negative $m^2$ the symmetry 
get spontaneously broken.
 Consider the latter case and write the potential as
\begin{align}
V(\phi)  &=  V_0 \left[ 1 + \frac{1}{2} \eta_0 \frac{\phi^2}{M_{\rm pl}^2} + \cdots \right]\ , \qquad {\rm where} \quad \eta \approx \eta_0 < 0\ . \label{equ:hill}
\end{align}
For small $\eta_0$, {\it hilltop inflation}~\cite{Boubekeur:2005zm} occurs (see fig.~\ref{fig:SR-examples}c).
The higher-order terms in (\ref{equ:hill}) become important for large values of~$\phi$. They define the precise value $\phi_{\rm end}$ at which inflation ends, and determine the value of the cosmological constant in the global vacuum after inflation. We will assume that $\phi_{\rm end} \lesssim M_{\rm pl}$.
If the higher-order terms in (\ref{equ:hill}) are irrelevant when the pivot scale exits the horizon, then
the spectral 
tilt and the tensor-to-scalar ratio are~\cite{Boubekeur:2005zm, Alabidi:2005qi}
\begin{align}
n_s - 1 &\, =\, 2 \hskip 1pt \eta_{0}\ ,\\
 r &\,=\, 2(1-n_s)^2 e^{-N_\star(1-n_s)} \, \left(\frac{\phi_{\rm end}}{M_{\rm pl}}\right)^2 \approx 10^{-3} \, \left(\frac{\phi_{\rm end}}{M_{\rm pl}}\right)^2 \ .
\end{align}
The model has two free parameters: the curvature of the hilltop, $\eta_0$, and the field value at the end of inflation, $\phi_{\rm end}$.

\begin{figure}[h!]
   \centering
     \includegraphics[scale=0.6]{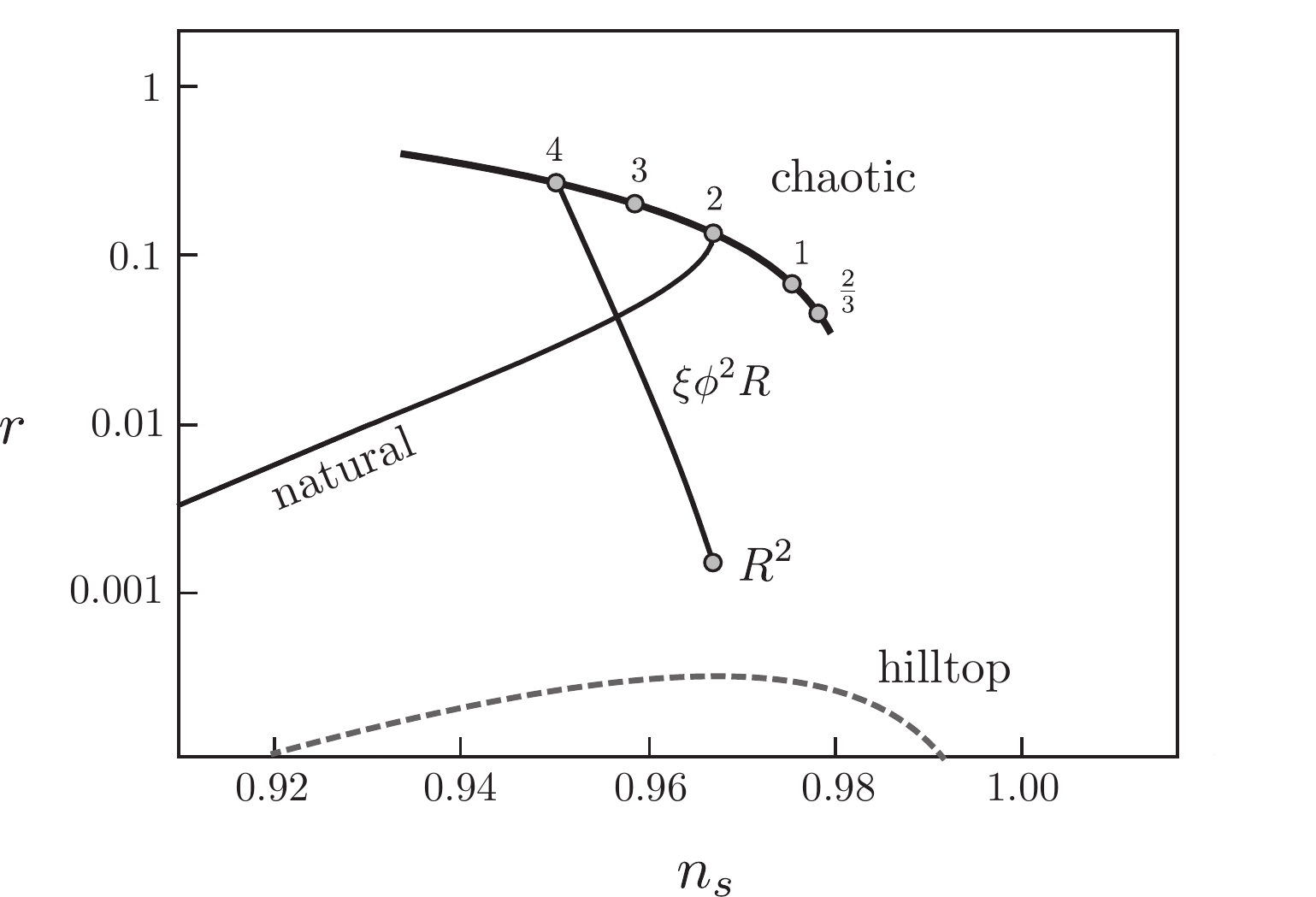}
   \caption{Some slow-roll predictions in the $n_s$-$r$ plane, assuming 60 $e$-folds of inflation.}
  \label{fig:nsrPredictions}
\end{figure}

\vskip 4pt
\noindent
{\it Inflection point inflation.}---Away from any symmetry points, a generic potential has the  expansion,
around $\phi=0$,
\beq
V(\phi) =  V_0 \left[ 1 + \lambda_0 \frac{\phi}{M_{\rm pl}} +  \frac{1}{2} \eta_0 \frac{\phi^2}{M_{\rm pl}^2} + \frac{1}{3!} \mu_0 \frac{\phi^3}{M_{\rm pl}^3} + \cdots \right] \ .
\eeq
Again, higher-order terms may become important towards the end of inflation, but are assumed to be irrelevant when the pivot scale exits the horizon.
To get enough $e$-folds of inflation, we require $|\eta_0| \ll 1$.
The special case $\eta_0 = 0$ corresponds to {\it inflection point inflation}~ ($V''_0 = 0$, see fig.~\ref{fig:SR-examples}d):
\beq
V(\phi) \approx  V_0 \left[ 1 + \lambda_0 \frac{\phi}{M_{\rm pl}} + \frac{1}{3!}\mu_0 \frac{\phi^3}{M_{\rm pl}^3} + \cdots \right] \ . \label{equ:INFLE}
\eeq
This type of potential arises in D-brane inflation~\cite{Baumann:2007np, Baumann:2007ah}  (see \S\ref{sec:dbrane}).
The spectral tilt 
derived from the potential (\ref{equ:INFLE}) is~\cite{BuenoSanchez:2006xk,  Baumann:2007ah}
\beq
n_s - 1 = - 4 \sqrt{\frac{\lambda_0 \mu_0}{2}} \cot \left( N_\star \sqrt{\frac{\lambda_0 \mu_0}{2}}\, \right) \ ,
\eeq
and
is uncorrelated with the value of the tensor-to-scalar ratio:
\beq
r = 16 \lambda_0^2\ .
\eeq
Constraints on the total number of $e$-folds and the scalar amplitude typically force $\lambda_0$ to be very small and the tensor signal to be unobservable, $r \ll 0.01$.

\begin{figure}[htbp]
    \centering
        \includegraphics[width=0.75 \textwidth]{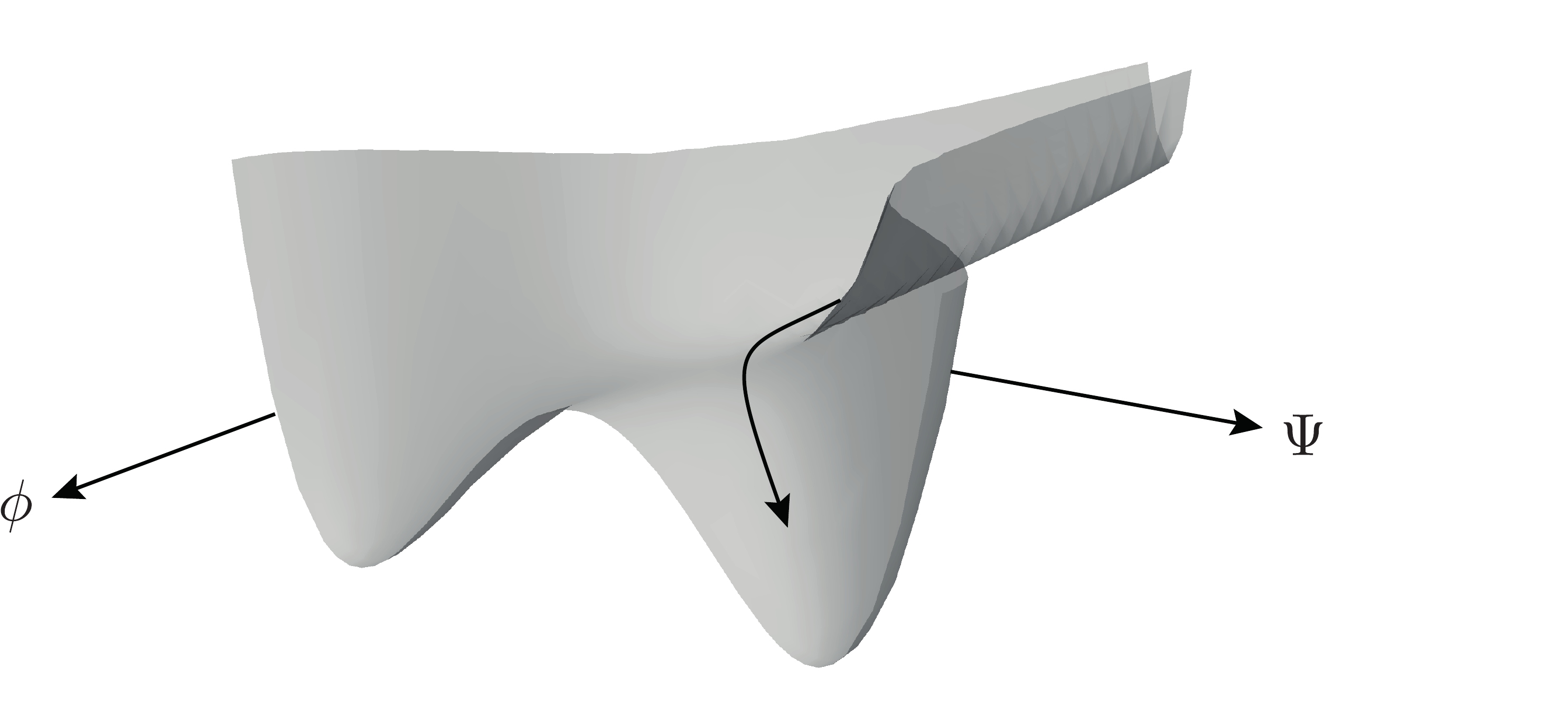}
\caption{A typical potential of hybrid inflation. A tachyonic instability ends inflation while the slow-roll parameter $\epsilon$ is still small.}   \label{fig:hybrid}
\end{figure}

\vskip 4pt
\noindent
{\it Hybrid inflation.}---Inflationary models with small-field potentials, such as (c) and (d) in fig.~\ref{fig:SR-examples}, often end through an instability induced by coupling the inflaton field $\phi$ to an additional `waterfall' field~$\Psi$.  The combination of a slow-roll potential and a waterfall instability is called {\it hybrid inflation}~\cite{Linde:1993cn}.
As a simple example, consider the 
two-field potential (see fig.~\ref{fig:hybrid})
\beq
V(\phi,\Psi) =  V(\phi) + V(\Psi) + \frac{1}{2} g\hskip 1pt \phi^2 \Psi^2\ ,
\eeq
where $V(\phi)$ is the slow-roll potential and $V(\Psi)$ is a potential of symmetry-breaking type,
\beq
V(\Psi) \equiv \frac{1}{4\lambda} \left(M^2-\lambda \hskip 1pt \Psi^2\right)^2\ .
\eeq
We assume that $V(\phi) \ll {M^4/4\lambda}$, so that the dominant contribution to the inflationary energy density comes from the false vacuum energy of the symmetry-breaking potential.
The coupling between $\phi$ and $\Psi$ induces an
effective mass for the waterfall field that depends on the value of the inflaton, 
\beq
M_\Psi^2(\phi) = - M^2 + g \phi^2\ .
\eeq
This vanishes at the special point $\phi = \phi_c \equiv M/\sqrt{g}$.
For $\phi > \phi_c$, the field $\Psi$ is stabilized at  $\Psi = 0$, and
can be integrated out, so that the theory reduces to that of single-field slow-roll inflation with $V_{\rm eff}(\phi) \approx M^4/4\lambda + V(\phi)$.
As $\phi$ approaches $\phi_c$ from above,  $\Psi$ becomes light and the effective description   involves both fields. Finally,
for $\phi < \phi_c$, the field $\Psi$ becomes tachyonic and ends inflation.
Notice that hybrid inflation requires a hierarchy between the masses of the two fields, 
$V_{,\phi\phi} \ll M^2$. This issue is discussed, and technically natural examples are constructed, in~\cite{Kaplan:2003aj, ArkaniHamed:2003mz}.

\vskip 4pt
\noindent
{\it Starobinsky's $R^2$ inflation.}---One of the earliest models of inflation was written down by Starobinsky in 1980~\cite{Starobinsky:1980te}.\footnote{The Starobinsky model has recently received renewed attention---see \cite{Kallosh:2013lkr}  for a superconformal generalization and \cite{Ellis:2013xoa, Ellis:2013nxa} for a no-scale supergravity version.
}
Motivated by \cite{Dowker:1975tf}, Starobinsky considered one-loop corrections to the Einstein-Hilbert action.  These lead to an effective action of the form of
(\ref{gravityexpansion}). Starobinsky's model considers only the $R^2$ correction,
\beq
S = \frac{M_{\rm pl}^2}{2} \int \d^4 x \sqrt{- g} \left(R + \frac{\alpha}{2 M_{\rm pl}^2} R^2 \right)\ .  \label{SStaro}
\eeq
For sufficiently large $\alpha$, this action leads to inflationary dynamics. The easiest way to see this is to perform a conformal transformation, $g_{\mu \nu} \mapsto \tilde g_{\mu \nu} = \Omega^2\hskip 1pt g_{\mu \nu}$, with $\Omega^2 \equiv 1 +  \alpha R/M_{\rm pl}^2$, to arrive at the action of a minimally coupled scalar field $\phi \equiv M_{\rm pl} \sqrt{\frac{2}{3}} \ln(1 +  \alpha R/M_{\rm pl}^2)$,
\beq
S = \int \d^4 x \sqrt{- \tilde g} \left( \frac{M_{\rm pl}^2}{2} \tilde R - \frac{1}{2}(\partial \phi)^2 - V(\phi) \right)\ ,
\eeq
with potential
\beq
V(\phi) = \frac{M_{\rm pl}^4}{4\alpha} \left( 1 - \exp\left[{- \sqrt{\frac{2}{3}} \frac{\phi}{M_{\rm pl}}}\right]\right)^2\ . \label{Vstaro}
\eeq
The slow-roll parameters associated with the potential
(\ref{Vstaro}) are
\beq
\eta = - \frac{4}{3} e^{- \sqrt{2/3}\hskip 1pt  \phi/M_{\rm pl}} \quad ,\quad \epsilon = \frac{3}{4} \eta^2\ .
\eeq
Inflation occurs for $\phi > M_{\rm pl}$.
The normalization of the scalar amplitude requires
\beq
\alpha = 2.2 \times 10^8\ . \label{ALPHA}
\eeq
Such a large parameter seems unnatural from a top-down perspective. Its bottom-up naturalness is discussed in~\cite{Burgess:2009ea,Hertzberg:2010dc}.
The scalar spectral
tilt and the tensor-to-scalar ratio are
\beq
n_s - 1 \approx - \frac{2}{N_\star}\quad , \quad r \approx \frac{12}{N_\star^2} \ . \label{equ:Staro}
\eeq

\vskip 4pt
\noindent
{\it Non-minimally coupled inflation.}---It is also interesting to consider
a scalar field $\varphi$ with a non-minimal coupling to gravity~\cite{Spokoiny:1984bd, Futamase:1987ua, Salopek:1988qh, Fakir:1990eg, Kaiser:1994vs, Komatsu:1999mt}.  The simplest such coupling is the
operator $\varphi^2 R$.\footnote{This interaction played a fundamental role in the
revival of Higgs inflation~\cite{Bezrukov:2007ep}. } Adding this to the action (\ref{equ:Vminimal}), we get
\beq
S = \int \d^4 x \sqrt{-g} \left[ \frac{M_{\rm pl}^2}{2}\left( 1 + \xi \frac{\varphi^2}{M_{\rm pl}^2} \right) R    - \frac{1}{2}(\partial \varphi)^2 - \frac{\lambda}{4} \varphi^4 \right]\ , \label{equ:SJordan0}
\eeq
where, for concreteness, we have chosen a quartic polynomial for the inflaton potential.
The parameter $\xi$ determines the strength of the non-minimal coupling to gravity.  Again, it is convenient to go to Einstein frame by performing a conformal rescaling, $\tilde g_{\mu \nu} = \Omega^2\hskip 1pt g_{\mu \nu}$, with $\Omega^2 \equiv 1 +  \xi \varphi^2/M_{\rm pl}^2$.
The action then takes the form
\beq
S = \int \d^4 x \sqrt{-\tilde g} \left[ \frac{M_{\rm pl}^2}{2}  \tilde R - \frac{1}{2} k(\varphi) (\partial \varphi)^2 - V(\varphi)\right]\ , \label{equ:NonC}
\eeq
where
\begin{align}
k(\varphi) &= \frac{1 + (6\xi +1) \psi^2}{(1+\psi^2)^2}\ , \\
 V(\varphi) &= \frac{\lambda M_{\rm pl}^4}{4 \xi^2} \frac{\psi^4}{(1+\psi^2)^2} \ , \qquad
\psi^2 \equiv \frac{\xi \varphi^2}{M_{\rm pl}^2}\ .
\end{align}
The canonically-normalized field, $\phi = \int \sqrt{k(\varphi)} \, \d \varphi$, is
\beq
\frac{\phi}{M_{\rm pl}} = \sqrt{\frac{6\xi +1}{\xi}} \sinh^{-1}\left(\sqrt{6\xi + 1}\, \psi \right) - \sqrt{6} \sinh^{-1} \left( \sqrt{6\xi} \frac{\psi}{\sqrt{1+\psi^2}} \right)\ .
\eeq
For $\xi \gg 1$, this can be approximated as
\beq
\frac{\phi}{M_{\rm pl}} \approx \sqrt{\frac{3}{2}} \, \ln(1+\psi^2)\ ,
\eeq
and the potential becomes
\beq
V(\phi) =  \frac{\lambda M_{\rm pl}^4}{4 \xi^2} \left( 1 - \exp\left[- \sqrt{\frac{2}{3}}\frac{\phi}{M_{\rm pl}} \right]  \right)^{2}\ . \label{equ:Vchi}
\eeq
This is identical to the potential (\ref{Vstaro}) in the Starobinsky model.
In the limit $\xi \gg 1$, the model (\ref{equ:SJordan0})
therefore
has the same phenomenology as (\ref{SStaro}).
The constraint (\ref{ALPHA}) translates into
\beq
\xi = 47000 \sqrt{\lambda} \ ,
\eeq
and the predictions for $n_s$ and $r$ are those of (\ref{equ:Staro}).
The predictions for general $\xi$ were derived in~\cite{Kaiser:1994vs, Komatsu:1999mt}:
\begin{align}
n_s - 1 &\,=\, - \frac{32\xi}{16 \xi N_\star - 1} \ \xrightarrow{\ \xi \gg 1\ }\ - \frac{2}{N_\star} \ , \\
 r &\, =\, +\frac{12}{N_\star^2} \frac{6\xi + 1 }{6\xi} \ \, \xrightarrow{\ \xi \gg 1\ }\ +\frac{12}{N_\star^2}\ .
\end{align}
These results are illustrated in fig.~\ref{fig:nsrPredictions}.
We see that the model interpolates between $\phi^4$ chaotic inflation (for $\xi = 0$) and the Starobinsky 
model (for $\xi \gg 1$).

\subsection{Non-Slow-Roll: K-Inflation}
\label{sec:PX}

So far, we have only considered slow-roll models with
canonical kinetic terms.
An alternative class of models---known as {\it k-inflation} \cite{ArmendarizPicon:1999rj, Garriga:1999vw} or $P(X)$ {\it theories} \cite{Chen:2006nt}---considers the possibility that inflation was driven by non-trivial kinetic effects rather than
by a flat potential.
An efficient way to model these effects is through the action
\beq
S =  \int \d^4 x \sqrt{-g} \, \left[  \frac{M_{\rm pl}^2}{2}R + P(X,\phi) \right] \ ,  \label{equ:PX0}
\eeq
where $P(X,\phi)$ is (so far) an arbitrary function of the inflaton field $\phi$ and of its kinetic energy $X \equiv - \frac{1}{2}(\partial \phi)^2$.   The stress-energy tensor arising from (\ref{equ:PX0}) corresponds to a perfect fluid with pressure $P$ and energy density $\rho = 2 X P_{,X} - P$,
 where $P_{,X}$ denotes a derivative with respect to $X$.
The Friedmann equation and the Klein-Gordon equation
are 
\beq
3 M_{\rm pl}^2 H^2 = 2 P_{,X} X - P \qquad {\rm and} \qquad \frac{d}{dt} \left( a^3 P_{,X} \dot{ \phi} \right) = a^3 P_{,\phi} \ ,
 \eeq
so the inflationary parameter (\ref{equ:eps}) becomes
 \beq
 \varepsilon = - \frac{\dot H}{H^2} = \frac{3 X P_{,X}}{2 X P_{,X} - P}\ .
 \eeq
 The condition for inflation is still $\varepsilon \ll 1$, 
 but it
is now a condition on the functional form of $P(X)$. The fluctuations in $P(X)$ theories propagate with a nontrivial speed of sound (see Appendix~B),
 \beq
 \label{equ:cs}
 c_s^2 = \frac{dP}{d\rho} = \frac{P_{,X}}{P_{,X} + 2 X P_{,XX}} \ .
 \eeq
 The predictions for $n_s$ and $r$ are the same as in \S\ref{sec:zeta}:
 \begin{align}
 n_s - 1 &\,=\, - 2 \varepsilon - \tilde \eta - \kappa\ , \\
 r &\,=\, 16 \varepsilon c_s\ ,
 \end{align}
 where $\tilde \eta$ and $\kappa$ were defined in (\ref{equ:Hubble2}).
We saw in \S\ref{sec:tests} that a small sound speed leads to observable equilateral-type non-Gaussianity.  We will discuss this further in \S\ref{ssec:NGUV}, where we also emphasize the need to UV-complete theories
such as (\ref{equ:PX0}).  In \S\ref{ssec:DBI}, we present {\it DBI inflation}~\cite{Silverstein:2003hf, Alishahiha:2004eh} as a specific example in string theory.

\subsection{Inflation in Effective Field Theory}
\label{ssec:UV}

The models that we have presented so far are toy models: they are decoupled from the rest of physics and lack ultraviolet completions.
The most conservative way to address these deficiencies is to work in effective field theory.
In the remainder of this chapter,
we will discuss the embedding of slow-roll inflation in the framework of effective field theory.

\vskip 4pt
The starting point is the EFT Lagrangian (\ref{equ:Leff3}) minimally coupled to gravity,
\beq
S_{\rm eff}[\phi] = \int \d^4 x \sqrt{-g} \left[ \,\frac{M_{\rm pl}^2}{2} R
+{\cal L}_l[\phi]
+ \sum_i c_i \hskip 2pt \frac{{\cal O}_i[\phi]}{\Lambda^{\delta_i -4}} \, \right]\ , \label{equ:Leff4}
\eeq   where ${\cal L}_l[\phi]$ includes the canonical kinetic term $- \frac{1}{2}(\partial \phi)^2$ as well as any renormalizable interactions.
As we explained at length above, the sum over non-renormalizable terms parameterizes the effects of massive fields on the EFT of the light fields.
When the UV theory is unknown, one can at best  make assumptions about the symmetry structure of the UV theory, and then include all higher-dimension operators~${\cal O}_i$ consistent with these symmetries.
Following the remarks in \S\ref{ssec:gravity}, the maximal cutoff of the EFT is the Planck scale, $\Lambda \lesssim M_{\rm pl}$.
In order for the effective theory (\ref{equ:Leff4}) to  remain valid  during
the freeze-out of cosmological perturbations, i.e.~when $\omega = H$, the minimal cutoff is the inflationary Hubble scale, $\Lambda \gtrsim H$.  Thus, all fields with masses $m \lesssim H$ are part of the EFT.
We will begin  by discussing the case where the only light degrees of freedom are the graviton and a  single real inflaton  scalar;  models with multiple light scalars  are discussed  in subsequent sections and in Appendix~C.

In most particle physics applications of effective field theory, higher-dimension operators only contribute small (`irrelevant') corrections to the leading dynamics. As the cutoff is pushed to the Planck scale,
these  contributions typically become negligible.
(One notable exception is gravity-mediated supersymmetry breaking.)
It is a special feature of the effective theory of inflation~(\ref{equ:Leff4}) that some irrelevant operators play a crucial role at low energies, not just for precision observables, but even for the zeroth-order dynamics.  Slow-roll inflation is sensitive even to Planck-suppressed operators.
  The next
  section is devoted to a careful discussion of this important fact.

\section{Ultraviolet Sensitivity}
\label{ssec:UV}

We will highlight four aspects of the UV sensitivity of inflation. The first two (eta problem I and II) are universal and apply to any slow-roll\footnote{Variations of these problems arise in most non-slow-roll models as well.  For example, when non-trivial kinetic terms  make a rapidly-varying potential innocuous (cf.~\S\ref{sec:PX}), one must still ensure that the  necessary kinetic terms  are not affected by Planck-suppressed contributions.   The general problem is to arrange that the {\it{action}}, not just the potential, changes slowly during inflation.} model of inflation. The last two (super-Planckian displacements and non-Gaussianity) only apply to specific classes of inflationary theories.

\subsection{Eta Problem I: Radiative Corrections} \label{etaI}

As we have explained in \S\ref{sec:EFT2}, the unknown heavy physics above the cutoff scale has two effects: {\it i}) it renormalizes the couplings of the light fields and {\it ii}) it introduces new non-renormalizable interactions. Both effects have to be addressed in a complete discussion of the inflationary dynamics.

\vskip 4pt
We have seen that
quantum corrections tend to drive scalar masses to the cutoff scale, unless the fields are protected by symmetries.
In the case of inflation, this implies the following  quantum correction to the inflaton mass:
\beq
\Delta m^2 \sim \Lambda^2\ . \label{equ:DML}
\eeq
Since consistency of the EFT treatment requires that $\Lambda > H$, we find a large renormalization of the inflationary eta parameter (\ref{equ:eta}),
\beq
 \Delta \eta \sim \frac{\Lambda^2}{H^2} \, \gtrsim\, 1\ ,
\eeq
and sustained slow-roll inflation appears to be unnatural.   This difficulty is known as the {\it eta problem}.  The eta problem in the context of supergravity was emphasized long ago in \cite{Copeland:1994vg}.   However, the issue is  actually far more general,  afflicting any construction of slow-roll inflation in effective field theory.

Two  strategies are available for addressing the eta problem: fine-tuning the potential,  or appealing to symmetries. The problem is a resilient one because  approaches based on symmetries
face serious limitations,
and have only occasionally been  successful.
The symmetry options are the same as discussed in \S\ref{sss:symmetry}: supersymmetry  and/or global internal symmetries.  We will discuss these  in turn,  but it is worth stating the upshot in advance: supersymmetry ameliorates  but cannot completely solve the problem,  while global symmetry arguments require  precise control of Planck-suppressed operators  breaking the symmetry,  motivating a treatment in quantum gravity.

\vskip 4pt
\noindent
{\it Supersymmetry.}---Even if the inflaton  is part  of a  supersymmetric action, the  inflationary background solution  spontaneously breaks SUSY, because  the energy density is necessarily positive.
Nevertheless, SUSY still  limits the size of radiative corrections, because  sufficiently high frequency modes are insensitive to the effects of the spacetime curvature during inflation.  The cancellation between boson and fermion loops therefore still applies in the high-energy regime, just as in flat space. On the other hand, modes with frequencies below the Hubble scale, $\omega \lesssim H$, do experience non-trivial effects from the expanding background.
Boson and fermion propagators are then modified by the coupling to the spacetime curvature, with
mass splittings within supermultiplets that are typically of order $H$,
and the corresponding loops no longer cancel.
Radiative corrections to the inflaton mass are therefore naturally of order of the Hubble scale,
\beq
\Delta m^2 \sim H^2  \ .
\eeq
This is smaller than the correction in (\ref{equ:DML}), but not small enough to evade the eta problem:
\beq
\Delta \eta \sim 1\ .  \label{etaestimate}
\eeq
This qualitative argument is confirmed in detail by  investigations of inflation in supergravity~\cite{Copeland:1994vg}  and in string theory (see \S\ref{etaproblem}).
Hence, although SUSY ameliorates the eta problem,
it does not solve it: successful inflation still requires fine-tuning of the mass term~\cite{Baumann:2011nk, Agarwal:2011wm}, although much less than in an EFT without SUSY.

The degree of fine-tuning implied by (\ref{etaestimate}) depends  to some  extent on the underlying model.
In small-field models with $\epsilon \ll \eta$, the value of $\eta$ at horizon crossing is related to the scalar spectral index, $\eta \approx \frac{1}{2}(n_s - 1)$.  For the Planck best-fit, $n_s \approx 0.96$, this implies $\eta \approx 0.02$, so the required fine-tuning is at the percent-level.

\vskip 4pt
\noindent
{\it Global symmetries.}---We have discussed global symmetries extensively in \S\ref{sss:symmetry}.
As we explained there, a small scalar mass is `bottom-up natural' if the renormalizable part of the Lagrangian (the `IR theory') respects an approximate shift symmetry
\beq
\phi \mapsto \phi + const. \label{equ:SHIFT}
\eeq
In other words, the theory contains no relevant or marginal operators that violate (\ref{equ:SHIFT}).
Then,  loops of the light fields do not drive  the scalar mass up to the cutoff:  quantum corrections to the scalar mass are suppressed by the parameter measuring the weak breaking of the symmetry.
Whether inflationary models based on (\ref{equ:SHIFT}) are `top-down natural' is an important question for a theory of quantum gravity.

\subsection{Eta Problem II: Higher-Dimension Operators}  \label{etaII}

We saw in \S\ref{sss:symmetry} that  not all desirable symmetries of the IR theory  can be  realized in a  consistent UV~theory: in particular, we recalled the common lore that quantum gravity breaks all continuous global symmetries.  Correspondingly,  although a  low-energy theory  with light scalars that respects~(\ref{equ:SHIFT}) is radiatively stable, such a theory is not necessarily `top-down natural':
{\it irrelevant operators} may spoil the desired symmetry.
Whether the symmetry survives is a question for the ultraviolet completion,  and cannot be addressed by studying the renormalizable Lagrangian.

\vskip 4pt
As an example, consider the dimension-six operator
\beq
 {\cal O}_6 = c \hskip 1pt V_{l}(\phi) \frac{\phi^2}{\Lambda^2} \ , \label{equ:L6}
\eeq
where $c$ is a constant, and $V_{l}(\phi)$ consists of the renormalizable terms in the potential, cf.~(\ref{equ:Leff4}).
Even if $V_l(\phi)$ respects an approximate shift symmetry,  this is broken by ${\cal O}_6$.
Provided that the inflaton vev is smaller than the cutoff, $\phi < \Lambda$, the operator ${\cal O}_6$ makes only a small correction to the inflationary potential, $\Delta V \ll V(\phi)$.
Nevertheless, its effect on the inflaton mass is  significant:
 \beq
 \Delta \eta \approx 2 c \hskip 1pt \left( \frac{M_{\rm pl}}{\Lambda}\right)^2  \ .  \label{equ:eta6}
\eeq
For $c \sim {\cal O}(1)$ and $\Lambda < M_{\rm pl}$, the theory again suffers from the eta problem.
(Notice that the overall scale of the potential  cancels in (\ref{equ:eta6}).)
Even if the operator in (\ref{equ:L6}) is Planck-suppressed, 
$\Lambda \to M_{\rm pl}$, it cannot be ignored in discussions of the inflationary dynamics.

In a theory with a single real scalar $\phi$, it is difficult to give a convincing argument for the absence of couplings of the form (\ref{equ:L6}).
Note in particular that if  one forbids  $\phi^2$  via a global symmetry under which $\phi$  transforms linearly,
one would simultaneously exclude the kinetic term $- \frac{1}{2}(\partial\phi)^2$.
An influential approach  is therefore to take $\phi$  to  transform nonlinearly under a global symmetry (e.g.~by taking $\phi$ to be an axion --- see \S\ref{sec:AxionInflation}),  and/or to be the phase of a complex scalar (cf.~e.g.~\cite{Baumann:2010nu}).

Although operators of the  simple form (\ref{equ:L6}) arise in  many ultraviolet completions (see \S\ref{etaproblem} and Chapter~\ref{sec:Examples}), more general
non-renormalizable interactions  are also common,  and can  give comparable (or larger)  effects.  Consider an operator of the form
\beq
 {\cal O}_{\delta} = c \hskip 1pt \langle V \rangle \left(\frac{\phi}{\Lambda}\right)^{\delta-4} \ , \label{equ:LDelta}
\eeq where $\langle V \rangle$ is the vacuum energy  at some stage of inflation.   The correction to $\eta$ is
\beq
 \Delta \eta \approx c \hskip 1pt (\delta-4)(\delta-5) \hskip 1pt  \left( \frac{M_{\rm pl}}{\Lambda}\right)^2 \left(\frac{\phi}{\Lambda}\right)^{\delta-6}  \ .  \label{equ:etaDelta}
\eeq
If $\Lambda=M_{\rm pl}$ and $\phi < \Lambda$, operators with $\delta \gg 6$  can be neglected.  For this reason,  addressing the eta  problem in small-field
inflation requires,  at a {\it{minimum}},  characterizing   Planck-suppressed interactions up to dimension six.   However, operators with $\delta$  slightly larger than six  are not strictly negligible unless $\phi \ll \Lambda$, while taking $\Lambda < M_{\rm pl}$ increases $\Delta \eta$ in (\ref{equ:etaDelta}).\footnote{Notice that for $4<\delta<6$, $\delta \neq 5$,  the correction $\Delta \eta$ increases for small $\phi/\Lambda$.   Irrelevant operators with non-integer dimensions $\delta < 6$  can  therefore dominate  the dynamics in small-field inflation.
For an example, see \S\ref{sec:dbrane}.}
The threshold beyond which non-renormalizable interactions can be neglected  therefore varies from model to model,  and depends on  $\Lambda/M_{\rm pl}$ and $\phi/M_{\rm pl}$.

Explaining the absence of operators like (\ref{equ:LDelta}) --- including the special  case (\ref{equ:L6}) --- requires an understanding of the leading high-energy corrections to the inflationary Lagrangian.  When a symmetry is assumed in the EFT, one must demonstrate, in the context of an ultraviolet completion, that the symmetry survives  non-renormalizable  corrections  such as  (\ref{equ:LDelta}).  This sensitivity to UV physics is the key challenge for realizing inflation in a theory of fundamental physics.
At the same time, the  fact that Planck-scale effects  do not decouple from inflation presents a striking opportunity:  one can hope to use cosmological observations as a laboratory for physics at the highest energy scales.

\subsection{Gravity Waves and Super-Planckian Fields}
\label{ssec:gwaves}

Inflationary models that predict a detectably-large primordial gravitational wave signal are extraordinarily  sensitive to ultraviolet physics.
To see this, we will derive the {\it Lyth bound}~\cite{Lyth:1996im}, which relates observable tensor modes to super-Planckian
displacements of the inflaton, $\Delta\phi \gtrsim M_{\rm pl}$.
We will  begin with a derivation of the Lyth bound in single-field slow-roll inflation, and then  present extensions to more general  scenarios.

\begin{figure}[htbp]
    \centering
        \includegraphics[width=0.55 \textwidth]{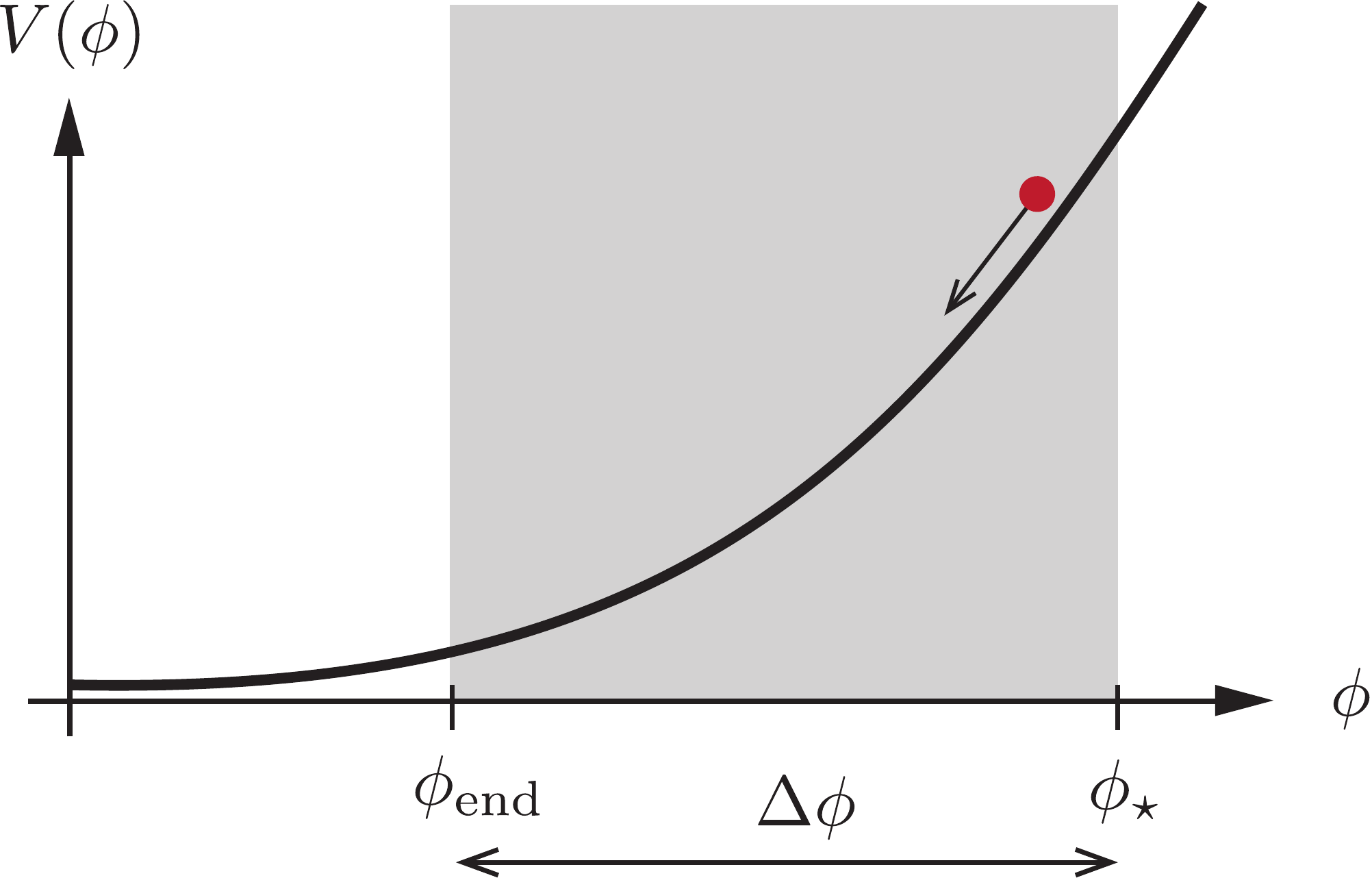}
\caption{Evolution of the inflaton field from the time when modes that are observable in the CMB exited the horizon, $\phi_\star$, to the end of inflation, $\phi_{\rm end}$. The total field displacement $\Delta \phi$ is related to the tensor-to-scalar ratio $r$
by (\ref{equ:LythBound}).}   \label{fig:Lyth}
\end{figure}

\vskip 4pt
\noindent
\subsection*{The Lyth Bound}

Substituting (\ref{equ:varE}) into $r= 16\varepsilon$, we can relate the tensor-to-scalar ratio $r$ to
the evolution of the inflaton field:
\beq
r =   8 \left(\frac{1}{M_{\rm pl}} \frac{\d \phi}{\d N} \right)^2\ , \qquad {\rm where} \quad \d N \equiv H \d t \ .  \label{Lythintegral}
\eeq
Integrating (\ref{Lythintegral}) from the time  $N_\star$ when modes that are observable in the CMB exited the horizon, until the end of inflation at $N_{\rm end} \equiv 0$ (see fig.~\ref{fig:Lyth}), we get~\cite{Lyth:1996im}
\beq
\frac{\Delta \phi}{M_{\rm pl}} \ =\, \int_{0}^{N_\star} {\rm d}  N\,  \sqrt{\frac{r(N)}{8}}\ . \label{equ:LythIntegral}
\eeq
To evaluate the integral in (\ref{equ:LythIntegral}), it is useful to define
\begin{equation}
N_{\rm{eff}} \equiv \, \int_{0}^{N_\star} {\rm d}  N\,  \sqrt{\frac{r(N)}{r_\star}}\ , \label{neffdef}
\end{equation} where $r_\star$ is the tensor-to-scalar ratio  measured in the CMB, so that
\begin{equation}
\frac{\Delta \phi}{M_{\rm pl}} = N_{\rm eff} \sqrt{\frac{r_\star}{8}} \ .  \label{LythwNeff}
\end{equation}
In slow-roll inflation, one can show that
\begin{equation}
\frac{\d \ln r}{\d N} = -\left[ n_s-1+\frac{r}{8} \right] \ ,
\end{equation}   Because both $n_s-1$ and $r$  are constrained by  observations, one can limit $N_{\rm{eff}}$ in slow-roll models:  a 
conservative estimate is $N_{\rm{eff}} \gtrsim 30$ \cite{Baumann:2006cd} (see also \cite{Easther:2006qu}), while  more typically $N_{\rm{eff}} \gtrsim 50$.
Taking $N_{\rm{eff}} \gtrsim 30$, we conclude that\footnote{One should not assume that simple models  will approximately saturate (\ref{equ:LythBound}): for example, chaotic inflation scenarios involve displacements roughly four times larger than required by the bound.}
\beq
\frac{\Delta \phi}{M_{\rm pl}} \gtrsim  
\left( \frac{r}{0.01} \right)^{1/2}\ . \label{equ:LythBound}
\eeq
To arrive at a maximally conservative bound in single-field slow-roll inflation,
one can assume that slow-roll is valid only while the {\it{observed}}  multipoles of the CMB exit the horizon,  corresponding to $N_{\rm eff} \approx 7$.
This leads to (cf.~\cite{Lyth:1996im}, which used a smaller  $N_{\rm eff}$ because fewer
multipoles had been observed in 1996)
\beq
\frac{\Delta \phi}{\Mp} \gtrsim 0.25 \times \left( \frac{r}{0.01} \right)^{1/2}\ . \label{conservativeLyth}
\eeq
It is quite a remarkable coincidence that the level of tensors that is experimentally accessible ($r \gtrsim 0.01$)
is tied to the fundamental scale of quantum gravity, $M_{\rm pl}$.

We strongly caution against viewing $\Delta\phi = M_{\rm pl}$ as an absolute dividing line:
the theoretical challenges of  models with $\Delta\phi > M_{\rm pl}$ are shared by models with slightly smaller displacements.
In particular, although gravity itself becomes strongly coupled
around the scale $\Mp$,
parametrically controlled ultraviolet completions of gravity  generally involve additional scales $\Lambda < \Mp$. For instance, the string scale and the Kaluza-Klein scale (see~\S\ref{overallgoal}) are typically well below the Planck scale. Field excursions that are large compared to those scales raise concerns similar to the super-Planckian issues we describe below.

Finally, let us emphasize that the Lyth bound (\ref{equ:LythBound}) is a purely kinematic statement, relating $r$ to the distance  in field space over which the inflaton moves.
Although the bound has profound consequences in the  context of  effective field theory reasoning about  natural Planck-suppressed interactions (see below),  the derivation of (\ref{equ:LythBound}) relied in no way on notions of naturalness, or on a Taylor expansion of the potential.

\subsection*{Super-Planckian Fields in Effective Field Theory}


The simplest  scenarios for large-field inflation involve a scalar field minimally coupled to gravity, with a monomial (or sinusoidal) potential  that varies slowly over super-Planckian distances  in field space.
To understand  the theoretical status of  these models, it is instructive  to examine them first from a purely bottom-up perspective, in  quantum field theory coupled to general relativity, {\it{without}} accounting for necessity of an ultraviolet completion of gravity.  The only degrees of freedom are then the graviton and a  single scalar inflaton.  From this perspective,  there are two issues that appear dangerous at first glance, but are in fact not at all problematic~\cite{Linde:2005ht}.

First, one might worry  that super-Planckian displacements of the inflaton  will lead to super-Planckian energy densities, and correspondingly large gravitational backreaction.   This concern is misplaced:  the normalization (\ref{naivespectrum}) of the scalar fluctuations requires that $V \ll \Mp^4$.
For instance, in $m^2 \phi^2$ chaotic inflation, (\ref{naivespectrum}) implies that the inflaton mass is small, $m \sim 10^{-5} \Mp$, so that the energy density never becomes significant even for super-Planckian fields.

A second concern  is radiative stability:  do  quantum corrections, from  graviton loops  and/or $\phi$ loops, destabilize the classical potential $V(\phi)$?  No: the small value of the inflaton mass\footnote{A parallel argument  applies to chaotic inflation with a non-quadratic  monomial potential.} $m$ is technically natural, because the theory  enjoys a shift symmetry in the limit $m \to 0$.  Quantum corrections therefore do not destabilize the potential.
 In particular, the one-loop correction from graviton loops is~\cite{Smolin:1979ca}
\beq
\frac{\Delta V}{V} = c_1 \frac{V''}{\Mp^2} + c_2 \frac{V}{\Mp^4} \ ,
\eeq
where $c_1$ and $c_2$ are order-one numbers.  Because $m \ll \Mp$ and $V \ll \Mp^4$,  this is a small correction.

To summarize,  in the low-energy theory of the inflaton and the graviton, potentials supporting large-field inflation can be radiatively stable, and  in particular free of significant  corrections from inflaton-graviton interactions.   Thus, from the bottom-up perspective, large-field inflation is not problematic.

The essence of the problem of large-field inflation is that  gravity requires an ultraviolet completion,  and  couplings  of the inflaton to the  degrees of freedom that provide this ultraviolet completion  do not {\it{necessarily}} respect  the symmetry structures  needed to protect the inflaton potential  in the low-energy theory.
The effects of classical gravity (i.e.~backreaction) and of semiclassical gravity (i.e.~graviton loops)  are not problematic, but full {\it{quantum gravity}} effects --- corresponding to integrating out  fields with Planck-scale or string-scale masses --- are subtle, and have the potential to be ruinous.\footnote{Some authors have argued that quantum gravity effects are necessarily small when all energy densities  are sub-Planckian.
As a general statement  about an arbitrary quantum gravity  theory, this is false: the fact that the eta problem appears in string theory, cf.~\S\ref{etaproblem}, is one simple counterexample, and the diverse failure modes of large-field models  in string theory discussed in \S\ref{sec:AxionInflation}
provide many more.  
Ignoring quantum gravity effects purely because all energy densities are small in Planck units  amounts to  attributing to the quantum gravity theory underlying our universe a property that is not seen in string theory.
This is a logically consistent position but is  very far from being agnostic about quantum gravity.}

From our  general discussion of  effective actions,  we know that integrating out fields of mass $\Lambda$ with order-unity couplings to the inflaton $\phi$ will lead to an effective theory of the form
\beq
{\cal L}_{\rm eff}[\phi] =   {\cal L}_{l}[\phi] +
 \sum_{i=1}^\infty \Biggl(
 \frac{c_i}{\Lambda^{2i}} \hskip 1pt \phi^{4+2i}
+ \frac{d_i}{\Lambda^{2i}} \hskip 1pt (\partial \phi)^2\phi^{2i}
+ \frac{e_i}{\Lambda^{4i}}\hskip 1pt (\partial \phi)^{2(i+1)} + \cdots
\Biggr) \ , \label{fullinfiniteexpansion}
\eeq  where the omitted terms involve additional derivatives acting on $\phi$  (or on the metric), and $c_i$, $d_i$, $e_i$  are dimensionless Wilson coefficients  that are  typically of order unity.

\begin{figure}[htbp]
    \centering
        \includegraphics[width=0.55 \textwidth]{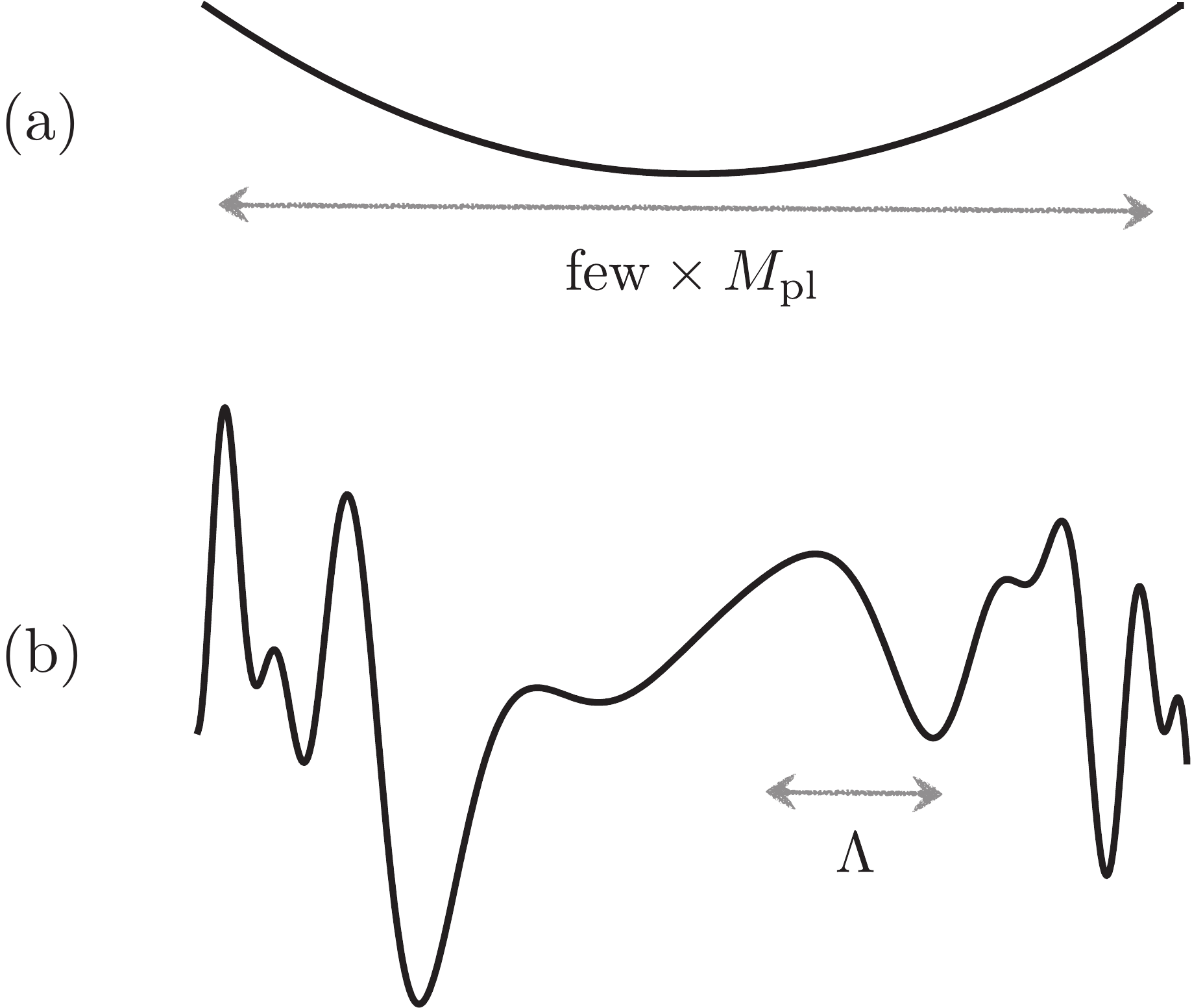}
\caption{(a): Observable tensor modes require a smooth inflaton potential over a super-Planckian range. (b): In the absence of symmetries, 
effective field theory predicts that generic potentials have structure on sub-Planckian scales, $\Lambda < M_{\rm pl}$. }   \label{fig:lyth}
\end{figure}

To begin, we discuss  contributions to the potential, i.e.~the terms  involving $c_i$ in (\ref{fullinfiniteexpansion}).
For $c_i \sim {\cal O}(1)$, we expect  that the dominant functional form of the potential will change when the field moves a distance of  order $\Lambda$: there is `structure'  in the potential  on scales of  order  $\Lambda$ (see fig.~\ref{fig:lyth}).
Even under the optimistic assumption that $\Lambda=M_{\rm pl}$, the potential (\ref{fullinfiniteexpansion}) will not support large-field inflation unless one effectively fine-tunes  the infinite set of Wilson coefficients $c_i$.  One might object that the  expansion (\ref{fullinfiniteexpansion}) is not a useful description over distances  $\gtrsim \Lambda$: as a practical matter one would not compute an infinite number of corrections.  This is true, but (\ref{fullinfiniteexpansion}) nevertheless serves to show how badly an expansion in low-dimension  operators can fail in large-field inflation;  the challenge is then to show that a more  sensible potential arises in some controlled setting.

As should be clear from the effective Lagrangian (\ref{fullinfiniteexpansion}), the problem of controlling super-Planckian displacements  is not simply a matter of protecting the potential:
other terms 
receive  equally  dramatic corrections  from higher-dimension operators.
The terms involving $d_i$ in (\ref{fullinfiniteexpansion}) correspond to  modifications of the two-derivative kinetic term for $\phi$, i.e.~corrections  to the metric on moduli space.
For $d_i \sim {\cal O}(1)$,  these corrections are large over distances of order $\Lambda$.

The leading idea for implementing large-field inflation is to use a symmetry to suppress the dangerous higher-dimension contributions in (\ref{fullinfiniteexpansion}). For example, an unbroken shift symmetry
\begin{equation}
\phi \mapsto \phi + const.  \label{shiftsymmetrylf}
\end{equation}
forbids all non-derivative operators in (\ref{fullinfiniteexpansion}),  including the desirable parts of the inflaton potential, while a suitable weakly-broken shift symmetry\footnote{Note that  to realize an approximate shift symmetry in the low-energy theory, it would suffice for the inflaton  to have weak couplings $g \ll 1$ to all  the degrees of freedom of the UV  completion:  the Wilson coefficients in (\ref{fullinfiniteexpansion})  would then be suppressed by powers of $g$.  Equivalently,  the effective cutoff scale  would become $\Mp/g \gg \Mp$.  The coupling of the inflaton to any additional degrees of freedom would be weaker than gravitational~\cite{ArkaniHamed:2006dz}.}
can give rise to a radiatively stable  model of large-field inflation.
Whether  such a  shift symmetry can be UV-completed is a  subtle and important question for a Planck-scale theory like string theory.
We will return to the problem of super-Planckian fields in Chapter~\ref{sec:StringInflation} (see also \S\ref{sec:AxionInflation}).

\subsection*{Evading the Lyth bound?}

Because the Lyth bound raises the specter of catastrophic quantum gravity corrections in all  models producing
detectable  primordial gravitational waves,  it is natural to pursue models that evade the bound by violating one or more of the assumptions that entered the derivation.  As we now review, the bound  is quite robust.

\vskip 4pt
Let us be precise about what it means to `evade the Lyth bound'.
In a model involving a single inflaton with canonical kinetic term, beginning in the Bunch-Davies vacuum, and with slow-roll unbroken throughout inflation, the bound (\ref{equ:LythBound}) generally applies.  If slow-roll  holds only  during the $N_{\rm eff} \approx 7$ directly-observed $e$-folds, and the subsequent evolution is arbitrary, the more conservative bound (\ref{conservativeLyth}) remains applicable.
Particularly because $\Mp = 2.4 \times 10^{18} {\rm{GeV}}$ is not a  precise and absolute  marker of  the realm where quantum gravity corrections are large, one should be wary of the claim that a marginal violation of  (\ref{equ:LythBound}), or even of  (\ref{conservativeLyth}), diminishes the problem of ultraviolet sensitivity in large-field inflation.\footnote{The Lyth bound is sometimes misunderstood as the statement that `detectable gravitational waves imply $\Delta\phi > 1.0 \times \Mp$', or equivalently ---  upon imposing a legalistic definition of small-field inflation as inflation with $\Delta\phi < 1.0 \times \Mp$ --- that detectable gravitational waves are impossible in `small-field' inflation.
Neither statement is true, so exhibiting  counterexamples  has limited utility.}
For this reason,  to truly evade the physical content of the Lyth bound,  one should alter an assumption entering the derivation  in such a way as to {\it{parametrically}}  violate the conservative bound (\ref{conservativeLyth}). In other words, the task is to  avoid the conclusion that `detectable gravitational waves imply displacements of order $\Mp$'.

\vskip 4pt
With this in mind, we comment on a few ideas for evasion of the Lyth bound:
\begin{itemize}

\item[$\triangleright$]  {\it Nontrivial evolution.}---A number of authors have proposed scenarios  in which nontrivial evolution after the horizon exit of the CMB fluctuation ---  for example, a steep drop in the potential ---  renders
(\ref{equ:LythBound}) inapplicable: see e.g.~\cite{BenDayan:2009kv,Shafi:2010jr,Hotchkiss:2011gz,Hebecker:2013zda}.
However, these models  do still satisfy (\ref{conservativeLyth}),  and correspondingly involve  displacements of order the Planck scale for $r\gtrsim 0.01$.  For a recent discussion see~\cite{Antusch:2014cpa}.

\item[$\triangleright$]  {\it Non-canonical kinetic terms.}---It is natural to ask whether the Lyth bound can be evaded if the inflationary phase  is supported by  kinetic energy.  As an explicit example, consider the $P(X)$ theories of \S\ref{sec:PX}. The naive  bound for the excursion of  $\phi$ becomes~\cite{Baumann:2006cd}
\beq
\frac{\Delta \phi}{\Mp}  =  (c_s P_{,X})^{-1/2} \sqrt{\frac{r}{8}}\,  \Delta N\ . \label{equ:PXLyth}
\eeq
This seems to suggest that the Lyth bound could be evaded by choosing $P_{,X} \gg 1$ for fixed~$c_s$.  However, when $P_{,X} \gg 1$ we are far from a canonical kinetic term for $\phi$, and must worry about corrections to the entire $P(X)$ action, not just to the potential.
In particular,  one should inquire about Planck-suppressed corrections of the form
\beq
\Delta {\cal L} = P\bigg(X - V(\phi) \frac{\phi^2}{\Mp^2}\bigg) = P(X) - P_{,X}  V(\phi) \frac{\phi^2}{\Mp^2} + \cdots\ .  \label{pxlyth}
\eeq
For $P_{,X} \gg 1$, the corrections (\ref{pxlyth}) are enhanced over potential corrections.
Thus, even though  taking $P_{,X} \gg 1$ does technically lead to models violating (\ref{conservativeLyth}), the  problem of Planck-suppressed corrections to the  effective Lagrangian is  undiminished, and merely moved from one class of terms to another.
A generalization of the Lyth bound to
a totally
general single-field Lagrangian~\cite{Cheung:2007st} was derived in~\cite{Baumann:2011ws} --- see Appendix~B
for further details.

\item[$\triangleright$]  {\it Multiple fields: arc length versus geodesic distance.}---The distance $\Delta\phi$ that enters (\ref{equ:LythBound}) is the arc length along the inflaton trajectory, not the geodesic distance between the starting and ending points of the trajectory.  The importance of this distinction is that large displacements appear unnatural when the inflaton travels outside the radius of convergence of a Taylor expansion of the low-energy potential.  If one can arrange that the inflaton trajectory winds or meanders to achieve a large arc length while remaining within the radius of convergence, then the problem of ultraviolet sensitivity is much diminished, even though the resulting models do obey (\ref{equ:LythBound}).
    This point was stressed in \cite{Berg:2009tg}, where monodromy in a two-axion system leads to a winding trajectory (see~\S\ref{ssec:AM}).

\item[$\triangleright$]
{\it Multiple fields: modified scalar perturbations.}---Because the scalar amplitude (\ref{equ:DZ}) entering the derivation of (\ref{equ:LythBound}) applies only to a single inflaton scalar, one can ask whether contributions to the scalar  perturbations by other light fields can  lead to a weaker bound.
     Let us first consider a multi-field inflation model in which as the observed CMB multipoles exit the horizon,  slow-roll is applicable, and moreover the field trajectory does not bend sharply.\footnote{Specifically, we require that the  usual slow-roll parameters are small, and moreover the parameter $\eta_{\perp}$  defined in (C.94)
      obeys $\eta_{\perp} \ll 1$.}
     Then, as explained in Appendix~C,
      the  fluctuations of fields transverse to the inflationary trajectory make strictly {\it{positive}}  contributions to $\Delta^2_{{\cal{R}}}$: see (C.107).
  As such, the contributions of additional fields  actually strengthen the Lyth bound,  increasing the displacement $\Delta\phi$  required to produce a given observed value of $r$.

     The bound can be (rather weakly)  violated if the slow-roll, slow-turn approximations  assumed above are invalid:  fluctuations of additional fields can then contribute negatively to $\Delta^2_{{\cal{R}}}$, increasing the effective value of $r$ --- see \cite{McAllister:2012am} for explicit examples.   However,  we are not aware of a plausible construction in which this effect is large enough to induce a meaningful weakening of the bound (\ref{conservativeLyth}).

     A more dramatic example of the effect of multiple fields arises if the inflaton contribution to the scalar perturbations is negligible in comparison to the perturbations arising from a curvaton  \cite{Lyth:2001nq}, or through modulated reheating \cite{Kofman:2003nx,Dvali:2003ar}.   In typical scenarios the inflaton does still fluctuate during inflation, but the modulated contributions imprinted later  are much larger, substantially increasing the power in scalar perturbations and  correspondingly strengthening the Lyth bound.   To evade the Lyth bound via a curvaton or modulated reheating one would have to suppress the inflaton fluctuations.

\item[$\triangleright$]  {\it Modifications of the initial state.}---The  tensor amplitude (\ref{equ:Dh})  is applicable when the initial state in which the two-point function is computed is the Bunch-Davies vacuum.   A  significant modification of the initial state may allow violations of (\ref{conservativeLyth}) \cite{Collins:2014yua,Ashoorioon:2014nta}, and the resulting  tensor spectrum can be expected to
    display significant scale-dependence \cite{Ashoorioon:2014nta,Aravind:2014axa}.

\item[$\triangleright$]  {\it  Other sources of gravitational waves.}---An alternative mechanism  for generating gravitational waves during inflation, as in e.g.~\cite{Cook:2011hg,Senatore:2011sp,Barnaby:2012xt}, can readily violate (\ref{conservativeLyth}), as the bound incorporates only the  primordial gravitational waves from quantum fluctuations of the gravitational field.
    A zeroth-order challenge  in such approaches is to ensure that the dynamics producing gravitational waves does not  render the scalar spectrum non-Gaussian~\cite{Carney:2012pk}.

\end{itemize}

\subsection{Non-Gaussianity}
\label{ssec:NGUV}

Single-field slow-roll inflation has
an approximate shift symmetry\footnote{This symmetry does not have to be fundamental, but may be the result of fine-tuning. Its presence is motivated by the observed scale-invariance of the primordial fluctuations.} (\ref{equ:SHIFT})
that constrains inflaton self-interactions in the potential and prevents large non-Gaussianity: $\fnl \sim {\cal O}(\epsilon, \eta) \ll 1$~\cite{Maldacena:2002vr}.
To generate observable levels of non-Gaussianity requires either higher-derivative interactions or couplings to extra fields.  Both options can be ultraviolet sensitive.



\vskip 4pt
\noindent
{\it Non-Gaussianity from higher derivatives.}---When higher-derivative interactions are important, the dynamics deviates significantly from slow-roll.  In \S\ref{sec:PX}, we presented $P(X)$ theories as a specific example.
We mentioned that fluctuations propagate with a nontrivial speed of sound,
 \beq
 \label{equ:cs}
 c_s^2= \frac{P_{,X}}{P_{,X} + 2 X P_{,XX}} \ .
 \eeq
However,
the effective theories corresponding to $c_s \ll 1$
cry out for ultraviolet completion.  In an EFT,  one thinks of the function $P(X)$ in a derivative expansion, cf.~eq.~(\ref{fullinfiniteexpansion}),
\beq
P = X + \frac{1}{2} \frac{X^2}{\Lambda^4} + \cdots\ , \label{equ:Crem}
\eeq
which truncates to a finite number of terms if $X \ll \Lambda^4$.
However, the condition $X \ll \Lambda^4$ also implies that the deviation from the slow-roll action, $P_{\rm s.r.} \equiv X - V(\phi)$, is a perturbative correction,
and the non-Gaussianity is  correspondingly small~\cite{Creminelli:2003iq, Chen:2006nt, Seery:2005wm}:
\beq
\big|\fnl^{\rm equil}\big| \sim \frac{1}{c_s^2} - 1 \, \approx\, \frac{X}{\Lambda^4} + \cdots \, \ll\, 1\ .
\eeq
On the other hand, $|\fnl^{\rm equil}| > 1$ can only arise for $X \gtrsim \Lambda^4$, in which case it is inconsistent to truncate the expansion in (\ref{equ:Crem}).  Instead, an infinite number of higher-derivative terms---those proportional to $e_i$ in (\ref{fullinfiniteexpansion})---become relevant.   Observably-large non-Gaussianity in single-field inflation is therefore UV sensitive.\footnote{This issue is also visible in the effective theory of fluctuations~\cite{Cheung:2007st} (see Appendix~B).
In the limit of observable non-Gaussianity, the theory of the fluctuations becomes strongly coupled {\it{below}} the symmetry-breaking scale $\dot \phi$~\cite{Baumann:2011su}, and must therefore be UV-completed below $\dot \phi$.  This is in contrast to the slow-roll limit, where questions about the UV completion are deferred to scales above $\dot \phi$.}
Special symmetries, such as the higher-dimensional boost symmetry of DBI inflation~\cite{Silverstein:2003hf} (see \S\ref{ssec:DBI}), are required to make sense of the UV completion of (\ref{equ:Crem}).\footnote{Another class of ghost-free, radiatively stable higher-derivative models is {\it Galileon inflation}~\cite{Burrage:2010cu}.
In these models the renormalization of the action is protected by the {\it Galilean symmetry} $\phi \ \mapsto \ \phi + b_\mu x^\mu + c$, which is a combination of the shift symmetry (\ref{equ:SHIFT}) and a spacetime translation.  No candidate  for an ultraviolet completion of a Galileon model in string theory has been proposed,  and whether one exists is an open question.}

\vskip 4pt
\noindent
{\it Non-Gaussianity from hidden sectors.}---As we will see in Chapters~\ref{sec:StringInflation} and \ref{sec:Examples}, ultraviolet completions of inflation invariably involve extra fields coupled to the inflaton. We will collectively denote these fields by $\psi$.
If these fields are sufficiently heavy ($m_\psi \gg H$), they can be integrated out and only affect the couplings of the single-field EFT. Light hidden sector fields ($m_\psi < H$), on the other hand, can affect the inflationary fluctuations and  may therefore leave imprints in cosmological observables.

Although the approximate shift symmetry (\ref{equ:SHIFT}) sharply limits the non-Gaussianity that can arise from self-interactions of the inflaton,
 the  couplings of hidden sector fields are much less constrained, and hidden-sector self interactions  can lead to visible non-Gaussianity, as we now explain.
Suppose that the shift symmetry of the inflaton is preserved by the coupling to a  hidden sector field~$\psi$.
Then the leading interaction between the hidden sector and the visible sector is the dimension-five operator~\cite{Assassi:2013gxa}
\beq
{\cal O}_5 = \frac{ \psi X}{\Lambda} \ .
 \label{dimensionfive}
\eeq
This coupling converts any non-Gaussianity in the hidden sector into observable non-Gaussianity in the inflaton sector.

Under rather natural circumstances, the fluctuations in the hidden sector can  be highly non-Gaussian.
For example, suppose that supersymmetry is spontaneously broken during inflation.   A generic hidden sector scalar field $\psi$ that is not sequestered from the inflationary supersymmetry breaking
will acquire a soft mass $m_\psi \sim H$  and cubic coupling (or $A$-term) $A \psi^3$, with $A \sim H$~\cite{Assassi:2013gxa, Kaplunovsky:1993rd}, by coupling to the inflationary vacuum energy.
Unless $\psi$  has a large supersymmetric mass,  it  can fluctuate during inflation,  and because $A \sim H$,  the  correlations of $\psi$ are order-one non-Gaussian.  Via the operator (\ref{dimensionfive}), this gets communicated to the visible sector.\footnote{Order-one
non-Gaussianity in the observed curvature perturbations would correspond to $\fnl \Delta_{\R} \sim 1$, not to $\fnl \sim 1$.}  The signal can be large while keeping the effective theory under perturbative control, with $X < \Lambda^4$.

Through the coupling (\ref{dimensionfive}), the Planck limits (\ref{equ:fl})--(\ref{equ:fo}) become
precision constraints on light hidden sector scalars~\cite{Green:2013rd, Assassi:2013gxa}.
For scalars with cubic couplings $\sim H\psi^3$, one finds the bound \cite{Assassi:2013gxa}
\beq
\Lambda \gtrsim 10^5 H \ . \label{equ:LB}
\eeq
This is a constraint on physics many orders of magnitude above the inflationary Hubble scale.
Using~(\ref{equ:HMp}), one can write the bound (\ref{equ:LB}) in terms of the Planck scale:
\beq
\Lambda \, \gtrsim\,  \left( \frac{r}{0.01}\right)^{1/2}\,  \Mp \ . \label{equ:EFTprec}
\eeq
It is a striking coincidence that a detection of primordial tensors, $r > 0.01$, would push the lower bound on $\Lambda$ to the Planck scale.
The bispectrum results of Planck would then imply constraints on Planck-suppressed couplings to hidden sectors.
Specifically,  we would learn that all hidden sector scalars  are either massive ($m_\psi \gg H$),  sequestered from  inflationary supersymmetry breaking ($A \ll H$),  or sequestered from  the inflaton itself ($\Lambda > \Mp$).

To understand the strength of a constraint of the form $\Lambda \gtrsim \Mp$, one should recognize that in parametrically controlled ultraviolet completions of gravity, the actual cutoff scale of an inflationary EFT is generally far below the Planck mass.
Thus, 
an  unambiguous detection of  
primordial tensors  would exclude  a broad  range of constructions involving light  hidden sector fields,
providing a powerful selection principle for ultraviolet completions of inflation.

\chapter{Elements of String Theory}
\label{sec:StringTheory}

String theory is the subject of a vast literature.\footnote{The fundamentals of the theory can be found in the classic textbooks~\cite{Green:1987sp, Green:1987mn, Polchinski:1998rq, Polchinski:1998rr}, as well as the lecture notes~\cite{Tong:2009np, Polchinski:1994mb, Kiritsis:1997hj, Bedford:2011uw, Forste:2001ah, Mohaupt:2002py, Sen:1998kr}. More recent advances are described in~\cite{Ibanez:2012zz, Becker:2007zj, Johnson:2003gi}. } Our aim in this section is to assemble the results that are most relevant for the study of string inflation (the subject of Chapters~\ref{sec:StringInflation} and \ref{sec:Examples}), making no pretense of completeness.  We will particularly focus on the four-dimensional effective actions arising in cosmologically realistic solutions of string theory.  Careful attention is paid to the problem of moduli stabilization, and de Sitter solutions are critically analyzed.

\section{Fundamentals}
\label{ssec:StringTheory}

\subsection{From Worldsheet to Spacetime}
\label{sssec:worldsheet}

An elementary starting point for string theory is the worldsheet action for a string, which defines a  (1+1)-dimen\-sional quantum field theory.
We will begin by describing bosonic string theory, and then turn to string theories whose worldsheet theories include fermionic fields.

\vskip 4pt
\noindent
{\it Bosonic string theory.}---The {\it{Polyakov action}} for a bosonic string propagating in $D$-dimensional Minkowski space~\cite{Deser:1976rb, Brink:1976sc} is
\begin{equation} \label{Polyakov}
S_{\rm P} = -\frac{1}{4\pi\alpha^{\prime}}\int \d^2\sigma \sqrt{-h} \hskip 1pt h^{a b} \hskip 1pt \partial_{a} X^{M}(\sigma) \partial_{b} X^{N}(\sigma)\hskip 1pt \eta_{MN}\ ,
\end{equation} where $X^{M}$, with $M = 0,\cdots, D-1$,
are the coordinates in the target spacetime;
$\sigma^{a}$, with $a=0,1$, are the coordinates on the string worldsheet; $h^{a b}$ is an independent metric on the worldsheet; and $2\pi\alpha^{\prime}$ is the inverse of the string tension.
The action~(\ref{Polyakov}) describes a two-dimensional field theory with $D$ scalar fields.  At the classical level, this theory is invariant under two-dimensional diffeomorphisms and under the Weyl symmetry $h_{a b} \mapsto e^{2\omega(\sigma)} \hskip 1pt h_{a b}$.  Famously, these classical symmetries are non-anomalous  if and only if $D=26$ \cite{Goddard:1973qh}.
The symmetries can be used  to set\footnote{This assumes that there  is no  topological obstruction to  the existence of a metric  that is flat everywhere.} $h_{a b} \mapsto \eta_{a b}$,  known as {\it conformal gauge}, so that the action takes the  more convenient form
\begin{equation} \label{PolyakovConf}
S_{\rm P} = -\frac{1}{4\pi\alpha^{\prime}}\int \d^2\sigma  \, \partial^{a} X^{M} \partial_{a} X_{M}\hskip 1pt \,,
\end{equation} in which the $X^M$ are  recognized as $D$ free fields  that respect a global $SO(D-1,1)$  symmetry.

Upon quantizing the string, one finds that the massless spectrum consists of a graviton $G_{MN}$, an antisymmetric tensor $B_{MN}$, and a scalar $\Phi$ known as the dilaton.  In addition, the spectrum contains massive excitations with scale set by $M_{\rm s} \equiv (\alpha^{\prime})^{-1/2}$.
The Polyakov action (\ref{Polyakov}) can be extended to a nonlinear $\sigma$-model action describing strings propagating in a target spacetime involving background profiles for the massless excitations:
\begin{align} \label{Polyakov2}
S_\sigma &\, =\, -\frac{1}{4\pi\alpha^{\prime}}\int \d^2\sigma \sqrt{-h}\, \Biggl( \left[h^{a b} G_{MN}(X) + \epsilon^{ab} B_{MN}(X) \right] \hskip 1pt \partial_{a} X^{M}\partial_{b} X^{N} \nonumber \\
& \hspace{4.5cm}+\, \alpha' \Phi(X) \, R(h) \Biggr) \ ,
\end{align} where $R(h)$ is the Ricci scalar constructed from $h_{ab}$.
Expanding the background fields around a given point, $X^{M} = X^{M}_{(0)} + \delta X^{M}$, one finds interaction terms such as
$h^{a b} \partial_P G_{MN}(X_{(0)}) \hskip 1pt \delta X^{P} \hskip 1pt \partial_{a} \delta X^{M}\partial_{b} \delta X^{N}$.
The nonlinear $\sigma$-model defined by (\ref{Polyakov2}) therefore describes an interacting quantum field theory.
When the gradients of the background fields are small in units of $\alpha^{\prime}$ --- and in particular, when all curvatures are small in string units --- these interactions can be treated perturbatively.
The corresponding expansion is known as the $\sigma$-model expansion or the $\alpha^{\prime}$ expansion.
Absence of anomalies in the quantum field theory defined by (\ref{Polyakov2})
requires that the background fields in the target spacetime obey certain differential equations that can be obtained order by order in the $\alpha^{\prime}$ expansion.
Consistency of string theory at the quantum level on the worldsheet therefore imposes equations of motion in the target spacetime~\cite{Friedan:1980jm}.
Remarkably, the equation of motion for $G_{MN}(X)$ at leading order in $\alpha^{\prime}$ is the Einstein equation!

The equations of motion for the background fields
can also be shown to follow from a $D$-dimensional spacetime action that parameterizes the interactions of the massless excitations of the bosonic string.
The idea is to construct an effective action in the sense described in Chapter~\ref{sec:EFT}: one imagines performing the path integral by first integrating out massive excitations of the string, leaving an effective action for the massless modes.  The theory that emerges at energies below the string scale $M_{\rm s}$
takes the form
(see \cite{Polchinski:1998rq} for details)
\begin{equation} \label{bosoniceffective}
S_{\rm B} = \frac{1}{2\kappa_D^2}\int \d^{D} X \sqrt{-G}\, e^{-2\Phi}\, \Biggl(R+ 4 (\partial \Phi)^2  -\frac{1}{2}|H_{3}|^2 - \frac{2(D-26)}{3\alpha^{\prime}} + {\cal{O}}(\alpha^{\prime})  \Biggr) \ ,
\end{equation}
where $\kappa_D$ is a coupling constant, and $H_3=\d B_2$ is the field strength of the antisymmetric tensor $B_{MN}$, or  equivalently of the two-form $B_2$.
Although the effective action (\ref{bosoniceffective}) lacks the good ultraviolet behavior of the full string theory (it violates perturbative unitarity at $E\sim M_{\rm s}$), it is nevertheless a convenient way to organize the interactions at energies below the cutoff, $E \ll M_{\rm s}$.  The omitted terms of higher order in $\alpha^{\prime}$ correspond to higher-dimension operators, including invariants constructed from the Riemann curvature of the target space.

In practice, the effective action  (\ref{bosoniceffective}) is obtained by computing scattering amplitudes for strings
via a path integral over worldsheets connecting initial and final states.
The path integral involves a sum over surfaces connecting the initial and final configurations, and the genus of the surface is a loop counting parameter: worldsheets of Euler number $\chi$ appear in the path integral with weight
\beq
e^{-\Phi\chi} = e^{-\Phi(2-2g)} \equiv g_{\rm s}^{2g-2}\ ,
\eeq
where $g$ is the genus of the worldsheet and $g_{\rm s} \equiv e^{\Phi}$ is the string coupling.
 Amplitudes are then defined order by order in the genus expansion, although except in special cases only one-loop results are available.
One can then ask which effective action in $D$-dimensional spacetime results in the same scattering amplitudes.
The amplitudes obtained at tree level in the genus expansion can be shown to follow from the effective action (\ref{bosoniceffective}), the very theory whose equations of motion enforce the absence of anomalies in the worldsheet theory (\ref{Polyakov2}).

In summary, the full $D$-dimensional action can   be expressed in a double expansion, in $g_{\rm s}$ and in $\alpha^{\prime}$.
The genus expansion corresponds to the
$\hbar$ expansion in the effective theory, while the $\alpha^{\prime}$ expansion controls the appearance of certain higher-dimension operators.
These expansions are controlled by vevs of dynamical fields, rather than by fundamental dimensionless parameters: the coupling `constant' in the genus expansion, $g_{\rm s}(\Phi)$, is the expectation value of the dilaton,
while
the expansion parameter of the $\sigma$-model is the curvature of the target spacetime  in units of $\alpha^{\prime}$.

\vskip 4pt
\noindent
{\it Superstring theories.}---The bosonic string theory defined by (\ref{Polyakov}) is unsuitable as a description of nature: the spacetime spectrum is devoid of fermions, and the theory suffers from a tachyonic instability~\cite{Polchinski:1998rq}.
Supersymmetric string theories are far more promising, and differ in important details.  Most fundamentally, the worldsheet actions involve additional fermionic terms: in the simplest case,
known as ${\cal N}=(1,1)$  worldsheet supersymmetry, the total action  in conformal gauge takes the form \cite{Green:1987sp}
\begin{equation}
S=S_{\rm P}+S_{\rm F} = -\frac{1}{4\pi\alpha^{\prime}}\int \d^2\sigma  \hskip 2pt \Bigl(\partial^{a} X^{M} \partial_{a} X_{M}-i\bar{\psi}^{M} \rho^{a}\partial_{a}\psi_{M} \Bigr)\ .
\end{equation}
Here, $\rho^a$ are two-dimensional Dirac matrices obeying the Dirac  (or Clifford) algebra
\begin{equation}
\{\rho^{a},\rho^{b}\}=-2\eta^{ab}\ ,
\end{equation} and $\psi^{M}$ is a Dirac spinor on the worldsheet that transforms as a  vector under Lorentz transformations  in the target space (which  correspond to global symmetry transformations of the worldsheet theory).
In terms of the two independent components of $\psi^{M}$,
\begin{equation}
\psi^{M} \equiv \binom{\psi_-^M}{\psi_+^M}\ ,
\end{equation}
the fermion action takes the form
\begin{equation}
S_{\rm{F}}=\frac{i}{2\pi\alpha^{\prime}}\int \d^2\sigma \Bigl(\psi_{-}^M\partial_{+}\psi_{-}^N + \psi_{+}^M\partial_{-}\psi_{+}^N \Bigr)\hskip 1pt \eta_{MN}\ , \label{rightmover}
\end{equation} where $\partial_{\pm} \equiv \frac{1}{2}(\partial_{\tau}\pm\partial_{\sigma})$, with
$\tau \equiv \sigma^0$ and $\sigma \equiv \sigma^1$.
The worldsheet fermions therefore separate into left-moving and right-moving modes.
The fermions $\psi^M_{\pm}$  contribute to the  central charge of the worldsheet field theory,  so that the  theory defined by $S= S_{\rm P} + S_{\rm F}$, with $S_{\rm P}$ given in (\ref{PolyakovConf}), has the critical dimension
$D=10$.

The action (\ref{rightmover}) for the worldsheet fermions does not completely determine the spacetime spectrum of the theory: one must also specify the periodicity of the fermions under transport around the closed string worldsheet.  Periodic fermions obeying
$\psi_{\pm}^{M}(\sigma  + \pi) = +\psi_{\pm}^{M}(\sigma)$ are said to be in the {\it{Ramond}} sector, while antiperiodic fermions with $\psi_{\pm}^{M}(\sigma  + \pi) = -\psi_{\pm}^{M}(\sigma)$ are said to be in the {\it{Neveu-Schwarz}} sector. This choice can be made separately for the left-moving and right-moving fermions, so that there are four possible sectors: NS-NS, R-R, R-NS, and NS-R.  The ten-dimensional effective actions describing the interactions of massless states of the superstring are supergravity theories involving additional fermionic and bosonic fields in comparison to (\ref{bosoniceffective}).  Bosonic fields in the target spacetime arise from  string states in the NS-NS and R-R sectors, while the R-NS and NS-R sectors give rise to spacetime fermions.

To construct a consistent closed string theory with spacetime fermions, it turns out to be necessary to impose a particular projection, the {\it{GSO  projection}},  on the spectrum.   This entails one further choice:  one can  perform identical GSO  projections in the R-NS and NS-R sectors, or opposite projections.   The former choice leads to  {\it{type IIB  string theory}}, which has a chiral spectrum in spacetime---in particular,  the two  gravitinos  have the same chirality.   The latter choice produces {\it{type IIA  string theory}}, which has a non-chiral spectrum.

Three other consistent superstring theories  are known.   To arrive at {\it{type I string theory}},  we consider the worldsheet parity operation $\Omega$, which reverses the orientation of the string worldsheet, and hence relates left-moving and right-moving modes.   In type IIB  string theory, the R-NS and NS-R sectors have the same spectra, so that worldsheet parity is a symmetry  of the theory, and it is consistent to project the spectrum onto states with $\Omega=+1$.   This operation,  which corresponds to gauging  the discrete symmetry of worldsheet parity,  leads to a theory of {\it{unoriented}} strings, because for any given string its  orientation-reversed image under $\Omega$  is also retained.   The projection removes one of the two gravitinos  from the spectrum,  yielding a theory with ${\cal N}=1$  supersymmetry in ten dimensions, the type I string.

The two remaining  theories also have ten-dimensional  ${\cal N}=1$  supersymmetry, but  have a different structure on the worldsheet.    While above we have discussed theories with left-moving and right-moving bosons, and left-moving and right-moving fermions,  it is  also consistent to take the left-moving sector to be that of the  bosonic string, and the right-moving sector to be that of the superstring.  Two  supersymmetric {\it{heterotic string theories}}  arise from this construction: the $SO(32)$  heterotic string, and the $E_8 \times E_8$  heterotic string.

\begin{figure}[h!]
   \centering
     \includegraphics[scale=0.7]{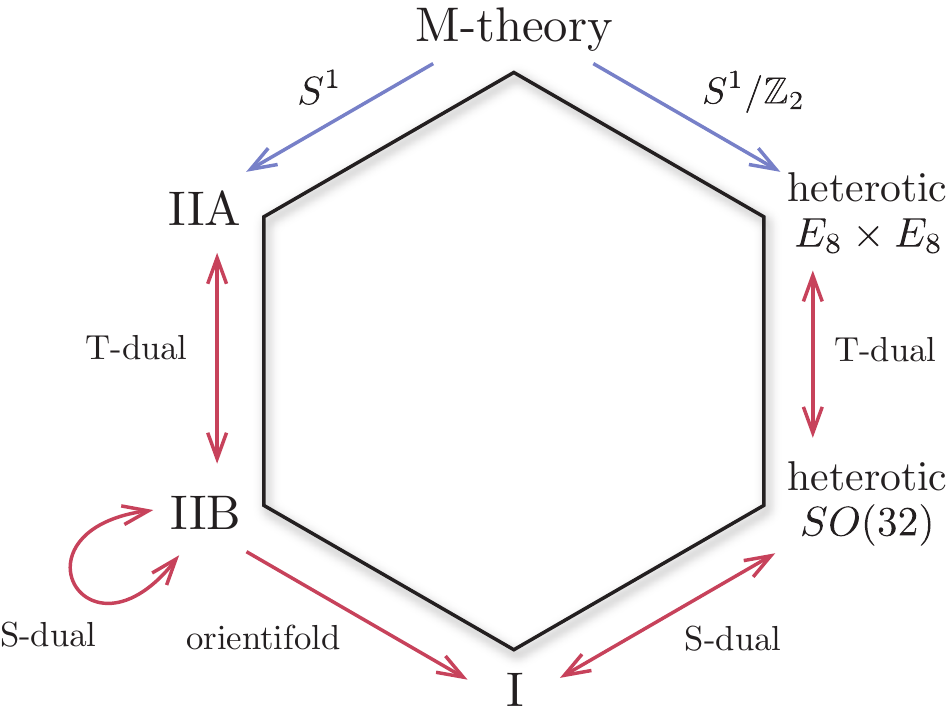}
   \caption{Dualities relating the supersymmetric string theories and M-theory.  S-duality exchanges strong coupling and weak coupling, while in compactification on a circle, T-duality exchanges momentum and winding.}
  \label{fig:Mtheory}
\end{figure}

The five superstring theories described above are  interrelated  by a number of dualities (see fig.~\ref{fig:Mtheory}), and correspond to different limits of an underlying theory which is sometimes called {\it M-theory}.

 \vskip 4pt
 \noindent
 {\it Supergravity limit.}---The low-energy limit of each of the consistent superstring theories is a ten-dimensional  {\it{supergravity}} theory.
We will now describe the corresponding effective actions for type~IIA and type~IIB string theory, focusing on the bosonic fields, which are directly relevant for obtaining classical solutions.  The actions for type I string theory and the heterotic string theories may be found in e.g.~\cite{Polchinski:1998rr}.

\vskip 4pt
The NS-NS sector of type II supergravity in ten dimensions contains the metric $G_{MN}$, the dilaton~$\Phi$ and the two-form~$B_2$. The action for these fields is
\beq
 S_{\rm NS}  = \frac{1}{2\kappa^2} \int \d^{10} X \sqrt{-G} \hskip 1pt e^{-2 \Phi} \left( R + 4 (\partial \Phi)^2 - \frac{1}{2} |H_3|^2 \right)\ , \label{equ:SNS}
\eeq
where $H_3 = \d B_2$.
The coupling constant~$\kappa^2$, corresponding to the Newton constant in ten dimensions, can be related to the string tension by comparing the worldsheet and supergravity actions: one finds~\cite{Polchinski:1998rq}
\beq
2 \kappa^2 = (2\pi)^7 (\alpha')^4\ .  \label{defkappa}
\eeq
In addition, type IIA supergravity has an R-R one-form $C_1$ and a three-form $C_3$.
The complete action then takes the form
\bea
S_{\rm IIA} =  S_{\rm NS} + S_{\rm R}^{\rm (IIA)} + S_{\rm CS}^{\rm (IIA)} \ ,
\label{equ:SIIAX}
\eea
where
\begin{align}
 S_{\rm R}^{\rm (IIA)}  &\, =\, - \frac{1}{4\kappa^2} \int \d^{10} X  \sqrt{-G}\, \left( |F_{2}|^2 + |\tilde F_{4}|^2\right)\ ,\\
 S_{\rm CS}^{\rm (IIA)} &\,=\, - \frac{1}{4 \kappa^2} \int B_{2} \wedge F_{4} \wedge F_{4}\ ,
\end{align}
with $F_{p} = \d C_{p-1}$ and $\tilde F_{4} = F_{4} + C_{1} \wedge H_{3}$.
The R-R fields in type IIB supergravity are a zero-form (scalar) $C_0$, a two-form $C_2$, and a four-form $C_4$ with self-dual field strength.  The complete action   is
\bea
S_{\rm IIB} =  S_{\rm NS} + S_{\rm R}^{\rm (IIB)} + S_{\rm CS}^{\rm (IIB)}\ , \label{equ:SIIBX}
\eea
where
\begin{align}
 S_{\rm R}^{\rm (IIB)}  &\,=\, - \frac{1}{4\kappa^2} \int \d^{10} X  \sqrt{-G}\, \left( |F_{1}|^2 + |\tilde F_{3}|^2 + \frac{1}{2}|\tilde F_{5}|^2\right)\ ,\\
 S_{\rm CS}^{\rm (IIB)} &\,=\, - \frac{1}{4 \kappa^2} \int C_{4} \wedge H_{3} \wedge F_{3}\ ,
\end{align}
with $F_{p} = \d C_{p-1}$, $\tilde F_{3} = F_{3} - C_{0} \wedge H_{3}$, and $\tilde F_{5} = F_{5} - \frac{1}{2}C_{2} \wedge H_{3}+ \frac{1}{2} B_{2} \wedge F_{3}$.
In addition, one must   impose the self-duality constraint
\beq
\tilde F_{5} = \star \tilde F_{5}\ .
\eeq

We have written the NS-NS sector (\ref{equ:SNS}) of  the actions (\ref{equ:SIIAX}) and (\ref{equ:SIIBX})  in `string frame', meaning that the Ricci scalar $R$ appears with the dilaton-dependent prefactor $e^{-2 \Phi}$.  This frame is convenient for comparing to the results of string perturbation theory.
However, for many questions involving gravity, it is more practical to work in {\it Einstein frame}, in which the dilaton prefactor is absent.
The  action can be written in the Einstein frame by performing the Weyl rescaling
$$G_{E,MN} \equiv e^{-\Phi/2}G_{MN}\,.$$
In type IIB string theory, it is convenient to define the combinations
\begin{align}
G_3 &\equiv F_3 - \tau H_3 \ , \label{g3def} \\
\tau &\equiv C_0 + i e^{-\Phi} \ ,
\end{align} in terms of which the action~(\ref{equ:SIIBX}), written in Einstein frame, takes the form
\begin{align}
S_{\rm IIB}  &\,=\, \frac{1}{2\kappa^2} \int \d^{10}X \sqrt{-G_{E}} \left[ R_E - \frac{|\partial \tau|^2 }{2({\rm Im}(\tau))^2} - \frac{|G_3|^2}{2\hskip 1pt  {\rm Im}(\tau)} - \frac{ |\tilde F_5|^2}{4}\right] \nonumber \\ & \ \  \ \ \ -\, \frac{i}{8 \kappa^2 } \int \frac{C_4 \wedge G_3 \wedge \bar G_3}{{\rm Im}(\tau)}  \ .  \label{SIIB}
\end{align}
The action (\ref{SIIB}) is the starting point  for our discussion of type IIB  flux compactifications in \S\ref{fluxcompactification}.

\subsection{D-branes}
\label{sec:Dbranes}

In addition to fundamental strings, string theory contains solitonic objects.  Most famous are D-branes, which are charged under the gauge symmetry of the R-R fields.  A D$p$-brane is an object with $p$ spatial dimensions that is charged under $C_{p+1}$ via the electric coupling\footnote{Type IIA string theory contains stable D$p$-branes with $p$ {\it even}, while type IIB string theory contains stable D$p$-branes with $p$ {\it odd}. Type I string theory has stable D$p$-branes with $p=1,5,9$.   D$p$-branes and D$(6-p)$-branes are charged under R-R potentials $C_{p+1}$ and $C_{7-p}$, whose field strengths $F_{p+2}$ and $F_{8-p}$ are dual to each other, $\star F_{8-p} = F_{p+2}$.  Thus, D$(6-p)$-branes carry magnetic charge under $C_{p+1}$.
In string theories with an NS-NS two-form $B_2$, fundamental strings carry charge under $B_2$ and are stable. Moreover, there is an additional soliton, the NS5-brane, which is magnetically charged under $B_2$.    }
\beq
S_{\rm CS}  = \mu_p \int_{\Sigma_{p+1}} C_{p+1}\ , \label{equ:CS}
 \eeq
where $\Sigma_{p+1}$ is the D$p$-brane worldvolume and $\mu_p$ is the brane charge.
The {\it Chern-Simons} action~(\ref{equ:CS}) is simply a higher-dimensional generalization of the
coupling of a charged point particle to
a
gauge potential, $\int \d x^\mu A_\mu \equiv \int A_1$.

A  defining characteristic  of D-branes  is that they are surfaces  on which strings can end.  The D in D-brane stands for Dirichlet, referring to the fact that open strings  ending on a D-brane  have Dirichlet boundary conditions in the directions transverse to the brane, i.e.~the open string endpoints cannot leave the D-brane.   Open strings have Neumann boundary conditions in the directions along  the spatial extent of a D$p$-brane  with $p>0$:  the endpoints are free to slide along the D-brane.

Quantization of the open strings residing on a D-brane yields a spectrum of bosonic and fermionic fields living on the worldvolume.  At the massless level, one finds scalar fields parameterizing fluctuations of the D-brane position, a worldvolume gauge field $A_a$ with field strength $F_{ab}$, and their superpartners.
The effective action for these fields  is an important object, because it encapsulates the low-energy dynamics of the D-brane.  Just as the low-energy effective action for the massless modes of a closed string could be determined by computing closed string scattering amplitudes in perturbation theory, the low-energy effective action that governs the massless fields on a D-brane can be derived by computing scattering amplitudes involving open strings ending on the D-brane.  Moreover, by computing amplitudes in which open strings on the D-brane interact with closed strings, one can determine the couplings of the D-brane to a closed string background.

\vskip 4pt
A general background  solution of type II or type I\hskip 1pt\footnote{The heterotic  string theories contain no R-R  fields, and correspondingly lack D-branes.} string theory will involve profiles for all the massless bosonic fields.
We would like to understand the  effective action for the light fields on a D$p$-brane in such a background.  For simplicity, we will restrict attention to the bosonic sector.

\vskip 4pt
\noindent
{\it D-brane action.}---An uncharged $p$-dimensional membrane  moving in a curved spacetime  with metric $G_{MN}$ can be described by the {\it Dirac} action, which is simply a higher-dimensional generalization of the Polyakov action~(\ref{Polyakov}):
\beq
S_{\rm D} = -  T_p \int \d^{p+1} \sigma \,  \sqrt{- \det(G_{ab} )}\ ,
\eeq
where
\beq
G_{ab} \equiv \frac{\partial X^M}{\partial \sigma^a} \frac{\partial X^N}{\partial \sigma^b} G_{MN}\ .
\eeq
Here, $G_{ab}$ is the pullback of the metric of the target spacetime, and $T_p$  is the tension of the membrane.

Next, we recall {\it{Born-Infeld}}  theory, a non-linear generalization of Max\-well's electromagnetism. The Born-Infeld  action in $p+1$  flat spacetime dimensions, for an Abelian gauge field $A_a$ with field strength strength $F_{ab}$, is
\begin{align}
S_{\rm BI} &\, =\, - Q_p \int  \d^{p+1} \sigma \,  \sqrt{- \det(\eta_{ab} + 2\pi \alpha' F_{ab})} \nonumber \\
&\,=\, - Q_p \int \d^{p+1} \sigma \,  \left( 1 + \frac{(2\pi \alpha')^2}{4} F_{ab} F^{ab} + \cdots \right) \ ,
\end{align}
where
$Q_p$ is a  constant with the dimensions of a $p$-brane  tension.

By computing open string amplitudes, and open+closed amplitudes,  one finds that  the action for a D$p$-brane  in a general closed string background involves a combination of the Dirac and Born-Infeld actions, the {\it Dirac-Born-Infeld} action
\beq
S_{\rm DBI} = - g_{\rm s} T_p \int \d^{p+1} \sigma \, e^{-\Phi} \sqrt{-\det(G_{ab} + {\cal F}_{ab})}\ , \label{equ:DBIXX}
\eeq
where
${\cal F}_{ab}$ is the gauge-invariant field strength
\beq
{\cal F}_{ab} \equiv B_{ab} + 2\pi \alpha' F_{ab}\ ,
\eeq and $B_{ab}$  is the pullback of $B_{MN}$  onto the D-brane worldvolume.
From the string amplitude computations one infers that the D$p$-brane tension is\footnote{In a background  with constant dilaton $\Phi$,  one has $g_{\rm s} e^{-\Phi}=1$.}
\beq
T_p \equiv \frac{1}{(2\pi)^p \hskip 1pt g_{\rm s}\hskip 1pt (\alpha^{\prime})^{ (p+1)/2}}\ ,  \label{deftp}
\eeq
leading to the important result that D$p$-branes are heavy at weak string coupling, $g_{\rm s} \ll 1$.
Next, the D-brane  tension $T_p$ can be related to the charge $\mu_p$ appearing in (\ref{equ:CS}).  The stable D-branes of type~I and type II string theories are {\it{BPS objects}}, and preserve half of the spacetime supersymmetries.  A BPS D$p$-brane ($p>0$) corresponds to a higher-dimensional generalization of an extremal black hole, with tension equal to its charge when expressed in appropriate units.  In our conventions, one finds $\mu_p = g_{\rm s} T_p$.

The Chern-Simons action in the presence of background fields in the target space and on the D-brane worldvolume takes the form
\beq
S_{\rm CS} = i\, \mu_p \int_{\Sigma_{p+1}}
\sum_{n} C_n \wedge e^{{\cal F}} \,
\ , \label{equ:NCS}
\eeq
where the sum runs over the R-R $n$-forms  of the  theory in question, and only $(p+1)$-forms contribute to  the integral in (\ref{equ:NCS}).
The complete bosonic action for D-branes in a supergravity background is then the  sum of the Dirac-Born-Infeld action (\ref{equ:DBIXX}) and the Chern-Simons action (\ref{equ:NCS}),
\begin{equation}
S_{{\rm D}p} = S_{\rm DBI} + S_{\rm CS}\ .  \label{DBIplusCS}
\end{equation}
When $N$~D$p$-branes  coincide, the worldvolume gauge theory  becomes non-Abelian, and the action becomes much more complicated, with a  potential involving commutators of the worldvolume scalars --- see \cite{Myers:1999ps}.

\vskip 4pt
\noindent
{\it D-branes as sources.}---The coupling of D-branes  to the background
fields has important consequences: in addition to responding to the closed string solution in which it is embedded,  a D-brane  contributes to the  profiles of the  massless closed string fields.  Specifically, a D-brane provides a  localized source of stress-energy  and R-R  charge, causing it to source curvature and R-R  fluxes in proportion to its  tension and charge.    Incorporating this `backreaction' is sometimes difficult, as explained in \S\ref{sss:approximations}.

The supergravity solution  sourced by one or more D$p$-branes corresponds to a spatially-extended extremal black hole, or {\it{extremal $p$-brane}}.
For $N$  coincident D$p$-branes, the  characteristic radius of curvature $r_+$  is given by \cite{Horowitz:1991cd}
\begin{equation}
r_+^{7-p} = d_p \hskip 2pt g_{\rm s} N (\alpha^{\prime})^{\frac{1}{2}(7-p)}\ ,  \label{curvaturelimit}
\end{equation}
where $d_p \equiv (4\pi)^{\frac{1}{2}(5-p)} \hskip 2pt \Gamma\left(\tfrac{1}{2}(7-p)\right)$.
The dilaton profile in the radial ($r$) direction takes the form
\begin{equation}
e^{\Phi} =g_{\rm s} \left(1+\left(\frac{r_+}{r}\right)^{7-p} \right)^{\frac{1}{4}(3-p)}\ . \label{dilatonlimit}
\end{equation}
A classical supergravity description  is valid if the curvatures are small in string units and moreover $g_{\rm s} \ll 1$.
For $p<7$, we find from (\ref{curvaturelimit}) that the curvature is small at large $g_{\rm s} N$ --- see \S\ref{sss:approximations} for further discussion of this point.
Moreover, for  the important special case $p=3$,  the dilaton is constant, and  can be small everywhere:  D3-branes decouple from the dilaton.
Thus,   if $p=3$  and
\begin{equation}
1 \ll g_{\rm s} N \ll N\ ,
\end{equation}  the $\alpha^{\prime}$  and $g_{\rm s}$ corrections to  leading-order classical supergravity  can be neglected everywhere.
On the other hand, for  $p \neq 3$ the non-trivial dilaton profile (\ref{dilatonlimit}) presents an obstacle  to  extending the classical supergravity solution over the entire spacetime.
See \cite{MAGOO} for a more  extensive  review of extremal $p$-brane solutions.

\section{Compactification}  \label{sec:compactification}

For our purposes, a {\it{solution}} of string theory is a configuration of the massless fields that solves the equations of motion of the effective theory, and hence leads to a worldsheet theory without anomalies.  
In
supersymmetric string theories, a geometric solution involves a ten-dimensional spacetime ${\cal M}_{10}$, and the solutions that are most
relevant for cosmology include a spacetime ${\cal M}_4$ with four large dimensions.  We therefore consider geometries of the form
\begin{equation} \label{vacconf}
{\cal M}_{10} = {\cal M}_4 \times X_{6}\ ,
\end{equation} where $X_{6}$ is a compact six-manifold.  This is referred to as a {\it compactification} of string theory on~$X_6$.

\vskip 4pt
\noindent
{\it Vacuum compactifications.}---We will distinguish vacuum configurations, i.e. solutions of the  ten-dimensional vacuum Einstein equations\footnote{The Einstein equations receive corrections in the $\alpha^{\prime}$ expansion, which can be important at large curvatures.} without sources of stress-energy, from solutions involving sources, and begin by considering vacuum solutions.
A suitable ansatz for vacuum configurations is
\begin{equation} \label{unwarped}
G_{MN}\hskip 1pt  \d X^{M} \d X^{N} =   \eta_{\mu\nu}\d x^{\mu} \d x^{\nu} +   g_{mn}\d y^{m} \d y^{n}\ ,
\end{equation}
where $y^m$, $m=1,\ldots, 6$ are coordinates on $X_{6}$, and  $g_{mn}$ is a metric on $X_{6}$. Such a geometry is an allowed vacuum configuration if and only if $G_{MN}$ solves the ten-dimensional vacuum Einstein equations, i.e.~if and only if $R_{\mu \nu} = R_{mn}=0$.  Thus, in vacuum solutions the compactification manifold must have vanishing Ricci tensor.
The best-understood non-trivial vacuum configurations take the form (\ref{vacconf}) with $X_6$ a Calabi-Yau three-fold.

\vskip 4pt
\noindent
{\it Warped compactifications.}---Vacuum configurations of the form (\ref{vacconf}), solving the vacuum Einstein equations, are a simple and well-understood starting point.  However, the ten-dimensional effective action involves fields beyond just the metric, and these fields can contribute stress-energy.  Furthermore, the extended objects described above (fundamental strings, D-branes, and NS5-branes) are local sources of stress-energy and of charge.
In non-vacuum solutions containing these sources --- or carrying the corresponding charges without any local sources --- the compactification manifold is generally not Ricci-flat.  For non-vacuum configurations with maximal symmetry in the noncompact spacetime, the product ansatz (\ref{warped}) is generalized to a `warped' product:
\begin{equation} \label{warped}
G_{MN}\hskip 1pt  \d X^{M} \d X^{N} =   e^{2A(y)} g_{\mu\nu}\d x^{\mu} \d x^{\nu} + e^{-2A(y)}  g_{mn}\d y^{m} \d y^{n}\ ,
\end{equation}
where now $g_{\mu\nu}$ is the metric of a maximally symmetric spacetime, the {\it{warp factor}} $A(y)$  is a function on $X_{6}$,
and the internal metric $g_{mn}$ is not necessarily Ricci-flat.

\vskip 4pt
\noindent
{\it Supersymmetric compactifications.}---The supergravity actions described in \S\ref{sssec:worldsheet} enjoy ${\cal N}=1$ or ${\cal N}=2$ supersymmetry in ten dimensions, but the solutions of the equations of motion need not preserve supersymmetry. Nevertheless, the best-understood solutions of string theory are supersymmetric,\footnote{Compactifications that break supersymmetry at the Kaluza-Klein scale (for  constructions with stabilized moduli, see e.g.~\cite{Dine:2006gx,Silverstein:2007ac,Haque:2008jz}) present an important alternative to supersymmetric compactifications.} for reasons that we briefly explain.  The first reason is geometric: the Ricci-flatness condition required in vacuum solutions is closely tied to reduced holonomy (see Appendix~A
 for details).
Suitably reduced holonomy leads to the existence of invariant spinors and hence to unbroken supersymmetry in four dimensions.  Most notably, Calabi-Yau three-folds have holonomy $SU(3)$ and correspondingly preserve one quarter of the ten-dimensional  supersymmetry:  a  Calabi-Yau compactification of type II  string theory has ${\cal N}=2$ supersymmetry in four dimensions.  The second reason is that unbroken supersymmetry provides unrivaled theoretical control,  by guaranteeing stability and by imposing intricate relations among the couplings in the effective theory.  A third reason is the historical and continuing interest in finding solutions of string theory with ${\cal N}=1$ supersymmetry broken near the electroweak scale, in order to address the hierarchy problem.

Inflationary configurations necessarily break supersymmetry, but a fruitful strategy is to study solutions with minimal (${\cal N}=1$) supersymmetry in four dimensions, and use these as the foundation for determining the effective action in ${\cal N}=0$ solutions.
Before  examining  supersymmetric compactifications, we will  first briefly  describe the structure of  the corresponding supergravity theories.

\vskip 4pt
\noindent
{\it ${\cal N}=1$ supergravity in four dimensions.}---The bosonic fields of a  general four-dimensional ${\cal N}=1$ supergravity theory  are the metric $g_{\mu \nu}$, gauge potentials $A_\mu^a$, and complex scalar fields $\phi^i$.   The low-energy interactions of the scalars
are encoded by the {\it{superpotential}} $W(\phi^i)$, which is a holomorphic function of the $\phi^i$, and by the {\it{K\"ahler potential}} $K(\phi^i, \bar \phi^i)$, which is a real analytic function of the fields.
In the absence of  gauge interactions, the Lagrangian for the scalar fields is
\beq
{\cal L}_\Phi = - K_{i \bar \jmath} \, \partial^\mu \phi^i \partial_\mu \bar \phi^j  - V_{F}\ ,  \label{L4dn1}
\eeq
where $K_{i \bar \jmath} \equiv \partial_i \partial_{\bar \jmath} K$ is the K\"ahler metric.
The
F-term potential $V_{F}$ appearing in (\ref{L4dn1}) is
\beq
V_{F}(\phi^i, \bar \phi^i) = e^{K/M_{\rm pl}^2} \left[ K^{i \bar \jmath} D_i W \overline{D_j W} - \frac{3}{M_{\rm pl}^2} |W|^2 \right] \ ,  \label{ftermpotdef}
\eeq
where $ K^{i \bar \jmath}$ is the inverse K\"ahler metric and $D_i W \equiv \partial_i W + \frac{1}{M_{\rm pl}^2} (\partial_i K) W$.

A primary task in studying a string compactification with ${\cal N}=1$ supersymmetry  is to compute the  superpotential and K\"ahler potential in terms of geometric data.   Through (\ref{L4dn1}) and (\ref{ftermpotdef}) these data determine the four-dimensional effective theory, to leading order in the low-energy (derivative) expansion.

\subsection{Dimensional Reduction}  \label{dimensionalred}

To compute the four-dimensional effective action of a string compactification, one begins with the appropriate ten-dimensional action and performs a  Kaluza-Klein reduction.
In order to develop intuition for this process, we will begin with a simple example.

\vskip 4pt
Consider  the ten-dimensional  geometry
\begin{equation} \label{breathing}
G_{MN}\hskip 1pt  \d X^{M} \d X^{N} =   e^{-6u(x)} g_{\mu\nu}\d x^{\mu} \d x^{\nu} + e^{2u(x)}\hat{g}_{mn}\d y^{m} \d y^{n}\ ,
\end{equation} where $\hat{g}_{mn}$ is a reference metric with fixed volume,
\begin{equation}
 \int_{X_6} \d^6 y\,\sqrt{\hat{g}} \equiv  {\cal V}\ ,
\end{equation}  while $e^{u(x)}$ is a `breathing mode'  that represents the variations in size of the internal space $X_6$ as a function of  the four-dimensional coordinate $x^\mu$.
The factor of $e^{-6u(x)}$ in the first term is a convenient choice  for which  the gravitational action in four dimensions  will  appear in Einstein frame.
We now examine the dimensional reduction of the Einstein-Hilbert term
  \beq
S_{\rm EH}^{(10)} = \frac{1}{2 \kappa^2} \int \d^{10} X \sqrt{-G}\hskip 2pt e^{-2\Phi} R_{10}  \ , \label{equ:SEH}
\eeq
where $R_{10}$ is the Ricci scalar constructed from $G_{MN}$.   We would like to express $R_{10}$ in terms of $R_{4}$ and $\hat{R}_{6}$, the Ricci scalars  constructed from  $g_{\mu\nu}$ and $\hat{g}_{mn}$, respectively.
For this purpose we note that if two D-dimensional metrics $g_{MN}$ and $\bar{g}_{MN}$ are related by  the conformal rescaling
\begin{equation}
\bar{g}_{MN} = e^{2\omega(x)} g_{MN} \ ,
\end{equation} then the corresponding Ricci scalars  are related by
\begin{align}
e^{2\omega} \bar{R} &= R - 2(D-1)\nabla^2 \omega - (D-2)(D-1)g^{MN}\nabla_M \omega \nabla_N \omega \ .
\end{align}
Similarly, the Laplacians constructed from $g_{MN}$ and $\bar{g}_{MN}$ are related by
\begin{equation}
e^{2\omega} \bar{\nabla}^2 = \nabla^2 + (D-2)g^{MN}\nabla_M \omega \nabla_N   \ .
\end{equation}
Using these results, we find
\begin{equation}
S_{\rm EH}^{(10)} = \frac{1}{2 \kappa^2} \int \d^{4} x\hskip 1pt \sqrt{-g}\hskip 2pt  \int_{X_6} \d^6 y \sqrt{\hat{g}}\, e^{-2\Phi} \left(R_{4}+e^{-8u}\hat{R}_{6} +
12 \partial_{\mu} u \partial^{\mu} u \right) \ . \label{dimred}
\end{equation}
If the string coupling $g_{\rm s} \equiv e^{\Phi}$  is constant over the internal space, then the four-dimensional Einstein-Hilbert  term can be written
\begin{equation}
S_{\rm EH}^{(4)} = \frac{M_{\rm pl}^2}{2} \int \d^{4} x \hskip 1pt  \sqrt{-g}\hskip 2pt R_{4} \ ,
\end{equation} with the four-dimensional Planck mass defined as
\begin{equation}
M_{\rm pl}^2 \equiv \frac{{\cal V}}{g_{\rm s}^2\hskip 1pt \kappa^2}\ . \label{equ:4dPlanck}
\end{equation}

We recognize the combination of derivatives of $u(x)$ appearing in (\ref{dimred}) as the kinetic term for a four-dimensional scalar field $u(x)$.   This field is a {\it{modulus}} corresponding to a spacetime-dependent deformation of the ten-dimensional solution.
As we will see below, in Calabi-Yau compactifications the breathing mode $u$ corresponds to one of the K\"ahler moduli: the kinetic term for $u$ in
(\ref{dimred}) follows from the K\"ahler potential
\begin{equation}
K = - 3\, \ln\, (T+\bar{T})\ ,
\end{equation} where we have set $\Mp =1$, and $T$ is a complex scalar field with ${\rm{Re}}(T) = e^{4u}$.
(The imaginary part of $T$  comes from the dimensional reduction of the  four-form potential: see \S\ref{sec:axions}.)

Notice that the Ricci scalar $\hat{R}_{6}$ yields a potential term for  the scalar $u$:
positive internal curvature ($\hat{R}_{6}>0$) contributes a negative potential term $V \propto -e^{-8u}$ in four dimensions, driving the compactification toward small volume,
while negative internal curvature contributes a positive potential term $V \propto +e^{-8u}$, leading to a decompactification instability.   In Ricci-flat  compactifications,  the  internal curvature term is absent and $u$ has vanishing potential in the classical theory.

More general Kaluza-Klein reductions involve both  more complicated ten-dimensional actions, for example involving $p$-form fields,  as well as  geometric deformations that generalize the very simple  breathing mode  described above.   However, the principles  underlying the general analysis are captured by the above example.

\subsection{Moduli}  \label{sss:moduli}

In the simple Kaluza-Klein reduction described above,  the breathing mode corresponding to an overall dilation of the internal space gave rise to a four-dimensional scalar field $u(x)$  parameterizing  spacetime-dependent changes in the compactification volume.  We will now describe the analogous moduli fields that arise in Calabi-Yau compactifications.
To simplify the presentation, we primarily discuss four-dimensional scalars, i.e.~moduli, leaving the actions for vector and tensor fields to the references.\footnote{A complete treatment can be found in \cite{Grimm:2004uq}, which we follow in this section.
See e.g.~\cite{Becker:2007zj} for  background on Calabi-Yau geometry.}

\vskip 4pt
\noindent
{\it Calabi-Yau compactifications with ${\cal N}=2$ supersymmetry.}---We begin by summarizing the effective theory  that results from Kaluza-Klein reduction in Calabi-Yau compactifications of type II string theory.  Consider the ten-dimensional geometry (\ref{unwarped}), with  $g_{mn}$ the Ricci-flat metric on a Calabi-Yau three-fold $X_{6}$.
Compactification of type II string theory on this background leads to a four-dimensional theory with ${\cal N}=2$ supersymmetry.\footnote{Calabi-Yau compactifications of type I string theory, or of the heterotic string, yield ${\cal N}=1$ supersymmetry.  We will primarily discuss type II compactifications.}
The geometric moduli of this compactification are scalar fields corresponding to deformations of the metric $g_{mn}$ that preserve the Calabi-Yau condition:
the {\it K\"ahler moduli} are deformations of the K\"ahler form
\begin{equation}
J \equiv i\,g_{i\bar \jmath} \, \d z^{i}\wedge \d\bar{z}^{\bar \jmath}\ ,
\end{equation} where $z^{i}$, $\bar{z}^{\bar \jmath}$, with $i,\bar \jmath =1,\ldots, 3$ are complex coordinates on  $X_{6}$, while {\it complex structure moduli} are deformations of the complex structure on $X_{6}$.
To parameterize the moduli, we introduce a set of harmonic (1,1)-forms $\omega^I$, $I=1,\ldots, h^{1,1}$,  comprising a basis for the Dolbeault cohomology group~$H^{1,1}$, as well as a set of harmonic (1,2)-forms $\chi^A$, $A=1,\ldots, h^{1,2}$, that form   a basis for $H^{1,2}$.  In terms of this basis, the
K\"ahler form is
\begin{equation}
J = t^I(x)\,  \omega_I\ ,
\end{equation}  where $t^I(x)$ are $h^{1,1}$ four-dimensional scalar fields, the K\"ahler moduli. Similarly, complex structure deformations may be written
\begin{equation}
\delta g_{ij} = \frac{i}{||\Omega||^2} \, \zeta^A(x) (\chi_{A})_{i\bar \imath \bar \jmath}\, \Omega^{\bar \imath \bar \jmath}_{~~j} \ ,
\end{equation}
where $\Omega$ is the holomorphic (3,0)-form of $X_{6}$ and $||\Omega||^2 \equiv \frac{1}{3!} \Omega_{ijk}\bar{\Omega}^{ijk}$.
The $h^{1,2}$ four-dimensional scalar fields $\zeta^A(x)$ are the complex structure moduli.

Additional scalar fields arise from expanding the NS-NS and R-R  potentials in the bases of harmonic forms.   We henceforth specialize to type IIB  string theory, where the relevant forms are
$B_2$, $C_2$, $C_4$, with the expansions
\begin{align}
B_2 &= B_2(x) + b^{I}(x) \hskip 1pt\omega_{I}\ ,\\
C_2 &= C_2(x) + c^{I}(x)\hskip 1pt\omega_{I}\ ,\\
C_4 &= \vartheta^I(x) \hskip 1pt \tilde{\omega}_{I}\ .  \label{c4}
\end{align}
Here, $B_2(x)$ denotes the  four-dimensional two-form $B_{\mu\nu}(x)\d x^{\mu}\d x^{\nu}$, to be distinguished from  the ten-dimensional two-form $B_2$, and similarly for $C_2$.
In (\ref{c4}),  we have suppressed vector field  contributions to the final equality (see \cite{Grimm:2004uq} for the complete expression), and have defined $\tilde{\omega}_{I}$, $I=1,\ldots, h^{1,1}$, as a basis for $H^{2,2}$.  Finally, the  dilaton $\Phi$ and the R-R zero-form $C_0$  give rise to two  more real scalars.

The scalar fields just described  appear in multiplets  of four-dimensional ${\cal N}=2$ supersymmetry.   The $4 h^{1,1}$ scalars $v^I,\vartheta^I,b^{I},c^{I}$  furnish the  bosonic content of $h^{1,1}$  hypermultiplets, while the $h^{1,2}$ real scalars $\zeta^A$ appear in ${\cal N}=2$ vector multiplets  (in combination with the vector fields $V^{\mu}_A \sim C^{\mu}_{ij\bar{k}} (\chi_{A})^{ij\bar k}$ from the dimensional reduction of $C_4$, which we have suppressed above).   Finally,  $\Phi$, $C_0$, $B_2$, $C_2$ form the `universal hypermultiplet',  after dualizing the two-forms to scalars in four dimensions --- see Appendix~A.

\vskip 4pt
\noindent
{\it Calabi-Yau orientifolds with ${\cal N}=1$ supersymmetry.}---Type II Calabi-Yau compactifications with unbroken ${\cal N}=2$ supersymmetry  do not yield realistic models of Nature: in particular, ${\cal N}=2$ supersymmetry does not allow fermions in chiral representations  of gauge groups.  More promising are type~II  compactifications that include local sources,  such as D-branes,  in addition to $p$-form fluxes.  The resulting  gauge theories  can be rich enough to include  the  Standard Model,  and  spontaneous breaking of supersymmetry in a metastable vacuum is plausibly achievable.

A fundamental consistency requirement  for flux compactifications with D-branes is cancellation  of all tadpoles associated with the charge  and tension of the sources.   Most dramatically,  the gravitational tadpole associated to the positive tension of a D-brane  requires the presence of a  negative-tension source \cite{Giddings:2001yu}.  The best-understood negative-tension objects are {\it{orientifold planes}}, which are non-dynamical  extended objects that appear at the fixed point loci of an  involution ${\cal O}$ that reverses the orientation of the string worldsheet.

We will describe  the essential aspects  of orientifolds here, referring the reader to \cite{Polchinski:1998rq} for a complete treatment.
An orientifold action  ${\cal O}$ is a symmetry that includes the worldsheet orientation reversal $\Omega_{ws}$. The orientifold  actions of primary interest here take the form
\begin{equation}
{\cal O} = (-1)^{F_L}\Omega_{ws}\hskip 1pt\sigma\ , \label{o3o7}
\end{equation}  where $(-1)^{F_L}$ is the worldsheet fermion number in the left-moving sector ---
cf.~the decomposition implied by (\ref{rightmover}) --- and the geometric involution $\sigma$ reverses the sign of the holomorphic $(3,0)$  form $\Omega$ of $X_6$, but leaves the metric  and complex structure invariant.  The fixed point loci  of an orientifold action  of the form (\ref{o3o7})  are  points or four-cycles in $X_6$. Because the  geometric action on the noncompact dimensions is trivial,  the resulting orientifold planes have three or seven  spatial dimensions, and are known as O3-planes  and O7-planes, respectively.

Under the action (\ref{o3o7}), the cohomology group $H^{1,1}$  can be decomposed as
\begin{equation}
H^{1,1} = H^{1,1}_+\oplus H^{1,1}_- \ ,
\end{equation}
with the subscript denoting the  parity  of the corresponding two-forms under the orientifold action.
Correspondingly, the basis $\omega^I$, $I=1,\ldots, h^{1,1}$  for $H^{1,1}$ decomposes into a  basis for the even eigenspace, $\omega^i$, $i=1,\ldots, h^{1,1}_+$, and a basis for the odd eigenspace $\omega^\alpha$, $\alpha=1,\ldots, h^{1,1}_-$.

To understand the effect of orientifolding  on the four-dimensional fields,  we  note that $v^I$, $\vartheta^I$, $\Phi$, $C_0$  are even under the orientifold action, while $\zeta^A$, $b^{I}$, $c^{I}$, $B_2(x)$, $C_2(x)$ are odd.   Invariant four-dimensional fields  arise from even ten-dimensional  fields  expanded in terms of even forms, or from odd ten-dimensional  fields  expanded in terms of odd forms.
The  K\"ahler  form can be written
 \begin{equation}
J = t^i(x) \hskip 1pt \omega_i\ ,
\end{equation}
so that the orientifold-invariant K\"ahler  moduli are the $h^{1,1}_+$ real scalars $t^i$, which measure the volumes  of two-cycles that are even under the involution.
Similarly, noting that the orientifold action projects out  the four-dimensional two-forms $B_2(x)$ and $C_2(x)$, we have the invariant fields  (again omitting vector contributions)
\begin{align}
B_2 &=   b^{\alpha}(x)\hskip 1pt\omega_{\alpha}\ , \\
 C_2 &=  c^{\alpha}(x)\hskip 1pt\omega_{\alpha}\ , \\
 C_4 &= \vartheta^i(x)\hskip 1pt \tilde{\omega}_{i}\ .  \label{c4o}
\end{align}   Likewise, the invariant complex structure moduli are $\zeta^{\alpha}$, for
$\alpha=1,\ldots, h^{1,2}_{-}$.
Finally, $\Phi$ and $C_0$ are automatically invariant.

It is important to  assemble the invariant scalars into the bosonic components of chiral multiplets of four-dimensional ${\cal N}=1$ supersymmetry,  i.e.\  to determine the proper K\"ahler coordinates on the moduli space.
First of all, the axion $C_0$ and dilaton $\Phi$ combine to form the complex axiodilaton,
\begin{equation}
\tau= C_0 + i e^{-\Phi}\,.  \label{theaxiodilaton}
\end{equation}
The complex structure moduli $\zeta^{\alpha}$ are automatically  good K\"ahler coordinates.
The `two-form scalars'
$b_{\alpha}$ and $c_{\alpha}$ form the complex combination
\begin{equation}
G_{\alpha} \equiv c_{\alpha}-\tau\,b_{\alpha} \,.  \label{gscalar}
\end{equation}
To go further, we note that the compactification volume ${\cal V}$ can be written in terms of the K\"ahler form $J$ as follows:
\beq
{\cal V} =  \frac{1}{6} \int_{X_6} J \wedge J \wedge J = \frac{1}{6} c_{ijk} t^i t^j t^k\ , \label{equ:calV}
\eeq
where $c_{ijk}$ are the triple intersection numbers of $X_6$.
Then, the  K\"ahler coordinates  describing complexified four-cycle volumes are \cite{Grimm:2004uq}
\begin{equation}
T_i \equiv \frac{1}{2} c_{i j k}\hskip 1pt t^{j}t^{k}  + i \vartheta_i + \frac{1}{4} e^{\Phi}\,c_{i \alpha\beta} \hskip 1pt G^{\alpha}(G-\bar{G})^{\beta}\ ,   \label{holomorphicvolume}
\end{equation}
The  expression (\ref{holomorphicvolume}) is not supposed to be obvious, but we can provide some intuition by dropping the contribution of
$G^{\alpha}$, so that
\begin{equation}
T_i  =  \frac{1}{2} c_{i j k}\hskip 1pt t^{j}t^{k} + i \vartheta_i\ .  \label{simpleholomorphicvolume}
\end{equation}
Now, we recall that the two-cycle volumes $t^i$ are related to the four-cycle volumes $\tau_i$ by
\beq
\tau_i = \frac{\partial {\cal V}}{\partial t^i} = \frac{1}{2} c_{ijk} t^j t^k\ , \label{equ:taut}
\eeq so that  (\ref{simpleholomorphicvolume})  can be recognized as
\begin{equation}
T_i  = \tau_i + i \vartheta_i\ . \label{equ:Ti}
\end{equation}
This is the familiar complexification of four-cycle volumes $\tau_i$ by $\vartheta_i$, i.e.~by the integral of $C_4$  over the corresponding four-cycle.
The more involved expression (\ref{holomorphicvolume}) shows that the  corresponding proper K\"ahler coordinate  depends on the vev of  the two-form $G^{\alpha}$.\footnote{This  fact  might seem to be an irrelevant technicality, but we will see in \S\ref{ssec:AM} that the mixing (\ref{holomorphicvolume}) is  the fatal flaw in one otherwise-compelling  scenario for inflation in string theory.}

In summary,  the K\"ahler coordinates on the moduli space
are the $h^{1,1}_+$ complexified  four-cycle volumes $T_i$ (\ref{holomorphicvolume}),  the $h^{1,1}_-$  two-form scalars $G^{\alpha}$ (\ref{gscalar}), the axiodilaton $\tau$ (\ref{theaxiodilaton}), and the $h^{1,2}_-$ complex structure moduli $\zeta^{\alpha}$.   All told,  a compactification of type IIB  string theory on an O3/O7 orientifold of  a Calabi-Yau manifolds leads to $h^{1,1}_++h^{1,1}_-+h^{1,2}_-+1=h^{1,1}+h^{1,2}_-+1$  complex moduli scalars in the four-dimensional  theory.   Further scalar fields can arise from the open string sector.

\subsection{Axions} \label{sec:axions}

One class of fields deserves  special discussion: these are {\it{axions}},  i.e.~pseudoscalar fields enjoying Peccei-Quinn (PQ)  shift symmetries of the form
\begin{equation}
a \mapsto a + const.
\end{equation}
The QCD axion is the original and most famous example of an axion,  and some authors reserve the word `axion'  for this field alone, but we stress that the  axionic fields  discussed here need not  couple to QCD.

\vskip 4pt
\noindent
{\it Axions from $p$-forms.}---Axions  arise in string compactifications from the integration of $p$-form gauge potentials over $p$-cycles of the compact space.
For example, in type IIB string theory, there are axions associated with the NS-NS two-form $B_2$, the R-R two-form $C_2$, and the R-R four-form $C_4$, integrated over suitable 2-cycles $\Sigma^{I}_2$ and 4-cycles $\Sigma^{I}_4$:
\beq
b_{I} = \frac{1}{\alpha^{\prime}} \int_{\Sigma^{I}_2} B_2 \ , \qquad   c_I = \frac{1}{\alpha^{\prime}} \int_{\Sigma^{I}_2} C_2 \ ,  \qquad \vartheta_I = \frac{1}{(\alpha^{\prime})^2}\int_{\Sigma^I_4} C_4\ ,  \label{IIBaxions}
\eeq
where we have chosen the following normalizations for the forms in~(\ref{c4}):
\begin{equation}
\int_{\Sigma^I_2} \omega^J = \alpha' \delta_I^{~J}\   ,  \qquad  \int_{\Sigma^I_4} \tilde{\omega}^J = (\alpha^{\prime})^2 \delta_I^{~J}\ .
\end{equation}
Finally, there are  three universal contributions: the R-R  axion $C_0$,  and two axions,
$b$ and $c$, from
dualizing $B_2(x)$  and $C_2(x)$, respectively.
In sum, a hypermultiplet  arising in ${\cal N}=2$  Calabi-Yau compactifications of type IIB  string theory  contains  three axions: for the $h^{1,1}$ `non-universal'  hypermultiplets,
the axions  descend from $B_2$, $C_2$,  and $C_4$, while  the axions in the universal hypermultiplet are $C_0$, $b_U$, and $c_U$.\footnote{The structure of shift symmetries arising in the universal hypermultiplet  has been  studied in \cite{Strominger:1997eb,Becker:1999pb,Antoniadis:2003sw}.}
We will collectively call these axions $a \equiv \{ b_I , c_I, \vartheta_I, C_0, b_U, c_U\}$.
Orientifolding  by an involution (\ref{o3o7}) with O3/O7 fixed planes  projects out some of the axions.
 Those that remain are $C_0$; $b_{\alpha}$  and $c_{\alpha}$, for $\alpha=1,\ldots, h^{1,1}_-$; and $\vartheta_{i}$, for $i=1,\ldots, h^{1,1}_+$.

\vskip 4pt
\noindent
{\it Axionic shift symmetries.}---At the classical level, each axion inherits a {\it{continuous}} shift symmetry, $a \mapsto a+ const.$,  from the corresponding $p$-form gauge invariances of the ten-dimensional theory.
Specifically, in a background with vanishing fluxes, the type IIB  action (\ref{SIIB}) is independent of $C_0, C_2, C_4, B_2$,
and involves only the associated field strengths.
The  continuous shift symmetry holds to all orders in perturbation theory,  but is broken nonperturbatively, by instanton effects. What remains is a {\it{discrete}} symmetry, $a \mapsto a + (2\pi)^2$.

\vskip 4pt
We now explain this important point in the concrete example of the $b$ axion, following the classic arguments by Dine, Seiberg, Wen, and Witten \cite{Wen:1985jz,Witten:1985bz,Dine:1986vd,Dine:1986zy} that established the shift symmetry to all orders in  the $g_{\rm s}$  and $\alpha'$  expansions.  The extension to axions from other $p$-forms is straightforward.
We start with (\ref{Polyakov2}), the worldsheet coupling of the  two-form $B_2$,
\begin{equation}
S_\sigma \, \supset\, -\frac{1}{4\pi\alpha^{\prime}}\int_{\Sigma_2} \d^2\sigma \,   \epsilon^{ab}   \hskip 1pt \partial_{a} X^{M}\partial_{b} X^{N} B_{MN}(X)    \ ,  \label{eq:bfull}
\end{equation}  or, equivalently,
\begin{equation}
S_\sigma \, \supset\, -\frac{1}{2\pi\alpha^{\prime}}\int_{\Sigma_2} B_2 \equiv - \frac{b}{2\pi} \ , \label{eq:btop}
\end{equation}  where the integral is taken over the string worldsheet.   We recognize (\ref{eq:btop}) as a topological coupling.
Expanding $B_{MN}(X)$  around a fiducial point $X_{(0)} \equiv 0$ yields
\begin{equation}
B_{MN}(X)=B_{MN}(X_{(0)}) + X^P \partial_{P}B_{MN}(X_{(0)}) + \cdots\ . \label{equ:BMN}
\end{equation}
The constant term $B_{MN}(X_{(0)})$ gives rise in (\ref{eq:bfull}) to a {\it{worldsheet total derivative}},
\beq
 -\frac{1}{4\pi\alpha^{\prime}}\int_{\Sigma_2} \d^2\sigma \,  \partial_a \left(  \epsilon^{ab}   \hskip 1pt X^{M}\partial_{b} X^{N} B_{MN}(X_{(0)})  \right) \ ,
\eeq
which  vanishes  unless the worldsheet either wraps a topologically non-trivial cycle, or has a boundary.   The remaining terms in (\ref{equ:BMN}) involving spacetime derivatives of $B_{MN}$ are nonvanishing in general, but correspond to finite-momentum couplings  (i.e.~derivative interactions  involving only $\partial_\mu b$
in the effective theory).  As derivative interactions do not break the shift symmetry, it suffices, for the purpose of ascertaining the symmetry structure, to consider the zero-momentum coupling arising from $B_{MN}(X_{(0)})$.

We  conclude
that the shift symmetry $b \mapsto b+ const$ can only be broken  if  the string worldsheet wraps a non-trivial cycle in the target spacetime,  or has a boundary.
Both sources of symmetry breaking play significant roles in model-building, and we will  discuss them in turn.
At any order in
$\sigma$-model perturbation theory, the string worldsheet wraps a topologically trivial cycle, but  the  fundamental nonperturbative contribution in the $\sigma$-model is a {\it{worldsheet instanton}}, i.e.~a worldsheet
wrapping a non-trivial cycle $\Sigma_2$. The corresponding  spontaneous breaking of the shift symmetry is nonperturbative in $\alpha'$, and is measured by the Euclidean action
\begin{equation}
S_{\rm{inst}} = {\rm{exp}}\left(-\frac{1}{2\pi \alpha^{\prime}}\int_{\Sigma_2} \big(J + i B_2\big)\right) \propto {\rm{exp}}\left(-i\, \frac{b}{2\pi}\right)\ ,
\end{equation} where $J$ is the K\"ahler form.
The result is a periodic potential for $b$, with periodicity\footnote{In this section we follow the conventions of \cite{McAllister:2008hb}.} $(2\pi)^2$.

Next, we consider the  string loop expansion.   The  preceding  arguments made no assumption about the genus  of the worldsheet,  and so must hold to any order in the  string loop expansion.  However, nonperturbatively in $g_{\rm s}$  a new possibility arises:  the  closed string worldsheet can  break open on a soliton (i.e.~a D-brane) and hence acquire a boundary.   Correspondingly, the shift symmetry can be broken by the presence of  spacetime-filling D-branes.

Finally, certain  types of Euclidean D-branes can break the shift symmetry, because $B_2$  appears in the Euclidean D-brane action.  Just as for worldsheet instantons, the resulting contribution to the potential is periodic, with scale
\begin{equation}
S_{{\rm ED}p} = {\rm{exp}}\Bigl(-T_{p}\, {\rm{Vol}}(\Sigma_p)\Bigr)\ ,
\end{equation} for a Euclidean D$p$-brane  wrapping a cycle $\Sigma_p$.

We conclude that  the axion  field $b$  in the  four-dimensional effective theory enjoys a continuous shift symmetry $b \mapsto b+ const$ that is spontaneously broken by worldsheet  and/or D-brane instantons to a discrete shift symmetry $b \mapsto b+ (2\pi)^2$, and may be  explicitly broken if D-branes are present in the compactification.

\vskip 4pt
\noindent
{\it Axion decay constants.}---The discrete  shift symmetry   $a \mapsto a+ (2\pi)^2$ constrains  the axion Lagrangian to take the form
\beq
{\cal L}(a) =  - \frac{1}{2} f^2 (\partial  a)^2 - \Lambda^4 \Big[ 1 - \cos(a/2\pi)\Big] + \cdots \ , \label{equ:La}
\eeq
where $\Lambda$ is a dynamically-generated scale; $f$ is  a
constant with dimensions of mass, known as the {\it{axion decay constant}}; and the omitted terms contain higher-derivative interactions and multi-instanton contributions.  In terms of the canonically-normalized field $\phi \equiv af$, the axion periodicity is $(2\pi)^2 f$.

After dimensional reduction, the decay constants can be deduced from the effective K\"ahler potential.  On the other hand, it is also instructive to compute them directly.
We again take the $b$ axions as an example.
The two-form $B_2$  can be expanded in terms of the four-dimensional  fields $b_{\alpha}(x)$ and the (1,1)-forms $\omega^{\alpha}$, $\alpha=1,\ldots, h^{1,1}_-$:
\beq
B_2 =   b_{\alpha}(x) \hskip 1pt \omega^{\alpha} \ .
\eeq
To determine the axion kinetic terms, and hence the decay constants, we dimensionally reduce the ten-dimensional action for the two-form,
\beq
\frac{1}{2(2\pi)^7 \, g_{\rm s}^2 (\alpha')^4} \int \d^{10} X\, |\d B_2|^2 \ \supset \ \frac{1}{2} \int \d^4 x \sqrt{-g}\, \gamma^{\alpha\beta} (\partial^\mu b_\alpha
\partial_\mu b_\beta)\ ,
\eeq
where
\beq
\gamma^{\alpha\beta} \equiv \frac{1}{6\hskip 1pt (2 \pi)^7 g_{\rm s}^2 (\alpha')^4} \int_{X_6}  \omega^{\alpha} \wedge \star_6 \hskip 2pt \omega^{\beta} \ .  \label{equ:intG}
\eeq
Performing the integral in (\ref{equ:intG}) and diagonalizing the result (i.e.~$\gamma_{\alpha\beta} \mapsto f_{\alpha}^2 \delta_{\alpha\beta}$), one can extract the axion decay constants $f_{\alpha}$.
For purposes of illustration, we consider an isotropic compactification with characteristic length $L$ and volume
${\cal V} = L^6/\alpha^{\prime 3}$. Using (\ref{equ:4dPlanck}) to relate the compactification volume to the four-dimensional Planck mass, we find
\beq
\frac{f^2}{M_{\rm pl}^2} \approx \frac{1}{6}\frac{\alpha^{\prime 2}}{L^4} \ .
\eeq
Since computational  control requires $L \gg \sqrt{\alpha^{\prime}}$, we infer that $f \ll M_{\rm pl}$.   Qualitatively similar  upper bounds on the decay constants  occur in all computable limits of string theory that have been explored  to date~\cite{Banks:2003sx,Svrcek:2006yi}.

\section{Moduli Stabilization} \label{sec:modulistabilization}

Generic Calabi-Yau compactifications come with many {\it moduli},\footnote{The word `moduli'  actually has  several different meanings in different contexts,  so a clarification is appropriate.
The geometric notion is that moduli parameterize  continuous families of solutions,  for example families of Ricci-flat  metrics.   In physics, a  modulus is a scalar field with gravitational-strength couplings that has vanishing potential at some level of approximation.  Some moduli have exactly vanishing potential before supersymmetry breaking, while others have vanishing classical potential but obtain a mass from quantum effects.
In some contexts, `moduli'  refers exclusively to parity-even real scalar fields, as distinguished from pseudoscalar axions,  but we will generally refer to complex moduli.} i.e.~zero-energy deformations arising from the  plethora of topologically distinct cycles in typical Calabi-Yau manifolds.
Understanding the dynamics of moduli is crucial for describing cosmological evolution.
During inflation, the positive vacuum energy tends to induce instabilities of massless scalar fields,  along directions that reduce the energy and  swiftly end inflation.
Moreover, quantum fluctuations of moduli during inflation  contribute to the primordial perturbations.   Furthermore, the impact of moduli  on cosmology after the time of inflation is profound and complex: moduli can  affect  Big Bang nucleosynthesis,  overclose the universe,  comprise some of the dark matter,  decay to dark radiation,  or mediate long-range interactions.
However, a modulus  that acquires a mass $m \gtrsim 30~\rm{TeV}$  decays before nucleosynthesis,  eliminating nearly all\footnote{Moduli  that decay early, but to fields  that themselves linger and affect late-time observables, are a very interesting exception: see e.g.~\cite{Cicoli:2012aq,Higaki:2012ar,Conlon:2013isa,Higaki:2013lra,Conlon:2013txa,Higaki:2013qka,Fairbairn:2013gsa}.} late-time effects.  A full treatment  of the cosmological moduli problem is  beyond the scope of this book, and we will content ourselves with describing the effects of moduli  during inflation.

A principal challenge in the search for cosmological models in string theory is the task of controlling instabilities  associated with the moduli, i.e.~finding  vacua in which  all the moduli have positive masses-squared:  this is known as {\it{moduli stabilization}}.  As we will explain in Section~\ref{sec:StringInflation},  giving non-zero masses to all moduli  does not suffice  to dispel the moduli problem --- for this purpose, the masses must be large compared to the scales accessed during inflation.   Even so, identifying  the leading contributions to the moduli potential is an essential first step  toward constructing realistic models.   We now turn to  a characterization of the moduli potential in the example of  flux compactifications of type IIB  string theory.

\subsection{Classical Solutions}\label{fluxcompactification}

In this section, we will review  the essential  features of type IIB  flux compactifications  on Calabi-Yau orientifolds, following the pioneering work by Giddings, Kachru, and Polchinski (GKP) \cite{Giddings:2001yu}.   Space limitations prevent us from detailing the many advances generalizing and extending  the analysis of \cite{Giddings:2001yu},  most notably to time-dependent backgrounds  and to solutions with strong warping (see \cite{Giddings:2005ff,Shiu:2008ry,Douglas:2008jx,Frey:2008xw}).
The  literature on flux compactifications beyond type~IIB  orientifolds  is so extensive that we  will not attempt  to summarize it: more complete  discussions of flux compactifications,  where the original references can be found, include \cite{Douglas:2006es,Denef:2007pq, Grana:2005jc}.

\vskip 4pt
\noindent
{\it Type IIB supergravity.}---At leading order in  $\alpha'$ and $g_{\rm s}$, the ten-dimensional action for the bosonic fields in Einstein frame~is given by (\ref{SIIB}).
In addition, there may be local sources, such as D-branes and orientifold planes, with corresponding action $S_{\rm loc}$.
We search for warped solutions with the ansatz\footnote{For time-dependent solutions, we would require a more general ansatz~\cite{Giddings:2005ff}.}
(\ref{warped}), but now taking $g_{\mu\nu} = \eta_{\mu\nu}$:
\begin{equation}
\d s^2 =   e^{2A(y)} \eta_{\mu\nu}\d x^{\mu} \d x^{\nu} + e^{-2A(y)}  g_{mn}\d y^{m} \d y^{n}\ . \label{minkowskiansatz}
\end{equation}
Four-dimensional Poincar\'e invariance requires that the three-form flux $G_3$ has no nonvanishing components in the noncompact spacetime,
while the self-dual five-form flux takes the form
\beq
\tilde F_5 = (1+\star_{10})\, \d \alpha(y) \wedge \d x^0 \wedge \d x^1 \wedge \d x^2 \wedge \d x^3\ , \label{equ:F5Ansatz}
\eeq
where $\star_{10}$ is the ten-dimensional Hodge star and $\alpha(y)$ is a scalar function on $X_6$.

\vskip 4pt
\noindent
{\it Equations of motion.}---The trace of the ten-dimensional Einstein equation yields
\beq
\nabla^2 e^{4A} = \frac{e^{8A}}{2 \hskip 1pt   {\rm Im}(\tau)} |G_3|^2 + e^{-4A} \left( |\partial \alpha |^2 + |\partial e^{4A}|^2\right) + 2 \kappa^2 e^{2A} {\cal J}_{\rm loc}\  , \label{EE}
\eeq
where $\nabla^2$ is the Laplacian on $X_6$, and  the effects of local sources are parameterized as
\beq
{\cal J}_{\rm loc} \equiv \frac{1}{4} \left( \sum_{M=4}^9 T^M{}_M - \sum_{M=0}^3 T^M{}_M \right)_{\rm loc} \,,
\eeq  with $T_{MN}$  the stress-energy tensor derived from $S_{\rm loc}$.
In the absence of local sources, i.e.~for $ {\cal J}_{\rm loc} = 0$, the solution is trivial, with constant $A$,  constant $\alpha$, and vanishing $G_3$. (To see this, note that the l.h.s.~of (\ref{EE}) integrates to zero on $X_6$, while the first three terms on the r.h.s.~are all non-negative.)
A non-trivial warped compactification requires one or more sources with $ {\cal J}_{\rm loc} < 0$~\cite{Maldacena:2000mw},  for example orientifold planes.

Next, the  Bianchi identity for the five-form flux is
\beq
\d \tilde F_5 = H_3 \wedge F_3 + 2 \kappa^2 T_3\hskip 1pt \rho_3^{\rm loc}\ , \label{BF}
\eeq
where $\rho_3^{\rm loc}$ is the D3-brane charge density  due to the local sources. Because $\tilde F_5$ is self-dual, (\ref{BF}) may also be thought of as an equation of motion.
Integrating (\ref{BF}) over $X_6$ leads to a tadpole-cancellation condition  (i.e.~Gauss's law  constraint)
\beq
\frac{1}{2\kappa^2 T_3} \int_{X_6} H_3\wedge F_3 + Q_3^{\rm loc} = 0\ ,
\eeq
where $Q_3^{\rm loc}$ is the total charge associated with $\rho_3^{\rm loc}$.
Substituting (\ref{equ:F5Ansatz}) into (\ref{BF}) and combining with (\ref{EE}), we get\footnote{This corrects the numerical factor appearing in \cite{Giddings:2001yu}, cf.~\cite{Baumann:2008kq}.}
\begin{align}
\nabla^2 \left( e^{4A} - \alpha \right) &= \frac{e^{8A}}{24\hskip 1pt {\rm Im}(\tau)} |i G_3 - \star_6 G_3|^2 + e^{-4A}|\partial(e^{4A} - \alpha)|^2 \nonumber \\[2pt]
& \ \ \ +\, 2 \kappa^2 e^{2A} \left( {\cal J}_{\rm loc} -
{\cal Q}_{\rm loc} \right) \ , \label{equ:EE2}
\end{align}
where $\star_6$ is the six-dimensional Hodge star and ${\cal Q}_{\rm loc} \equiv T_3 \rho_3^{\rm loc}$.
The l.h.s.~of (\ref{equ:EE2}) integrates to zero on $X_6$, while the non-localized sources on the r.h.s.~are  non-negative.
As for the localized contribution ${\cal J}_{\rm loc} - {\cal Q}_{\rm loc}$, many well-understood localized sources satisfy the BPS-like condition
\beq
{\cal J}_{\rm loc} \ge {\cal Q}_{\rm loc} \ . \label{BPS}
\eeq
The condition (\ref{BPS}) is saturated by D3-branes and O3-planes, and by D7-branes wrapping four-cycles (in such a way as to
respect the ${\cal N}=1$  supersymmetry preserved by D3-branes).
It is satisfied, but not saturated, by anti-D3-branes and by D5-branes wrapped on collapsed two-cycles.
However, $\overline{O3}$-planes and $O5$-planes violate (\ref{BPS}),  because they are incompatible with the supersymmetry preserved by D3-branes.

Consider a compactification in which all sources satisfy (\ref{BPS}).  Integrating (\ref{equ:EE2}) reveals that we must in fact demand that all sources saturate (\ref{BPS}) ---  i.e.~only D3-branes, O3-planes,  and D7-branes  are allowed --- and that the three-form flux is
imaginary self-dual (ISD),
\beq
\star_6 G_3 = i G_3\ , \label{equ:ISD}
\eeq while the warp factor is equal to the four-form potential
\beq
e^{4A} = \alpha\ \,.
\eeq
A configuration  meeting these criteria is called an {\it{ISD  solution}}.

\vskip 4pt
To recapitulate,  the Einstein equation and  five-form Bianchi identity
can be combined to give, at leading order in $\alpha'$ and $g_{\rm s}$,  the key relations  (\ref{EE}) and (\ref{equ:EE2}).  These expressions are parallel in form: the l.h.s.~expressions integrate to zero, while the r.h.s.~in each case  involves  a sum of non-localized (`bulk') terms  that are everywhere non-negative,  as well as a localized contribution.
If the localized contribution  is non-negative, it must  in fact be zero, and then the  bulk terms  must be identically zero.   In the case of the Einstein equation (\ref{EE}),  this implies that {\it{in the absence of negative tension sources, only unwarped  solutions (without positive tension sources) are allowed.}}    From the Einstein equation  minus Bianchi identity  (\ref{equ:EE2}), we learn that {\it in the absence of  sources violating} (\ref{BPS}), {\it only ISD solutions are allowed.}   Because well-understood supersymmetric configurations of O3-planes  and  O7-planes (as well as D3-branes  and  D7-branes) yield negative tension  without violating  (\ref{BPS}),  it is straightforward to exhibit ISD warped solutions.   Non-ISD solutions are much less studied at  present, because of the difficulty of controlling the comparatively exotic orientifold  planes that violate (\ref{BPS}).

\vskip 4pt
A significant property of compactifications with three-form flux, including ISD solutions, is that the complex structure moduli $\zeta^{\alpha}$ and the axiodilaton $\tau$  experience a potential.  To see this, we note that the ten-dimensional type IIB  action (\ref{SIIB}) contains the term
\begin{equation}
V_{\rm flux} =    \frac{1}{2\kappa^2} \int \d^{10}X \sqrt{-G_{E}} \left[  - \frac{|G_3|^2}{2\hskip 1pt  {\rm Im}(\tau)} \right]\ ,
\end{equation} which involves the complex structure moduli via the metric contraction, and the axiodilaton both through the denominator and through the definition (\ref{g3def}) of $G_3$.  As a result, for a generic choice of quantized fluxes, $\tau$ and all of the $\zeta^{\alpha}$ receive masses at the classical level, i.e.~at leading order in $\alpha^{\prime}$.

\vskip 4pt
\noindent
{\it Effective supergravity.}---The data of the four-dimensional effective theory of an ISD compactification can be usefully repackaged in terms of a K\"ahler potential and superpotential of ${\cal N}=1$ supergravity.  At leading order in the $\alpha'$ and string loop expansions, the K\"ahler potential is
\beq
K_0 = - 2\ln ({\cal V}) - \ln\left(-i( \tau - \bar \tau)\right) - \ln \left(-i \int \Omega \wedge \bar \Omega \right)\ . \label{K0}
\eeq
Here, the volume ${\cal V}$ and the holomorphic three-form $\Omega$ depend implicitly on the K\"ahler moduli $T_i$ and the complex structure moduli $\zeta_\alpha$, respectively.
The ISD condition (\ref{equ:ISD}) can be derived from the Gukov-Vafa-Witten flux superpotential~\cite{Gukov:1999ya}
\beq
W_0 = \frac{c}{\alpha^{\prime}}\, \int G_3 \wedge \Omega\ , \label{GVW}
\eeq where $c$ is a constant (see~\cite{Conlon:2006gv}).
Since $G_3$ depends on the dilaton and $\Omega$ involves the complex structure moduli, the superpotential~(\ref{GVW}) leads to a non-trivial potential for these moduli.
The scalar potential associated with $K_0$ and $W_0$ is
\beq
V_F = e^{K_0} \left[ K_0^{I \bar J} D_I W_0 \overline{D_J W_0} -  3 |W_0|^2 \right]\ , \label{VF}
\eeq
where $I,J$ run over all the moduli ($T_i$, $\zeta_\alpha$ and $\tau$).
Supersymmetry is preserved if all F-terms vanish,\footnote{When gauge multiplets  are present in the effective theory, D-term contributions are an important alternative source of supersymmetry breaking, but our present discussion is confined to the moduli sector.} i.e.~if
\beq
D_I W_0 \equiv \partial_I W_0 + (\partial_I K) W_0 = 0\ ,
\eeq
where $I$ runs over all the moduli.

\vskip 4pt
\noindent
{\it No-scale structure.}---The K\"ahler potential (\ref{K0}) is of a specific form that satisfies
\beq
\sum_{I, J = T_i} K^{I\bar J}_0 \partial_I K_0 \partial_{\bar J} K_0 =3\ .  \label{noscaleK}
\eeq
Since the superpotential (\ref{GVW})  is independent of the K\"ahler moduli,
the scalar potential (\ref{VF}) is of the {\it no-scale} type, i.e.~it is  independent of the F-terms of the K\"ahler moduli,
\beq
V_F =e^{K_0} \sum_{I,J\ne T_i} K^{I\bar J}_0 D_I W_0 \overline{D_JW_0} \ . \label{noscaleV}
\eeq
This potential is positive semi-definite, and $V_F = 0$ when $D_{I \ne T_i} W_0 = 0$.
The minimum is not necessarily supersymmetric, as in general
we may have $D_{T_i} W_0 \ne 0$.

\vskip 4pt
\noindent
{\it No-scale structure and D3-branes.}---Thus far we have discussed the effective action for massless closed string fields, but the positions of D-branes provide an important additional class of {\it{open string moduli}}.
Consider a D3-brane that fills spacetime and sits at a point in a flux compactification on
a Calabi-Yau manifold.
Evaluating the DBI+CS action (\ref{DBIplusCS}) in an ISD background, one finds that the potential energy for D3-brane motion vanishes identically: the complex scalars $z_{\alpha}$, $\alpha=1,2,3$, that parameterize the D3-brane position are massless moduli.
The four-dimensional action derived from the dimensional reduction of (\ref{DBIplusCS}) can be expressed in ${\cal N}=1$ supergravity via the {\it{DeWolfe-Giddings K\"ahler potential}}, which for a compactification with a single K\"ahler modulus $T$
takes the form~\cite{DeWolfe:2002nn}
\beq
K(T, \bar T, z_\alpha, \bar z_\alpha) = - 3 \ln\Big[T + \bar T - \gamma k(z_\alpha, \bar z_\alpha) \Big] \ , \label{equ:DeWolfeG}
\eeq
where $\gamma$ is a constant, and $k(z_\alpha, \bar z_\alpha)$ is the K\"ahler potential for the metric on the Calabi-Yau manifold.
The K\"ahler potential (\ref{equ:DeWolfeG}) is of no-scale type: if the superpotential $W$ is independent of $T$ and of the $z_{\alpha}$, then the F-terms of
these fields
do not appear in the F-term potential.
The mixing between the K\"ahler modulus $T$ and the D3-brane position moduli implied by (\ref{equ:DeWolfeG}) has significant ramifications for inflationary model building with D3-branes: see \S\ref{sec:dbrane}.

\vskip 4pt
In summary, in a `no-scale' compactification with imaginary self-dual fluxes, one finds, at leading order in $\alpha'$ and $g_{\rm s}$, that the vacuum energy vanishes,\footnote{Having a non-supersymmetric vacuum with vanishing vacuum energy seems too good to be true, and it is: no-scale  structure on its own is not a solution to the cosmological constant problem, because it does not survive quantum corrections.} the complex structure moduli  and axiodilaton are stabilized, the K\"ahler moduli and D3-brane position moduli have vanishing potential.

\subsection{Quantum Effects}  \label{quantumeffects}

Perturbative and nonperturbative corrections to the effective action  are known to break the no-scale symmetry, lifting or destabilizing the flat directions and  altering the vacuum energy.
We will begin by discussing perturbative corrections to the K\"ahler potential, in both the $\alpha'$ and $g_{\rm s}$ expansions, and then discuss nonperturbative corrections to the superpotential.

\subsection*{Perturbative Corrections}

The most famous perturbative correction to the  K\"ahler potential descends from an $(\alpha^{\prime})^3$ curvature correction in ten dimensions, namely the quartic  invariant ${\cal R}^4$  appearing in (\ref{IIBgravityexpansion}).   This term is part of the classical, higher-curvature ten-dimensional supergravity theory:  it arises via a four-loop correction to the $\beta$-function of the
worldsheet $\sigma$-model \cite{Gross:1986iv},  rather than from a loop in spacetime.
In the four-dimensional effective theory,  the  result takes the form \cite{Becker:2002nn}
\beq
K = - 2 \ln \left[ {\cal V} + \frac{\xi}{2 g_{\rm s}^{3/2}}\right] \ , \qquad \xi \equiv - \frac{\chi(X_6) \zeta(3)}{2(2\pi)^3} \ , \label{equ:Ka}
\eeq where $\chi(X_6)$ is the Euler  characteristic of $X_6$, and $\zeta(3) \approx 1.202$ is Ap\'ery's constant.
The K\"ahler potential (\ref{equ:Ka}) does not satisfy the no-scale condition  (\ref{noscaleK})  (unless $\chi=0$).


\vskip 4pt
Perturbative  corrections  from loop effects in spacetime, i.e.~from  higher-genus string worldsheets,  will also  generically spoil the no-scale structure (\ref{noscaleK}).  The only explicit results available are for ${\cal N}=1$ compactifications on the
toroidal orientifold $T^6/(\mathbb{Z}_2 \times \mathbb{Z}_2 )$~\cite{Berg:2005ja, Berg:2005yu}.
To give a concrete picture of string loop corrections,  we  now sketch this specific result.
The correction to the K\"ahler potential takes the form
\beq
\delta K_{(g_{\rm s})} = \delta K_{(g_{\rm s})}^{\rm KK} + \delta K_{(g_{\rm s})}^{\rm W} \ , \label{equ:KKgs}
\eeq  where the term $\delta K_{(g_{\rm s})}^{\rm KK}$ comes from the exchange of closed strings with Kaluza-Klein (KK) momentum  between D7 and D3-branes, while
$\delta K_{(g_{\rm s})}^{\rm W}$ comes from the exchange of closed strings with nonvanishing winding~(W).
The former  is given by
 \beq
 \delta K_{(g_{\rm s})}^{\rm KK} = - \frac{1}{128 \pi^2} \sum_{i=1}^3 \frac{{\cal E}_i^{\rm KK}(\zeta,\bar \zeta)}{{\rm Re}(\tau) \hskip 1pt \tau_i} \ ,\label{equ:Kgs1}
\eeq
where $\tau_i$ stands for the K\"ahler modulus associated with the four-cycle wrapped by the $i$-th D7-brane.
The second term in (\ref{equ:KKgs})  takes the form
\beq
\delta K_{(g_{\rm s})}^{\rm W} =
\left. - \frac{1}{128 \pi^2} \sum_{i=1}^3 \frac{{\cal E}_i^{\rm W}(\zeta,\bar \zeta)}{\tau_j \tau_k} \right|_{j \ne k \ne i} \ .\label{equ:Kgs2}
\eeq
These results have a complicated dependence on the complex structure moduli $\zeta$ (encoded by the functions ${\cal E}_i^{\rm KK}(\zeta,\bar \zeta)$ and ${\cal E}_i^{\rm W}(\zeta,\bar \zeta)$ given in \cite{Berg:2005ja, Berg:2005yu}), but
have
a simple scaling with the K\"ahler moduli~$\tau_i$.  A conjectural  generalization of the results of \cite{Berg:2005ja, Berg:2005yu} to general Calabi-Yau three-folds appears in~\cite{Berg:2007wt} (see also \cite{vonGersdorff:2005bf} for related earlier work), but giving an  explicit characterization of this leading string loop correction  remains an open problem.

Even though the perturbative corrections (\ref{equ:Kgs1}) and (\ref{equ:Kgs2}) manifestly violate no-scale structure,  the corresponding contributions to the scalar potential  cancel to some  extent: see the discussion in \S\ref{lvscpct}.

\subsection*{Nonperturbative Effects}

Although the K\"ahler potential for the K\"ahler moduli receives perturbative corrections in the $\alpha'$ and $g_{\rm s}$ expansions, the superpotential receives no corrections  in either expansion, to any order in perturbation theory, as we now explain.

The fact that the superpotential of a supersymmetric field theory receives no perturbative corrections in the ordinary $\hbar$ expansion, corresponding to the $g_{\rm s}$ expansion in string theory, was originally established directly \cite{Grisaru:1979wc}.
Elegant non-renormalization theorems in string theory \cite{Wen:1985jz,Witten:1985bz,Dine:1986vd,Dine:1986zy}  arrived at the same end by combining holomorphy and shift symmetry  arguments.
In the heterotic string setting emphasized in \cite{Wen:1985jz,Witten:1985bz,Dine:1986vd,Dine:1986zy}, the argument for non-renormalization in $g_{\rm s}$ is more straightforward than in type IIB flux compactifications,  because the classical superpotential in the heterotic string is independent of the dilaton, whereas
the classical GVW flux superpotential (\ref{GVW}) involves the dilaton through the definition (\ref{g3def}) of $G_3$.  A careful demonstration of the absence of string loop corrections to (\ref{GVW}) appears in \cite{Burgess:2005jx}.

Next, to address $\alpha^{\prime}$  corrections, we recall that the axionic imaginary parts of the K\"ahler moduli (\ref{equ:Ti}) are protected by shift symmetries, $\vartheta_i \mapsto \vartheta_i + const.$, which hold to all orders in perturbation theory, as explained in \S\ref{sec:axions}.  (These shift symmetries rely in no way on supersymmetry.)
Holomorphy dictates that the superpotential can only depend on $T_i$, rather than on $T_i + \bar{T_i}$, but no non-trivial polynomial in $T_i$ is invariant under the shift of the axion.
Thus, the superpotential can depend on $T_i$ only nonperturbatively.  Because corrections in the $\alpha'$ expansion  must change in magnitude as the  $T_i$ are varied,  but the superpotential is independent of the  $T_i$  to all orders, it follows that $W$ receives no  perturbative $\alpha^{\prime}$  corrections.

\vskip 4pt
Let us now discuss nonperturbative contributions to the superpotential.  Consider a compactification in which a stack of
$N_c$  D7-branes  wraps a four-cycle $\Sigma_4$.   The  worldvolume theory  of the D7-branes  includes a  Yang-Mills action for four-dimensional gauge fields $A_{\mu}$, of the form
\begin{equation}
S = \frac{1}{2g_{7}^2} \int_{\Sigma_4} \d^4 \sigma \sqrt{g_{\rm ind}\vphantom{-g}}\, e^{-4A(y)}\cdot \int \d^4 x \sqrt{-g}\,\, {\rm{Tr}}\big[F_{\mu \nu}F^{\mu \nu}\big]\ ,
\end{equation}  where the indices are raised with  the unwarped metric $g_{\mu \nu}$, and
$g_{7}$ is the gauge coupling of the (7+1)-dimensional  Yang-Mills theory,
\begin{equation}
g_{7}^2 = 2(2\pi)^{5}(\alpha')^2\ .
\end{equation}
The gauge coupling  of the four-dimensional  Yang-Mills theory is
\begin{equation}
\frac{1}{g^2} = \frac{T_3 {\cal V}_{4}}{8\pi^2}\ ,
\end{equation}
where we have defined the volume of $\Sigma_4$ as
\begin{equation}
{\cal V}_{4} \equiv \int_{\Sigma_4} \d^4 \sigma \sqrt{g_{\rm ind}\vphantom{-g}}\, e^{-4A(y)}\ , \label{warpedvolume}
\end{equation}
and $g_{\rm ind}$  is the induced metric on the D7-brane.
Because of  the appearance of $e^{-4A(y)}$, ${\cal V}_{4}$ as defined in (\ref{warpedvolume})  is sometimes called the `warped volume'.

Given certain topological conditions  on $\Sigma_4$, discussed further below ---  heuristically, one asks that $\Sigma_4$  have no  deformations  that could correspond to charged matter fields ---  the four-dimensional gauge theory  arising upon dimensional reduction is {\it{pure glue}} ${\cal N}=1$  super Yang-Mills theory.   At low energies, this field theory  generates a nonperturbative superpotential from gaugino condensation~\cite{Novikov:1983ek,Novikov:1983ee,Ferrara:1982qs,Dine:1985rz,Derendinger:1985kk,Shifman:1987ia} (cf.~\cite{Gorlich:2004qm}):
\begin{equation}
|W_{\lambda\lambda}| = 16\pi^2 M_{\mathsmaller{\rm UV}}^3 \, {\rm{exp}}\left(-\frac{1}{N_c}\frac{8\pi^2}{g^2}\right) \propto {\rm{exp}}\left(-\frac{T_3 {\cal V}_{4}}{N_c}\right)\ . \label{gauginocondensate}
\end{equation}
The volume ${\cal V}_{4}$ is proportional to the real part\footnote{Supersymmetry requires that the  superpotential is a holomorphic function of the moduli, but verifying that ${\cal V}_{4}$ is the real part of a holomorphic function is highly non-trivial \cite{Baumann:2006th,Baumann:2007ah}. When D3-branes are present, their backreaction on the  volume ${\cal V}_{4}$ must be incorporated  in order to maintain holomorphy \cite{Baumann:2006th}.  This effect was first understood in the open string channel,  as a threshold correction to the gauge coupling $g$ \cite{Berg:2004ek}.} of  a corresponding K\"ahler modulus $T$, so the  gaugino condensate superpotential  may be written as
\begin{equation}
W_{\lambda\lambda} = {\cal A} \, e^{-a T}\,,  \label{gauginocondensation}
\end{equation}  where $a=\frac{2\pi}{N_c}$
and  the prefactor ${\cal A}$ is independent of all the K\"ahler moduli, but generally depends on the complex structure moduli and on the positions of any D-branes.  One might  suspect from~(\ref{gauginocondensate}) that ${\cal A} \propto \MKK^3$, but because $\MKK/M_{\rm pl}$ depends on $T+\bar{T}$,  such a dependence would not be holomorphic.
Instead, for typical  complex structure moduli vevs and D-brane positions,  one has ${\cal A} \sim M_{\rm pl}^3$: see \cite{Conlon:2006gv,McAllister:2008hb}.

A very similar superpotential contribution arises if $\Sigma_4$ is wrapped not by spacetime-filling D7-branes, but by {\it{Euclidean D3-branes}}, also known as {\it{D3-brane instantons}} \cite{Witten:1996bn} (see \cite{Blumenhagen:2009qh} for a review).  A Euclidean D$p$-brane is an instantonic  contribution to the path integral  whose  Euclidean action has a real part that is proportional to the volume of the $(p+1)$-cycle wrapped by the Euclidean brane, while the imaginary part is determined by the corresponding Chern-Simons action.
For a Euclidean D3-brane wrapping $\Sigma_4$, the resulting superpotential term is
\begin{equation}
W_{\rm{ED3}} = {\cal A} \, e^{-a T}\ , \label{EuclideanD3brane}
\end{equation}  where $a=2\pi$,  and as in (\ref{gauginocondensation})  the prefactor ${\cal A}$ can depend on the complex structure moduli and on D-brane positions, but is independent of the K\"ahler moduli.

We now  turn to the necessary and sufficient topological conditions for the generation of a nonperturbative superpotential, focusing on the case of Euclidean D3-branes.
These conditions  can be expressed most simply in terms of an auxiliary eight-dimensional geometry $Y$, in which the axiodilaton $\tau$ parameterizes an elliptic fibration over the  six-dimensional  manifold $X$ on which the type~IIB  theory is compactified.\footnote{An elliptic fibration is a fibration  in which almost all fibers are non-singular and have the topology of two-tori, but a finite number of singular fibers can appear.  The possible singular fibers have been classified by Kodaira in \cite{MR0187255}.}
This construction is known as {\it{F-theory}} \cite{Vafa:1996xn}: one says that F-theory has been  compactified on $Y$, which is an elliptically-fibered four-fold over the base $X$.

Witten observed in \cite{Witten:1996bn} that  a necessary condition for a non-vanishing Euclidean D3-brane superpotential term associated with a four-cycle $\Sigma_4 \subset X$  is that $\Sigma_4$ is the projection of a six-cycle $D \subset Y$ obeying
\begin{equation}
\chi({\cal{O}}_D) \equiv \sum_{i=0}^{3}(-1)^{i}h^{0,i}(D) = 1\ . \label{arithmetic}
\end{equation}
where ${\cal{O}}_D$ denotes the trivial line bundle defined on $D$ (see e.g.~\cite{griffiths2011principles} for the relevant mathematical background).
The number $\chi({\cal{O}}_D)$ is known as the  holomorphic Euler characteristic of $D$ \cite{griffiths2011principles}, or the  arithmetic genus of $D$.\footnote{In the mathematics literature, some authors define the arithmetic genus $p_a(D)$  so that $p_a(D) = 1 - \chi({\cal{O}}_D)$, for $D$  a six-manifold \cite{griffiths2011principles}.  Here, as in most of the string theory literature,  the arithmetic genus and the holomorphic Euler characteristic are  both equal to $\chi({\cal{O}}_D)$, cf.~\cite{Witten:1996bn,hirzebruch2009concept}.
In (\ref{arithmetic}), the notation $\chi({\cal{O}}_D)$ for the holomorphic Euler characteristic  is used instead of $\chi(D)$ because the latter can be confused with the  more familiar topological Euler characteristic.}
Next, a sufficient condition  for a non-vanishing Euclidean D3-brane superpotential is \cite{Witten:1996bn}
\begin{equation}
h^{0,1}(D) = h^{0,2}(D) = h^{0,3}(D) = 0 \ . \label{rigidity}
\end{equation}
A six-cycle $D$ obeying (\ref{rigidity}) is said to be {\it{rigid}}: the  Hodge numbers in (\ref{rigidity}) count the independent deformations of $D$.

The sufficient condition (\ref{rigidity}) is unmodified by the presence of flux, but in flux  backgrounds the necessary condition (\ref{arithmetic}) is modified and becomes less restrictive \cite{Gorlich:2004qm,Kallosh:2005yu,Saulina:2005ve,Tripathy:2005hv,Kallosh:2005gs,Bergshoeff:2005yp,Martucci:2005rb}.  Couplings to flux can give mass to (some of the) deformations of Euclidean D3-branes, and of D7-branes, counted by $h^{0,2}$.  Generalizations of (\ref{arithmetic}) to backgrounds with flux,  and further consistency conditions, are  described in \cite{Robbins:2004hx,Gorlich:2004qm,Kallosh:2005yu,Saulina:2005ve,Tripathy:2005hv,Kallosh:2005gs,Bergshoeff:2005yp,Martucci:2005rb,Blumenhagen:2007sm} and reviewed in \cite{Denef:2008wq,Blumenhagen:2009qh}.

\subsection{Volume Stabilization}
\label{modulistab}

Having assembled the known perturbative and nonperturbative corrections to the potential for the K\"ahler moduli in type IIB  flux compactifications,  we are in a position to ask whether the quantum-corrected theory has cosmologically interesting metastable vacua, even though the classical theory has unstabilized K\"ahler moduli.

\begin{figure}[h!]
   \centering
     \includegraphics[width=0.97 \textwidth]{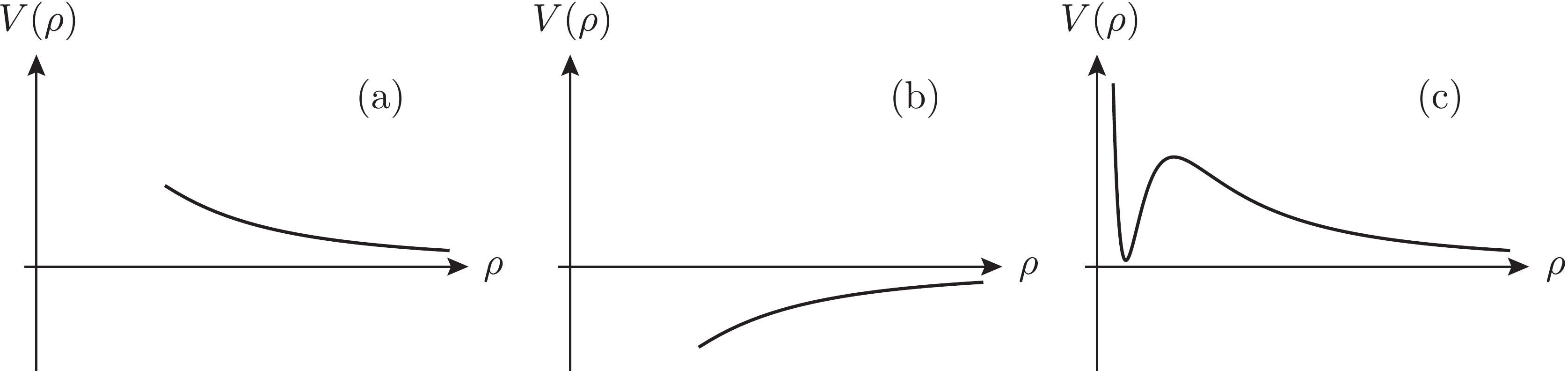}
   \caption{The Dine-Seiberg problem \cite{Dine:1985he} for a modulus $\rho$.   In case (a),  there is a runaway to $\rho = \infty$,  where the theory is free.   In case (b),  the leading correction drives the theory toward small $\rho$, where it is strongly coupled.   The  existence of the minimum in case (c) requires competition among  at least three terms.}
  \label{fig:DS}
\end{figure}

There is a very general problem \cite{Dine:1985he} underlying any search for string compactifications in which flat directions are stabilized by perturbative or nonperturbative corrections.
The {\it{Dine-Seiberg problem}} \cite{Dine:1985he} can be summarized  as follows: when corrections  are important, they are not computable, and when they are  computable, they are not important \cite{Denef:2008wq}.  To understand the observation of \cite{Dine:1985he}  in more detail, let $\rho$ be a modulus that controls a weak coupling expansion,  such that $\rho \to \infty$  is the free limit.  Concretely, $\rho$  could be the K\"ahler modulus  that measures the compactification volume, $\rho = T+\bar T$, so that $\rho \to \infty$ corresponds to decompactification to ten  dimensions; or, for the string loop expansion, $\rho = g_{\rm s}^{-1} = e^{-\Phi}$.
We now ask whether  perturbative or nonperturbative corrections  generate a potential for $\rho$  that has a minimum at finite  $\rho$.
Because the leading-order classical action  is valid for $\rho \to \infty$, the potential $V(\rho)$ generated by perturbative and nonperturbative corrections must vanish for $\rho \to \infty$.   In particular,
$V(\rho)$  must approach zero  from above or from below as $\rho \to \infty$ (see fig.~\ref{fig:DS}).  If $V(\rho)$  is positive for large $\rho$, then the leading correction term in $V(\rho)$, which dominates for $\rho \to \infty$, creates an instability that drives the theory toward  $\rho = \infty$.  If instead $V(\rho)$  is negative for large $\rho$, then the leading correction to the free theory creates an instability that drives the theory toward smaller $\rho$, and hence toward stronger coupling.
Either way,  the leading correction term  creates an instability, and a (meta)stable vacuum can arise only if higher-order corrections  make comparably important contributions that counterbalance the instability.   But once two\footnote{When $V>0$ for large $\rho$, three separate terms are required --- see \cite{Silverstein:2004id,Denef:2008wq}.} consecutive  terms in the weak coupling expansion  are comparable,  one expects that the entire series must be included.  While it could happen that the first and second non-vanishing terms are competitive because the second is accidentally large,  verifying that this leads to a consistent solution requires examining  higher terms in the series  to rule out  unanticipated accidental enhancements  at higher orders.  Thus,  metastable vacua are quite generally found at points in moduli space where the weak coupling expansions break down.
This fact presents a major obstacle to the search for metastable string vacua, because in nearly all cases, at most the {\it{first}} non-vanishing correction
in each expansion ($\alpha'$ or $g_{\rm s}$) is known explicitly.

In the case of K\"ahler moduli stabilization in type IIB flux compactifications, no-scale structure ensures  that the  classical potential for the K\"ahler moduli vanishes,  so the leading  correction to the potential is  in fact the dominant  potential energy term  overall.
At {\it{generic}} points in the parameter space, one expects that the correction of leading importance will  come from the  first non-vanishing perturbative correction, which
is necessarily the first
correction\footnote{Whether the leading perturbative correction to the potential comes from the first $\alpha'$ correction to $K$, or  instead from the first
$g_{\rm s}$ correction to $K$,  is not obvious a priori, and can depend on parameter values --- see \S\ref{sec:LVS} for a detailed discussion.} to $K$,  because the superpotential is not renormalized  in perturbation theory.
Following \cite{Dine:1985he}, we conclude that  vacua at generic points in the parameter space  are the result of competition among terms at different perturbative orders.
Because of the absence of perturbative computations beyond leading order, it has proved very difficult to find controllable vacua in this regime (however, see~e.g.~\cite{Bobkov:2004cy,Berg:2005yu,Parameswaran:2006jh}).

The two leading ideas  for K\"ahler moduli stabilization, the KKLT scenario~\cite{Kachru:2003aw} and the Large Volume Scenario (LVS) \cite{Balasubramanian:2005zx}, succeed by targeting regions of parameter space where  vacua result from competition among {\it{known}}  correction terms.   To anticipate slightly,  the KKLT mechanism involves competition  between a classical
flux superpotential (\ref{GVW}),  made small by fine-tuning fluxes,  and the nonperturbative superpotential (\ref{gauginocondensation}).  The LVS construction works in a region of K\"ahler moduli space where some cycles  are exponentially larger than others,  so that the leading $\alpha^{\prime}$ correction~(\ref{equ:Ka}) involving the large overall volume ${\cal V}$  competes with nonperturbative superpotential terms~(\ref{gauginocondensation}) involving the small cycles.  In both cases,  one can argue that the unknown higher  corrections do not spoil the vacuum structure.
We  now turn to explaining these mechanisms in more detail.

\subsection*{KKLT Scenario}

The seminal KKLT proposal~\cite{Kachru:2003aw} for constructing stabilized vacua bypasses all perturbative corrections and  instead makes use of nonperturbative contributions to the superpotential.

\vskip 4pt
In the presence of three-form flux the complex structure moduli and dilaton acquire supersymmetric masses via the classical superpotential (\ref{GVW}), cf.~\S\ref{fluxcompactification}.
If we denote the  typical mass scale by $m_{\rm flux}$, then at energies $E \ll m_{\rm flux}$ the complex structure moduli and dilaton can be integrated out (see the discussion below),  and the classical superpotential $W_0$ becomes a constant.
The  fields remaining in the low-energy effective theory are the K\"ahler moduli,\footnote{If  spacetime-filling D3-branes are present, their  positions are also light fields in the  effective theory, as explained  in detail in \S\ref{sec:dbrane}.} which do not appear in the classical superpotential.

\begin{figure}[h!]
   \centering
     \includegraphics[width=0.9 \textwidth]{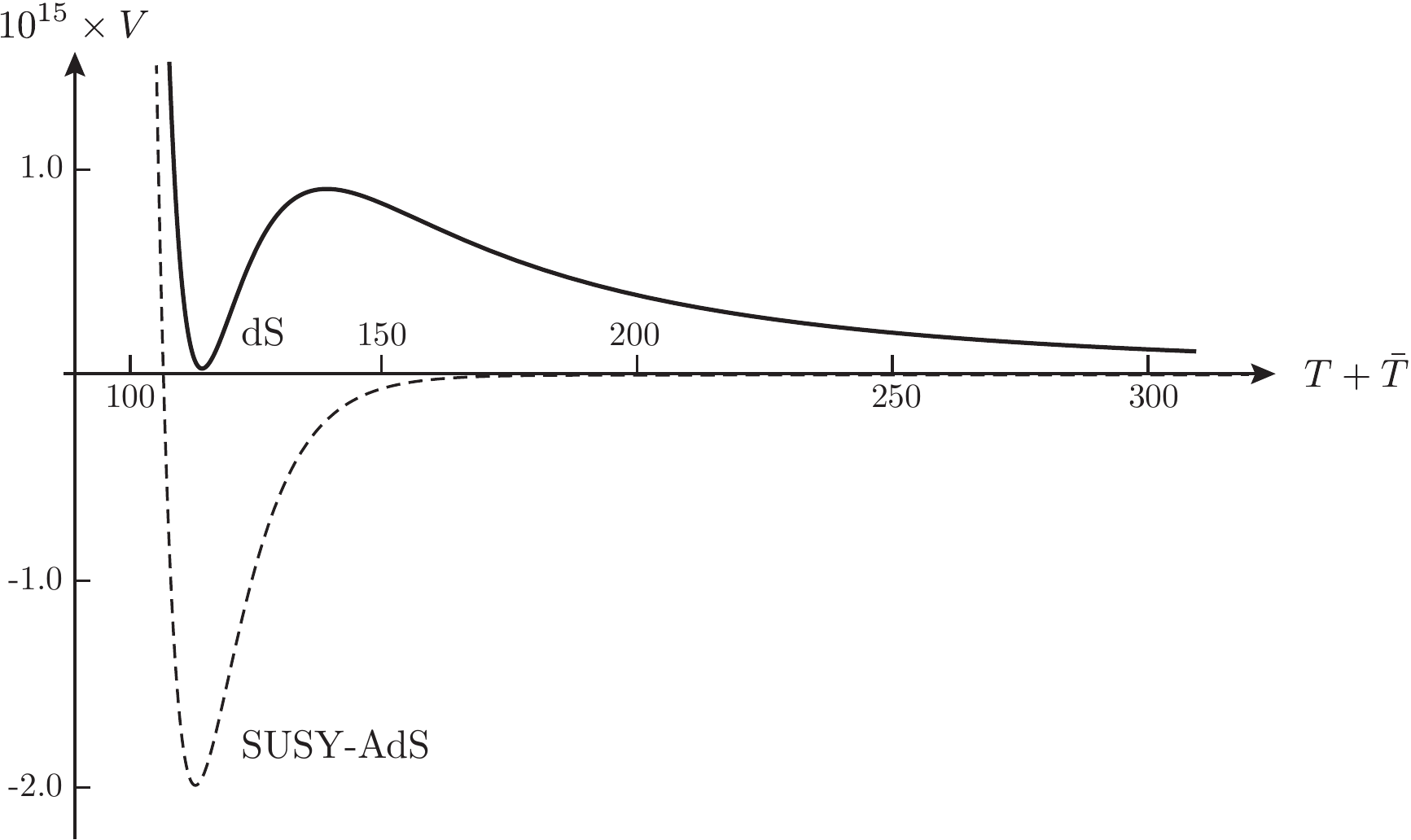}
   \caption{Potential for the K\"ahler modulus $T$ in a KKLT scenario with $h^{1,1}_{+}=1$.  The dashed line shows the potential in the absence of a supersymmetry-breaking anti-D3-brane. The figure was generated for ${\cal A} = 1$, $a=0.1$, and $W_0 = - 10^{-4}$.}
  \label{fig:KKLT}
\end{figure}

As shown in \S\ref{quantumeffects}, nonperturbative effects  can generate superpotential  interactions for the K\"ahler moduli, either through strong gauge dynamics (such as gaugino condensation) on D7-branes,  or  through instanton contributions from Euclidean D3-branes.
The combination of the constant flux superpotential (\ref{GVW}) with the nonperturbative terms (\ref{gauginocondensation}) or (\ref{EuclideanD3brane}) leads to
\beq
W = W_0 + \sum_{i=1}^{h^{1,1}_+} {\cal A}_i\, e^{-a_i T_i} + \cdots\ ,  \label{KKLTW}
\eeq
where  the ellipses denote higher-order nonperturbative effects.
In writing (\ref{KKLTW}),  we have assumed that there is a  nonperturbative term for each of the K\"ahler moduli $T_i$.
The  status of this important assumption  is not completely understood: while examples do exist  in which there is a nonperturbative term for each K\"ahler modulus \cite{Denef:2004dm,Denef:2005mm}, it has not been shown that this situation is generic.\footnote{The stabilization scenario of \cite{Bobkov:2010rf}  is very similar to the KKLT construction, but requires only one nonperturbative term, arising on  a four-cycle $\Sigma_4$ that is {\it{ample}}.  Roughly speaking, $\Sigma_4$ is ample if it is a positive linear combination of a basis of four-cycles  of positive volume --- see \cite{Bobkov:2010rf} for further background and a precise definition.}

For an arbitrary K\"ahler potential $K$, the superpotential (\ref{KKLTW}) leads to the scalar potential
\begin{align}
V_{(np)} &\, =\, e^{K} K^{j \bar \imath} \left[ a_j {\cal A}_j a_{\bar \imath} \bar {\cal A}_{\bar \imath}\, e^{-(a_j T_j + a_i \bar T_i)} \right. \nonumber \\
& \left. \hspace{1cm}-\, \left(a_j {\cal A}_j \, e^{-a_j T_j} \bar W \partial_{\bar \imath} K + a_{\bar \imath} \bar {\cal A}_{\bar \imath} \, e^{- a_i \bar T_i } W \partial_j K \right)\right]\ . \label{equ:Vnp}
\end{align}
Taking $K= K_0 = - 2\ln ({\cal V})$, cf.~eq.~(\ref{K0}),  and considering the single-modulus case ($h^{1,1}_+ =1$), ${\cal V} = (T+\bar T)^{3/2}$, one finds
\beq
V_{(np)} = \frac{a {\cal A} \,e^{-a (T+\bar T)}}{2(T+\bar T)^2} \left[ \left(1 + \frac{T+\bar T}{3} \right) a {\cal A} \, e^{-a(T+\bar T)} + W_0 \right] \ . \label{equ:Vnp2}
\eeq	
This potential is plotted in fig.~\ref{fig:KKLT} (dashed line).  It is easy to see that the vacuum solution is supersymmetric anti-de Sitter space.
Letting $(T+\bar T)_\star$ be the value of the K\"ahler modulus at the minimum, we find $(\partial_T V_{(np)})_\star = (D_T W)_\star = 0$ and
\beq
W_0 = - {\cal A} \, e^{- a (T+\bar T)_\star} \left( 1 + \frac{2}{3} a (T+\bar T)_\star \right)\ . \label{W0}
\eeq
Control over the instanton  expansion of the superpotential, corresponding to neglecting the ellipses in (\ref{KKLTW}), requires that  $a (T+\bar T)_\star \gg 1$.
Moreover, perturbative ($\alpha'$ and $g_{\rm s}$) corrections to the K\"ahler potential~(\ref{K0}) may be neglected if $(T+\bar T)_\star \gg 1$.\footnote{String loop corrections to $K$  are suppressed at large volume,  and not only by factors of  $g_{\rm s}$, because $K_0$ involves ${\cal V}$,  so that any additive correction to $K$ is subleading in volume; see \S\ref{sec:LVS}.}
We see from (\ref{W0}) that the volume is stabilized in a controlled limit only for an exponentially small value of the flux superpotential, $W_0 \ll {\cal A}$.
This can be achieved through a fine-tuned choice of quantized flux, following~\cite{Bousso:2000xa}.

A number of authors have critically examined the two-step procedure of integrating out the complex structure moduli and dilaton,  and then studying the effective theory for the K\"ahler moduli, instead of analyzing  all moduli on the same footing \cite{Choi:2004sx,deAlwis:2005tf,Lust:2005dy,Curio:2006ea,Abe:2006xi}.
The underlying justification for a two-step procedure is that the mass scale $m_{\rm flux}$  is set by the flux quantization condition,  and  does not diminish as $W_0$ is fine-tuned to be small,  whereas the  mass of the K\"ahler modulus $T$  at the minimum is proportional to $W_0$.
To understand this, we  expand the flux superpotential  around the supersymmetric minimum,\footnote{To be precise,  we mean the minimum determined by $D_{\zeta_{\alpha}}W_0 = D_{\tau}W_0 = 0$, where we stress that $W_0$ is the flux superpotential (\ref{W0}), not the  full superpotential (\ref{KKLTW}).}
\begin{equation}
W_0 = W_0|_{Z=0} + \ell_A Z^A + m_{AB} Z^{A}Z^{B}  + \cdots \ ,
\end{equation} where $Z^{A} \equiv \{\tau,\zeta^{\alpha}\}$, and $W_0|_{Z=0},\ell_A,m_{AB}$ are constants dictated by the quantized  three-form fluxes.
Via a fine-tuned choice of  fluxes,  one can arrange for $W_0|_{Z=0}$ to be small,  and this contrivance does {\it{not}}  render $m_{AB}$ atypically small at the same time ---  in fact, a further  fine-tuning would  be needed to reduce $m_{AB}$.
This fact is  true in generic configurations, but can fail  in simple examples with special structures.
For example, because $W_0$ involves the dilaton only through the definition (\ref{g3def}) of $G_3$, which  is linear in $\tau$,  we see that $m_{\tau\tau}=0$.
As a result, the dilaton acquires a mass from $W_0$ only by mixing with the complex structure moduli $\zeta^{\alpha}$, through couplings $m_{\tau\zeta^{\alpha}}$.
In a compactification with $h^{1,2}_-=0$, where there are no complex structure moduli, no such coupling is possible, the dilaton does not acquire a mass of order $m_{\rm flux}$, and it is not consistent to integrate out $\tau$  before studying the  K\"ahler moduli \cite{Choi:2004sx,deAlwis:2005tf}.
However, this example is of limited interest because the  mechanism of~\cite{Bousso:2000xa} is inoperative there.
In summary (see \cite{Abe:2006xi}),  integrating out the complex structure moduli and dilaton is consistent when these fields have large supersymmetric masses,  which is the generic situation.\footnote{Some care is needed to ensure that holomorphy is maintained in this process,  as explained in \cite{deAlwis:2005tf} and described in an explicit example in \cite{Curio:2006ea}.}
One can therefore  treat $W_0$  as a constant,  taking $W_0=W_0|_{Z=0}$, as we have done  in  the remainder of this section.

\subsection*{Large Volume Scenario}

The Large Volume Scenario \cite{Balasubramanian:2005zx} achieves stabilization of the K\"ahler moduli  by balancing the leading $\alpha^{\prime}$ correction~(\ref{equ:Ka}) to $K$ against the nonperturbative superpotential~(\ref{gauginocondensation}).  The success of this approach rests on  stabilizing the overall volume ${\cal V}$ at such large values that one can consistently neglect the (unknown) $\alpha^{\prime}$ and $g_{\rm s}$  corrections that are formally subleading in ${\cal V}$ compared to~(\ref{equ:Ka}).

Combining the constant flux superpotential (\ref{GVW}) with the $\alpha'$-corrected K\"ahler potential (\ref{equ:Ka}) leads to the following contribution to the scalar potential:
\beq
\delta V_{(\alpha')} \, =\, 3 \hskip 1pt \hat \xi e^{K} \frac{\big( \hskip 2pt  \hat \xi^2 + 7 \hat \xi {\cal V} +{\cal V}^2  \hskip 1pt  \big) }{\big(  \hskip 2pt  {\cal V} - \hat \xi \hskip 2pt \big) \big(  \hskip 2pt  2{\cal V}+ \hat \xi  \hskip 2pt  \big)^2} \, W_0^2\, \approx \, \frac{3 }{4} \hskip 1pt \hat \xi \hskip 1ptW_0^2\, \frac{1}{{\cal V}^3}\ , \label{equ:Valpha}
\eeq
where we have defined $\hat \xi \equiv \xi/g_{\rm s}^{3/2}$, cf.~(\ref{equ:Ka}).
Adding (\ref{equ:Valpha}) to (\ref{equ:Vnp}),
one finds\footnote{For the $\alpha'$ and string loop expansions to be valid, we require ${\cal V} \gg \hat \xi \gg 1$, as discussed further below; see \cite{Conlon:2005ki} for a systematic exposition  of the $\alpha^{\prime}$  expansion in this setting.}
\begin{align}
V_{(np)} + \delta V_{(\alpha')} &= e^{K} \Bigg\{ K^{j \bar \imath} \left[ a_j {\cal A}_j \,a_{\bar \imath} \bar {\cal A}_{\bar \imath}\, e^{-(a_j T_j + a_i \bar T_i)} \right. \nonumber \\
&\hspace{1cm} \left.
 - \, \left(a_j {\cal A}_j \, e^{-a_j T_j} \bar W \partial_{\bar \imath} K + a_{\bar \imath} \bar {\cal A}_{\bar \imath} \, e^{- a_i \bar T_i } W \partial_j K \right) \right] \nonumber  \\
&\hspace{1cm}  +\,  \frac{3 }{4} \hskip 1pt \hat \xi \hskip 1ptW_0^2\, \frac{1}{{\cal V}}\Bigg\}\ .  \label{lvsfullv}
\end{align}
At very large volume,
the perturbative term (\ref{equ:Valpha}) dominates over the nonperturbative terms~(\ref{equ:Vnp}).
Competition between  (\ref{equ:Valpha}) and (\ref{equ:Vnp}) can occur if one  or more  cycles are exponentially smaller than the largest cycles.
Denoting the small cycle volumes by $\tau_s \equiv \frac{1}{2}(T_s+\bar{T_s})$,  the idea is to  take the limit
\begin{equation}
{\cal{V}} \to \infty \ , \quad {\rm with}\ \ a_s \tau_s = \ln{\cal V} \ .  \label{lvslim}
\end{equation}
Along the ray in the K\"ahler moduli space defined by (\ref{lvslim}), the exponentials $e^{- a_s T_s }$ in (\ref{equ:Vnp}) are proportional to $1/{\cal{V}}$, and all terms in (\ref{lvsfullv}) are of the same order in $1/{\cal{V}}$.  Notice that the hierarchy~(\ref{lvslim}) is only possible for $h^{1,1}_{+}>1$ --- we will therefore take $h^{1,1}_{+}>1$ for the remainder of this discussion.

The sign of $\hat\xi$ is determined by the topology of the compactification, with $\hat\xi >0$ corresponding to $\chi(X_6) < 0$. In this section we will assume that $\hat\xi >0$,  which implies
that the contribution (\ref{equ:Valpha}) is positive, so that the potential~(\ref{lvsfullv}) approaches zero from below at large  ${\cal{V}}$ along the ray~(\ref{lvslim}).
To establish the existence of a minimum, one then needs to argue, first, that  the potential along (\ref{lvslim})  becomes positive at  sufficiently  small ${\cal{V}}$, so that  by continuity the potential restricted to (\ref{lvslim}) is minimized at an intermediate point ${\cal{V}}_{\star}$.
Second, one must show that at ${\cal{V}}_{\star}$, (\ref{lvsfullv}) is non-decreasing in the $h^{1,1}_{+}-1$  directions in the K\"ahler moduli space that are perpendicular to the ray (\ref{lvslim}).

A  useful heuristic argument that is valid in certain simple cases (with provisos enumerated below) goes as follows. If the term (\ref{equ:Valpha})  is dominant over the exponential terms  at small volume,  this establishes
that (\ref{lvsfullv}) likewise becomes positive at small volume.  Next, if the leading exponential terms in (\ref{lvsfullv}) are  positive, and all $h^{1,1}_{+}-1$ K\"ahler moduli appear in the nonperturbative superpotential,  the potential increases in the directions transverse to the ray~(\ref{lvslim}).   In combination, these  assumptions imply the existence of a minimum  at exponentially large volume.
This minimum has negative vacuum energy, so the spacetime solution is $AdS_4$.   Because the F-terms are non-vanishing  in the minimum, supersymmetry is  spontaneously broken \cite{Balasubramanian:2005zx}.

Let us now discuss  the conditions for a minimum in more detail, following \cite{Cicoli:2008va}.
We divide the K\"ahler moduli into two classes; those corresponding to big and small cycles,
\beq
\{T_i\}=\{T_{b}^{\rho}\}\cup\{T_{s}^{r}\}\ ,
\eeq
where
$r=1,\ldots, N_{s}$ and
$\rho=1,\ldots, N_{b} = h^{1,1}_+ - N_s$.
We consider  the large volume limit
\begin{equation}
{\cal{V}} \to \infty \ , \quad {\rm with}\ \ a_s^{r} \tau_s^{r} = \ln{\cal V}\, ~~\mbox{for all} \, ~r=1,\ldots , N_{s}\ .  \label{lvslimN}
\end{equation}
To check for the  existence of a minimum  in the limit (\ref{lvslimN}), one needs to examine in detail the inverse K\"ahler metric $K^{i\bar{\jmath}}$, and in particular the block corresponding to the small cycle moduli~$T_{s}^{(a)}$.
A systematic treatment for
$N_{s}=1$ and $N_{s}=2$ appears in \cite{Cicoli:2008va}.

To understand the results of \cite{Cicoli:2008va},  one piece of geometrical background is necessary.
Suppose that ${\cal M}$  is a complex manifold (potentially containing singularities) of complex dimension $n$,  and let $p$  be a point in ${\cal M}$.
The {\it{blowup}} of ${\cal M}$ at a non-singular point $p$ replaces $p$  with a copy of $\mathbb{P}^{n-1}$,  known as the {\it{exceptional divisor}}.
The blowup of a singular point can  result in more general exceptional divisors.  When the blowup of ${\cal M}$ is a Calabi-Yau threefold, the exceptional divisor is a four-cycle,  with size parameterized by one of the K\"ahler moduli.
When the exceptional divisor satisfies the rigidity condition (\ref{rigidity}), the corresponding Euclidean D3-brane superpotential term is non-vanishing~\cite{Witten:1996bn}.

A necessary condition  for an LVS minimum is that at least one of the $N_{s} \ge 1$ small cycles is a rigid
exceptional divisor arising from blowing up a singular point~\cite{Cicoli:2008va}.
When $N_{s}=1$,  this condition guarantees  that (\ref{lvsfullv}), {\it{restricted to the ray}} (\ref{lvslimN}), has a minimum at exponentially large volume.
Whether this is a minimum of the full potential depends on the curvature in the $N_{b}-1$  directions  perpendicular to (\ref{lvslimN}), as we discuss further below.
For the case $N_{s}=2$, if  the two small cycles  correspond to blowups of distinct points, then (\ref{lvsfullv}) restricted to (\ref{lvslimN}) again has a minimum at exponentially large volume, with the same caveat  about transverse directions.   If instead the two small cycles are two independent resolutions of the same  singular point,  then an LVS  minimum along~(\ref{lvslimN}) exists only if there is a basis in which the volume ${\cal V}$  is symmetric in the two K\"ahler moduli $T_{s}^{1}$ and $T_{s}^{2}$.
For a discussion of the necessary conditions on $K^{i\bar{\jmath}}$ in  the context of a survey of a class of Calabi-Yau manifolds,
see \cite{Gray:2012jy}.

A canonical class of examples  of LVS vacua arise in  what are known as `Swiss-cheese' Calabi-Yau  manifolds, whose volumes can be written as\footnote{See \cite{Gray:2012jy} for a study of the incidence of the form (\ref{equ:weakSwiss}) in a class of Calabi-Yau manifolds.}
\beq
{\cal V} = \alpha\hskip 1pt  \tau_b^{3/2} - p^{(3/2)}(\tau_s^{r})  \ , \label{equ:weakSwiss}
\eeq
where $\alpha > 0$, and $p^{(3/2)}$ is a homogeneous polynomial of degree $3/2$ in  the small cycle moduli $\tau_s^{r}$, $r=1,\ldots , N_{s}$.
A proper subset of Swiss-cheese   Calabi-Yau manifolds  take the  `strong' form
\beq
{\cal V} = \alpha \left( \tau_b^{3/2} - \sum_{r=1}^{N_s} \lambda_r (\tau_s^{r})^{3/2} \right)\ , \label{equ:Swiss}
\eeq
with $\lambda_r > 0$.
This compactification has a single large four-cycle, with volume $\tau_b$, and $N_s = h^{1,1}_+-1$ small  four-cycles, with volumes $\tau^r_s$.
Increasing one of the $\tau^r_s$ with all else fixed {\it{decreases}} ${\cal V}$, so the small cycles act like holes in a large cheese.
The structure (\ref{equ:Swiss}) can arise if  the $N_s$ small cycles  correspond to the  blowups  of $N_s$ distinct singular points.
In the case of a  compactification of strong Swiss cheese form (\ref{equ:Swiss}), the necessary conditions described in \cite{Cicoli:2008va} are readily met, for any~$N_s >0$.

The final, critical question  is whether the potential is stable or unstable in the $N_{b}-1$  directions  perpendicular to~(\ref{lvslimN}).  In fact, (\ref{lvsfullv}) per se,  which includes only the leading $\alpha^{\prime}$  correction to $K$, viz.~(\ref{equ:Ka}), has $N_{b}-1$ flat directions.  The exact moduli potential, incorporating all perturbative and nonperturbative effects  in $g_{\rm s}$ and $\alpha^{\prime}$, very plausibly depends on the $N_{b}-1$ fields that are unlifted  by~(\ref{lvsfullv}).\footnote{As we will explain in \S\ref{lvscpct}, it has been suggested \cite{Cicoli:2008va,Cicoli:2008gp} that the leading $g_{\rm s}$  correction to $K$, with form conjectured in \cite{Berg:2007wt} following computations in \cite{Berg:2005ja, Berg:2005yu},  can stabilize the $N_{b}-1$ flat directions.  However, a  more detailed demonstration of  stability  would be valuable.}
However, appealing to an unknown  and uncomputable potential to lift these  remaining moduli  is problematic, not least because  there is no evidence that the resulting masses-squared will all be positive.   That is,  further perturbative corrections beyond (\ref{equ:Ka}) could well introduce instabilities along one or more of the $N_{b}-1$ flat directions of (\ref{lvsfullv}), leading to an LVS saddle point rather than a minimum.\footnote{However, if it can be established that the potential increases as one moves toward each of the boundaries of the moduli space, then one can again make a continuity argument for the existence of a minimum.  We thank Joe Conlon and Fernando Quevedo  for discussions of this point.}  Indeed, as we argue in \S\ref{sec:RMTsupergravity} below, in certain ensembles of supergravity theories it is overwhelmingly  improbable that  all $N_{b}-1$ flat directions are stabilized rather than destabilized: the probability of stability is exponentially small in  $N_{b}$.  Whether the assumptions of  \S\ref{sec:RMTsupergravity} are applicable to the moduli potential in LVS is an important open question (see \cite{Rummel:2013yta} for recent work).

\vskip 4pt
In summary,  the necessary conditions for an LVS  minimum are the following: $\hat\xi >0$; $h^{1,1}_{+} \equiv N_{s} + N_{b} >1$; $N_{s}\ge 1$ K\"ahler moduli corresponding to the blowups of points.
For $N_{s}>1$, further conditions on the blowups are necessary \cite{Cicoli:2008va}, while for $N_{b}>1$, it is necessary that further corrections, beyond~(\ref{lvsfullv}), render stable the $N_{b}-1$ flat directions of (\ref{lvsfullv}).   Explicit examples with $N_{b}=1$ that meet all other necessary criteria  are now well-known \cite{Balasubramanian:2005zx,Conlon:2005ki,Cicoli:2008va}.

Several differences between LVS and the KKLT scenario  should be emphasized.  In LVS, some cycles  are exponentially larger than others, while in KKLT the cycles  are not hierarchically different in size.  In KKLT,  the classical flux superpotential $W_0$ is fine-tuned to be  exponentially small,  while in LVS $W_0$  is of order unity.
In KKLT,  the $AdS_4$  vacuum is supersymmetric,  whereas in LVS  the $AdS_4$  vacuum is non-supersymmetric.   However, in both scenarios some form of `uplifting'  effect is required  to achieve a de Sitter  vacuum, as we now explain.

\section{De Sitter Vacua}
\label{sec:dSVacua}

The KKLT and LVS vacua just described are fully stabilized,  in the sense that there are no remaining instabilities and no flat directions of the potential.  Even so, these vacua have negative energy and are therefore unsuitable for a realistic cosmology. To describe the early universe (inflation) and the late universe (dark energy)
requires vacua with positive energy, i.e.~de Sitter solutions.
Constructing metastable de Sitter vacua in string theory turns out to be far more difficult than constructing stable anti-de Sitter vacua.
As a first step toward appreciating the problem, one can ask what it is about AdS vacua that makes them a natural endpoint of  the moduli stabilization procedure.
In the KKLT scenario, supersymmetry guarantees the stability of  the AdS solution.
In LVS, the AdS vacua are not supersymmetric, but their stability can be established by asymptotic arguments, in particular by the fact that $V \rightarrow 0$ from below for ${\cal V} \rightarrow \infty$.
In contrast, dS vacua are much more susceptible to instabilities. This becomes apparent when one tries to construct explicit de Sitter solutions in string compactifications.

\subsection{Uplifting to De Sitter}

The leading paradigm for constructing metastable de Sitter solutions from stabilized AdS
solutions is known as {\it uplifting}.
The stable vacuum is interpreted as a background solution to which the effects of supersymmetry breaking in some new sector, not considered in the original stabilization, may be added.
Although the steps of stabilization and uplifting are conveniently described as sequential, in reality the full set of equations of motion, for all fields, must of course be solved simultaneously.
This presents a difficulty, because the vacuum energy contribution from the uplifting sector cannot be a perturbatively small correction to the original vacuum energy.
In most approaches the stabilization in AdS is analyzed in a supersymmetric effective action, and one must take care that the large supersymmetry breaking from the uplifting sector does not invalidate this treatment.
In summary, the task in uplifting is to identify a sector that breaks supersymmetry dynamically, in a parametrically controlled manner, and makes a positive contribution to the vacuum energy without disrupting the physics that led to a stabilized AdS vacuum.
As we explain in the following, these requirements are very challenging, even taken in isolation.

\vskip 4pt
First of all, one must  engineer a sector of fields that breaks supersymmetry.
As a concrete example, consider
placing multiple D-branes at the singular apex of a Calabi-Yau cone, leading to a supersymmetric gauge theory in four dimensions.   Some of the  resulting  gauge theories have metastable vacua in which supersymmetry is dynamically broken~\cite{Franco:2006es,Argurio:2006ew,Florea:2006si,Argurio:2006ny,Diaconescu:2007ah,GarciaEtxebarria:2007vh,Buican:2007is},  while in other cases, such as \cite{Berenstein:2005xa,Franco:2005zu,Bertolini:2005di}, there are runaway instabilities in  directions parameterized by K\"ahler moduli~\cite{Franco:2005zu,Intriligator:2005aw,Brini:2006ej}.
But even if  one finds a configuration of D-branes on a noncompact Calabi-Yau cone leading to a flat space gauge theory that dynamically breaks supersymmetry,  establishing that metastability survives compactification is highly non-trivial (but see \cite{Diaconescu:2007ah}).  The essential issue is that in  the low-energy Lagrangian of a compactification, all parameters are determined by the
vevs of fields, and  are therefore dynamical  at sufficiently high energies.  Any gauge theory construction relying on a  non-dynamical parameter --- for example, the mass of a quark flavor,  as in \cite{Intriligator:2006dd} ---  is potentially vulnerable,  upon compactification, to an instability along which this parameter evolves.
Often a second stage of model-building is required  in which one generates the desired vev  dynamically and establishes the absence of instabilities --- see  e.g.~\cite{Argurio:2006ny}.

After identifying  a supersymmetry-breaking sector,  one must compute  the effects of supersymmetry breaking on the remaining fields.  A pervasive but  potentially deceptive picture for uplifting is that the uplifting sector exists `somewhere else' in the compactification: stabilization in AdS is imagined to result from sources and fields in one region, while supersymmetry breaking arises in another region, and the vacuum energy contributions are therefore approximately additive, by locality in the extra dimensions.
One problem with this modular picture, as we explain in detail in \S\ref{etaproblem}, is that geometric separation does not imply complete decoupling of two sectors.
At the very least, the supersymmetry-breaking sector interacts with the remaining fields by its coupling to the overall compactification volume ${\cal V}$: any source ${\cal S}$ of positive energy\footnote{A ten-dimensional cosmological constant would be an exception, but this is excluded by ten-dimensional supersymmetry.} in the four-dimensional theory must be negligible in the limit  ${\cal V} \to \infty$, and so must enter the Lagrangian as
\begin{equation}
\rho_{{\cal S}} = \frac{D}{{\cal{V}}^{\alpha}}\ ,  \label{upliftingterm}
\end{equation}  asymptotically at large ${\cal V}$, with $D$ and $\alpha$ being positive constants.
The potential (\ref{upliftingterm}) contributes to the equation of motion for the K\"ahler modulus parameterizing the volume ${\cal V}$: because $D>0$, there is a force toward larger volume, cf.~\S\ref{dimensionalred}.
This force can substantially change the vev of ${\cal V}$, or even drive runaway decompactification.  The net result is that a computation of physical parameters in the original AdS vacuum will not necessarily give an accurate prediction for these quantities in the dS solution.
Accurate determination of the effective action in  metastable de Sitter solutions remains a core challenge  for inflationary model-building in string theory, as we  discuss further in Section~\ref{sec:StringInflation}.

Many constructions of uplifting to de Sitter vacua along the lines of \cite{Kachru:2003aw}, as well as alternatives to uplifting, have been proposed: see e.g.~\cite{Burgess:2003ic,Escoda:2003fa,Brustein:2004xn,Saltman:2004sn,Saltman:2004jh,Parameswaran:2006jh,Lebedev:2006qq,Lebedev:2006qc,Achucarro:2006zf,Dudas:2006gr,Westphal:2006tn,Cremades:2007ig,Covi:2008zu,Polchinski:2009ch,Krippendorf:2009zza,Dong:2010pm,Rummel:2011cd,Louis:2012nb,Cicoli:2012fh}.
Analyses in type IIA  string theory include~\cite{Davidse:2005ef,Saueressig:2005es,Silverstein:2007ac,Caviezel:2008ik,Haque:2008jz,Danielsson:2009ff,Wrase:2010ew,Danielsson:2010bc,Danielsson:2011au,Shiu:2011zt,Danielsson:2012by,Danielsson:2012et}, while for proposals in the heterotic string, see~\cite{Buchbinder:2003pi,Becker:2004gw,Buchbinder:2004im,Curio:2006dc,Parameswaran:2010ec}.
See \cite{Maloney:2002rr} for an  early construction  of de Sitter vacua in  supercritical string theory, i.e.~for total spacetime dimension $D>10$.  Discussions of de Sitter vacua of M-theory,  and of supergravity theories with $N>1$  supersymmetry in four dimensions,  can be found in e.g.~\cite{Hull:2001ii,Kallosh:2001gr,Gibbons:2001wy,Kallosh:2001tm,Chamblin:2001dx,Fre:2002pd,GomezReino:2008bi,Roest:2009tt}.

\subsection{SUSY Breaking from Antibranes} \label{d3b}

The archetypal  configuration  \cite{Kachru:2002gs} for uplifting to de Sitter space consists of $p$  anti-D3-branes placed at the tip of a {\it{Klebanov-Strassler (KS) throat}} \cite{Klebanov:2000hb},  which is a smooth, asymptotically conical supergravity solution described in detail in \S\ref{ssec:warped}.
The tip of the KS throat  is a three-sphere threaded by three-form flux:
\begin{equation}
\frac{1}{(2\pi)^2 \alpha'} \int_{S^3} F_3 \equiv M \ ,
\end{equation} with $M$ an integer.
The KS solution preserves ${\cal N}=1$  supersymmetry in four dimensions, but the anti-D3-branes  are incompatible with these supersymmetries,  so the total  configuration is non-supersymmetric.

Because the anti-D3-branes carry negative D3-brane charge, cf.~eq.~(\ref{equ:CS}),   while the fluxes in the  KS solution carry positive D3-brane charge,  annihilation of anti-D3-branes and flux is possible in some circumstances.
For a given background there is a critical value  $p_{\star} \approx 0.08 \hskip 1pt M$ such that for $p>p_{\star}$, rapid classical annihilation can occur, while for $p<p_{\star}$ the leading
annihilation instability involves quantum tunnelling~\cite{Kachru:2002gs}, and is nonperturbatively slow.
It was therefore argued in \cite{Kachru:2002gs} that a collection of $p<p_{\star}$ anti-D3-branes in a KS throat is  a metastable, supersymmetry-breaking configuration.
In \cite{Kachru:2003aw}, and in many subsequent works, this configuration was used as a module effecting uplifting: see fig.~\ref{fig:KKLT}.

The idea of antibrane uplifting has recently been challenged~\cite{Bena:2009xk}.
In particular, it was observed that the known, approximate solutions for $p$ anti-D3-branes in a KS background  are singular.
If one could establish that  the  corresponding full,  exact solution  manifests {\it{unphysical}}  singularities,  this would imply that anti-D3-branes in a KS throat do not
provide a consistent metastable super\-symmetry-breaking  configuration.
To discuss this important point
\cite{Bena:2009xk,Bena:2010ze,Dymarsky:2011pm,Bena:2011hz,Bena:2011wh,Blaback:2012nf, Bena:2012tx,Kutasov:2012rv,Bena:2012bk,Bena:2012vz,Bena:2012ek,Junghans:2013xza,Dymarsky:2013tna,Junghans:2014xfa,Junghans:2014wda}, we first have to explain the sense in which the known solutions are approximations (see also~\S\ref{sss:approximations}).

\vskip 4pt
\noindent
{\it The meaning of exact and approximate.}---By an exact solution of string theory, we mean a configuration of the massless fields that
solves the exact equations of motion, i.e.~the equations of motion that incorporate all perturbative and nonperturbative corrections in the $\alpha^{\prime}$  and $g_{\rm s}$  expansions.
In contrast, an exact solution of  classical, two-derivative\footnote{The `two-derivative'  qualifier refers to omission of higher-curvature contributions, and is usually assumed implicitly.} supergravity ---  generally abbreviated as an `exact supergravity solution' --- solves the equations of motion  expressed to leading order in  $\alpha^{\prime}$  and $g_{\rm s}$.  These are the equations of motion  determined by the two-derivative, ten-dimensional actions (\ref{equ:SIIAX}), (\ref{equ:SIIBX}) for type IIA and type IIB  string theory, respectively.
Next,  we recall exact and approximate solutions involving D-brane sources (see \S\ref{sec:Dbranes} for more details).
Consider, for example, a stack of $N$ coincident D3-branes placed in ten-dimensional Minkowski space.
The D3-branes  warp the space: comparing to (\ref{curvaturelimit}), the  characteristic radius of curvature~$R$~is
\begin{equation}
R^{4} = 4\pi \hskip 1pt g_{\rm s} N (\alpha^{\prime})^2\ ,
\end{equation} so that corrections in the $\alpha^{\prime}$  expansion can be ignored for $g_{\rm s} N \gg 1$, while as usual corrections in the string loop expansion can be ignored for $g_{\rm s} \ll 1$.  Thus, the exact supergravity solution determined by the D3-brane sources is an approximation to an underlying exact string theory solution,  and the small expansion parameters governing the approximation are  $g_{\rm s} \ll 1$ and $(g_{\rm s} N)^{-1} \ll 1$.  Notice that for  any fixed $N$,  the string loop and $\alpha^{\prime}$  expansions cannot both be arbitrarily accurate.
In particular, if one imagines sending  $g_{\rm s} \rightarrow 0$ for $N$  fixed, a curvature singularity develops, and the $\alpha^{\prime}$  expansion  becomes invalid  near the source.
For  one or more D3-branes in flat space,  this singularity  is not  surprising, and is not indicative of any sickness: at  weak string coupling a D3-brane is a heavy  source whose transverse thickness is of order $\sqrt{\alpha'}$.  This system is well-behaved and can be defined by referring to the conformal field theory describing open strings ending on the D-branes.
Less practically, one could imagine  incorporating all $\alpha^{\prime}$ corrections in order to  obtain a solution that does not break down near the source.
Summarizing, a single D3-brane in flat space  is a singular source in supergravity: this is the expected and allowable singularity that arises from a localized source,  just as for a point charge  in  electromagnetism.  For $N$  coincident D3-branes,  the  curvature of the supergravity solution is  small at large $g_{\rm s} N$.

\vskip 4pt
\noindent
{\it Singular antibranes.}---In view of the above remarks, it should come as no surprise that a single anti-D3-brane placed in a KS throat is a singular source in supergravity.
More generally, for $p$  anti-D3-branes, there is no reason to expect a smooth supergravity solution if $g_{\rm s} p \ll 1$: this would amount to better behavior than  that of D3-branes, which are supersymmetric in the KS background  and hence are `maximally innocuous'.
On the other hand, for $g_{\rm s} p \gg 1$ it is reasonable to expect that a smooth, exact supergravity solution  exists, but none has been constructed to date: only singular approximate solutions have been obtained.   The important question is whether the singularities are a  signal of unphysical behavior,  or instead merely  reflect our technical limitations.

In the following, we will discuss two aspects of  the  singularity problem: first, we will ask whether the singularities could be artifacts of the approximations involved in the analysis.
Even if one can argue that approximations are not the cause of the singularity,  one still has to ask whether the singularities are unexpected and signal an inconsistency for antibranes in KS throats.

\vskip 4pt
\noindent
{\it Approximate treatments.}---Determining the supergravity solution for antibranes in KS is extremely complicated,  and two further approximations,  beyond the fundamental expansions described above,  have been employed  to simplify the task: these are {\it{linearization}}  and {\it{smearing}}.
Linearization refers to an expansion of the supergravity equations of motion to first order in the strength of the source.
The smearing  approximation
 replaces anti-D3-branes at  a specific location on the $S^3$  with an equivalent charge and tension uniformly distributed over the $S^3$.  This reduces the equations of motion from PDEs to ODEs.   One may wonder whether either of these approximations could be the source of the apparent singularity.

\begin{itemize}
\item[$\triangleright$] {\it Linearization.}---The linearized supergravity solution for $p$  anti-D3-branes  smeared around the $S^3$ has been obtained in \cite{Dymarsky:2011pm,Bena:2011hz} (for related earlier work see \cite{DeWolfe:2008zy,McGuirk:2009xx}), and passes non-trivial consistency checks~\cite{Bena:2010ze,Dymarsky:2013tna}.
The characteristic  radius of curvature  near the source is
$R_{p}= (4\pi g_{\rm s} p)^{1/4} \alpha^{\prime 1/2}$,  so the linearized solution can be trusted at radial distances $r \gg R_{p}$  away from the tip: nearer to the tip it is inconsistent to neglect $\alpha^{\prime}$  corrections, and some of the background fields become singular.
In particular, the  three-form fluxes  are singular near the source.
It has been argued in \cite{Bena:2012bk} that the singularity in the flux is not a consequence of linearization:  the  nonlinearly backreacted, but still smeared, solution  displays singularities.
This leaves smearing as  perhaps the most  plausible  cause of the singularities.  (See e.g.~\cite{Blaback:2011nz,Bena:2012tx,Junghans:2013xza} for related work on the problem of singularities from localized sources.)

\item[$\triangleright$] {\it Smearing and brane polarization.}---What sort of smooth supergravity solution  might one expect for $p \gg g_{\rm s}^{-1}$ non-smeared anti-D3-branes?
As noted in  \cite{Kachru:2002gs}, anti-D3-branes  that are initially coincident are driven to redistribute themselves along an $S^2$ in the $S^3$,   manifestly breaking some of the symmetries  preserved by a configuration  smeared on the $S^3$.  This process can be viewed as {\it polarization} of the branes \cite{Myers:1999ps} by the flux  background, as in the related solution found by Polchinski and Strassler~\cite{Polchinski:2000uf}, where brane polarization resolves the singularity present in the unpolarized configuration.
In \cite{Dymarsky:2011pm}, it was conjectured that a smooth anti-D3-brane solution can be modelled on the system in \cite{Polchinski:2000uf},  with polarization of the anti-D3-branes along an $S^2 \subset S^3$ being responsible for removing the singularities.\footnote{D-branes can  generally polarize  in  multiple ways, and an alternative to the polarization identified in \cite{Kachru:2002gs}, where the anti-D3-branes spread along an $S^2 \subset S^3$, is for the anti-D3-branes to spread along a different $S^2$, namely the $S^2$  that shrinks toward the tip of the throat (see \S\ref{ssec:warped}).   This process moves the anti-D3-branes radially outward, away from the tip, a direction of motion that is opposed by the classical potential from fluxes.  It was shown in \cite{Bena:2012vz} that this  alternative, radial polarization is not possible, but this does not exclude the expected polarization of \cite{Kachru:2002gs}.}
Such a solution is  clearly incompatible with a smearing approximation, but solving the equations of motion in this setting is a formidable technical challenge, and at present it is not known whether brane polarization will resolve the singularities.
\end{itemize}

\vskip 4pt
\noindent
{\it Expected and unexpected singularities.}---One further issue in the study of singularities from anti-D3-branes concerns the nature of the singular behavior.
No one should be surprised  by the fact that the electric field sourced by a pointlike electron in classical Maxwell theory  is singular near the  electron.
The corresponding potential $\Phi$ obeys
\begin{equation}
\nabla^2 \Phi = 4\pi e \, \delta(\x) \ , \label{electrons}
\end{equation}
for an  electron at position $\x$,  which is  solved by
\begin{equation}
\Phi(\x') = -\frac{e}{|\x -\x'|} \ .  \label{emphi}
\end{equation}  The singularity of (\ref{emphi}), and of the corresponding electric field, is expected, because the electron  is a singular,  perfectly localized source for the  electric field.
Of course, the divergence  in the energy of the electric field is removed in the  quantum theory.

The question, then, is whether the singularities seen in \cite{Bena:2009xk,Dymarsky:2011pm,Bena:2011hz} are expected,  and therefore  plausibly resolved in the exact solution.
A central concern raised by \cite{Bena:2009xk} is that the singularities in three-form flux ``do not appear to have a distinct physical origin'' \cite{Bena:2009xk}.  That is, according to~\cite{Bena:2009xk} it is not obvious how the anti-D3-brane can serve as a source for singular three-form flux,  and correspondingly these singularities are unexpected.

It is certainly true that  the only flux  sourced by an anti-D3-brane in empty flat space is  five-form flux $F_5$, just as for a D3-brane: see the Chern-Simons coupling~eq.~(\ref{equ:CS}).
For an anti-D3-brane in a classical flux background, the problem is more subtle:  the supergravity equations of motion are nonlinear, and the various fluxes are coupled to each other, as we will explain below.
To understand this case, we begin by developing intuition in a simpler example.

Let us see how a  point source of one field ${\cal A}$,  in a classical background of a second field ${\cal B}$,  can source a singular profile of a third field ${\cal C}$, even if the source does not have a direct coupling to ${\cal C}$ in the Lagrangian.
Consider classical  four-dimensional electromagnetism coupled to an axion $\phi$,  with Lagrange density
\begin{equation}
{\cal L} = - \frac{1}{2}(\partial\phi)^2-\frac{1}{4}F_{\mu\nu}F^{\mu\nu} - \frac{\phi}{f}F_{\mu\nu}F_{\rho\sigma}\epsilon^{\mu\nu\rho\sigma}\ ,
\end{equation}
where $f$ is the axion decay constant.
Suppose that  there is a  constant classical background magnetic field ${\boldsymbol B} = B\hskip 1pt \hat{\boldsymbol z}$, and place an electron in this field,  at rest at the origin.  The electron  sources  an electric field ${\boldsymbol  E}=-e\hskip 1pt \hat{\boldsymbol r}/r^2$,  so that  in
spherical polar coordinates $(r,\theta,\varphi)$ one has
\begin{equation}
{\boldsymbol E}\cdot {\boldsymbol B} = -\frac{e B}{r^2} \cos \theta \ .
\end{equation}
The equation of motion for $\phi$  is therefore
\begin{equation}
\nabla^2\phi =  \frac{e B}{f r^2} \cos \theta \ .
\end{equation}
Thus, the axion $\phi$ effectively has a local source, even though the electron alone does not couple to $\phi$.
This example illustrates that in theories with  local sources and multiple coupled fields, not every singular field profile  arises from  a `one field, one source'  coupling in the Lagrangian:  classical background fields can also play a role.

In the case of an anti-D3-brane in a KS throat, the classical background field analogous to ${\boldsymbol B}$ is the three-form flux $G_3$ of the KS solution.  Schematically, the anti-D3-brane  sources  singular five-form flux, which couples  to the  non-singular background three-form flux,  and thereby sources a singular three-form flux.   To see this more explicitly, we consider the equation of motion of the three-form flux. In the KS solution the dilaton is constant and the imaginary anti-self-dual component of the flux, $G_- \equiv (\star_{6} - i)G_3$, satisfies
\begin{equation} \label{intertwined}
{\rm d} G_- = - {\rm{d}}\left(\frac{\Phi_- G_+}{\Phi_+} \right) \ ,
\end{equation} where $G_+ \equiv (\star_{6} + i)G_3$ and $\Phi_\pm \equiv e^{4A} \pm \alpha$.
The anti-D3-brane is a singular source for $\Phi_-$, while $G_+ \neq 0$ and $\Phi_+ \neq 0$ in the KS background.
Thus,  solving (\ref{intertwined}) requires that $G_-$ be singular.
As a result, divergences in three-form flux are to be expected when  anti-D3-branes are placed in a KS~throat.

\vskip 4pt
\noindent
{\it Summary.}---Let us summarize
the key facts  and questions about singularities from anti-D3-branes.  The linearized solution describing $p$ anti-D3-branes smeared around the tip of  a Klebanov-Strassler throat  has been obtained in \cite{Bena:2009xk,Dymarsky:2011pm,Bena:2011hz}, and passes multiple consistency checks \cite{Bena:2010ze,Dymarsky:2013tna}.
The  three-form flux in this solution is singular near the source \cite{Bena:2009xk}.  The singularity is not an artifact of linearization \cite{Bena:2012bk},  but it is not known  whether  the smearing approximation is responsible for the singularity.
We have argued that  singularities in flux should  in fact be expected  in this setting,
but  to show definitively that the  singularities found in the solutions of \cite{Bena:2009xk,Dymarsky:2011pm,Bena:2011hz} are  (or are not) physical,  the most  compelling course is to exhibit the  corresponding non-singular solution  (or show that none exists).
A leading proposal for a non-singular resolution, for $g_{\rm s}p \gg 1$, involves the anti-D3-branes
polarizing \cite{Myers:1999ps},  as in  the Polchinski-Strassler  solution \cite{Polchinski:2000uf}, but this necessarily breaks the symmetries used to smear the anti-D3-branes, and obtaining the corresponding solution is a difficult open problem.


\section{Statistics of String Vacua} \label{landscapestatistics}

\vspace{0.3cm}
\begin{quote}
{\footnotesize I would be happy personally if the multiverse interpretation is not correct, in part because it potentially limits our ability to understand the laws of physics.
But none of us were consulted when the universe was created.}

\hfill {\footnotesize Edward Witten~\cite{Witten}.}
\end{quote}
\vspace{0.2cm}

At the fundamental level, string theory contains no continuously-adjust\-able
dimensionless parameters, but the theory has an astronomical number of solutions, or vacua.
These solutions are distinguished from each other by the vevs of continuously-adjustable moduli fields,
and also by discrete data, consisting of topological  invariants of the compactification itself, such as Hodge numbers;  topological properties of any branes wrapping internal cycles, and of gauge bundles on these branes;  and the number of units of quantized flux threading each cycle.
The number of distinct choices of integer data is extremely large, because many compactifications have  hundreds of independent cycles  on which flux can be placed.

\subsection{Landscape of Stabilized Vacua}

For the purposes of cosmology, it is important to understand solutions whose effective theories contain no massless scalar fields, i.e.\ solutions without moduli.  Such vacua are necessarily isolated: classical transitions from one to another require energy input.  The {\it{string landscape}} is the collection of all consistent solutions of string theory that have four large spacetime dimensions\footnote{The four-dimensional spacetime is often assumed to be maximally symmetric, i.e.~de Sitter space, Minkowski space, or anti-de Sitter space.  Isolated solutions without four large spacetime dimensions could also be considered to be part of the landscape, but we will focus on the class of vacua that are directly relevant for cosmology.}
and do not have moduli.  The expectation that string theory has a vast array of isolated solutions dates back to the early days of the theory \cite{Strominger:1986uh,Lerche:1986cx}, but detailed understanding of flux compactifications in recent years has brought the landscape into focus and has made explicit investigation possible.
In the same period, the discovery of dark energy \cite{Perlmutter:1998np, Riess:1998cb}
has
made understanding de Sitter solutions of string theory an urgent question.

What are the prospects for understanding the structure of the string landscape? There are two overarching challenges: accurately characterizing the effective theories whose isolated solutions comprise the landscape, and then exploring their innumerable vacua.  At present, there is some degree of understanding of the effective theories resulting from Calabi-Yau compactifications of type~II, type~I, and heterotic strings, in the regime of weak coupling and large volume.  Certain compactifications of M-theory and F-theory are likewise understood.   However, despite prolonged study,  non-Calabi-Yau compactifications (even if supersymmetric) are less understood, in part because fewer geometric and topological tools are applicable.
It would be premature to declare that the properties of the effective theories of presently-understood compactifications are in fact general characteristics of string theory.  Indeed, we find it plausible that most of the landscape remains to be discovered.  Even so, in the absence of an alternative, one can begin by surveying the part of the landscape that rests on known compactifications.

This brings us to the second difficulty, of working out the characteristics of the set of vacua of a
fully specified
ensemble of effective theories.  Understanding through enumeration is inconceivable for systems with $10^{500}$ vacua, which strongly motivates a {\it{statistical}} approach,  initiated by Douglas in~\cite{Douglas:2003um}.
Instead of computing all physically relevant quantities (a.k.a.~`observables') --- such as gauge groups, coupling constants, and mass spectra --- in a small number of actual vacua, one can instead determine the
statistical distribution of a given observable, or the correlations among observables, in a broad class of vacua.
We stress that the motivation for a statistical treatment of observables in the landscape goes beyond the practical difficulty of computing observables in explicit examples.  Few  now believe that string theory has a unique vacuum consistent with all observations, and the pressing task is not so much to find `the' vacuum describing our universe, but rather to understand the characteristic properties of realistic vacua.  Solving the Schr\"odinger equation for one single microstate of the ocean is of much less practical use than understanding thermodynamic and hydrodynamic quantities: the statistical description is simpler, but also more important as a description of the phenomena of the system.  Equally, in the landscape, the distributions of observables can display emergent simplicity.
Examples of simple patterns seen in the distributions of observables can be found in \cite{Grana:2005jc,Kumar:2006tn,Douglas:2006es,Denef:2007pq}.
For a comprehensive account of the statistics of flux vacua, we refer the reader to the excellent review \cite{Douglas:2006es}.

\subsection{Counting Vacua}

There is a general consensus\footnote{Limitations and weaknesses of the current evidence have been described in e.g.~\cite{Dine:2004fw,Banks:2004xh,Bena:2009xk,Bena:2012vz,Bena:2012ek,Banks:2012hx}.}
that the number of vacua in the landscape is immense, but it will be worthwhile to review key aspects of the argument.  For concreteness, we will consider type IIB flux compactifications on Calabi-Yau orientifolds (or more generally, compactifications of F-theory).

Consider an orientifold of a Calabi-Yau threefold, with a specified choice $\mathfrak{F}$ of quantized three-form fluxes: that is, for each independent three-cycle $\Sigma_3$, one chooses $\int_{\Sigma_3} F_3 \in (2\pi)^2\alpha'\,\mathbb{Z}$ and $\int_{\Sigma_3} H_3 \in (2\pi)^2\alpha'\,\mathbb{Z}$.
The result is a potential on the complex structure moduli space ${\cal M}_{{\cal{C}}}$,
\beq
V=V_{\mathfrak{F}}(\zeta_1,\, \ldots\, , \zeta_{h^{2,1}})\ .
\eeq
As reviewed above, this flux-induced potential is responsible for the stabilization of the complex structure moduli: the local minima of $V_{\mathfrak{F}}$ are generally isolated points $\{p_1,\ldots, p_{K}\}$ in ${\cal M}_{{\cal{C}}}$, and the complex structure moduli masses are generically nonvanishing at such minima.  However, the number $K$ of local minima of $V_{\mathfrak{F}}$ is {\it{not}} the primary large number responsible for the scope of the landscape: instead, the large number of choices ${\cal N}_{\mathfrak{F}}$ of quantized flux $\mathfrak{F}$, corresponding to distinct possibilities for the elementary topological data of the compactification, is the origin of the diversity of vacua.  As explained in \cite{Douglas:2006es}, ${\cal N}_{\mathfrak{F}}$ is large in Calabi-Yau compactifications because there are many --- typically, hundreds --- of independent three-cycles that the two fluxes can thread.
Each choice $\mathfrak{F}$ creates a distinct potential $V_{\mathfrak{F}}$ on ${\cal M}_{{\cal{C}}}$, and the number ${\cal N}_{\mathfrak{F}}$ of such choices is inarguably stupendous, at least of order $10^{500}$.

Let us now describe more carefully how the  number of choices of flux ${\cal N}_{\mathfrak{F}}$ is related to the number of vacuum  solutions.  As a first step  toward understanding the statistics of string vacua,  one can count {\it{supersymmetric}} vacua in type IIB flux compactifications.   More precisely,  following \cite{Ashok:2003gk},
we will discuss configurations in which the F-terms\footnote{The F-terms  described  here are those due to the classical flux superpotential $W_0$,  but nonperturbative contributions to the superpotential ---  for example, from Euclidean D3-branes ---  introduce further dependence on the complex structure moduli.} $D_{\zeta_i}W_0$ of the complex structure moduli $\zeta_i$ vanish.   At this stage the K\"ahler moduli sector is  ignored completely, so one must bear in mind that what we term `vacua' here  are  merely  solutions to the equations of motion in one sector, not full-fledged solutions of the total theory.

To count vacua, a natural object to consider is the  density of vacua as a function of  the location $\zeta$ in moduli space:
\begin{equation}
\d{\cal N}_{\rm{min}}(\zeta) \equiv \sum_i \delta(\zeta-\zeta_i) \ .
\end{equation}
In practice, $\d{\cal N}_{\rm{min}}$ is  far more challenging to study than  the related {\it{index density}} $\d{\cal I}_{\rm{min}}$, defined by
\begin{equation}
\d{\cal I}_{\rm{min}}(\zeta) \equiv \sum_i \delta(\zeta-\zeta_i)\, (-1)^{F_i} \ ,
\end{equation}
where $(-1)^{F_i}$  is the sign of the determinant  of the fermion mass matrix (see \cite{Douglas:2006es}).   The integral of $\d{\cal I}_{\rm{min}}$  over the moduli space is  manifestly not  the total number of vacua: it is instead a sum weighted by signs.
The advantage of considering $\d{\cal I}_{\rm{min}}$ is that it is computable: one can obtain the elegant Ashok-Douglas formula \cite{Ashok:2003gk}\footnote{Evidence supporting the result (\ref{exponentialgrowth}) in explicit flux compactifications on Calabi-Yau three-folds was obtained in \cite{Giryavets:2004zr,DeWolfe:2004ns}, building on \cite{Giryavets:2003vd}.}
\begin{equation}
\sum_{L \le L_{\rm{max}}} \d{\cal I}_{\rm{min}} = \frac{(2\pi L_{\rm{max}})^{b_3}}{\pi^{b_3/2}b_3!}\, {\rm{det}}(-{\cal R}-\omega)\ ,  \label{exponentialgrowth}
\end{equation}  where $\omega$  is the K\"ahler form on the moduli space, ${\cal R}$  is the curvature two-form, $b_3$ is the third Betti number
of the compactification, and the number $L_{\rm{max}}$  represents a tadpole constraint on the flux.  Equipped with (\ref{exponentialgrowth}), one can estimate the actual number of vacua by  attempting to place bounds on the  degree of difference between $\d{\cal I}_{\rm{min}}$ and $\d{\cal N}_{\rm{min}}$.
One  pivotal observation  is that the  number of vacua  is {\it{exponential in $b_3$}}.

There are two critical caveats that prevent one from concluding at this stage that type IIB  string theory compactified on a Calabi-Yau manifold with large $b_3$  has an exponentially large number of metastable de Sitter vacua.  First, we have thus far described only the complex structure moduli, and a local minimum of the potential on ${\cal M}_{{\cal{C}}}$ may or may not correspond to a local minimum of the exact potential on the full moduli space ${\cal M}_{\rm{total}}$, which also includes the K\"ahler moduli and the positions of D-branes.  Second, ${\cal M}_{{\cal{C}}}$ is {\it{noncompact}}, as is ${\cal M}_{\rm{total}}$: in particular, the K\"ahler moduli space ${\cal M}_{{\cal{K}}}$ can be continued toward infinite volume, where one recovers ten-dimensional flat space.  Noncompactness of ${\cal M}_{{\cal{C}}}$ implies
that  $V_{\mathfrak{F}}$ may not have a minimum in ${\cal M}_{{\cal{C}}}$.\footnote{For ${\cal M}_{{\cal{K}}}$, one manifestation of the corresponding fact is that the potential can have its minimum at infinite compactification volume.}  Thus, one is not strictly guaranteed {\it{any}} vacua for a given choice of flux.  Equation counting does certainly suggest that $V_{\mathfrak{F}}$ will generically have one or more minima inside ${\cal M}_{{\cal{C}}}$, but topology does not necessitate this.

With this background, we emphasize that
the celebrated counting of $10^{500}$ vacua in the landscape (cf.~\cite{Douglas:2006es}) does not refer to a counting of metastable vacua of the full potential for all moduli (at any level of approximation): it is a counting of {\it{supersymmetric vacua of the complex structure moduli sector}}, neglecting the K\"ahler moduli and postponing the question of  metastable supersymmetry breaking.

Let us therefore ask whether one can extrapolate from this result to estimate the number of de Sitter vacua in type IIB flux compactifications.  One might be tempted to argue as follows: suppose that one single metastable de Sitter vacuum is found, e.g.\ a KKLT solution on a particular Calabi-Yau with a particular choice $\mathfrak{F}_{\star}$ of quantized flux.  As famously explained by Bousso and Polchinski~\cite{Bousso:2000xa}, the many possible choices of quantized $p$-form flux in compactifications with many $p$-cycles lead to a `discretuum' of closely-spaced vacuum energy densities.  Can one then apply this logic and appeal to the existence of many fluxes $\mathfrak{F}_{\star}', \mathfrak{F}_{\star}'',\cdots$ that differ (by discrete quanta) from $\mathfrak{F}_{\star}$, but lead to a very similar cosmological constant, in order to replicate the single de Sitter vacuum into ${\cal O}({\cal N}_{\mathfrak{F}})$ de Sitter vacua?
No: the fact that $V_{\mathfrak{F_{\star}}}$ has a metastable local minimum in no way implies that $V_{\mathfrak{F_{\star}'}}$ has a local minimum.  This fact can also be understood in concrete examples: a change of quantized fluxes that leads to a small change in the cosmological constant generally involves large changes in the individual flux quanta, and correspondingly makes an order-unity change to the effective action, entirely changing the distribution of extrema (if any exist).

One must therefore be
cautious when using the vast number of supersymmetric (or `no-scale' supersymmetry-breaking) vacua in the complex structure moduli sector, cf.~(\ref{exponentialgrowth}), to argue for the existence of a comparable number of metastable de Sitter vacua of the full potential on the total moduli space: ${\cal N}_{dS} \neq {\cal N}_{\mathfrak{F}}$ in general.
We will now discuss this
issue in detail.

\subsection{Random Supergravity}  \label{sec:RMTsupergravity}

As a practical matter,  it is far easier to find {\it{critical points}} of $V_{\mathfrak{F}}$, i.e.\  points where $\partial_a V_{\mathfrak{F}}=0$, than it is to find  minima of $V_{\mathfrak{F}}$.  For the problem of counting metastable vacua, one  can therefore employ a strategy of counting the number ${\cal N}_{\rm{c.p.}}$ of critical points  and estimating ${\cal N}_{dS}$ via
\begin{equation}
{\cal N}_{dS} = {\cal N}_{\rm{c.p.}} \times f_{dS}\ ,  \label{cpanddS}
\end{equation}  where $f_{dS}$, defined by (\ref{cpanddS}), is the  fraction of all critical points that are in fact metastable de Sitter vacua.
(Precisely analogous logic  applies for  vacua with any other property---for example,  one could estimate the number of vacua with Standard Model gauge group by computing ${\cal N}_{\rm{c.p.}}$ and the associated fraction $f_{\rm SM}$.)
To further simplify the analysis, one can first ask what fraction $f_{\rm{min}}$ of all critical points are local minima, without demanding that the cosmological constant at the minimum be positive:
\begin{equation}
{\cal N}_{\rm min} =  {\cal N}_{\rm{c.p.}} \times f_{\rm{min}}  \ .
\end{equation}
The number of local minima, ${\cal N}_{\rm{min}}$, obviously provides an upper bound on   ${\cal N}_{dS}$.

The problem of counting de Sitter vacua therefore hinges on determining the probability that a randomly-chosen critical point is in fact a metastable minimum.  Let us be very precise about the notion of probability that is relevant here.  The intent is to begin with a compactification of fixed topology --- for example, a Calabi-Yau with specified Hodge numbers --- and consider all consistent choices of quantized flux $\mathfrak{F}$.  For each choice $\mathfrak{F}_{\star}$, one imagines finding all the critical points $\{p_i^{(\mathfrak{F_{\star}})}\}$ of $V_{\mathfrak{F_{\star}}}$ in the moduli space ${\cal M}$ (rather than in its compactification $\overline{\cal M}$), and assembling the ensemble~$\mathfrak{C}$ of all critical points,
\begin{equation}
\mathfrak{C} \equiv \bigcup_{\mathfrak{F}_{\star}}\, \Bigl\{p_i^{(\mathfrak{F}_{\star})}\Bigr\}\ ,
\end{equation} for any choice of flux.  Equation counting suggests that for a generic choice of flux, there will be at least one critical point, so we expect\footnote{In some circumstances one can show that the number of critical points  per choice of flux is exponentially large.   We thank Edward Witten for this observation.}
\begin{equation}
{\cal N}_{\rm{c.p.}} \gtrsim {\cal N}_{\mathfrak{F}} \ .
\end{equation}

However, it  still remains to estimate $f_{dS}$.   In \cite{Marsh:2011aa}  it was shown that for  broad classes of supergravity theories  with $N \gg 1$  scalar fields, $f_{dS}$ is spectacularly small, and can even be smaller than $1/{\cal N}_{\mathfrak{F}}$.   We will
now
summarize the argument of \cite{Marsh:2011aa}.

Consider an ${\cal N}=1$  supergravity theory with $N$  chiral superfields.   The F-term potential, in units with $\Mp=1$, is
\bea
 V = e^{K} \big( F_{a} \bar{F}^{a} - 3 |W|^2 \big) \ .
\eea
The object of primary interest is the {\it{Hessian matrix}} ${\cal H}$ at a critical point $p$ of the potential,
\bea
{\cal H} = \left(
\begin{array}{c c}
\partial^2_{a \bar{b}} V & \partial^2_{a b} V \\
\partial^2_{\bar{a} \bar{b}} V & \partial^2_{\bar{a} b} V
\end{array}
\right) \ .
\eea
At a local minimum of the potential,  the eigenvalues $\lambda_1 \le \lambda_2 \ldots \le \lambda_N$ of  ${\cal H}$ are nonnegative, so
\begin{equation}
f_{\rm{min}} = P(\lambda_1>0)\ ,
\end{equation}  where as explained above, the probability $P$  is computed in the ensemble consisting of the Hessian matrices at each of the critical points in $\mathfrak{C}$.

To express ${\cal H}$ in a convenient form \cite{Denef:2004cf,Marsh:2011aa}, we perform a coordinate transformation to  set $K_{a\bar{b}} = \delta_{a\bar{b}}$ at $p$, and a K\"ahler transformation to set $K=0$ at $p$.
We denote the geometrically-covariant  and  K\"ahler-covariant derivative by ${\cal D}_a$, and define the first three covariant derivatives of the superpotential as
\begin{equation}
F_a \equiv {\cal D}_a W\ , \quad
Z_{ab} \equiv {\cal D}_a {\cal D}_b W \ , \quad
U_{abc} \equiv {\cal D}_a {\cal D}_b {\cal D}_c W\ .
\end{equation}
The Hessian then takes the form \cite{Denef:2004cf,Marsh:2011aa}
\begin{align}
{\cal H} &=
\left(
\begin{array}{c c}
 Z_{a}^{~\bar c}\ \bar{Z}_{\bar{b} \bar{c}} -  F_a \bar{F}_{\bar{b}} - R_{a \bar{b} c \bar d} \bar{F}^c F^{\bar d}  & U_{a b c} \bar{F}^c - Z_{a b} \overline{W} \\
\overline{U}_{\bar{a} \bar{b} \bar c} F^{\bar c} - \bar{Z}_{\bar{a} \bar{b}} W & \bar{Z}_{\bar{a}}^{~c}\ Z_{b c} -  F_b \bar{F}_{\bar{a}} - R_{b \bar{a} c \bar d} \bar{F}^c F^{\bar d}
\end{array}
\right) \nonumber \\[4pt]
&\ \ \ \ \, +\,
\mathds{1}\,  \Big(F^2 - 2 |W|^2 \Big) \  ,  \label{fullHessian}
\end{align}
where indices are raised  with $\delta^{a \bar{b}}$, $\mathds{1}$ is the $2N \times 2N$ identity matrix, and $R_{a \bar{b} c \bar d}$ is the Riemann tensor of the  metric on field space.

The idea at this stage is to recognize that the large dimension $N$ of the field space need not remain an obstacle, but can instead be an expansion parameter!  The Hessian is a large matrix, and {\it{random matrix theory}} \cite{mehta2004random} provides a powerful tool for determining its eigenvalue spectrum.  The foundational insight in random matrix theory \cite{Wigner:1951:SDW} is that one can make sharp predictions about the statistical properties of the eigenvalues of a large ($N \times N$) diagonalizable matrix given very limited information about the actual entries of the matrix.  The guiding principle here is {\it{universality}}, which states that for $N \gg 1$, the statistics of the eigenvalues have little dependence on the statistics of the matrix entries.  This may be thought of as central limit behavior for matrices.

Universality will be essential to the argument, so we pause for a brief illustration; see \cite{Erdos,2011arXiv11035922K,2006math.ph...3038D} for in-depth discussions.
Consider a real, symmetric $N \times N$ matrix $M$, whose independent entries $M_{ij}$ ($i \ge j$) are independent stochastic variables drawn from a normal distribution~${\cal{N}}(0,\sigma)$
with mean zero and  standard deviation $\sigma$.  Compare to this a real, symmetric $N \times N$ matrix $\tilde{M}$ that has the same symmetries as $M$, but whose independent entries $\tilde{M}_{ij}$~($i \ge j$) are stochastic variables that are not necessarily independent---i.e.~the entries may have some correlations---and are drawn from diverse non-Gaussian distributions.  The magic of universality is that for $N \gg 1$, $M$ and $\tilde{M}$ have the same eigenvalue spectrum: the correlations and non-Gaussianities disappear\footnote{The fine print is that the correlations cannot be too numerous \cite{2005math.ph...5003S}, and the statistical distributions must have appropriately bounded moments.
Universality has been formulated and established rigorously in many settings --- cf.~\cite{Erdos,TaoVu} --- but to simplify the discussion we will continue to omit the associated technicalities.  More details can be found in \cite{Marsh:2011aa}.} at large $N$.

To apply random matrix theory to vacuum statistics, following \cite{Douglas:2003um,Ashok:2003gk,Denef:2004cf,Marsh:2011aa}, we first define a {\it{random supergravity}} as a four-dimensional ${\cal N}=1$ supergravity theory whose superpotential $W$ and
K\"ahler potential $K$ are random functions, in the sense that the components of their covariant derivatives, such as $F_a, Z_{ab}$, and $U_{abc}$, are stochastic variables drawn from one or more statistical distributions.  (See \cite{Distler:2005hi} for related work.)
In view of universality, it will suffice to take all the independent tensor components to be drawn from normal distributions (though this choice is not central to the analysis), but with the possibility of distinct {\it{scales}} $F_{\rm{rms}}, Z_{\rm{rms}}$, and $U_{\rm{rms}}$ for the components of $F_a, Z_{ab}$, and $U_{abc}$,  respectively.
These relative scales control the degree of supersymmetry breaking: the soft supersymmetry-breaking masses are of order $F/M_{\rm pl}$,  while the size $m_{\rm{susy}}$ of the supersymmetric mass terms is determined by the eigenvalues of $Z_{ab}$.
Strictly unbroken supersymmetry  would imply vacuum stability,  while for
\begin{equation}
F \ll m_{\rm{susy}} M_{\rm pl}  \label{wastelandalmostsusy}
\end{equation} supersymmetry breaking is a small effect, and supersymmetry  may be expected to increase the likelihood of stability.
If instead
\begin{equation}
F \gtrsim m_{\rm{susy}} M_{\rm pl}\ ,  \label{wastelandgeneric}
\end{equation} the supersymmetry-breaking masses are at least as large as the supersymmetric masses, and supersymmetry has little protective effect. When the scales appearing in the input statistical distributions are  taken to be comparable, i.e.~when
\begin{equation}
F_{\rm{rms}}\sim Z_{\rm{rms}}\sim U_{\rm{rms}}\ , \label{defofgeneric}
\end{equation} then (\ref{wastelandgeneric}) holds  for a typical member of the ensemble,  while approximate supersymmetry as in (\ref{wastelandalmostsusy})  can occur via a rare fluctuation.  We will refer to  the ensemble of critical points  generated via (\ref{defofgeneric}) as {\it{generic}} critical points; it was argued in \cite{Marsh:2011aa} that  the overwhelming majority of critical points  are in fact of this form.

In a random supergravity theory with $N$ chiral superfields, the Hessian (\ref{fullHessian}) is a $2N \times 2N$ matrix whose entries are stochastic variables.  Considerable structure is evident in (\ref{fullHessian}), and the next step, following \cite{Marsh:2011aa}, is to decompose (\ref{fullHessian}) into a sum of constituent random matrices with simple properties.  The eigenvalue spectrum of ${\cal H}$ --- which is the quantity controlling the probability of metastability --- can then be obtained by appropriately convolving the spectra of the constituents.

To this end, we briefly outline the properties of two classic ensembles of random matrices.  The (complex) {\it{Wigner ensemble}}, also known as the {\it{Gaussian Unitary Ensemble}}, consists of $N\times N$ Hermitian matrices $M$ of the form
\begin{equation}
M = A + A^{\dagger}\ ,
\end{equation}   where the entries $A_{ij}$ are stochastic complex variables with uniformly-distri\-buted phase and normally-distributed magnitude, $|A_{ij}| \in {\cal{N}}(0,\sigma)$.
The eigenvalue density $\rho(\lambda)$ of a typical member of the Wigner ensemble is given by the Wigner semicircle law,
\begin{equation}
\rho(\lambda) =  \frac{1}{2\pi N \sigma^2}\sqrt{4N \sigma^2-\lambda^2} \ .  \label{Wignerspectrum}
\end{equation}
Next, the {\it{complex Wishart ensemble}} consists of matrices of the form
\begin{equation}
M = A A^{\dagger}\,, \label{wishform}
\end{equation} where $A$ is a complex $N \times P$ matrix ($P \ge N$), and
again $A_{ij}$ are stochastic variables with magnitude drawn from~${\cal{N}}(0,\sigma)$.  From the form (\ref{wishform}) it is clear that the eigenvalues of a Wishart matrix are necessarily nonnegative.  The eigenvalue spectrum of a typical member of the Wishart ensemble is given by the Mar\v{c}enko-Pastur law,
\begin{equation}
\rho(\lambda) =  \frac{1}{2\pi N \sigma^2 \lambda}\sqrt{(\eta_+-\lambda)(\lambda-\eta_-)} \ , \label{Wishartspectrum}
\end{equation}
with $\eta_\pm \equiv N \sigma^2 (1\pm\sqrt{P/N} \hskip 2pt)^2$.

\begin{figure}
\begin{center}
$\begin{array}{l c r}
\includegraphics[width=10.2cm]{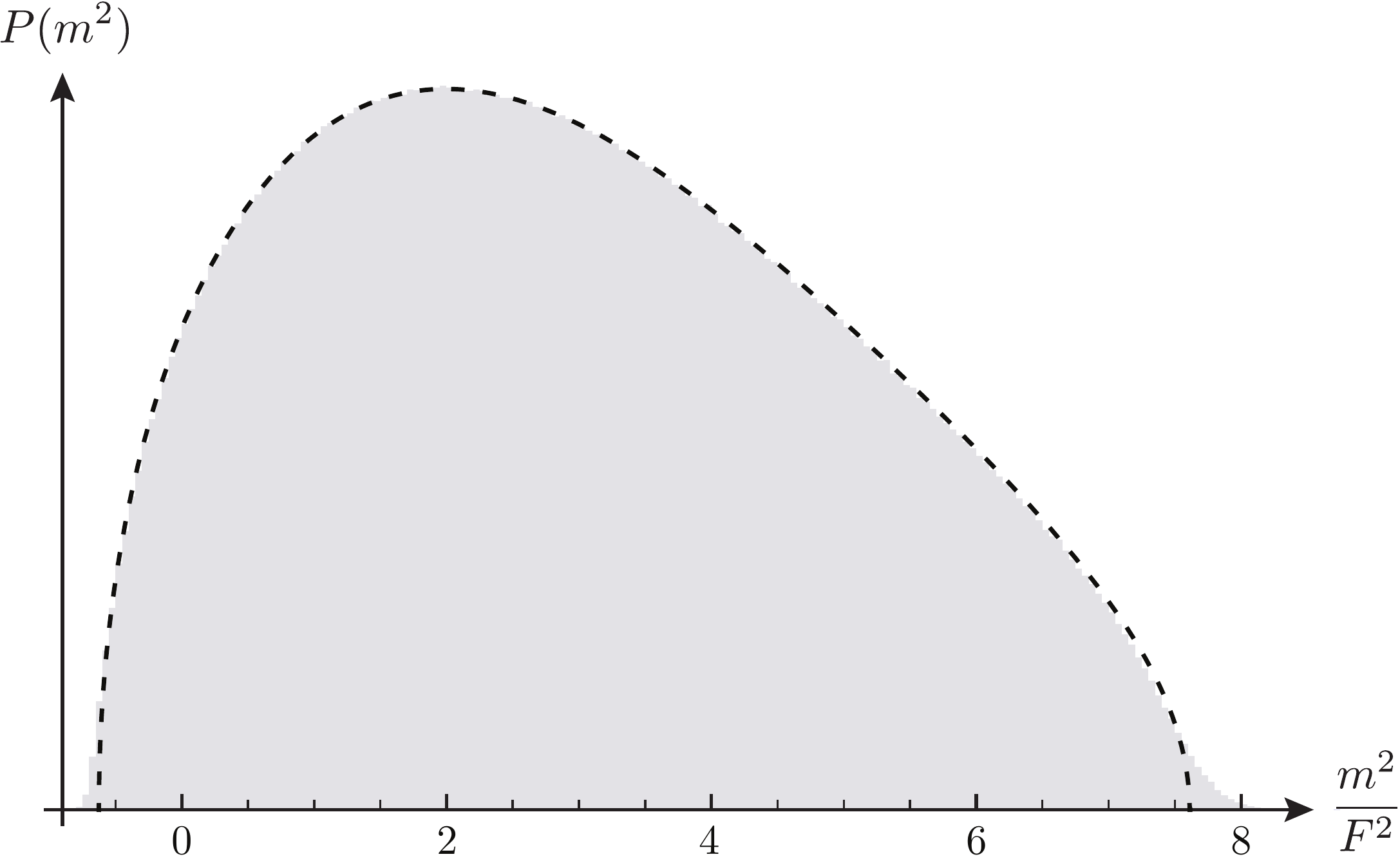}
\end{array}$
\caption{The histogram shows the spectrum of eigenvalues of the full Hessian matrix in random supergravity (for $N=200$ fields), while the curve gives the analytic result from the $WWW$ model~(\ref{WWWmodel})~\cite{Marsh:2011aa}.   The curve is a  parameter-free prediction of the model, not a fit.  (Figure adapted from \cite{Marsh:2011aa}.)}
\label{fig_WWW}
\end{center}
\end{figure}

We are now prepared to use random matrix theory to analyze the eigenvalue spectrum of the supergravity Hessian matrix (\ref{fullHessian}), in a random supergravity theory.  We will begin by studying generic critical points, as defined by (\ref{defofgeneric}).
One recognizes (\ref{fullHessian}) as the sum of constituent matrices with simple structures: for example,
\begin{align}
{\cal H}_{Z} &\equiv
\left(
\begin{array}{c c}
 Z_{a}^{~\bar c}\ \bar{Z}_{\bar{b} \bar{c}}    & 0 \\
 0 & \bar{Z}_{\bar{a}}^{~c}\ Z_{b c}
\end{array}
\right) \,
\end{align} is manifestly positive-definite, and is well-approximated by a Wishart matrix.
By continuing  along these lines, one finds \cite{Marsh:2011aa} that the eigenvalues of  the Hessian (\ref{fullHessian}) are well-approximated  by those of
\begin{equation}
{\cal{H}}_{WWW} \equiv  {\cal{H}}_{\rm{Wigner}} + {\cal{H}}_{\rm{Wishart}}^{(I)} + {\cal{H}}_{\rm{Wishart}}^{(II)}  \ ,  \label{WWWsetup}
\end{equation} where ${\cal{H}}_{\rm{Wigner}}$ is a Wigner matrix, and ${\cal{H}}_{\rm{Wishart}}^{(I),(II)}$ are Wishart matrices.
To obtain the spectrum of ${\cal{H}}_{WWW}$, one convolves the spectra of the constituents, which are given in (\ref{Wignerspectrum}) and (\ref{Wishartspectrum}).
Because  the matrices in question do not commute  with each other, this must be what is known as a {\it{free convolution}} \cite{voiculescu1992free},  denoted by $\boxplus$:
\begin{equation}
\rho({\cal{H}}_{WWW}) =  \rho({\cal{H}}_{\rm{Wigner}}) \boxplus \rho({\cal{H}}_{\rm{Wishart}}^{(I)}) \boxplus \rho({\cal{H}}_{\rm{Wishart}}^{(II)})   \ .  \label{WWWmodel}
\end{equation}   An analytic  expression for $\rho({\cal{H}}_{WWW})$ was obtained in \cite{Marsh:2011aa}  (we omit it here for brevity).   In fig.~\ref{fig_WWW} we compare a histogram of the eigenvalues of the full Hessian matrix (\ref{fullHessian}) in random supergravity, making no approximation, to the analytic result of the Wigner $\boxplus$ Wishart $\boxplus$ Wishart ($WWW$) model (\ref{WWWmodel}).
The model has no freely-adjustable parameters: we take $N=200$ fields  in  both the simulations and  the analytic model.  The agreement is  excellent; the slight tail at  the right edge  is a consequence of  finite $N$. Although formal results in the subject often require the limit $N \to \infty$, fig.~\ref{fig_WWW} makes it clear that $N=200$, which is an entirely reasonable number of fields in Calabi-Yau compactification, is a sufficiently large value of $N$.
We conclude that the Hessian matrix (\ref{fullHessian}) at a generic critical point of a random supergravity theory is very  well approximated by the analytic model (\ref{WWWmodel}).

At first glance, the eigenvalue spectrum (\ref{WWWmodel}) depicted in fig.~\ref{fig_WWW} may appear to determine  $f_{\rm{min}}$  as follows: according to the probability density $\rho(\lambda)$, a single eigenvalue is positive with probability
\begin{equation}
f_{>} \equiv \frac{\int_{0}^{\infty} \rho(\lambda)}{\int_{-\infty}^{\infty} \rho(\lambda)}\ ,
\end{equation} suggesting that  $f_{\rm{min}} = (f_{>})^N$.  This is {\it{not}} correct: $\rho(\lambda)$ describes the eigenvalue density for a typical Hessian matrix in the ensemble, and can be used to compute the probability that a single eigenvalue $\lambda_i$ falls in some given interval, provided that the remaining $N-1$ eigenvalues are unconstrained.
However, the eigenvalues of a random matrix are {\it{strongly correlated}}, and manifest eigenvalue repulsion (see \cite{mehta2004random}): if $N-1$ eigenvalues happen to be positive, the probability density for the final eigenvalue $\lambda_i$ is very different from (\ref{WWWmodel}).

What is needed, beyond knowledge of the typical eigenvalue spectrum, is a means of computing the probability of finding a very atypical matrix ${\cal H}_{WWW}$ in the
$WWW$ ensemble (\ref{WWWsetup}), one that has only positive eigenvalues.
Fortunately, there is a well-developed theory, pioneered by Tracy and Widom \cite{tracy1994level}, describing fluctuations of the extreme (i.e.~largest and smallest) eigenvalues of a random matrix.  For a large class of random matrix ensembles,
including the  $WWW$ ensemble \cite{Bachlechnertoappear}, one finds \cite{BenArous,Dean:2006wk,PhysRevE.77.041108}
\begin{equation}
P(\lambda_1>\zeta) = \exp \left(-N^2\hskip 2pt \Psi(\zeta)\right)\ ,
\end{equation} where the function
$\Psi(\zeta)$, which depends on the particular ensemble and is computable in simple cases, is $N$-independent at leading order in large $N$.

In summary, eigenvalue repulsion dictates that the probability that a Hessian matrix in the ensemble defined by (\ref{defofgeneric})
has only positive eigenvalues is given by
\begin{equation}
f_{\rm{min}} \equiv P(\lambda_1>0) = \exp \left(-c\hskip 2pt N^2 \right)\,,  \label{N2scaling}
\end{equation} for a constant $c$.  Thus, at large $N$, an overwhelming fraction of
generic critical points are unstable saddle points, not metastable minima.\footnote{The  extreme scarcity  of minima in a landscape whose Hessian  matrices are governed by Wigner's Gaussian Orthogonal Ensemble  was first discussed in the context of cosmology in \cite{Aazami:2005jf}.
An analysis of uplifting of supersymmetric AdS  vacua of type IIA  string theory, leading to the same conclusion, appears in \cite{Chen:2011ac}.}
This is to be contrasted to the naive estimate $f_{\rm{min}} = (f_{>})^N$, which is `only' exponentially small at large $N$.
Furthermore, we recall from (\ref{exponentialgrowth}) that the number of choices of flux is exponential in $N$, ${\cal N}_{\mathfrak{F}} \propto e^{d N}$, for a positive constant $d$.
Comparing to (\ref{N2scaling}), we conclude that when the assumptions of the above random supergravity analysis hold to good approximation, there are  hardly any vacua at large $N$ --- at least, none that correspond to  critical points that are generic in the sense of (\ref{defofgeneric}), with soft supersymmetry-breaking masses at least as large as the supersymmetric masses.

Because metastability  is so improbable in the absence of supersymmetry, it is natural to examine the sub-population of critical points that are
{\it{approximately supersymmetric}}, obeying (\ref{wastelandalmostsusy}).
Unbroken supersymmetry would guarantee stability,\footnote{Unbroken supersymmetry in AdS  does not guarantee the absence of tachyons allowed by the Breitenlohner-Freedman  bound \cite{Breitenlohner:1982bm}, but the techniques described above can be used to compute the probability that there are no tachyons \cite{Bachlechner:2012at}.  See also \cite{Achucarro:2007qa}.} and
one expects greatly increased likelihood of stability in the approximately-supersymmetric regime \cite{Denef:2004cf}.\footnote{The principal instability corresponds to the scalar  partner of the Goldstino --- see \cite{Covi:2008ea} for an analysis of geometric conditions that ensure stability  in this direction, and \cite{Covi:2008cn,Borghese:2012yu} for discussions of inflation along this  direction.} Detailed investigation \cite{2012arXiv1208.2691B} shows that at an approximately-supersymmetric critical point,
\begin{equation}
f_{\rm{min}} \equiv P(\lambda_1>0) = \exp \left(-d\hskip 2pt N\hskip 1pt \right)\ ,  \label{N2scalingSUSY}
\end{equation} where $d \approx 0.35$ is a constant.
We conclude that most metastable flux vacua  arise in regions of approximate supersymmetry.

An important proviso is that the $WWW$ model (\ref{WWWmodel}) describes the spectrum of eigenvalues of ${\cal H}$ in a random supergravity theory, in which $K$ and $W$ are random functions (as defined above).  Although universality blunts the effect of non-random correlations on the eigenvalue spectrum, there are well-motivated supergravity theories in which $K$ and $W$ have so much structure that (\ref{N2scaling}) must be modified.  The simplest example\footnote{A second important example is the  Large Volume Scenario, cf.~\S\ref{modulistab}, which leads to non-supersymmetric $AdS_4$  vacua that have been argued to be automatically tachyon-free in certain cases \cite{Balasubramanian:2005zx,Cicoli:2008va}.  The incidence of instabilities in LVS  has been analyzed in \cite{Rummel:2013yta}.} consists of two sectors, heavy and light, with additively-separable $K$ and $W$:
\begin{align}
K &= K_l(\phi,\bar{\phi}) + K_h(\Sigma,\bar{\Sigma}) \ , \\
W &= W_l(\phi) + W_h(\Sigma)\ .
\end{align}
If the $N_h$ heavy fields $\Sigma$ receive large supersymmetric masses at a high scale $\Lambda_h$, and supersymmetry is spontaneously broken in the sector of $N_l$ light fields $\phi$ at a much lower scale $\Lambda_l$, then only the light fields will be vulnerable to instabilities caused by supersymmetry breaking.  One therefore finds
\begin{equation}
f_{\rm{min}} \equiv P(\lambda_1>0) = \exp \left(-c\hskip 2pt N_l^2 \right)\ ,
\end{equation} which for $N_h \gg N_l$ is a vastly increased probability of stability, compared to the estimate $P(\lambda_1>0)=\exp (-c\hskip 2pt (N_l+N_h)^2)$ that overlooks the fact that the heavy fields are robustly stabilized.
The lesson is that the number $N$ appearing in (\ref{N2scaling}) is the number of fields that are {\it{dynamically accessible}} at the energy scale of the critical point in question.\footnote{See also \cite{Bachlechner:2014rqa}, where the probability of metastability in a Gaussian landscape was shown to be anticorrelated with the  magnitude of the vacuum energy.}

The principal reason for using caution in applying the results (\ref{N2scaling}) and (\ref{N2scalingSUSY}) was already noted above: we have only a rudimentary understanding of the array of effective theories that emerge from string theory, so it is too early to give a complete account of the vacua of string theory, by any means.  Nevertheless, we would like to stress that the assumption of a random superpotential and K\"ahler potential that underpinned the discussion of random supergravity is {\it{not}} tantamount to assuming that the supergravities arising in string theory have `no structure'.  Instead, universality ensures that in the large $N$ limit, the eigenvalue spectrum of ${\cal H}$ takes the universal form, unless the correlations in $K$ and $W$ are extremely strong.  In other words, many sorts of underlying patterns in the ${\cal N}=1$ data are compatible with (\ref{N2scaling}) and (\ref{N2scalingSUSY}): such patterns are obscured in the eigenvalue spectrum, and the only patterns that do survive in the spectrum are those determined by the macroscopic structure of (\ref{fullHessian}), not by the statistical properties of $K$ and $W$ themselves.  For this reason, random matrix theory is actually a conservative approach to the problem of counting vacua: it serves to expose structure that is inherent in supergravity, through the form of (\ref{fullHessian}), while blurring out detailed --- and presently unknown --- microphysics.

The techniques described above have a wealth of applications, most notably to the problem of characterizing inflation in a potential with many fields \cite{Aazami:2005jf,Easther:2005zr,Battefeld:2012qx,Westphal:2012up,Pedro:2013nda,Battefeld:2013xwa,Marsh:2013qca}, which we will briefly discuss in \S\ref{ssec:multi}, \S\ref{sec:potential},  \S\ref{warpedphenomenology}, and \S\ref{nflationsection}.

\chapter{What is String Inflation?}
\label{sec:StringInflation}

Inflationary scenarios constructed in effective field theory have limitations stemming from incomplete knowledge of the ultraviolet completion.  Because the inflationary dynamics is extraordinarily
sensitive to Planck-suppressed operators in the effective theory, merely parameterizing our ignorance of quantum gravity is untenable:
predictions obtained in this approach
amount to reflections of
implicit or explicit assumptions about the characteristics of quantum gravity.  This fundamental problem motivates pursuing a more complete understanding of inflation in the context of string theory.

In Chapter~\ref{sec:Examples}, we will discuss an array of  attempts  to derive inflation in string theory.
Before grappling with model-dependent  details, however, it is worthwhile to have a broad overview of the subject.
Many of the technical challenges that arise in string inflation are cognate across a range of models, and the  phenomenological characteristics are likewise parallel.  In this chapter, we  provide a schematic account of the  essential aspects of inflation in string theory.
We will sharpen these considerations with detailed case studies in Chapter~\ref{sec:Examples}.

\section{From Strings to an Inflaton}  \label{overallgoal}

The aim of most work on the subject can be summarized by the simple expression
\beq
S_{10}[{\cal C}] \, \mapsto\, S_4[\Phi(t)] \ .
\eeq
where the configuration ${\cal C}$ refers to the ten-dimensional data of geometry, fluxes,  localized sources, and quantum effects,
while $\Phi(t)$  represents a time-dependent configuration of scalar fields in the four-dimensional effective theory.
The task is to specify compactification data ${\cal C}$ that lead, upon dimensional reduction, to an effective theory $S_4$ with interesting cosmology.
To describe inflationary solutions, we require that $S_4$ has a positive vacuum energy contribution and one or more light moduli $\Phi$ whose time-dependent vevs describe a controlled instability of the vacuum.

\subsection{Energy Scales}

Understanding the primary energy scales that are involved provides a useful perspective  on the problem. Observations of the CMB  directly probe energies of order the inflationary expansion rate $H$ when modes cross the horizon and freeze (see Chapter~\ref{sec:dS}).
However, as shown in Chapter~\ref{sec:EFT}, inflation is sensitive to physics at  higher energy scales.
When inflation is formulated in effective field theory, these scales parameterize unknown  ultraviolet physics, but in string theory they  are computable and have specific meanings,  as we now explain.

The fundamental scale of string theory is the string scale $M_{\rm s} = (\alpha')^{-1/2}$. At energies below $M_{\rm s}$, only the massless states of the string are excited, and the theory reduces to an effective supergravity in ten dimensions. Most models of inflation in string theory are formulated in this island of theoretical control. The drawback of being in the regime $H \ll M_{\rm s}$ is that truly stringy effects are highly suppressed as far as CMB observables are concerned. To describe situations with $H > M_{\rm s}$, one would need to use the full string theory: the time dependence of the background would create excited string states.  Quantitative analysis of such a regime is out of reach at present.

Compactification on an internal space of volume ${\cal V} M_{\rm s}^{-6}$ introduces one or more additional scales, the Kaluza-Klein scales $\MKK \sim  M_{\rm s}\hskip 1pt {\cal V}^{-1/6}$.
We will usually work in the regime where $\MKK \ll M_{\rm s}$, so that  the theory is a ten-dimensional supergravity for intermediate energies, $\MKK < E < M_{\rm s}$, while it reduces to a  four-dimensional effective theory at low energies, $E < \MKK$.   The  four-dimensional theory  will itself be supersymmetric if the compactification preserves some of the ten-dimensional supersymmetries,  for example by having suitably reduced holonomy.  Most models of string inflation satisfy $H < \MKK$,  and as a result it is hardly surprising that  many such models reduce to well-known EFT models.  Formulating inflation (or alternatives to inflation) as a truly higher-dimensional phenomenon would be interesting, but requires rethinking many of the fundamental aspects of the problem, such as the horizon problem and the generation of primordial perturbations.

The four-dimensional Planck scale becomes a derived scale in string theory. It is related to the string scale, the Kaluza-Klein scale and the string coupling via (\ref{equ:4dPlanck}), which has the schematic form
\beq
M_{\rm pl} \sim g_s^{-1} \left({M_{\rm s}/\MKK} \right)^3 M_{\rm s} \gg M_{\rm s} \ .
\eeq
We note that applying the standard inflationary slow-roll analysis requires that one works in the four-dimensional Einstein frame and normalizes all fields with respect to the fixed Planck scale.

Finally, we have the scale of supersymmetry breaking  in the early universe, $\MSUSY$,
by which we mean the  highest scale of supersymmetry breaking that is unrelated to inflation.
The fact that  no superpartners  have been observed to date  plausibly puts the scale  of supersymmetry breaking in the present vacuum at or above the TeV scale, but  the breaking of supersymmetry may well have been different at the time of inflation.
For $\MSUSY < H$, supersymmetry is only spontaneously broken during inflation, and can partially protect against radiative corrections (see \cite{Baumann:2011nk} for a recent discussion).
The associated theoretical control provides crucial underpinning for most models of inflation in string theory.
However, supersymmetry could be much more badly broken: indeed, in non-supersymmetric compactifications $\MSUSY \gtrsim \MKK$.

\vskip 4pt
To sum up, most controlled treatments of string compactifications, and of inflation within it,
rely on the hierarchy of scales
\beq
\MSUSY \ < \ H \ <\ \MKK \ < \ M_{\rm s}\ <\ M_{\rm pl}\ .  \label{desiredhierarchy}
\eeq
As our understanding of string theory improves, it may  be  possible to move away from the comfort of this particular hierarchy of scales and explore a wider parameter space of string cosmologies.

\begin{figure}[h!]
   \centering
     \includegraphics[scale=0.4]{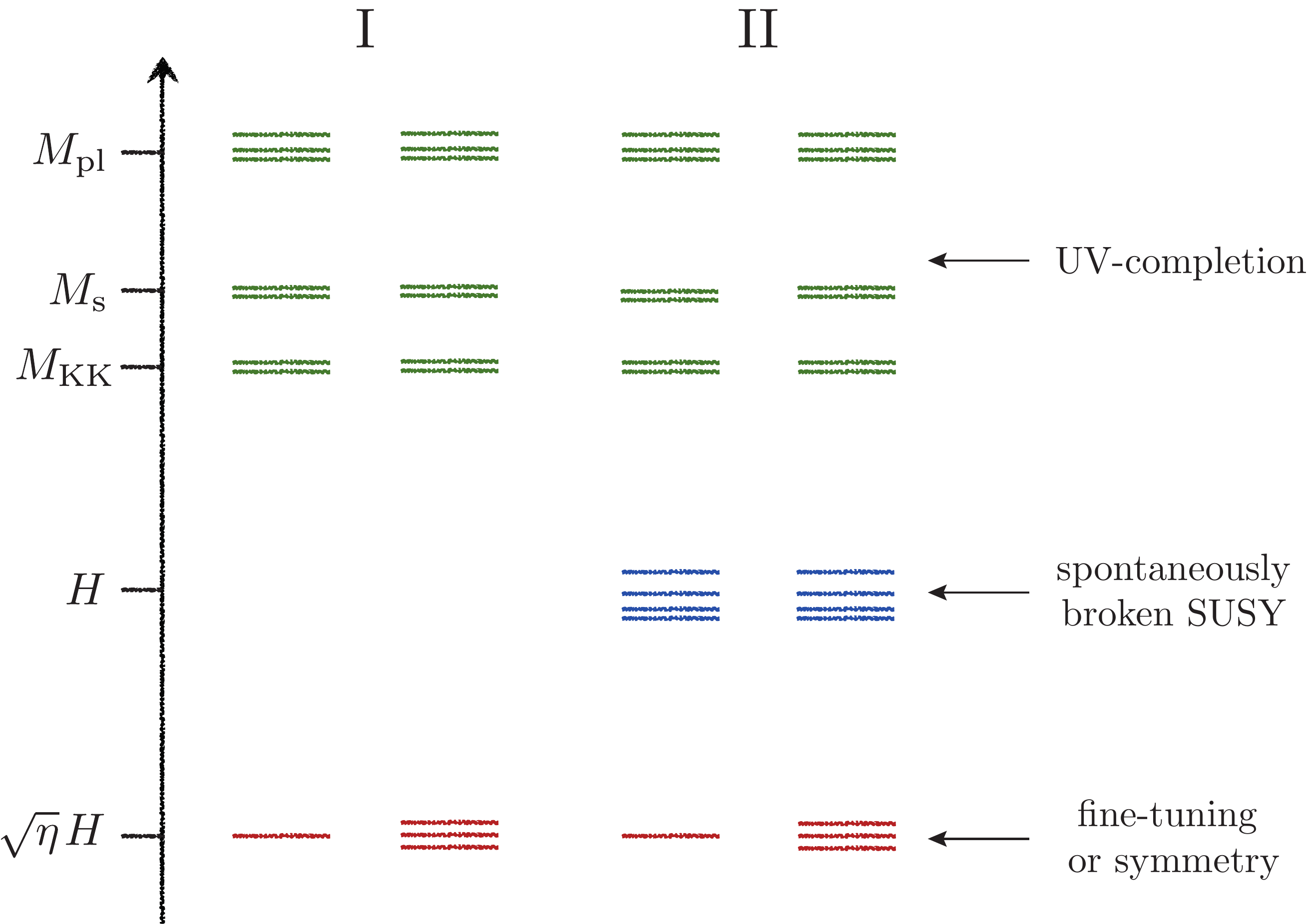}
   \caption{Mass spectra of inflationary models.  Phenomenological models of inflation
    frequently
    assume a large hierarchy between one or more light inflaton fields and the extra states of the UV completion (I).  On the other hand, concrete examples of inflation in string theory often contain fields with masses of order the Hubble scale (II) arising from the spontaneous breaking of supersymmetry.  Robust symmetries, or fine-tuning, are required to explain the presence of scalars with masses $m \sim \sqrt{\eta} \hskip 1pt H$.}
  \label{fig:ParticleSpectrum}
\end{figure}

\subsection{Spectrum of States}

Fields with masses that are smaller than the Hubble scale, $m < \frac{3}{2} H$, are both classically and quantum-mechanically active during inflation. To characterize effective theories of inflation, we need to determine the spectrum and the interactions of these light fields.
The simplest toy models of inflation assume that only one field is light and the rest have masses far above the Hubble scale (see fig.~\ref{fig:ParticleSpectrum}): the heavy fields can then be integrated out, and one is left with a model of {\it single-field inflation}.
If several fields are light one speaks of {\it multi-field inflation}.  In both scenarios, the heavy fields only affect the couplings of the low-energy theory, but do not participate actively in the generation of the primordial perturbations.

However, this hierarchy of mass scales is rarely the situation one encounters in actual constructions of inflation in string theory.
Even if one manages to arrange for one or more very light fields, it is typically hard to avoid having extra fields with intermediate masses.
In particular, most models of inflation in string theory constructed to date involve spontaneously broken supersymmetry~($\MSUSY < H$),  which  generally leads to moduli fields with masses of order $H$~\cite{Baumann:2011nk}.\footnote{Lighter moduli, with $m \ll H$, may be  natural in certain circumstances: see e.g.~\cite{Burgess:2010sy}.}
These fields fluctuation quantum-mechanically during inflation and therefore
have to be included
in the computation of
the primordial perturbations. The phenomenology of these models of {\it quasi-single-field inflation}~\cite{Chen:2009zp} has been explored in~\cite{Baumann:2011su, Baumann:2011nk, McAllister:2012am, Assassi:2012zq, Assassi:2013gxa, Noumi:2012vr, Burgess:2012dz, Sefusatti:2012ye, Avgoustidis:2012yc, Achucarro:2012sm, Cespedes:2012hu, Shiu:2011qw, Achucarro:2010da, Cremonini:2010ua}.

\subsection{Inflaton Candidates}

Models of string inflation can be classified by the nature of the field that serves as the inflaton.   A few of the leading candidates are:
\begin{itemize}
\item[$\triangleright$]  {\it Brane moduli.}---The positions of mobile,  spacetime-filling branes\footnote{The primary  examples are D$p$-branes with $p \ge 3$, NS5-branes, or M5-branes, wrapping suitable cycles.  Orientifold planes, in contrast, are non-dynamical:  their positions are not parameterized by light scalars.} in the internal space can be moduli in the four-dimensional effective theory.
Many leading models of string inflation are built on the time dependence of these brane position moduli.
The complex interactions of a brane with other sources in the compactification create the inflaton potential.
If the forces on the brane are weak enough, it moves non-relativistically and may source slow-roll evolution in the four-dimensional spacetime.  In \S\ref{sec:dbrane} and \S\ref{ssec:brane2}, we will analyze
several examples of this sort.
If instead the brane moves relativistically, kinetic effects dominate the dynamics. This leads to the interesting possibility of inflation being driven not by a flat slow-roll potential, but by the non-linear interactions in the kinetic part of the DBI action (\ref{equ:DBIXX}) for the brane.  Models of DBI inflation  are explored in \S\ref{ssec:DBI}: we will see that string theory plays an important role in explaining the radiative stability of these theories.

\item[$\triangleright$]  {\it K\"ahler moduli.}---Models of K\"ahler moduli inflation
identify the inflaton with time-dependent deformations of the  volumes of even-dimen\-sional cycles.
 Some of the most promising configurations involve  changes in the volume of one or more four-cycles, keeping the overall volume fixed.
 In constructions in type IIB string theory, the inflaton potential  arises from the leading effects  that violate no-scale structure,  typically a combination of $\alpha'$ corrections, string loop corrections and nonperturbative effects.  We describe various realizations of K\"ahler moduli inflation in \S\ref{sec:LVS}.

\item[$\triangleright$]  {\it Complex structure moduli.}---In the best-understood moduli stabilization scenarios in type IIB  string theory, the complex structure moduli are stabilized at a high scale by flux and  can be integrated out  at the time of inflation.   Unsurprisingly, time dependence of complex structure moduli  has played a limited role in models of inflation in type IIB  flux compactifications.
    In stabilized flux compactifications of type IIA string theory \cite{Villadoro:2005cu,DeWolfe:2005uu,Ihl:2006pp}, in contrast, complex structure moduli could a priori be natural inflaton candidates.   However, in the best-understood classes of type IIA compactifications,  there are no-go theorems for inflation \cite{Hertzberg:2007wc,Flauger:2008ad} (see also \cite{Hertzberg:2007ke}).  As a result, it appears difficult to construct explicit scenarios for inflation driven by complex structure moduli.

\item[$\triangleright$]  {\it Axions.}---String compactifications typically contain a plenitude of axion fields. These axions are particularly attractive inflaton candidates because they enjoy shift symmetries to all orders in perturbation theory.  Such symmetries are a key ingredient in technically natural models of inflation~\cite{Freese:1990rb},
including the large-field models required for
significant primordial gravitational waves.  String theory offers the opportunity to
determine which low energy shift symmetries are compatible with quantum gravity.  Inflation driven by a single axion field requires a super-Planckian decay constant, a feature that  is difficult to realize in controlled string compactifications~\cite{Banks:2003sx}.  In \S\ref{sec:AxionInflation}, we show how models of axion inflation in string theory can overcome this  obstacle,  and we discuss the rich phenomenology of axion inflation.
\end{itemize}


\subsection{Approximations}
\label{sss:approximations}

In  an ideal world, one would derive the inflaton action from first principles, beginning with fundamental  integer data ${\cal C}$ for a compactification,  solving the equations of motion of the ten-dimensional effective supergravity theory $S_{10}$, order by order in $\alpha'$  and in $g_{\rm s}$,  and then integrating out massive degrees of  freedom (including the Kaluza-Klein modes  of the compactification) to arrive at a four-dimensional effective theory $S_4$.\footnote{More ambitiously, one  might pursue solutions that are exact in $\alpha'$  and $g_{\rm s}$, but there  has been very little progress in this direction.}
Unfortunately, computing the effective action in a metastable non-supersymmetric  compactification  is a formidable technical challenge: a direct approach, making no approximations,  would require unforeseen  advances in our understanding of string theory.
Indeed, even in compactifications that preserve ${\cal N}=2$  supersymmetry in four dimensions,  such as  compactifications of type II  string theory on Calabi-Yau three-folds,  the metric on the internal space cannot be computed  analytically; in non-supersymmetric solutions, the  difficulties are far greater.   In practice, a four-dimensional effective theory  is deduced based on a partial specification of the compactification data ${\cal C}$, and an arsenal of approximation schemes  is used in  place of a  complete calculation.

\vskip 4pt
We will outline some of the most important expansion parameters, systematic  (and non-systematic!) approximation schemes, and simplifying assumptions  that are used in determining four-dimensional effective theories and extracting their dynamics.  It will be important to remember these limitations when we present our case studies in Chapter~\ref{sec:Examples}.

\begin{itemize}
\item[$\triangleright$]
{\it $\alpha^{\prime}$ expansion.}---The $\alpha^{\prime}$ expansion is reliable  when the gradients of the background fields are small in units of $\alpha^{\prime}$.   However, the compactification volume is finite, and is typically restricted by the desire to achieve the hierarchy $H < \MKK$, so that inflation is inherently four-dimensional. As a result, the $\alpha^{\prime}$ expansion is often a barely-controlled approximation scheme, rather than a convergent parametric expansion, in the regions of interest.  This issue is particularly severe in models of high-scale (equivalently, large-field) inflation, because the lower limits on the Kaluza-Klein mass become more stringent.

\item[$\triangleright$]
{\it String loop expansion.}---The weak coupling approximation retains only the leading terms in $g_{\rm s} \ll 1$.
One might  hope to make $g_{\rm s} = e^{\Phi}$ small in a stabilized vacuum through appropriate choices of flux, but because the dilaton $\Phi$ couples to most fields and  localized sources,  taking
$g_{\rm s} \ll 1$ very often disrupts the delicate balance of energies responsible for moduli stabilization.
As a result,  arbitrarily weak coupling is rarely achievable in practice.

\item[$\triangleright$]
{\it Probe approximation.}---Localized objects, such as D-branes or orientifold planes, are often treated as probes, meaning that they are not included as sources in the ten-dimensional equations of motion.
For sources that respect some of the supersymmetries preserved by the background, the backreaction can be restricted to a limited set of fields, and sometimes obeys a superposition principle, e.g.\ in the case of D3-branes in a background of ISD flux.  However, for non-supersymmetric sources this approximation is tenable only at considerable distances, which are not always available in a compact space.  For lack of an alternative, an unjustified probe approximation is occasionally made for low-codimension objects such as D7-branes and D8-branes, whose effects can be felt at arbitrarily large distances.  Many instances have been found in which the probe approximation misses crucial aspects of the inflationary dynamics.

\item[$\triangleright$]
{\it Large charge approximation.}---The polar opposite of the probe approximation takes the number of backreacting localized sources, or the total corresponding charge, to be so large that the radii of curvature of the resulting geometry are large in units of $\alpha^{\prime}$.  While this limit has proved very fruitful in noncompact geometries, e.g.\ taking a large D3-brane charge leads to the large $N$ limit in the AdS/CFT correspondence~\cite{Maldacena:1997re}, Gauss's law presents difficulties in compact models.  In many compact examples the large charge approximation is used despite being marginally valid at best.

\item[$\triangleright$]
{\it Smeared approximation.}---When computing the backreaction of localized sources is unmanageably complex, considerable simplifications can be achieved by imagining that the sources are distributed in a highly symmetric manner, or are distributed throughout the entire space.  For example, D-brane sources in Calabi-Yau cones are often treated as smeared over one or more angular directions of the cone, so that a problem that is properly posed as a system of PDEs is approximated by a system of ODEs.  Similarly, negative tension contributions from orientifolds are sometimes taken to be uniformly distributed in the compactification, postponing the question of possible singularities near an actual localized orientifold.   This approximation is very effective at reducing the difficulty of a problem, but its accuracy is poorly characterized.

\item[$\triangleright$]
{\it Linear approximation.}---Linearization in the strength of a source for some of the supergravity fields is very common, but not always self-consistent.
For example, in \cite{Baumann:2010sx} it was shown that the leading contribution to the action for a D3-brane in a warped region arises at quadratic order in an expansion in the strength of sources in the bulk of the compactification.

\item[$\triangleright$]
{\it  Noncompact approximation.}---Although analytic expressions for metrics on compact Calabi-Yau three-folds remain unavailable,  metrics are known for many noncompact Calabi-Yau manifolds, i.e.~Calabi-Yau cones.  The noncompact approximation attempts to represent a region in a Calabi-Yau compactification as a finite portion of a noncompact Calabi-Yau cone, subject to boundary conditions in the ultraviolet that represent the effects of compactification.

\item[$\triangleright$]
{\it Large volume expansion.}---Although the $\alpha^{\prime}$ expansion corresponds to an expansion in inverse volumes, there is a special sort of large volume expansion that deserves separate mention.
When the compactification volume is {\it{exponentially}} large, as in the large volume scenario~\cite{Balasubramanian:2005zx, Conlon:2006gv}, then a considerable number of corrections to the effective action in the $\alpha^{\prime}$ and string coupling expansions can be ignored.  In particular, it has been argued that only a subset of the terms arising at order $(\alpha')^{3}$ make leading contributions.

\item[$\triangleright$]
{\it Adiabatic approximation.}---It is often assumed that heavy fields adiabatically follow their instantaneous minima as a light field evolves.
The detailed form of the  inflaton potential can depend on the precision with which heavy fields are integrated out (see \S\ref{sec:dbrane}  for an example).
As discussed in \S\ref{ssec:Time},  a heavy mode of frequency~$\omega$ can be integrated out in this way only when the adiabatic condition $\dot{\omega}/\omega^2 < 1$ holds: more rapid evolution leads to  excitation of the heavy modes.  (Analyses of  the effects of heavy fields include \cite{Achucarro:2010jv,Achucarro:2010da,Achucarro:2012sm,Achucarro:2012yr}.
See also the discussion of resonance  in \S\ref{axionphenomenology}.)

\item[$\triangleright$]
{\it Truncation.}---Often the low-energy effective theory is truncated by omitting
one or more fields that  would be integrated out in a more  sophisticated analysis.
Omitting fields with  $m \ll H$, which evolve and fluctuate during inflation, is widely understood to be inconsistent.
A more reasonable --- but not always justified ---  approximation omits fields with $m \gtrsim H$.

\item[$\triangleright$]
{\it Moduli space approximation.}---An unjustified and misleading oversimplification asserts that the existence of a moduli space for some field $\varphi$ in a supersymmetric compactification `suggests' that $\varphi$ will have a relatively flat potential even after supersymmetry breaking.  This serves only to mask the actual problem, which is understanding the effective action of the non-supersymmetric theory.

\end{itemize}

Zealous application of the  approximation schemes described above can eventually lead to a well-characterized four-dimensional effective action,  but inflation is not an automatic consequence.  In the following sections, we will describe the most common obstacles that  arise after the effective theory has been determined.

\section{The Eta Problem}
\label{etaproblem}

The eta problem is omnipresent in realizations of inflation in string theory, but it takes various guises in different models.  In this section, we will summarize the causes of the eta problem in string inflation at a conceptual level, to provide a framework for understanding the detailed incarnations of the problem in the examples of Chapter~\ref{sec:Examples}.

\subsection{Compactness and Non-Decoupling}
\label{sec:Non-Dec}

A  pivotal insight about inflation in string theory is that the effects of  compactification and moduli stabilization do {\it not} decouple from the inflationary dynamics.
The problem of stabilizing the moduli and the problem of computing the inflaton potential cannot be treated independently, and
the inflaton sector cannot be
understood in isolation  from the other sectors  of the theory.  The importance of moduli stabilization in string inflation is widely appreciated in  the recent literature, but achieving control of the moduli potential remains one of the main technical challenges of the subject.  Moreover, the failure of decoupling of different sectors in string compactifications has important consequences for the dynamics of inflation in the presence of moduli-stabilizing ingredients.

\vskip 4pt
An instructive picture of decoupling is possible in compactifications with D-branes, where one often constructs  distinct sectors of the theory on collections of D-branes located in different parts of the compactification.
These sectors are said to {\it{decouple}}  if the details of one sector are irrelevant for the dynamics in another, i.e.~if  the sectors serve as non-interacting modules  for the purpose of computing some four-dimensional observable.  Complete decoupling is not  always desirable:  the interaction between two sectors could be responsible for inflationary  evolution,  as in the example of a  well-separated brane-antibrane pair --- see \S\ref{sec:dbrane} and \S\ref{ssec:brane2}.
 However, the problem is that hardly any sector decouples from inflation,  so
detailed understanding of all hidden sectors is necessary \cite{Hardeman:2010fh}.

A common but dangerous assumption is that sufficient geometric separation of two sectors, $A$ and $B$, makes the couplings between the sectors negligible.  (More refined criteria involve separation  along a warped direction, or separation without any branes  stretched between the sectors,  but the principle is the same.)   To check this assumption, one has to compute the couplings between the sectors by integrating out massive fields that couple to both $A$ and $B$.   In particular, open strings with one end on $A$ and another end on $B$ lead to massive fields in the four-dimensional theory.  Integrating out these strings leads to operators of the form\footnote{In many settings it is
more efficient to compute the couplings between two separated sectors  by working in supergravity,  rather than by integrating out stretched open strings.
In
this closed string approach, one finds a supergravity solution that incorporates the backreaction of sector $A$, and then evaluates the probe action for sector~$B$ at the appropriate location in this solution, in order to determine the effect of sector $A$ on sector $B$.  This method  has been used, for example, to determine the coupling between D3-branes and quantum effects stabilizing the K\"ahler moduli \cite{Baumann:2006th}.}
\begin{equation}
\Delta {\cal L} \, \supset\, \frac{1}{M_{AB}^{\delta_A+\delta_B-4}} \hskip 1pt {\cal{O}}_A^{(\delta_A)} {\cal{O}}_B^{(\delta_B)} \ ,
\end{equation}
where ${\cal{O}}_A^{(\delta_A)}$ is an operator of dimension $\delta_A$ consisting of the fields of sector $A$, and similarly for ${\cal{O}}_B^{(\delta_B)}$, while $M_{AB}$ is the mass of the strings stretched between the sectors.
For example, if ${\cal{O}}_B^{(4)} \equiv V_0$ is a constant contribution to the vacuum energy originating in sector $B$, and taking $\phi$ to be a scalar field in sector $A$, then with ${\cal{O}}_A^{(2)} \equiv \phi^2$ we find the coupling
\begin{equation}
\Delta {\cal L}\, \supset\, \frac{V_0}{M_{AB}^{2}}\hskip 1pt \phi^2\ .  \label{etadecoupling}
\end{equation}
This is precisely the dimension-six ultraviolet-sensitive inflaton mass term discussed in \S\ref{ssec:UV}.
The  problematic interaction (\ref{etadecoupling}), and kindred couplings, will be negligible if $M_{AB} \gg M_{\rm pl}$,  but will otherwise alter the inflationary dynamics.
Notice that two sectors decouple, for the purposes of inflation, if  the interactions  between the sectors are {\it{more than Planck-suppressed}}.
The general expectation in effective field theory  is that  Planck-mass  degrees of freedom that participate in the ultraviolet completion of gravity will induce Planck-suppressed interactions:  the absence of such couplings requires a special structure or symmetry in the quantum gravity theory.   We will see that this  expectation is borne out in string theory.\footnote{If $B$  is taken to be a  supersymmetry-breaking sector, and $A$  is the visible sector, then  the notion of decoupling described here corresponds to what is called {\it{sequestering}}  \cite{Randall} in the literature on supersymmetry breaking.  Investigations of sequestering in string theory \cite{Anisimov,Kachru2,Kachru,Blumenhagen:2009gk,BergMarsh,BergConlon} have confirmed that complete decoupling is extremely rare, but partial suppression of some couplings can occur in  certain cases \cite{Blumenhagen:2009gk}.}

\vskip 4pt
The erroneous  intuition that supports decoupling  is that $M_{AB}$  is dictated by the distance $d$ between the sectors, via $M_{AB} \sim d/\alpha^{\prime}$,  so that the effects of $A$  on $B$  can be made negligible by taking $d$ to be large.
Of course, in a compactification, the distance $d$ is bounded by the diameter $L$ of the compact space.  Moreover, in a roughly isotropic compactification, the total volume scales as ${\cal V} \propto L^6$, so that $M_{\rm pl} \propto L^3/(g_{\rm s}(\alpha')^2)$ and hence\footnote{We display only the parametric scaling:
factors of $2\pi$  can be important for this relation, but  depend on the precise geometry  and must be analyzed on a case-by-case basis.}
\beq
\frac{M_{AB}}{\Mp} \lesssim g_{\rm s}  \left(\frac{\ell_{\rm s}}{L} \right)^2\ .
\eeq
Thus, $M_{AB}/M_{\rm pl} < 1$ when the volume is controllably large: the stretched string mass cannot parametrically exceed the Planck mass in an isotropic compactification.  Consequently, the couplings between spatially separated D-brane sectors will  generically be at least gravitational in strength: the corresponding operators will be suppressed by no more than the Planck mass.

Isotropy is a strong assumption, and it is important to check whether
decoupling  arises automatically
in suitably anisotropic compactifications.  If the compactification has $p$ large directions of size $L$ and $6-p$ small directions of size $S$, then
\begin{equation}
\frac{M_{AB}}{M_{\rm pl}} \lesssim g_{\rm s} \left( \frac{\ell_{\rm s}}{L}\right)^{\frac{1}{2}p-1} \left( \frac{\ell_{\rm s}}{S} \right)^{\frac{1}{2}(6-p)} \ ,
\end{equation}
so that for $p>1$ the coupling is again at least gravitational in strength at large volume.
(For the case $p=1$, see \S\ref{superPlanckian}.)
A significant example consists of a warped throat geometry: a warped cone over an angular manifold $X_5$ is an example of a highly anisotropic space, if $X_5$ is chosen appropriately --- for example, one might consider $X_5 = S^5/\mathbb{Z}_k$ for $k \gg 1$. Rather surprisingly, it was shown in \cite{Baumann:2006cd} that for {\it{any}} $X_5$, the stretched string mass is less than the Planck mass (see \S\ref{ssec:warped}). Thus, `slender' warped throats do not evade the general argument that gravitational-strength couplings are unavoidable.

\vskip 4pt
The fact that compactness prevents decoupling leads to important constraints on the
interactions between localized sources.  We will illustrate the issues  in the example of a D3-brane/anti-D3-brane pair in a general unwarped six-manifold~$X_6$, though the problem is more general (see \ref{sec:dbrane}).

The Coulomb potential of a D3-brane/anti-D3-brane pair separated by a distance $r$ is
\begin{equation}
V(r) = 2\hskip 1pt T_3 \Biggl(1 - \frac{1}{2\pi^3}\frac{T_3 g_{\rm s}^2 \kappa^2}{r^4} \Biggr)\ ,  \label{coulombunwarped}
\end{equation} where $T_3$  is the D3-brane tension (\ref{deftp}), and $\kappa$ is the gravitational coupling defined in (\ref{defkappa}).
The  canonically-normalized field $\phi$ is related to $r$ by $\phi = \sqrt{T_3}\hskip 1pt r$.   Computing the slow-roll parameter $\eta$,  we find
\begin{equation}
\eta \approx -\frac{10}{\pi^3} \hskip 1pt \frac{{\cal{V}}}{r^6}\ ,
\end{equation}
where we have used (\ref{equ:4dPlanck}) and $T_3^2 g_{\rm s}^2 \kappa^2 = \pi$.
The Coulomb potential (\ref{coulombunwarped}) is evidently steep at small separations  and grows flatter at large separations.
However, the brane-antibrane pair cannot be separated by a  distance greater than the  diameter of the compactification,  so unless $X_6$ is highly anisotropic,  the potential (\ref{coulombunwarped}) is too steep  to support inflation \cite{Burgess:2001fx}.\footnote{When the background is warped,  the Coulomb potential  (\ref{coulombunwarped}) takes  the modified form (\ref{equ:VC}) \cite{Kachru:2003sx}, and is extremely flat  even at modest separations: see \S\ref{sec:dbrane}.}  This is  one of the simplest  examples  of the phenomenon of non-decoupling described above.

\subsection{Compactness and Backreaction}
 
In a warped background, the version of the eta problem that we just discussed seems to disappear~\cite{Kachru:2003sx}.
However, a more subtle issue arises: the backreaction of the D3-branes on the compact geometry leads to instabilities and to a recurrence of the eta problem. We will briefly sketch the argument.

\vskip 4pt
Going beyond the probe approximation,
a D3-brane located at the position $y_b$ in a six-dimensional space
with coordinates $y$ acts as a point source for a perturbation
$\delta e^{-4A}$ of the geometry (\ref{warped}):
\beq \label{equ:laplace}
  \nabla_y^2 \hskip 1pt \left(  \delta e^{-4A(y_b;y)} \right)= - \hskip 1pt{\cal C}
\left(\frac{\delta(y_b -y)}{\sqrt{g(y)}} - \bar \rho(y)\right)
\ ,
\eeq
with ${\cal C} \equiv 2 g_{\rm s}^2 \kappa^2  T_{3} = (2\pi)^4 g_{\rm s}
(\alpha')^2$.
In order to satisfy Gauss's law on the compact
space \cite{Giddings:2001yu},
we have included a background charge density
 $\bar \rho(y)$, with $\int {\rm d}^6 y \sqrt{g}\, \bar \rho(y) = 1$.
To be precise, the tadpole in question is gravitational, so that $\bar \rho(y)$ corresponds to a  negative tension source, as in \S\ref{fluxcompactification}.
The solution to (\ref{equ:laplace}) can be written as \cite{Baumann:2006th} \beq \label{equ:solvedlaplace} \delta e^{-4A(y_b;y)} = {\cal C} \left( {\cal G}(y_b;y) - \int \d^6 y' \sqrt{g}\,
{\cal G}(y;y') \, \bar \rho(y') \right) \ , \eeq
where the function ${\cal G}(y;y')$ satisfies
\beq
\label{equ:easylaplace} \nabla_{y'}^2 {\cal G}(y;y') = \nabla_{y\vphantom{'}}^2
{\cal G}(y;y') = - \frac{\delta(y-y')}{\sqrt{g}}+ \frac{1}{{\cal V}} \
 . \eeq
  Acting with $\nabla_{y_b}^2$ on (\ref{equ:solvedlaplace}), we find
 \beq \label{equ:modulicharge}  \nabla_{y_b}^2 \left( \delta e^{-4A(y_b;y)}\right) =
- \hskip 1pt{\cal C} \left({\delta(y_b-y) \over \sqrt{g(y_b)}} -
\frac{1}{{\cal V}} \right) \ , \eeq which does not depend on the background charge distribution\footnote{For discussions of the effects of the background charge, see \cite{Shandera:2003gx}.} $\bar \rho(y)$.
The leading term in the scalar potential  for a D3-brane
is
therefore
(see \S\ref{sec:dbrane} for more details)
\begin{equation}
V(y_b) = 2 T_3  e^{4A(y_b)} \approx 2 T_3 \left(1 - \delta e^{- 4A(y_b)} \right) \ .
\end{equation}
Computing the trace of the Hessian,  we find
\begin{equation}
{\rm Tr}(\eta) \approx -\frac{M_{\rm pl}^2}{T_3}\, \nabla^2_{y_b} \left( \delta e^{-4A(y_b;y)}\right) = -2\ ,
\end{equation}  where we used (\ref{equ:modulicharge}) and (\ref{equ:4dPlanck}).
Thus, the potential for a D3-brane
in the presence of an anti-D3-brane, with no other sources beyond those required by tadpole cancellation,
necessarily has a steep unstable direction, preventing sustained inflation \cite{Kachru:2003sx}.

\vskip 4pt
Although we have presented the problem in the example of D3-branes, parallel considerations apply  to any scenario in which  the backreaction of a source creates a potential for the motion of some object within the compactification:  the instabilities that arise will quickly end inflation.  On the other hand, all realistic models  involve additional sources of stress energy --- at the very least, to stabilize the moduli --- and the moduli-stabilizing contributions can in principle lead to a potential suitable for inflation.  This almost always requires some degree of fine-tuning.  To make this fine-tuning {\it{explicit}}, and thus to obtain a complete inflationary scenario in string theory, rather than a plausibility argument for inflation, requires computing the moduli potential in extraordinary detail.

\subsection{The Eta Problem in Supergravity}
\label{susyetaproblem}

Most contemporary scenarios for string inflation preserve supersymmetry down to the scale $H < \MKK$, and hence can be described in four-dimensional ${\cal N}=1$ supergravity. The positive vacuum energy during inflation spontaneously breaks supersymmetry.
Inflation then often suffers from a particular form of the eta problem that arises from couplings in supergravity~\cite{Copeland:1994vg}.

\vskip 4pt
We sketched the basics of ${\cal N}=1$ supergravity in four dimensions in \S\ref{sec:compactification}.
Let us take the inflaton $\varphi$ to be a complex\footnote{The  actual inflationary instability will generally involve one real component of $\varphi$, e.g.~the real or imaginary part, phase,  or magnitude of $\varphi$.}  scalar in the chiral multiplet.
Assuming that $\varphi$ is a gauge singlet, its interactions are determined by the K\"ahler potential $K(\varphi,\bar \varphi)$ and the superpotential $W(\varphi)$. The Lagrangian  for the inflaton is
 \beq
{\cal L}\ =\ - K_{\varphi \bar \varphi} \hskip 1pt \partial_\mu \varphi \partial^\mu \bar \varphi - e^{K/M_{\rm pl}^2} \left[ K^{\varphi \bar \varphi} D_\varphi W \overline{D_\varphi W} - \frac{3}{M_{\rm pl}^2} |W|^2 \right] \ . \label{equ:LF}
\eeq
In (\ref{equ:LF}) we have omitted the F-terms $D_\chi W$ of  additional moduli $\chi$: including these terms is straightforward and does not change our conclusions.
We have also omitted  a possible D-term  contribution, which we will comment on below.

Expanding the K\"ahler potential around a reference location $\varphi \equiv 0$,
\begin{equation}
K = K(0) + K_{\varphi \bar \varphi}(0) \varphi \bar \varphi + \cdots \ ,   \label{equ:KF2}
\end{equation} the Lagrangian (\ref{equ:LF}) becomes
\begin{align}
{\cal L} &\ \approx\  - \partial_\mu \phi \partial^\mu \bar \phi  - V(0) \left( 1 + \frac{\phi \bar \phi }{M_{\rm pl}^2} + \cdots \right)\ , \label{equ:LF2}
\end{align}
where we have defined the canonically-normalized field $\phi \bar \phi \equiv K_{\varphi \bar \varphi}(0)\hskip 1pt \varphi \bar \varphi$. The ellipses in (\ref{equ:LF2}) correspond to terms arising from the expansion of $K$ and $W$ inside the square brackets in (\ref{equ:LF}). These terms are model-dependent and can be of the same order as the model-independent term that we have shown explicitly. However, without fine-tuning the model-dependent terms against the universal term, we get a large contribution to the inflaton mass and hence to the eta parameter:
\beq
m_\phi^2 = \frac{V(0)}{M_{\rm pl}^2} + \cdots = 3 H^2 + \cdots \qquad \Rightarrow \qquad \eta = 1 + \cdots\ .
\eeq
Thus, a generic inflationary model in ${\cal N}=1$ supergravity suffers from the eta problem \cite{Copeland:1994vg}.

An instructive special case  is the theory of a spacetime-filling D3-brane in a compactification with a single K\"ahler modulus $T$.
Parameterizing the D3-brane position  in the compact space with three complex scalars $z_{\alpha}$, $\alpha=1,2,3$, the K\"ahler potential takes the DeWolfe-Giddings~\cite{DeWolfe:2002nn} form (\ref{equ:DeWolfeG}):
\beq
K= - 3 \ln\Big[T + \bar T - \gamma k(z_\alpha, \bar z_\alpha) \Big] \equiv -2 \ln {\cal{V}} \ , \label{equ:DeWolfeGin4}
\eeq
in units where $\Mp \equiv 1$.
In the second equality we have indicated the dependence on the
{\it physical volume} ${\cal V}$, as contrasted with the {\it{holomorphic volume}} $T$.  The latter is the proper  K\"ahler coordinate on the moduli space, and can appear in the superpotential.  On the other hand, the rescaling to  four-dimensional Einstein frame entering (\ref{breathing}) involves a power of ${\cal V}$, and so  all sources of positive energy in four dimensions contribute to a runaway potential for ${\cal V}$.

In the absence of a superpotential for $T$ and $z_{\alpha}$, all four fields have vanishing potential.
One might hope that $T$  could be stabilized by superpotential interactions, leaving $z_{\alpha}$ as flat directions.
However, in the presence of a superpotential for $T$, the F-term potential (\ref{equ:LF}) depends both on $T$, through the superpotential,  and on ${\cal V}$, through the prefactor
$e^K$.
Displacement of the D3-brane changes $k(z_\alpha, \bar z_\alpha)$, and hence alters either $T$ or ${\cal V}$.  As a result,  superpotential stabilization of $T$  leads to a mass for $z_{\alpha}$, through the mixing  in (\ref{equ:DeWolfeGin4}).  This is another manifestation of the eta problem.

\vskip 4pt
It has been suggested that the eta problem in supergravity may be evaded if inflation is driven by a D-term  potential \cite{Binetruy:1996xj}: the argument given above is then inapplicable.
Moreover, the D-term  potential has been argued to be less sensitive  than the F-term  potential to inflaton-dependent corrections to the K\"ahler potential.   A significant difficulty\footnote{A criticism of  D-term  inflation based on consistency conditions in supergravity can be found in \cite{Seiberg2}.} is that all known scenarios for  complete moduli stabilization involve  some F-term potential for the moduli,  and in general $V_{F} \gtrsim V_D$.   Expanding $V_F$  as in (\ref{equ:LF2}), the eta problem reappears,  because of the inflaton dependence of the F-term  contribution to moduli stabilization.  See \cite{McAllister:2005mq}  and \S\ref{d3d7section} for discussions of this effect in an  explicit string inflation scenario.

\vskip 4pt
By particle physics standards, the fine-tuning required to go from $\eta \sim {\cal O}(1)$ to $\eta \sim {\cal O}(0.01)$ is not extreme.
Nevertheless, it would certainly be preferable if a symmetry principle made inflation technically natural.
A simple way to achieve this in the present context \cite{Kawasaki:2000yn}\footnote{See also \cite{Gaillard:1995az}, in which an assumed  Heisenberg symmetry protects the flatness of the potential.} is to impose a shift symmetry on one of the real components of the complex scalar $\phi$, e.g.~$(\phi + \bar \phi) \mapsto  (\phi + \bar \phi) + const.$
If this symmetry is exact, then the superpotential is independent of $\phi$ and the K\"ahler potential can only be a function of the imaginary part $\phi - \bar \phi$, i.e.~at lowest order we have
\beq
K = (\phi - \bar \phi)^2 \ .
\eeq
Now the $e^K$
factor in (\ref{equ:LF}) is independent of $\phi + \bar \phi$, and the real part of $\phi$ is protected from a dangerous mass term.
This time only the unprotected field $\phi - \bar \phi$ receives a mass of order $H$.
Examples of supersymmetric inflation models with these structures can be found in~\cite{ArkaniHamed:2003wu, ArkaniHamed:2003mz, Kaplan:2003aj, Baumann:2010ys, Baumann:2010nu, Baumann:2011nk}.
Further work on inflation in supergravity includes~\cite{Roest:2013aoa, Kallosh:2010xz, Kallosh:2010ug, Ferrara:2010in, Ferrara:2010yw, AlvarezGaume:2011xv, Borghese:2012yu}.

A fundamental  limitation of simply assuming a shift symmetry in the low-energy  supergravity
is  that couplings to Planck-mass degrees of freedom can readily spoil the symmetry (see \S\ref{ssec:gravity}).
Thus, asserting an exact shift symmetry in supergravity is untenable,\footnote{In certain field theories with special structures, it is possible to suppress all dangerous symmetry breaking terms to the necessary level. For example, in \cite{Baumann:2010nu}, it was shown that if the inflaton is the phase of a baryonic operator in SUSY QCD with gauge groups $SU(N\ge 5)$, symmetry breaking operators only arise at dimension seven or larger. In this case, the inflaton shift symmetry is an accidental symmetry and symmetry breaking effects are controlled by gauge symmetry. (The same mechanism controls proton decay in the Standard Model.) Similarly, coupling the inflaton to a conformal field theory can suppress the Wilson coefficients of the dangerous operators by RG flow~\cite{Baumann:2010ys}.} and the question is how badly the symmetry is lifted in string theory.
In \S\ref{d3d7section} and \S\ref{sec:AxionInflation},  we will encounter examples of inflation in string theory that try to exploit shift symmetries to construct natural models of slow-roll inflation. This is a  prime example of the utility of string theory in assessing ultraviolet-sensitive questions: the nature of the remnant symmetry can be determined by direct calculation within string theory.
A fair summary is that  approximate symmetries are ubiquitous in string theory,  but symmetries that are powerful enough to resolve the eta problem and make inflation natural are considerably less common.

\section{Super-Planckian Fields}  \label{superPlanckian}

The recent
BICEP2 detection of primordial B-modes makes it essential to understand inflationary scenarios involving super-Planckian inflaton displacements, $\Delta\phi \gtrsim M_{\rm pl}$.
As we explained in \S\ref{ssec:UV}, such large-field models are exquisitely sensitive to Planck-scale physics: at least naively, an infinite series of  non-renormalizable  terms
should be incorporated  in the inflaton action.
Examining large-field inflation in string theory sharpens and refines the problem: the  task of understanding and controlling
the effective inflaton action
 becomes a  matter of explicit computation.

\vskip 4pt
It is useful to divide constraints on super-Planckian displacements  into two classes, kinematic and dynamic.
Kinematic constraints  on the field range are purely geometric: if the field space has a finite diameter,  then by definition there is a maximum possible geodesic distance  between two points,  although the path length between an initial and final configuration can still be arbitrarily large.
For  fields in string theory that have restricted ranges for purely geometrical reasons,
one can make very strong statements about the impossibility of using those fields to construct inflationary models with observable tensors.
Even for the fields that kinematically allow super-Planckian vevs,  one must consider  the dynamical question of whether inflation  can persist over such a displacement, i.e.~whether controllably flat potentials can extend over such large distances in field space. This requires careful study of corrections to the inflaton potential.

In this section we will  describe  some of the general aspects of the kinematic and dynamic problems.
A definitive treatment of dynamics requires detailed information about  the geometry and potential energy in a metastable compactification, and is therefore deferred to the examples of Chapter~\ref{sec:Examples}.

\subsection{Geometric Constraints}

First,  we will examine the size of the moduli space for a D$p$-brane in a simple toroidal compactification.
Consider a D$p$-brane that fills the four-dimensional spacetime and wraps a $(p-3)$-cycle of volume ${\cal V}_{p-3}=(2\pi L)^{p-3}$ on an isotropic six-torus  of volume ${\cal V} = (2\pi L)^6$.  The dynamics of the brane is then that of a point particle in $9-p$ compact dimensions. We will derive a kinematic constraint on the canonical range of this particle.
Suppose that the D$p$-brane moves along one of the circles in the $T^6$, with coordinate $y$; the maximum possible distance from its starting point is then $\Delta y = \pi L$.
Dimensional reduction of the DBI~action defines the canonically-normalized field as $\phi^2 = T_p {\cal V}_{p-3}\, y^2$, so  that the maximal displacement is
\beq
\Delta \phi^2 \ <\ \frac{1}{8\pi} \frac{M_{\rm s}^2}{g_{\rm s} } \left( \frac{L}{\ell_{\rm s}}\right)^{p-1}\ .
\eeq
It may appear that we can make this field range arbitrarily large by choosing $L \gg \ell_{\rm s}$ and/or $g_{\rm s}\ll 1$. However, what is relevant for the Lyth bound is the canonical field range in units of the four-dimensional Planck mass (\ref{equ:4dPlanck}),
\beq
M_{\rm pl}^2 =  \frac{1}{\pi} \frac{M_{\rm s}^2}{g_{\rm s}^2} \left(\frac{L}{\ell_{\rm s}}\right)^6\ .
\eeq
We find
\beq
\frac{\Delta \phi^2}{M_{\rm pl}^2} \ <\ \frac{g_{\rm s}}{8} \left( \frac{\ell_{\rm s}}{L}\right)^{7-p}\ .  \label{isotorus}
\eeq
For $p < 8$, the Planck mass grows faster with $L$ than $\Delta \phi$
does, so that in the limit of theoretical control ($L > \ell_{\rm s}$ and $g_{\rm s} < 1$),  the field excursion is sub-Planckian.

The constraint (\ref{isotorus}) is clearly weakest for a high-dimensional brane on an anisotropic compactification with one large dimension.
Consider the spacetime $\mathbb{R}^{1,3} \times S^1/\mathbb{Z}_2 \times X_5$, where $X_5$ is a compact manifold of volume ${\cal V}_{5}$ and the interval $S^1/\mathbb{Z}_2$  has length $\pi L$.
A D8-brane that fills $\mathbb{R}^{1,3}$ and wraps $X_5$  is then a point particle on $S^1/\mathbb{Z}_2$. 
Going through the same logic as above, one finds
\beq
\frac{\Delta \phi^2}{M_{\rm pl}^2} < \frac{g_{\rm s}}{4\pi} \frac{L}{\ell_{\rm s}}\ ,  \label{anisorange}
\eeq with no dependence on ${\cal V}_{5}$.
The field range now becomes parametrically
large for $L \gg \ell_{\rm s}$.  This result closely parallels the finding in \S\ref{sec:Non-Dec}  that stretched string masses can become super-Planckian in compactifications with one large dimension and five small dimensions.



\subsection{Backreaction Constraints}

Although the kinematic range (\ref{anisorange}) accessible to a {\it{probe}} D8-brane can be very large,  the low codimension of the D8-brane makes backreaction  a serious problem.   In fact, backreaction by the D8-brane restricts the range to be sub-Planckian.\footnote{We thank Juan Maldacena for discussions of this point.}


First, we note that the D8-brane charge and tension  lead to tadpoles that must be canceled.  For a consistent compactification on $\mathbb{R}^{1,3} \times S^1/\mathbb{Z}_2 \times X_5$,  we  introduce a pair of O8-planes that sit at each end of the interval and wrap $X_5$,  and take the total number of D8-branes to be 16,  initially situated in two groups of eight  on top of the orientifold planes.\footnote{This is known  as a compactification of the
type ${\rm{I}}^\prime$ theory --- see for example the discussion in \cite{Becker:2007zj}.}   Now the inflaton candidate is the position $y$ of a single D8-brane, leaving the remaining D8-branes at the  endpoints of the interval.
The backreaction problem is that the  moving D8-brane  has charge and tension, and sources corrections to the metric and dilaton once it is removed from the O8-plane: cf.~(\ref{dilatonlimit}).  Because of the low codimension of the source,  it turns out that the dilaton {\it{diverges}}  before the D8-brane  can be displaced by $\Delta\phi=\Mp$.  Thus, consistently incorporating backreaction prevents super-Planckian displacements.

\vskip 4pt
A rather different example where  the would-be inflaton  induces corrections  that limit its own field range arises in N-flation~\cite{Dimopoulos:2005ac}, as detailed in \S\ref{nflationsection}.
The  essential idea is that, as in assisted inflation~\cite{Liddle:1998jc},  the inflaton $\Phi$ is a collective excitation of $N \gg 1$  elementary fields $\phi_i$.  If the $\phi_i$  each have kinematic range $\Delta\phi$,  the total range is
\begin{equation}
\Delta \Phi = \sqrt{N} \Delta \phi\ . \label{nflationdisplacement}
\end{equation}
The  backreaction problem in this scenario is that  the Planck mass is renormalized by loops of the $N$ light  fields.  Without detailed knowledge of the ultraviolet completion, one can estimate this correction as
\begin{equation}
\delta   M_{\rm pl}^2 \sim \frac{N}{16\pi^2}\, \Lambda_{\mathsmaller{\rm UV}}^2 \ , \label{renormalizationmp4}
\end{equation}  in terms of  an  ultraviolet cutoff $\Lambda_{\mathsmaller{\rm UV}}$.
Because the correction (\ref{renormalizationmp4}) has the same scaling with $N$  as the displacement (\ref{nflationdisplacement}),   taking $N$  large does not  parametrically increase the field range  in a theory where the quantum corrections take the form (\ref{renormalizationmp4}).
Overcoming this problem requires replacing the estimate (\ref{renormalizationmp4}) with a precise computation in an ultraviolet completion,  and then identifying  circumstances in which the scaling differs from (\ref{renormalizationmp4}): see \S\ref{nflationsection}.

\subsection{Stability Constraints}

Given a field space in which super-Planckian displacements  are possible,  sustained large-field inflation
requires a potential energy source that varies slowly over this distance.
A significant obstacle to constructing a gently sloped potential in a string compactification is that the inflationary energy itself backreacts on the geometry, and can disrupt the stabilization of the moduli, as we now explain.

\vskip 4pt
Many sources contribute to the moduli potential  in a general string compactification: $p$-form fluxes,  localized D-branes and orientifold planes,  and perturbative and nonperturbative quantum effects are among the  best-studied examples.
A single source generally induces an instability, as explained in \S\ref{modulistab}, and the characteristic of solutions with stabilized moduli is a delicate ---  and often precarious --- balance  among multiple contributions to the  potential energy,  leading to a moduli potential $U_{\rm{mod}}$  with a  local minimum.
The inflationary potential energy itself is one such contribution,  but, crucially, this energy $V$ necessarily diminishes as inflation proceeds, with initial and final energies  differing by $V_i - V_{f} \equiv \Delta V$.
In scenarios where $V \ll U_{\rm{mod}}$ and $\Delta V \ll U_{\rm{mod}}$, the inflationary energy poses a limited risk to stability.
When instead $V \gtrsim U_{\rm{mod}}$, the initial inflationary energy  may overcome the barriers in the moduli potential,  driving runaway evolution.   Even worse,  when $\Delta V \gtrsim U_{\rm{mod}}$  the inflationary contribution changes so dramatically during the course of inflation that  instabilities are unavoidable unless the remaining  sources for the moduli potential  provide  precisely compensating energies with  just the right time-dependence.

Destabilization is a particular difficulty for large-field inflation in string theory, because the inflationary energy density $V$ is necessarily large, of order $M_{\rm{GUT}}^4$,
and changes significantly  during inflation.\footnote{While the precise change is model-dependent,  the ratio of initial to final energies  is generally sizable:  for example, $V_i/V_f \sim 10^2$ in $m^2\phi^2$  chaotic inflation.}
With only  two decades of energy between the inflationary energy and the Planck scale, there is little
room for a hierarchy of the form
\beq
V^{1/4} \ \ll\  \MKK \ \ll\ M_{\rm s} \ \ll  \ M_{\rm pl} \   \label{newdesiredhierarchy}
\eeq
that would underpin theoretical control, as in (\ref{desiredhierarchy}).
The Kaluza-Klein scale $\MKK$ sets the maximal scale of the moduli potential, $U_{\rm{mod}} \lesssim \MKK^4$, so
the first relation in (\ref{newdesiredhierarchy}), $V \ll  \MKK^4 $, indicates  the separation of scales that could be compatible with $V \ll U_{\rm{mod}}$.

While destabilization that leads to runaway  decompactification is ruinous, more controllable backreaction of the inflationary energy on the moduli potential can  alter the character of an inflationary model without  preventing sustained inflation.  In particular,  given sufficiently high  barriers around a local minimum of the moduli potential, a time-dependent inflationary energy can induce  evolution of the moduli within the basin of attraction  of the minimum.
Incorporating the motion of the moduli  can then change the form of the inflaton potential,
as in the  rather general {\it{flattening}} mechanism of~\cite{Dong:2010in}.\footnote{See also \cite{Baumann:2007ah,Panda:2007ie}, where the slight shift of the overall volume  induced by motion of a D3-brane leads to important corrections to the D3-brane potential.}
Thus,  although shifts of the moduli  do not necessarily end inflation, their  effects must be  taken into account.

\vskip 4pt
The twin issues of limited parametric separation and of backreaction by the inflationary energy are common to all scenarios for large-field inflation in compactifications of string theory: the problem is simply an outcome of the high energy scale (\ref{equ:HMp}), combined with the existence of extra dimensions with radii greater than the Planck length.  Even so, these fundamental problems take many different guises in explicit constructions, and can be subtle to identify and extirpate.
In Chapter~\ref{sec:Examples}, we will encounter these challenges explicitly: e.g.~backreaction by relativistic D-branes in the DBI model (\S\ref{ssec:DBI}), and by induced charge on NS5-branes in axion monodromy models (\S\ref{ssec:AM}).

\section{Multi-Field Dynamics}
\label{ssec:multi}

Moduli fields are ubiquitous in string compactifications, as we explained in \S\ref{sec:compactification}.
After integrating out ultraviolet degrees of freedom, incorporating  the effects of fluxes,  localized sources, and quantum  corrections  to the action, one generally finds a complicated potential for the moduli.
Although a subset of the moduli may acquire large supersymmetric masses, $m \gg H$ --- e.g.~complex structure moduli in type IIB  flux compactifications, cf.~\S\ref{sec:modulistabilization} --- the generic outcome is that a significant number of moduli have masses $m \lesssim H$, and are therefore dynamically  active during inflation.
The resulting inflationary models are quite complex, and are just beginning to be explored in detail.


The challenge of analyzing a model with  multiple light moduli can be divided into two  principal tasks:  {\it i})
determining the effective Lagrangian,
and {\it ii}) computing the observational signatures.
We will address these issues in turn.

 
\subsection{Ensembles of Effective Theories}

As explained in \S\ref{sec:EFT2}, the effective Lagrangian for $N$ scalar fields $\Phi \equiv \{\phi_1,\ldots, \phi_N\}$ can be written in the form (\ref{equ:Leff3}),
\beq
{\cal L}_{\rm eff}[\Phi] = {\cal L}_l[\Phi] + \sum_i c_i \hskip 2pt \frac{{\cal O}_i[\Phi]}{\Lambda^{\delta_i-4}} \ , \label{equ:ensemble}
\eeq
where 
${\cal O}_i[\Phi]$ stands for operators of dimension $\delta_i$ constructed from $\phi_1,\ldots, \phi_N$ and their derivatives,\footnote{Curvature invariants  are also allowed in  principle, but can usually be neglected during an inflationary phase  with $H \ll M_{\rm{pl}}$.} and $c_i$ are the associated Wilson coefficients.\footnote{Symmetries  of the high-scale theory  may forbid certain operators, or  suppress different Wilson coefficients  to varying degrees,  as detailed in~\S\ref{sec:EFT2}.   Incorporating  these effects  in the ensemble is straightforward, cf.~e.g.~\cite{Baumann:2010sx,Agarwal:2011wm,Gandhi:2011id}.}
The Wilson coefficients depend on unknown details of the compactification, and computing them is impractical.
Moreover, we lack any principle that could select a single compactification, and are therefore obliged to
marginalize over the unconstrained details of the bulk.  Said differently, scenarios for inflation in flux compactifications of string theory lead not to one fully specified Lagrangian, but to  an {\it{ensemble}} of possible inflationary Lagrangians, each with the same operator content $\{{\cal{O}}_i\}$ but with different sets of associated Wilson coefficients~$\{c_i\}$.  Fine-tuning the parameters of a model ---  implicitly, by adjusting quantized fluxes and other integer data --- ultimately  involves  selecting an appropriate  Lagrangian from the ensemble.

How can  anything be  learned if the Wilson coefficients are unknown?
One strategy is
to take the $c_i$ to be elements of some statistical distribution $\Omega$, and then determine only the statistical properties of the
ensemble of effective Lagrangians.
A natural concern is that
 the conclusions
 might depend on $\Omega$,  which, just like the values of the individual $c_i$, is usually not computable.
Fortunately, in effective theories with {\it{many fields}} --- and  therefore a large number of  operators with\footnote{See \S\ref{etaII} for an explanation of the cutoff value $\delta \sim 6$  in small-field inflation.} $\delta \lesssim 6$ --- the potential is a sum  of many terms,  with the consequence that central limit behavior can wash out most of the dependence
on the shape
of $\Omega$.  Universality therefore restores
some degree of
predictivity.
Concretely, one can  approximate $\Omega$  by a Gaussian distribution with zero mean and standard deviation\footnote{The standard deviation $\sigma$  controls the rms size of non-renormalizable contributions to the potential, and is therefore physical; one can  estimate $\sigma$  by the general logic of \S\ref{sec:EFT2}.} $\sigma$, even if the true distribution of the individual Wilson coefficients $c_i$ is highly non-Gaussian.

In this approach,  inflation can arise  from accidental cancellations among two or more terms in the potential.
A primary goal  for a statistical analysis
is then to determine how frequently inflation occurs, and when it does, what the characteristic  properties  of the evolution are.
  
Although numerical experiments in the particular example of  warped D-brane inflation (see~\S\ref{sec:dbrane}) give strong evidence that $six$  fields can be large enough  for universality to take hold \cite{Agarwal:2011wm,Dias:2012nf}, much remains to be learned about the statistics of general multi-field models --- see \cite{Tegmark:2004qd,Tye:2008ef,Tye:2009ff,Battefeld:2008qg,Frazer:2011tg, Frazer:2011br, McAllister:2012am, Frazer:2013zoa,Marsh:2013qca} for related work.

\subsection{Multi-Field Perturbations}

Extracting the cosmological signatures of an effective theory with multiple light fields is challenging.
We will briefly describe the qualitative problems (and opportunities), deferring details to
Appendix~C.

\vskip 4pt
\noindent
{\it Super-horizon evolution.}---The essential difference between a model with one light field and a model with two or more light fields is that in the former case there is only one clock, so that the evolution of the perturbations is captured by the  Goldstone action (\ref{equ:piAction}) for $\pi$.  The resulting curvature perturbations, $\R = - H \pi$, are purely adiabatic, and are conserved outside the horizon.  In multi-field models, the vevs of additional fields $\psi$ provide additional clocks, whose fluctuations correspond to {\it{entropy}} perturbations.

Entropy fluctuations can evolve outside the horizon, and also couple
to the
curvature perturbations
in such a way as to permit the latter to evolve outside the horizon, so that the late-time curvature perturbation can be a complicated function of all the fluctuations at horizon crossing,
\beq
\R = f(\pi_\star, \psi_\star)\ . \label{equ:Rmulti}
\eeq
In some cases, the entropy perturbations eventually decay and the evolution reaches an {\it{adiabatic limit}}, where the curvature perturbation can again be expressed as $\R = - H \pi$.
After that time, the superhorizon curvature perturbations are conserved.  If instead reheating occurs before an adiabatic limit is reached, the late-time curvature perturbations are extremely sensitive to the details of reheating,  leading to a loss of predictivity.

Single-field slow-roll models automatically predict curvature perturbations that are adiabatic, approximately scale-invariant, and approximately Gaussian, in excellent agreement with observations.  None of these properties is automatic in a general multi-field model.
For $m_\psi \sim H$, the entropy fluctuations have a strongly scale-dependent spectrum.
If these fluctuations give the dominant contribution in (\ref{equ:Rmulti}), this
can
destroy the scale-invariance of the spectrum of curvature perturbations.
Scale-invariance can be preserved, however, if the  couplings of the inflaton to the additional fields preserve the approximate shift symmetry of the inflaton~\cite{Chen:2009zp, Baumann:2011nk}.


\vskip 4pt
\noindent
{\it Alternative sources for curvature perturbations.}---
Models with multiple light fields offer alternative mechanisms for generating the observed density perturbations,  including {\it{modulated reheating}}~\cite{Dvali:2003em, Dvali:2003ar, Zaldarriaga:2003my, Kofman:2003nx} and the {\it{curvaton scenario}}~\cite{Linde:1996gt, Lyth:2001nq,Moroi:2001ct, Enqvist:2001zp}.

In modulated reheating, superhorizon fluctuations in one or more light spectator fields $\psi$ modulate the end of inflation, or the decay rate of the inflationary energy density.
In other words, if the decay rate~$\Gamma$ is a function of the fields,  $\Gamma=\Gamma(\psi)$,  then the decay rate inherits the spatial variations of the $\psi$ fields.   This converts fluctuations of $\psi$ to density fluctuations in the post-inflationary universe.  The  observed curvature perturbations can be non-Gaussian if the function $\Gamma(\psi)$ is non-linear.

In the curvaton scenario, a light spectator field $\psi$, the `curvaton', 
survives until after reheating. Once the Hubble rate drops below the mass $m_{\psi}$, the  curvaton  begins  to oscillate,  evolving as non-relativistic matter. The energy density associated with the curvaton therefore redshifts more slowly than the post-inflationary radiation background,  and eventually the curvaton  makes a significant contribution to the total energy density of the universe.  When $\psi$ ultimately decays, its superhorizon fluctuations are imprinted into density fluctuations in the visible sector. The fluctuations may be non-Gaussian if the potential $V(\psi)$ is anharmonic and/or if the decay rate $\Gamma(\psi)$ is non-linear.

\section{Reheating}

Any complete model of inflation must explain how the energy stored in the inflaton eventually reaches the visible sector and initiates the hot Big Bang.  There are two basic requirements for the process of {\it reheating}: Standard Model degrees of freedom must be heated to a temperature sufficient for baryogenesis, and the cosmic history must not be spoiled by overproduction of relic particles in other sectors.  The rich structure of inflationary models in string theory leads to significant challenges for successful reheating, as well as a range of novel phenomena, as we now review.\footnote{See \cite{Allahverdi:2010xz} for a review of reheating in field theory.}

\subsection{Heating the Visible Sector}

The universal feature of string constructions that complicates reheating is the existence of fields beyond the inflaton, the Standard Model fields, and the four-dimensional graviton.  Light, long-lived hidden-sector fields, such as moduli, have long been known to threaten the successes of the standard thermal history.  Moduli decays occurring after baryogenesis can dilute the baryon asymmetry, while decays occurring during or after Big Bang nucleosynthesis can photodissociate the light elements, ruining the prediction of their abundances.
On the other hand, cosmologically long-lived relic particles can yield too much dark matter or even overclose the universe.  String theory provides a plethora of candidates for dangerous relics, including compactification moduli, Kaluza-Klein modes, excited strings, axions, as well as hidden sector matter and radiation.\footnote{Dark radiation,  corresponding to relativistic species in a hidden sector, can have distinctive signatures: see  e.g.~\cite{Cicoli:2012aq,Higaki:2012ar,Conlon:2013isa,Conlon:2013txa}.}

In conventional field-theoretic studies of reheating, as well as of the related nonperturbative process known as {\it preheating}, a primary question is the efficiency with which the inflaton transfers its energy into other degrees of freedom.   The
difficulty in many string-theoretic constructions is rather different \cite{Kofman:2005yz}: the inflaton readily liberates its energy into hidden sector fields, and the question is whether a sufficiently large fraction ends up in the Standard Model rather than in harmful relics.
The challenge of reheating  after inflation in string theory can be compared to that of keeping a house warm
in a cold winter:
a furnace alone is insufficient, and
one must also have insulation to direct a large fraction of the energy output  to the desired region.

Reheating crucially involves the Standard Model, so to discuss reheating in a string construction one cannot remain agnostic about how the visible sector is realized. In D-brane models in type II and type I string theory, as well as in the strongly-coupled heterotic string, the visible sector is generally localized on one or more branes (or at the intersections of branes).
When the inflationary energy is also localized on a brane, one can take a modular approach, in which the inflationary sector and the visible sector constructed separately, in local geometries approximating regions of some unspecified compactification, and their interactions are then computed or parameterized.
This strategy has been fruitful in extensive explorations \cite{Barnaby:2004gg,Kofman:2005yz,Frey:2005jk,Chialva:2005zy,Barnaby:2006cq,Langfelder:2006vd,Chen:2006ni,Brandenberger:2007ca,Berndsen:2008my,Dufaux:2008br,Panda:2009ji,Frey:2009qb} in the context of warped D-brane inflation\cite{Kachru:2003sx}, as we review in \S\ref{sec:dbrane}.  Reheating in other models involving D-branes  has been studied in e.g.~\cite{Brandenberger:2008kn,Brandenberger:2008if,Bouatta:2010bp}.

In models where the inflaton is a closed string modulus (see \S\ref{sec:LVS}), new challenges arise, as described in \cite{Green:2007gs,Cicoli:2010ha}.  Investigations in models where the inflaton is a closed string modulus include \cite{Barnaby:2009wr,Burgess:2010bz,Braden:2010wd,Cicoli:2010yj,Cicoli:2012cy}.

The phenomenology of reheating  in string theory is quite rich.  A violent end to inflation, e.g.~through brane-antibrane annihilation, provides a setting in which fields that can otherwise be omitted from the effective theory play a role: strong violations of the adiabatic approximation allow very massive fields to contribute to the dynamics, as further discussed in \S\ref{sec:Dissipation}.  Furthermore, the existence of multiple light fields can lead to `modulated reheating', in which  the dominant contribution to the temperature anisotropies arises from spatial variations in the couplings between the inflaton and the visible sector,  or to  the conversion of  entropic perturbations to curvature perturbations: investigations of these effects in string theory include
\cite{Burgess:2010bz,Cicoli:2012cy} and \cite{Brandenberger:2007ca,Dias:2012nf,McAllister:2012am},  respectively.   Finally, condensation of a complex tachyon produces a network of topological defects, {\it{cosmic strings}},  which have striking signatures, as we now explain.

\subsection{Cosmic Strings}  \label{cosmicstrings}

Symmetry-breaking phase transitions can lead to the formation of topological defects classified by the topology of the vacuum manifold.
Cosmologically important examples include zero-dimensional defects, such as magnetic monopoles; one-dimensional defects, known as {\it cosmic strings}; and two-dimensional defects, i.e.~domain walls.
Magnetic monopoles from a GUT phase transition could overclose the universe \cite{Preskill:1979zi}, and one of the early successes of inflation was explaining how monopoles could be diluted~\cite{Guth:1980zm}.  Domain walls likewise come to dominate the energy density of the universe, and are ruled out.  Cosmic string networks, on the other hand, evolve so that their density tracks the density of the dominant component (radiation or matter): this is called {\it scaling}.
As a result, cosmic strings are constrained, but not excluded, and they produce spectacular, unmistakable signatures that could be detected in coming experiments.  Moreover, cosmic strings arise very naturally in constructions of inflation in string theory.  Here, we will review key facts about cosmic strings, referring the reader to the textbook \cite{vilenkin2000cosmic} and the reviews \cite{Hindmarsh:1994re,Polchinski:2004ia,Kibble:2004hq,Copeland:2009ga} for many more details.  We will begin with generalities that apply to all cosmic strings, and then describe the special aspects of the cosmic superstrings that arise in string theory, following \cite{Polchinski:2004ia}.

Cosmic strings arise whenever a $U(1)$  symmetry is broken: the winding number of the $U(1)$ around the core of the string is the topological conserved quantity  responsible for stability.
The minimum cosmic string density produced in a cosmological phase transition  in which a $U(1)$  symmetry  is broken  is set  by the {\it Kibble mechanism} \cite{Kibble:1976sj}: causality prevents  the phase of the  complex scalar  order parameter  from  being correlated on super-horizon distances, so that  at least one  horizon-spanning string defect is produced per horizon volume.

Cosmic string evolution involves: stretching along with the  expansion of the universe;  intersection and reconnection,  including loop formation;  and  energy loss through emission of gravitational radiation.    Reconnection  of a string after intersection is known as {\it  intercommutation},  and the probability $P$  of intercommutation  is a key phenomenological parameter.    The self-intersection  and intercommutation of a long string  leads to the formation of a loop,  which breaks off of the long string and gradually decays by emitting gravitational waves.   Thus, the  network of strings involves a number of long, horizon-crossing strings, as well as populations of loops in different stages of decay.

The signatures of cosmic strings  are distinctive.   A string produces a conical defect geometry: denoting the string tension by $\mu$, the deficit angle is $8\pi G \mu$, where $G$ is  Newton's constant.  The associated gravitational lensing of background objects can lead to double images.\footnote{Most searches  for cosmic string lensing  involve extragalactic objects (cf.~e.g.~\cite{Christiansen:2010zi}),  but microlensing of stars within the galaxy~\cite{Chernoff:2007pd} (rather than of distant quasars \cite{Kuijken:2007ma}) could probe very low tensions, particularly if the string loops cluster substantially \cite{Chernoff:2009tp}.}
Moreover, a moving cosmic string generates a temperature contrast in the CMB \cite{Kaiser:1984iv}---this is known as the Kaiser-Stebbins effect (for  related signatures in 21 cm  radiation, see \cite{Brandenberger:2010hn}). Stochastic contributions to the CMB  anisotropy are also important:  cosmic strings  with   high tension could produce  density perturbations sufficient to seed large-scale structure. However,  the  corresponding anisotropies lack  phase coherence,  and so do not  manifest acoustic peaks.  Thus, cosmic strings can at most  contribute a subdominant component  \cite{PhysRevD.65.021301,Jeong:2004ut,Pogosian:2004ny,PhysRevD.72.023513,PhysRevLett.100.021301} of  the primordial perturbations.  The  continual emission of gravitational radiation  produces a stochastic background of gravitational waves,  which could be  detected directly by LIGO or Virgo \cite{Abadie:2011fx},  or indirectly by inducing stochastic fluctuations in the  arrival of pulsar  signals (see e.g.~\cite{PhysRevLett.98.111101}).  Finally, smooth loops of string  develop one or more sharp {\it cusps}  in each period of oscillation.   Near the cusp, the string is extremely relativistic, and emits  an intense  burst of gravitational waves in a cone pattern \cite{Damour:2000wa,Damour:2001bk,Damour:2004kw}.   A cusp  event directed toward  a gravitational wave detector such as LIGO  could allow detection of strings with comparatively low tension.\footnote{Cosmic strings of even lower tension might be detectable if they passed through the Earth \cite{Motohashi:2013kka}, causing devastating earthquakes while simultaneously providing a window on Planck-scale physics.}  Bursts can also occur if strings break following the formation of monopole-antimonopole pairs \cite{Martin:1996cp,Leblond:2009fq}.

For many years, the study of cosmic strings  focused exclusively on strings arising in quantum field theory --- Nielsen-Olesen  strings \cite{Nielsen:1973cs}, also called vortex lines ---  rather than on the fundamental strings of superstring theory.  Witten had  observed in \cite{Witten:1985fp} that  in perturbative constructions, the tension of fundamental strings was  large enough so that cosmic F-strings were excluded by the isotropy of the CMB.\footnote{The isotropy of the CMB gives an upper limit on the inflationary scale,  and hence on the  tension of cosmic strings that could be produced in a phase transition after inflation.  Moreover,  high-tension strings  can be excluded by  searches for lensing  and for the Kaiser-Stebbins effect.}
Furthermore, he showed that heterotic  cosmic strings form the boundary  for axion domain walls, whose tension causes the strings to contract rapidly and disappear.

A renewed study of cosmic superstrings  was initiated by
Tye and collaborators in \cite{Jones:2002cv,Sarangi:2002yt,Jones:2003da}.  The essential  new insight was that if the Standard Model arises on D-branes, the  visible sector couplings and the string tension in Planck units can be adjusted independently,  by changing the string coupling $g_{\rm s}$  and the compactification volume.  Thus, the string tension can be low enough to  satisfy observational constraints.  (A similar argument applies in the strongly coupled heterotic string \cite{Becker:2005pv}.)
Moreover, in inflationary scenarios involving  moving D-branes,  reheating typically proceeds by the condensation of a complex tachyon \cite{Sen:1998tt}, leading to cosmic string defects via the Kibble mechanism.

Cosmic superstrings have several  important characteristics that distinguish them from strings arising as topological defects in perturbative quantum field theories \cite{Dvali:2003zj,Copeland:2003bj,Jackson:2004zg}.
In type IIB  string theory, there are two  elementary one-dimensional objects:  the  fundamental string, or `F-string',  and the D1-brane, or `D-string'.
These strings  can form bound states  involving $p$  F-strings and $q$  D-strings, if $p$  and $q$  are relatively prime \cite{Harvey:1995rn,Schwarz:1995dk}.   The resulting $(p,q)$  string has tension \cite{Schwarz:1995dk}
\begin{equation}
\mu_{p,q} = \frac{1}{2\pi\alpha^{\prime}}\sqrt{(p-C_0\hskip 1pt q)^2+e^{-2\Phi}q^2}\ .  \label{pqtension}
\end{equation}
Networks of $(p,q)$ strings yield scaling  solutions \cite{Tye:2005fn},  just like simpler cosmic strings.
The intercommutation probabilities of cosmic F-strings and D-strings can be much  smaller than those for  field theory cosmic strings, as carefully examined in \cite{Jackson:2004zg}.
In particular, a colliding  pair of strings can miss each other in the compact dimensions \cite{Jones:2003da,Dvali:2003zj,Jackson:2004zg}, and the string coupling $g_{\rm s}$  also suppresses  the intercommutation probability.

Perhaps the most compelling setting  for cosmic superstring production  is warped D-brane inflation~\cite{Kachru:2003sx}, in which annihilation of a D3-brane/anti-D3-brane pair via condensation of a complex tachyon  automatically produces  a collection of cosmic strings, and  warping  provides a natural parametric mechanism through which  the tension can  be small enough to obey observational bounds.  The stability and tension of these strings depend on the  the details of the model \cite{Copeland:2003bj,Leblond:2004uc,Firouzjahi:2005dh}, and we defer further discussion to \S\ref{warpedphenomenology}.

One might hope that  cosmic superstrings can be distinguished from strings arising as topological defects in field theory --- see \cite{Polchinski:2004ia} for a thorough discussion of this point. This hope is not entirely unjustified: cosmic superstrings  with $P<1$  can be told apart from  strings in a {\it perturbative} field theory,  which have $P \approx 1$.
Furthermore,  the spectrum of tensions (\ref{pqtension}) appears distinctive.
On the other hand, a field theory  with $SL(2,\mathbb{Z})$  invariance  would reproduce  (\ref{pqtension}).  More generally, the duality between  string theory and field theory makes  it difficult, even in principle, to distinguish F-strings, D-strings, or $(p,q)$
strings
of string theory from corresponding defects  in  strongly-coupled field theories \cite{Polchinski:2004ia}: for example,
the $(p,q)$ strings produced in warped D-brane inflation can also be viewed as strings of the dual gauge theory.
Even so, the detection of a network of cosmic $(p,q)$  strings would be an unsurpassed opportunity to probe high-scale physics!

\section{Inflation in String Theory: a Checklist}

An ideal  model of the early universe in quantum gravity would begin from fundamental topological data, arrive at an effective theory  via an  explicit and well-controlled  computation,  and make definitive, distinctive predictions that are consistent with current data but could be falsified  or verified  with future experiments.
There is little prospect of deriving such a model in the near future.
A more realistic hope is to specify  some integer data (for example, the topology of a Calabi-Yau  orientifold)  and  explicitly solve some equations of motion (e.g.~those of the K\"ahler moduli)  while appealing to the existence  of generic solutions for the remaining equations (e.g.~the complex structure moduli and dilaton equations of motion given a choice of quantized three-form flux).

Let us
 summarize the essential requirements for a  successful model of inflation derived in string theory:
\begin{itemize}
\item[$\triangleright$] The inflaton  action  should be computed in an expansion around a meta\-stable de Sitter  vacuum, with all  approximations under good control.
\item[$\triangleright$]  For  every physical effect contributing  to the moduli potential, one must know  the corresponding correction to the inflaton potential.
\item[$\triangleright$]  All assertions about ultraviolet-sensitive quantities must be justified  through controlled calculations.
\item[$\triangleright$]  If a dimensionless parameter  needs to be large or small in order for inflation to succeed,  one should know whether the required value can be achieved in a consistent compactification.
\item[$\triangleright$]  For  each field  with a mass $m \ll H$,  the small mass should be explained either by  fine-tuning of explicitly known,  fully specified operators in the effective theory,  or by a symmetry that can be shown to survive in string theory.

\item[$\triangleright$] All quantum-mechanically active fields, i.e.~fields with $m < \frac{3}{2} H$, must be included in the phenomenology.

\item[$\triangleright$] The model should contain a mechanism to produce
density fluctuations that are nearly scale-invariant, Gaussian and adiabatic.

\item[$\triangleright$]  The inflationary phase must end, and then  transition to successful reheating of the Standard Model, without overproduction of relics.

\end{itemize}
Distinctive  observational signatures, while obviously desirable, are ultimately optional.

\vskip 4pt
In the next chapter, we will review some of the leading examples of string inflation.
We will see that no model is completely successful on all points of the above checklist. 
\chapter{Examples of String Inflation}
\label{sec:Examples}

In this chapter, we will survey a number of representative examples of inflation in string theory.
We will try to be reasonably complete in our discussion of inflationary mechanisms, within the limitations of space and expertise,
but we will not be able to present all the results in the subject.
Our focus will be on extracting  a few important lessons from the collective works of many researchers.

\vskip 4pt
Some of the scenarios that we will discuss  make predictions that appear incompatible with the  observational bounds described in  Chapter \ref{sec:dS}.   At the time of writing, the predictions of  most models of inflation in string theory  are works in progress,  because the inflaton Lagrangian depends on details of the compactification  for which we
currently have only zeroth-order approximations.  Indeed, we will argue below that theoretical uncertainties  in determining the {\it{scalar}} power spectrum (and its tilt $n_s - 1$) are systematically understated in  most of the literature on inflation in string theory.  On the other hand, because  the {\it tensor} amplitude (or the tensor-to-scalar ratio $r$)  is  linked to  a purely kinematic quantity, the length of the inflaton trajectory,  one can sometimes determine with high confidence whether $r$  is large  or small in a given model.  Then, in view of the  detection of primordial gravitational waves reported by BICEP2~\cite{Ade:2014xna},  small-field models may be rejected as candidates for the history of our universe.  Even so, we  provide details of a number of small-field scenarios in this chapter, because they serve as  comparatively simple building blocks from which more realistic models can be  developed.   This
is in the same spirit as  the study of string compactifications with  unbroken supersymmetry, which certainly do not describe our universe, but facilitate the construction of more complete models.

\vskip 4pt
The individual sections are largely self-contained and can be read in any order.
In \S\ref{sec:dbrane}, we consider the motion of a D3-brane~\cite{Dvali:1998pa} in a warped throat region~\cite{Kachru:2003sx} as a source for inflation.
We present a number of interrelated perspectives on the potential energy of a D3-brane in a warped flux compactification, and then discuss the challenge of achieving slow-roll behavior in this setup.
In~\S\ref{ssec:brane2}, we study a few examples of brane inflation in unwarped compactifications,  including D3/D7 inflation~\cite{Dasgupta:2002ew,Hsu:2003cy,Dasgupta:2004dw,Hsu:2004hi,Haack:2008yb}, fluxbrane inflation~\cite{Hebecker:2011hk,Hebecker:2012aw} and M5-brane inflation~\cite{Buchbinder:2004nt, Becker:2005sg}.
In \S\ref{ssec:DBI}, we discuss relativistic brane motion as a source of non-slow-roll inflation.  We  describe DBI inflation~\cite{Silverstein:2003hf} as an effective field theory and highlight microphysical constraints imposed by  compactification.
In \S\ref{sec:AxionInflation}, we argue that string axions are promising inflaton candidates.
We  give detailed analyses of N-flation~\cite{Dimopoulos:2005ac} and axion monodromy inflation~\cite{McAllister:2008hb}.
In \S\ref{sec:LVS}, we describe models in which the inflaton is a  K\"ahler modulus (or the associated axion),  including racetrack inflation~\cite{BlancoPillado:2004ns, BlancoPillado:2006he} and inflationary scenarios in large volume compactifications~\cite{Conlon:2005jm, Cicoli:2008gp, Cicoli:2011ct}.  Finally, in \S\ref{sec:Dissipation}, we look at dissipative effects as a source of inflation and  critically assess the prospects  for dissipative inflation in string compactifications~\cite{Green:2009ds, DAmico:2012sz, DAmico:2012ji, Martinec:2012bv}.

\section{Inflating with Warped Branes}
\label{sec:dbrane}

The positions of localized sources in a string compactification  correspond to scalar fields in the four-dimensional effective theory.
In \cite{Dvali:1998pa}, Dvali and Tye proposed that the separation between two branes could serve as an inflaton candidate.
This idea was made more precise in \cite{Dvali:2001fw,Burgess:2001fx},  where the two branes  were taken to be a D3-brane and an anti-D3-brane, respectively.
These  objects  attract each other gravitationally, and also through the R-R four-form potential $C_4$, under which they carry opposite charges; moreover, at small separations a
tachyon appears in the spectrum, and the brane and antibrane annihilate, providing a natural end to inflation.  (See \cite{Alexander:2001ks} for a  proposal  in which the annihilation itself drives inflation.)

In \cite{Dvali:2001fw,Burgess:2001fx}, the Coulomb interaction (\ref{coulombunwarped}) of the brane-antibrane pair was computed and identified with the inflaton potential.  The Coulomb  force diminishes with increasing distance, suggesting that for sufficiently large separations,  the Coulomb  interaction could drive slow-roll inflation.
However, Burgess et al.~\cite{Burgess:2001fx} demonstrated that the branes would have to be separated by a distance that is larger than the size of the compact space to give a potential that can source slow-roll inflation~(see~\S\ref{etaproblem}).

The character of the problem changed when Kachru et al.~(KKLMMT)~\cite{Kachru:2003sx} made two pivotal observations about D-brane inflation.
First, they  established that {\it warping} of the extra dimensions suppresses the Coulomb force between the brane-antibrane pair, flattening the potential even for modest brane separations.
However,  building on  advances in moduli stabilization (cf.~\S\ref{modulistab}), they also showed that the inflaton potential for a D-brane system is {\it{not}}  given by the Coulomb potential alone: the leading  contributions to the curvature of the inflaton potential come from the physical effects that stabilize the moduli.  This was the first of many manifestations of the eta problem in the context of stabilized string compactifications.   The task is therefore to  specify the moduli-stabilizing effects and  derive the complete inflaton potential.  We pick up the story at this stage.\footnote{This section is based mostly on refs.~\cite{Kachru:2003sx, Baumann:2007np, Baumann:2010sx}. }

\subsection{D3-branes and Warped Geometries}
\label{ssec:warped}

The scenario of \cite{Kachru:2003sx} operates in the context of flux compactifications of type IIB string theory (see~\S\ref{modulistab}), which
can naturally contain warped throat regions.
In this section, we will introduce some geometrical facts about these spacetimes.  We will first approximate the warped region by five-dimensional anti-de Sitter space, and then upgrade to the warped deformed conifold geometry~\cite{Klebanov:2000hb, Klebanov:2000nc}.
In \S\ref{sec:potential}, we will derive the D3-brane potential in these warped backgrounds.

\subsection*{D3-branes in Anti-de Sitter Space}

Consider a stack of $N$ D3-branes in ten-dimensional Minkowski space. The D3-branes source a non-trivial background for the massless fields of type IIB supergravity.
In string frame, the solution for the metric is
\beq
\d s^2 = e^{2A(r)} \eta_{\mu \nu} \d x^\mu \d x^\nu + e^{-2A(r)} \left(  \d r^2 + r^2 \d \Omega_{S_5}^2 \right)\ , \label{equ:ND3metric}
\eeq
where  $\d \Omega_{S_5}^2$  is the metric on a five-sphere and $e^{4A(r)}$ is a harmonic function of the transverse coordinates,
\beq
e^{-4A(r)} =  1 + \frac{L^4}{r^4} \ , \qquad {\rm with}  \qquad
\frac{L^4}{(\alpha')^2} = 4\pi g_{\rm s} N \ .   \label{adswarp}
\eeq
This is a  simple example  of a warped solution, as in (\ref{warped}).
The solution has constant dilaton\footnote{Recall from \S\ref{sec:Dbranes} that D3-branes decouple from fluctuations of the dilaton.
Moreover, their backreaction on the metric of an ISD  compactification (cf.~\S\ref{fluxcompactification}) is completely captured by an overall warp factor, as in (\ref{equ:ND3metric}).
D3-branes are therefore considerably simpler to treat than branes of other dimensionality.} and a non-trivial four-form potential
\beq
\alpha(r) \equiv (C_4)_{t x^i} =  e^{4A(r)} \ . \label{CAdS}
\eeq
Eq.~(\ref{CAdS}) corresponds to the self-dual five-form flux $\tilde F_5 = (1+\star_{10}) \d C_4$.
Recalling the line element of five-dimensional anti-de Sitter space, $AdS_5$, in Poincar\'e coordinates,
\beq
\d s^2_{AdS_5} =  \frac{L^2}{r^2} \d r^2 + \frac{r^2}{L^2}\, \eta_{\mu \nu} \d x^\mu \d x^\nu  \ ,
\eeq
we see that (\ref{equ:ND3metric}) reduces to $AdS_5 \times S^5$ for $r \ll L$.

\begin{figure}[h!]
   \centering
     \includegraphics[scale=0.75]{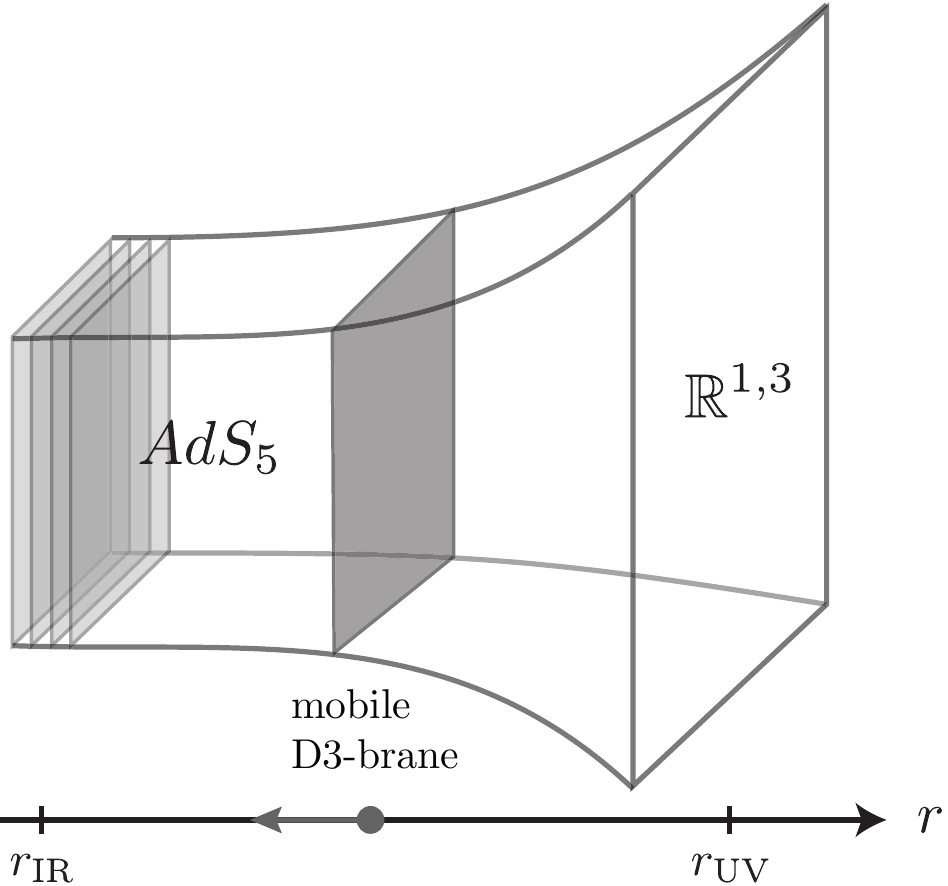}
   \caption{Brane inflation in $AdS_5$. A mobile D3-brane fills four-dimensional spacetime and is pointlike in the extra dimension.}
  \label{fig:AdS-BraneInflation}
\end{figure}

We now consider the dynamics of a mobile D3-brane in the $AdS_5 \times S^5$ background (see fig.~\ref{fig:AdS-BraneInflation}).
The action for a D3-brane in Einstein frame is\footnote{We have  taken the  gauge field strength ${\cal F}_2$  on the D3-brane worldvolume to vanish, which corresponds to considering a D3-brane  without  dissolved D1-brane charge.}
\beq
S_{{\rm D}3} = - T_3 \int \d^{4} \sigma\, \sqrt{-\det (G_{ab}^E)} + \mu_3 \int C_4\ . \label{equ:D3action}
\eeq
To preserve four-dimensional Poincar\'e symmetry, the D3-brane is spacetime-filling, i.e.~its worldvolume coordinates $\sigma^a$ coincide with the spacetime coordinates $x^\mu$. The brane is pointlike in the extra dimensions.
We denote its radial location in anti-de Sitter space by $r$.
Since the angular isometries of $S^5$ are unbroken, we can (for now) assume that the D3-brane has a fixed location along the angular coordinates.
Evaluating the action (\ref{equ:D3action}) in the background (\ref{equ:ND3metric}) gives the following Lagrangian for the brane position:
\beq
{\cal L} \, =\, -  T_3 e^{4A(r)} \sqrt{1 + e^{-4A(r)} g^{\mu \nu} \partial_\mu r \partial_\nu r} \, +\, T_3\hskip 1pt \alpha(r)\ . \label{equ:LDBI}
\eeq
For small velocities, $\dot{r}^2 \ll e^{4A(r)}$, we can expand the square root to get
\beq
{\cal L} \, \approx\, - \frac{1}{2} (\partial \phi)^2 -  T_3 \left(e^{4A(\phi)} - \alpha(\phi)\right)\ , \label{equ:LD3}
\eeq
where we have defined the canonically-normalized field $\phi^2 \equiv T_3 r^2$.
From (\ref{CAdS}), we see that a single D3-brane experiences no force in the anti-de Sitter background: electrostatic repulsion from the four-form background exactly cancels the gravitational attraction.

\subsection*{D3-branes on the Conifold}

An anti-de Sitter background  is not a
realistic setting for D-brane inflation.
First of all, the spacetime is not compact, but ranges from $r=0$ to $r=\infty$.  Furthermore, the metric becomes singular, with infinite redshift, at $r = 0$.
A more  promising scenario for D-brane inflation\footnote{{\it{Mirage  cosmology}} \cite{Kehagias:1999vr} is an alternative to inflation  in which the  spacetime metric  is the induced metric on a D-brane moving through a background supergravity solution.  Discussions   of mirage  cosmologies involving  D3-branes in warped throat regions include \cite{Kachru:2002kx,Germani:2006pf,Easson:2007fz}.} involves a D3-brane  in a finite warped throat region  of a flux compactification \cite{Kachru:2003sx}.  We will now review a few geometric prerequisites for a discussion of this model.

\vskip 4pt
\noindent
{\it Singular conifold.}---The singular conifold is a six-dimensional Calabi-Yau cone $X_6$ that can be presented as the locus in $\mathbb{C}^4$  defined by
\beq
\sum_{A=1}^4 z_A^2 = 0\ , \label{equ:KT}
\eeq
where $A \in \{1,2,3,4\}$.   This describes a  cone over a base $Y_5$,  which is topologically ---  but not metrically ---  equivalent to $S^2 \times S^3$.
To see this, note that if $z^A$ is a solution to (\ref{equ:KT}) then so is $\lambda z^A$, with  $\lambda \in \mathbb{C}$.  Writing $z^A = x^A + i y^A$, the complex equation (\ref{equ:KT}) may be recast as three real equations,
\beq
x \cdot x = \frac{1}{2}\rho^2 \ , \qquad y \cdot y = \frac{1}{2}\rho^2\ , \qquad x \cdot y = 0\ .
\eeq
The first equation defines a three-sphere $S^3$ with radius $\rho/\sqrt{2}$, while the last two equations describe a two-sphere $S^2$ fibered over the $S^3$.
More precisely, the base $Y_5$ of the cone is the Einstein manifold\footnote{An Einstein manifold
satisfies $R_{ab} \propto g_{ab}$.}~$T^{1,1}$, which is the coset space
\begin{equation}
T^{1,1} = [SU(2) \times SU(2)]/U(1)\ ,
\end{equation} with isometry group  $SU(2) \times SU(2) \times U(1)$.
The  metric on $T^{1,1}$ is
\beq
d\Omega_{T^{1,1}}^2 \equiv \frac{1}{9} \left( \d \psi + \sum_{i=1}^2 \cos \theta_i \d \phi_i \right)^2 + \frac{1}{6} \sum_{i=1}^2 \Big( \d \theta_i^2 + \sin^2 \theta_i \d \phi_i^2 \Big)\ ,
\eeq
where $ \theta_i \in [0,\pi]$, $\phi_i \in [0, 2\pi]$ and $\psi  \in [0,4\pi]$.  The metric  on the conifold can then be written as
\beq
\d s^2 = \d r^2 + r^2 \d \Omega_{T^{1,1}}^2\ , \label{equ:ConMet}
\eeq
where $r \equiv \sqrt{3/2}\hskip 2pt \rho^{2/3}$.
To express (\ref{equ:ConMet}) as a manifestly
K\"ahler metric, we introduce three complex coordinates $z^{\alpha}$, $\alpha \in \{ 1,2,3\}$.
The Ricci-flat K\"ahler metric on the singular conifold,
\beq
\d s^2 = k_{\alpha \bar \beta} \, \d z^\alpha \d \overline{z^\beta}\ ,
\eeq then follows from the K\"ahler potential \cite{Candelas:1989js}
\beq
k(z_\alpha, \bar z_\alpha) = \frac{3}{2}\left( \sum_{A=1}^4 |z^A|^2 \right)^{2/3}  \ , \label{equ:smallk}
\eeq via $k_{\alpha \bar \beta} = \partial_\alpha \partial_{\bar \beta} k$.

\vskip 4pt
\noindent
{\it Deformed conifold.}---In  the singular conifold, the base manifold $T^{1,1}$  shrinks to zero size at  $z_A=0$,  and the metric on the cone has a curvature singularity.
To remove the singularity, we consider a small modification of the embedding condition (\ref{equ:KT}),
\beq
\sum_{A=1}^4 z_A^2 = \varepsilon^2\ . \label{equ:KS}
\eeq
This defines the deformed conifold.
The deformation parameter $\varepsilon$ can be made real by an appropriate phase rotation.
Eq.~(\ref{equ:KS}) can then be written as
\begin{align}
x \cdot x - y \cdot y = \varepsilon^2 \ ,\\
x \cdot x + y \cdot y = \rho^2\ .
\end{align}
At the tip of the cone, $\rho^2 = \varepsilon^2$, the $S^3$ remains finite ($x \cdot x = \varepsilon^2$), while the $S^2$ shrinks to zero size~($y \cdot y = 0$).
Sufficiently far from the tip, the right-hand side of (\ref{equ:KS}) can be ignored and the metric of the deformed conifold is well-approximated by that of the singular conifold.
Most models of D-brane inflation operate in this regime.

\vskip 4pt
\noindent
{\it D3-branes on the conifold.}---Now consider placing a stack of $N$ D3-branes at the singular tip, $z_A = 0$, of the singular conifold.
As before, the branes backreact on the geometry,  producing the warped ten-dimensional line element~\cite{Klebanov:2000nc}
\beq
\d s^2 = e^{2A(r)} \eta_{\mu \nu} \d x^\mu \d x^\nu + e^{-2A(r)} \left(  \d r^2 + r^2 \d \Omega_{T^{1,1}}^2 \right)\ , \label{equ:KTmetric}
\eeq
where
\beq
e^{-4A(r)} = 1+ \frac{L^4}{r^4}  \qquad {\rm with} \quad L^4 \equiv \frac{27\pi}{4} g_{\rm s} N(\alpha')^2\ .  \label{equ:KWw}
\eeq
For $r\ll L$, the  solution is $AdS_5 \times T^{1,1}$ \cite{Klebanov:1998hh}.

\vskip 4pt
\noindent
{\it Warped deformed conifold.}---Finally, we describe the  warped deformed conifold, or {\it{Klebanov-Strassler}} (KS)  geometry \cite{Klebanov:2000hb}.
This is a noncompact, smooth solution of type IIB  supergravity in which  warping is supported by background fluxes.
The KS solution can be obtained by considering the backreaction of $N$  D3-branes  at the tip of the  singular conifold,  together with  the backreaction of $M$  D5-branes  wrapping the collapsed $S^2$  at the tip,  but  we will find it useful to give an alternative presentation in which all D-branes are replaced by fluxes carrying  the associated charges (cf.~\cite{Vafa:2000wi}).

The geometric substrate for the solution is the deformed conifold (\ref{equ:KS}), which contains two independent three-cycles:  the $S^3$  at the tip, known as the {\it{A-cycle}}, and the Poincar\'e dual  three-cycle,  known as the {\it{B-cycle}}.  The  background three-form fluxes of the KS solution are quantized
\begin{equation}
\frac{1}{(2\pi)^2 \alpha'} \int_A F_3 = M \qquad {\rm and}
\qquad \frac{1}{(2\pi)^2 \alpha'} \int_B H_3 = K\ ,  \label{KSflux}
\end{equation} where $M \gg 1$ and $K \gg 1$ are  integers.   These fluxes give rise to  non-trivial warping.
The line element for the KS solution takes the form
\beq
\d s^2 = e^{2A(r)} \eta_{\mu \nu} \d x^\mu \d x^\nu + e^{-2A(r)} \d \tilde s^2\ , \label{equ:KSmetric}
\eeq  where $\d \tilde s^2$ is the metric of the deformed conifold defined by (\ref{equ:KS}).
As in the deformed conifold, the infrared geometry is smooth: the A-cycle is finite in size,  with radius $r_{A} = \sqrt{g_{\rm s} M \alpha'}$,  so  the supergravity approximation remains valid near the tip provided that $g_{\rm s} M \gg 1$.   For our purposes, it will suffice to cut off the radial coordinate at a minimum value $r_{\mathsmaller{\rm IR}}$, and work at $r \gg r_{\mathsmaller{\rm IR}}$  (but see \cite{Klebanov:2000hb} for a precise description of the tip geometry).
Far from the tip, the line element is well-approximated by (\ref{equ:KT}), with
\beq
e^{-4A(r)} = \frac{L^4}{r^4}\Biggl(1+ \frac{3 g_{\rm s} M}{8\pi K} +  \frac{3 g_{\rm s} M}{2\pi K}\hskip 1pt \ln \frac{r}{r_{\mathsmaller{\rm UV}}}\Biggr)\ ,  \label{kswarp}
\eeq
where
\beq L^4 \equiv \frac{27\pi}{4} g_{\rm s} N(\alpha')^2\ ,  \quad N \equiv MK\ .
\eeq
Here, $r_{\mathsmaller{\rm UV}}$ is an ultraviolet cutoff, discussed further below.   The logarithmic running of the warp factor corresponds to that seen in the singular warped conifold solution of \cite{Klebanov:2000nc}.
The warp factor $e^{A(r)}$ in (\ref{equ:KSmetric})  reaches a minimal value  $e^{A(r_{\mathsmaller{\rm IR}})} \equiv e^{A_{\mathsmaller{\rm IR}}}$ at the tip, and is given in terms of the flux  quanta by \cite{Giddings:2001yu}
\begin{equation}
e^{A_{\mathsmaller{\rm IR}}}= {\rm exp}\left(-\frac{2\pi K}{3 g_{\rm s} M}\right)\ .
\end{equation}  The exponential hierarchy is a consequence of the logarithmic running in (\ref{kswarp}).
The KS solution given in (\ref{equ:KSmetric}) and (\ref{kswarp}) is the canonical example of  a {\it warped throat} geometry, and provides the basis for the most explicit studies of warped D-brane inflation.

\begin{figure}[h!]
   \centering
     \includegraphics[width=0.7\textwidth]{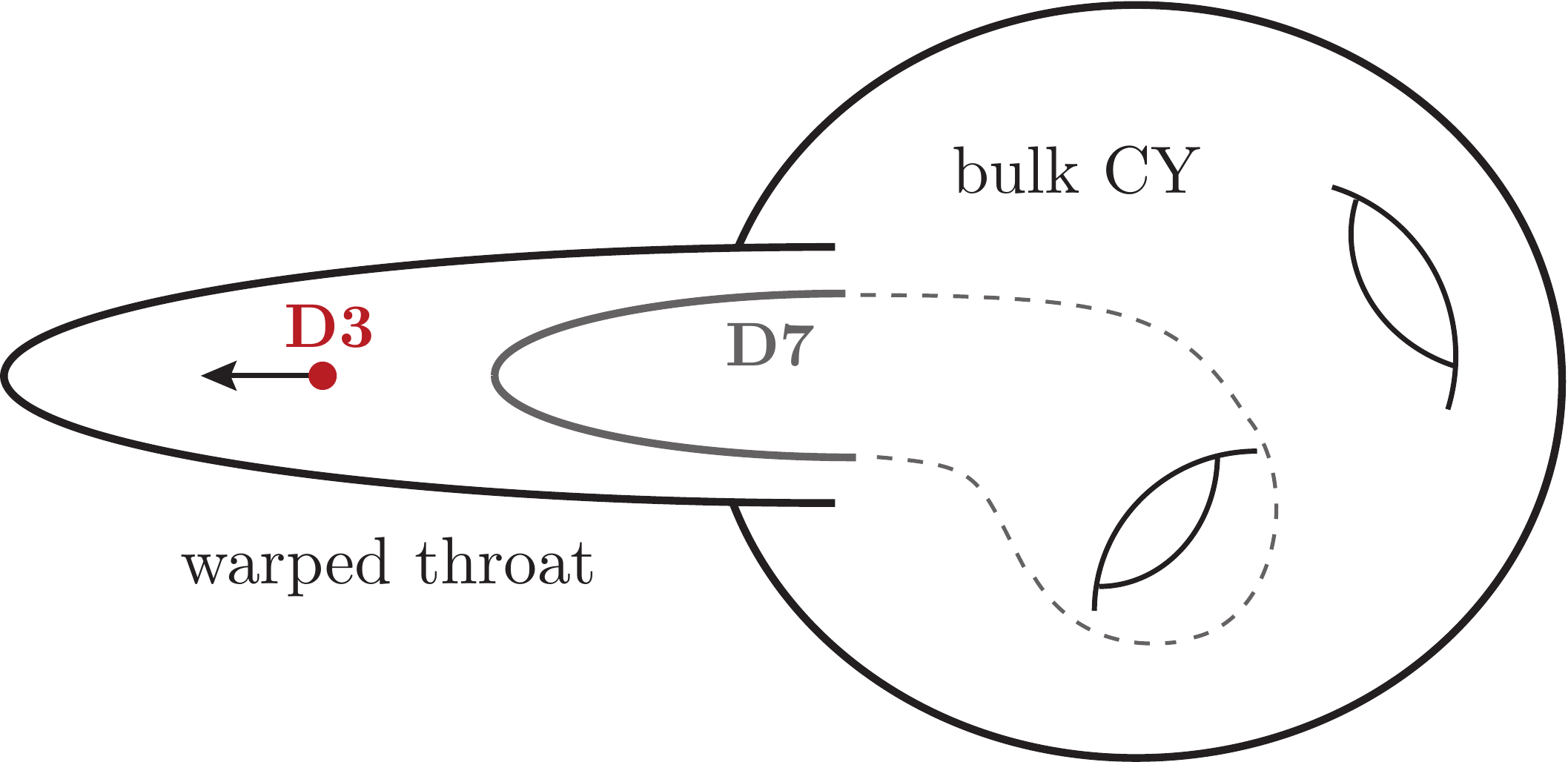}
   \caption{Schematic of a finite  warped throat containing D7-branes wrapping a compact four-cycle.
A portion of the four-cycle  extends into the throat region.
Gaugino condensation on the D7-branes leads to a D3-brane potential.}
  \label{fig:BraneInflation}
\end{figure}

Before proceeding,  we should emphasize that the ten-dimensional KS solution, involving a noncompact  warped deformed conifold,  does not give rise to dynamical gravity  upon dimensional reduction to four dimensions:  the compactification volume,  and hence the four-dimensional Planck mass, are infinite.   For model-building  purposes, one considers  instead a flux compactification containing a finite warped throat region that is well-approximated by a  finite portion of the KS solution, from the tip $r=r_{\mathsmaller{\rm IR}}$ to some ultraviolet cutoff  $r=r_{\mathsmaller{\rm UV}}$.  Beyond this, the throat attaches to a {\it bulk}  space, corresponding to the remainder of the  compactification (see fig.~\ref{fig:BraneInflation}).
The metric of the bulk is poorly characterized in general, but the influence of the bulk  supergravity solution  on dynamics in the throat region can be parameterized very effectively.
The validity of the  finite throat approximation was  systematically  investigated in \cite{Baumann:2010sx} --- see \S\ref{sec:potential}.

\subsection*{A Field Range Bound}

The total compactification volume is the sum of the throat volume,
\beq
{\cal V}_{\cal T} \equiv \int \d \Omega_{T^{1,1}}^2 \int_{r_{\mathsmaller{\rm IR}}}^{r_{\mathsmaller{\rm UV}}} r^5 \d r\, e^{-4A(r)} \ =\ 2\pi^4 g_{\rm s} N (\alpha')^2 \, r_{\mathsmaller{\rm UV}}^2\ ,
\eeq
and the volume ${\cal V}_{\cal B}$ of the bulk space.
The Planck mass  (\ref{equ:4dPlanck}), $M_{\rm pl}^2 = {\cal V}/g_{\rm s}^2 \kappa^2$, is finite,
with ${\cal V} \equiv {\cal V}_{\cal T} + {\cal V}_{\cal B}$. Ignoring the bulk volume gives a
lower bound on the Planck mass,
\beq
M_{\rm pl}^2 > \frac{N}{4} \frac{r_{\mathsmaller{\rm UV}}^2}{(2\pi^3) g_{\rm s} (\alpha')^2} \ .\label{equ:DMp}
\eeq
The amount of canonical field range available to a D3-brane in the throat region (the region of controlled evolution) is bounded from above by
\beq
\Delta \phi^2 < T_3 r_{\mathsmaller{\rm UV}}^2 = \frac{r_{\mathsmaller{\rm UV}}^2}{(2\pi^3) g_{\rm s} (\alpha')^2}\ . \label{equ:Dphi}
\eeq
Combining (\ref{equ:DMp}) and (\ref{equ:Dphi}), we  arrive at the remarkably simple formula~\cite{Baumann:2006cd}
\beq
\frac{\Delta \phi}{M_{\rm pl}} \le \frac{2}{\sqrt{N}} \ . \label{equ:BMbound}
\eeq
Since the validity of the supergravity approximation requires $N \gg 1$,
this result precludes super-Planckian field ranges in models of inflation based on D3-branes  in warped throats.
The  geometric bound (\ref{equ:BMbound}) implies that warped D3-brane inflation does not allow for observable gravitational waves.   Note that this argument is purely kinematic, and does not involve the D3-brane potential.

\subsection{The D3-brane Potential}
\label{sec:potential}

Eq.~(\ref{equ:LD3}) gives the potential for a D3-brane in the warped backgrounds (\ref{equ:ND3metric}), (\ref{equ:KT}), and (\ref{equ:KSmetric}) as
\beq
V(\phi) = T_3 \left(e^{4A(\phi)} - \alpha(\phi)\right)\ . \label{equ:Vphi}
\eeq
This vanishes for compactifications with imaginary self-dual (ISD) fluxes \cite{Giddings:2001yu}.  However, generic string compactifications contain various sources that break the ISD condition and generate a non-trivial potential for the D3-brane.

\vskip 4pt
\noindent
{\it Coulomb potential.}---In~\cite{Kachru:2003sx}, an anti-D3-brane was added to the compactification, following \cite{Kachru:2002gs,Kachru:2003aw}.
The antibrane minimizes its energy in regions of maximal warping,  and is therefore stabilized at the tip of the conifold, $r=r_{\mathsmaller{\rm IR}}$.
The anti-D3-brane perturbs the background supergravity solution, and the D3-brane experiences a  corresponding force.
This is described by the
Coulomb potential~\cite{Kachru:2003sx, DeWolfe:2008zy, Bena:2010ze}
\beq
V_{\cal C}(\phi) = D_0 \left( 1 -\frac{27 }{64\pi^2} \frac{D_0}{\phi^4} \right)\ , \label{equ:VC}
\eeq
where the scale of the potential, $D_0 \ll 2 T_3$, is set by the warped tension of the antibrane
\beq
D_0 \equiv  2 T_3 e^{4A(r_{\mathsmaller{\rm IR}})} \ .
\eeq
The potential (\ref{equ:VC}) is extremely flat, even for small values of the field $\phi$. If this were the end of the story,
warped D-brane inflation would be a strikingly natural scenario,
but life is not so simple.

\vskip 4pt
\noindent
{\it Curvature coupling.}---To source inflation, the system has to be coupled to dynamical gravity.
Besides the Einstein-Hilbert term, the four-dimensional effective action contains a
curvature coupling~\cite{Kachru:2003sx}
\beq
V_{\cal R}(\phi)  = \frac{1}{12} R\, \phi^2\ . \label{equ:VR}
\eeq
In de Sitter space, the four-dimensional spacetime curvature $R$ equals $12H^2$.
During inflation, the coupling in (\ref{equ:VR}) therefore induces a dangerous mass term for the inflaton
\begin{align}
V(\phi) &= V_{\cal C}(\phi)+ V_{\cal R}(\phi)  + \cdots \nonumber \\
&\approx V_0 + H^2\, \phi^2+ \cdots \qquad \Rightarrow \qquad \eta \approx \frac{2}{3} + \cdots\ . \label{equ:ETAprob}
\end{align}
This is an incarnation of the eta problem. The flatness of the Coulomb potential has been completely destroyed by the curvature coupling.
However, this is still not the final answer~\cite{Kachru:2003sx}.
In all stabilized string compactifications there are additional contributions to the D3-brane action, and these must
be included  in order to determine whether inflation can occur.

\vskip 4pt
\noindent
{\it Beyond the probe approximation.}---To compute these corrections we have to go beyond the probe approximation and allow the D3-brane to backreact on the geometry.  In fact, the curvature coupling~(\ref{equ:VR}) can be interpreted as such a backreaction effect~\cite{Baumann:2010sx}.
The presence of the D3-brane perturbs the overall volume of the compactification, ${\cal V}$. Moreover, this perturbation will depend on the position of the brane.  As the brane moves through the warped region, its effect on the volume varies. The compactification volume therefore develops a dependence on the brane position,
${\cal V}={\cal V}(\phi)$.
As a result, a potential that is flat in string frame need not stay flat in Einstein frame, since the transformation between the frames involves a factor of the volume.  The eta problem in (\ref{equ:ETAprob}) arises
from precisely
this effect: see \S\ref{susyetaproblem}.

However, it is easy to see that there will be  further corrections. In  \S\ref{modulistab}, we  explained that  K\"ahler moduli stabilization  in the KKLT scenario involves nonperturbative effects on D7-branes (or from Euclidean D3-branes) wrapping certain four-cycles.
The volumes ${\cal V}_4$ of these four-cycles will also depend on the D3-brane position, ${\cal V}_4(\phi)$.
As the D3-brane moves, the four-cycle volume adjusts. This changes the gauge coupling on the wrapped D7-branes (or the Euclidean D3-brane action) and hence the strength of the nonperturbative effects. This leads to important corrections to the D3-brane potential.

\vskip 4pt
In the following, we will describe the complete D3-brane potential from two different perspectives: first we will derive the potential in four-dimen\-sional supergravity, and then we will provide an equivalent treatment in ten-dimensional supergravity.

\subsection*{4D Perspective} \label{sec:D34d}

The four-dimensional effective theory can be described by the F-term potential of ${\cal N}=1$ supergravity,
\beq
V_F = e^K \left[ K^{I \bar J} D_I W \overline{D_J W}- 3 |W|^2 \right]\ ,
\eeq
where $I, J$ runs over all moduli.  We make the standard KKLT assumption that the complex structure moduli and the dilaton are stabilized at sufficiently high energies.
The remaining moduli are then the K\"ahler moduli $T_i$ and the brane position moduli $z^\alpha$ ($\alpha =1,2,3$). For simplicity of presentation, we restrict to compactifications with only a single K\"ahler modulus $T$, but all our considerations generalize to $h_{+}^{1,1}>1$.  We define $Z^I \equiv \{ T, z^\alpha\}$.
The tree-level K\"ahler potential is the logarithm of the compactification volume
\beq
K = - 2\ln ({\cal V})\ , \label{equ:KV}
\eeq
where ${\cal V}$ is an implicit function of the $Z^I$.  Corrections to (\ref{equ:KV}) are important in many other contexts, cf.~\S\ref{sec:LVS}, but can be neglected in D3-brane inflation.

 \vskip 4pt
\noindent
{\it Backreaction on the volume.}---As mentioned above, a D3-brane with finite energy density backreacts on the overall compactification volume, which therefore depends on the brane position $z^\alpha$~\cite{DeWolfe:2002nn, Baumann:2007ah}:
\beq
{\cal V} = \Bigl( T + \bar T - \gamma k(z_\alpha, \bar z_\alpha) \Bigr)^{3/2}\ , \label{equ:vol}
\eeq
where $k(z_\alpha, \bar z_\alpha)$ is the K\"ahler potential (\ref{equ:smallk}) and $\gamma$ is a constant. In Appendix~B of \cite{Baumann:2007ah}, the parameter $\gamma$ was related to the stabilized value of the K\"ahler modulus,
\beq
\gamma \equiv \frac{T_3}{6} \left(T + \bar T \right)_{\mathsmaller{\rm IR}}\ .
\eeq
Here, $T_{\mathsmaller{\rm IR}} \equiv T(r_{\mathsmaller{\rm IR}})$ stands for the value of the K\"ahler modulus when the
D3-brane is near the tip of the throat.
In \cite{Baumann:2007ah}, it was shown that the  minimum of the potential for the K\"ahler modulus $T$  shifts slightly  as the D3-brane moves,
and the effect of this  shift was further examined in \cite{Panda:2007ie}.

\vskip 4pt
\noindent
{\it F-term potential.}---Combining (\ref{equ:KV}) and (\ref{equ:vol}), we find that the K\"ahler potential is of the form postulated by DeWolfe and Giddings~\cite{DeWolfe:2002nn}, cf.~(\ref{equ:DeWolfeG}):
\beq
K(Z^I, \bar Z^I) = - 3 \ln\Big[T + \bar T - \gamma k(z_\alpha, \bar z_\alpha) \Big] \equiv - 3 \ln \Big[ U(Z^I, \bar Z^I) \Big]\ . \label{equ:KKK}
\eeq
The F-term potential for (\ref{equ:KKK}) combined with a general superpotential $W(Z^I)$ was determined in \cite{Baumann:2007ah,Burgess:2006cb, Krause:2007jk}
\begin{align} V_F(T, z_\alpha) \ =\ &
\frac{1}{3 U^2} \Biggl[ \bigg (T + \bar{T}+
\gamma \left( k_{\gamma} k^{\gamma
\overline{\delta}} k_{\overline{\delta}}-k \right) \bigg)
|W_{,T}|^2 - 3 \left(\overline{W} W_{, T} + c.c.\right)  \nonumber \\
& \hspace{1cm}\mbox{} + \underbrace{ \left( k^{\alpha \overline{\delta}}k_{\overline{\delta}}
\overline{W_{,T}}  W_{,\alpha} + c.c.\right) +
\frac{k^{\alpha \overline{\beta}}}{\gamma}  W_{,\alpha}
\overline{W_{,\beta}} }_{\Delta V_F} \ \ \Biggr]\ ,
\label{equ:Fterm}
\end{align}
where $k_{\alpha} \equiv \partial_\alpha k$ and $k_{\alpha \bar \beta} \equiv \partial_\alpha \partial_{\bar \beta} k$.
The label $\Delta V_F$ has isolated terms that arise exclusively from
the dependence of the superpotential on the brane position $z^\alpha$.  The remainder is the standard KKLT F-term potential~\cite{Kachru:2003aw}.

First consider the situation in which the superpotential does not depend on the brane coordinate, $W = W(T)$.
In this case, $\Delta V_F = 0$ and the remaining terms in the square bracket in~(\ref{equ:Fterm}) depend only weakly on the inflaton.
The potential can therefore be written as
\beq
V_F(r) \approx \frac{V_0}{(1- \frac{1}{6}\phi^2)^2} \approx V_0 + \frac{1}{3} \frac{V_0}{M_{\rm pl}^2} \phi^2\ , \label{equ:Veta}
\eeq
where in the second equality we have made the dependence on the Planck mass explicit.
We see that the inflaton has a mass of order the Hubble scale, $H^2 \approx V_0/(3M_{\rm pl}^2)$.
This is how the curvature coupling (\ref{equ:VR}) arises in the effective supergravity description.

\vskip 4pt
\noindent
{\it Backreaction on D7-branes.}---Gaugino condensation on a stack of $N_c$ D7-branes
leads to
\begin{equation}
|\Delta W| \propto {\rm exp}\Bigl(-\frac{2\pi}{N_c}\hskip 1pt {\cal{V}}_4\Bigr)\ ,
\end{equation} where ${\cal{V}}_4$ is the `warped volume' (\ref{warpedvolume}) wrapped by the D7-branes.    Changing the position $\phi$ of a spacetime-filling D3-brane alters the warp factor
$A(\phi)$,
and hence ${\cal V}_4(\phi)$,
so that  $\Delta W=\Delta W(\phi)$.
To quantify this effect, one computes
 the backreaction of the D3-brane on the four-cycle wrapped by the D7-branes~\cite{Baumann:2006th}.
For a four-cycle
 defined by a holomorphic embedding
\beq
f(z_\alpha) = 0\ , \label{equ:emb}
\eeq the result can be written as
\beq
\label{equ:W}
 W(T,z_\alpha) = W_0 + {\cal A}(z_\alpha) e^{- a T}\ , \quad \quad \quad
a \equiv \frac{2 \pi}{N_c}\ ,
\eeq
where the function ${\cal A}(z_\alpha)$ is defined in terms of the embedding~(\ref{equ:emb}),
\beq
{\cal A}(z_\alpha) = {\cal A}_0 \left( \frac{f(z_\alpha)}{f(0)}\right)^{1/N_c}\ .
\eeq

\begin{figure}[htbp]
    \centering
        \includegraphics[width=0.97 \textwidth]{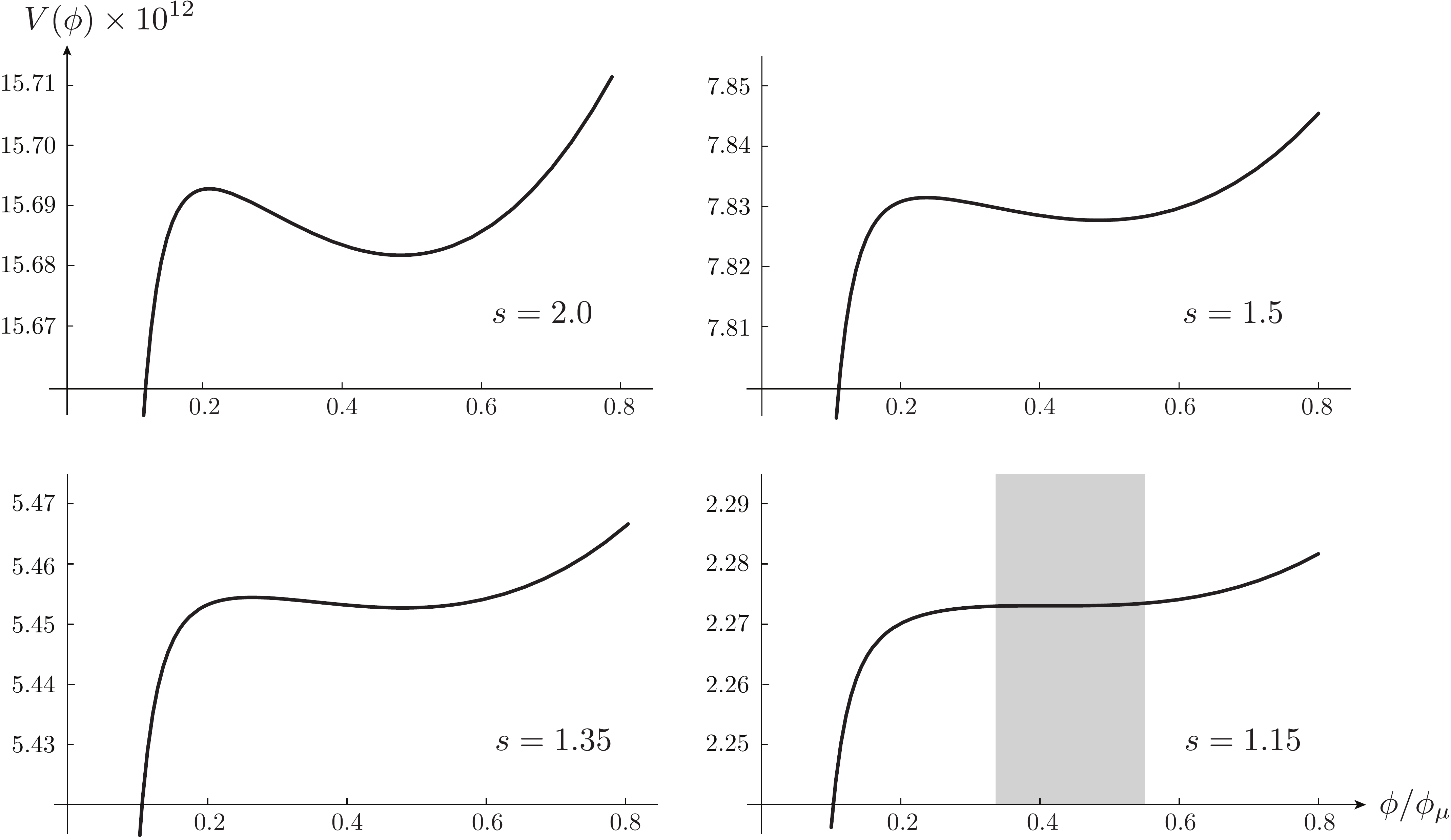}
\caption{Example scan through the parameter space of warped D3-brane inflation (figure adapted from~\cite{Baumann:2007ah}). The scan parameter $s$ is the ratio of the antibrane energy to the F-term energy before uplifting. Successful inflation occurs in the gray shaded region.}   \label{fig:Scan}
\end{figure}

\noindent
{\it Fine-tuning to produce a flat potential.}---Which embedding functions $f(z_\alpha)$ lead to forces that can balance the curvature coupling? This question was addressed in a number of papers~\cite{Baumann:2007np, Baumann:2007ah, Krause:2007jk, Burgess:2006cb}.
An important no-go result was proven in~\cite{Burgess:2006cb, Baumann:2007ah}.
The infinite class of embeddings studied in~\cite{Arean:2004mm} does not allow any inflationary solutions.
In fact, to date only a single explicit embedding is known in which inflation can occur~\cite{Krause:2007jk,Baumann:2007ah}.
This is the
so-called Kuperstein embedding~\cite{Kuperstein:2004hy}
\beq
f(z_1) = \mu - z_1 \ .
\eeq
In this example, the scalar potential (\ref{equ:Veta}) receives a correction scaling as $\phi^{3/2}$ ($\propto z_1$),
\beq
V_F(\phi) \approx V_0 + \cdots + \lambda \phi^{3/2}   + \frac{V_0}{M_{\rm pl}^2} \phi^2 + \cdots\ . \label{equ:V32}
\eeq
The
last
two terms
shown in (\ref{equ:V32}) contribute to $\eta$ with opposite signs.
Let $\phi_0$ be the point in field space where the second slow-roll parameter vanishes, $\eta(\phi_0) = 0$.  Near this point we have $|\eta| \ll 1$.  This is not even a fine-tuning, but arises dynamically. What does involve fine-tuning is the requirement that $\phi_0$ is in the region of control (i.e.~inside the warped throat) and that the potential is monotonic and has a small first derivative (small $\epsilon$) at the same point.
If this can be arranged, then we get inflation near an approximate {\it inflection point}. Fig.~\ref{fig:Scan} shows an example of a successful scan in the parameter space of warped D3-brane inflation~\cite{Baumann:2007ah}.

\subsection*{10D Perspective} \label{sss:D310d}

The example above provides an existence proof for inflation in warped throat geometries, but the setup is too special to provide a good sense for the range of possibilities.
Moreover, the above analysis implicitly assumed that the physics inside the throat decouples completely from the physics of the bulk, which  as we stressed in \S\ref{etaproblem}  is rarely the case.
Finally, we have modeled the warped throat region by a finite portion of a noncompact warped Calabi-Yau cone.
This approximation fails where the finite throat is attached to the remainder of the compactification.
In this section, we describe a more general analysis that addresses these deficiencies.

\vskip 4pt
The essential idea is that all `compactification effects' --- i.e.~all information about moduli stabilization and supersymmetry breaking in the remainder of the compactification --- can be expressed as non-normalizable perturbations of the noncompact solution~\cite{Baumann:2010sx, Baumann:2009qx},
\begin{equation}
\delta \Phi(r) = \delta\Phi(r_{\mathsmaller{\rm UV}}) \left(\frac{r}{r_{\mathsmaller{\rm UV}}}\right)^{\Delta} \ .
\end{equation}
Here, $\delta\Phi$ is the deviation of some supergravity field $\Phi$ from its value in the noncompact solution, $r_{\mathsmaller{\rm UV}}$ is the radial location of the ultraviolet end of the throat, and $\Delta$ is the scaling dimension of~$\delta \Phi$.\footnote{In AdS/CFT, this corresponds to the dimension of the operator dual to the perturbation $\delta\Phi$.}  By determining the spectrum of perturbations of the warped conifold, we will be able to identify the leading corrections to the D3-brane potential.

Locality in the internal space dictates that the effective action for a D-brane probe at some point is specified by the supergravity fields at that point. This suggests the following strategy: find the most general supergravity solution for a finite warped throat that asymptotes in the infrared to the Klebanov-Strassler solution, by classifying all possible perturbations $\delta\Phi$.  Far from the ultraviolet region, the solution is given to good approximation by retaining the subset of modes with the lowest values of $\Delta$, i.e.~the modes dual to the most relevant perturbations of the dual field theory Lagrangian.

In a general six-dimensional cone, it would be challenging to determine the spectrum of dimensions~$\Delta$.  However, the conifold is a cone over the coset space $T^{1,1}$, which is amenable to harmonic analysis via group theory techniques.  Thus, by approximating a finite warped region as a portion of the warped conifold and using the spectroscopy of $T^{1,1}$, one can determine the leading non-normalizable modes.  Correspondingly, one obtains the form of the leading contributions to the potential  of a D3-brane in a KS throat.   We now give a few details of this analysis.

\begin{figure}[h!]
   \centering
     \includegraphics[width=0.9\textwidth]{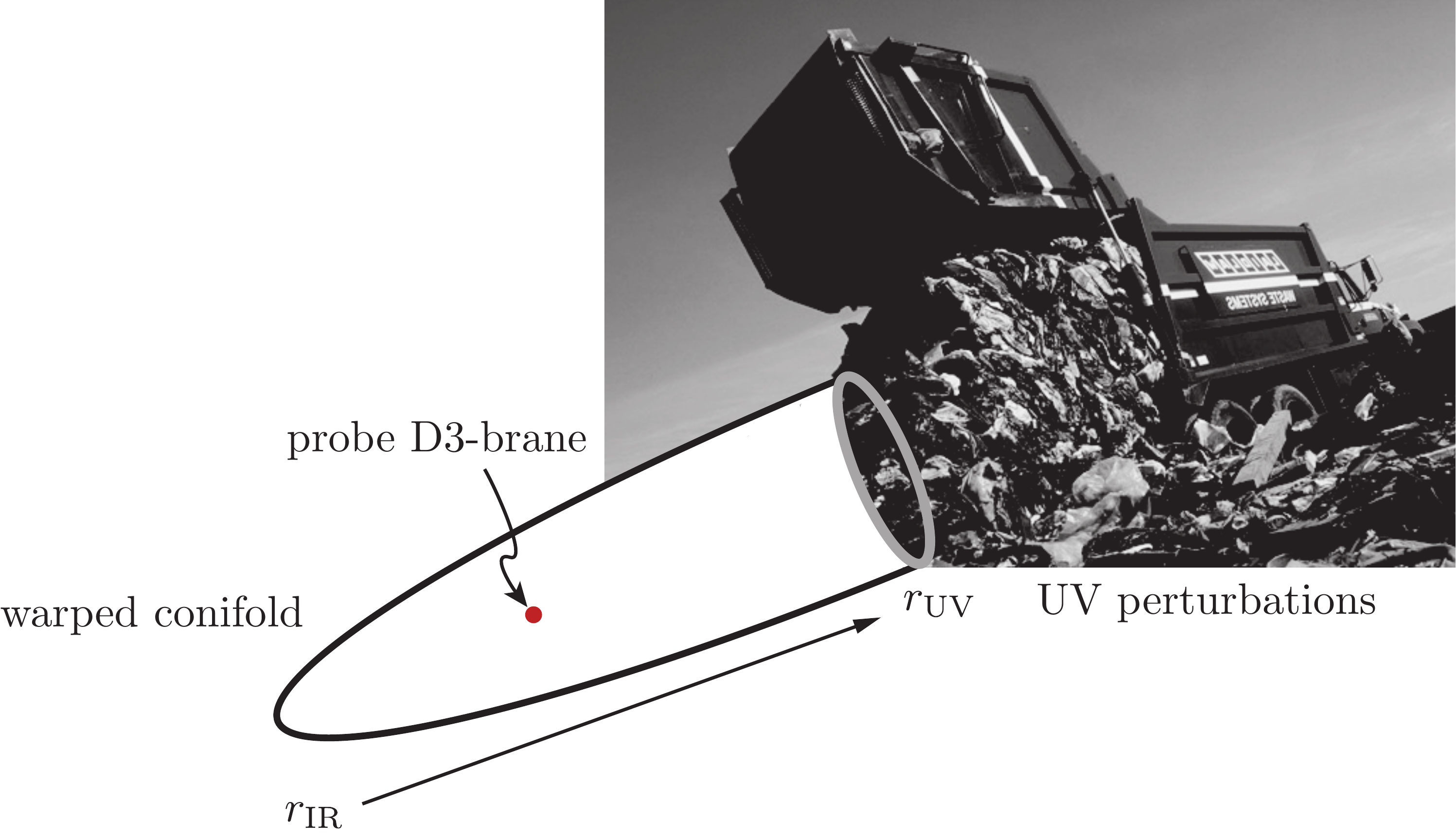}
   \caption{Compactification induces UV perturbations to the warped conifold solution.
In the IR the lowest-dimension perturbations dominate in the D3-brane
potential.}
  \label{fig:Truck}
\end{figure}

\vskip 6pt
\noindent
{\it 10D Supergravity.}---To determine the D3-brane potential (\ref{equ:Vphi}), we need solutions for the warp factor $e^{4A(r)}$ and the four-form potential $\alpha(r)$. In particular, we will be interested in the solution for the field
\beq
\Phi_- \equiv e^{4A} - \alpha \  . \label{phiminusdef}
\eeq
Taking the metric ansatz
\begin{equation}
\d s^2 =   e^{2A(y)} g_{\mu\nu}\d x^{\mu} \d x^{\nu} + e^{-2A(y)}  g_{mn}\d y^{m} \d y^{n}\ ,  \label{msansatz}
\end{equation} where $g_{\mu\nu}$ is  the metric of a maximally symmetric  four-dimensional spacetime,
the field equations of ten-dimensional type IIB supergravity imply the master equation\footnote{In comparison to (\ref{equ:EE2}), we have now allowed the four-dimensional curvature ${R}_{4}$ to be nonvanishing: compare (\ref{minkowskiansatz}) and (\ref{msansatz}).}
\beq
\nabla^2 \Phi_- =  {R}_{4} + \frac{g_{\rm s}}{96} |\Lambda|^2  + e^{-4A}|\nabla\Phi_-|^2 + {\cal S}_{\rm loc} \ , \label{equ:master}
\eeq
where $\nabla^2$ is the Laplacian constructed using the conifold metric (\ref{equ:ConMet}),
${\cal S}_{\rm loc}$ is a localized source due to anti-D3-branes, and
\begin{equation}
\Lambda \equiv \Phi_+ G_- + \Phi_- G_+ \ ,
\end{equation} with
\begin{equation}
G_{\pm} \equiv (\star_{6}\pm i)G_3 \qquad {\rm and} \qquad \Phi_\pm \equiv e^{4A} \pm \alpha\ .  \label{gpm}
\end{equation}
At the same time, the three-form flux  must satisfy the equation of motion
\begin{equation}  \label{lambdafluxeom}
{\rm d}\Lambda + \frac{i}{2}\frac{{\rm d}\tau}{{\rm Im}\tau}\wedge(\Lambda+\bar{\Lambda}) = 0 \ .
\end{equation}

The solutions to (\ref{equ:master}) can be organized as follows:
\beq
V(x, \Psi) = V_0 + V_{\cal C}(x) + V_{\cal R}(x) + V_{\cal B}(x,\Psi)\ , \label{equ:Vx}
\eeq
where $x \equiv r/r_{\mathsmaller{\rm UV}}$ and $\Psi$ stands collectively for all five angular coordinates. We will describe each of the terms in (\ref{equ:Vx}) in turn.

\vskip 4pt
\noindent
{\it Constant contributions.}---The constant $V_0$ represents possible contributions from
distant sources of supersymmetry breaking --- in  the bulk of the compactification, or in other throats ---  that  exert negligible forces on the D3-brane, and only contribute to the inflationary vacuum energy.
This situation corresponds to  maximal decoupling of the source of supersymmetry breaking  from the D3-brane action:  the two sectors communicate only through four-dimensional curvature.
As explained in~\S\ref{etaproblem},  complete decoupling  of this sort is  very rare.
We have in fact made an  artificial but convenient division, using $V_0$  to represent  the sum  of all\footnote{In fact, one  constant contribution is grouped in $V_{\cal C}$  rather than in $V_0$:  this is the vacuum energy contributed by the brane-antibrane pair,  denoted $D_0$ in (\ref{eq:vCoul}).} constant contributions to the potential,  from diverse sources,  each of which will in general also contribute non-constant terms  in other categories described below.

\vskip 4pt
\noindent
{\it Local sources.}---As before, $V_{\cal C}(x)$ is the Coulomb potential sourced by ${\cal S}_{\sf local}$,
\beq  \label{eq:vCoul}
V_{\cal C}(x) = D_0 \left( 1 - \frac{27}{64 \pi^2} \frac{D_0}{T_3^2 r_{\mathsmaller{\rm UV}}^4} \frac{1}{x^4}\right)\ .
\eeq
In the inflationary regime (far from the tip), the dependence on the D3-brane position $x$ is a subdominant effect.  This is a restatement of the fact that warping --- captured by the smallness of $D_0$ in (\ref{eq:vCoul}) ---  makes the Coulomb potential extremely flat.

\vskip 4pt
\noindent
{\it The eta problem revisited.}---The Friedmann equation relates the
Ricci curvature in four dimensions, $R_{4} = 12 H^2$, to the inflationary energy density, $V \approx V_0 + D_0$.  Integrating (\ref{equ:master}), we find a curvature-induced mass term
\beq
V_{\cal R}(x) =  \frac{1}{3}\mu^4 x^2 + \cdots\ , \qquad {\rm where} \quad \mu^4 \equiv (V_0 + D_0) \frac{T_3r_{\mathsmaller{\rm UV}}^2}{M_{\rm pl}^2} \ .
\eeq
This is how the curvature-coupling aspect of the eta problem arises in ten-dimensional supergravity.

\vskip 4pt
\noindent
{\it Bulk contributions.}---Finally, we have a term that characterizes all possible contributions from stress-energy in the bulk of the compactification,
\beq
V_{\cal B}(x,\Psi) = \mu^4 \sum_{LM} c_{LM} \, x^{\Delta(L)}\, f_{LM}(\Psi)\ , \label{equ:Vbulk}
\eeq
where $c_{LM}$ are constant coefficients, $L \equiv (j_1, j_2, R)$ and $M \equiv (m_1, m_2)$ label the $SU(2)\times SU(2)\times U(1)$ quantum numbers under the isometries of $T^{1,1}$, and the functions  $f_{LM}(\Psi)$
 are angular harmonics on $T^{1,1}$ (whose explicit forms can be found in~\cite{Baumann:2010sx}). The exponents $\Delta(L)$ have been computed in detail in~\cite{Baumann:2010sx}, building on a spectroscopic analysis of perturbations on $AdS_5 \times T^{1,1}$~\cite{Ceresole:1999zs}. We briefly summarize the results.
We split the bulk contributions into homogeneous solutions of the six-dimensional Laplace equation~\cite{Baumann:2008kq}
\beq
\nabla^2 \Phi_{h} = 0\ , \label{equ:homo}
\eeq
and inhomogeneous contributions sourced by flux~\cite{Baumann:2009qx},
\beq
\nabla^2 \Phi_{f} =  \frac{g_{\rm s}}{96} |\Lambda|^2\ . \label{equ:fluxeq}
\eeq
The solutions are characterized by their scaling dimensions $\Delta(L)$. Solutions to (\ref{equ:homo}) satisfy
\beq
\Delta_h(L) \equiv - 2 \sqrt{H(j_1,j_2,R) + 4} \ ,
\eeq
where
\beq
H(j_1,j_2,R)  \equiv 6 \left[ j_1(j_1+1) + j_2(j_2+1) - \frac{1}{8} R^2\right]\ .
\eeq
Taking into account selection rules~\cite{Ceresole:1999zs, Baumann:2009qx} for the angular quantum numbers, the first few scaling dimensions are
\beq
\Delta_h \ = \ \frac{3}{2}\ , \ 2\ , \ 3\ , \ \sqrt{28} - 2\ ,\ \cdots   \label{homdim}
\eeq
The flux contributions in (\ref{equ:fluxeq}) lead to the following solutions:
\beq
\Delta_f(L) = \delta_i(L) + \delta_j(L) - 4\ ,
\eeq
where
\begin{align}
\delta_1(L) &\equiv - 1 + \sqrt{H(j_1,j_2,R+2) + 4}\ ,\\
\delta_2(L) &\equiv  \sqrt{H(j_1,j_2,R) + 4}\ ,\\
\delta_3(L) &\equiv 1 + \sqrt{H(j_1,j_2,R-2) + 4}\ .
\end{align}
Incorporating
the selection rules, we find~\cite{Baumann:2009qx, Gandhi:2011id}
\beq
\Delta_f \ =\ 1\ ,\ 2\ ,\  \frac{5}{2}\ ,\ \sqrt{28} - \frac{5}{2}\ , \ \cdots  \label{inhomdim}
\eeq
The total bulk potential (\ref{equ:Vbulk}) therefore contains terms with the scaling dimensions
\beq
\Delta \ =\ \left\{ \Delta_h , \Delta_f \right\}\ = \ 1\ ,\ \frac{3}{2}\ ,\ \sqrt{28} - \frac{5}{2}\ , \ 3\ , \ \sqrt{28} - 2\ ,\ \frac{7}{2}\ , \ \sqrt{28} - \frac{3}{2} \ , \ \cdots
\label{finalspectrum}
\eeq

A few remarks about the analysis leading to (\ref{finalspectrum}) are necessary.   One should recognize that (\ref{equ:fluxeq}) is non-linear in perturbations  of the background:  a linear treatment would capture  only the homogeneous solutions solving (\ref{equ:homo}), with dimensions given in (\ref{homdim}), while the {\it{leading}}  term at small $r$,  corresponding to $\Delta_f=1$ in (\ref{inhomdim}),  actually arises at quadratic order in  perturbations of three-form flux.   This is possible because the perturbations corresponding to various supergravity fields  do not enter  on equal footing:  some perturbations  are allowed by the ISD  background,  and hence have order-unity perturbations $\delta \Phi$  at $r=r_{\mathsmaller{\rm UV}}$, while other perturbations are forbidden in the ISD  solution, and have perturbations $\delta \Phi \sim e^{-a T}$  at $r=r_{\mathsmaller{\rm UV}}$.  These hierarchies can be captured by a careful spurion analysis \cite{Baumann:2010sx, Gandhi:2011id}.

Notice that we again have a contribution scaling as $\phi^{3/2}$, just as in the four-dimensional analysis.
This suggests that the basic phenomenology is again that of inflection point inflation, and a number of numerical investigations~\cite{Ali:2008ij,Ali:2010jx,Agarwal:2011wm, McAllister:2012am, Dias:2012nf} have confirmed this  expectation.

\subsection{Multi-Field Dynamics}
\label{sec:MD}

The  effective theory describing an  inflating D3-brane in a conifold region attached to a stabilized compactification has a natural mass scale:  the inflationary Hubble parameter, $H$.   Moreover, all continuous global symmetries are broken by the compactification.  The general arguments reviewed in \S\ref{sec:EFT2}   then suggest that the six  real scalar  fields  parameterizing the D3-brane position  should have masses $m \sim {\cal{O}}(H)$.
Inflation will not occur naturally, and some accidental  cancellations  among terms in the potential are required  in order for one of the scalars to have a mass $m \ll H$.
Once such a cancellation has occurred, it is quite unlikely that all five  of the other   fields will have masses $m \gg H$:  a more probable outcome is that one or more of these fields will be light enough to evolve and fluctuate during inflation.  Thus,  the warped D3-brane inflation  scenario  generically gives rise to models of multi-field inflation, or more precisely of {\it{quasi-single-field}} inflation~\cite{Chen:2009zp}.

To  understand the phenomenology of these models,  neither a slow-roll approximation  nor a single-field truncation  is appropriate, and one must solve the
equations of motion for the perturbations  numerically, without  making any approximations.
The exact power spectra for more than $10^4$  realizations from the ensemble of \cite{Agarwal:2011wm} were  obtained in \cite{McAllister:2012am},  with key results summarized in \S\ref{warpedphenomenology}.

\vskip 4pt
One intriguing finding of  \cite{McAllister:2012am} is that the spectrum of scalar masses  is predicted to good accuracy by a  very simple  matrix model inspired by \cite{Marsh:2011aa}, cf.~\S\ref{landscapestatistics}.
The model for the $6\times 6$ mass matrix ${\cal M}$ takes the form\footnote{The physical relevance of the matrix model (\ref{abc}) can be understood by  comparing it to  the Wigner+Wishart+Wishart  model (\ref{WWWmodel}) of \cite{Marsh:2011aa}.  The positive-definite blocks $A\bar A$ and $B\bar B$ are  consequences of spontaneously broken four-dimensional supersymmetry:  in the limit of unbroken supersymmetry  the  mass matrix must be positive definite. The  methods used in \cite{Baumann:2010sx} to construct the ensemble of  effective Lagrangians were inherently ten-dimensional, and  made no direct connection to the  structure of  four-dimensional ${\cal N}=1$  supersymmetry.  The fact that  the stability properties enjoined by  four-dimensional supersymmetry  nevertheless emerge after the  intricate analysis  described above  is  encouraging evidence  that the entire construction is  self-consistent.}
\begin{equation} \label{abc}
{\cal M} =
\left(
\begin{array}{c c}
A\bar A + B\bar B&  C \\
 \bar C & \bar A A +  \bar B B \end{array}
\right)\,,
\end{equation} in terms of $3\times 3$ complex symmetric matrices $A$, $B$, and $C$ whose entries are assumed to be  random complex numbers drawn from a Gaussian distribution.
The eigenvalue spectrum of ${\cal M}$  agrees surprisingly well  with the empirical mass spectrum found in \cite{McAllister:2012am}, even though the methods of random matrix theory are  formally applicable only  to large matrices:  evidently 3  is a sufficiently large number in the present context.

\vskip 4pt
The procedure described above led to an EFT  for six  real fields,  the coordinates of the D3-brane. This captures completely general contributions to the action for these fields that stem from {\it heavy} degrees of freedom in the remainder of the compactification.   However, the  open string EFT  constructed in this way can differ from the complete EFT  that arises from dimensional reduction  of all open and closed string fields.   We have implicitly truncated  the spectrum (see \S\ref{sss:approximations}), assuming that the  closed string moduli have masses $m \gg H$.   For complex structure moduli  and the axiodilaton,  which  acquire large supersymmetric masses from three-form flux, truncation is  generally justified; but  without special model building (cf.~\cite{Kallosh:2004yh}) the  typical mass scale of the  K\"ahler moduli  is $m \sim H$.   As a result,  the EFT  may  include  a number of relatively light K\"ahler moduli, in addition to the six open string fields studied above, and there is comparatively little hope of determining the precise form  of the potential for these  closed string moduli.\footnote{As noted above, some effects  of a single light K\"ahler modulus were  considered in \cite{Panda:2007ie}, and  the response of the overall volume to the displacement of the D3-brane played a key role in the stability analysis of \cite{Baumann:2007ah}.}  However, in view of the successes of universality  and random matrix theory  in characterizing the six-field effective theory  \cite{Agarwal:2011wm, McAllister:2012am}, and bearing in mind that having more fields makes these methods more robust, we find it plausible that  the statistical signatures of scenarios with dynamical K\"ahler moduli can be obtained  in like manner.

\subsection{Reheating}

The reheating stage of warped D-brane inflation was carefully examined in \cite{Barnaby:2004gg,Kofman:2005yz,Chialva:2005zy,Chen:2006ni,Berndsen:2008my}, revealing a complex cascade of energy from the inflaton to the visible sector and to invisible relics.
To set the stage, we remark that a modular approach to reheating is very natural in this context: because the inflaton sector involves a D3-brane in a local geometry, it is reasonable to identify the warped throat where inflation occurs as one module, and to situate the Standard Model on D-branes in a different region of the geometry, either in another warped throat or in the unwarped bulk region.  These model-building choices critically affect the success of reheating.
We will not review all possibilities here, and will emphasize the interesting `two-throat' scenario in which the visible sector resides in a warped throat distinct from that in which inflation occurs. This choice affords much latitude in model-building, as well as leading to novel phenomenology for reheating.  Moreover, if the Standard Model D-branes were inside the inflationary throat, any relic cosmic strings would quickly disintegrate  through  contact with these D-branes; in the bulk or in another throat, the D-branes are at a safe distance, and long-lived cosmic strings, with the associated interesting signatures, are possible.

The outline of the end of inflation, and of reheating, is as follows.  Inflation occurs while the D3-brane passes through the vicinity of an inflection point in its potential, and accelerated expansion ends once the D3-brane reaches a steeper portion of the potential. The D3-brane then falls rapidly toward the anti-D3-brane at the tip of the throat.  Eventually, the separation of the brane-antibrane pair becomes small enough that a tachyon develops.  The tachyonic instability causes the D3-brane pair  to fragment, and to decay into  highly-excited, non-relativistic closed string modes \cite{Sen:2002nu,Sen:2002in,Lambert:2003zr}, which quickly
decay into massive Kaluza-Klein excitations of the  supergravity fields (i.e.~massless string modes) in the inflationary throat.

A few words about  interactions in warped throats are necessary. The Kaluza-Klein modes of a warped throat  have wavefunctions that peak  exponentially in the infrared,  and their mutual interactions are suppressed by the infrared scale $m_{\mathsmaller{\rm IR}} \sim e^{A_{\mathsmaller{\rm IR}}}\,M_{\rm pl} \ll M_{\rm pl}$, where  $e^{A_{\mathsmaller{\rm IR}}}$ is the warp factor at the tip of the throat.  On the other hand, their couplings  to Kaluza-Klein {\it zero modes},  including the graviton,  are suppressed by $M_{\rm pl}$.
The warping creates a  gravitational potential barrier  that confines massive  particles to the infrared region:  access to  other throats is via tunnelling\footnote{See \cite{Mukohyama:2007ig} for an analysis of energy transfer in warped reheating via induced motion of D-branes.} through the bulk  of the compactification, which is very slow compared to perturbative decays \cite{Dimopoulos:2001ui,Barnaby:2004gg,Kofman:2005yz,Chialva:2005zy,Langfelder:2006vd}.
As a result, the  characteristic timescales  typically obey
\begin{equation}
\tau_{\rm therm} \ll \tau_{{\rm graviton}} \ll \tau_{{\rm tunnel}}\,,
\end{equation}  where $\tau_{\rm therm}$ denotes the thermalization time for Kaluza-Klein modes of the inflationary throat, $\tau_{\rm{graviton}}$ is the timescale for decay to gravitons, and
$\tau_{\rm{tunnel}}$ is the tunnelling timescale.

Shortly after the decay of excited strings to excited Kaluza-Klein modes, the energy previously stored in the inflaton condensate is  still largely confined to the inflationary throat. The success of reheating depends on channeling a sufficiently large fraction of this energy into Standard Model degrees of freedom, rather than into four-dimensional gravitons; long-lived relic particles protected by approximate isometries; or matter or radiation in other sectors.  We now discuss these challenges in turn.

\begin{figure}[h!]
   \centering
     \includegraphics[width=0.95\textwidth]{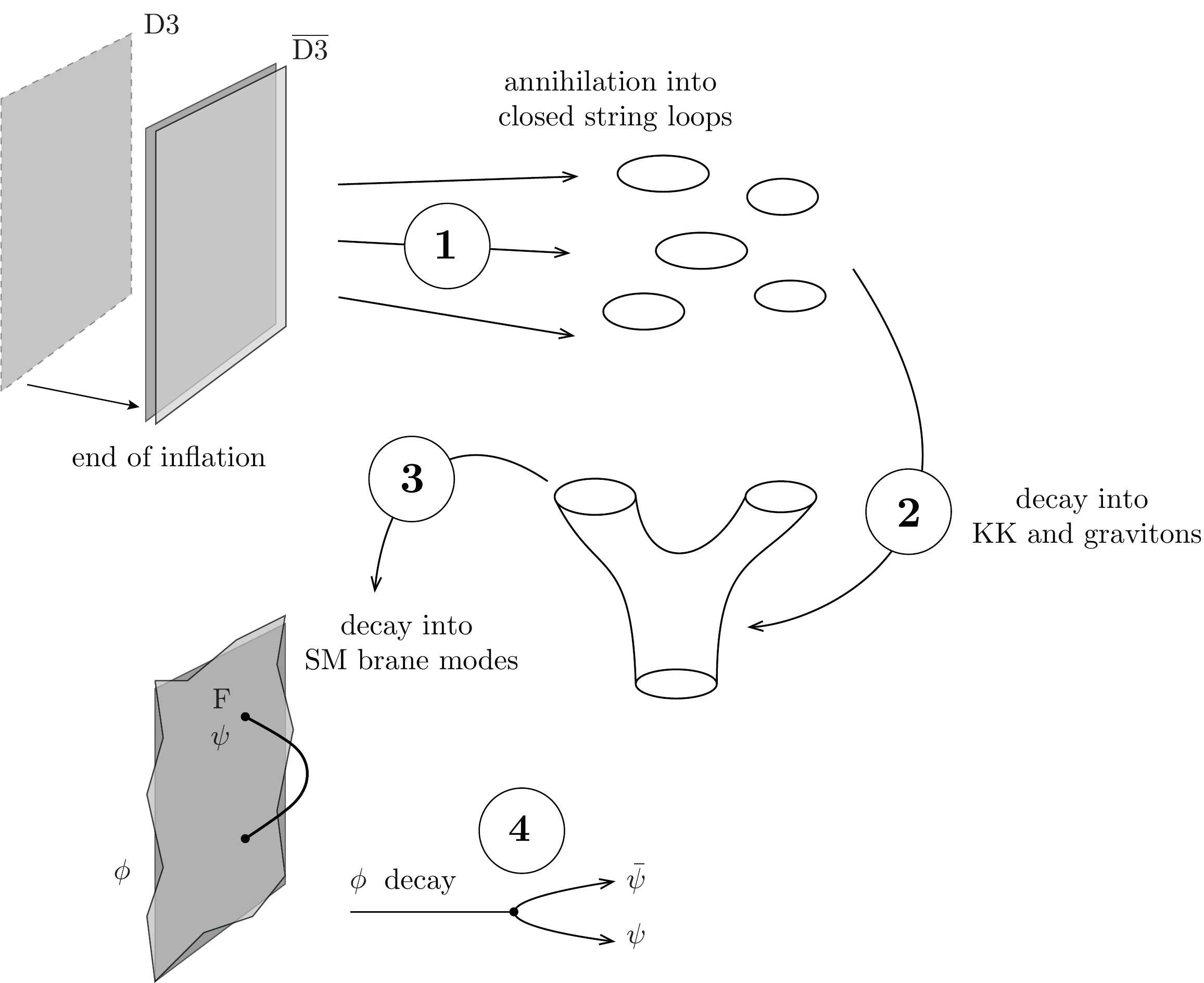}
   \caption{The stages of reheating after warped D-brane inflation (figure adapted from \cite{Kofman:2005yz}).}
  \label{fig:reheating}
\end{figure}

\begin{itemize}

\item[$\triangleright$]  {\it Overproduction of gravitons.}---KK modes decay to four-dimensional gravitons with a rate set by $M_{\rm pl}$.
If no other channels extract energy  more quickly from the inflationary throat,  the universe will be dominated by gravitational radiation, ruining  Big Bang nucleosynthesis.
Tunneling  can transfer energy to other throats, but because generically $\tau_{\rm{graviton}} \ll \tau_{\rm{tunnel}}$ (cf.~\cite{Barnaby:2004gg}), additional mechanisms may be needed to dilute  the graviton abundance.

The heaviest  Kaluza-Klein modes have the wavefunctions that reach  farthest into the ultraviolet, and so have the largest tunnelling probability. Efficient tunnelling therefore requires that
the lifetime $\tau_{\rm KK}$ of the heaviest Kaluza-Klein modes obeys $\tau_{\rm KK} \gtrsim \tau_{\rm{tunnel}}$.  This presents a further constraint on the parameters \cite{Barnaby:2004gg}.

\item[$\triangleright$]  {\it Kaluza-Klein relics from angular isometries.}---Suppose that one of the throats  in the compactification enjoys approximate angular isometries,  such as the $SU(2)\times SU(2)$  isometry of the Klebanov-Strassler  solution.   The associated angular momentum is  approximately conserved, and Kaluza-Klein modes  carrying  this charge can only decay\footnote{Annihilation  can in principle reduce the  relic density \cite{Chen:2006ni}, but only  for problematically small values of the warp factor \cite{Berndsen:2008my}.} through symmetry-violating interactions.  Charged Kaluza-Klein modes  produced  during  reheating  will be long-lived Kaluza-Klein relics \cite{Kofman:2005yz}, and can readily overclose the universe.

    To  determine whether Kaluza-Klein relics decay sufficiently quickly  for successful cosmology,  one can
  examine the isometry-breaking perturbations sourced by the compactification \cite{Aharony:2005ez,Kofman:2005yz,Berndsen:2008my}, as explained in \S\ref{sss:D310d}.  A detailed analysis of this issue appeared in \cite{Berndsen:2008my}, where it was concluded that
       for the relics to decay before nucleosynthesis, irrelevant\footnote{The restriction to  irrelevant perturbations in \cite{Berndsen:2008my} rests on the requirement that the background throat solution  is a good approximation in the infrared.  However, \cite{Baumann:2010sx,Gandhi:2011id}  showed that certain {\it relevant} perturbations are necessarily present in Klebanov-Strassler regions of KKLT compactifications.   Approximate no-scale symmetry ensures that these perturbations have exponentially small coefficients and do not destroy the throat.   The effects of such perturbations on Kaluza-Klein relic decays have not been assessed.} perturbations that  break supersymmetry at a sufficiently high scale  must be introduced.
The conclusion obtained in \cite{Chen:2006ni} is more positive: the effects of warping and of  the compact bulk  were argued to lead to a much smaller relic abundance  than that found in \cite{Kofman:2005yz,Berndsen:2008my}.

\item[$\triangleright$]  {\it Excitation of other sectors.}---Some scenarios consider an additional `intermediate' throat (for example, where supersymmetry is broken)  whose warp factor falls between those of the inflationary and visible-sector throats. In that case, tunneling leads to Kaluza-Klein  excitations of this throat.  These are only very slowly depopulated by transfer to the
visible-sector throat, presenting a serious problem \cite{Kofman:2005yz}.  More generally, if light moduli  associated with other sectors become populated, these can come to dominate the energy density of the universe,  with consequences discussed in e.g.~\cite{Cicoli:2012aq,Higaki:2012ar,Conlon:2013isa,Higaki:2013lra,Conlon:2013txa,Higaki:2013qka}.

\item[$\triangleright$]  {\it Reheating above the local string scale.}---The reheating temperature can exceed the warped string scale $e^{A_{\mathsmaller{\rm IR}}}/\sqrt{\alpha^{\prime}}$ in a throat that is much more strongly warped than the inflationary throat ---  for example, if the  electroweak hierarchy is addressed by warping  of the visible-sector throat.
Reheating can then induce copious production of excited strings in the strongly
warped throat \cite{Frey:2005jk}.   Analyzing this process in detail remains challenging.
\end{itemize}

\subsection{Fine-Tuning}

Considerable effort has been directed at finding  mechanisms that can alleviate the fine-tuning of the potential in  warped D-brane inflation --- see \cite{Silverstein:2003hf,Firouzjahi:2003zy,Iizuka:2004ct,Cline:2005ty,Hoi:2008gc,Cline:2009pu,Easson:2009kk}.   Here, we will  outline a few of the leading  approaches.
The {\it{DBI mechanism}}, which  turns a steep potential from a liability into an asset, will be discussed in \S\ref{ssec:DBI}.  Discrete symmetries can be used to  forbid  problematic mass terms \cite{Iizuka:2004ct}, and  in some cases have been shown to be compatible with moduli stabilization \cite{Kachru:2009kg}.
Dynamical mechanisms have also been found: it  was shown in \cite{Cline:2005ty, Cline:2009pu} that if $N$ D3-branes become trapped in a metastable minimum of the potential  in the throat, and sequentially tunnel out,  the  barrier diminishes with each tunneling event.   For  favorable parameter values, the  potential for the final D3-brane is an  inflationary inflection point.

The second fine-tuning problem of warped D-brane inflation --- indeed,  of most  scenarios for inflation in string theory  ---  is that  rather special initial conditions are required for successful inflation  to occur.  When the potential is approximately flat in a small fraction of the field space, and is steep elsewhere, then generic trajectories passing through the  would-be inflationary region  will {\it{overshoot}}  the flat portion without initiating an inflationary phase, as  emphasized long ago in~\cite{Brustein:1992nk}.
The DBI kinetic term (see \S\ref{ssec:DBI}) has been argued to ameliorate the overshoot problem \cite{Underwood:2008dh}, though this conclusion was challenged  by \cite{Bird:2009pq}.
Negative spatial curvature  resulting from tunneling entirely removes overshooting in certain classes of potential, and reduces its severity in general \cite{Freivogel:2005vv, Dutta:2011fe}.
Finally,  it was argued in \cite{Itzhaki:2007nk}  that the overshooting of an inflection point is ameliorated by particle production near points in  field space where new species become light  \cite{Kofman:2004yc,Watson:2004aq,Greene:2007sa,Battefeld:2010sw}.

A different perspective on overshooting was  given in \cite{Agarwal:2011wm}, in which inflationary solutions were found by Monte Carlo sampling of the ensemble of potentials obtained in \cite{Baumann:2010sx},  followed by numerical solution of the six-field equations of motion.  In  this setting, the potential was fine tuned by chance, rather than by hand.   Surprisingly, the overshoot problem was absent: for each  inflationary trajectory that was found for a given potential $V$
and for some fixed initial conditions, an ${\cal O}(1)$ fraction of the space of possible initial  positions likewise led to prolonged inflation.   Thus, while inflation was not a generic outcome in the  joint space of Lagrangians and initial conditions,  for each successful Lagrangian that was found, inflation  occurred for generic  initial  positions\footnote{The initial kinetic energy  of the D-brane  was required to be somewhat smaller than the initial potential  energy.} of the D3-brane.

\subsection{Phenomenology} \label{warpedphenomenology}

The phenomenology of warped D-brane inflation has been the subject of intense investigation (e.g.~\cite{Shandera:2006ax, Baumann:2007ah, Krause:2007jk, Panda:2007ie,Peiris:2007gz, Chen:2008ai, Agarwal:2011wm, McAllister:2012am, Dias:2012nf}).  In this section, we will summarize some of the main conclusions.\footnote{We will emphasize  the original scenario \cite{Kachru:2003sx}  in which a D3-brane falls  toward the tip  of a Klebanov-Strassler throat.   Scenarios involving D-branes  moving  on the $S^{3}$ at the tip  of the throat include \cite{DeWolfe:2004qx,Pajer:2008uy}.}
We will start with the simplified single-field treatment~\cite{Baumann:2007ah, Krause:2007jk} in which the angular degrees of freedom are integrated out using an adiabatic approximation.
This is not always consistent, as the angular fields can have masses that are smaller than the inflationary Hubble scale,  but serves to develop intuition for the more complex multi-field dynamics studied in~\cite{Panda:2007ie,Chen:2008ada,Chen:2010qz,Agarwal:2011wm, McAllister:2012am, Dias:2012nf}.
We will then present the multi-field results of~\cite{Agarwal:2011wm, McAllister:2012am, Dias:2012nf}, which incorporate the  complete potential  derived in \cite{Baumann:2010sx},  and follow the full six-field\footnote{As explained above, light K\"ahler moduli may also  evolve during inflation.} dynamics numerically, making no approximation.

\subsection*{Single-Field Expectations}

In \cite{Baumann:2007ah, Krause:2007jk}, the six-dimensional field space was analyzed analytically. The potential was minimized in the angular
directions and an effective potential for the radial direction was determined.
As expected, the potential for the effective radial coordinate has an inflection point.
Near the inflection point, we can write the potential as
\beq
V(\phi) \approx  V_0 \left[ 1 + \lambda_0 \frac{\phi}{M_{\rm pl}} + \frac{1}{3!}\mu_0 \frac{\phi^3}{M_{\rm pl}^3} + \cdots \right] \ , \label{equ:Inflection}
\eeq
where the constants $V_0$, $\lambda_0$ and $\mu_0$ can be related to microscopic parameters of the model~\cite{Baumann:2007ah}.
A slow-roll analysis of this potential leads to the following predictions~\cite{Baumann:2007ah}:
\begin{itemize}
\item[$\triangleright$]  {\it Power spectrum.}--- The spectral index derived from (\ref{equ:Inflection}) has
the analytic solution~\cite{BuenoSanchez:2006xk,  Baumann:2007ah}
\beq
n_s - 1 \approx - \frac{4\pi}{N_{\rm tot}} \cot \left(\pi  \frac{N_\star}{N_{\rm tot}}\, \right) \approx - \frac{4}{N_\star} \left( 1 + {\cal O}\left( \frac{N_\star^2}{N_{\rm tot}^2}\right) \right) \ , \label{equ:nsBrane}
\eeq
where $N_\star$ corresponds to the number of $e$-folds between the horizon exit of the pivot scale and the end of inflation and $N_{\rm tot}$ denotes the total number of $e$-folds, defined as
\beq
N_{\rm tot} = \int_{-\infty}^\infty \frac{1}{\sqrt{2\epsilon}} \frac{\d \phi}{M_{\rm pl}} = \pi \sqrt{\frac{2}{\lambda_0 \lambda_1}} \ .
\eeq
The number of $e$-folds from some initial vev $\phi$  until the end of inflation at $\phi_{\rm{end}}$  is
 \beq
N_{e}(\phi) = \int_{\phi_{\rm{end}}}^\phi \frac{1}{\sqrt{2\epsilon}} \frac{\d \phi}{M_{\rm pl}} = \frac{N_{\rm tot}}{\pi}{\rm{arctan}}\Biggl(\frac{\eta(\phi)N_{\rm tot}}{2\pi}\Biggr)\Biggr|_{\phi_{\rm{end}}}^\phi \ .
\eeq
For $N_{\rm tot}$ not much greater than $N_\star \approx 60$ the spectrum is strongly blue and the model is hence ruled out by observations (see fig.~\ref{fig:ns-brane}). For $N_{\rm tot} \approx 2 N_\star$,
the spectrum on CMB scales is exactly scale-invariant, while for $N_{\rm tot} > 2 N_\star$, the spectrum is red and asymptotes to the lower limit $n_s \to 1 - 4/N_\star \approx 0.93$ for $N_{\rm tot} \gg 2N_\star$.
\begin{figure}[h!]
   \centering
     \includegraphics[width=0.75\textwidth]{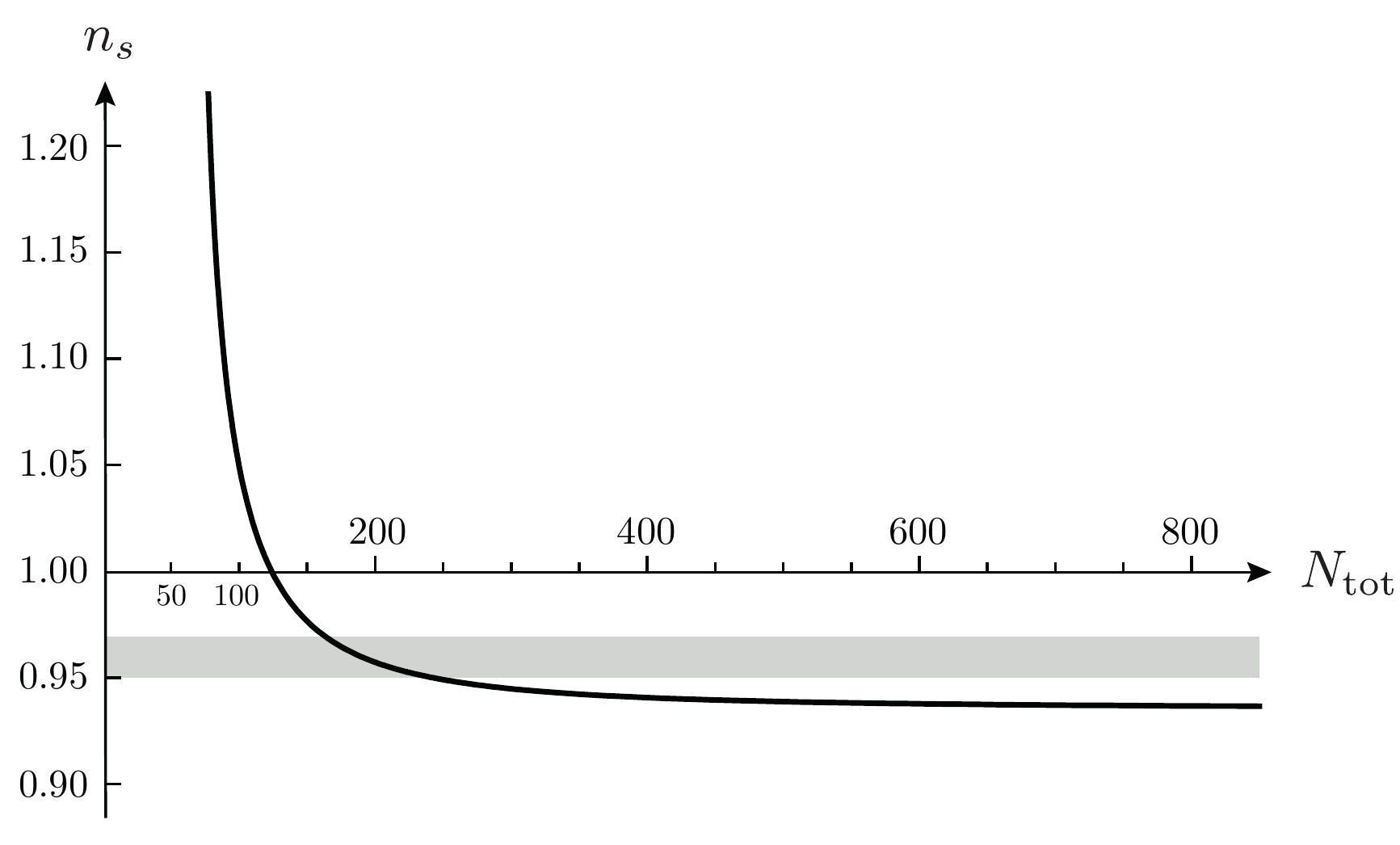}
   \caption{Prediction for $n_s$ as a function of the total number of $e$-folds. The gray band shows the range of $n_s$ allowed by Planck.}
  \label{fig:ns-brane}
\end{figure}

The running of the spectral index follows from (\ref{equ:nsBrane}),
\beq
\alpha_s = - \frac{4\pi^2}{N_{\rm tot}^2} \sin^{-2}  \left(\pi  \frac{N_\star}{N_{\rm tot}}\, \right)  \approx - \frac{4}{N_\star^2} \left( 1 + {\cal O}\left( \frac{N_\star^2}{N_{\rm tot}^2}\right) \right) \ .
\eeq
Notice that both the tilt $n_s$ and the running $\alpha_s$ are determined by $N_{\rm tot}$ alone (for fixed $N_\star$).

\item[$\triangleright$]  {\it Absence of tensors.}---Combining the geometric bound (\ref{equ:BMbound}) on the inflaton field range\footnote{See \cite{Chen:2006hs}  for another discussion of geometric constraints in warped D-brane inflation, with implications for eternal inflation.} with the Lyth bound (\ref{equ:LythBound}), we find~\cite{Baumann:2006cd}
\beq
r < \frac{4}{N} \times 0.01 \ \ll\ 0.01\ .
\eeq
This is a conservative bound
that assumes that inflation occurs over the entire length of the throat, and that the bulk makes a negligible contribution to the total compactification volume.
In all known examples, inflation is confined to a small part of the throat (the part where the potential is tuned to be flat) and the tensor amplitude is much smaller than the maximum allowed by the geometric bound.  This implies that gravitational waves are unobservable in warped brane inflation.\footnote{If inflation is driven by the motion of a D$p$-brane  wrapping a $(p-3)$-cycle, the field range can be larger than that for a D3-brane \cite{Kobayashi:2007hm,Becker:2007ui}.   However, arranging for a nearly-flat potential is challenging, and backreaction of the moving brane can be important.}
\end{itemize}

\subsection*{Multi-Field Effects}

As explained in \S\ref{sec:MD}, a proper description of warped D-brane inflation  involves all six D3-brane coordinates (and  ultimately any light K\"ahler moduli).
Here, we summarize a few key phenomenological results that emerge from an  intensive Monte Carlo  investigation of the  dynamics  and signatures of the six-field effective theory~\cite{Agarwal:2011wm, McAllister:2012am, Dias:2012nf}.  A few words about the methodology are necessary: in \cite{Agarwal:2011wm, McAllister:2012am,Dias:2012nf}, scalar potentials  were drawn at random from the ensemble  described in \S\ref{sec:MD}, and the equations of motion were solved numerically beginning from a random initial condition.  The cosmological signatures were then evaluated in the  subset of trials that led to $N_e \ge 60$ $e$-folds of inflation.

\begin{itemize}
\item[$\triangleright$]  {\it Inflationary probabilities.}---First,  one  can  compute the relative probability $P(N_e)$ of $N_e$  $e$-folds  of inflation in the ensemble.
In \cite{Agarwal:2011wm}, it was shown that
\begin{equation}
P(N_e) = P(N_{\star}) \left(\frac{N_{\star}}{N_e}\right)^3\ , \label{pne}
\end{equation}  where $N_{\star} \gtrsim 10$ is a reference value encoding the absolute probability.
Thus, the probability of $N_e$  $e$-folds  of inflation is proportional to $1/N_e^3$.
This  result  can be derived analytically in an inflection point model \cite{Agarwal:2011wm} (see \cite{Freivogel:2005vv} for earlier work in a slightly different model), and is consistent with the simpler analytic arguments of \S\ref{sss:D310d}, which focused on the appearance of the term $\phi^{3/2}$  in an effective single-field description.  Of course, the  total number of $e$-folds  is not itself an observable, but whether  or not $N_e \gg 60$  strongly influences the likelihood of observing relics of a pre-inflationary stage, such as traces of  bubble collisions~\cite{Chang:2007eq, Chang:2008gj, Feeney:2010dd, Feeney:2010jj, Osborne:2013hea}.

\item[$\triangleright$]  {\it Violations of slow roll.}---A useful measure of violations of the slow-roll approximation is the ratio $m_{\sigma}^2/H^2$, where $m_{\sigma}$ is the mass of fluctuations  in the {\it{adiabatic}} direction (see
Appendix~C
for a precise definition and further discussion).   Slow-roll violations are strongly correlated with the total number of $e$-folds of inflation: realizations with $N_e \gg 100$ have $m_{\sigma}^2 \approx -0.1 \hskip 1pt H^2$  at the moment  when the CMB exits the horizon,  agreeing with  the  analytic result for  single-field inflection point inflation.  However, realizations with $N_e \approx 60$ have $m_{\sigma}^2 \approx H^2$, so that the slow-roll approximation  is marginally valid at best.  The  effect on the spectrum  is a slight  increase in $n_s$  compared to the slow-roll result \cite{McAllister:2012am}, see fig.~\ref{multifieldobservables1}.

\begin{figure}[h!]
   \centering
     \includegraphics[width=0.75\textwidth]{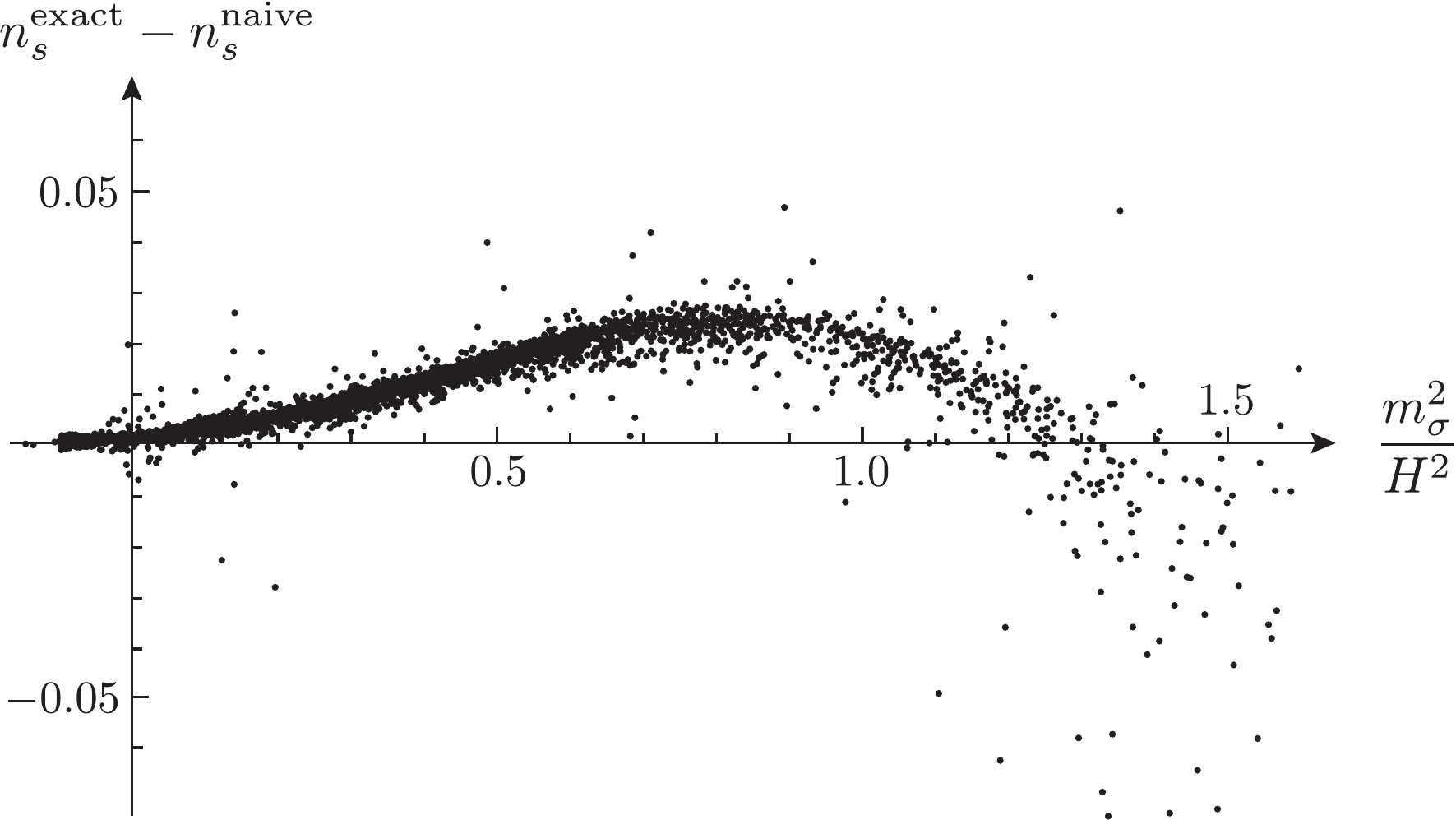}
 \hspace{0.3cm}  \caption{Multi-field effects on the spectral index in warped D-brane inflation, versus the mass $m_{\sigma}$ of the adiabatic fluctuation (figure adapted from~\cite{McAllister:2012am}). The exact tilt, $n_{s}^{\rm exact}$, is the result of a six-field numerical calculation making no slow-roll approximation, while the naive tilt, $n_{s}^{\rm naive}$, follows from simply evaluating (\ref{naivetilt}) at horizon exit.  }
  \label{multifieldobservables1}
\end{figure}

\item[$\triangleright$]  {\it Bending of the trajectory.}--- Characteristic trajectories  leading to prolonged inflation begin  by spiraling in the angular directions, and then  settle  down to an inflection point that is  approximately parallel  to the radial direction.    As a result,  multi-field effects are generically significant during the first 5 -- 10  $e$-folds of inflation, but are  subsequently exponentially suppressed.  See
Appendix~C
for background on multi-field effects  from bending trajectories.

\item[$\triangleright$]  {\it Decay  of entropic perturbations.}---Although all  six open string scalars have masses  that are very roughly ${\cal{O}}(H)$,  the precise distribution of masses  is important.   By directly evaluating the Hessian matrix, or using the matrix model given in \S\ref{sss:D310d}, one can show that  in nearly all realizations  the  lightest field  is tachyonic,  the second-lightest field has $m^2 \sim H^2$,  and the four remaining fields  have $m^2 > \frac{9}{4}H^2$. Thus,  there is at most one instability,  and only two fields fluctuate.   Moreover, the five entropic perturbations decay exponentially after exiting the horizon \cite{McAllister:2012am}: i.e.~an `adiabatic limit' \cite{Elliston:2011dr} is reached.
    This is important, for if one or more entropic perturbations  were to persist until the time of reheating, predicting the scalar power spectrum  would become extremely difficult \cite{Elliston:2011dr}.  A detailed treatment of related multi-field effects at the end of D-brane inflation  appears in \cite{Chen:2008ada}.

\item[$\triangleright$]  {\it Scalar power spectrum.}---In models producing $N_e \lesssim 60$ -- $70$ $e$-folds in total, multi-field effects dictate the observable anisotropies,  while in models yielding $N_e \gg 70$ $e$-folds,  a single-field approximation is  valid and the analytic treatment  given above  applies without  modification.  In light of (\ref{pne}), models with multi-field effects are much more common than approximately single-field models, within the class of all realizations  yielding $N_e \ge 60$ $e$-folds.  However --- see fig.~\ref{fig:ns-brane}  and the related discussion ---  the scalar power spectrum  computed {\it{in the single-field approximation}}  is  unacceptably blue in models producing $N_e \lesssim 120$ $e$-folds.  Multi-field effects quite generally shift the spectrum toward the red, i.e.~$n_{s}^{\rm{exact}}-n_{s}^{\rm{naive}}<0$, but the magnitude of the effect is only occasionally large enough to produce models consistent with observations,  which fall in the gray band in fig.~\ref{multifieldobservables2}.

\begin{figure}[h!]
   \centering
        \includegraphics[width=0.8\textwidth]{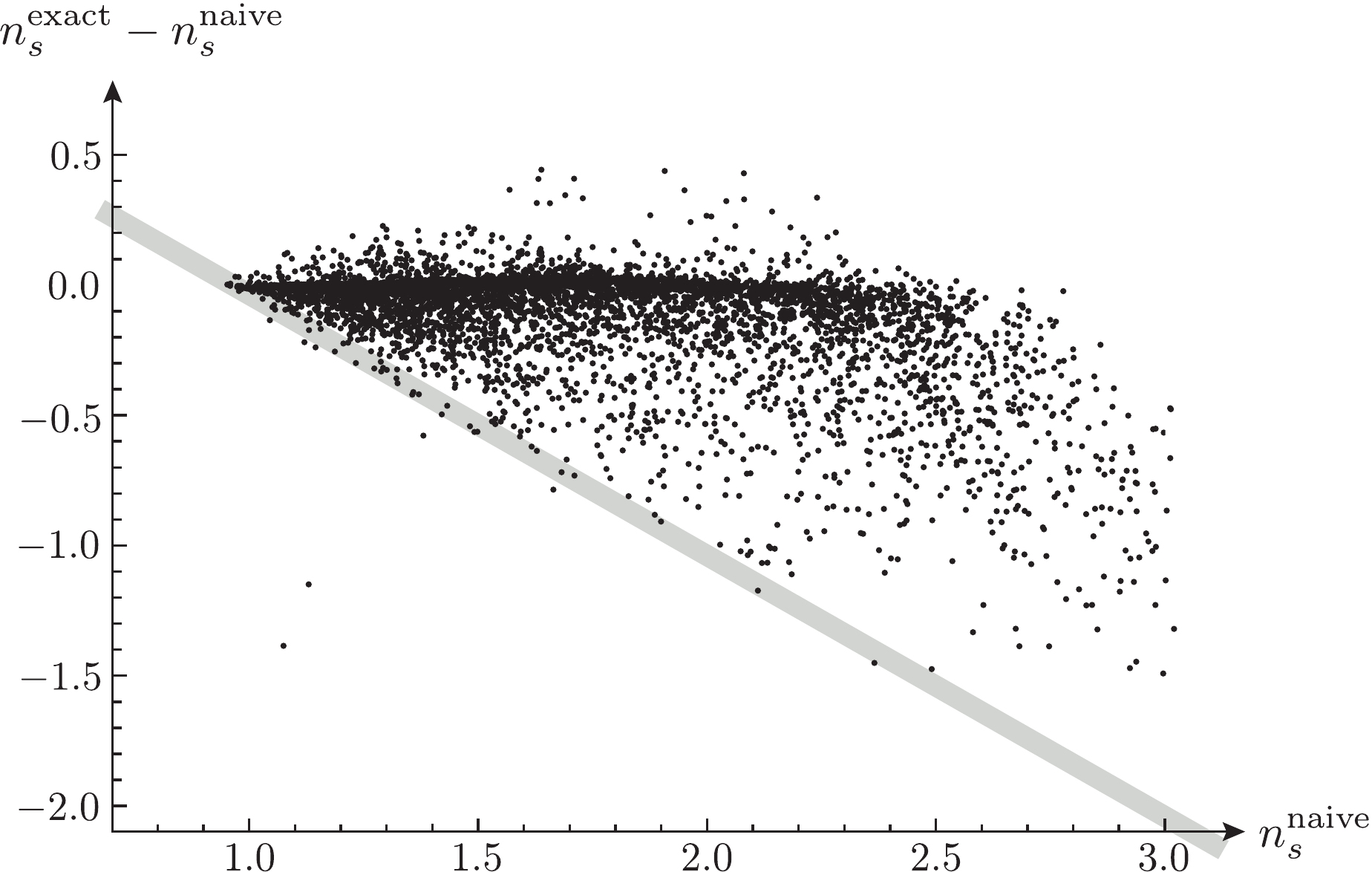}
   \caption{The spectral index in realizations of warped D-brane inflation with significant multi-field effects  (figure adapted from~\cite{McAllister:2012am}).  The gray band shows the region allowed at 2$\sigma$ by WMAP7; the Planck constraints are slightly more stringent.}
  \label{multifieldobservables2}
\end{figure}

\item[$\triangleright$]  {\it Tensor amplitude.}---The  inflationary inflection points  arising in the ensemble are  extremely small in Planck units: for the parameters explored, $r \lesssim 10^{-12}$,  which is far below the upper limit allowed by the Lyth bound (\ref{equ:LythBound}) combined with the geometric bound~(\ref{equ:BMbound}).
\item[$\triangleright$]  {\it Non-Gaussianity.}---Although conversion of entropic perturbations to curvature perturbations is commonplace in models yielding $N_e \lesssim$ 60 -- 70 $e$-folds, this does not  automatically lead to large non-Gauss\-ian\-ity, because the cubic couplings in the D3-brane potential  can be quite small~\cite{McAllister:2012am}.  More importantly, in the subset of models allowed by constraints on the tilt,  multi-field effects, including non-Gaussianity,  are extremely  rare.\footnote{Sharp features in the radial profile of the warp factor were argued in \cite{Bean:2008na}  to produce observable signatures in the power spectrum and bispectrum.}
\end{itemize}

\subsection*{Cosmic Strings}

Cosmic superstrings  are one of the most striking signatures of D-brane inflation.
Following \cite{Polchinski:2004ia}, we recall  the conditions for cosmic strings to be cosmologically relevant: the strings must be produced after inflation,  remain stable over cosmological times,  and be observable  without already being excluded. Finally, one  may also hope that the strings have distinctive signatures revealing their origin in string theory.
All four conditions  can be met  in warped D-brane inflation, as we now explain.

\vskip 4pt
Condensation of the D3-brane/anti-D3-brane  tachyon at the end of inflation automatically produces a population of cosmic F-strings and D-strings, as well as the more general $(p,q)$  string bound states.  Whether these strings are stable  depends on whether there are D-branes in the warped throat where inflation occurs --- see \cite{Copeland:2003bj}  for a detailed treatment.
First of all,  $(p,q)$  strings  (including  the $(1,0)$ F-string and  $(0,1)$ D-string) are not BPS  in this setting: the two-forms $B_{\mu\nu}$ and $C_{\mu\nu}$  whose charges the strings carry are projected out  by the orientifold action \cite{Copeland:2003bj}.  Correspondingly, a string can  break apart by coming into contact with its orientifold image.   However,  in the generic situation in which there are no orientifold  fixed planes within the throat itself, a string  has to fluctuate out of the throat to meet its image in the image throat.   This is an  exponentially slow process,  as the  potential due to the warp factor confines the strings to the bottom of  their respective throats, and for practical purposes breakage via the orientifold image can be ignored \cite{Copeland:2003bj}.  A more significant  risk comes from D3-branes or anti-D3-branes  in the inflationary throat,  which could serve as the substrate for the  Standard Model \cite{Burgess:2004kv}
or as a source of supersymmetry-breaking  energy \cite{Kachru:2002gs}.   If any D3-branes or anti-D3-branes  are present, cosmic strings fragment  immediately and are cosmologically irrelevant.  If D7-branes  are present but D3-branes  and anti-D3-branes  are not,  the D-string  remains stable \cite{Copeland:2003bj}.

The spectrum of  tensions of $(p,q)$  strings in a warped throat was obtained in \cite{Firouzjahi:2006vp}:
\begin{equation}
T_{(p,q)} \approx \frac{e^{2A_{\mathsmaller{\rm IR}}}}{2\pi\alpha^{\prime}}\sqrt{\frac{q^2}{g_{\rm s}^2} +\left(\frac{b M}{\pi}\right)^2{\rm{sin}}^2 \left(\frac{\pi(p-q C_0)}{M}\right)} \ ,
\end{equation} where $e^{A_{\mathsmaller{\rm IR}}}$ is the warp factor at the tip of the throat, $M$  is the  flux  on the $A$-cycle, and $b \approx 0.93$  is a constant arising in the Klebanov-Strassler solution.
This result is primarily governed by the warp factor, which can be exponentially small. Hence, if the warp factor were a free parameter, it would be easy to  ensure that the cosmic string tension is low enough to satisfy any  conceivable observational bound.   However, the warp factor in the inflationary throat  determines the scale of the inflaton potential, and is  therefore constrained  by the
normalization of the scalar fluctuations.  Recalling from (\ref{naivespectrum})  that the  amplitude of the scalar power spectrum involves both $V$  and $\epsilon$,   we conclude that once $\epsilon$ is known,  the warp factor and hence the cosmic string tension are predicted.  The distribution  of values of $\epsilon$ in a simple model  for the potential was studied in \cite{Firouzjahi:2005dh}.  Tensions  that  satisfy present constraints but can be detected in the coming generation of observations  are achievable, but the associated fine-tuning has not yet been quantified completely.

\section{Inflating with Unwarped Branes}
\label{ssec:brane2}

We have just seen, in \S\ref{sec:dbrane}, that D3-branes in warped throat regions  of type IIB flux compactifications
lead to a class of highly computable inflationary models with rich phenomenology.  At the same time, D-branes  in more general geometries --- in which warping may be present but is not a dominant effect --- provide an array of interesting models  with some theoretical advantages.  In this section, we will discuss a few examples of  inflation driven by branes in unwarped regions.

\subsection{D3/D7 Inflation}   \label{d3d7section}

An interesting and uniquely explicit scenario for D-brane inflation in an unwarped compactification is the D3/D7 model \cite{Dasgupta:2002ew,Hsu:2003cy,Dasgupta:2004dw,Hsu:2004hi,Haack:2008yb,Burgess:2008ir}.   This model has close parallels to the warped brane inflation scenario  detailed in the previous section,  so we will be brief, emphasizing the  distinctive features of the D3/D7  construction.

\vskip 4pt
The background geometry is a compactification of type IIB string theory on the orientifold $K3 \times T^2/\mathbb{Z}_2$.  At each of the four fixed points there are four D7-branes atop an O7-plane, all of which wrap the $K3$ manifold. This configuration corresponds to M-theory on $K3 \times T^4/\mathbb{Z}_2$, with the $T^4/\mathbb{Z}_2$ being the orbifold limit of $K3$.  Displacing the D7-branes from the orientifold planes leads to a geometry that lifts to M-theory on $K3 \times K3$ (see \cite{Aspinwall:2005ad} for an analysis of moduli stabilization in this regime).

\begin{figure}[h!]
   \centering
     \includegraphics[scale=0.5]{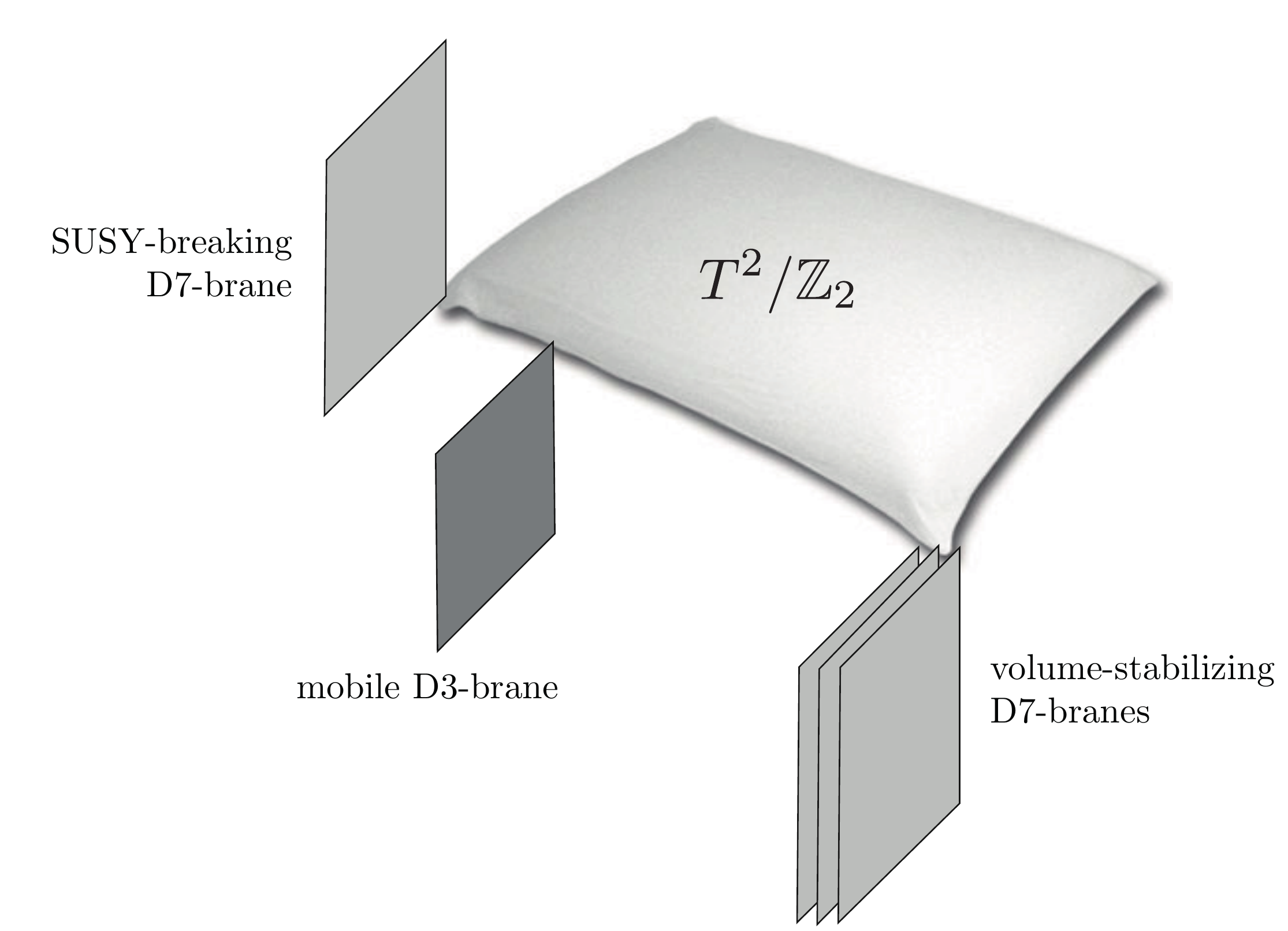}
   \caption{Schematic of D3/D7 inflation (figure adapted from~\cite{Haack:2008yb}).}
  \label{fig:D3D7}
\end{figure}

Now we add a spacetime-filling D3-brane, which sits at a point in the internal space, and in particular on $T^2/\mathbb{Z}_2$ (see fig.~\ref{fig:D3D7}).
The position of the D3-brane on $T^2/\mathbb{Z}_2$, relative to the stack of D7-branes, was proposed to be the inflaton~\cite{Dasgupta:2002ew}.  The inflaton sector therefore consists of two real fields describing the D3-brane location on the torus.

A spacetime-filling D3-brane in the $K3 \times T^2/\mathbb{Z}_2$ orientifold actually preserves ${\cal N}=2$ supersymmetry in four dimensions, so there is no potential for D3-brane motion, and the would-be inflaton is perfectly massless.  However, introducing two-form flux ${\cal F}_2$ in the D7-brane worldvolume modifies the situation: when the flux is not self-dual in the worldvolume (i.e.~$\star_{4}{\cal F}_2\neq {\cal F}_2$, with $\star_4$ the Hodge star in the four compact directions of the D7-brane worldvolume), then supersymmetry is broken, and the D3-brane feels a force.  As explained in \cite{Dasgupta:2002ew}, the worldvolume flux corresponds to a field-dependent Fayet-Iliopoulos term\footnote{See \cite{Binetruy:2004hh,Komargodski:2009pc,Dienes:2009td} for analyses of consistency conditions for Fayet-Iliopoulos terms in supergravity, and \cite{Gwyn:2011tf} for the implications for cosmic strings.} $\xi$, so the D3/D7 model described so far is a model of D-term inflation.\footnote{An important alternative  means of breaking supersymmetry  is the addition of an anti-D3-brane in a warped region --- see \cite{Burgess:2008ir}  for a comprehensive discussion of the D3/D7  model with antibrane supersymmetry breaking.}   Specifically, D-term supersymmetry breaking by D7-brane fluxes introduces mass splittings in the supermultiplets of strings stretched between  the D3-brane and the D7-brane, and integrating out these fields leads to the Coleman-Weinberg potential (cf.~\cite{Haack:2008yb} for an updated  discussion)
\begin{equation}
V_D(\phi)= \frac{g^2\xi^2}{2}\Bigl(1+ \frac{g^2}{16\pi^2}U(x)\Bigr) \ ,  \label{VDphi}
\end{equation} where  we have defined $x\equiv \phi/\sqrt{\xi}$,  and
\begin{equation}
U(x) \equiv  (x^2+1)^2 \ln(x^2+1)+(x^2-1)^2\ln(x^2-1)-4x^4 \ln(x)-4 \ln(2)\ ,
\end{equation}  with $g$  the coupling of the $U(1)$  gauge field.

\vskip 4pt
At this stage we have specified the geometric data of a non-super\-symmetric compactification, but without further ingredients this configuration will be unstable to decompactification: the ${\cal N}=2$ supersymmetric compactification has unfixed moduli, and supersymmetry-breaking positive energy from the D3-brane potential creates an instability.
Fixing all the K\"ahler moduli of the supersymmetric compactification may be achievable \cite{Aspinwall:2005ad}, but we will first focus on a scenario in which a single overall volume modulus is stabilized by gaugino condensation in the gauge theory living on the D7-branes.

As explained in detail in \S\ref{sec:D34d}, mixing of D3-brane position moduli with K\"ahler moduli, via the DeWolfe-Giddings K\"ahler potential (\ref{equ:DeWolfeG}), implies that stabilization of the volume generically leads to stabilization of the D3-brane position.\footnote{A very different perspective on this fact was recently given in \cite{Komargodski:2010rb}.}  We will now examine this crucial point more closely and determine whether there are non-generic exceptions.

The K\"ahler potential takes the form
\beq
K(Z^I, \bar Z^I) = - 3 \ln\left[T+ \bar T - \gamma k(z_\alpha, \bar z_\alpha) \right]\ ,
\eeq
with $k(z_\alpha, \bar z_\alpha)$ the K\"ahler potential for the metric on the internal space.  Let us restrict attention to a single complex field $z_1 \equiv x + i y$,
corresponding to the D3-brane position on $T^2/\mathbb{Z}_2$.  Suppose for the moment that the superpotential $W$ is independent of $z_1$, so that the only dependence of the potential energy on $x$ and $y$ comes through the appearance of these fields in $K$.  If $k(x,y)$ depends non-trivially on both $x$ and $y$, then both real fields will have non-trivial masses in the stabilized vacuum, rather than corresponding to flat directions.

An influential proposal is to invoke a shift symmetry in the K\"ahler potential \cite{Kawasaki:2000yn}, so that $k$ --- and hence $K$ --- is independent of one of the fields.
For example, if
\beq
k=\frac{1}{2}(z_1 + \bar z_1)^2 = x^2\ ,
\eeq
then $y$ corresponds to a flat direction of the  F-term potential  for the moduli.   (The dependence of the D-term potential (\ref{VDphi}) on $z_1$ is mild enough to be suitable for inflation.)  Although this approach appears reasonable in supergravity, in a string construction one is not free to write down an effective action with a desired form: the action follows from dimensional reduction of a specified configuration.  Moreover, some shift symmetries do not admit ultraviolet completions.  It is therefore essential to determine whether the D3/D7 model actually enjoys a shift symmetry that allows K\"ahler modulus stabilization to coexist with a flat direction for D3-brane motion.

It was shown in \cite{Hsu:2003cy,Hsu:2004hi} that the tree-level K\"ahler potential is indeed shift-symmetric, so that before accounting for additional terms in the effective action, the D3-brane potential takes the form of a nearly flat trough oriented along the symmetry direction.  However, the nonperturbative superpotential, which is critical in the stabilization of the volume, necessarily depends on the D3-brane position, contrary to our assumption above.  The one-loop correction to the gauge kinetic function for the D7-brane gauge theory was computed explicitly in \cite{Berg:2004ek}, and was found to depend on the D3-brane position, so that the gaugino condensate superpotential likewise depends on the D3-brane location.
For D7-branes  with gauge group $SU(N_c)$ at position $z_{D7}=\mu$  in $T^2/\mathbb{Z}_2$, one finds~\cite{Berg:2004ek}
\begin{equation}
W=W_0 + \Bigl[\vartheta_1\bigl(\sqrt{2\pi}(z_1+\mu),\zeta\bigr)\, \vartheta_1\bigl(\sqrt{2\pi}(z_1-\mu),\zeta\bigr)\Bigr]^{-1/N_c}e^{-2\pi T/N_c}\ ,  \label{BHKW}
\end{equation}
where $\vartheta_1$ is a Jacobi theta function, and $\zeta$ is the complex structure of the $T^2$~\cite{Berg:2004ek}.  It was then shown in \cite{Berg:2004sj,McAllister:2005mq} that the appearance of the D3-brane position in the nonperturbative superpotential (\ref{BHKW}) spoils the shift symmetry and prevents inflation from occurring naturally.

The gauge theory description of this effect is simple and instructive: strings stretching from the D3-brane to the D7-branes (`3-7 strings')
correspond to flavors in the condensing theory, and their masses depend on the D3-brane's separation from the D7-branes.  The dependence of the low-energy condensate on the mass of the flavors implies that the superpotential depends on the D3-brane position.
As a simple example, consider ${\cal N}=1$ supersymmetric Yang-Mills theory with gauge group  $SU(N_c)$ (for $N_c >2$) and a single flavor $Q$ with mass parameter $m$.
The gaugino condensate superpotential below the scale $m$, which results from integrating out $Q$, takes the form
\begin{equation}
W = \Lambda^{3-1/N_c} \hskip 1pt m^{1/N_c}\ ,
\end{equation} where $\Lambda$ is the dynamical scale of the high-energy theory.
In a string theory realization of this gauge theory, $m = m_{37}$ is the mass of the stretched strings, which depends on the D3-brane position $\phi$: for sufficiently small separations, $m \propto \phi$.  Thus, $W \propto \phi^{1/N_c}$, and the gaugino condensate superpotential depends on the D3-brane position.

It is worthwhile to recognize that one of the virtues attributed to models of D-term inflation is the absence of inflaton mass terms from K\"ahler potential couplings.
The D3/D7 model is arguably the best-studied model of D-term inflation in string theory, and an important lesson from this model is that moduli stabilization by superpotential terms introduces F-term energy, which itself may depend on the inflaton, even if the `intended' inflaton potential comes from a D-term.  In other words, a model of D-term inflation in a compactification stabilized by superpotential terms for the moduli is not purely a D-term scenario, and the moduli sector introduces masses in the inflaton sector.

Although the  global form of the moduli potential  is readily computed from (\ref{BHKW}), in practice one can expand in $\phi$:  the leading contribution to the inflaton potential from moduli stabilization is an inflaton mass term.  The total potential $V=V_F+V_D$ then  takes the form
\begin{equation}
V(\phi) = V_D(\phi) -\frac{m^2}{2}\phi^2 +\frac{\lambda}{4}\phi^4\ , \label{D3D7V}
\end{equation}  where $V_D(\phi)$ is given in (\ref{VDphi}).   The resulting phenomenology is discussed in \S\ref{D3D7ph}.

\subsection{Fluxbrane Inflation}

An influential idea  for achieving inflation with D-branes is to consider a pair of branes  that are separated in the compact space, and are almost parallel, but misaligned by a small  relative angle~$\theta$~\cite{GarciaBellido:2001ky,Blumenhagen:2002ua,GomezReino:2002fs}.\footnote{A T-dual configuration, in which the inflationary coordinate is a Wilson line,  has been investigated in~\cite{Avgoustidis:2006zp} (see also \cite{Avgoustidis:2008zu}).}  The small angle leads to  controllably small breaking of supersymmetry, resulting in a force that  draws the branes together, at which point they  merge and reheat the universe.   Brane-antibrane  inflation \cite{Dvali:2001fw,Burgess:2001fx} can be viewed as a special case  in which the branes are precisely antiparallel.

\vskip 4pt
Just as in the cases of brane-antibrane inflation  and the D3/D7 model, the approach taken  in the literature was to begin by analyzing the  interaction potential $V_{\rm int}$ of the misaligned D-brane pair,  assuming that the closed string moduli were stabilized by some mechanism,  and then later attempt to incorporate  (or minimize) the effects of the moduli potential $V_F$, which we may  take to be an F-term  potential.
This approach was a  pragmatic one, because methods for computing the interaction potential  were developed long before techniques for computing the moduli potential.
However,  from the present perspective we must emphasize that  the division into interaction potential and moduli potential is  somewhat arbitrary, and  is often very misleading: the  essence of the eta problem  described in \S\ref{etaproblem}  is that  the moduli potential {\it{is not subleading}}  as a contribution to the inflationary  dynamics.  Bearing this in mind, we will nevertheless  briefly describe the properties of the interaction potential $V_{\rm int}$.

The interaction potential for a brane-antibrane pair in an  unwarped compact space  is generally too steep for successful inflation \cite{Burgess:2001fx}, except possibly for  certain antipodal configurations  (see  e.g.~\cite{Jones:2002cv}).  The proposal of \cite{GarciaBellido:2001ky}  was that weak  supersymmetry breaking by a small angle  $\theta \ll 1$ would diminish the Coulomb force  to the extent that $V_{\rm int}$ could drive slow-roll inflation.
Compactness  of the internal space introduces a crucial difficulty: the  potential between branes with $\theta \ll 1$  is indeed small (compared to the vacuum energy)  if the  computation is performed  with the internal directions taken to be noncompact, but the result is quite different for compact internal dimensions \cite{Kachru:2003sx}.  As explained in \S\ref{etaproblem}, the effect of compactification  is to make the  interaction potential for the branes be of the same order as the vacuum energy,  ruining the favorable hierarchy obtained by taking the branes to be noncompact.

More recently, the  relative position of two D7-branes has been proposed  as an inflationary direction \cite{Hebecker:2011hk,Hebecker:2012aw}.  Consider type IIB string theory compactified on an O3/O7  orientifold of a Calabi-Yau three-fold $X_6$.   Suppose that there is a continuous family $\Sigma_4$ of four-cycles in $X_6$, on any representative of which a D7-brane can be wrapped.
Wrap two  D7-branes $a$, $b$  on distinct  representatives in $\Sigma_4$: the D7-branes can  then be  separated to some  extent,  although they generally intersect along a  two-cycle.   If gauge flux ${\cal F}$ is introduced on the shared two-cycle,  the D7-branes  feel a force  that tends to make them coincide.\footnote{This setup is T-dual to a configuration of branes at angles:  to see this, take the compactification to be a torus  and T-dualize along a  circle in the two-cycle threaded by the flux ${\cal F}$.}   Because a  key part of the inflaton potential arises from worldvolume flux, this scenario is called {\it{fluxbrane inflation}}.

\begin{figure}[h!]
   \centering
     \includegraphics[scale=0.5]{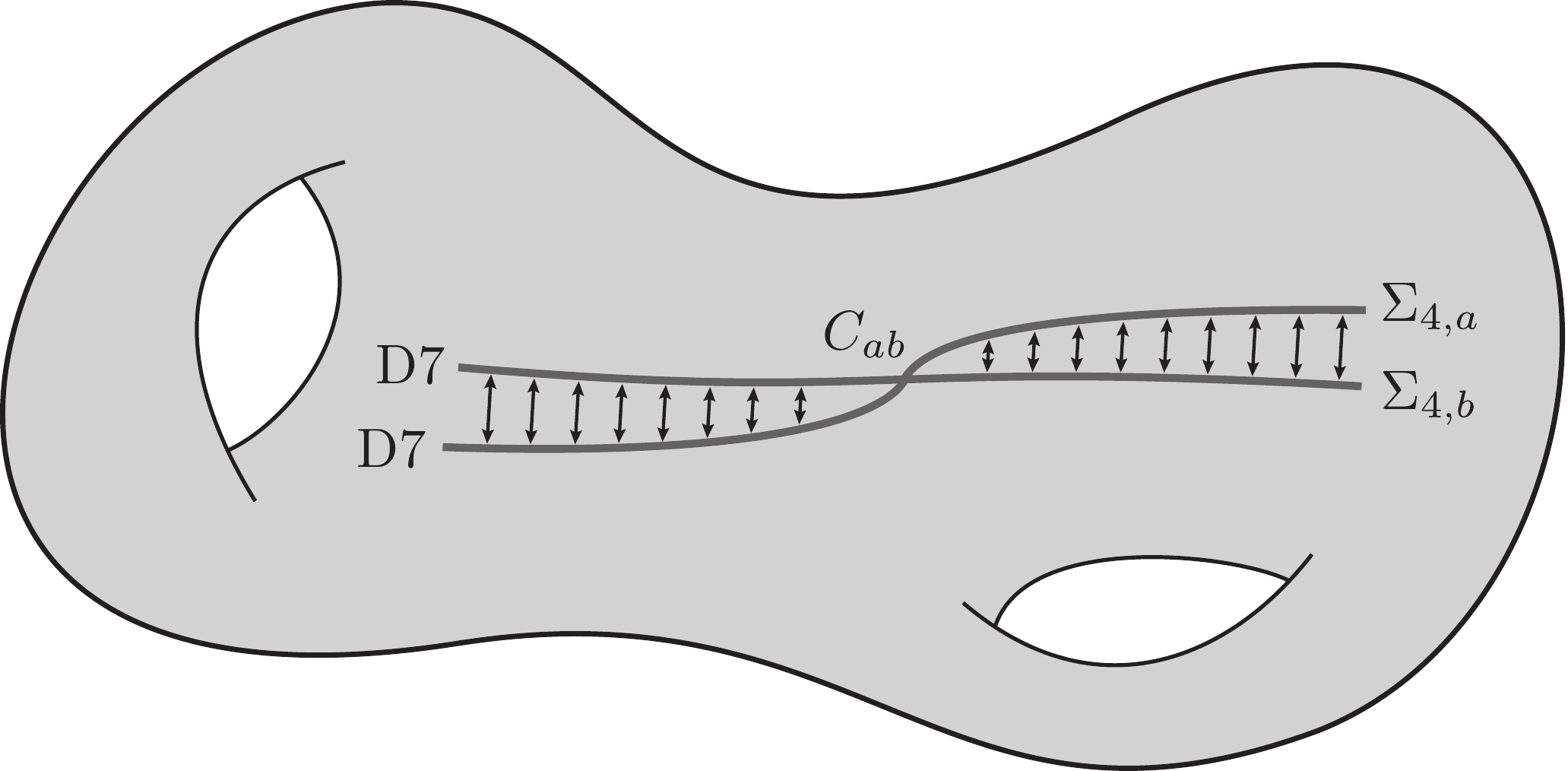}
   \caption{In fluxbrane inflation \cite{Hebecker:2011hk,Hebecker:2012aw}, the inflaton coordinate is the effective separation of
   a pair of intersecting D7-branes (figure adapted from~\cite{Hebecker:2011hk}).}
  \label{fig:fluxbrane}
\end{figure}

The potential  for the  canonically-normalized D7-brane  coordinate takes the form
\begin{equation}
V(\phi) = V_D(\phi) + V_F(\phi)\ ,
\end{equation} where\footnote{In an alternative parameter regime for fluxbrane inflation, the dominant  non-constant term in the potential is sinusoidal \cite{Hebecker:2013zda},  instead of  logarithmic as in (\ref{vdflux}).}
\begin{equation}
V_D(\phi) = V_0\Bigl(1+\alpha \ln(\phi/\phi_0) \Bigr)\ ,  \label{vdflux}
\end{equation}
and $V_0$, $\alpha$, $\phi_0$ are constants.   The D-term potential $V_D(\phi)$ is a consequence of the supersymmetry-breaking flux,  and can be thought of as a Coulomb potential, while the F-term potential $V_F(\phi)$ results from moduli stabilization and has not been computed in detail.

Although the similarities to D3/D7  inflation are apparent, several distinctive features of fluxbrane inflation  were identified in \cite{Hebecker:2011hk}.
First of all, the D7/D7 interaction potential due to flux can be made flat enough for inflation, evading the well-known difficulty \cite{Burgess:2001fx} of  achieving a  sufficiently flat Coulomb interaction  within an unwarped compact space.   Of course, as in warped D-brane inflation \cite{Kachru:2003sx}, the Coulomb potential  is only a small part of the final story:  one must compute the moduli-stabilizing potential $V_F(\phi)$ ---  which generically contributes $\Delta \eta \gtrsim 1$ ---  and determine  whether it spoils inflation.
Moduli stabilization in the type IIB  orientifold (and F-theory)  context is comparatively well understood,  so detailed  study of the moduli potential is possible. Careful investigations of this issue appear in \cite{Hebecker:2011hk,Hebecker:2012aw}, but the issue is not yet settled.

The second  notable feature of fluxbrane inflation  is that the range of the canonically-normalized inflaton  corresponding to a wrapped D7-brane can be much larger than in D3-brane inflation, as first recognized in \cite{Kobayashi:2007hm,Becker:2007ui}.  Moreover, the D-term potential  is readily adjusted to avoid cosmic string constraints,  by  arranging for a hierarchy in the  stabilized values of the K\"ahler moduli \cite{Hebecker:2011hk}.

A fundamental challenge in fluxbrane inflation
is to determine whether
there might be an approximate shift symmetry protecting the potential for D7-brane motion.
At first sight this appears implausible, because {\it{generic}}  choices of three-form flux  lift  the D7-brane moduli, giving large supersymmetric masses to the D7-brane scalars.\footnote{D3-branes, in contrast, do enjoy a moduli space in the leading order no-scale compactifications of \cite{Giddings:2001yu}, but at this same order the K\"ahler moduli are unstabilized.  The challenge described in \S\ref{sec:dbrane} is that the  nonperturbative effects that lift the K\"ahler moduli  inevitably spoil the flatness of the D3-brane potential.}
On the other hand, Appendix E  of \cite{Hebecker:2011hk} gives a plausibility argument for the existence of fluxes that stabilize all closed string moduli while leaving D7-brane flat directions, at least at the level of the classical flux superpotential. It then remains to be shown that perturbative ($g_{\rm s}$ and $\alpha^{\prime}$) corrections to $K$,  and nonperturbative contributions to $W$,  respect this approximate symmetry.  This is an open question:  Euclidean D3-branes carrying worldvolume flux may  introduce a D7-brane potential \cite{Hebecker:2012aw} (cf.~\cite{Grimm:2011dj}), and there are intricate interactions with the stabilization of other moduli, including the dilaton and  the K\"ahler moduli.  Determining the moduli potential in detail will be  a crucial step toward obtaining the phenomenological signatures of the model.

\subsection{M5-brane Inflation}

Although until now we have discussed D3-branes in type IIB string theory, closely-related proposals exist in the context of M-theory compactifications on $S^1/\mathbb{Z}_2 \times X_6$, with $X_6$ a Calabi-Yau threefold.
In this case, it was proposed that the inflaton could correspond to the position of one or more M5-branes along the interval, with inflation ending as the M5-branes collide with and dissolve into the `end-of-the-world' brane.  The single-M5-brane scenario was proposed in \cite{Buchbinder:2004nt}, while a multiple-brane model appeared in \cite{Becker:2005sg}.\footnote{It was argued in \cite{Krause:2007jr} that the tensor-to-scalar ratio $r$ can be large in multi-M5-brane inflation.}

A fundamental difficulty in realizing inflation along these lines is the absence of parametrically controlled constructions of de Sitter vacua in
heterotic string theory, at weak or at strong coupling.  Extensive efforts building on sophisticated studies of heterotic compactifications have led to scenarios for stabilization of the geometric\footnote{Vector bundle moduli are not necessarily stabilized, but are sometimes assumed to be absent.} moduli in anti-de Sitter vacua --- see for example \cite{Anderson:2011cza}.  Even so, de Sitter constructions remain challenging (but see the recent work \cite{Cicoli:2013rwa}).
A general obstacle to parametric control, in both supersymmetric and non-supersymmetric vacua, is that the  quantized three-form flux $H_3$ in
heterotic string theory is {\it{real}}, and hence cannot be adjusted in the same manner as the complex flux $G_3$ of type IIB string theory.\footnote{The difficulties inherent in constructing parametrically controlled heterotic vacua with $H_3$ were appreciated many years ago \cite{Dine:1985rz,Derendinger:1985kk}, and have been only partially overcome: see \cite{Gukov:2003cy,Cicoli:2013rwa}.}

Furthermore, the eta problem appears in a predictable manner in models with moving M5-branes.  The K\"ahler potential for the volume modulus $S$ of the Calabi-Yau, the length modulus $T$ of the interval, and the position $Y$ of a single M5-brane includes the term~\cite{Becker:2005sg}
\begin{equation}
K \supset - \ln \left(S+\bar{S} -\frac{(Y+\bar{Y})^2}{T+\bar{T}} \right)\ .
\end{equation}
This is evidently analogous to the DeWolfe-Giddings K\"ahler potential for a D3-brane, and leads to a mass term for the M5-brane position in the presence of effects stabilizing $S$.  More generally, it is difficult to arrange for an M5-brane to have a potential suitable for inflation while simultaneously stabilizing the geometric moduli.  It was suggested in \cite{Becker:2005sg} that the M5-brane potential would be satisfactory if the effects of gaugino condensation and $H_3$ flux could be neglected during inflation, becoming relevant only later in cosmic history, but it is not clear
that such a scenario, if realizable, can avoid destabilization of the geometric moduli.

\subsection{Phenomenology} \label{D3D7ph}

Once moduli stabilization  is properly incorporated, inflation in the D3/D7  model remains possible, but necessarily involves fine-tuning.
Equipped with  the global form of the nonperturbative superpotential thanks to the worldsheet calculation of \cite{Berg:2004ek}, the authors of \cite{Haack:2008yb,Burgess:2008ir}  systematically analyzed the potential in search of inflationary regions.  Two qualitatively  different scenarios were found:
\begin{itemize}
\item[$\triangleright$]  {\it Saddle-point inflation.}---If the  
condensate responsible for K\"ahler moduli stabilization  is assumed to arise exclusively on a stack of D7-branes near a single fixed point of  $T^2/\mathbb{Z}_2$, then after  fine-tuning of the parameters, the potential for a D3-brane at an  approximately antipodal location in the torus  can develop an unstable saddle point.  (For this scenario, it is essential that the primary  source of supersymmetry breaking  is an anti-D3-brane in a warped region.)  The resulting model has $r \ll 1$  and $n_s \lesssim 0.95$.  The characteristic redness of the spectrum in saddle-point models of this form is discussed in \cite{Brax:2007fz}.

\item[$\triangleright$]  {\it Inflection point inflation.}---If the dominant force on the D3-brane comes from interactions with supersymmetry-breaking fluxes on a D7-brane,  as in \cite{Dasgupta:2002ew}, then the potential can be fine-tuned to have an inflationary inflection point, with phenomenology  broadly similar to that described in \S\ref{warpedphenomenology}.  The potential takes the form (\ref{D3D7V}), incorporating a Coleman-Weinberg term, as well as  quadratic  and quartic terms from moduli stabilization.
    When the quartic terms are significant, the fine-tuning for inflation is extreme,  and was argued in \cite{Burgess:2008ir} to be  at the level of one part in $10^6$.   On the other hand, \cite{Haack:2008yb} exhibit parameter ranges in which the moduli contribution is approximately quadratic and the fine-tuning is milder.
\end{itemize}

The  kinematical  field range $\Delta\phi$ of the canonically-normalized inflaton in D3/D7 inflation can be  super-Planckian, $\Delta\phi > M_{\rm pl}$, if the
$T^2/\mathbb{Z}_2$ is highly anisotropic \cite{Haack:2008yb}.  For a rectangular torus  with side lengths $L_1$ and $L_2$,  we have
\begin{equation}
 M_{\rm pl} \propto \sqrt{{\rm{Vol}}(K3)L_1 L_2}\ ,
\end{equation}
while the field range  along the side of length $L_1$  has the parametric dependence
\begin{equation}
\Delta\phi_1 \propto \sqrt{L_1/L_2}\ .
\end{equation}
In \cite{Haack:2008yb},  it was argued that one can take $L_1/L_2$ to be  large enough  so that $\Delta\phi_1 > M_{\rm pl}$, without compromising computability.
This  fact is consistent with the general arguments made in \S\ref{superPlanckian},  where anisotropy of the compactification was the only plausible route to a parametrically controlled  super-Planckian field range for a D-brane.   Even so, this observation has not led to a full-fledged inflationary scenario  with large tensor-to-scalar ratio $r$, because of the difficulty of arranging that the potential remains flat over a distance $\Delta\phi > M_{\rm pl}$.

Cosmic strings in models of D-term inflation have been a subject of much  discussion --- see \cite{Dasgupta:2004dw,Haack:2008yb} for  summaries with original references.
In early versions of the D3/D7 model,  long-lived cosmic strings  were thought to be present in problematic numbers,  in conflict with upper limits from  measurements of the CMB temperature power
spectrum.  This problem has been avoided in two ways:  first, in extensions of the scenario with additional charged fields \cite{Urrestilla:2004eh,Dasgupta:2004dw}, the  vacuum manifold is simply connected and the  resulting cosmic strings are `semilocal',  i.e.~non-topological,  leading to  weaker constraints.   Second, it was recognized in \cite{Haack:2008yb} that contributions to the inflaton potential  from moduli stabilization introduce an additional parameter,  allowing separation of the amplitude of inflationary density perturbations from the  density perturbations due to cosmic strings.

\vskip 4pt
The signatures of  fluxbrane  inflation are broadly similar to those of D3/D7 inflation, although as noted above  the observational constraints from cosmic strings are readily satisfied through hierarchical stabilization of the K\"ahler moduli.  Detailed statements, for example about the spectral index, will have to await a complete computation of the moduli potential.

\vskip 4pt
The phenomenology of M5-brane  inflation is more difficult to characterize,  because it depends on  presently-unknown properties of the  moduli potential,  as well as  on the intricacies  of the M5-brane collision with the end-of-the-world brane.  We will leave the signatures of this scenario as a question for the future.

\section{Inflating with Relativistic Branes}
\label{ssec:DBI}

As we have learned in the previous two sections,
inflation in systems driven by slowly moving D-branes suffers from the eta problem.
In~\cite{Silverstein:2003hf}, Silverstein and Tong observed in that this problem may be alleviated if the D-branes move relativistically.
The model relies on the non-trivial structure of the DBI action for the D-branes and is called {\it DBI inflation}.
In this section, we will explore this idea.\footnote{This section is based mostly on \cite{Silverstein:2003hf, Alishahiha:2004eh}.}
In \S\ref{dbii}, we present the DBI mechanism, followed, in \S\ref{sec:DBIEFT} and \S\ref{sec:DBICFT}, by explanations  of the  symmetries that control quantum corrections.
Microphysical constraints arising
in the embedding in string theory are presented in \S\ref{sec:DBImicro}, and the observational signatures of DBI inflation are summarized in \S\ref{dbiph}.

\subsection{Dirac-Born-Infeld Inflation}  \label{dbii}

We begin with the same system as in \S\ref{sec:dbrane}:
a spacetime-filling D3-brane probing $AdS_5 \times S^5$,
\begin{equation}
\d s^2 = \Big(\frac{r}{L}\Big)^2\eta_{\mu \nu }\d x^{\mu}\d x^{\nu} + \Big(\frac{L}{r}\Big)^2\Bigl(\d r^2+r^2 \d \Omega_{S^5}^2 \Bigr)\ , \label{equ:AdS5S5}
\end{equation} where $L^4 = 4\pi g_{\rm s} N (\alpha')^2$, with $N$  the total D3-brane charge of the background.
To arrive at a simple model with four-dimensional gravity, we cut off the AdS space  in the infrared and ultraviolet, taking $r_{\mathsmaller{\rm IR}} < r < r_{\mathsmaller{\rm UV}}$ (as in Randall-Sundrum  scenarios \cite{Randall:1999ee}).  The D3-brane  Lagrangian takes the form
\beq
{\cal L}\,=\, - \frac{\phi^4}{\lambda} \left( \sqrt{1 + \frac{\lambda}{\phi^4}  \hskip 1pt (\partial \phi)^2} - 1 \right) - V(\phi)\ , \label{equ:DBI}
\eeq
where $\phi \equiv \sqrt{T_3}\hskip 1pt r$ and\footnote{The constant $\lambda$ is proportional, but not equal, to  the standard 't Hooft  coupling in $AdS_5 \times S^5$, $\lambda_{t} \equiv g_{\rm{YM}}^2 N = 4\pi g_{\rm{s}} N$.}
\beq
\lambda \equiv T_3 L^4 = \frac{1}{2\pi^2}N\ .
 \eeq
The potential $V(\phi)$ in (\ref{equ:DBI})  requires some explanation.
In $AdS_5 \times S^5$ a probe D3-brane feels  no force, but  physical effects associated with the  infrared and ultraviolet deformations of the spacetime typically generate a potential for D3-brane motion.   Indeed, as we have seen in \S\ref{sec:dbrane}, the cutoff AdS  geometry  is most naturally viewed as an  approximation to a region of a  finite Klebanov-Strassler throat\footnote{Such a region actually corresponds to a section of $AdS_5 \times T^{1,1}$, up to logarithmic  corrections, but the  angular manifold is immaterial at present.} that is attached to a flux compactification ---  the  IR cutoff at $r_{\mathsmaller{\rm IR}}$ corresponds to the tip of the throat,  while the UV cutoff at $r_{\mathsmaller{\rm UV}}$ corresponds to the remainder of the compactification.   Supersymmetry breaking in the infrared (e.g.~by an anti-D3-brane),  as well as supersymmetry breaking  and moduli stabilization in the ultraviolet then lead to a potential for the D3-brane position,  as explained in detail in \S\ref{sss:D310d}.   Any such effect  can be captured by a suitable change to the supergravity background, cf.~eq.~(\ref{equ:Vphi}).

The action (\ref{equ:DBI}) is adapted straightforwardly to more general warped backgrounds
\beq
{\cal L} \, =\,  - T(\phi) \left( \sqrt{1 + \frac{(\partial \phi)^2}{T(\phi)}} - 1 \right) - V(\phi) \ , \label{equ:DBIX}
\eeq
where we introduced the warped tension of the brane, $T(\phi) \equiv T_3\hskip 1pt e^{4A(\phi)}$.
The precise functional forms of the warp factor $e^{4A(\phi)}$ and the potential $V(\phi)$ determine the details of the phenomenology of DBI inflation (see~\cite{Shandera:2006ax, Bean:2007hc}).
As a generalization of the AdS background (\ref{equ:AdS5S5}), we will use Calabi-Yau cones that approximate finite warped throat regions attached to
type IIB  flux compactifications. Although the
KS throat provides a rare example of a warped throat that is smooth in the infrared,  here we will also consider more general warped  Calabi-Yau cones, with  ten-dimensional line element
\beq
\d s^2 = e^{2A(r)} \eta_{\mu \nu} \d x^\mu \d x^\nu + e^{-2A(r)} \left(  \d r^2 + r^2 \d \Omega_{X_5}^2 \right)\ , \label{equ:CYcone}
\eeq where $X_5$ is an arbitrary Einstein manifold.   Just as the KS solution can be approximated by  $AdS_5 \times T^{1,1}$ for $r_{\mathsmaller{\rm IR}} \ll r \ll r_{\mathsmaller{\rm UV}}$, up to logarithmic corrections,  many solutions of the form (\ref{equ:CYcone})  can be approximated by $AdS_5 \times X_5$: the warp factor is then\footnote{For a generalization of DBI inflation to arbitrary warp factor, assuming an appropriate potential, see \cite{Spalinski:2007dv,Spalinski:2007qy}.}
\beq
e^{-4A(r)} \approx  \frac{L^4}{r^4}  \qquad {\rm with} \quad L^4 \equiv \frac{4\pi^4 g_{\rm s}}{{\rm{Vol}}(X_5)}  N(\alpha')^2\ ,  \label{equ:cycw}
\eeq where ${\rm{Vol}}(X_5)$  denotes the volume of $X_5$ (in string units).
For a throat of the form  (\ref{equ:CYcone}), the range of the canonically-normalized  D3-brane position is  given,  as in (\ref{equ:BMbound}), by \cite{Baumann:2006cd}
\beq
\frac{\Delta \phi}{M_{\rm pl}} \le \frac{2}{\sqrt{N}} \ , \label{equ:BMboundX5}
\eeq and in particular is independent of $X_5$.  We will see in \S\ref{dbiph} that  if one manages to achieve a DBI phase, (\ref{equ:BMboundX5}) provides a  stringent upper bound on the tensor-to-scalar ratio.

\vskip 4pt
\noindent
{\it The UV model.}---We will refer to the UV model as the situation in which a D3-brane   moves {\it into} the warped region, i.e.~toward small $\phi$, from the ultraviolet \cite{Silverstein:2003hf, Alishahiha:2004eh}.

\vskip 4pt
\noindent
{\it The IR model.}---An  interesting alternative is the IR model, in which  inflation occurs as the D3-brane  leaves the tip region and moves  toward the ultraviolet end of the throat \cite{Chen:2004gc, Chen:2005ad, Chen:2005fe}.
The initial conditions  for  the IR model are very  appealing \cite{Chen:2004gc}.
Suppose that $p$ anti-D3-branes  are introduced into a  KS throat region.
If $p \lesssim 0.08 \hskip 1pt M$, with $M$ the flux quantum number defined in (\ref{KSflux}),  the  anti-D3-branes  form a metastable configuration  at the tip
\cite{Kachru:2002gs}.  Over an exponentially long timescale,  this state can decay:  the anti-D3-branes  annihilate against flux,  liberating $(M-p)$  D3-branes.
The observation of \cite{Chen:2004gc} is that the D3-brane potential arising from moduli stabilization may  drive  some or all of these D3-branes  to move out of the throat region,   and during this process a phase of DBI inflation can occur.
A simple model of the  radial potential is
\begin{equation}
V(\phi) = V_0 - \frac{1}{2}\beta H^2 \phi^2\ ,  \label{IRV}
\end{equation} where $V_0$  is a constant, and in generic configurations,\footnote{Potentials that tend to expel a D3-brane from the infrared are quite common in the ensemble obtained in \cite{Baumann:2010sx}, but  are  far more complicated than (\ref{IRV}), involving all five angular directions and an  array of competing terms (see~\cite{Agarwal:2011wm}).} $-1 \lesssim \beta \lesssim 1$.  The IR  model corresponds to $\beta >0$.  For a discussion of obstacles to a  computable realization of the IR  model in string theory, see \S\ref{sec:DBImicro}.

\vskip 4pt
\noindent
{\it Relativistic dynamics.}---For a spatially-homogeneous D3-brane, i.e.~for $\phi=\phi(t)$, it is natural to define a `Lorentz factor', by analogy to relativistic particle dynamics:
\beq
\gamma\, \equiv \, \left(1 - \frac{\dot{\phi}^2}{T(\phi)}  \right)^{-1/2} \ .  \label{gammadefT}
\eeq
The requirement that $\gamma$ be real enforces a speed limit on the motion of the probe D3-brane:
\beq
\dot{ \phi}^2\, < \,  T(\phi) \ . \label{dbiL}
\eeq
  Notice that the bound is independent of the  properties of the potential and
becomes stronger in regions of strong warping, where $e^{4A(\phi)} \ll 1$ and hence $\dot{\phi}^2 \ll T_3$.

In \S\ref{sec:dbrane}, we studied non-relativistic D3-brane motion, corresponding to $\gamma \approx 1$: expanding the square root in (\ref{equ:LDBI}) then led to the two-derivative action
(\ref{equ:LD3}).
DBI inflation
operates in the regime of
 relativistic brane dynamics, with $\gamma \gg 1$,  where higher-derivative terms in (\ref{equ:DBI}) cannot be neglected.
Varying (\ref{equ:DBI}) with respect to the four-dimensional metric gives the stress-energy tensor sourced by the D3-brane.
This corresponds to the stress-energy of a perfect fluid, with energy density and pressure given by
\begin{align}
\rho &\, =\,  \bigl(\gamma -1 \bigr) \hskip 1pt T + V \ ,  \label{DBIrhoeq} \\
P &\, =\,  \bigl(\gamma- 1 \bigr) \hskip 1pt \frac{T}{\gamma} - V \ .
\end{align}
Coupling to gravity gives the Friedmann equation
\beq
3 M_{\rm pl}^2 H^2 = \bigl(\gamma -1 \bigr) \hskip 1pt T + V(\phi) \ , \label{DBIFRW}
\eeq
and the continuity equation
\beq
 \dot{\phi} = - \frac{2M_{\rm pl}^2 H'}{\gamma}\ ,  \label{DBIcont}
\eeq
where $H' \equiv dH/d\phi$.  Using (\ref{DBIcont}) in (\ref{gammadefT}), we find
\beq
\gamma = \sqrt{1+ \frac{(2 M_{\rm pl}^2 H')^2}{ T}}\ .   \label{gammahp}
\eeq
The Hubble slow-roll parameters are~\cite{Silverstein:2003hf}
\begin{align}
\varepsilon &\, =\, - \frac{\dot H}{H^2} = \frac{2 M_{\rm pl}^2}{\gamma} \left( \frac{H'}{H}\right)^2 \ , \\
\tilde \eta &\,=\,\,\, \frac{\dot \varepsilon}{H \varepsilon}\,\, \, =  \frac{2 M_{\rm pl}^2}{\gamma} \left[ 2 \left( \frac{H'}{H}\right)^2  -  2 \frac{H''}{H} +  \frac{H'}{H} \frac{\gamma'}{\gamma} \right]   \ .
\end{align}
Notice the factors of $\gamma^{-1}$  in both $\varepsilon$ and $\tilde \eta$. For large $\gamma$, the slow-roll parameters are therefore {\it{suppressed}} relative to the expectation derived from the non-relativistic limit. This leads to the intriguing possibility of achieving inflation even for potentials that naively seem  to be too steep to drive prolonged inflation.

Accelerated expansion occurs if  the potential energy dominates over the kinetic energy.
Demanding that $V(\phi)$  is the leading term on the
right-hand side of (\ref{DBIrhoeq}) gives the condition
\begin{equation}
\frac{V}{ \gamma \hskip 1pt T}  \gg 1\ .  \label{DBIaccelerate}
\end{equation}
Thus, DBI inflation can occur near the location $\phi$  only if the potential is {\it{large in local string units}}.
Next, a defining requirement for a
DBI phase is that the D3-brane is relativistic.\footnote{In the remainder of this section, we will work to leading order in $\gamma^{-1} \ll 1$.}
Using~(\ref{DBIaccelerate}) in (\ref{DBIFRW}) and (\ref{gammahp}), we find
\begin{equation}
\gamma^2 = \frac{2}{3}\hskip 1pt \epsilon \hskip 1pt \frac{V}{T}  \gg  1   \ ,  \label{DBIgammacondition}
\end{equation} where $\epsilon \equiv \frac{1}{2}\Mp^2 (V'/V)^2$.
We observe that although  (\ref{DBIgammacondition})  involves $\epsilon$ and can plausibly be  satisfied by making the potential very steep, the condition (\ref{DBIaccelerate}) is independent of the functional form of the potential.  We will see below that (\ref{DBIaccelerate})  presents a serious obstacle in the search for a consistent embedding in string theory.

Analytical~\cite{Silverstein:2003hf, Alishahiha:2004eh} and numerical~\cite{Shandera:2006ax, Bean:2007hc} studies have shown that  for suitable potentials $V(\phi)$, the DBI Lagrangian (\ref{equ:DBI}) can indeed  support an  inflationary phase  in which  the non-trivial kinetic term plays a crucial role.    However, before  describing the phenomenology of DBI inflation, one should first ask whether the  DBI Lagrangian (\ref{equ:DBI}) gives an  accurate and consistent  representation of the physics of a relativistic D3-brane.    There are several  important questions:  first,  do quantum  corrections, either in EFT  or from Planck-scale physics,  lead to significant changes to the  very special kinetic terms in (\ref{equ:DBI})?   Second, do potentials $V(\phi)$  satisfying the necessary conditions  (\ref{DBIgammacondition}) and (\ref{DBIaccelerate}) arise naturally,  in the same setting where the kinetic terms  take the necessary form?   Finally,  does  backreaction of the D3-brane energy, which is large when (\ref{DBIaccelerate}) holds, modify the  background or the dynamics?
We will discuss these issues in turn.

\subsection{DBI as an EFT}
\label{sec:DBIEFT}

The action in (\ref{equ:DBI}) may be viewed as an effective theory with cutoff $\Lambda^4(\phi) \equiv \phi^4/\lambda$. The relativistic limit corresponds to the limit $(\partial \phi)^2 \to \Lambda^4$.
Naively, this suggests a breakdown of the effective theory and a loss of predictivity.
In particular, one might worry that the form of the DBI action in the ultra-relativistic limit is unstable to quantum corrections.
Moreover, one may ask  whether it is consistent to  work to arbitrary order in single derivatives $\partial_\mu\phi$  and yet neglect all terms involving higher derivatives, such as $\Box \phi$.
In this section, we address both of these concerns.

\vskip 4pt
\noindent
{\it Quantum corrections.}---We will first  explain why the action in (\ref{equ:DBI}) does not receive large quantum corrections in the ultra-relativistic limit. In particular, we will show that the DBI action in anti-de Sitter space is uniquely fixed by the nonlinearly realized conformal symmetries inherited from the isometries of the background spacetime.\footnote{This argument was first made in \cite{Maldacena:1997re} and further elaborated in~\cite{deRham:2010eu, Goon:2011qf}.}

The isometry group of five-dimensional anti-de Sitter space, $SO(2,4)$, contains Poincar\'e transformations, spacetime dilatations and special conformal transformations (SCTs).
The D3-brane action is invariant under the four-dimensional Poincar\'e  subgroup $ISO(1,3) \subset SO(2,4)$, but some of the five-dimensional symmetries are only nonlinearly realized.
In particular, the D3-brane position modulus $\phi$ transforms nonlinearly  under the conformal symmetries of the AdS spacetime~\cite{Maldacena:1997re}:
\begin{align}
\mbox{dilatation} : \qquad x^\mu &\mapsto \tilde x^\mu \equiv (1+c)x^\mu \ ,  \nonumber\\[2pt]
 \phi(x) &\mapsto \phi(\tilde x) + c \ , \label{equ:LG1} \\[4pt]
\mbox{SCTs} : \qquad x^\mu &\mapsto \tilde x^\mu \equiv x^\mu +  (b \cdot x) x^\mu - \frac{1}{2} \left( x^2 + \frac{\lambda}{\phi^2} \right) b^\mu\ ,\nonumber \\
\phi(x) &\mapsto \phi(\tilde x) \big(1 - (b \cdot x) \big)\ , \label{equ:LG2}
\end{align}
where $c$ and $b^\mu$ are infinitesimal transformation parameters.
We see that dilatations and SCTs shift the field value and its gradient, respectively.
After gauge fixing, these transformations of $\phi$ become global symmetries~\cite{Goon:2011qf},  which constrain  the form of the action.
First, we note that the unbroken four-dimensional Lorentz symmetry and the nonlinearly realized dilatation symmetry (\ref{equ:LG1}) imply
\beq
S = \int \d^4 x\,\, \phi^4 f\Bigl((\partial \phi)^2/\phi^4 \Bigr) + \cdots\ ,
\eeq
where $f$ is an arbitrary function and the ellipses denote corrections involving at least two derivatives acting on $\phi$. In order for this action to be invariant under the SCTs (\ref{equ:LG2}), the function $f$  must take the form \cite{Maldacena:1997re}
\beq
f(z) = \alpha \left[ \sqrt{1 + \lambda \, z} - \beta  \right]\ ,
\eeq where $z$ is shorthand for $(\partial \phi)^2/\phi^4$.  The coefficients $\alpha$ and $\beta$  can be fixed using supersymmetry:\footnote{Here, we assume that supersymmetry is only broken spontaneously during inflation, so it still constrains the form of the action.}
first, recall that in the absence of a supersymmetry-breaking potential, a D3-brane feels no force in an AdS background (it is BPS). This implies that $\beta=1$.
Second, the kinetic term $(\partial \phi)^2$ is not renormalized in the supersymmetric limit.
This fixes the normalization, $\alpha = - \lambda^{-1}$.
Purely on the basis of symmetries, we have therefore arrived at the DBI action in the form (\ref{equ:DBI}), implying that the action is protected by symmetry.  Moreover, the argument is nonperturbative and so applies to all orders in the quantum theory.  This is the famous non-renormalization theorem of the DBI action:
quantum corrections can only arise at higher order in derivatives. Similar non-renormalization arguments apply to the generalized backgrounds (\ref{equ:DBIX}).

\vskip 4pt
\noindent
{\it Higher-derivative corrections.}---But what about the higher-derivative terms?
In the limit $(\partial \phi)^2 \to \Lambda^4$, the full square-root structure of the kinetic term is important and the dynamics {\it cannot} be described by the first few orders in an expansion in $(\partial \phi)^2/\Lambda^4$.  This is somewhat unconventional from a low-energy effective field theory point of view, so it deserves a bit more discussion.  For example, is it really consistent to go to all orders in $(\partial \phi)^2$, but ignore all operators with higher derivatives such as those involving $\Box \phi$?
Notice that we do the same when we study a point particle in the ultra-relativistic limit,
$\dot x^2 \to 1$.
In that case, we also trust the full square-root action, $L = - m^2 \sqrt{1-\dot x^2}$, but neglect higher time derivatives such as terms involving the acceleration $\ddot x$.  The justification
is that the equations of motion enforce all higher derivatives to vanish in the ultra-relativistic limit,
i.e.~as $\dot x^2 \to 1$.  In the absence of warping, an identical argument holds for the DBI action, i.e.~the DBI action is valid for arbitrarily high velocities $\dot{\phi}$, as long as the proper acceleration is smaller than the string scale.
One expects  the same  conclusion to hold in a  warped background,  provided that the warp factor changes sufficiently slowly.
As shown in~\cite{Silverstein:2003hf}, this is indeed the case in the AdS background as long as $\lambda \gg 1$,   which is precisely the limit in which the supergravity description is a good approximation.

\subsection{DBI as a CFT}
\label{sec:DBICFT}

The AdS/CFT correspondence~\cite{Maldacena:1997re,Gubser:1998bc,Witten:1998qj} (for a review, see \cite{MAGOO})
provides an alternative viewpoint on the inflaton action.  To describe this,
we will briefly  sketch the essential elements of the correspondence in its simplest incarnation,  which is the duality between ${\cal N} = 4$ super-Yang-Mills (SYM) theory with gauge group $U(N)$,  in flat four-dimensional spacetime,  and type IIB  string theory in $AdS_5 \times S^5$~\cite{Maldacena:1997re}.

The essential idea of the correspondence is that there are two equivalent descriptions of the region near to a stack of $N$  D3-branes in flat ten-dimensional spacetime:  the gauge theory description involving open strings on the D-branes, and  the gravitational description  involving the curved spacetime sourced by the branes.
Near a stack of $N$ D3-branes, the background takes the $AdS_5 \times S^5$  form (\ref{equ:AdS5S5}).
The asymptotic symmetry group of this spacetime (at the boundary, $r \to \infty$) is $SO(4,2)\times SO(6)$.
We recognize $SO(4,2)$ as the {\it{conformal group}} in four spacetime dimensions.   Recalling that ${\cal N} = 4$ super-Yang-Mills theory is a conformal theory and has an R-symmetry group $SO(6) \simeq SU(4)$,  one finds a  perfect match between the  global symmetries of the gauge theory and the asymptotic symmetries of the gravitational solution.

For the present purpose, the relevant  application of the duality is to the {\it{Coulomb branch}}\footnote{The terminology  is arguably more appropriate to ${\cal N} = 2$ theories,  where the Coulomb branch  is distinguished from the Higgs  branch because the former is parameterized by  scalars in vector multiplets and the latter is parameterized by scalars in hypermultiplets.   The important point is just that  motion on the Coulomb branch does not change the rank of the gauge group, but can change the rank of the non-Abelian part of the gauge group, i.e.~$U(N) \to U(N-1) \times U(1)$, while motion on the Higgs  branch can change the total rank.} of  ${\cal N} = 4$ SYM.  This is the moduli space corresponding to the positions  of D3-branes along the radial coordinate $r$  of $AdS_5$,  and in the five angles on $S^5$.  The D3-brane positions are parameterized by scalars that transform  in the adjoint  of $SU(N)$.
Let us give a vev to one eigenvalue $\phi$ of the adjoint scalar.
This induces the symmetry breaking $U(N) \to U(N-1) \times U(1)$. (In the bulk this corresponds to separating one of the branes from the stack.)
Modes $\Psi$ that are charged under the $U(1)$ symmetry obtain masses proportional to $\phi$. (In the bulk picture these correspond to the masses of strings stretching from the mobile brane to the stack.)
Integrating out the fields $\Psi$ generates higher-dimension operators suppressed by powers of $\phi$ itself.
The vev of $\phi$ can be extremely small compared to the string scale and,  thus,  higher-dimension contributions are more important in the DBI model than one might naively expect.
The first correction is protected by supersymmetry and takes the form
$\lambda/\phi^4 (\partial \phi)^4$. The CFT is strongly coupled, so all higher-order terms are important and need to be resummed. This is difficult to do in the field theory, but the AdS/CFT correspondence tells us that the answer will be the DBI action~(\ref{equ:DBI}).

\subsection{Microphysical Constraints}
\label{sec:DBImicro}

In \S\ref{sec:DBIEFT}, we have seen that DBI inflation is natural in the bottom-up sense, in that loop corrections in the EFT are under control. We now turn to a discussion of top-down naturalness,  including the question of whether DBI inflation  arises in a  consistent string compactification.
For concretness, we will focus our discussion on relativistic branes in warped Calabi-Yau cones, cf.~eq.~(\ref{equ:CYcone}).\footnote{Identifying an alternative setting for DBI  inflation in a  string compactification would be very interesting, and would undoubtedly  lead to modified microphysical constraints, but we are not aware of any complete example.}

\begin{itemize}
\item[$\triangleright$]
{\it Achieving a steep potential.}---Slow-roll inflation requires a potential that is flat in Planck units, with $\eta \ll 1$,   whereas DBI inflation  requires a potential that is steep enough to drive the moving D-brane to have large kinetic energy, i.e.~to obey (\ref{DBIgammacondition}).
These two options are {\it{not exhaustive}}: a  potential can easily be too steep for slow-roll inflation and yet too gentle for DBI inflation.

To  understand whether (\ref{DBIgammacondition}) is readily satisfied  for D3-branes in Calabi-Yau cones,  we can make use of the potential derived in \cite{Baumann:2010sx}, which describes the forces on a D3-brane in a KS throat attached to a KKLT  compactification (see \S\ref{sec:dbrane}).  This potential  incorporates the Coulomb  interaction with an anti-D3-brane,  as well as the full spectrum of contributions from moduli stabilization.  In a Monte Carlo study \cite{Agarwal:2011wm} based on \cite{Baumann:2010sx}, DBI inflation did not occur by chance:  in  the full set of  more than $10^7$ trials, $\gamma -1$ never exceeded $10^{-8}$.   The reason for this finding is that the  D3-brane potential from moduli stabilization  is small in local string units, $V(\phi) \ll T(\phi)$, and is not parametrically steep,  so that $\epsilon \lesssim 1$.  It follows from (\ref{DBIgammacondition})  that $\gamma-1 \ll 1$.
While it is  plausible that  somewhat larger Lorentz  factors  could arise very near the tip of the throat,  which was not directly studied in~\cite{Agarwal:2011wm}, the problem of backreaction  becomes severe in this regime, as discussed below.
The earlier analysis \cite{Bean:2007hc} worked with a simpler model of the potential,  but arrived at  compatible conclusions:  in realizations consistent with the field range bound (\ref{equ:BMboundX5}) of \cite{Baumann:2006cd}, and with observational  constraints on $n_s$,  the Lorentz factor was bounded by $\gamma -1 < 10^{-7}$.

A remark about the Coulomb  potential is relevant here.
The potential $V_{\cal C}(\phi)$ given in (\ref{equ:VC}) applies to a KS throat  (in the $AdS_5 \times T^{1,1}$ approximation).
For a cone over  $X_5$, one finds instead
\beq
V_{\cal C}(\phi) = D_0 \left( 1 -\frac{\pi}{4\hskip 1pt {\rm{Vol}}(X_5)} \frac{D_0}{\phi^4} \right)\ . \label{equ:VCX}
\eeq  For small ${\rm{Vol}}(X_5)$ --- for example, if $X_5 = S^5/Z_p$ with $p \gg 1$ ---  the Coulomb force is increased in strength: heuristically, the field lines are collimated  along a narrow throat.   In extreme cases,  this increased force can compel the D3-brane to be relativistic:  \cite{Bean:2007hc} found that a DBI  phase arose for ${\rm{Vol}}(X_5) \lesssim 10^{-17}$.   However, it is highly implausible that such a throat can be embedded in a consistent compactification, and moreover, even if this issue is overlooked, the examples of \cite{Bean:2007hc}  are incompatible either with observations or with the microphysical bound of \cite{Baumann:2006cd}.

\item[$\triangleright$]  {\it Backreaction.}---The requirement (\ref{DBIaccelerate}) for accelerated expansion, $V(\phi)  \gg \gamma T(\phi)$, implies that the D3-brane potential energy must substantially exceed the local string scale.   This  carries the risk that whatever physics generates the potential will simultaneously distort the background supergravity solution (i.e.~the AdS geometry~(\ref{equ:AdS5S5}) or a related warped throat geometry).   We will offer two  related perspectives on this problem:

First, we  present an observation due to Maldacena (also described in \cite{McAllister:2007bg}).
Suppose that the D3-brane potential  arises from coupling the  warped throat sector to a hidden sector  that breaks supersymmetry,  and take the potential to be  quadratic, $V(\phi)  = \frac{1}{2} m^2\phi^2$, so that
\begin{equation}
\frac{V(\phi)}{\gamma \hskip 1pt T(\phi)} = \frac{\lambda}{2\gamma} \frac{m^2}{\phi^2}\ .   \label{vtratio}
\end{equation}
Consider the effect of hidden sector supersymmetry breaking on the Kaluza-Klein spectrum of the throat.
Barring an efficient sequestering mechanism,  the  Kaluza-Klein modes will acquire masses $\MKK \sim m$.
Since the lightest  Kaluza-Klein modes in an undistorted throat  have
\begin{equation}
\MKK \sim \frac{1}{L} e^{A_{\mathsmaller{\rm IR}}} \sim \frac{r_{\mathsmaller{\rm IR}}}{L^2}\ ,
\end{equation} supersymmetry breaking  will typically cut off the throat at
\begin{equation}
r_{\mathsmaller{\rm IR}} \sim m L^2\ .  \label{rcutoff}
\end{equation}
Thus,  the  canonical field $\phi$ parameterizing D3-brane motion  obeys
\begin{equation}
\phi^2 \gtrsim \phi_{\mathsmaller{\rm IR}}^2 \sim  T_3 (m L^2)^2 = \lambda m^2\ .  \label{minimumcanonical}
\end{equation}
Combining (\ref{vtratio}) and (\ref{minimumcanonical}), we find that
\begin{equation}
\frac{V(\phi)}{\gamma T(\phi)} \lesssim \frac{1}{2\gamma} \ll 1\ .  \label{equ:Juan}
\end{equation}
We conclude that  unless the source of supersymmetry breaking couples much more strongly to the D3-brane than to the  Kaluza-Klein modes of the  background,  the  throat is truncated in the infrared, excluding the region where DBI inflation could occur.

One can argue for the cutoff (\ref{rcutoff}) in a slightly different way, beginning with the potential for a D3-brane probe  of a general supergravity  background.
From (\ref{equ:Vphi}), we have $V(\phi) = T_3 \Phi_-$, where the scalar $\Phi_-$, defined in (\ref{gpm}), involves the warp factor and the four-form potential.
Thus,
\begin{equation}
\frac{V(\phi)}{\gamma T(\phi)} =   \frac{1}{\gamma}  \frac{e^{4A}-\alpha}{e^{4A}}\ .
\end{equation}
DBI inflation therefore requires
\begin{equation}
|\alpha| \gg \gamma e^{4A}\ . \label{alphae}
\end{equation}
In the noncompact, supersymmetric Klebanov-Strassler background, $e^{4A}=\alpha$, i.e.~$\Phi_-=0$, but in a finite warped throat, supersymmetry breaking and moduli stabilization in the bulk source perturbations of $\Phi_-$ and of $G_-$ \cite{Baumann:2008kq,Baumann:2010sx} --- see (\ref{gpm}).  In turn, these perturbations source corrections to the metric.
The condition (\ref{alphae}) requires strong deviations from the ISD background, which lead to correspondingly large corrections to the metric that eventually terminate the throat in the infrared.  One can then show \cite{Gandhi:2011id,Gandhitoappear} that $V(\phi) \lesssim T(\phi)$ in the accessible region, in agreement with (\ref{equ:Juan}).

\item[$\triangleright$]  {\it Super-Planckian fields.}---Another important microphysical difficulty is that in many simple examples (e.g.~for a quadratic potential) a field range $\Delta\phi$ of order $M_{\rm pl}$ is necessary \cite{Alishahiha:2004eh}.
In view of (\ref{equ:BMboundX5}), this is at best marginally possible, but only if $N \lesssim 4$: at large $N$, $\Delta\phi \ll M_{\rm pl}$.
On the other hand, the requirement of a nearly-Planckian range in DBI inflation is a common finding, not a theorem, and it may be possible to find a compactification in which the potential satisfies (\ref{DBIgammacondition}) and (\ref{DBIaccelerate}) without violating the immutable bound (\ref{equ:BMboundX5}).

\item[$\triangleright$]  {\it  Bremsstrahlung.}---A further consistency requirement is that the speed limit felt by the D3-brane is dictated by the non-trivial kinetic term, not by other modes of dissipation.\footnote{While alternative means of shedding energy could give rise to interesting cosmological scenarios, the dynamics would not be governed by (\ref{equ:DBI}) alone, and much further analysis would be required.  See \S\ref{sec:Dissipation} for several examples of dissipative models.}  Because the D3-brane is necessarily relativistic, and is accelerated by the potential, it will emit gravitational and scalar synchrotron radiation into the compact dimensions.  This process was analyzed in \cite{Bachlechner:2013fja}, where it was shown that losses due to bremsstrahlung dominate the dynamics in a significant fraction of parameter space --- including the regime of weak string coupling and large volume --- but can be neglected in the remainder.

\item[$\triangleright$]  {\it  Excitation of massive strings.}---A very significant  obstacle to predictivity  in the IR model is that in the initial stage of expansion,  the Hubble scale exceeds  the local string tension, and massive open strings can be excited \cite{Chen:2004gc, Chen:2005ad, Chen:2005fe} (see the summary in \cite{Bean:2007eh}).   One can try to estimate  the corresponding perturbations \cite{Chen:2005ad, Chen:2005fe},  but a reliable computation is  inaccessible with current tools.   This problem cannot be  relegated to unobservably  large angular scales: the maximum number of $e$-folds $N_{e}^{\mathsmaller{\rm EFT}}$  that can be produced after the EFT computation of the perturbations becomes valid is \cite{Chen:2005ad}
\begin{equation}
N_{e}^{\mathsmaller{\rm EFT}} \approx \frac{N^{1/8}}{\sqrt{\beta}}\ ,
\end{equation}
where  as usual $N$  is the D3-brane charge of the throat and $\beta$ was defined in (\ref{IRV}).   Unless $N_{e}^{\mathsmaller{\rm EFT}} \gg 60$, the observed CMB  perturbations will be dictated by uncomputable fluctuations of massive strings.  While this is a fascinating possibility  that escapes the confines of the effective field theory of a finite number of fields,  no meaningful predictions are possible at present.
For generic  potentials with $\beta \sim 1$,  solving the horizon problem without  encountering uncomputable perturbations requires $N \gtrsim 10^{14}$;  moreover, because $\beta N_e \gg 1$ (with $N_e$ the number of $e$-folds) is required for a consistent DBI phase \cite{Bean:2007eh}, we must have $N \gtrsim 10^{7}$ regardless of the value of~$\beta$.   D3-brane charges of this magnitude are difficult to  realize in  compact spaces.

\end{itemize}

\subsection{Phenomenology}  \label{dbiph}

Despite the many apparent obstacles to realizing DBI inflation in a consistent string compactification, the DBI scenario is a leading example of a field-theoretic inflationary mechanism
that is underpinned by the symmetries of an ultraviolet theory.
Many authors have deferred the question of explicit ultraviolet completion, and directly investigated the rich phenomenology that follows from (\ref{equ:DBIX}) --- or a generalization, e.g.~with additional fields --- in the regime where $\gamma \gg 1$.
We will briefly sketch the signatures of DBI inflation that emerge from this approach.

\vskip 4pt
DBI inflation is a special case of so-called $P(X)$ theories (see \S\ref{sec:PX}) whose action is given by
\beq
  S = \int \d^4 x \sqrt{-g} \left[ \frac{\Mp^2}{2}R + P(X,\phi)  \right]\ ,
\eeq
where $P(X,\phi)$ is an arbitrary function of $X\equiv -\frac{1}{2}(\partial \phi)^2$ and $\phi$.  DBI inflation is recovered for
\beq
P(X,\phi) = - T(\phi) \left( \sqrt{1-\frac{2X}{T(\phi)}} - 1 \right) - V(\phi) \ . \label{equ:PXDBI}
\eeq
The phenomenology of $P(X)$ theories has been explored in \cite{Chen:2006nt}.

\begin{itemize}
\item[$\triangleright$]  {\it Scalar modes.}---The theory for the fluctuations in
$P(X)$ theories
maps to the effective Goldstone action of \S\ref{sec:Gold} with
sound speed given by
\beq
c_s^2 = \frac{P_{,X}}{P_{,X} + 2 X P_{,XX}} = \frac{1}{\gamma^2(\phi)} \ .
\label{soundspeed}
\eeq
The scalar power spectrum therefore follows from (\ref{equ:DZ}).
Substituting (\ref{equ:PXDBI}) into (\ref{soundspeed}), we get
\beq
c_s^2 = 1 -\frac{2X}{T(\phi)}\ .
\eeq
We see that a non-trivial warp factor can lead to a field dependence of the sound speed, $c_s(\phi)$. On CMB scales, the scale-invariance of the spectrum constrains the variation of the sound speed. On smaller scales, a significant evolution of the spectrum and hence
of the sound speed is still allowed.  Large-scale structure constraints on the running of the spectrum were analyzed in~\cite{Sefusatti:2009xu}.

\item[$\triangleright$]  {\it Tensor modes.}---The tensor-to-scalar ratio  in $P(X)$ theories is given by
\begin{equation}
r = 16 c_s \varepsilon \ ,  \label{DBItensor}
\end{equation}
and a generalized Lyth bound can be derived~\cite{Baumann:2006cd}:
\beq
\frac{\Delta \phi}{\Mp} = \int_{0}^{N_\star} \sqrt{\frac{r(N)}{8} \frac{1}{c_s P_{,X}}} \, \d N \ .
\eeq
We notice a non-trivial generalization of the slow-roll result (\ref{equ:LythIntegral})
 through the factor $c_s P_{,X}$.
However, the Lagrangian of DBI inflation (\ref{equ:PXDBI}) is algebraically special, satisfying
\beq
c_s P_{,X} = 1\ .
\eeq
The correspondence between $\Delta \phi$ and $r$ is therefore the same as for slow-roll inflation.
The geometric field range bound (\ref{equ:BMbound}) therefore also forbids large tensors in DBI inflation~\cite{Baumann:2006cd}.\footnote{The bound $r<10^{-7}/{\rm{Vol}}(X_5)$  can be derived by  combining (\ref{equ:BMbound})  with the {\it{assumption}}  that $\fnl^{\rm equil} \gtrsim 1$.  In other words, DBI inflation in a Calabi-Yau cone cannot  have both detectable non-Gaussianity and detectable tensors~\cite{Lidsey:2007gq}.}

\item[$\triangleright$]  {\it Equilateral non-Gaussianity.}---The cubic Goldstone action takes the form of (\ref{equ:S3}), with
\beq
A = c_s^2 \left( -1 - \frac{2}{3} \frac{X P_{,XXX}}{P_{,XX}} \right)\ . \label{equ:PXA}
\eeq
Substituting (\ref{equ:PXDBI}) into (\ref{equ:PXA}), we get\footnote{This result is  independent of the  potential and the warp factor, unlike other observables such as $n_s$.}
\beq
A = - 1\ .
\eeq
Mapped onto the equilateral and orthogonal templates (see \S\ref{sec:tests}), the size of the bispectrum for $P(X)$ theories is
\begin{align}
\fnl^{\rm equil} &= \big(-0.27 + 0.08 A \big)\, \frac{1-c_s^2}{c_s^2}  \ , \\
\fnl^{\rm ortho} &= \big(+0.02 + 0.02 A \big)\, \frac{1-c_s^2}{c_s^2}\ .
\end{align}
For the DBI Lagrangian, the orthogonal component is small and the amplitude of equilateral non-Gaussianity becomes
\beq
\fnl^{\rm equil} = - \frac{35}{108} \left( \frac{1}{c_s^2}  - 1\right) \approx  - \frac{35}{108}\gamma^2 \ ,  \label{DBIfnl}
\eeq
where in the second equality we have assumed the relativistic limit.
The Planck constraint,
$- 117 < \fnl^{\rm equil} < 33$
(68\% C.L.)~\cite{PlanckNG}, implies
\beq
\gamma \lesssim 24 \quad (95\%~{\rm C.L.})\  \label{equ:GCon}
\eeq

\item[$\triangleright$]  {\it Joint constraints for a quadratic potential.}---In  the important special case where  $V(\phi)  = \frac{1}{2} m^2\phi^2$, the above results  can be combined to  place strong limits on  the model parameters~\cite{Baumann:2006cd} (see also \cite{Lidsey:2007gq}).   For a quadratic potential, one has \cite{Silverstein:2003hf}
\begin{equation}
 2\left(\frac{M_{\rm pl}}{\phi}\right)^2 = \varepsilon\gamma =  \frac{1}{16} r \gamma^2\ ,   \label{quadraticDBI}
\end{equation}
where in the final equality we used (\ref{soundspeed}) and (\ref{DBItensor}).
From (\ref{DBIfnl}), one then finds \cite{Baumann:2006cd}
\begin{equation}
 2\left(\frac{M_{\rm pl}}{\phi}\right)^2 = \frac{27}{140} \, r\, \big|\fnl^{\rm equil} \big|\ .  \label{DBIcombined}
\end{equation}
Incorporating the geometric bound (\ref{equ:BMboundX5}), we  find the upper limit
\begin{equation}
N \, <\,  \frac{27}{70} \,r\, \big|\fnl^{\rm equil}\big|\, \lesssim\, 9 \quad (95\%~{\rm C.L.})\ .
\end{equation}
This small value of $N$  is in tension with the required amplitude of the scalar perturbations:  using (\ref{DBIcombined}) in (\ref{equ:DZ}) leads to \cite{Baumann:2006cd}
\begin{equation}
\Delta_{\cal{R}}^2(k_{\star}) = \left(\frac{32}{3\pi}\right)^2 \frac{3}{r^4 ({\fnl^{\rm equil}})^2} \frac{{\rm{Vol}}(X_5)}{N} \, \gtrsim\, 15 \hskip 1pt  \frac{{\rm{Vol}}(X_5)}{N} \ ,
\end{equation} where the inequality  uses the Planck upper limits on $r$ and $\fnl^{\rm equil}$.
Because $\Delta_{\cal{R}}^2(k_{\star})=2.2\times 10^{-9}$, we conclude that
\begin{equation}
N \, \gtrsim\, 7\times 10^9~{\rm{Vol}}(X_5)\ .
\end{equation}
Quadratic DBI inflation  in a Calabi-Yau cone is therefore virtually excluded.  One would need an extremely small angular  manifold $X_5$, e.g.~by orbifolding  by a large discrete group, while also keeping $N \lesssim 9$,  which almost certainly renders the supergravity  approximation  invalid.

\item[$\triangleright$]  {\it Multi-field effects.}---So far, our discussion of DBI inflation has been restricted to purely radial evolution of the D3-brane,  but in general the potential will depend on the angular coordinates. Including these effects leads to multi-field models of DBI inflation,\footnote{For an analysis of DBI inflation with $N$ D3-branes, see \cite{Ward:2007gs}.} whose phenomenology has been studied comprehensively in~\cite{Easson:2007dh,Huang:2007hh, Langlois:2008qf, Langlois:2008wt, Langlois:2009ej, RenauxPetel:2009sj, Langlois:2008mn}.
     When one or more of the angular fields  are light during inflation, their  quantum fluctuations lead to entropy perturbations, which propagate with the same speed of sound $c_s$ as the adiabatic mode.  In the limit $c_s \ll 1$, the amplitude of the entropy perturbations  is boosted relative to the adiabatic fluctuations.  The amplitude of the bispectrum can be suppressed,
\beq
\fnl^{\rm equil}  \approx  - \frac{35}{108}\gamma^2 \cos^2 \Theta\ ,
\eeq
where the angle $\Theta$ parameterizes how much of the final curvature perturbation
 arises from entropy perturbations,\footnote{The transfer from entropy perturbations to curvature perturbations can depend on the physics of reheating.
Investigations of the distinctive features of reheating in DBI inflation include \cite{Leblond:2006cc,Bouatta:2010bp,Zhang:2013asa}.}
with $\Theta =0$ if there is no transfer of entropy modes and $\Theta = \pi/2$ if the final curvature perturbation is mostly of entropic origin. This can relax the constraint (\ref{equ:GCon}) on $\gamma$.

\end{itemize}

\section{Inflating with Axions}
\label{sec:AxionInflation}

String axions are promising inflaton candidates.
Equipped with a continuous shift symmetry to all orders in perturbation theory,
the axion potential is stable against radiative corrections.
Weakly breaking the symmetry---either spontaneously by nonperturbative  effects, or explicitly through the presence of branes---can lead to inflationary theories that are natural in the bottom-up sense of~\S\ref{ssec:natural}.
In {\it natural inflation}~\cite{Freese:1990rb,Adams:1992bn}, the role of  the inflaton is played by a single axion $\phi$  with the Lagrangian~(\ref{equ:La}),
\beq
{\cal L}(\phi) =  - \frac{1}{2} (\partial \phi)^2 - \Lambda^4 \left[ 1 - \cos\left(\frac{\phi}{f}\right)\right] + \cdots \ ,  \label{naturalinflationLagrangian}
\eeq where $f$ is the axion decay constant, and $\Lambda$  is a dynamically-generated scale. To facilitate comparison with the literature we have redefined the decay constant by a factor of $2\pi$ compared to \S\ref{sec:axions}, with $f_{\rm here} = 2\pi f_{\rm there}$.

For sufficiently large  values of $f/M_{\rm pl}$, the model (\ref{naturalinflationLagrangian}) gives rise to prolonged inflation, and for
$f \gtrsim 10 \hskip 1pt M_{\rm pl}$,  the  spectral index $n_s$   is compatible with the constraints from the Planck  mission.\footnote{The constraint on the decay constant depends on the choice of prior. The result we have cited here is for a uniform prior on $\log(f)$~\cite{PlanckInflation}.}
However, as we have reviewed in \S\ref{sec:axions}, super-Planckian decay constants  have not  been obtained  to date  in a controlled string compactification.
Natural inflation is therefore an interesting example of the importance of  explicit ultraviolet completion. Although inflation with a single axion and super-Planckian decay constant is natural from the bottom-up perspective, it does not seem to be natural from the top down, unless additional structures are present.
In the rest of this section, we will discuss  a few of the leading proposals for what these extra structures might be.

\subsection{Inflation with Multiple Axions}
\label{nflationsection}

One strategy is to extend (\ref{naturalinflationLagrangian}) by introducing one or more additional axions \cite{Kim:2004rp,Dimopoulos:2005ac}, each with a sub-Planckian decay constant,  and arrange that a  combination of these fields  effectively enjoys a super-Planckian  decay constant.   We will describe two  mechanisms: `alignment'  of two axions \cite{Kim:2004rp}, and assisted inflation \cite{Liddle:1998jc} with $N \gg 1$  axions \cite{Dimopoulos:2005ac}.

\subsection*{Two axions}

Consider two axion fields $\phi_1$ and $\phi_2$  with decay constants $f_1$ and $f_2$, respectively.
Suppose that these axions couple to  linear combinations of two confining non-Abelian gauge groups $a$ and $b$,  with the following Lagrangian density \cite{Kim:2004rp}
\begin{equation}  \label{twogroups}
{\cal L} \ \supset\ \sum_{i=1}^2\,  \frac{\phi_i}{f_i} \left( \frac{c_{ia}}{32\pi^2}\, {\rm Tr}\left[F^{(a)}\wedge F^{(a)}\right] + \frac{c_{ib}}{32\pi^2}\,{\rm Tr}\left[F^{(b)}\wedge F^{(b )}\right]\right) \ ,
\end{equation}
where the coefficients $c_{ia} = \{c_{1a}, c_{2a}\}$ and $c_{ib} = \{c_{1b}, c_{2b}\}$ are dimensionless.
In terms of the dynamical scales $\Lambda_a$ and $\Lambda_b$ of the two gauge groups, the potential for the axions is 
\begin{equation}
V = \Lambda_a^4 \left[ 1 - \cos\left( c_{1a}\frac{\phi_1}{f_1} + c_{2a}\frac{\phi_2}{f_2}\right)\right] + \Lambda_b^4 \left[ 1 - \cos\left( c_{1b}\frac{\phi_1}{f_1} + c_{2b}\frac{\phi_2}{f_2}\right)\right] \ .
\end{equation}
The central observation of \cite{Kim:2004rp} is that if
\begin{equation}
\frac{c_{1a}}{c_{2a}} = \frac{c_{1b}}{c_{2b}} \ ,   \label{cratio}
\end{equation} then one linear combination $\xi$ of the axions is unlifted,  and effectively has infinite range.\footnote{A related mechanism  is used in racetrack constructions---see \S\ref{racetrack}.}
When (\ref{cratio}) is approximately  satisfied, $\xi$ has  a decay constant $f_{\xi}$  that can be much larger than $f_1$ and $f_2$.  In particular, for sufficiently precise alignment,  one can have $f_{\xi} > M_{\rm pl}$ with $f_1,~f_2 \ll M_{\rm pl}$.  For the simple case
where $c_{1a}=c_{1b}=1$ and $\Lambda_a^4 \gg \Lambda_b^4$, one finds
\begin{equation}
f_{\xi} =   \frac{\left(c_{2a}^2 f_1^2 + f_2^2 \right)^{1/2}}{\big|c_{2b}-c_{2a}\big|} \ , \qquad {\rm with} \quad \ \
\xi = \frac{\phi_2 f_2 - c_{2a} \phi_1 f_1}{c_{2a}^2 f_1^2 + f_2^2} \ .
\end{equation}

The relation (\ref{cratio}) is plausibly radiatively stable.
However, one should bear in mind that it only ensures a flat direction if we assume that the particular linear combination of axions $\xi$ is unlifted by the leading  instanton effects in the gauge groups given in~(\ref{twogroups}).  Whether other effects lift the flat direction is an important question: in particular, it would be  valuable to construct an explicit example in a stabilized string compactification  and ascertain whether any nonperturbative effects  associated with moduli stabilization, which might be unrelated to the confining gauge groups in (\ref{twogroups}), spoil the flatness  of the inflaton direction.  (For a closely related idea, see \cite{Berg:2009tg}, as discussed in \S\ref{ssec:AM} below.)

\subsection*{Many axions: N-flation}

A different idea  for axion inflation takes advantage of the fact that string compactifications often come with large numbers of axion fields: perhaps the collective excitations of hundreds of axions, each with a sub-Planckian decay constant, can yield an effective decay constant of super-Planckian size.\footnote{This section is based on~\cite{Dimopoulos:2005ac, Easther:2005zr, Grimm:2007hs}.}
This idea is called {\it N-flation}~\cite{Dimopoulos:2005ac} and is based on an earlier proposal of {\it assisted inflation}~\cite{Liddle:1998jc}.\footnote{A different approach to assisted inflation  in string theory is M-flation \cite{Ashoorioon:2009wa,Ashoorioon:2009sr}.}

\vskip 4pt
Consider $N$ axions whose Lagrangian is simply $N$ copies of (\ref{naturalinflationLagrangian}),
\beq
{\cal L} =  \sum_{i=1}^N \left[  - \frac{1}{2}(\partial \phi_i)^2 - V_i( \phi_i) \right] \ ,  \label{NflationLfull}
\eeq
where  
$V_i( \phi_i) \equiv \Lambda_i^4 \big[ 1- \cos(\phi_i/f_i) \big]$.
We have assumed that  cross-couplings in the axion potential are negligible, as  discussed below.
Each individual axion therefore experiences a force only from its own potential~$V_i$, but Hubble friction from the sum of all potentials $\sum_i V_i$,
\beq
\ddot{\phi}_i + 3 H \dot{\phi}_i  = - \partial_i V_i \ , \qquad {\rm where} \quad 3 M_{\rm pl}^2 H^2 \approx \sum_{i=1}^N V_i \ .
\eeq
Compared to the single axion case, each individual axion feels enhanced Hubble friction, suggesting that one might be able to achieve a friction-dominated situation even without any axion being at a super-Planckian distance from the minimum.

Let us begin by considering the simple case in which all the axions have equal masses $m_i  = \Lambda_i^2/f_i \equiv m$.
Moreover, let us consider small displacements from the minimum, 
$\phi_i \ll f_i$.
The collective excitation $\Phi^2 \equiv \sum_i \phi_i^2$
then has the
potential
\beq
V(\Phi) = \frac{1}{2} m^2 \Phi^2 \ .
\eeq
Successful inflation requires $\Phi > M_{\rm pl}$, but this does not mean that the individual axion vevs $\phi_i$ have to be large:
for sufficiently large $N$, the individual displacements can be sub-Planckian, $\phi_i \approx \Phi /\sqrt{N} < M_{\rm pl}$.  In typical examples the required number of axions is
$N \gtrsim {\cal O}(10^3)$ \cite{Dimopoulos:2005ac}.

One might object that  quantum gravity constraints should  restrict the range of $\Phi$  to be sub-Planckian,  just like the ranges of the $\phi_i$, because in moving from  individual displacements $\phi_i$ to the collective field $\Phi$, one has merely changed from Cartesian to  spherical polar coordinates,  which should be immaterial  unless some physical effect  is sensitive to the change of coordinates.  This concern is unfounded:  the $N$  discrete axionic shift symmetries 
$\phi_i \mapsto \phi_i + 2\pi f_i$ that persist  at the nonperturbative level do in fact define a preferred coordinate system,  the Cartesian one.  (After the periodic identifications,  the axion field space is an $N$-torus,  with  individual radii $f_i <  M_{\rm pl}$.)
In string theory constructions (see below),  only the $f_i$  are directly constrained.

A  much graver concern is that loops of the $N$ light axion fields renormalize the Planck mass:  on general grounds one expects \cite{Dimopoulos:2005ac}\footnote{A counterpoint to this finding appears in  \cite{Ashoorioon:2011aa}, where it is argued that the portion of the eta problem  arising from loops of light fields  is actually suppressed  at large $N$.}
\begin{equation}
\delta   M_{\rm pl}^2 \sim \frac{N}{16\pi^2}\, \Lambda_{\mathsmaller{\rm UV}}^2 \ , \label{renormalizationmp}
\end{equation}  where $\Lambda_{\mathsmaller{\rm UV}}$ is the ultraviolet cutoff.
Because the collective field displacement scales as $\Phi \propto \sqrt{N} \phi$, with $\phi$ the mean of the individual displacements, while the correction to the Planck mass  in (\ref{renormalizationmp})  also scales as $\sqrt{N}$,  we conclude that  one
cannot  obtain a parametrically
super-Planckian displacement  purely by working at large $N$.  Instead, one must grapple with the ultraviolet-sensitive details, e.g.~by refining the field-theoretic estimate (\ref{renormalizationmp})  through a computation in string theory.
To learn more, we now turn to a string theory realization of  N-flation~\cite{Dimopoulos:2005ac, Easther:2005zr, Grimm:2007hs}.

\vskip 4pt
\noindent
{\it N-flation in type IIB string theory.}---As an explicit realization of N-flation  in string theory, we consider a KKLT compactification in which $h^{1,1}_{+} \equiv N \gg 1$ complexified K\"ahler moduli $T_i$ are stabilized by nonperturbative effects.
As explained in \S\ref{sec:compactification}, the associated axions $\vartheta_i$, $i=1,\ldots, N$, correspond to the integrals of $C_4$  over orientifold-even four-cycles, cf.~eq.~(\ref{IIBaxions}).
For  simplicity of presentation, we take $h^{1,1}_{-}=0$, so that by (\ref{equ:Ti}), $T_i =\tau_i + i\vartheta_i$, with $\tau_i$ a real four-cycle volume.\footnote{Another interesting and rather explicit construction with similar qualitative features  works with $h^{1,1}_{-} \equiv N \gg 1$, taking the  inflationary axions to  arise from the dimensional reduction of the R-R two-form potential, cf.~eq.~(\ref{gscalar})~\cite{Grimm:2007hs}.
One advantage of the  model of~\cite{Grimm:2007hs} is that it is  comparatively straightforward to arrange that the K\"ahler moduli masses are larger than the inflaton mass.}

The superpotential takes the form (\ref{KKLTW}),
\beq
W = W_0 + \sum_{i=1}^N {\cal A}_i\hskip 1pt e^{-a_i T_i}  \ ,  \label{NflationW}
\eeq
where $a_i = 2\pi$ (for Euclidean D3-branes) or $a_i = 2\pi/N_i$ (for gaugino condensation in a gauge group with dual Coxeter number $N_i$).
Note that the axions  $\vartheta_i$ appear in the phase of each nonperturbative term.
In view of (\ref{K0}), the K\"ahler potential has a complicated dependence on the K\"ahler moduli~$T_i$:
\beq
K = -2 \ln\left[ {\cal V}(T_i,\bar T_i)\right]\ , \label{NflationK}
\eeq
where ${\cal V}$ is the volume of the compact Calabi-Yau manifold.
The ${\cal N} =  1$ supergravity theory defined by (\ref{NflationW}) and (\ref{NflationK}) has $N$ light chiral superfields,\footnote{At the energy scales of interest, the dilaton and the complex structure moduli are already stabilized, and $W_0$ and ${\cal A}_i$ are constants.} and generically admits supersymmetric $AdS_4$  solutions.   Introducing an anti-D3-brane  following \cite{Kachru:2003aw},  one  can find a solution with a small positive cosmological constant.   The   anti-D3-brane potential  energy has negligible dependence on the axions~$\vartheta_i$,  and will be treated as a constant for the purpose of axion inflation.

\begin{figure}[h!]
   \centering
     \includegraphics[scale=0.45]{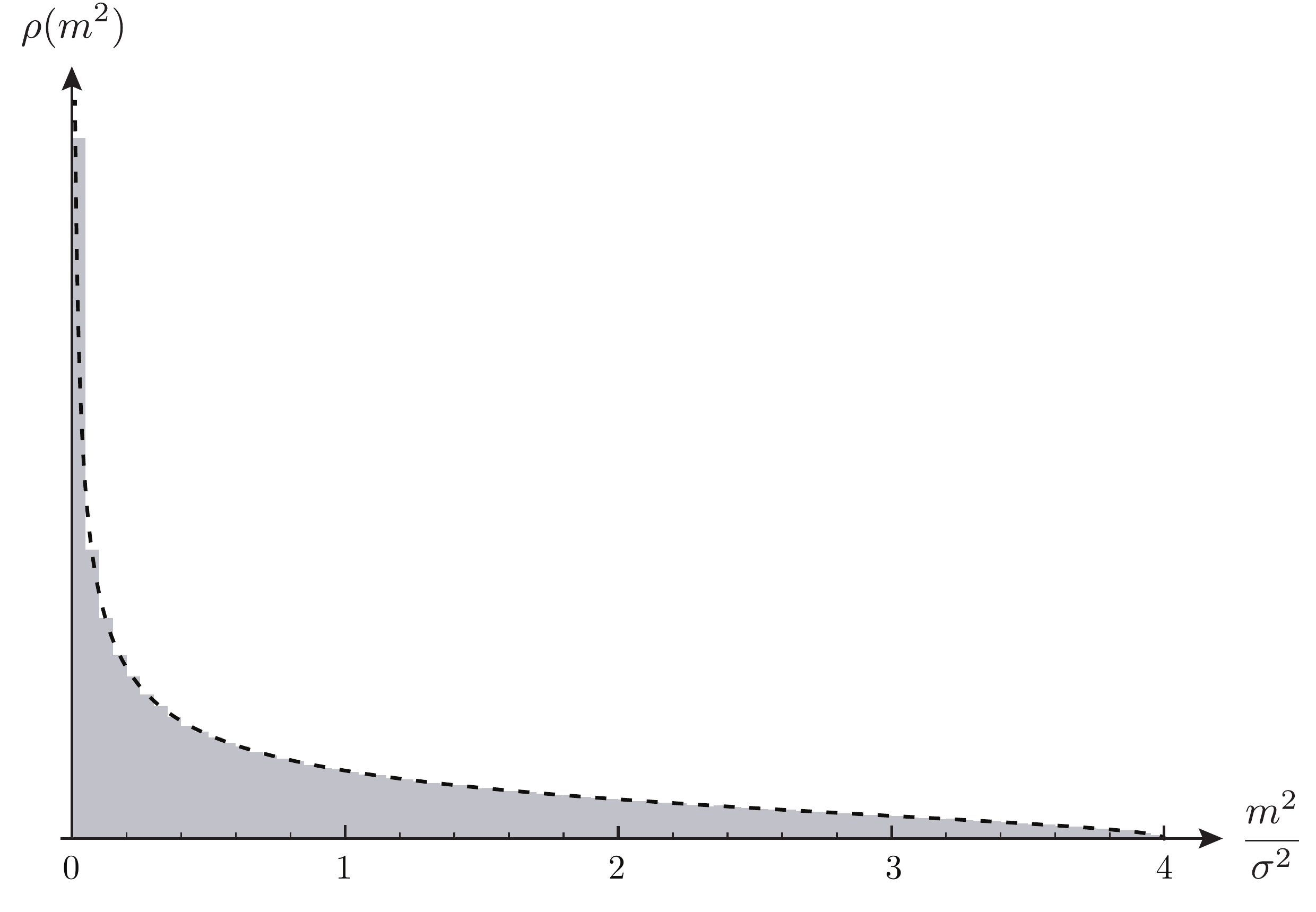}
     \vspace{-0.3cm}
   \caption{The eigenvalue spectrum of a Wishart matrix, given by the Mar\v{c}enko-Pastur law (\ref{MPspectrum})  (figure adapted from \cite{Marsh:2011aa}).
   The  spectrum of axion masses  around a KKLT minimum  is well-described  by this law when $N \equiv h^{1,1}_{+} \gg 1$.
   The curve  is the analytic result, while the histogram is the result of simulations,  both for $N=P=200$.}
  \label{fig:MP}
\end{figure}

We now expand around the minimum, in small fluctuations of the (dimensionless) axions $\vartheta_i$:
\beq
{\cal L} = - \frac{1}{2} M_{\rm pl}^2  K_{ij} \partial_\mu \vartheta^i\partial^\mu \vartheta^j -  \frac{1}{2} M_{ij} \vartheta^i \vartheta^j + \cdots \ . \label{equ:LN}
\eeq
The mass matrix   $  M_{ij}$ is determined by the superpotential (\ref{NflationW}), the K\"ahler potential (\ref{NflationK}) and their derivatives---see \cite{Easther:2005zr}  for the explicit expression---and depends on the vevs of the (real) K\"ahler moduli $\tau_i$. In a generic basis, the axions will be cross-coupled both in their kinetic terms and in the potential: neither $K_{ij}$  nor  $  M_{ij}$  will be diagonal.
Because of the axion shift symmetry, the K\"ahler metric $K_{ij}$ is independent of the $\vartheta^i$, up to nonperturbatively small corrections.
One can always perform a change of coordinates to set $K_{ij} \mapsto \delta_{ij}$,  which rotates and rescales the axion fields.
Denoting the new, canonically-normalized   axion fields  by $\phi_i$ and performing a  further orthogonal rotation to diagonalize the mass matrix, we arrive at
\beq
{\cal L} = \sum_{i=1}^N \left[ - \frac{1}{2} (\partial \phi_i)^2 - \frac{1}{2}m_i^2\,\phi_i^2\right] \ .  \label{NflationL}
\eeq
The masses $m_i$ of the decoupled fields $\phi_i$  have a rather complicated dependence on $W_0, {\cal A}_i, \tau_i$.
Fortunately, at large $N$, random matrix theory  yields a simple expression for the axion mass spectrum~\cite{Easther:2005zr}.  The mass matrix belongs to the
Wishart ensemble described in \S\ref{sec:RMTsupergravity}, and the mass spectrum is given by the  Mar\v{c}enko-Pastur law (\ref{Wishartspectrum}),
\begin{equation}
\rho(m^2) =  \frac{1}{2\pi N \sigma^2 m^2}\sqrt{(\eta_+-m^2)(m^2-\eta_-)} \ . \label{MPspectrum}
\end{equation}
Here, $\eta_\pm \equiv N \sigma^2 (1\pm\sqrt{P/N} \hskip 2pt)^2$, where, as above, $N=h^{1,1}_{+}$, while $P=h^{2,1}_{-}$  is the number of complex structure moduli,
and $\sigma$  controls the typical scale of terms in the moduli potential---see \cite{Easther:2005zr} for a detailed explanation.\footnote{One further assumption  implicit in obtaining (\ref{MPspectrum}) in a KKLT compactification is that the gravitino mass $m_{3/2}$  is small compared to the scale of supersymmetric masses: see \cite{Marsh:2011aa,Bachlechner:2012at} for discussions of how this could be achieved.}  The  eigenvalue spectrum (\ref{MPspectrum}) is shown in fig.~\ref{fig:MP}.

Because the axion decay constants $f_i$  are proportional to the eigenvalues of the K\"ahler metric $K_{ij}$, the distribution of decay constants  in a class of string compactifications could be determined in much the same way that the distribution of masses $m_i$  was obtained above.

\vskip 4pt
\noindent
{\it Microscopic challenges.}---Two  significant issues make it difficult to realize N-flation  in a stabilized string compactification.
\begin{itemize}
\item[$\triangleright$]  {\it Light K\"ahler moduli.}---One obstacle faced by any  realization of N-flation  in a string compactification with spontaneously broken supersymmetry is that the axions are partnered with real scalar fields, by ${\cal N}=1$  supersymmetry:  in the KKLT example given above, these are the  four-cycle volumes $\tau_i$.
Arranging that the real moduli have masses above the Hubble scale, $m \gtrsim H$,  and hence are (barely)  frozen during inflation, is challenging \cite{Kallosh:2007cc}.
Highly contrived configurations, e.g.~involving many racetrack  terms in the superpotential,  could in principle  give large masses to these fields (see the discussion in \cite{Marsh:2011aa}),  but no examples have been constructed.
If supersymmetry is not spontaneously broken during inflation, but is instead broken at a higher scale,  then the real partners of the axions can be decoupled, but control of the potential becomes more difficult \cite{Cicoli:2014sva}.

\item[$\triangleright$]  {\it Renormalization of the Planck mass.}---The problem of the renormalization of the Planck mass (\ref{renormalizationmp})  is not automatically alleviated in string theory, but it takes a more precise form, as we now explain in the example of a KKLT realization of N-flation.
 The leading correction to the four-dimensional Planck mass  that  scales with the number of axions is (\ref{IIBgravityexpansion}), the four-loop $\sigma$-model correction  to the ten-dimensional action.
Upon dimensional reduction, one finds the Einstein-Hilbert term \cite{Dimopoulos:2005ac}
\begin{equation}
{\cal{L}} = \frac{M_{\rm pl}^2}{2}\Biggl(1+\frac{\zeta(3) \chi(X_6)}{(2\pi)^3} \, \frac{(\alpha^{\prime})^3}{ {\cal V}}+ \cdots\Biggr)~R_4 \ ,  \label{leadingrenorm}
\end{equation}
where $\chi(X_6)$ is the Euler  characteristic of the compactification manifold $X_6$,
\begin{equation}
\chi(X_6)= 2 h^{1,1}-2 h^{2,1} \ ,
\end{equation} and the omitted terms are higher order in $\alpha^{\prime}$  and/or in $g_{\rm s}$.
The renormalization of the Planck mass  encoded in (\ref{leadingrenorm})   indeed has the same scaling with $N$
as the field theory result\footnote{A potential confusion is that the leading correction in (\ref{leadingrenorm})  arises as a loop effect on the worldsheet,  not from loops  of light  moduli in spacetime,  but nevertheless involves the number of moduli.   As such, it  corresponds to a non-renormalizable term in the effective theory, as in \S\ref{etaII},  rather than a radiative correction as in (\ref{renormalizationmp}), cf.~\S\ref{etaI}.  However, corrections at higher order in $g_{\rm s}$  may be expected to involve actual loops of the light fields,  in closer analogy to (\ref{renormalizationmp}).}  (\ref{renormalizationmp}), {\it{unless}} $h^{1,1}-2 h^{2,1} \ll h^{1,1}$.
Thus,  for suitable Hodge numbers,  the correction to the Planck mass from (\ref{leadingrenorm}) can be neglected.   However, there are higher-order  corrections in $\alpha^{\prime}$, and in $g_{\rm s}$, whose form is not known:  if these are
also proportional to $\chi(X_6)$,  then renormalization of the Planck mass is harmless in compactifications with $h^{1,1}-2 h^{2,1} \ll h^{1,1}$, but otherwise the problem plausibly reappears.\footnote{One cannot  work at arbitrarily weak coupling and large volume, because in this limit the individual decay constants~$f_i$  are parametrically small compared to $M_{\rm pl}$: see \S\ref{sec:axions}.}
More detailed understanding of  ultraviolet-sensitive quantum corrections  would be needed to resolve this issue.
\end{itemize}

\vskip 4pt
\noindent
{\it Phenomenology.}---A number of authors have studied  the  signatures of the phenomenological model (\ref{NflationLfull}),  anticipating a microphysical realization: see e.g.~\cite{Easther:2005zr,Kim:2006ys,Piao:2006nm,Kim:2006te,Battefeld:2006sz,Kim:2007bc,Battefeld:2007en,Battefeld:2008bu,Adshead:2008gk,Kim:2010ud,Kim:2011jea}.
If a quadratic approximation is applicable, then the tensor-to-scalar ratio is given by $r \approx 0.13$  (for 60 $e$-folds of inflation).

\subsection{Axion Monodromy Inflation}
\label{ssec:AM}

Another idea to extend the effective axion range uses the phenomenon of {\it monodromy}.\footnote{This section is based on \cite{Silverstein:2008sg, McAllister:2008hb}.}
We speak of monodromy when a system reaches a new configuration after being transported around a closed loop in the (naive) configuration space.
The classic example is a spiral staircase, where the naive configuration space is a circle, but the system changes upon transport by $2\pi$: after each circuit we reach a higher level on the staircase.   Something very similar occurs in the scalar potential for axions in  certain string compactifications:  the potential energy  continues to increase as the axion  traverses multiple  circuits of its  fundamental domain.   The basic idea of monodromy inflation \cite{Silverstein:2008sg}  is that inflation can persist  through many cycles  around the configuration space.
The effective field range is then  much larger than the fundamental period, but the axion shift symmetry protects the structure of the potential  over each  individual cycle \cite{Silverstein:2008sg, McAllister:2008hb}.

Monodromies are widespread in string compactifications,  but constructing an explicit model of axion monodromy inflation is delicate, as we now explain.

\vskip 4pt
\noindent
{\it Axion potentials.}---We will begin with a simple example: a D5-brane in type IIB string theory~\cite{McAllister:2008hb}.\footnote{The first example of monodromy inflation constructed in string theory involves a D4-brane in a nilmanifold  compactification of type IIA  string theory~\cite{Silverstein:2008sg}, and relies on a scenario for moduli stabilization  that is rather different from
that presented in \S\ref{modulistab}.  We will discuss  the  scenario of \cite{Silverstein:2008sg} in \S\ref{sec:trapped2}.}
The brane fills the four-dimensional spacetime and wraps a two-cycle $\Sigma_2$ in the compact space. The axion $b \equiv \frac{1}{\alpha'} \int_{\Sigma_2} B_2$ will exhibit monodromy in the potential energy, i.e.~the potential energy of the wrapped brane is {\it not} a periodic function of the axion.
To see this, consider the DBI action~(\ref{equ:DBIXX}) for the D5-brane,
\beq
S_{{\rm D}5} = \frac{1}{(2\pi)^5 g_{\rm s} (\alpha')^3} \int\limits_{{\cal M}_4 \times \Sigma_2} \hspace{-0.2cm} \d^6 \sigma \, \sqrt{-{\rm det}(G_{ab} + B_{ab})} \ .
\eeq
Performing the integral over the two-cycle,
we obtain the potential for the axion in the four-dimensional effective theory:
\beq
V(b) = \frac{ \varrho}{(2\pi)^6  g_{\rm s} (\alpha')^2}\, \sqrt{(2\pi)^2 \ell^4 + b^2} \ ,
\eeq
where $\ell$ is the size of $\Sigma_2$ in string units, and the dimensionless number $\varrho$ encodes a possible dependence on the warp factor.  The presence of the brane has broken the axion shift symmetry, $b \mapsto b + (2\pi)^2$.
We say that the brane has generated a monodromy for the axion.  For large values of the axion vev,
$b \gg \ell^2$, the potential is linear, $V(b) \propto b$.

A similar effect occurs if the D5-brane is replaced by an NS5-brane.
The wrapped NS5-brane now produces a monodromy for the axion $c \equiv \frac{1}{\alpha'} \int_{\Sigma_2} C_2$. Dimensional reduction of the action for the NS5-brane introduces the following potential for the $c$ axion
\beq
V(c) =  \frac{\varrho}{(2\pi)^6  g_{\rm s}^2 (\alpha')^2} \, \sqrt{(2\pi)^2 \ell^4 + g_{\rm s}^2 c^2} \ .  \label{ns5dbi}
\eeq

\vskip 4pt
\noindent
{\it Axion monodromy inflation.}---In both cases, inflation
can occur if the axion ($b$ or $c$) has a large initial vev.
The Lagrangian for the canonically-normalized field is
\beq
{\cal L} = - \frac{1}{2}(\partial \phi)^2 - \mu^3 \phi\ , \qquad {\rm with} \quad \mu^3  \equiv \frac{1}{f}\, \frac{\varrho}{(2\pi)^6 g_{\rm s} (\alpha')^2} \ ,  \label{axionlinear}
\eeq
where $f$ is the decay constant of the corresponding axion, as defined in \S\ref{sec:axions}.
During inflation the worldvolume flux on the wrapped fivebrane decreases, and the axion vev drops.
For large  initial vev, the axion moves a large effective distance in field space.
Provided that the axion shift symmetry  continues to protect the potential  across super-Planckian displacements,  the result is a  natural model of chaotic inflation  in string theory.  (We critically discuss the stability of the potential in the remainder of this section.)  Inflation ends  at a small axion vev,  at which point the  inflaton starts to oscillate around the minimum of the potential.  Couplings between the axion and other fields will drain energy from the inflaton sector. If this energy transfer happens predominantly to the visible-sector degrees of freedom, then the system successfully reheats and the hot Big Bang is initiated (see \S\ref{sec:AxionReheating}).

Although axion monodromy from an NS5-brane source  yields the asymptotically linear  potential~(\ref{axionlinear}),  many other variants of chaotic inflation  can arise via monodromy.
One can parameterize the  resulting theories in terms of an  exponent $p$  as
\beq
{\cal L} = - \frac{1}{2} (\partial \phi)^2 - \mu^{4-p} \phi^p \ .
\eeq
The nilmanifold monodromy scenario of~\cite{Silverstein:2008sg} yields $p=2/3$, while versions of axion  monodromy inflation with  $p=2$
can arise from an appropriate coupling of an
axion to a four-form field strength~\cite{Kaloper:2008fb},\footnote{For related field-theoretic constructions, see \cite{Kaloper:2011jz,Dubovsky:2011tu}.} or on a pair of sevenbranes \cite{Palti:2014kza}.

A general mechanism  known as {\it{flattening}}~\cite{Dong:2010in} can affect the asymptotic form of the scalar potential for a light field $\phi$ in the presence of additional heavy fields $\Psi$.   Given appropriate couplings of  $\phi$ to $\Psi$, integrating out  $\Psi$ flattens $V(\phi)$  at large  $\phi$, in the sense of reducing the exponent~$p$.
It was argued in \cite{Dong:2010in} that the linear potential of (\ref{axionlinear}) is an example of flattening:
the type IIB  action includes terms proportional to
\begin{equation}
S \supset \int \d^{10} X\, |C_2 \wedge H_3|^2 \ ,
\end{equation}  which  naively  give rise to an energy that is quadratic in $\phi \propto \int C_2$,  but the  actual potential (\ref{axionlinear}) is linear.
The claim of \cite{Dong:2010in} is that backreaction  of localized D3-brane charge, which shifts the moduli vevs, is responsible for the flattening from $p=2$  to $p=1$.

\subsection*{Compactification and Tadpole Cancellation}

The attentive reader will appreciate by now that it is essential to check that a proposed inflationary mechanism in string theory---and any symmetries that underlie it---survives compactification and moduli stabilization.   We turn to a critical discussion of these issues.

\begin{figure}[h!]
   \centering
     \includegraphics[scale=0.25]{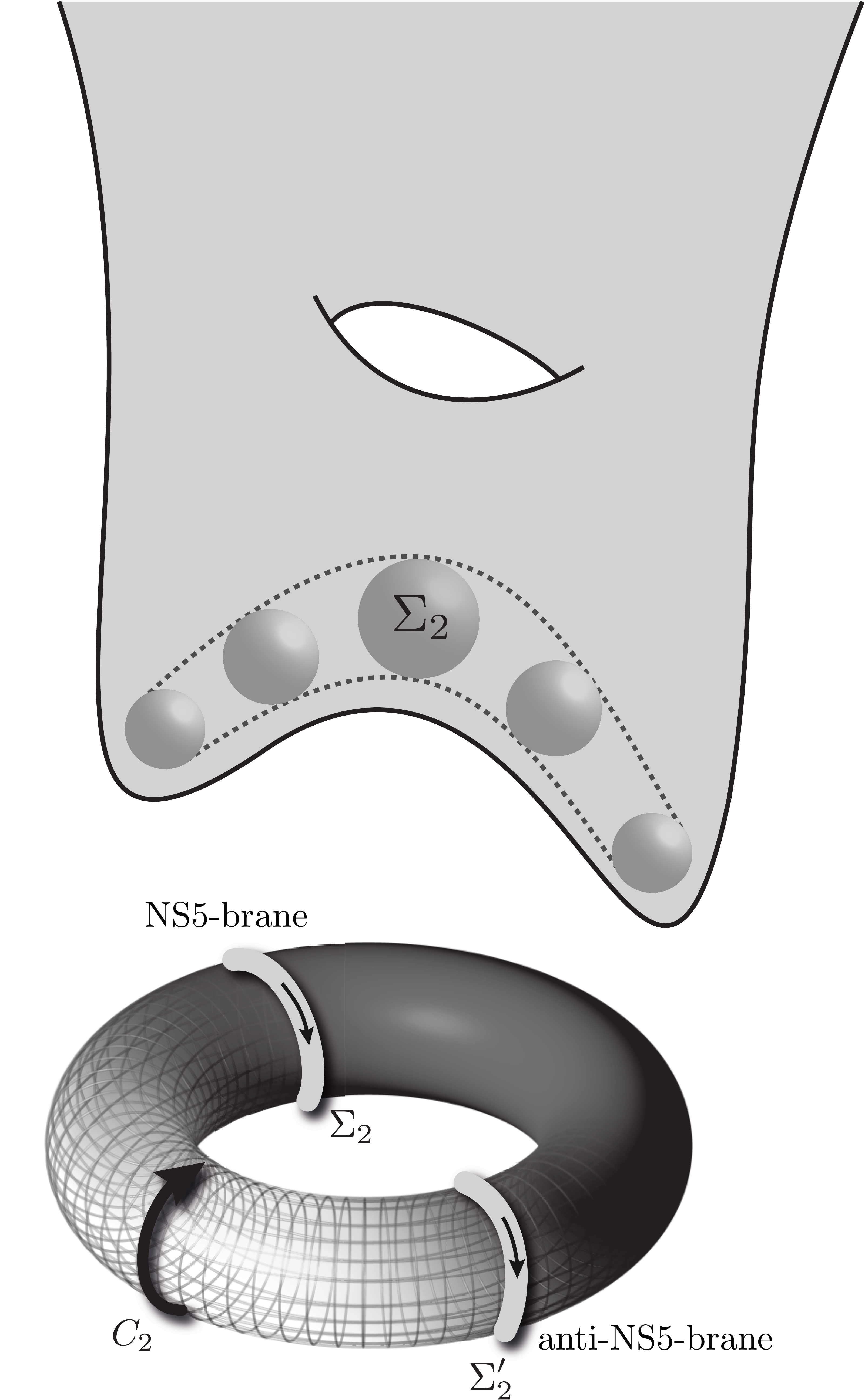}
     \vspace{-0.3cm}
   \caption{The integral of the two-form $C_2$ over a two-cycle $\Sigma_2$ defines the $c$ axion. In the presence of a wrapped NS5-brane this develops a monodromy. An anti-NS5-brane is required by Gauss's law on the compact space.  The entire configuration should be situated in a warped region, and have a distant orientifold image (not shown). In the lower figure, the two-cycles are represented by circles.}
  \label{fig:Tadpole}
\end{figure}

A fundamental  consistency requirement in a compact model is cancellation of all tadpoles.    Changing the axion vev, $b$,  in the presence of a D5-brane  alters the D3-brane charge induced on the  D5-brane, because of the Chern-Simons coupling (\ref{equ:NCS}):
\begin{equation}
S_{\rm{CS}} \, \supset\, i \mu_{5} \int\limits_{{\cal{M}}_4 \times \Sigma_2} C_4 \wedge {\cal{F}}_{2}\ .
\end{equation}
That is, because a D3-brane has the Chern-Simons coupling
\begin{equation}
S_{\rm CS}^{(\rm D3)} = i \mu_3 \int_{{\cal{M}}_4} C_4\ ,
\end{equation}
a D5-brane wrapping $\Sigma_2$, with
\begin{equation}
\int_{\Sigma_2} {\cal{F}}_{2} \neq  0\ ,
\end{equation} carries non-vanishing D3-brane charge.
Without a  mechanism for absorbing or canceling this induced charge,  Gauss's  law would fix $b$  to  one definite value.
However, there is a natural  configuration in which the induced charge is canceled automatically: instead of a  single D5-brane  on a two-cycle $\Sigma_2$, consider a D5-brane and an  anti-D5-brane, each wrapping $\Sigma_2$, but at different locations in the compact space,\footnote{That is, the D5-brane and anti-D5-brane wrap distinct, well-separated representatives of the same homology class.} as in the constructions of \cite{Aganagic:2006ex}.  A similar construction applies for an  NS5-brane pair (see fig.~\ref{fig:Tadpole}),  which, as we will see, yields a more promising inflationary model.

\subsection*{Symmetry Breaking from Nonperturbative  Effects}
\label{sec:CE}

To analyze the impact of moduli stabilization on axion monodromy inflation, it is necessary to specify a scenario for stabilization.  Here, we will discuss axion monodromy inflation on fivebranes in type IIB compactifications with nonperturbatively-stabilized K\"ahler moduli.

\vskip 4pt
\noindent
{\it Eta problem for the $b$ axion.}---To assess axion monodromy inflation with $b$  as the inflaton, we consider a KKLT compactification with $h^{1,1}_+=h^{1,1}_-=1$, and correspondingly a single K\"ahler modulus $T$ and a single complex `two-form scalar' $G = c -\tau b$.
The ${\cal N}=1$ supergravity data obtained from dimensional reduction takes the form
\begin{align}
W &= W_0 + {\cal A} \hskip 1pt e^{-2\pi T}\ , \label{wnp1} \\
K &= - 3 \ln \left(T + \bar T + \gamma b^2\right)\ , \label{kp1}
\end{align}
where $\gamma  = e^{-\Phi}c_{+--}$,
 with $c_{+--}$ the triple intersection number  of the even and odd cycles.

Stabilization of $T$ via the nonperturbative superpotential (\ref{wnp1}) breaks the shift symmetry for~$b$, because of the way that  $b$ and $T$  mix in the  K\"ahler potential (\ref{kp1}).  One way to understand this effect is by considering the distinction between `physical volumes'  and `holomorphic volumes'.  The proper  K\"ahler coordinates  on the moduli space  are the complex scalar $G$  and the holomorphic four-cycle volume~$T$   given in (\ref{holomorphicvolume}), but the  physical volume of the compactification  is  ${\cal V} \propto (T + \bar T + \gamma b^2)^{3/2}$, which involves a non-trivial mixture of $T$ with the inflaton $b$.
If $T$ is unstabilized, then the potential has a flat direction along which $T$ and $b$  shift, but $T + \bar T + \gamma b^2$ is invariant. Along  this direction, the  physical volume of the compactification is unchanged, but the holomorphic volume $T$  is altered.

K\"ahler modulus stabilization through superpotential interactions  involves the introduction of a  potential for the holomorphic coordinate~$T$.  On the other hand, the presence of sources of positive or negative energy (such as fluxes  and D-branes)  leads to terms in the scalar potential that depend on physical volumes, and are proportional to powers of $e^{K} \propto {\cal V}^{-2}$.    Because the potential energy therefore depends on both $T + \bar T + \gamma b^2$ and $T$,   the flat direction along which $b$  shifts is lifted.  One finds that  the canonically-normalized field obtained from $b$  has $\eta \sim 1$, so that inflation does not generically occur.

Notice that with  the identification $b \to \phi$,
$\gamma b^2 \to -\gamma k(\phi,\bar{\phi})$ the discussion above  is exactly analogous to the eta problem in D3-brane inflation (see \S\ref{sec:D34d}). The basic point is that a coordinate that appears in the K\"ahler potential  almost certainly receives a  potential on moduli stabilization.

\vskip 4pt
\noindent
{\it Perturbative shift symmetry for the $c$ axion.}---The NS-NS two-form $B_2$ is related to the R-R  two-form $C_{2}$ by the S-duality subgroup  of the full $SL(2,\mathbb{Z})$  duality, which  also exchanges D5-branes and NS5-branes.   This immediately suggests  an S-dual  of the above scenario, in which an NS5-brane  is wrapped on a two-cycle $\Sigma_2$,  and the inflaton is proportional to $c \equiv \frac{1}{\alpha'} \int_{\Sigma_2} C_{2}$.   The background  flux compactification breaks $SL(2,\mathbb{Z})$, so the difficulties of the $b$ axion model do not necessarily have to arise in the $c$ axion model.
The natural motivation  for considering a $c$ axion model is that $c$  does not appear in the tree-level K\"ahler potential, so that the eta problem  observed for $b$ is absent.
This is a  direct consequence of the
PQ symmetry for $c$, which is unbroken at the  perturbative level,  and constitutes one of the simplest realizations in string theory of the shift symmetry proposal of \cite{Kawasaki:2000yn}.

\vskip 4pt
\noindent
{\it Nonperturbative symmetry breaking for the $c$ axion.}---Let us now discuss  the leading effects  that break the shift symmetry of $c$.
We begin at the level of the ${\cal N} = 1$  data, i.e.~we first consider spontaneous breaking of the shift symmetry in the supersymmetric theory, before incorporating the explicit breaking due to the NS5-brane.
In view of our general discussion in \S\ref{sec:axions} of the breaking of axionic symmetries in string theory,  the only possibilities involve nonperturbative effects:
\begin{itemize}

\item[$\triangleright$]   {\it Euclidean strings.}---For a two-form symmetry, the first place to look is a two-dimensional worldvolume, corresponding to  some form of Euclidean string.  Although  ordinary worldsheet instantons (Euclidean F-strings) do break the shift symmetry of $b$,  fundamental strings do not carry charges under Ramond fields, so even worldsheet instantons will not  break the shift symmetry of $c$.   On the other hand, Euclidean D1-branes---or more generally, Euclidean $(p,q)$  strings---do couple to $C_2$, and can  break the symmetry.   However, no such effect can be present in the superpotential, because of holomorphy:  the  real part of the action of any Euclidean string involves the volume of a two-cycle, which is not
    the real part of a holomorphic coordinate  and therefore cannot appear in the superpotential \cite{Witten:1996bn}.   (The corresponding superfields are  linear superfields, not chiral superfields.)
We conclude that symmetry-breaking terms from  Euclidean D1-branes are confined to the K\"ahler potential. These contributions can easily be made negligible: they are nonperturbatively small,  and---unlike Euclidean D3-brane terms in the superpotential---are unrelated  to, and can be much smaller than, the terms appearing in the moduli potential.   Indeed, recall that in a KKLT vacuum,  perturbative corrections to $K$  can be consistently neglected,  and nonperturbative corrections to $K$  are even smaller.  See \cite{Flauger:2009ab} for a more detailed discussion.

\item[$\triangleright$] {\it Euclidean D3-branes.}---Because  Euclidean D3-branes can make important contributions to the moduli potential, it is important to ask whether they also affect the potential for $c$. The  action for a Euclidean D3-brane  wrapping  a four-cycle $\Sigma_4$ contains a term
\begin{equation}
S \, \supset\,  \int_{\Sigma_4} C_2 \wedge {\cal F}_2 \ , \label{sfc}
\end{equation}  where ${\cal F}_2$  is the worldvolume  flux.   The path integral  includes a sum  over `magnetizations', i.e.\ over topologically distinct choices of ${\cal F}_2$, and if  there exists a choice of magnetic flux such that (\ref{sfc}) is nonvanishing,  this will generally give rise to an  eta problem for $c$.  In  short,  un-magnetized Euclidean D3-branes do not affect $c$, but magnetized Euclidean D3-branes that intersect the NS5-brane can break the shift symmetry and lead to an eta problem.

\item[$\triangleright$] {\it Gaugino condensation on D7-branes.}---Perhaps surprisingly, breaking of the shift symmetry of $c$ by gaugino condensation on D7-branes is negligibly small.   To see this, recall that the nonperturbative superpotential (\ref{gauginocondensation}) from $N_c$ D7-branes can be written in the form
\begin{equation}
W_{\lambda\lambda} = {\cal{A}} \hskip 1pt e^{-f/N_c} \ ,
\end{equation}  where  the Wilsonian gauge kinetic function $f$---not to be confused with the axion decay constant---is a holomorphic function of the moduli,  with
${\rm{Re}}(f)=8\pi^2/g^2$.
The gauge kinetic function is renormalized  only at one loop and nonperturbatively:
\begin{equation}
f=f_0 + f_1 + f_{np}\ ,  \label{fWilpoly}
\end{equation}  with $f_0 = 2\pi T$.
The one-loop correction $f_1$  is independent of $c$,  because perturbative strings do not carry R-R  charge, so $W_{\lambda\lambda}$ can depend on $c$  only through $f_{np}$.  However,  on general grounds $f_{np} \lesssim e^{-S}$, where $S$  denotes the Euclidean action (for some appropriate extended object that couples to $c$,  for example a magnetized Euclidean D3-brane) evaluated at  a dominant saddle point.  Thus,
\begin{equation}
W_{\lambda\lambda} = {\cal{A}} \hskip 1pt e^{-(f_0+f_1)/N_c} \Bigl[1+ {\cal{O}}(e^{-S}) \hskip 2pt g(c)  \Bigr]\ ,
\end{equation} where $g(c)$  is some function of $c$.   We conclude that the dependence of the gaugino condensate superpotential  on $c$  is exponentially weaker than its dependence on $T$,  so  one can arrange for stabilization of $T$  without inducing a large potential for $c$.
\end{itemize}

Let us summarize the  nonperturbative effects  that spontaneously break the shift symmetry of $c$.  Euclidean $(p,q)$ strings make innocuous periodic contributions to the K\"ahler potential, and do not introduce  dangerously large masses for $c$.   Gaugino condensation  yields negligibly small  dependence on $c$.   Magnetized Euclidean D3-branes are problematic, and as a model-building criterion, one must ensure that any four-cycles  stabilized by Euclidean D3-branes do not intersect the NS5-brane.\footnote{The precise condition on the triple intersection form is $c_{i \varphi\varphi}=0~\forall\, i$, where $\varphi$  denotes the orientifold-odd  two-form corresponding to the inflaton, i.e.\ $c=G^{\varphi}\omega_{\varphi}$ (no sum),  and $i$ indexes all four-cycles stabilized by  Euclidean D3-branes.}

\subsection*{Symmetry Breaking from Backreaction}

The arguments above establish that given appropriate topology of the four-cycles involved in K\"ahler moduli stabilization, and taking the NS5-brane to be a {\it{probe}} of a fixed background compactification, the axion $c$ has an all-orders shift symmetry for which the leading spontaneous breaking comes from Euclidean D1-brane contributions to the K\"ahler potential, and is easily made negligibly small.   The leading
explicit breaking originates in the NS5-brane DBI action (\ref{ns5dbi}): this is the candidate inflaton potential.

However, assuming that the NS5-brane is purely a probe of the compactification is not consistent, as originally noted in \cite{McAllister:2008hb}:

\begin{itemize}

\item[$\triangleright$] {\it Backreaction from D3-brane charge.}---In the
configuration with $c \gg 1$, a large D3-brane charge is induced on the NS5-brane.  Upon solving the supergravity equations of motion with this charge as a source, one finds significant corrections to the warp factor\footnote{This is easily understood if one recalls that the backreaction of D3-branes in flat space leads to the $AdS_5 \times S^5$ geometry.} of the background that depend on $c$.\footnote{The corresponding calculation for  a  D3-brane  was performed in \cite{Baumann:2006th} --- see also the discussion following eq.~(\ref{equ:laplace}).}
The critical question is whether a $c$-dependent warp factor leads to additional $c$-dependence in the potential energy, beyond that captured by the DBI action of the NS5-brane.   We recall that the K\"ahler moduli $T_i$ are assumed to be stabilized by a combination of Euclidean D3-branes and gaugino condensation on D7-branes, and the Euclidean D3-brane action (or D7-brane gauge coupling) depends on the
warped volume of the corresponding four-cycle: cf.~eq.~(\ref{warpedvolume}).  Thus, a $c$-dependent modification of the warp factor entails a $c$-dependent correction to the exponentials appearing in the moduli potential.   Schematically, one has
\begin{equation}
V(c)  \, =\, \mu^3 f c +  \widehat V[T_i(c)]\ ,
\end{equation} where the first term comes from the DBI action and $\widehat V$ stands for the moduli potential.
The  K\"ahler moduli $T_i$  depend on $c$  via the warp factor,
\begin{equation}
{\rm{Re}}(T_i) = \int_{\Sigma_{4}^i} \d^4 y \sqrt{g} \, e^{-4A(y;c)} \ .
\end{equation}
In generic configurations, the resulting $c$-dependence of the moduli potential is large enough to invalidate the derivation made in the probe approximation.

Fortunately,  there is a mechanism that  provides parametric suppression of this  problematic backreaction effect.
In order to satisfy Gauss's  law, it was necessary to introduce  an  anti-NS5-brane, in addition to the  NS5-brane:  the induced charges  on the brane and antibrane are equal and opposite.
Thus, if the NS5-brane and anti-NS5-brane are relatively near to each other compared to their distance from the relevant four-cycles,  the  net flux of $F_5$  past the four-cycle will be suppressed,  and the correction to the warp factor will be correspondingly small.  Instead of  seeing a  monopole  D3-brane charge,   the four-cycles  see only a dipole.  A concrete realization  of this protective mechanism involves placing an NS5-brane and an anti-NS5-brane in a common warped throat, as discussed in detail in \cite{Flauger:2009ab}.\footnote{A  precise computation  of the axion decay constant in such a setting is an open problem.  If the decay constant is  significantly reduced by warping,  the requisite  induced charge increases,  intensifying the problem of backreaction.}

To recap, backreaction of the  induced D3-brane charge  that is ultimately responsible for the inflationary energy  leads to a correction to the warp factor that  affects the scalar potential by modifying the Euclidean D3-brane action.  This is  consistent with the general arguments,  in that the breaking of the axionic shift symmetry by backreaction  effects  is, strictly speaking, `nonperturbatively small',  being proportional to ${\rm{exp}}(-T_{p}\, {\rm{Vol}}(\Sigma_p))$.   However, it is essential to understand that the moduli potential,  and the inflationary vacuum energy,  are necessarily nonperturbatively small in the same sense.   For the backreaction to be a small correction, the geometry must be arranged to respect an
additional
approximate symmetry,  e.g.\  by situating the fivebrane pair  at the bottom of a warped throat, as noted above.  The original  axion shift symmetry,  on its own,  does not suffice to guarantee a flat potential.

\item[$\triangleright$]  {\it Backreaction from the NS5-branes.}---There is  another backreaction effect  that presents a possible concern \cite{Conlon:2011qp}.
The NS5-brane/anti-NS5-brane pair explicitly breaks supersymmetry,  and moreover  either member of the pair, in isolation, sources a fivebrane tadpole that is canceled by its counterpart.
A toroidal orbifold computation  presented in \cite{Conlon:2011qp} shows that---in that  unwarped setting---the backreaction of the five-branes themselves, not including the  induced D3-brane  charge described above, grows logarithmically with their separation.  The scale of the potential from backreaction is set by the mass of strings stretched between the branes.
The conclusion of \cite{Conlon:2011qp} is that for an NS5-brane/anti-NS5-brane pair that occupy  well-separated throats,  the potential from  fivebrane backreaction is of order the unwarped  bulk scale.  It would be worthwhile to understand whether the findings of \cite{Conlon:2011qp},  especially the claim that  the backreaction of a homologous fivebrane pair  has real codimension two,  are applicable in general warped backgrounds.   One should bear in mind that the geometric configuration required to address the problem of D3-brane backreaction---namely, situating  the fivebrane pair  in a common warped region---also  ameliorates the fivebrane backreaction  described in \cite{Conlon:2011qp}.
\end{itemize}

The difficulties  described above are  substantially alleviated in the two-field  generalization of axion monodromy known as {\it{Dante's Inferno}}~\cite{Berg:2009tg}, in which the  inflaton trajectory  is a gradual spiral in the two-dimensional  axion field space.
A hierarchy between the two axion decay constants ---  which is
plausibly radiatively stable --- provides the large number that serves to  enlarge the  effective axion field range,  and correspondingly to suppress backreaction.
One striking feature of this scenario is that the length of the inflationary trajectory can be parametrically larger than the diameter of the region in field space traversed during inflation.  This  makes it possible  to produce a large  primordial tensor signal without a super-Planckian displacement.

\subsection*{Reheating}
\label{sec:AxionReheating}

Reheating in a string theory model with a shift-symmetric inflaton poses particular difficulties, as explored in \cite{Green:2007gs,Brandenberger:2008kn,Barnaby:2009wr,Braden:2010wd}.  The  general issue is that the shift symmetry that protects the inflaton simultaneously
limits the couplings of the inflaton to the visible sector.   In a  naive model  containing only the inflaton,  the  visible sector,  and general relativity ---  with no additional sectors associated with the ultraviolet completion of gravity ---  the primary consequence would be  slow reheating,  which is not necessarily fatal.  However, if the inflaton couples  at least as strongly to hidden sector fields as it does to the visible sector, which is  frequently the case in models of closed string inflation,  then  problematic reheating of the hidden sectors  is difficult to avoid.

At present, no concrete results  concerning reheating are available for the specific  scenario of axion monodromy inflation with a $c$ axion coupled to an NS5-brane pair,  but we find it reasonable to expect that reheating of hidden sectors, including fields associated with the NS5-brane pair and the enveloping warped geometry,  will present a challenge  for the model.

\subsection{Phenomenology}  \label{axionphenomenology}

The phenomenology of axion inflation can be extremely rich.\footnote{See~\cite{Pajer:2013fsa} for a recent review.}
Besides the model-independent gravitational wave signal, a host of additional  model-dependent signatures have been explored,  including oscillations in the power spectrum~\cite{Flauger:2009ab}, deviations from scale-invariance~\cite{Barnaby:2011qe}, non-Gaussianity~\cite{Flauger:2010ja, Barnaby:2010vf}, chiral gravitational waves~\cite{Barnaby:2011qe, Cook:2011hg}, and primordial black holes~\cite{Linde:2012bt}.
Since all of these effects are tied to the underlying axion shift symmetry, one has the hope of finding correlated signatures  across different observational channels.

\vskip 4pt
\noindent
{\it Signatures of nonperturbative effects.}---At leading order in the instanton expansion, the Lagrangian for axion monodromy inflation
takes the form 
\beq
{\cal L} = - \frac{1}{2} (\partial \phi)^2 - V_0(\phi)  -\Lambda^4 \cos \left( \frac{\phi}{f} \right) \ . \label{axionmL}
\eeq
where $V_0(\phi) \equiv \mu^{4-p} \phi^p$.
Since the scale $\Lambda$ is  generated nonperturbatively, the modulations can  quite naturally be exponentially small,  but it is also possible  that the modulations could be large enough to spoil the monotonicity of the inflaton potential.  Here, we assume that $\Lambda$ happens to be large enough to be phenomenologically interesting, but not  large enough to dominate the evolution.
The monotonicity constraint is
\beq
b_\star \equiv \frac{\Lambda^4}{V_0'(\phi_\star) f}  < 1\ .
\eeq
This parameter depends on the inflaton vev, unless the potential is linear.
Here, we have evaluated it at $\phi = \phi_\star$, the value of the inflaton when the pivot scale $k=k_\star$ exits the horizon.

Before  proceeding,  we  point out that $f/\Mp$ is  bounded from below,  for two reasons.
First, requiring that the $\alpha^{\prime}$  expansion  is under good control (in the NS5-brane  construction \cite{McAllister:2008hb} with $p=1$) implies~\cite{Flauger:2009ab}
\beq
\frac{f^2}{\Mp^2} > \frac{\sqrt{g_{\rm s}}}{(2\pi)^3 {\cal{V}}}  \ . \label{equ:lowf}
\eeq
Another bound on $f$ arises from requiring that the effective theory of the fluctuations~is weakly coupled
for the parameter values of interest~\cite{Behbahani:2011it}. Control over the observational predictions requires that the oscillation frequency $\omega = (-2\Mp^2 \dot H)^{1/2}/f$ is much smaller than the unitarity bound $4\pi f$. This implies
\beq
f \, \gg\, (2\Delta_\R)^{-1/2}\, \frac{H}{2\pi}\ .
\eeq
Using (\ref{HMp}), we get
\beq
\frac{f}{\Mp} \, \gg\, \sqrt{\frac{r}{16}} \hskip 1pt \Delta_\R^{1/2}  \, \approx\, 4.5 \times 10^{-4} \left( \frac{r}{0.07} \right)^{1/2}\ . \label{equ:funi}
\eeq
The predictions for cosmological observables depend on the parameters $b_\star$ and $f$, as well as the parameters of the potential $V_0$ (e.g.~$V_0(\phi_\star)$, $\epsilon_\star$ and $\eta_\star$).

\vskip 4pt
Striking signatures of axion monodromy inflation  arises from  the periodic modulation in~(\ref{axionmL}).
In a general  inflationary model, the  solutions for the primordial perturbations  inside the horizon are oscillatory.   The  periodic potential in (\ref{axionmL}) introduces a periodic driving force, which  can resonate with  the freely-oscillating  perturbations, modulating their amplitude  as a function of wavenumber~$k$ \cite{Chen:2008wn, Flauger:2009ab}.
This resonance leads  to modulations of the power spectrum, as well as to a specific type of non-Gaussianity:

\begin{itemize}
\item[$\triangleright$]  {\it Modulated power spectrum.}---For $b_\star \ll 1$, the oscillatory term in the potential can be treated as a perturbation.
At first order in $b_\star$, the power spectrum is~\cite{Flauger:2009ab}
\beq
\Delta_\R^2(k) = \Delta_\R^2(k_\star) \left(\frac{k}{k_\star}\right)^{n_s-1} \left[ 1 + {\cal A}\, \cos \left( \frac{\phi_k}{f} \right) \right] \ ,
\eeq
where $\phi_k \simeq \phi_\star - \sqrt{2\epsilon_\star} \ln(k/k_\star)$ is the field value at horizon exit of the mode $k$.
We have defined $n_s = 1 + 2 \eta_\star - 6 \epsilon_\star$ and
\beq
{\cal A} \, \equiv\,   3 b_\star \left( \frac{2\pi f}{\sqrt{2 \epsilon_\star}} \right)^{1/2}\ ,
\eeq in units where $\Mp \equiv 1$, and have worked  to leading order in $f/\sqrt{2 \epsilon_\star}$.
For the special case $p=1$,  i.e.~the linear potential  that arises from NS5-branes, we have $1/\sqrt{2 \epsilon_\star} = \phi_\star$, and $\Delta_\R^2(k)$  is known for arbitrary $f \phi_\star$:
\beq
{\cal A}_{p=1} \, =\, \frac{6 b_\star}{\sqrt{1+ (3 f \phi_\star)^2}} \, \sqrt{\frac{\pi}{2} \coth \left( \frac{\pi}{2 f \phi_\star } \right) f \phi_\star}  \ .
\eeq
These oscillations in the power spectrum have recently been searched for in  the WMAP data~\cite{Peiris:2013opa} and the Planck data~\cite{FlaugerPrivate}.  So far, no signal has been detected.\footnote{Using WMAP9, it was found that a modulation with $\log_{10}(f/\Mp) = -3.38$ improves the fit by $\Delta \chi^2 = 19$~\cite{Peiris:2013opa}, but the frequency of this signal coincides with the unitarity bound  (\ref{equ:funi}). Moreover, the signal does not seem to be present in the Planck data~\cite{FlaugerPrivate,Easther:2013kla}.}

\item[$\triangleright$]  {\it Resonant non-Gaussianity.}---At first order in $b_\star$,
the bispectrum is~\cite{Flauger:2010ja}
\begin{align}
B_\R(k_1,k_2, k_3) &\, =\,  \fnl^{\rm res} \times  \frac{(2\pi \Delta_\R)^4}{k_1^2 k_2^2 k_3^2} \Bigg[ \sin\left(\frac{\sqrt{2\epsilon_\star}}{f} \ln \frac{K}{k_\star} \right) \nonumber \\
& \ \ \ \ \ \ \ \ +\,  \frac{f}{\sqrt{2\epsilon_\star}} \sum_{i \ne j} \cos \left(\frac{\sqrt{2\epsilon_\star}}{f} \ln \frac{K}{k_\star} \right) + \cdots \Bigg] \ , \label{equ:Bres}
\end{align}
where $K \equiv k_1 + k_2 + k_3$ and
\beq
 \fnl^{\rm res} \equiv \frac{3 \sqrt{2\pi} }{8} \, b_\star\, \left( \frac{\sqrt{2\epsilon_\star}}{f}\right)^{3/2} \ .
\eeq
The ellipses in (\ref{equ:Bres}) stand for terms that are suppressed by higher powers of the slow-roll parameters or by positive powers of $f/\sqrt{2\epsilon_\star}$.
Observable non-Gaussianity requires $f/\sqrt{2\epsilon_\star} \ll 1$.
In this case, the second term in (\ref{equ:Bres}) is suppressed relative to the first term, except in the squeezed limit (where it ensures that Maldacena's consistency relation (\ref{equ:CONSIST}) holds).
The unitarity bound (\ref{equ:funi}) implies an upper bound on $\fnl^{\rm res}$,
\beq
\fnl^{\rm res} \ll \frac{3 \sqrt{\pi}}{2} (2\Delta_\R)^{-3/4} \approx 3 \times 10^3\ .
\eeq

Because the bispectrum is oscillating, it is nearly orthogonal to the standard bispectrum templates.
It is therefore  barely constrained by the present bispectrum results (\ref{equ:fl})--(\ref{equ:fo}),
and a dedicated analysis  is required to put  meaningful constraints on the parameter $\fnl^{\rm res}$.
\end{itemize}

\vskip 4pt
\noindent
{\it Signatures of gauge field production.}---In order to reheat, the axion has to be coupled to extra fields, and one can ask whether these
couplings
affect the perturbations during inflation.
Particularly interesting is the following dimension-five operator that couples the inflaton to a gauge
field:\footnote{We will assume $\alpha \le 1$. The case $\alpha \gg 1$ is discussed in~\cite{Anber:2009ua}, while estimates for $\alpha$  in type IIB  string theory appear in \cite{Barnaby:2011qe}.
Both bottom-up and top-down naturalness of the regime $\alpha \gg 1$ remain to be established.}
\beq
{\cal L} \, \supset\,  - \frac{\alpha}{4} \frac{\phi}{f} F \tilde F \ , \label{equ:pFF}
\eeq
where $F$ is the gauge field strength.
This  coupling respects the shift symmetry of the inflaton, as for constant $\phi$ the operator is a total derivative.
When the inflaton has a time-dependent vev, $\phi(t)$, the conformal invariance of the gauge field is broken.   This  leads to production  of gauge field  quanta during inflation.
To see this, consider the equation of motion for the two polarization modes of the gauge field (in Coulomb gauge):
\beq
\left( \frac{\partial^2}{\partial \tau^2} + k^2 \mp 2aH k \hskip 1pt \xi \right) A_\pm(\tau,k) = 0  \ , \qquad {\rm where} \quad \xi \equiv \frac{\alpha \dot \phi}{2fH}\ .
\eeq
We see that one of the helicities of the gauge field experiences tachyonic growth for $k/(aH) < 2\xi$. For $\xi > 0$, the unstable mode is $A_+$.
Most of the power in the produced gauge field is in modes with $(8\xi)^{-1} < k/(aH) < 2\xi$.
In this regime, the solution can be written as~\cite{Anber:2009ua}
\beq
A_+(\tau, k) \simeq \frac{1}{\sqrt{2k}} \left( \frac{k}{2\xi aH}\right)^{1/4} e^{\pi \xi- 2\sqrt{2\xi k/(aH)}}\ .
\eeq
The coupling of this solution to the inflaton, via (\ref{equ:pFF}), leads to a number of observational signatures:

\begin{itemize}
\item[$\triangleright$]  {\it Equilateral non-Gaussianity.}---The gauge field non-linearities in the operator (\ref{equ:pFF}) source non-Gaussian inflaton fluctuation (this may be thought of as an inverse decay, $\delta A + \delta A \to \delta \phi$).
The bispectrum is of equilateral type and has the amplitude~\cite{Barnaby:2010vf}
\beq
\fnl^{\rm equil} \simeq \frac{\Delta_{\R,0}^6}{\Delta_\R^4} \, f_3(\xi) e^{6\pi \xi}\ ,
\eeq
where $\Delta_{\R,0}$ stands for $\Delta_\R|_{\xi=0} = H^2/(2\pi \dot \phi)$.  The function $f_3(\xi)$ is determined numerically, but has the following limits:
\begin{align}
f_3(\xi) &= 2.8 \times 10^{-7} \, \xi^{-9}\quad \ \hskip 1pt  {\rm for} \quad \xi \gg 1\ , \\
f_3(\xi) &\approx 7.4 \times 10^{-8} \xi^{-8.1} \quad {\rm for} \quad 2 < \xi < 3\ .
\end{align}
We see that $\fnl^{\rm equil}$ is exponentially sensitive to the model parameter $\xi$.
Let us denote by $\xi_\star$ the value of $\xi$ at the pivot scale $k_\star = 0.002$ Mpc${}^{-1}$. Using the seven-year WMAP data, ref.~\cite{Meerburg:2012id} found
$\xi_\star < 2.45$ (95\% CL). In terms of the axion decay constant, this corresponds to
\beq
f > \frac{\alpha}{10 \pi} \frac{H}{\Delta_\R}\ .
\eeq
Using (\ref{HMp}), this can be written as
\beq
\frac{f}{\Mp} \, >\,  \frac{\alpha}{10} \sqrt{\frac{r}{2}} \, \approx\, 2\times 10^{-2} \left(\frac{\alpha}{1} \right) \left( \frac{r}{0.07} \right)^{1/2}\ .
\eeq
For $\alpha \sim {\cal O}(1)$, this provides a strong constraint on the axion decay constant.  This bound is similar in spirit to the bound in (\ref{equ:EFTprec}).
\item[$\triangleright$] {\it Non-scale-invariance.}---The power spectrum of curvature perturbation receives contributions both from the vacuum fluctuations of the inflaton and the fluctuations sourced by the gauge field,
\beq
\Delta_\R^2(k) = \Delta_{\R,0}^2(k)\left[ 1+ \Delta_{\R,0}^2(k) \, f_2(\xi) e^{4\pi \xi} \right]\ , \label{equ:DA}
\eeq
where
\begin{align}
f_2(\xi) &= 7.5 \times 10^{-5} \, \xi^{-6} \quad \ \hskip 1pt {\rm for} \quad \xi \gg 1\ , \\
f_2(\xi) &\approx 3.0 \times 10^{-5} \xi^{-5.4} \quad {\rm for} \quad 2 < \xi < 3\ .
\end{align}
For large $\xi$, the sourced fluctuations can dominate over the vacuum fluctuations.
Notice that $\xi \propto \sqrt{\epsilon}$ grows during inflation. Although
$\xi$ grows slowly, it appears in the exponent
in (\ref{equ:DA}), and
can therefore lead to
significant scale-dependence of the power spectrum. Although this effect is constrained mainly by small-scale CMB and LSS data, it remains convenient to express the
constraint as a bound on the parameter $\xi_\star$ (evaluated at $k_\star = 0.002$~Mpc${}^{-1}$).  Assuming a quadratic inflaton potential and using WMAP and ACT data, ref.~\cite{Meerburg:2012id} found  $\xi_\star < 2.41$ (95\% CL).

\item[$\triangleright$]  {\it Primordial black holes.}---The growth of $\xi$ can lead to the formation of primordial black holes.  This can disturb the standard cosmology. Estimating the effects of strong backreaction at the end of inflation, ref.~\cite{Linde:2012bt} find $\xi_\star < 1.5$. Although this is the strongest constraint on the parameter $\xi_\star$, it  is  plausibly subject to the largest theoretical uncertainties.

\item[$\triangleright$]  {\it Chiral gravitational waves.}---The stress tensor associated with the produced gauge fields sources gravitational waves.
An interesting  property of the  resulting tensor signal is that it is chiral~\cite{Sorbo:2011rz}, essentially because only one chirality of the gauge field is unstable.  This parity violation could, in principle, be tested using the TB and EB correlators of the CMB.  However, to achieve a sufficiently large tensor amplitude requires values of $\xi_\star$ that are already ruled out by the bound on non-Gaussianity.
On the other hand, the tensor modes become large on small scales, just like the scalars.
It is therefore conceivable for the signal to be small on CMB scale, but detectable on
scales accessible to terrestrial interferometers.
In~\cite{Crowder:2012ik} it was estimated that $\xi_\star > 2.2$ could be probed with Advanced LIGO.
\end{itemize}

\section{Inflating with K\"ahler Moduli}
\label{sec:LVS}

A very early idea for inflation in string theory was that a modulus could be the inflaton \cite{Binetruy:1986ss,Banks:1995dp}.
In particular,  the compactification volume is invariably present in the four-dimensional effective theories of string compactifications, and it is natural to ask if the volume modulus could be the field driving slow-roll inflation.  In this section, we will describe scenarios in which  the inflaton is a K\"ahler modulus,  or an axion paired with a K\"ahler modulus.\footnote{This section is based mostly on~\cite{Balasubramanian:2005zx, Conlon:2005jm, Cicoli:2008gp}. For a recent review see~\cite{Cicoli:2011zz}.}

\subsection{Racetrack Inflation}
\label{racetrack}

Modular inflation in the context of flux compactifications was first realized in the {\it{racetrack inflation}} scenario \cite{BlancoPillado:2004ns}.
As a concrete example,
we consider a KKLT compactification with a single K\"ahler modulus $T$, and a superpotential of the
`racetrack' form \cite{BlancoPillado:2004ns}
\begin{equation}
W= W_0 + {\cal A}\,e^{-aT} + {\cal B}\,e^{-bT}\ , \label{equ:Wrace}
\end{equation} where $W_0$ is the constant flux superpotential, ${\cal A}$ and ${\cal B}$ are prefactors that depend on the vevs of the stabilized complex structure moduli, and $a$, $b$ are constants.  A superpotential of this form can be generated by gaugino condensation in a product gauge group: for gauge group $SU(N) \times SU(M)$ one has $a=2\pi/N$ and $b=2\pi/M$.
The classical K\"ahler potential takes the form
\begin{equation}
K=-3 \ln (T+\bar{T})\ ,
\end{equation} up to $\alpha'$ corrections that will be discussed momentarily.
To complete the specification of the effective theory, we incorporate supersymmetry breaking by an anti-D3-brane in a warped throat region, which leads to a term in the potential of the form
\begin{equation}
\delta V = \frac{\varrho}{(T+\bar{T})^2}\ ,  \label{delV2}
\end{equation} with $\varrho$ a constant that depends on the warp factor at the location of the anti-D3-brane.\footnote{For an explanation  of the exponent $2$ in (\ref{delV2}), which differs from the result given in \cite{Kachru:2003aw}, see \cite{Kachru:2003sx}.}

The authors of \cite{BlancoPillado:2004ns} showed that for suitable values of the parameters $W_0,{\cal A},{\cal B},a,b,\varrho$, the potential for $T$ develops a saddle point\footnote{Racetrack-type superpotentials for K\"ahler moduli have also been argued to yield inflection point inflation \cite{Badziak:2008gv}.}
 that is suitable for inflation.  The evolution is primarily in the direction of the axion ${\rm Im}(T)$, but ${\rm Re}(T)$, corresponding to the compactification volume, does also evolve.
In a related construction, ref.~\cite{Linde:2007jn}
followed the Kallosh-Linde  scenario \cite{Kallosh:2004yh} for  K\"ahler moduli stabilization via a racetrack, and  showed that  for special values of the parameters, the axion ${\rm Im}(T)$ is stabilized, but a single-field inflection point appears in the ${\rm Re}(T)$ direction.   The volume modulus  then serves as the inflaton.

The parameter values that can lead to racetrack inflation  in type IIB  string theory are quite contrived.\footnote{The example given in \cite{BlancoPillado:2004ns} has $N=90$ and $M=100$, as well as fine-tuned values for $W_0,{\cal A},{\cal B},\varrho$. In the more explicit construction of \cite{BlancoPillado:2006he} (building on moduli stabilization results of \cite{Denef:2004dm}),
successful inflation was found  for $N=40$ and $M=258$, again with fine-tuning of the remaining parameters, while the scenario of \cite{Linde:2007jn} used $N=58$ and $M=60$, with $B$  specified to 11  decimal places.} This is not a fatal objection: because $W_0$, ${\cal A}$, ${\cal B}$  are determined by the vevs  of complex structure moduli,  which are in turn dictated by quantized fluxes,  it is possible in principle to adjust their values rather precisely (as originally noted for the cosmological constant problem  in \cite{Bousso:2000xa}).  On the other hand, large values of $M,N$  require stacks of many D7-branes,  which can be difficult to construct in  explicit compactifications.

It is important to recognize that fine-tuning the leading-order classical effective action does not necessarily lead to a self-consistent model. Quantum effects, as well as curvature corrections  from the $\alpha'$  expansion,  inevitably contribute to the action,  and  it is not  consistent to study solutions of the leading-order theory that require precision comparable to the size of these corrections.   Indeed, it was shown in \cite{Greene:2005rn}  that the higher-curvature term of (\ref{IIBgravityexpansion}) generically  destabilizes ${\rm Re}(T)$,  and  also renders the potential more steep in the ${\rm Im}(T)$  direction, spoiling inflation.   Although it is conceivable that  new parameter values could be found for which the corrected action yields inflation, the complete set of contributing terms has not yet been determined,  so that it is difficult to make the model more explicit and predictive.

A related idea is that the volume modulus $T$  can serve as the inflaton if  finely-tuned  combinations of corrections to the K\"ahler potential, cf.~\S\ref{quantumeffects}, lead to the  appearance of an inflection point in the potential \cite{Conlon:2008cj}.  The scenario of \cite{Conlon:2008cj}  alleviates the tension between low-energy supersymmetry and high-scale inflation identified in \cite{Kallosh:2004yh} --- see also \cite{Buchmuller:2004tz,Graesser:2006ft,Badziak:2008yg,He:2010uk,Linde:2011ja}.

\subsection{Large Volume Compactifications}  \label{lvscpct}

The most concrete models of K\"ahler moduli inflation have been constructed in the Large Volume Scenario (LVS)~\cite{Conlon:2006gv}.\footnote{We thank  Michele Cicoli,  Joe Conlon, and Fernando Quevedo  for helpful discussions of the material in this section.}
We reviewed LVS compactifications in \S\ref{modulistab}.
For convenience, we will quickly summarize the aspects of that discussion that will be relevant for inflationary model-building.

\vskip 4pt
The tree-level K\"ahler potential in the presence of the leading sigma model corrections, at order $(\alpha^{\prime})^3$~\cite{Becker:2002nn} is
\beq
K = K_0 + \delta K_{(\alpha')} = - 2\ln({\cal V}) - \frac{\hat \xi}{ {\cal V}} \ . \label{equ:KLVS}
\eeq
The contribution $\delta K_{(\alpha')}$ breaks the no-scale structure, but
only lifts the volume modulus ${\cal V}$, leaving $(h^{1,1}_+ -1)$ complexified K\"ahler moduli  massless (at this level of approximation).
These are natural inflaton candidates.
To assess whether such scenarios are viable, it is important to consider
string loop corrections to the K\"ahler potential,
\beq
K = K_0 + \delta K_{(\alpha')} + \delta K_{(g_{\rm s})}\ ,
\eeq
but these corrections are  difficult to compute.
The only explicit results available are those obtained by Berg, Haack, and K\"ors for ${\cal N}=1$ compactifications on the toroidal orientifold $T^6/(\mathbb{Z}_2 \times \mathbb{Z}_2 )$~\cite{Berg:2005ja, Berg:2005yu}, cf.~eqs.~(\ref{equ:Kgs1}) and (\ref{equ:Kgs2}).
Some progress has been made on extending these results to general Calabi-Yau manifolds~\cite{Berg:2007wt, Cicoli:2007xp}, and a specific functional form of the string loop corrections has been conjectured by Berg, Haack, and Pajer in \cite{Berg:2007wt} (see also \cite{vonGersdorff:2005bf}  for a general discussion).
Recall from \S\ref{modulistab} that these corrections can be separated into two types: those associated with the exchange of closed strings with Kaluza-Klein momentum,  and those  associated with the exchange of  strings that wind  non-contractible cycles.\footnote{The cycles in question are one-cycles within  the curve of intersection of two D7-branes: the parent Calabi-Yau manifold  does not have any non-contractible (non-torsion)  one-cycles.}
The correction $\delta K_{(g_{\rm s})}^{\rm KK}$ is conjectured to be
\beq
 \delta K_{(g_{\rm s})}^{\rm KK} \,\sim\, g_{\rm s}\, \sum_{i=1}^{h^{1,1}} \frac{{\cal C}_i^{\rm KK}(\zeta,\bar \zeta) M_{\rm KK}^{-2}}{ {\cal V}} \, \sim\, g_{\rm s}\, \sum_{i=1}^{h^{1,1}}  \frac{{\cal C}_i^{\rm KK}(\zeta,\bar \zeta) (a_{ij} t^j)}{ {\cal V}}\ , \label{equ:KGS1}
 \eeq
where $a_{ij} t^j$ is some linear combination of the two-cycle size moduli $t^j$.
The conjectured  result for $\delta K_{(g_{\rm s})}^{\rm W}$ is
\beq
\delta K_{(g_{\rm s})}^{\rm W} \, \sim\, \sum_i \frac{{\cal C}_i^{\rm W}(\zeta,\bar \zeta) M_{\rm W}^{-2}}{{\cal V}} \,\sim\,   \sum_i \frac{{\cal C}_i^{\rm W}(\zeta,\bar \zeta) }{(b_{ij} t^j){\cal V}}\ , \label{equ:KGS2}
\eeq
where the two-cycle $b_{ij} t^j$ corresponds to the curve of intersection of two D7-branes (see \cite{Berg:2007wt} for details).
The unknown complex structure dependence has been absorbed into the functions ${\cal C}_i^{\rm KK}(\zeta,\bar \zeta)$ and ${\cal C}_i^{\rm W}(\zeta,\bar \zeta)$. Since we assume that the complex structure moduli are stabilized at higher energies, we will treat them as constants, focusing on the dependence on the K\"ahler moduli.

The supergravity approximation  holds when $t_i \gg 1$, in which case
\beq
\delta K_{(g_{\rm s})}^{\rm KK} \, \sim\, \sum_i \frac{t_i}{{\cal V}} \, > \, \delta K_{(\alpha') } \,\sim\, \frac{1}{{\cal V}}\ .
\eeq
One might then worry that  at large volume the $g_{\rm s}$ corrections (\ref{equ:KGS1}) and (\ref{equ:KGS2}) overwhelm the $\alpha^{\prime}$  corrections $\delta K_{(\alpha')}$ and  drastically change the vacuum structure, to say nothing of the inflationary phenomenology.
However, what happens is a bit more subtle \cite{Berg:2007wt,Cicoli:2007xp}.
Although the $g_{\rm s}$ corrections dominate over the $\alpha'$ corrections in the K\"ahler potential, they cancel to a certain degree in the scalar potential, so that the dominant contribution to
the scalar potential actually comes from the $\alpha'$ corrections.  This important phenomenon is called {\it extended no-scale structure} \cite{Cicoli:2007xp}.
It arises because $\delta K_{(g_{\rm s})}^{\rm KK}$ is a homogeneous function, of degree $-2$,  in the two-cycle volumes $t_i$ \cite{Cicoli:2007xp}.

To illustrate extended no-scale structure, we consider an example with a single modulus $\tau$~\cite{Cicoli:2011zz}.
Schematically, we can write the K\"ahler potential as
\beq
K = - 2 \ln({\cal V}) - \frac{\hat \xi}{{\cal V}} + \frac{\sqrt{\tau}}{{\cal V}} \ . \label{equ:KKK}
\eeq
Taking the superpotential to be a constant, $W=W_0$, we find
\beq
V = \frac{W_0^2}{{\cal V}^3} \left[ 0 + \hat \xi + 0 \cdot \sqrt{\tau} + \frac{1}{\sqrt{\tau} } + \frac{1}{\tau^{3/2} }\right] \ .\label{equ:Vgs}
\eeq
The first zero in (\ref{equ:Vgs})  corresponds to the famous no-scale structure, while the second vanishing contribution (namely, $0\cdot \sqrt{\tau}$) is  the consequence of what we just referred to as extended no-scale structure.
We see that the leading $g_{\rm s}$ contribution to the scalar potential scales as $1/\sqrt{\tau}$, and is {\it{smaller}} than the  leading $\alpha'$~contribution proportional to $\hat \xi$.  The $g_{\rm s}$ contribution to the scalar potential is therefore smaller than naively expected.   Even so, we will find that $g_{\rm s}$  corrections can still make dangerously large contributions to the inflaton potential.

Extended no-scale structure can be understood from an alternative point of view~\cite{Cicoli:2007xp}.
In the low-energy effective field theory, we can interpret the $g_{\rm s}$ contribution as the
one-loop Coleman-Weinberg potential
\beq
\delta V_{\mathsmaller{\rm CW}} \simeq 0 \cdot \Lambda^4 + \Lambda^2 {\rm STr} (M^2) + {\rm STr} \left[ M^4 \ln\left( \frac{M^2}{\Lambda^2}\right)\right] \ , \label{equ:VCW}
\eeq
where
\beq
\Lambda = M_{\rm KK} \simeq \frac{\Mp}{{\cal V}^{2/3}} \quad , \quad {\rm STr}(M^2) \simeq \frac{\Mp^2}{{\cal V}^2} \ , \label{equ:CUT}
\eeq and ${\rm STr}$ denotes the supertrace.\footnote{The supertrace is defined by ${\rm STr}(M^2) \equiv \sum_s (2s+1)(-1)^{2s}{\rm Tr}(M^2_{s})$, where $s$  is the  spin and
$M^2_{s}$ is the matrix of masses squared for particles of spin $s$.  Note that bosonic and fermionic contributions enter with opposite sign.}
The first term in (\ref{equ:VCW}) vanishes by supersymmetry. Substituting (\ref{equ:CUT}) into (\ref{equ:VCW}), we get
\beq
\delta V_{\mathsmaller{\rm CW}} \simeq 0 \cdot \frac{1}{{\cal V}^{8/3}} + \frac{1}{{\cal V}^{10/3}} + \frac{1}{{\cal V}^4}\ , \label{equ:VCW2}
\eeq
which, upon using $\tau \sim {\cal V}^{2/3}$, can be written as
\beq
\delta V_{\mathsmaller{\rm CW}} \simeq \frac{1}{{\cal V}^3} \left[ 0 \cdot \sqrt{\tau} + \frac{1}{\sqrt{\tau}} + \frac{1}{\tau^{3/2}} \right] \ .
\eeq
This  precisely matches the scaling in (\ref{equ:Vgs}).  We have therefore related the extended no-scale feature of the potential to supersymmetry~\cite{Cicoli:2007xp}.

\vskip 4pt
In the rest of this section, we will use these results to construct inflationary solutions in LVS.
We will describe three ways in which the inflaton potential is generated:
i) via nonperturbative effects~\cite{Conlon:2005jm} (\S\ref{sec:blow}), ii) via string loops~\cite{Cicoli:2008gp}~(\S\ref{sec:fibre}),
and iii) via poly-instanton effects~\cite{Cicoli:2011ct,Blumenhagen:2012ue}~(\S\ref{sec:poly}).

\subsection{Blow-up Inflation}
\label{sec:blow}

\begin{figure}[h!]
   \centering
     \includegraphics[scale=0.5]{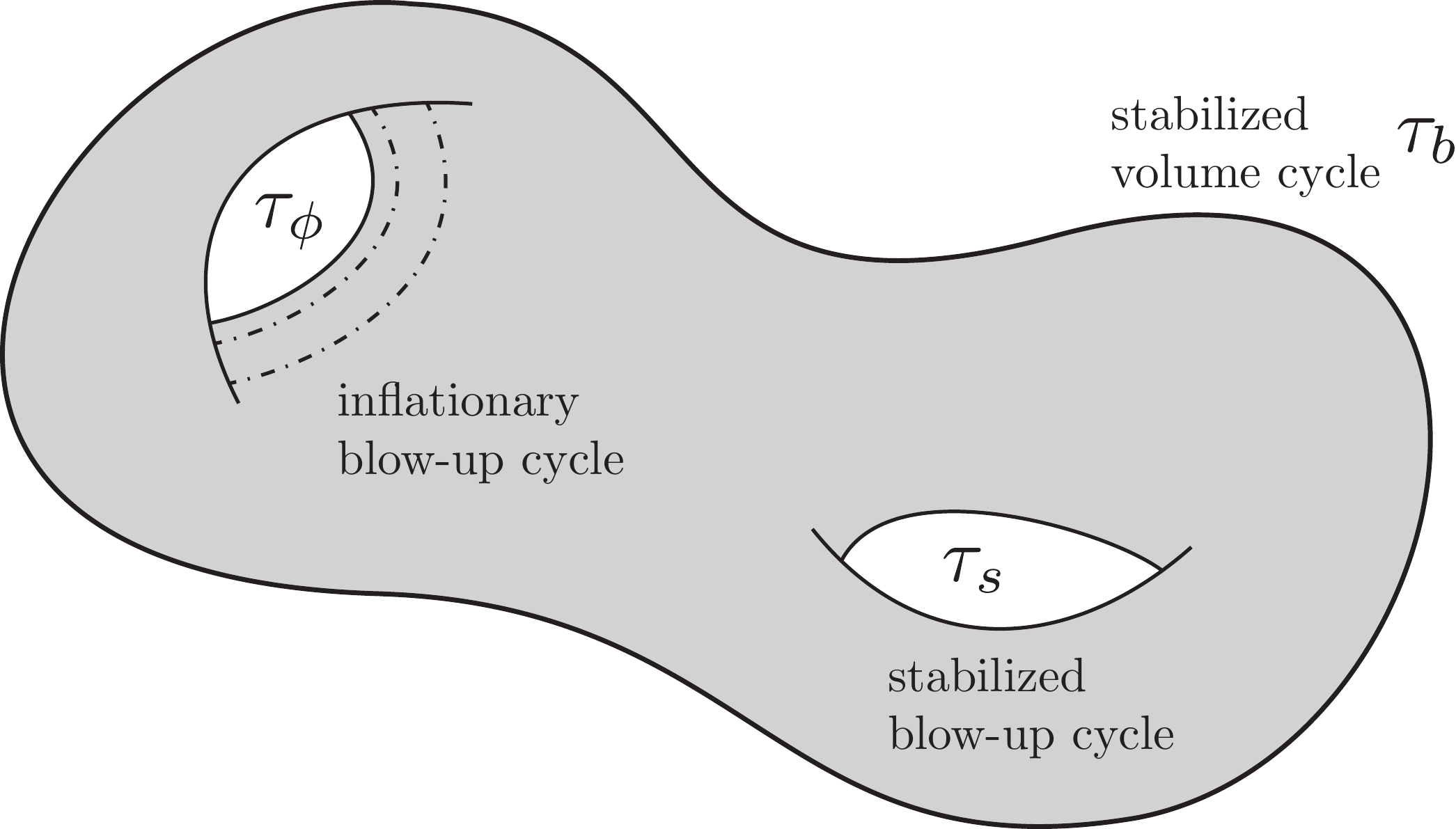}
   \caption{Blow-up inflation  in a three-modulus Swiss-cheese compactification. An evolving blow-up cycle $\tau_\phi(t)$ drives inflation, while a second blow-up cycle $\tau_s$ stabilizes the overall volume ${\cal V} \sim \alpha \tau_b^{3/2}$.}
  \label{fig:Kahler}
\end{figure}

The first models of K\"ahler moduli inflation~\cite{Conlon:2005jm} were constructed in Swiss-cheese compactifications of the Large Volume Scenario (see \S\ref{modulistab}.)    In order for one of the blow-up cycles to play the role of the inflaton, while keeping the overall volume fixed, at least three K\"ahler moduli are required (see fig.~\ref{fig:Kahler}).
The compactification volume is  then
\beq
{\cal V} = \alpha \left(\tau_b^{3/2} - \lambda_\phi \tau_\phi^{3/2} - \lambda_s \tau_s^{3/2} \right)\ .
\eeq
We will look at the part of the moduli space satisfying the hierarchies $\tau_b \gg \tau_\phi \gg \tau_s$.
The field $\tau_b$ then determines the overall volume, while $\tau_\phi$ and $\tau_s$ are blow-up cycles.

\vskip 4pt
\noindent
{\it Inflaton potential.}---We first assume that string loop corrections can be ignored, so that the K\"ahler potential is given by (\ref{equ:KLVS}). For the superpotential, we take
\beq
W = W_0 + {\cal A}_\phi \hskip 1pt e^{-a_\phi T_\phi} + {\cal A}_s \hskip 1pt e^{-a_s T_s} \ .
\eeq
This structure stabilizes the moduli\footnote{The axionic partner $\vartheta_\phi$ of the inflaton $\tau_\phi$ is not necessarily stabilized, and allowing $\vartheta_\phi$ to have a nonvanishing initial velocity  leads to the rich dynamics known as {\it{roulette inflation}} \cite{Bond:2006nc}.} $\tau_b$ and $\tau_s$, with ${\cal V} \sim \alpha \tau_b^{3/2}\sim e^{a_s \tau_s}$ and $\tau_s \gg 1$.
Integrating out $\tau_b$ and $\tau_s$ leads to the potential \cite{Conlon:2005jm}\footnote{This corrects a misprint in the corresponding formula (41) in~\cite{Cicoli:2011zz}.}
\beq
V = W_0^2 \left( \mathfrak{a}\, \frac{\sqrt{\tau_\phi}\hskip 2pt e^{-2 a_\phi \tau_\phi}}{{\cal V}} - \mathfrak{b}\, \frac{\tau_\phi \hskip 2pt e^{-a_\phi\tau_\phi}}{{\cal V}^2} + \mathfrak{c}\, \frac{\hat \xi}{{\cal V}^3}  \right) \ , \label{equ:V1Kahler}
\eeq
where $\mathfrak{a}$, $\mathfrak{b}$ and $\mathfrak{c}$ are order-one coefficients~\cite{Conlon:2005jm}.  While ${\cal V}$ is fixed\footnote{See \cite{BlancoPillado:2009nw} for numerical evidence  supporting the validity of the  single-field approximation.} during inflation, $\tau_\phi$ will evolve, playing the role of the inflaton.
For large $\tau_\phi$, the last two terms in (\ref{equ:V1Kahler}) dominate and determine the inflaton potential
\beq
V(\phi) \simeq V_0 \left( 1 - c_1 \hskip 1pt {\cal V}^{5/3} \hskip 1pt \phi^{4/3} \exp\left[- c_2\hskip 1pt {\cal V}^{2/3} \hskip 1pt \phi^{4/3}\right] \right) \ , \label{equ:Vblow}
\eeq
where we have defined the canonically-normalized inflaton,
\beq
\phi \equiv \sqrt{4\lambda_\phi/(3{\cal V})}\, \tau^{3/4}_\phi\ ,
\eeq
as well as the parameters $V_0 \equiv {\cal O}(1) \times W_0^2 \, \hat \xi \hskip 1pt {\cal V}^{-3}$, $c_1 \equiv {\cal O}(1) \times \hat \xi^{-1}$ and $c_2 \equiv {\cal O}(1) \times a_\phi$.
It is  instructive to expand (\ref{equ:Vblow}) around the vev of $\phi$ in the minimum of (\ref{equ:V1Kahler}), where
\beq
a_\phi \langle \tau_{\phi} \rangle \approx {\cal O}(1) \times \ln({\cal V})\ \qquad \Leftrightarrow \qquad  \langle \phi \rangle \approx {\cal O}(1) \times \frac{\ln({\cal V})^{3/4}}{ {\cal V}^{1/2}}\ .  \label{equ:Vblowvev}
\eeq
Writing $\phi = \langle \phi \rangle + \hat \phi$ and using ${\cal V} \gg 1$, we find
\beq
V(\hat \phi) \simeq V_0 \left( 1 - \kappa_1 \hskip 1pt e^{- \kappa_2 \hat \phi} \right)\ , \label{equ:Vblow2}
\eeq
where\footnote{Our volume scalings of $\kappa_1$ and $\kappa_2$ differ from~\cite{Burgess:2013sla}.}
\begin{align}
\kappa_1 &\equiv c_1 \hskip 1pt {\cal V}^{5/3}  \hskip 1pt \langle \phi \rangle^{4/3} \approx {\cal O}({\cal V}  \ln({\cal V}))\ , \\
\kappa_2 &\equiv \frac{4}{3} c_2 {\cal V}^{2/3} \langle \phi \rangle^{1/3} = {\cal O}({\cal V}^{1/2} \ln({\cal V})^{1/4})\ .
\end{align}

\vskip 4pt
\noindent
{\it Eta problem from string loops.}---The region of interest for inflation is given by
\begin{equation}
{\cal V}^{2/3} \hskip 1pt \phi^{4/3} \gg 1 \qquad {\rm{and}} \qquad \phi \ll 1\ ,  \label{blowupregime}
\end{equation}  where the first condition renders the potential (\ref{equ:Vblow}) exponentially flat, while the second ensures that $\tau_{\phi} \ll \tau_b$, i.e.~that the inflationary blow-up cycle makes a negligible  contribution to the overall volume ${\cal V} \approx \alpha\hskip 1pt \tau_b^{3/2}$.
However, at this point one should remember that we have not yet included string loop corrections.
Using (\ref{equ:KKK}), we can estimate the string loop correction to the inflaton potential,
\beq
\delta V_{(g_{\rm s})} \sim \frac{1}{\sqrt{\tau_\phi}\, {\cal V}^3} \sim \frac{1}{\phi^{2/3} {\cal V}^{10/3}} \ .  \label{lvsloopestimate}
\eeq
The associated correction to the $\eta$ parameter is
\beq
\delta \eta \sim \frac{\delta V_{(g_{\rm s})}''}{V_0} \sim \frac{1}{\phi^{8/3} {\cal V}^{1/3}}  \sim \frac{{\cal V}}{\tau_\phi^2}\ ,  \label{lvsdeltaeta}
\eeq
where we are still using units with $\Mp \equiv 1$.
Using $\tau_\phi \approx \langle \tau_\phi \rangle$ and inserting (\ref{equ:Vblowvev}) in (\ref{lvsdeltaeta}), we find that \beq
\delta \eta \approx a_{\phi}^2\,  \frac{{\cal V}}{\ln({\cal V})^2} \gg 1\ .
\eeq
Thus, the leading string loop corrections to the K\"ahler potential ---  even after incorporating the cancellation of extended no-scale structure ---  lead to {\it{parametrically large}} values of $\eta$.

\vskip 4pt
In conclusion,  K\"ahler moduli inflation from a blow-up cycle  suffers from a severe eta  problem induced by string loop corrections associated with D7-branes wrapping the inflationary cycle.
Note that nonperturbative effects on this cycle are required in order to generate the exponential in (\ref{equ:Vblow}), which is central to the mechanism.
One suggestion for evading the  eta problem \cite{Cicoli:2008gp} is to arrange that only Euclidean D3-branes, not D7-branes, wrap the inflationary cycle: the desired superpotential is then generated, while the associated quantum corrections to the K\"ahler potential are not obviously determined by known and conjectured results \cite{vonGersdorff:2005bf, Berg:2005ja, Berg:2005yu,Berg:2007wt, Cicoli:2007xp}.  However, Euclidean D3-branes  and gaugino condensation  involve closely related physics.  Indeed, there are examples where a quantum correction  that was first computed as an open string loop effect in the D7-brane case, and appeared  inaccessible  in the corresponding Euclidean D3-brane case,  was shown (by a closed string computation) to  take precisely the same form for Euclidean D3-branes \cite{Baumann:2006th}.
Whether the substitution of Euclidean D3-branes will address the eta problem of blow-up inflation  remains an open question that could be resolved by direct computation.

\subsection{Fibre Inflation}
\label{sec:fibre}

A fundamental feature of the Large Volume Scenario is that the leading $\alpha^{\prime}$  correction depends only on the overall volume,  leaving the remaining K\"ahler moduli as flat directions.
As we explained above, in blow-up inflation~\cite{Conlon:2005jm} nonperturbative effects  generate an exponentially flat term in the potential; but it has proven difficult to prevent perturbative quantum corrections to the  K\"ahler potential from introducing a parametrically larger --- and unacceptably steep ---  contribution.
Faced with this situation, it is natural to  ask whether  there exist compactifications in which the perturbative  contributions, which are almost invariably significant, actually  drive inflation.

The first proposal of this sort is {\it fibre inflation} \cite{Cicoli:2008gp}.
The setting is a Calabi-Yau manifold that is a K3  fibration over a $\mathbb{P}^1$ base.
In  the simplest  explicit example,\footnote{The simplest example is  a Calabi-Yau hypersurface in $\mathbb{P}_{4}^{(1,1,2,2,6)}$ --- see \cite{Candelas:1993dm} for details.}
the volume can be written as
\beq
{\cal V} = \frac{1}{2} \hskip 1pt \sqrt{\tau_1} \tau_2 \  , \label{equ:VolFibre1}
\eeq
where we have chosen a convenient basis  in which $\tau_1$  is the volume of the K3 fiber \cite{Cicoli:2008gp}.
As in blow-up inflation, a third blow-up cycle (whose volume we again denote by $\tau_s$) turns out to be necessary.
The volume is therefore assumed to take the form  \cite{Cicoli:2008gp}
\beq
{\cal V} = \alpha \left( \sqrt{\tau_1} \tau_2 - \lambda_s \tau_s^{3/2} \right) \ , \label{equ:VolFibre}
\eeq where $\alpha$  and $\lambda_s$ are model-dependent constants.

\vskip 4pt
\noindent
{\it Inflaton potential.}---Before including string loop corrections, the K\"ahler potential is given by (\ref{equ:KLVS}).
If $\tau_1 , \tau_2 \gg 1$, nonperturbative effects involving $\tau_1$ and $\tau_2$ can be neglected, and the superpotential  takes the form
\beq
W = W_0 + {\cal A}_s e^{-a_s T_s} \ .
\eeq
The scalar potential is then
\beq
V = a_s^2 {\cal A}_s^2\, \frac{\sqrt{\tau_s}}{{\cal V}} e^{-2 a_s \tau_s} - a_s {\cal A}_s W_0\, \frac{\tau_s}{{\cal V}} e^{-a_s \tau_s} + \hat \xi W_0^2 \, \frac{1}{{\cal V}^3}\ . \label{fibrenoloop}
\eeq
The potential (\ref{fibrenoloop}) depends only on $\tau_s$ and ${\cal V}$, which are stabilized at $\tau_s \sim g_{\rm s}^{-1}$ and ${\cal V} \sim W_0 \sqrt{\tau_s} \hskip 1pt e^{a_s \tau_s}$. This leaves a flat direction in the $(\tau_1, \tau_2)$ plane --- namely, the direction along  which ${\cal V}$  remains constant.
This flat direction is plausibly lifted by string loop corrections to the K\"ahler potential.  The main idea of fibre inflation  is that these quantum corrections will provide the leading (non-constant) terms in the inflaton potential.   Before proceeding, we must emphasize that the string loop corrections in question, eqs.~(\ref{equ:KGS1}) and (\ref{equ:KGS2}),
are those {\it{conjectured}} in \cite{Berg:2007wt} (see also~\cite{vonGersdorff:2005bf, Cicoli:2007xp}) as generalizations of the explicit computations of~\cite{Berg:2005ja, Berg:2005yu} for the toroidal orientifold $T^6/(\mathbb{Z}_2 \times \mathbb{Z}_2 )$.  The viability of fibre  inflation rests on the specific form assumed in \cite{Berg:2007wt}, and it would be valuable to obtain more direct and detailed understanding of quantum corrections to the K\"ahler potential.
Without further apologies, the potential from  the conjectured string loop corrections is
\beq
\delta V_{(g_{\rm s})} =  \frac{W_0^2}{{\cal V}^2} \left(\mathfrak{a}\, \frac{g_{\rm s}^2}{\tau_1^2} - \mathfrak{b}\, \frac{1}{\sqrt{\tau_1}\, {\cal V}} + \mathfrak{c}\, \frac{g_{\rm s}^2 \tau_1}{{\cal V}^2}\right)\ ,  \label{delvgs}
\eeq
where $\mathfrak{a}$, $\mathfrak{b}$ and $\mathfrak{c}$ are unknown order-one constants. This fixes the
fiber modulus
$\tau_1$ at $\tau_1 \sim g_{\rm s}^{4/3} {\cal V}^{2/3} $.
An inflationary phase  can arise if $\tau_1$ is displaced far from this minimum, i.e.~if the K3  fiber is initially large compared to the base, and then relaxes to smaller values.

\begin{figure}[h!]
   \centering
     \includegraphics[scale=0.7]{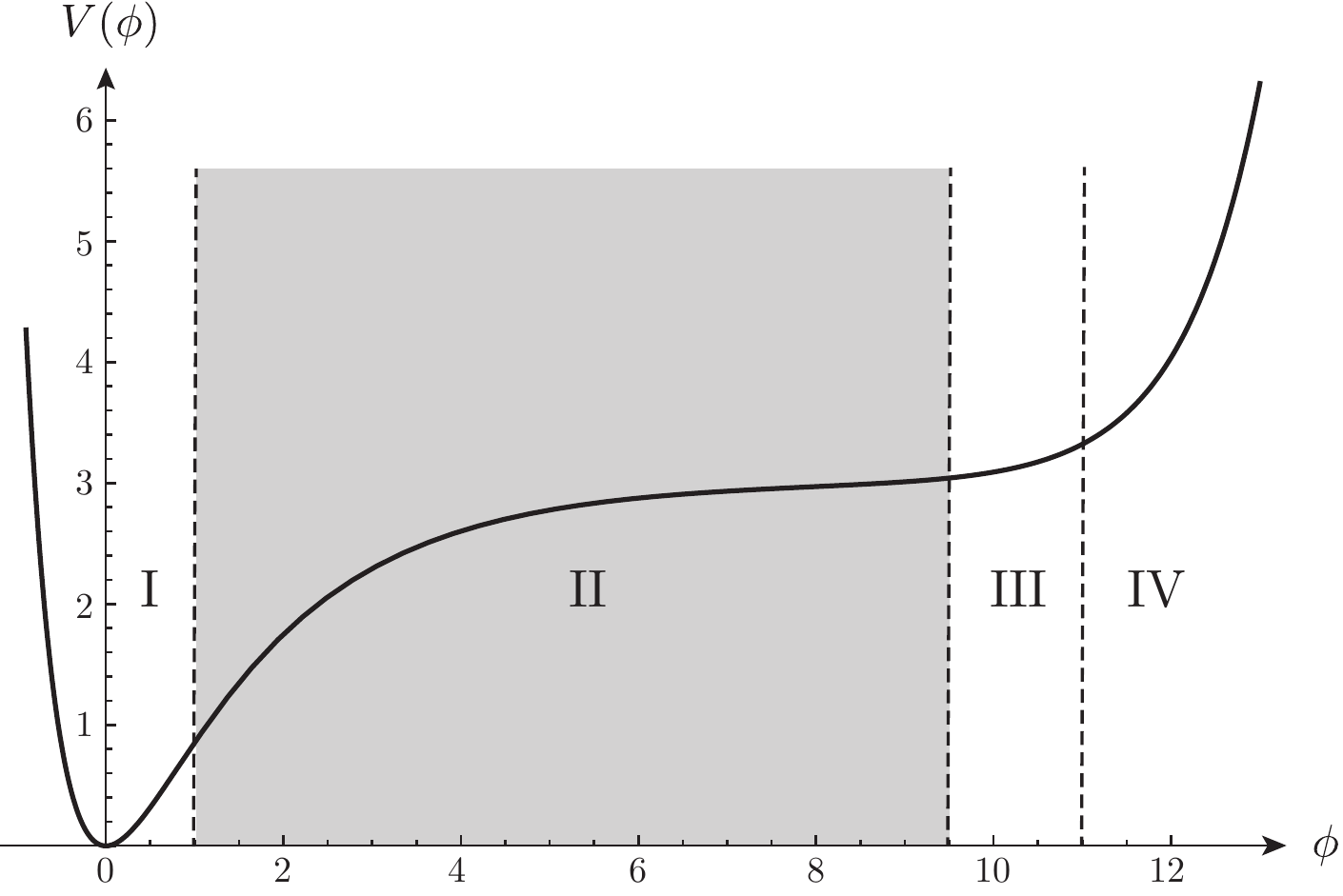}
   \caption{Sketch of the potential for fibre inflation (figure adapted from~\cite{Cicoli:2011zz}). The phenomenologically viable inflationary regime is the gray shaded region II. The slow-roll conditions are also satisfied in region~III, but the spectrum of fluctuations is blue.}
  \label{fig:Fibre}
\end{figure}

As a simple first step,  we suppose that $\tau_s$ and ${\cal V}$  remain fixed at their minima  while $\tau_1$ evolves, and can  be integrated out.
The resulting single-field potential takes the form
\beq
V(\phi) = V_0 \left( 1 - \frac{4}{3} e^{-\phi/\sqrt{3}} + \frac{1}{3} e^{-4\phi/\sqrt{3}} + \frac{\mathfrak{C}}{3}\hskip 1pt e^{2\phi/\sqrt{3}} \right)\ ,  \label{vfibre1}
\eeq
where $V_0\equiv {\cal O}(1) \times{\cal V}^{-10/3} $, $\mathfrak{C} \equiv 16 \hskip 1pt \mathfrak{ac/b}^2\sim g_{\rm s}^4 \ll 1$ and
\beq
\phi \equiv \frac{\sqrt{3}}{2} \ln \tau_1\ . \label{equ:phiFibre}
\eeq
The potential is plotted in fig.~\ref{fig:Fibre}. Successful inflation occurs in region II,\footnote{The slow-roll conditions are also satisfied in region III, but constraints on the spectral index are violated there; see \S\ref{Kahlerphenomenology}.} where
the potential can be approximated as
\beq
V(\phi) \simeq V_0 \left( 1 - \frac{4}{3}\, e^{-\phi/\sqrt{3}} \right) \ . \label{equ:Vfib}
\eeq
Interestingly, this form of the potential is similar to that obtained in the Starobinsky model and in Higgs inflation (see \S\ref{sec:SRM}).

The single-field treatment presented above  is not a priori justified, because the compactification volume ${\cal V}$  is light enough to evolve during inflation.
Even so, ref.~\cite{Cicoli:2008gp}  presents extensive  numerical and analytical evidence showing that the single-field potential (\ref{vfibre1}) gives an accurate picture of the two-field evolution.   The motion of ${\cal V}$ is slow until the end of inflation, and moreover upon incorporating its evolution, i.e.~setting ${\cal V}={\cal V}(\phi)$,  one finds negligible corrections to the slow-roll parameters of the effective single-field model.  It would be interesting to know whether fluctuations of  ${\cal V}$ contribute to the primordial perturbations in fibre  inflation, along the lines of \cite{Burgess:2010bz}.

\vskip 4pt
\noindent
{\it Naturalness and higher corrections.}---The structure of  the potential (\ref{vfibre1}) is ultimately dictated by the leading $\alpha^{\prime}$  correction (\ref{equ:KLVS}), which depends only on ${\cal V}$, and by the string loop corrections (\ref{equ:KGS1}) and (\ref{equ:KGS2}), which enter in (\ref{delvgs}) and lift all flat directions.
Corrections  from higher string loops,  and at higher order in $\alpha^{\prime}$,  have not been computed, but are suppressed by additional (possibly fractional) powers of $g_{\rm s}$  and ${\cal V}^{-1}$.

To understand whether  all such corrections can be neglected, we  recall that the remarkable resilience of LVS rests  in large part on the fact that  at {\it{exponentially}}  large ${\cal V}$,  any unknown or unwanted corrections that are suppressed by any reasonable power (including a fractional power) of ${\cal V}$ are effectively negligible.
As an example, a possible higher-derivative correction  to the ten-dimensional action at order  $(\alpha^{\prime})^4$ would on dimensional grounds be suppressed compared to the  leading term (\ref{equ:Ka}) by a factor ${\cal V}^{-1/3}$.   In `traditional' LVS  constructions, ${\cal V}^{-1/3}$ is a very small number, justifying the omission of higher-derivative terms in ten dimensions.\footnote{See \cite{Conlon:2005ki} for
the details of the $\alpha^{\prime}$  expansion in LVS.}

In fibre inflation, however, reproducing the normalization of the scalar power spectrum compels the volume to be modest in size.
This follows  because the potential (\ref{equ:Vfib}) has only one free parameter: the scale is set by $V_0  \propto {\cal V}^{-10/3}$ and the slow-roll parameter $\epsilon$  is not parametrically adjustable. For the benchmark parameters given in \cite{Cicoli:2008gp}, the  scalar power spectrum has the right amplitude for ${\cal V} \approx 1700$.
Neglecting terms suppressed by integer powers of ${\cal V}$  is clearly safe,  but suppressions by ${\cal V}^{-1/3}$ are marginal,  particularly in cases where the dimensionless prefactor is entirely unknown.\footnote{However,  because the  leading $\alpha^{\prime}$  correction (\ref{equ:Ka}) does not depend on the inflaton,  one might conjecture that subleading $\alpha^{\prime}$  corrections are  likewise inflaton-independent,  and hence unimportant.  We thank Michele Cicoli  for discussions of this point.}

Higher-loop corrections are a  potentially important issue in a model driven by one-loop  corrections.
To understand higher-loop corrections  in fibre inflation, we consider the limit $\tau_1 \to \infty$  at constant ${\cal{V}}$,  corresponding to a K3  fiber that is large compared to the base $\mathbb{P}^1$, which becomes singular in the limit.  The  geometric singularity is reflected in a  (power law) divergence  in the one-loop corrections involving $\tau_2$,  which vanishes in the  large fiber limit \cite{Cicoli:2008gp}.  An  obvious concern is that higher-loop corrections to the K\"ahler potential will become important in this regime.   However, it was shown in \cite{Cicoli:2008gp}  that  slow-roll inflation  also breaks down at small base volume,  at a value of the base volume that is large  enough so that higher-loop corrections are still small.   As a result, higher-loop corrections are argued to be negligible  during the inflationary phase.

Finally, the authors of~\cite{Cicoli:2008gp} have argued that fibre  inflation is robust because of a  hidden symmetry  that emerges in the  limit of infinite volume.
In four-dimensional terms,  the problematic corrections to the inflaton potential are suppressed by  powers of ${\cal V}$,  and vanish  in the  decompactification limit  ${\cal V} \to \infty$.
This limit enjoys  additional symmetries, most notably ten-dimensional general covariance, that could be related to cancellations in the four-dimensional action.  Further analysis of this interesting possibility would be worthwhile.

\vskip 4pt
In conclusion, fibre  inflation is a promising  inflationary scenario in LVS  compactifications.  If one grants the conjectured string loop corrections (\ref{equ:KGS1}) and (\ref{equ:KGS2}), which are the key to the inflaton potential, and also omits  higher-order corrections, the dynamics  is quite robust.   We have argued above that at the comparatively small values of ${\cal V}$ relevant for fibre  inflation,  the validity of the approximations made to arrive at the inflationary potential merits further scrutiny.  To be clear, the possible corrections are not  parametrically large (as they are in  blow-up inflation),  but they could be important, and it would be interesting to  have a sharper picture.

\subsection{Poly-Instanton Inflation}
\label{sec:poly}

In an effort to evade the eta problem of blow-up inflation, ref.~\cite{Cicoli:2011ct} (see also \cite{Blumenhagen:2012ue}) constructed a model in which {\it poly-instanton} terms in the superpotential
make a critical contribution to the scalar potential.  A poly-instanton\footnote{Poly-instantons should not be confused with {\it{multi-instantons}}, which are well-known in field theory and correspond in string theory to multiple Euclidean branes wrapping the same cycle.} effect arises when the Euclidean action $S_a$ of an instanton $a$  receives corrections from a second instanton $b$ \cite{Blumenhagen:2008ji} (see also the earlier work \cite{GarciaEtxebarria:2007zv}), so that
\begin{equation}
W =   {\cal{A}}_a \hskip 1pt {\rm{exp}}\Bigl(-S_a + {\cal{A}}_b \hskip 1pt e^{-S_b} \Bigr)\ ,
\end{equation} where the moduli-dependent prefactors ${\cal{A}}_a$ and ${\cal{A}}_b$ are one-loop determinants.\footnote{Precisely this structure arises in axion monodromy inflation through instanton corrections to the holomorphic gauge coupling function $f$, cf.~eq.~(\ref{fWilpoly}).}

\vskip 4pt
\noindent
{\it Setup.}---The compactification geometry assumed in \cite{Cicoli:2011ct} is the same as in \S\ref{sec:fibre}; in particular, the compactification volume is given by (\ref{equ:VolFibre}),
\beq
{\cal V} = \alpha \left( \sqrt{\tau_1} \tau_2 - \lambda_s \tau_s^{3/2} \right) \ . \label{equ:VolPoly}
\eeq
Building on  explicit poly-instanton constructions in \cite{Blumenhagen:2012kz}, ref.~\cite{Blumenhagen:2012ue}  considered a slightly different model with
\beq
{\cal V} = \tau_b^{3/2} - \tau_s^{3/2} - (\tau_s+\tau_w)^{3/2} \ . \label{equ:VolPolyB}
\eeq
We will focus on (\ref{equ:VolPoly}), as realized in  the explicit constructions of \cite{Lust:2013kt}, but the issues and most of the phenomenology are very similar with the choice (\ref{equ:VolPolyB}) \cite{Blumenhagen:2012ue}.

We consider a stack of D7-branes wrapping the four-cycle associated with the modulus $\tau_s$. We assume that the field theory on the D7-branes can be broken into two sectors that separately undergo gaugino condensation.  The superpotential is then of the racetrack\footnote{The racetrack  is unrelated to the existence of poly-instanton effects:  it is a further model-building requirement.
By introducing  another adjustable parameter, the racetrack superpotential  allows one to evade constraints that arise in a single-condensate model \cite{Cicoli:2011ct,Blumenhagen:2012ue}.}
form~(\ref{equ:Wrace}),
\beq
W = W_0 + {\cal A}\hskip 1pt \exp[-a T_s] - {\cal B}\hskip 1pt \exp[- b T_s]\ ,
\eeq where the sign of the final term is a convenient phase choice.
In addition, a Euclidean D3-brane is taken to wrap the fiber associated with $\tau_1$. This leads to nonperturbative corrections to the gauge kinetic functions of the two condensing gauge groups.
The poly-instanton corrected superpotential then takes the form
\beq
W = W_0 + {\cal A}\hskip 1pt \exp\left[-a \left(T_s + c_1 e^{-2\pi T_1}\right)\right] - {\cal B} \hskip 1pt \exp\left[-b\left(T_s + c_2 e^{-2\pi T_1}\right)\right]\ , \label{equ:Wpoly}
\eeq
where $c_1$ and $c_2$ are constants.

\vskip 4pt
\noindent
{\it Inflaton potential.}---In the absence of the poly-instanton corrections,  i.e.~for $c_1=c_2=0$, the fields ${\cal V}$ and $\tau_s$ are stabilized as before. Again, one is left with a flat direction in the $(\tau_1,\tau_2)$ plane.  This time, the flat direction is lifted by the poly-instanton contributions in (\ref{equ:Wpoly}).
As before,  one can consistently integrate out ${\cal V}$ and $\tau_s$, as well as the axion partners of all K\"ahler moduli (see \cite{Cicoli:2011ct} for details).  The scalar potential for the distance from the minimum in the $\tau_1$ direction, i.e.~$\hat \tau_1 \equiv \tau_1 - \langle \tau_1 \rangle$, is found to be
\beq
V = \frac{F_{\rm poly}}{{\cal V}^{3+p}} \left( 1 - (1+ 2\pi \hat \tau_1) e^{-2\pi \hat \tau_1} \right) \ ,
\eeq
where $F_{\rm poly} \equiv {\cal O}(1) \times W_0$ and $p={\cal O}(1)$. The order-one factors in the parameters $F_{\rm poly}$  and $p$ depend in a complicated way on the microscopic parameters of the theory, and the precise expressions can be found in \cite{Cicoli:2011ct}.
Using (\ref{equ:phiFibre}), we can write the potential in terms of the canonically-normalized inflaton field,
\beq
V(\hat \phi) \simeq V_0 \left( 1 - \kappa_2 \hskip 1pt \hat \phi \hskip 1pt e^{-\kappa_2 \hat \phi} \right)\ , \qquad \hat \phi \approx \frac{\sqrt{3}}{2} \frac{\hat \tau_1}{\langle \tau_1 \rangle} \ , \label{equ:Vpoly}
\eeq
where $V_0 \equiv F_{\rm poly} {\cal V}^{-(3+p)}$ and $\kappa_2 \simeq {\cal O}(1) \times \ln({\cal V})$.

\vskip 4pt
\noindent
{\it Corrections.}---Because the setting (\ref{equ:VolPoly}) for  poly-instanton inflation in \cite{Cicoli:2011ct} is precisely that of fibre inflation, while  the geometry (\ref{equ:VolPolyB}) in \cite{Blumenhagen:2012ue} is similar to that in blow-up inflation, one  should ask about the string loop corrections to the K\"ahler potential that were  crucial in \S\ref{sec:blow} and \S\ref{sec:fibre}.
In \cite{Cicoli:2011ct}, it is argued that because D7-branes only wrap $\tau_s$, not ${\tau_1}$ or $\tau_2$, with only  Euclidean D3-branes wrapping~${\tau_1}$, one does not expect open string loop corrections that depend on the inflaton ${\tau_1}$.  However, as we remarked in \S\ref{sec:blow}, it  has not  actually been shown that dangerous open string loop corrections  are absent in this setting.  Instead, a fair summary is that  the calculation of \cite{Berg:2005ja, Berg:2005yu} that led to the conjecture \cite{Berg:2007wt} is not immediately applicable, and  no first-principles computation of the quantum corrections has been presented.  The absence of  (a certain sort of) quantum corrections to an  unprotected quantity such as the K\"ahler potential would be quite striking, and further investigation of this point is warranted.

In  addition to  corrections from loops of open strings ending on D7-branes,   the K\"ahler potential can also be corrected by loops of closed strings.
This quantum correction  was estimated in \cite{Cicoli:2011ct},  where it was found that closed string loops can
significantly affect the  shape of the inflaton potential.   The size of the effect depends on an undetermined amplitude ${\cal C}_{\rm{loop}}$ that  depends on the complex structure moduli, and
may be assumed to be of order unity in generic situations.  In \cite{Cicoli:2011ct},  it was assumed that for appropriate choices of flux one has ${\cal C}_{\rm{loop}} \lesssim 0.1$,  in which case  the loop corrections can be neglected.

\subsection{Phenomenology}
\label{Kahlerphenomenology}

In the truncation to a single-field description, the models of  inflation in LVS  described in this section
can all be written in terms of the approximate potential
\beq
V(\phi) \approx  V_0 \left( 1 -  \kappa_1 \hskip 1pt e^{-\kappa_2 \phi}\right)\ . \label{equ:VKahler}
\eeq
In blow-up inflation, $\kappa_1 = {\cal O}({\cal V} \ln {\cal V})$ and $\kappa_2 = {\cal O}({\cal V}^{1/2} (\ln {\cal V})^{1/4})$, while in fibre inflation $\kappa_1 \sim \kappa_2  ={\cal  O}(1)$, and in
poly-instanton inflation $\kappa_1 \sim \kappa_2 = {\cal  O}(\ln ({\cal V}))$.
The slow-roll parameters derived from (\ref{equ:VKahler}) are
\beq
\eta \simeq - \kappa_1 \kappa_2^2\,  e^{- \kappa_2 \phi} \qquad {\rm and} \qquad \epsilon \simeq \frac{1}{2}\frac{\eta^2}{\kappa_2^2} \ .
\eeq
This class of models therefore satisfies $\epsilon \ll \eta$ and hence  (\ref{naivetilt}) becomes
\beq
n_s \simeq 1 + 2\eta \ .
\eeq
Given $n_s$, one predicts
the tensor-to-scalar ratio:
\beq
 r \simeq \frac{2}{\kappa_2^2}(n_s-1)^2\  \xrightarrow{\ n_s = 0.96\ } \ \frac{3\times 10^{-3}}{\kappa_2^2}\ . \label{equ:rP}
\eeq
This prediction for $r$ depends on the parameter $\kappa_2$, which differs for the different classes of K\"ahler moduli inflation  scenarios:
\begin{itemize}
\item[$\triangleright$]  {\it Blow-up inflation.}---Because of the parametrically large string loop correction (\ref{lvsdeltaeta}) to $\eta$ in blow-up inflation,
it is not necessarily well-motivated to derive predictions from the uncorrected model of the form (\ref{equ:VKahler}).\footnote{For the same reason,
in \S\ref{sec:dbrane} we did not analyze the phenomenology that would arise in warped D3-brane inflation driven by a Coulomb potential with no  corrections from moduli stabilization:  although these predictions are widely  quoted in the literature,  they have little meaning.}  Here, we will only point out that in blow-up inflation without string loop corrections, $\kappa_2 = {\cal O}({\cal V} (\ln {\cal V})^{1/4}) \gg 1$, cf.~(\ref{equ:Vblow2}).
Thus, by (\ref{equ:rP}), the tensor-to scalar ratio $r$  is  extremely small.   However, this  feature relies on exponential flatness of the potential,  which as explained above is very vulnerable to corrections.

\item[$\triangleright$]  {\it Fibre inflation.}---Eq.~(\ref{equ:Vfib}) is of the form (\ref{equ:VKahler}), with $\kappa_2 = 1/\sqrt{3}$.  This leads to a direct correlation between the scalar spectral index and the tensor-to-scalar ratio,
\beq
r \simeq 6 (n_s - 1)^2 \  \xrightarrow{\ n_s = 0.96\ } \  0.01\ .
\eeq
A word about  predictions for $n_s$ in fibre inflation is necessary.   From (\ref{vfibre1})  one readily sees that slow-roll inflation can occur in both regions II and III depicted in fig.~\ref{fig:Fibre}.  In region III $\eta >0$,  so that $n_s > 1$:  the spectrum has a blue tilt,  which is  strongly disfavored by observations.
The approach of \cite{Cicoli:2008gp} is to consider only inflationary dynamics in region II,  but in fact the full model (\ref{vfibre1}) can produce a blue {\it{or}} a red spectrum, depending on where on the potential the large-angle CMB  exits the horizon: see also \cite{Cicoli:2013oba,Pedro:2013pba}.  The situation is similar to that in inflection point inflation (cf.~\S\ref{warpedphenomenology}),  which is unsurprising given the shape of the potential in fig.~\ref{fig:Fibre}.

\item[$\triangleright$]  {\it Poly-instanton inflation.}---In (\ref{equ:Vpoly}), we found $\kappa_2 = {\cal O}(\ln({\cal V}))$.  A typical model~\cite{Cicoli:2011ct} has $\kappa_2 = \ln (10^3) \sim 10$ and hence
\beq
r \sim 10^{-5}\ .
\eeq
Such a low tensor amplitude is unobservable.

\end{itemize}

It seems quite generic that the inflaton field in K\"ahler  moduli inflation couples to additional light degrees of freedom.  This can modify the above results, which were based on a truncation to single-field inflation, and may lead to additional signatures. For example,
variations of blow-up inflation have been proposed~\cite{Burgess:2010bz} that allow for large local non-Gaussianity via the curvaton mechanism~\cite{Linde:1996gt, Lyth:2001nq,Moroi:2001ct, Enqvist:2001zp}: $\fnl^{\rm loc} \sim {\cal O}({\rm few}) \times 10$.   Similarly, extensions of the simplest fibre inflation models have been constructed~\cite{Cicoli:2012cy}
that produce relatively large local non-Gaussianity from modulated reheating~\cite{Dvali:2003em, Dvali:2003ar, Zaldarriaga:2003my, Kofman:2003nx}: $\fnl^{\rm loc} \sim {\cal O}({\rm few})$.
Both possibilities are strongly constrained by the Planck bound~(\ref{equ:fl}).

\section{Inflating with Dissipation}
\label{sec:Dissipation}

In systems  where the potential energy  function is too steep to support slow-roll inflation,  dissipation  can provide an alternative source of accelerated expansion.
Microscopically, dissipative effects arise if the inflaton is coupled to, and excites, additional degrees of freedom during inflation.
To model this we add a direct coupling between the inflaton and the extra fields, collectively denoted~${\psi}$:
\beq
S = \int \d^4 x \sqrt{-g} \, \left[  \frac{M_{\rm pl}^2}{2}R - \frac{1}{2}(\partial \phi)^2 - V(\phi) + {\cal O}(\phi, \psi)\, \right]\ .
\eeq
Suitable couplings can lead to the production of ${\psi}$-particles, which drains energy from the inflaton sector and leads to an enhanced effective friction that slows the evolution of the inflaton field.  However, since the density of particles is diluted exponentially during inflation, it  is difficult to maintain friction-dominated evolution. In this section, we present a few ideas for how this might nevertheless be achieved.\footnote{An effective field theory of dissipative inflation was constructed in~\cite{LopezNacir:2011kk}.  Related work on {\it warm inflation} \cite{Berera:1995ie} is reviewed in \cite{WarmInflation} (see also \cite{Anber:2009ua, Battefeld:2011yj}).}

\vskip 4pt
In \S\ref{sec:trapped}, we describe {\it trapped inflation}~\cite{Green:2009ds},  in which dissipative dynamics  arises from repeated production of particles or strings.
We  explain how trapped inflation  could plausibly arise in the  class of string  compactifications discussed in  the context of axion monodromy inflation in \S\ref{ssec:AM}, albeit in a slightly different parameter regime.  Then, in \S\ref{ssec:FluxCascades} and \S\ref{ssec:magnetic}, we present two very recent ideas: inflation via flux cascades~\cite{DAmico:2012sz} and via magnetic drift~\cite{Adshead:2012kp}.  Both are imaginative additions to the string inflation literature,  so we include them here even though,  at the time of writing,
the models still lack explicit embeddings into fully specified string compactifications including moduli stabilization.
We hope our discussion will inspire the reader
to determine whether these ideas can be realized in concrete compactifications.  Finally,
we close, in \S\ref{sec:TrapPheno}, with a summary of the phenomenology of trapped inflation.

\subsection{Trapped Inflation}  \label{sss:trappedI}

A good place to learn about quantum-mechanical particle production during inflation is the pioneering work of Kofman, Linde and Starobinsky~\cite{Kofman:1997yn}.  Here, we describe the basic elements of that analysis and then apply
them to trapped inflation.\footnote{This section is based mostly on \cite{Kofman:2004yc, Green:2009ds}.}

\subsection*{Particle Production}

We start by computing the particle production for a simple field theory model~\cite{Kofman:1997yn,Kofman:2004yc, Barnaby:2009mc}.
The result of this computation will feed into the dynamics of the inflationary model.
Consider a scalar field $\psi$ coupled to the inflaton $\phi$ via the   interaction
\beq
{\cal L}_{\rm int} \, = \, - \frac{1}{2}g^2 (\phi - \phi_0)^2 \psi^2\ . \label{equ:cop}
\eeq
Notice that the field $\psi$ becomes massless at a specific point in field space, $\phi = \phi_0$. This is where the $\psi$ particles are produced.
Near this point, we can approximate the homogeneous inflaton evolution as
\beq
\phi(t) \approx \phi_0 + \dot \phi_0 (t-t_0) \ ,
\eeq
which implies a time-dependent effective mass for the $\psi$ particles,
\beq
m_\psi^2(t) \equiv g^2(\phi- \phi_0)^2 \approx k_\star^4 (t-t_0)^2\ ,
\eeq
where $k_\star^2 \equiv g|\dot \phi_0 |$.
The evolution equation for a Fourier mode of the
$\psi$ field
is then
\beq
\ddot \psi_k + 3 H \dot \psi_k + \underbrace{\left(\frac{k^2}{a^2} + k_\star^4 (t-t_0)^2 \right)}_{\equiv \, \omega_k^2(t)}\, \psi_k = 0 \ \ . \label{equ:chiEOM}
\eeq
Particles are produced when the evolution becomes non-adiabatic,
\beq
|\dot \omega_k| > \omega_k^2\ .
\eeq
This occurs in the time interval $|t-t_0| < k_\star^{-1}$ and for momenta $k < k_\star$.
Solving (\ref{equ:chiEOM})
gives the occupation number of the
$\psi$ particles~\cite{Kofman:1997yn,Kofman:2004yc}\footnote{This result assumes $k_\star > H$.}
\beq
n_k = e^{-\pi k^2/k_\star^2}\ .
\eeq
Shortly after $t=t_0$, the number density of $\psi$ particles is
\beq
n_\psi(t_0) = \int \frac{\d^3 k}{(2\pi)^3} \, n_k \approx \frac{k_\star^3}{(2\pi)^3} \ . \label{equ:dense}
\eeq
This depends on a combination of the coupling constant $g$ and the inflaton speed $|\dot \phi_0|$.
We assume that the $\psi$ particles become sufficiently massive after the production event so that they can be treated as non-relativistic matter.
The density of $\psi$ particles then dilutes as $a^{-3}$,
\beq
n_\psi(t) = \frac{k_\star^3}{(2\pi)^3}  \frac{a^3(t_0)}{a^3(t)}\, \Theta(t-t_0)\ ,
\eeq
where $\Theta$ is the Heaviside function.
The energy density of the $\psi$ particles is $\rho_\psi(t) = m_\psi n_\psi(t)$.

\begin{figure}[h!]
   \centering
      \includegraphics[scale=0.6]{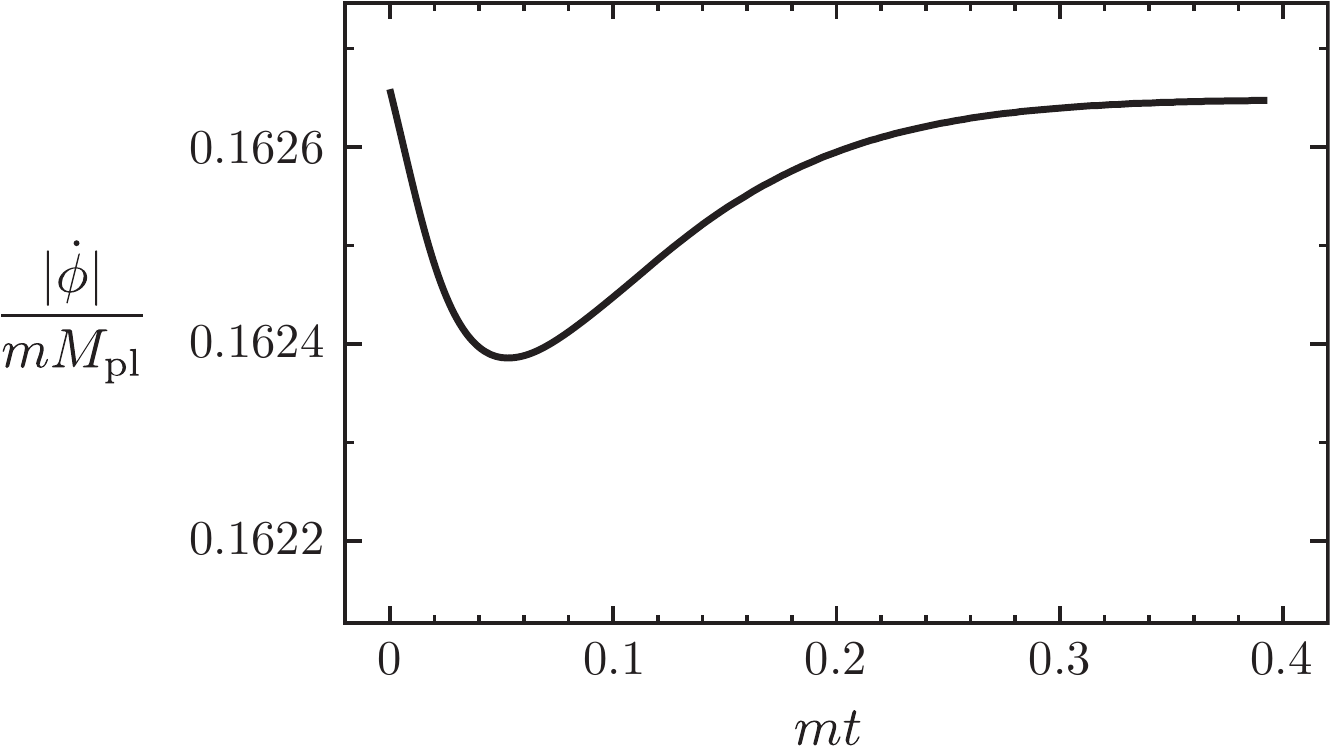}
   \caption{Decay of the inflaton velocity due to particle production (figure adapted from \cite{Barnaby:2009mc}). The time $t=0$ corresponds to the production event, the coupling is $g^2=0.1$, and $m^2 \equiv V''$.}
  \label{fig:trapped}
\end{figure}

\vskip 4pt
Next, we determine how the density of $\psi$ particles affects the evolution of the inflaton field $\phi$.
The effect can be estimated by using the following mean-field equation~\cite{Kofman:1997yn,Kofman:2004yc}
\beq
\ddot{\phi} + 3 H \dot {\phi} + V' = - g^2(\phi - \phi_0) \langle \psi^2\rangle  \ , \label{equ:back}
\eeq
where\footnote{For a derivation of (\ref{equ:Psi2}) see~\cite{Kofman:1997yn,Kofman:2004yc}
.}
\beq
\langle \psi^2 \rangle \approx \frac{n_\psi(t)}{g|\phi-\phi_0|} \ . \label{equ:Psi2}
\eeq
Fig.~\ref{fig:trapped} shows a numerical solution of eq.~(\ref{equ:back}). We see that the inflaton velocity $\dot{\phi}$ decays after the production event, but then returns almost to its initial value as the effect of the particles gets diluted away.  The cosmological evolution is affected only temporarily by the particle production.\footnote{Particle production can be continuous if the inflaton is coupled to a gauge field (see \S\ref{ssec:magnetic}).}
In other words, a single particle production event does not lead to many $e$-folds of dissipative dynamics.
Achieving
inflation from dissipation requires that the density of particles is kept high by repeated particle production.  The resulting inflationary model is called {\it trapped inflation}.\footnote{Trapped inflation  was first proposed in \cite{Kofmanunpublished,Kofman:2004yc}. The inflationary mechanism,  the spectrum and bispectrum,  and possible microphysical embeddings were systematically analyzed in \cite{Green:2009ds}.  See also the related work \cite{Barnaby:2009mc}.}
  We first describe an effective field theory construction of trapped inflation and then present a string theory realization.

 \subsection*{Trapped Inflation in Effective Field Theory}
\label{sec:trapped}

To get repeated particle production, we  replicate the coupling (\ref{equ:cop}) at
$N$ points $\phi_i$,
\beq
{\cal L}_{\rm int} \, = \, - \frac{1}{2} g^2\sum_{i=1}^N (\phi - \phi_i)^2 \psi^2_i \ , \label{equ:Ltrap}
\eeq
We take these points to be evenly spaced, with $\phi_{i+1} -\phi_i \equiv \Delta $, both to simplify the presentation  and because uniform spacing  is natural in microphysical models involving monodromy (see below).
Particles are now produced periodically with densities given by (\ref{equ:dense}),
\beq
n_{\psi_i}(t_i) \simeq \frac{\big(g \dot{\phi}(t_i)\big)^{3/2}}{(2\pi)^3}  \equiv   \frac{\dot{\varphi}^{3/2}(t_i)}{(2\pi)^3} \ .
\eeq
We have ignored the effects of any finite pre-existing particle density on the particle production, and we have defined $\varphi \equiv g\phi$  for later convenience.
Replacing $\psi_i^2$ by its expectation value $\langle \psi_i^2\rangle $, we get
an equation of motion for the inflaton:
\beq
\ddot{\phi} + 3 H \dot{\phi} + V' + \sum_i  \frac{g \dot{\varphi}^{3/2}(t_i)}{(2\pi)^3} \frac{a^3(t_i)}{a^3(t)} = 0 \ . \label{equ:phiEOM}
\eeq
If the production events are spaced densely enough\footnote{The necessary condition is $\Delta \ll \{\, \dot {\phi}/H\, , \, \dot{\phi}^2/\ddot{\phi}\, \}  $~\cite{Green:2009ds}.}, then we can replace the sum by an integral
\beq
 \sum_i  \frac{g \dot{\varphi}^{3/2}(t_i)}{(2\pi)^3} \frac{a^3(t_i)}{a^3(t)}  \approx \int^t \frac{\d t'}{\Delta}\, \frac{\dot{\varphi}^{5/2}(t')}{(2\pi)^3} \frac{a^3(t')}{a^3(t)} \ \approx\ \frac{1}{3H\Delta}\frac{\dot{\varphi}^{5/2}(t)}{(2\pi)^3} \ , \label{equ:INT}
 \eeq
and (\ref{equ:phiEOM}) becomes
\beq
 \ddot{\phi} + 3 H \dot{\phi} + V' +  \frac{1}{24 \pi^3} \frac{g^{5/2}}{H\Delta} \, \dot{\phi}^{\hskip 1pt 5/2} = 0 \ .
\eeq
Notice the extra friction term proportional to $\dot{\phi}^{\hskip 1pt 5/2}$ provided by the finite density of $\psi$ particles.
Assuming slow-roll ($|\ddot{\phi}| \ll 3 H |\dot{\phi}|$), and taking the damping  to be dominated by  particle production ($3H|\dot{\phi}| \ll V'$), we find
\beq
\dot \varphi = g \dot{\phi} \approx - \big( 24 \pi^3 \hskip 1pt H  \Delta\hskip 1pt V' \big)^{2/5}\ . \label{equ:DotPhi}
\eeq

Let us estimate the conditions for this solution to correspond to inflation.
We assume that the Hubble parameter is dominated by the potential energy of the inflaton,
\beq
3 M_{\rm pl}^2 H^2 = \rho_\phi + \rho_\psi \approx V(\phi) \ ,
\eeq
while its evolution is sourced by the $\psi$ particles,
\beq
2 M_{\rm pl}^2 \dot H \approx - \rho_\psi \ ,
\eeq
where we have used $\dot \rho_\psi \simeq - 3 H \rho_\psi \gg \dot \rho_\phi$. The Hubble slow-roll parameter is then
\beq
\varepsilon = - \frac{\dot H}{H^2} \approx \frac{3}{2} \frac{\rho_\psi}{V} \ , \label{equ:VarEps}
\eeq
where
\beq
\rho_\psi(t) = \sum_i g | \phi - \phi_i|\, n_{\psi_i}(t)\approx \int^t \frac{\d t'}{\Delta}\, |\phi(t) -  \phi(t')|\, \frac{\dot{ \varphi}^{5/2}(t')}{(2\pi)^3} \frac{a^3(t')}{a^3(t)} \ . \label{equ:integral}
\eeq
Using $| \phi(t) -  \phi(t')| \approx \dot{\phi} (t-t')$, we can approximate the integral in (\ref{equ:integral}) in the same way as in~(\ref{equ:INT}),
\beq
\rho_\psi(t) \approx \frac{1}{(3H)^2} \frac{1}{g \Delta} \frac{\dot{ \varphi}^{7/2}(t)}{(2\pi)^3} \ .
\eeq
Using the solution (\ref{equ:DotPhi}) to replace $ \dot{ \varphi}$, we can write (\ref{equ:VarEps}) as
\beq
\varepsilon \,\sim\, \frac{\epsilon^{7/10}}{g} \left( \frac{H}{M_{\rm pl}} \frac{\Delta^2}{M_{\rm pl}^2}\right)^{1/5} \ , \label{equ:VAReps}
\eeq
where $\epsilon$ is the potential slow-roll parameter (\ref{equ:eta}) and we have dropped some unimportant numerical factors.
We see that inflation can occur ($\varepsilon < 1$) even for a steep potential ($\epsilon > 1$). The parametric scaling of the answer in (\ref{equ:VAReps}) is as expected:
particle production is more efficient for larger coupling $g$ and smaller spacing $\Delta$; both of these effects correspond to smaller $\varepsilon$ for fixed $\epsilon$.
Consistency conditions and further constraints on $g$ and $\Delta$ were studied in~\cite{Green:2009ds}.

 \subsection*{Trapped Inflation in String Theory}
\label{sec:trapped2}

The  core requirement for trapped inflation is a closely-spaced series of  particle production events along the inflationary trajectory, as in the toy Lagrangian (\ref{equ:Ltrap}).
This structure appears contrived in four-dimensional EFT,  but  readily arises in string theory as a consequence of monodromy (cf.~\S\ref{ssec:AM}).
We will describe two approaches to a string theory embedding of trapped inflation~\cite{Green:2009ds, Silverstein:2008sg, McAllister:2008hb}.

\vskip 4pt
\noindent
{\it Wrapped brane monodromy.}---We first examine a D4-brane in a nilmanifold compactification \cite{Silverstein:2008sg}, where the replication responsible for serial particle production  is most easily visualized.
Consider the three-dimensional nilmanifold
(or `twisted torus') ${\cal N}_3$ defined by coordinates $u_1, u_2, x$
identified by
\begin{align}
t_x\, &: \quad (x,u_1, u_2) \mapsto (x+1,u_1,u_2) \\
t_{u_1}\, &: \quad (x,u_1,u_2) \mapsto (x-Mu_2, u_1+1,u_2)  \label{sltwist} \\
t_{u_2} \, &: \quad (x,u_1,u_2) \mapsto (x,u_1,u_2+1)\ ,
\end{align} with the  line element
\beq
\frac{\d s^2}{\alpha'} = L_{u_1}^2 \d u_1^2  + \underbrace{L_{u_2}^2 \d u_2^2 + L_x^2 \left( \d  x + M u_1 \d u_2 \right)^2}_{T^2} \ ,
\eeq where $L_{u_1}$, $L_{u_2}$, and $L_{x}$ are  dimensionless constants.
This geometry corresponds to a $T^2$  fibration over a circle  parameterized by $u_1$, which we denote by $S^1_{u_1}$: for each value of $u_1$ there is a $T^2$ in $u_2$ and $x$,
\beq
\frac{\d s^2_{T^2}(u_1)}{\alpha'} = L_{u_2}^2 \d u_2^2 + L_x^2 \left(\d x + M u_1 \d u_2\right)^2\ .
\eeq
The identification (\ref{sltwist}) shows that the fiber $T^2$ at $u_1=1$ is twisted by an $SL(2,\mathbb{Z})$ transformation  before being glued to the fiber at $u_1=0$.   More precisely,
the complex structure of the torus shift by $M$ units, i.e.~$\tau \mapsto \tau + M$ as $u_1 \mapsto u_1 + 1$.
These equivalent tori are identified by the projection $t_{u_1}$.
At $M$ special locations around $S^1_{u_1}$, $M u_1 = j \in \mathbb{Z}$, the tori are rectangular:
\beq
\frac{\d s_{T^2,{\perp}}^2}{\alpha'} = L_x^2 \d y_1^2 + L_{u_2}^2 \d y_2^2\ .
\eeq
We have defined coordinates $y_1 \equiv x + j u_2$ and $y_2 \equiv u_2$ obtained from an $SL(2,\mathbb{Z})$ transformation of $x$ and $u_2$.

The configuration of interest is type IIA string theory compactified on an orientifold of the product space ${\cal N}_3  \times \tilde {\cal N}_3$,
with $\tilde {\cal N}_3$ a second nilmanifold.
For the moment it suffices to consider a single ${\cal N}_3$.
We consider a D4-brane wrapped on the one-cycle defined by $u_2 = \lambda$, or equivalently by $(y_1,y_2) = (j \lambda, \lambda)$. The role of the inflaton is played by the $u_1$ coordinate of the D4-brane.
The key point is that if the D4-brane  is transported in the
$u_1$ direction, the fiber torus returns to an
equivalent torus, but the one-cycle does {\it not}: e.g.~at $u_1=0$, the brane wraps $(y_1,y_2) = (0,\lambda)$, while at $u_1 = 1$, the brane wraps $(y_1,y_2) = (M\lambda, \lambda)$.
\begin{figure}[h!]
   \centering
     \includegraphics[width=0.9\textwidth]{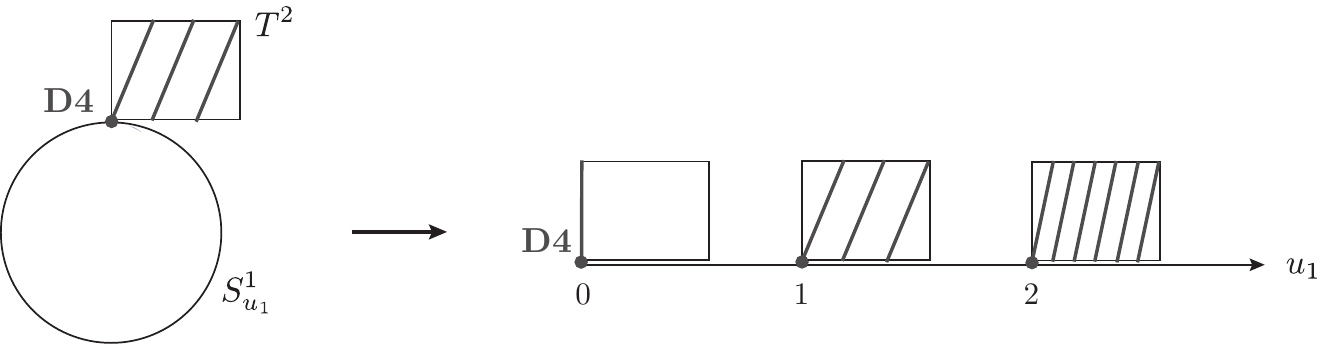}
   \caption{Monodromy of a wrapped D4-brane on a nilmanifold.}
  \label{fig:Nil1}
\end{figure}
The D4-brane  undergoes monodromy  upon transport around $S^1_{u_1}$ \cite{Silverstein:2008sg}.

To derive the dynamics of the wrapped D4-brane, we consult
the DBI action
\beq
S_{{\rm D}4} = - \frac{1}{(2\pi)^4 g_{\rm s} (\alpha')^2} \int \d^4 x \sqrt{-g} \, \sqrt{L_{u_2}^2 + L_x^2 M^2 u_1^2} \left(1 - \frac{1}{2}  \alpha' L_{u_1}^2 \dot u_1^2 \right)\ .
\eeq
For $L_x M u_1 \gg L_{u_2}$,
we get
\beq
S_{{\rm D}4} = \int \d^4 x \sqrt{-g} \left( \frac{1}{2} \dot \phi^2 - \mu^{10/3} \phi^{2/3} \right) \ ,
\eeq
where
\begin{align}
\frac{\phi^2}{\Mp^2} &= \frac{2}{9} (2\pi)^3 g_{\rm s} \frac{M}{L^3} \frac{L_{u_1}}{L_{u_2}} u_1^3 \ , \\
\frac{\mu}{\Mp} &= \frac{M_{\rm s}}{\Mp} \left( \frac{9}{4} \frac{M^2}{(2\pi)^8 g_{\rm s}^2} \left( \frac{L_x}{L}\right)^3 \frac{L_{u_2}}{L_{u_1}} \right)^{1/10}\ .
\end{align}
Here, we have defined $L^3 \equiv L_{u_1} L_{u_2} L_x$.
The field range can be super-Planckian if $L_{u_1} \gtrsim L_{u_2}$ and
\beq
\Delta u_1^3 \gg \frac{L^3}{M} \ .
\eeq
This corresponds to moving around the $S^1$ many times (see fig.~\ref{fig:Nil1}).

\begin{figure}[h!]
   \centering
     \includegraphics[scale=1.4]{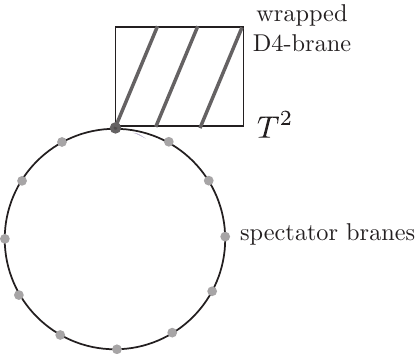}
   \caption{Trapped inflation on a nilmanifold.}
  \label{fig:Nil2}
\end{figure}

\vskip 4pt
\noindent
{\it Trapped inflation from wrapped branes.}---A small modification of the scenario of \cite{Silverstein:2008sg} allows for repeated particle production events, as in
 trapped inflation (see fig.~\ref{fig:Nil2}).
 In addition to the inflationary brane, we consider $N$  D4-branes wrapping the $SL(2,\mathbb{Z})$ transforms of the one-cycle associated with the inflaton brane.
The $j$th brane has a potential $(u_1 - j/M)^2 + \cdots$ and therefore minimizes its energy at $u_1 = j/M$. As the inflaton brane unwinds, it comes close to each of these spectator branes. The strings stretching between the mobile brane and the lattice of stationary branes play the role of the extra fields $\psi_j$.
Before accounting for moduli stabilization, the effective potential is of the form given in (\ref{equ:Ltrap}).

A serious obstacle to realizing trapped inflation in a  nilmanifold compactification is that  the presence of a large number  of D4-branes  tends to destabilize the moduli \cite{Green:2009ds}. For the moduli stabilization constructions described in \cite{Silverstein:2007ac,Haque:2008jz}, the D4-brane energy exceeds the scale of the moduli potential barriers when
$N_{{\rm D}4} \gtrsim {\cal{O}}(10)$, which does not  allow enough particle production events for trapped inflation.

\vskip 4pt
\noindent
{\it Trapped inflation from axions.}---A closely related scenario  in which instabilities are under better control \cite{Green:2009ds} is the axion monodromy model \cite{McAllister:2008hb} described in \S\ref{ssec:AM}.
Let us denote by $\ell^2$ the volume of the two-cycle $\Sigma_2$ wrapped by the NS5-brane, cf.~(\ref{ns5dbi}).  A D3-brane wrapping $\Sigma_2$, with worldvolume flux
\begin{equation}
\int_{\Sigma_2} {\cal{F}}_2 = n \in \mathbb{Z}\ ,  \label{fonD3}
\end{equation}
 gives rise in four dimensions to  a string with tension 
\beq
T_{{\rm D}3/\Sigma_2} = T_3 \sqrt{\ell^2 + (c\hskip 1pt g_{\rm s} +n)^2} \ ,  \label{wrappedD3tension}
\eeq
where  $c \equiv \int_{\Sigma_2} C_2$ measures induced D1-brane charge in the D3-brane. For $c g_{\rm s} \in \mathbb{Z}$, this induced charge can be canceled by  the quantized flux ${\cal F}_2$ in (\ref{fonD3}), if $n=-c g_{\rm s}$.
We now notice that if $\ell \to 0$,
this configuration gives rise to a {\it{tensionless string}}  in four dimensions whenever integer values of $c g_{\rm s}$ are canceled by appropriate flux ${\cal F}_2$.  These  strings play the role of the $\psi$ particles in the EFT discussion of \S\ref{sec:trapped}:  as $c$  diminishes from a large initial vev,  tensionless strings are produced at  regularly-spaced intervals.

A necessary  condition for trapped inflation in this setting  is a field range \cite{Green:2009ds} $\Delta\phi/\Mpl \gtrsim \sqrt{60}/10$, which is a bit milder than the requirement for chaotic inflation in \S\ref{ssec:AM}.
On the other hand, a systematic study of  moduli stabilization and backreaction  would be necessary to determine sufficient conditions for trapped inflation in the context of axion monodromy, and to characterize corrections to the simple model of~(\ref{equ:Ltrap}).

\vskip 4pt
\noindent
{\it Weak coupling limit of DBI inflation.}---Trapped inflation is closely related to DBI inflation.  Consider a D3-brane in the background of a stack of  $N-1$ D3-branes, corresponding to a location on the Coulomb branch of  ${\cal N} = 4$ super-Yang-Mills theory (see \S\ref{sec:DBICFT}).
At large $g_{\rm s} N$,  the D3-brane is a probe of an $AdS_5 \times S^5$  geometry.
Taking $\phi$  to be the canonical field  representing  the radial position of the D3-brane,  one easily sees  that strings stretching between the  isolated D3-brane  and the stack  have masses proportional to $\phi$.  By analogy to  the Higgs mechanism,  the fields with masses $m \propto \phi$  are sometimes called `W-bosons' (though  some of the relevant  fields are fermions).

Now suppose that the D3-brane moves toward the stack, breaking supersymmetry  by virtue of its kinetic energy.
There are two  important effects that can change its trajectory: {\it{virtual}} W-bosons  induce quantum corrections  to the action for $\phi$, while pair production of {\it{on-shell}}  W-bosons ---  caused by the time-dependence of their mass --- drains energy from  the
$\phi$ sector.   The former effect  leads to the DBI action~(\ref{equ:DBI}):  indeed, it is the fact that  the W-bosons  have  $m \propto \phi$ that causes the non-renormalizable terms in (\ref{equ:DBI}) to be suppressed by $\phi$,  rather than by a fixed cutoff scale.   The latter effect  is precisely the particle production process  described in \S\ref{sss:trappedI}.
At large 't Hooft coupling,  so that the gauge theory is strongly coupled  but the supergravity background probed by the D3-brane is weakly curved,  the dominant effect on the D3-brane dynamics comes from  virtual W-bosons \cite{Silverstein:2003hf}:  the  evolution is governed by the DBI action,  with negligible particle production.    If  instead the gauge theory is weakly coupled,  particle production dominates.
In the sense, trapped inflation is the weak-coupling analogue of DBI inflation, even though --- as usual with  strong-weak dualities ---  one  rarely has  control of both sides in the same setting.
Indeed, we saw above that the most plausible string theory realizations of trapped inflation do  not involve taking the  weak coupling limit  of  the configurations  studied in \S\ref{ssec:DBI}
(namely, D3-branes  in Calabi-Yau cones):  instead,  compactifications involving monodromy  are a more fruitful setting.

\vskip 4pt
The phenomenology of trapped inflation will be discussed in \S\ref{sec:TrapPheno}.

\subsection{Flux Cascades}
\label{ssec:FluxCascades}

The {\it{unwinding inflation}} scenario \cite{DAmico:2012ji, DAmico:2012sz} combines  bubble nucleation, dissipation/trapping, monodromy, the DBI effect, oscillations in the potential, and a hybrid exit through brane-antibrane annihilation. The basic setup is the following:
a $(p+2)$-form flux $F_{p+2}$ fills the noncompact spacetime and threads a $(p-2)$-cycle in the compact space. Initially there are $Q_0 \gg 1$ units of $F_{p+2}$, but the flux can be discharged by the nucleation of a $p$-brane/anti-$p$-brane pair, followed by ${\cal O}(Q_0)$ `unwindings',  in which the brane and antibrane move  in opposite directions around the compact cycle,  reducing the flux and colliding with each  other in every circuit.\footnote{The first proposal to use self-collisions of a bubble in compact extra dimensions to drive inflation  appears in \cite{Brown:2008ea}.  Cascades following nucleation events were discussed in \cite{Easther:2009ft,Kleban:2011cs}.  A closely related scenario in which D-brane motion around the compact cycle discharges a flux is \cite{Shlaer:2012by}.}
In this section, we will give some of the details of  unwinding inflation, and comment on the prospect of realizing this idea in string theory.\footnote{This section is based on  \cite{DAmico:2012ji, DAmico:2012sz}.}

\vskip 4pt
The unwinding mechanism is applicable in a  broad class of higher-dimen\-sional gravity theories involving  suitable fluxes,  but, anticipating a UV completion in string theory, we will  limit our discussion to string compactifications. Consider string theory in the ten-dimensional spacetime
\begin{equation}
{\cal M}_{10} = dS_4 \times X_6\ ,
\end{equation}
for $X_6$ a compact manifold,  and take $Q_0 \gg 1$ units of the R-R $(p+2)$-form flux $F_{p+2}$ to fill the  noncompact spacetime and thread a $(p-2)$-cycle $\Sigma_{p-2}$ in $X_6$, 
\begin{equation}
\int\limits_{dS_4 \times \Sigma_{p-2}} \hskip -10pt F_{p+2} \, =\, Q_0 \gg 1\ .
\end{equation}
The flux induces an effective cosmological constant in four dimensions: this will play the role of the inflationary energy density.
A D$p$-brane  carries electric charge under $F_{p+2}$,  and nucleation of a  bubble bounded by a D$p$-brane creates a region (the bubble interior) in which the flux is reduced by one unit compared to the exterior,  as in \cite{Brown:1988kg}: this is a higher-dimensional analogue of the  Schwinger process in QED.   The background flux creates a force  on the bubble,  driving it to expand in $\Sigma_{p-2}$.\footnote{As a simple analogy \cite{DAmico:2012ji},
one can picture the flux as a rubber sheet  that wraps 
repeatedly around $\Sigma_{p-2}$. The initial bubble nucleation corresponds to  the appearance of an  approximately spherical hole in one layer of the wrapped sheet.   The tension of the rubber  causes the hole to expand,  unwinding  layer after layer of the sheet.}

\begin{figure}[h!]
   \centering
     \includegraphics[scale=0.65]{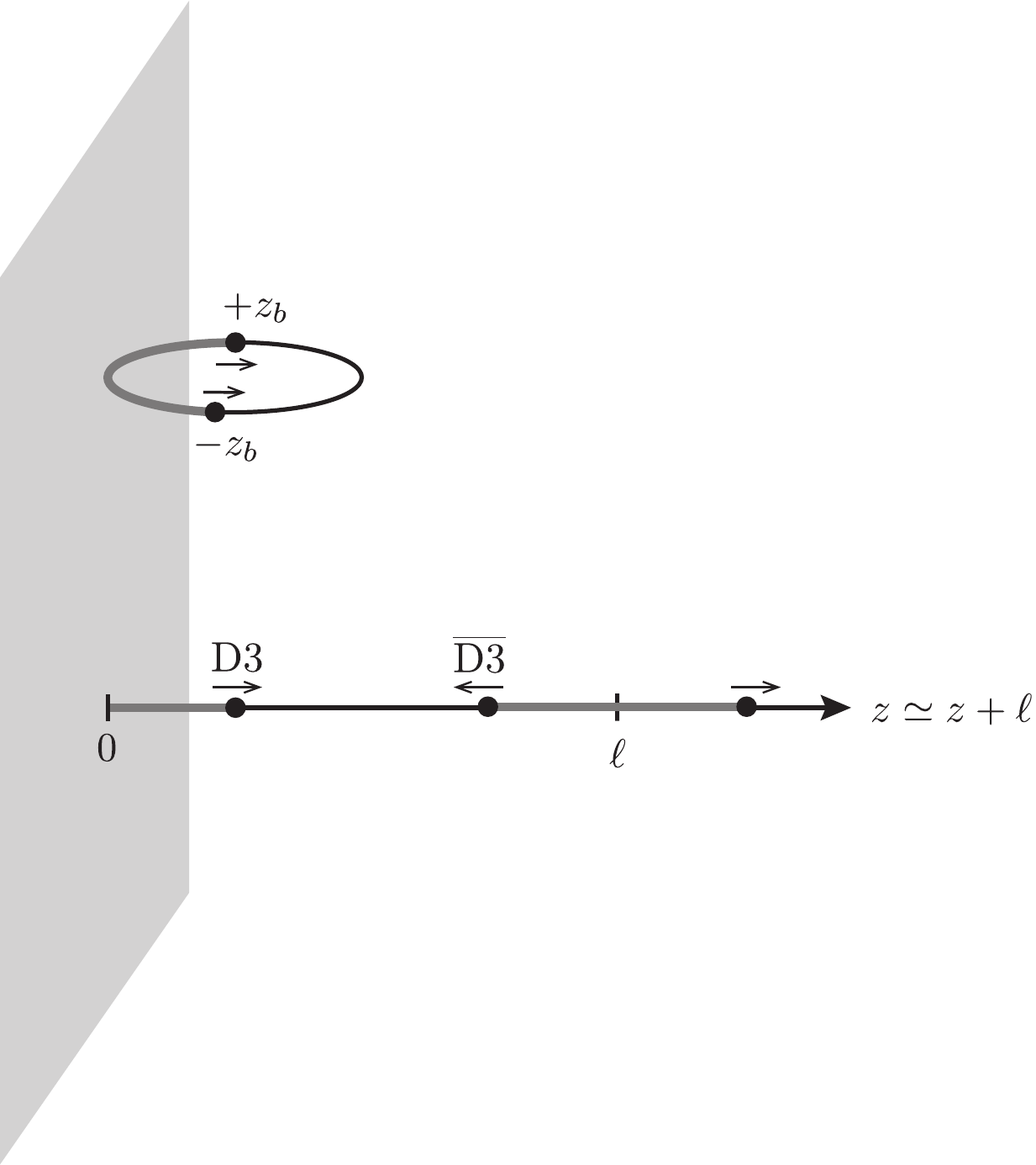}
   \caption{A cascade of five-form flux on $dS_4 \times S^1$. The compact $S^1$ and its covering space are both shown. }
  \label{fig:Unwrapping}
\end{figure}

Eventually the D$p$-brane  bubble  becomes so large that it `unwraps' $\Sigma_{p-2}$ and collides with itself,  dissipating energy into  open string degrees of freedom.
Because the portions of the bubble that collide have opposite orientation,  this is locally a  D$p$-brane/anti-D$p$-brane collision.
Provided that the collision happens so rapidly that the brane-antibrane tachyon does not have time to condense, and provided that the dissipation is not strong enough to stall the unwinding process --- see below  for discussions of these important points --- the brane and the antibrane pass through each other  and continue to unwind (see fig.~\ref{fig:Unwrapping} for a five-dimensional example).
Each subsequent collision reduces the flux by a further unit.  In the four-dimensional effective theory, this appears as a slow reduction of the effective cosmological constant, mimicking the evolution during slow-roll inflation.\footnote{This is similar in spirit to {\it chain inflation}~\cite{Freese:2004vs,Feldstein:2006hm,Huang:2007ek,Chialva:2008xh,Ashoorioon:2008pj,Ashoorioon:2008nh,Cline:2011fi}, although the microscopic details are quite different, and is also very similar to the unwinding of a wrapped D4-brane in monodromy inflation in nilmanifold  compactifications \cite{Silverstein:2008sg}, and to the reduction of induced D3-brane charge during axion monodromy inflation \cite{McAllister:2008hb}, cf.~\S\ref{ssec:AM}.}
When the flux has dropped sufficiently, the branes stop moving relativistically.  Tachyon condensation can then be efficient when the branes approach each other, and brane-antibrane annihilation provides a natural  hybrid exit from inflation.

A number of important questions arise at this stage.
What sort of bubble nucleation event leads to a flux discharge cascade?   What is the  four-dimensional effective action for the unwinding branes?   Is dissipation a small correction  to the background evolution,  and to the  scalar and tensor perturbations?
What is the  dynamics of the  D-brane pair in the  compact directions  perpendicular to the flux?   Are the requirements of  unwinding inflation compatible with compactification and moduli stabilization?  We will  briefly address the first three points, following \cite{DAmico:2012ji, DAmico:2012sz}, and then  review some of the difficulties  involved in embedding these ideas in string theory.

To discuss the effective action, it  will be instructive to  examine the  simplified example of $dS_4 \times S^1$ with five-form flux $F_5$~\cite{DAmico:2012sz}.  In this case, bubble nucleation leads to the situation depicted in fig.~\ref{fig:Unwrapping}. The bubble is bounded by D3-branes at $+z_b$ and $-z_b$:
because these branes have opposite charges, one can think of them as a brane-antibrane pair.
The role of the inflaton is played by $z_b$, the radius of the bubble in the extra dimension.\footnote{For D$p$-branes in a  compactification of critical string theory, there will be additional scalars  describing the remaining coordinates of the D-branes (at least if $p<8$,  which  includes all examples of interest).    One should bear in mind that these  fields could  be crucial for the background evolution and the perturbations,  so this five-dimensional toy model  may not give a faithful representation of  unwinding in a string compactification.}
To determine  the  effective action  for $z_b$, ref.~\cite{DAmico:2012sz}
examined bubble nucleation in the Euclidean spacetime  with metric
\begin{equation}
\d s_{\rm E}^2 = H^{-2}\Bigl(\d\xi^2 +{\rm{sin}}^2\xi \hskip 1pt \d\Omega_3^2\Bigr) + \d z^2\ ,
\end{equation}  where $\d\Omega_3^2$ is the line element on $S^3$  and $z$  is the coordinate for an $S^1$ of circumference $\ell$.
The bubbles  of  primary interest have initial size\footnote{A bubble of size $\Delta z \gtrsim \ell$  would correspond to an ordinary  bubble  of reduced flux  in $dS_4$,  and would  expand in $dS_4$ without initiating a cascade.}  $\Delta z \ll \ell$ and have the maximum possible symmetry (as this is characteristic of  dominant instantons).
Solving the Euclidean equations of motion and then continuing back to Lorentzian signature, ref.~\cite{DAmico:2012sz} obtained the action
\beq
S  = \int \d z \int \d {\cal H}_3 \hskip 1pt \d t \,\, \frac{\sinh^3(Ht)}{H^3}  \left( - 2 \sigma\, \delta(z-z_b) \sqrt{1- (\partial z_b)^2} \, - \, \frac{F_5^2}{2 \cdot 5!}  \right)\ , \label{equ:ActionBubble}
 \eeq
where $\d {\cal H}_3 \equiv \sinh^2 (\rho)\hskip 1pt  \d \rho \hskip 1pt \d \Omega_2$ is the integration measure on a three-hyper\-boloid, $\sigma$ is the tension of the wall (a D3-brane),
and
\beq
\frac{F_5^2}{5!} = \mu^5 Q^2 = \mu^5 \left( Q_0 + \sum_{j=-\infty}^\infty \Big[ \Theta(z-z_b+ j \ell) - \Theta(z+z_b + j \ell) \Big] \right)^2 \ . \label{equ:flux}
\eeq
Here, $Q_0$ stands for the initial flux before bubble nucleation, $\mu^{5/2}$ is the charge of the brane, and the sum is over all the image branes (cf.~fig.~\ref{fig:Unwrapping}).
Eq.~(\ref{equ:flux}) is a consequence of Gauss's law, which requires that the flux changes by one unit of the brane charge across the bubble wall.
Once the radius of curvature of the bubble becomes much larger than the Hubble radius, the four-dimensional spacetime inside the bubble can be approximated by flat de Sitter space. Integrating over the fifth dimension, the action~(\ref{equ:ActionBubble}) becomes
\beq
S = \int \d t \hskip 1pt {\rm d}^3 x\, a^3(t)\left( - 2 \sigma \sqrt{1-(\partial z_b)^2} - V(z_b) \right)\ , \label{5daction}
\eeq
where $a(t) \equiv e^{Ht}$.

Upon solving the equations of motion that follow from (\ref{5daction}) ---  still neglecting dissipation during collisions ---  one finds that the  D3-brane  velocity $\dot{z}_b$ is relativistic, and approximately constant.
The branes  collide with image branes when $z_b = n \ell/2$, for $n \in \mathbb{Z}$.
This leads to discrete jumps in the vacuum energy $V$ perceived by an observer in the four-dimensional spacetime.
On timescales that are large relative to the Kaluza-Klein scale, this reduction in the inflationary energy density appears continuous and can be approximated as
\beq
V(z_b)  \sim \mu^5 \left( Q_0 - \frac{z_b}{\ell} \right)^2 \ .
\eeq
Tachyon condensation ends unwinding inflation, so it is crucial that the brane and antibrane can collide ${\cal O}(Q_0)$ times without slowing down so much from the resulting dissipation that the tachyon condenses prematurely.  Tachyon condensation is suppressed when the brane-antibrane collision is relativistic, and a priori one could  construct a configuration in which the electric force from the flux accelerates the brane to an arbitrarily large $\gamma$, allowing a correspondingly large number of cycles.  In a realistic cosmology, however, $\gamma$ is bounded from above: the DBI kinetic term of the moving brane leads to equilateral non-Gaussianity,  as described in \S\ref{ssec:DBI}, and the Planck upper limit requires that $\gamma \lesssim 24$, cf.~eq.~(\ref{equ:GCon}).  This limits the degree to which tachyon condensation can be  deferred.

Computing the production of open and closed strings in a relativistic brane-antibrane collision --- particularly in the most singular case of zero impact parameter --- is highly nontrivial.
If the branes are taken to be homogeneous, some aspects of the calculation can be performed in the four-dimensional EFT involving the lightest string modes, but there are subtleties in this approach.  The naive EFT obtained by dimensional reduction of a compactification containing a stationary brane-antibrane pair does not  correctly capture the spectrum of masses of stretched strings between a brane and an antibrane in {\it{relativistic}} relative motion \cite{McAllister:2004gd}.  To obtain the  correct rate of open string production, one
applies the optical theorem to the annulus amplitude for the moving branes \cite{Bachas:1995kx}, which reveals that the effective tension of the stretched string diminishes at large $\gamma$, increasing the pair production rate \cite{McAllister:2004gd, Bachlechner:2013fja}.  Building on the results of \cite{Bachas:1995kx}, ref.~\cite{DAmico:2012sz}  argued that  for the velocities allowed by (\ref{equ:GCon}), only the lowest few massive string modes are produced  during the collision.  One limitation of this approach is that the annulus amplitude provides information about pair production in a  constant-velocity scattering process, while in practice the dissipation from a head-on collision may substantially (albeit temporarily) decelerate the brane-antibrane pair.\footnote{This deceleration leads to bremsstrahlung, which is dramatically enhanced at large $\gamma$ \cite{Bachlechner:2013fja}.  The values of $\gamma$ allowed by limits on non-Gaussianity are not large enough for the results of \cite{Bachlechner:2013fja}, where an ultrarelativistic limit was assumed, to be directly applicable, but the losses  to closed string radiation during unwinding inflation may nevertheless be significant, and deserve further study.}  A direct calculation of the production of excited open strings in a series of relativistic scattering processes with varying velocity (and perhaps with inhomogeneities) would be a major technical challenge.

Even if the dissipation in each collision is a mild correction to the background evolution, dissipation could have a major impact on the perturbations.
One possibility is that the periodic modulations of the Hubble constant will induce resonant contributions to the spectrum and bispectrum,  as described for  axion monodromy inflation in
\S\ref{ssec:AM}.  More dramatically, the repeated production of  open strings could source the dominant component of the scalar power spectrum, as for trapped inflation in~\S\ref{sss:trappedI}.

The evolution described above assumes that only a single coordinate (the bubble radius) is relevant, and that production of particles and strings, as well as the eventual tachyon condensation, are not strongly inhomogeneous.  These issues are linked: a fluctuation of the  moving brane in a transverse direction changes the impact parameter of the collision, and both particle production and the tachyon mass depend sensitively on the impact parameter.  Although a number of related consistency checks were performed in the toy models of \cite{DAmico:2012sz}, the geometries considered in \cite{DAmico:2012sz} may be too simple to capture the dynamics of unwinding in a realistic compactification, and further investigation is warranted.

\vskip 4pt
The essential mechanism  of unwinding inflation is fairly simple, and appears to arise naturally in toy flux compactifications, with simple unwarped geometries and with the moduli stabilized by fiat. Above we have highlighted some limitations of the calculations of \cite{DAmico:2012sz}, as well as some ways in which the presence of extra compact dimensions --- still stabilized by hand --- could complicate, or prematurely terminate, the unwinding process.  In closing, we will reemphasize the importance of  complete and calculable moduli stabilization.  For a proper perspective, one should recognize that {\it{all}} of the mechanisms for inflation in string theory that we have described thus far appear to succeed naturally in unstabilized toy compactifications, but (we would argue) {\it{none}} has been automatically successful after moduli stabilization and careful implementation of microphysical constraints.
Because unwinding inflation tends to occur at a high scale \cite{DAmico:2012sz}, it faces the very general problem of achieving adequate barriers to destabilization and decompactification, which in various guises plagued the large-field axion models of \S\ref{sec:AxionInflation}.   Addressing this issue by realizing unwinding inflation in a fully stabilized string compactification is an interesting problem for the future.

\subsection{Magnetic Drift}
\label{ssec:magnetic}

Another class of scenarios for dissipative inflation invokes couplings between the inflaton and  one or more gauge fields.
Suppose first \cite{Anber:2009ua} that the inflaton is an axion $\phi$ that couples to $N$ $U(1)$ gauge fields $A_i$ via the standard axionic coupling
\beq
{\cal L} \, \supset\, - \,  \sum_{i=1}^N  \alpha_i\, \frac{\phi}{f} \hskip 1pt  F_i \tilde F_i\ , \label{equ:gauge}
\eeq
where $F_i = \d A_i$ is the gauge field strength  of the $i$th  gauge group, and $\tilde F_i $ is its dual.
A time-dependent axion vev, $\phi(t)$, breaks the conformal invariance of the action for the gauge field, leading to the production of quanta of the gauge field.
It is  natural to ask whether dissipation through production of gauge fields
can slow $\phi$ sufficiently to give inflation. In~\cite{Anber:2009ua}, it was shown  a successful inflationary period with phenomenologically viable perturbations requires large couplings, $\alpha_i \sim {\cal O}(100)$, to a large number of gauge fields, $N \sim 10^5$.
The top-down naturalness of this mechanism  remains to be established,  and in particular  no complete string theory realization has been constructed.\footnote{One way to achieve large $\alpha$ is to fine-tune two axion decay constants to be nearly coincident~\cite{Anber:2009ua}.}

Another  recent proposal, which we will now describe in some detail, is that inflation  can be achieved by coupling an axion to non-Abelian gauge fields  with suitable vevs \cite{Adshead:2012kp, Martinec:2012bv} (see \cite{Maleknejad:2011sq, Maleknejad:2011jw,SheikhJabbari:2012qf}  and the review article \cite{Maleknejad:2012fw} for  the related idea of {\it gauge-flation}).
A large Chern-Simons coupling between the axion and the gauge fields transfers energy from the inflaton sector to the
gauge fields --- without dissipation --- and allows slow-roll to occur even in the presence of a steep potential.  The basic dynamics is similar to that of a charged particle in a magnetic field.\footnote{This section is based mostly on \cite{Adshead:2012kp, Martinec:2012bv}. We thank Peter Adshead and Emil Martinec for helpful discussions.}

\vskip 4pt
As a concrete example, let us consider a stack of $N$ D3-branes, with $SU(N)$ gauge theory on their   worldvolume. The Chern-Simons coupling (\ref{equ:NCS}) includes the term
\beq
S_{\rm CS} = \frac{i}{2\pi}  \int_{{\cal M}_4} C_0\, {\rm Tr}\left[ {\cal F}_2 \wedge {\cal F}_2 \right]\ .
\eeq
This is a topological term, and will not appear in the stress tensor.
Combining this with the standard kinetic terms for the axion and the gauge field, and
evaluating the action in a homogeneous FRW background, we find
\begin{align}
S_{\rm eff} &= \int \d^4 x\, \Bigg\{ a^3 \left[   \gamma_C \dot C_0^2  - V(C_0) + \gamma_A \frac{{\rm Tr}(\dot A^2)}{a^2}   + \gamma_A \frac{{\rm Tr}([A,A]^2)}{a^4}  \right] \nonumber \\ & \hspace{2cm}   +\, \kappa\, C_0\, {\rm Tr}(\dot A[A,A]) \Bigg\} \ , \label{equ:Schromo}
\end{align}
where $\gamma_A \equiv 1/((2\pi)^2 g_{\rm s})$, $\gamma_C \equiv g_{\rm s}^2 M_{\rm pl}^2$, and $\kappa = 1$ (in D7-brane generalizations  discussed below, we will have $\kappa \in \mathbb{Z}$).
We have left the axion potential, $V(C_0)$, unspecified.
We introduce the canonically-normalized inflaton
\beq
C_0(t) \equiv \frac{\phi(t)}{\sqrt{\gamma_C}}\ , \label{equ:Cansatz}
\eeq
and choose an initial gauge field configuration with a rotationally invariant vacuum expectation value,
\beq
A_0 \equiv 0 \ , \qquad A_i(t) \equiv \frac{a(t)\psi(t)}{\sqrt{\gamma_A \nu}} \hskip 2pt J_i\ , \label{equ:Aansatz}
\eeq
where $J_i$ are the generators of $SU(2)$ in the $N$-dimensional representation.\footnote{Any $SU(N)$ group has
an $SU(2)$ subgroup, and here we have identified the global part of this $SU(2)$ with the group $SO(3)$ of spatial rotations.}
Substituting (\ref{equ:Cansatz}) and (\ref{equ:Aansatz}) into (\ref{equ:Schromo}), we find
\begin{align}
S_{\rm eff} &= \int  \d^4 x \, a^3 \left[ \, \frac{1}{2}\dot \phi^2 -V(\phi)  +\, \frac{3}{2}(\dot \psi+H\psi)^2 - \frac{3}{2} g^2 \psi^4   \right. \nonumber \\
&\hspace{2.3cm} \left.  - \, \frac{3g \lambda}{f} \phi\hskip 2pt  \psi^2(\dot \psi+H\psi)\, \right]\ , \label{equ:Schromo2}
\end{align}
where $f \equiv \sqrt{\gamma_C}$, $\lambda \equiv \kappa/\gamma_A$, and $g \equiv 1/\sqrt{\gamma_A \nu}$. The same action arises in phenomenological models of {\it chromo-natural inflation}~\cite{Adshead:2012kp}.

\begin{figure}[h!]
   \centering
     \includegraphics[width=0.95\textwidth]{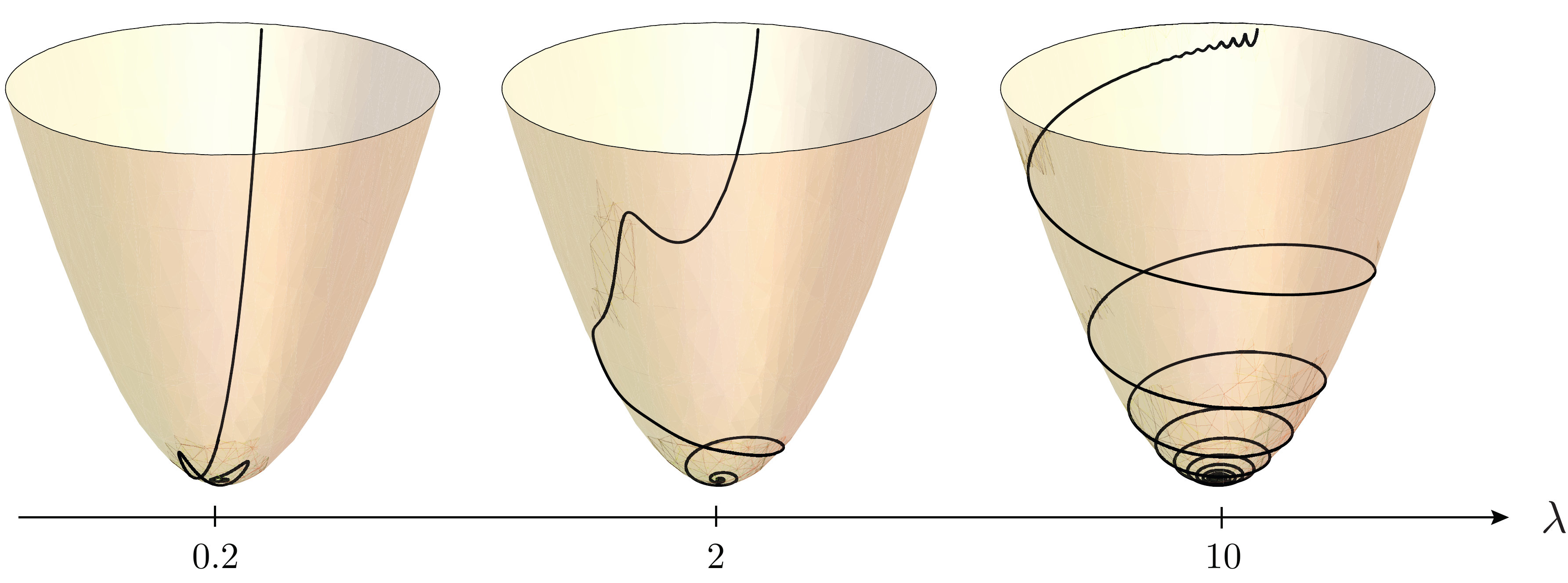}
   \caption{Evolution of a charged particle in two dimensions with quadratic external potential and a coupling to a homogeneous magnetic field. For large enough coupling to the magnetic field, the particle experiences slow magnetic drift. (The numerics for these figures was kindly provided by Peter Adshead.)}
  \label{fig:Magnetic}
\end{figure}

The equation of motion for the inflaton is
\beq
\ddot \phi + 3 H \dot \phi + V_{,\phi} = - 3 \frac{g \lambda}{f} \psi^2(\dot \psi + H\psi) \ . \label{equ:InfEOM}
\eeq
In addition to the force from the bare axion potential, the field experiences the analogue of a {\it magnetic drift}
force proportional to the coupling $\lambda$.  For large $\lambda$, the two forces balance each other and hence allow a slow evolution of the inflaton.
 This is closely related to the magnetic drift phenomenon of a charged particle coupled to a magnetic field; see fig.~\ref{fig:Magnetic}.

For sufficiently large $\lambda$, the effective action (\ref{equ:Schromo2}) leads to inflation.
The maximum number of $e$-folds that can be achieved while the axion rolls to the minimum of its potential is found to be~\cite{Adshead:2012kp}
\beq
(N_{e})_{\rm max} \approx \frac{3}{5} \lambda\ .
\eeq
Thus, successful chromo-natural inflation requires a large Chern-Simons coupling, $\lambda \gtrsim {\cal O}(100)$.

\vskip 4pt
The crucial question is whether such a large coupling can be achieved in a controlled string compactification.
In fact, it is easy to see that this cannot be achieved for a stack of D3-branes at weak coupling, as in that case the Chern-Simons coupling is fixed by the string coupling, $\lambda \sim g_{\rm s} \ll 1$.
A possible alternative is a stack of D7-branes wrapping a four-cycle $\Sigma_4$, with Euler number $\chi(\Sigma_4) = \int_{\Sigma_4} \hat \chi(R)$, and with worldvolume  gauge field
instanton number $c_2=K$.
The Chern-Simons interaction is then
\beq
S_{\rm CS} = \frac{i}{(2\pi)^3}  \int\limits_{{\cal M}_4 \times \Sigma_4} \frac{C_0}{24}\, \left({\rm Tr}[{{\cal F}_2} \wedge {{\cal F}_2} \wedge {{\cal F}_2} \wedge {{\cal F}_2}]  + \frac{1}{2} {\rm Tr}[{\cal{F}}_2 \wedge {\cal{F}}_2] \, \hat\chi(R) \right)\ ,
\eeq
where we have included the curvature coupling proportional to the Euler density $\hat\chi(R)$.
The effective coupling in (\ref{equ:Schromo2}) is then found to be \cite{Martinec:2012bv}
\beq
\lambda = \left[ K + \frac{\chi(\Sigma_4)}{24} \right] \times  g_{\rm s} \times \frac{\ell_{\rm s}^4}{{\cal V}_4}\ .
\eeq
It appears difficult, but not impossible, to achieve $\lambda \gtrsim {\cal O}(100)$ in this setting while retaining control of the $g_{\rm s}$ and $\alpha^{\prime}$  expansions. Further attempts to obtain large magnetic couplings in string theory are discussed in \cite{Martinec:2012bv}.

\subsection{Phenomenology}
\label{sec:TrapPheno}

The study of the  primordial perturbations arising in dissipative models is comparatively new, and because  realizations in string theory are also a work in progress,  a definitive characterization of the phenomenology is not available at present.  In this section, we will  briefly describe some of the  more robust signatures.

\vskip 4pt
\noindent
{\it Trapped inflation.}---The perturbations in trapped inflation, while understood in some detail~\cite{Green:2009ds, LopezNacir:2011kk}, are not easily described analytically.
We therefore present only the main results, referring the reader to the original literature for derivations~\cite{Green:2009ds, LopezNacir:2011kk}.

\begin{itemize}

\item[$\triangleright$]  {\it Power spectra.}---Particle production affects the spectrum of primordial perturbations.  The inflaton fluctuations satisfy
\begin{align}
\ddot{\delta \phi} &+ \left(M^2 + \frac{k^2}{a^2}\right) \delta \phi  \nonumber \\[4pt]
&+ \int^t \d t' \, M^2 \left( \frac{5}{2} \dot{\delta \phi}(t') - 3H \delta \phi(t') \right) \frac{a^3(t')}{a^3(t)} = - g\hskip 1pt \Delta n_\psi(k, t)\ , \label{equ:tra}
\end{align}
where we have defined  the time-dependent effective mass of inflaton fluctuations,
\beq
M^2 \equiv \frac{g^{5/2}}{(2\pi)^3} \frac{\dot{\phi}^{\hskip 1pt 3/2}}{\Delta}\ , \label{equ:Mtra}
\eeq
and the variance in the number of produced $\psi$ particles, $\Delta n_\psi \equiv g \sum_i \left( \psi_i^2 - \langle \psi_i^2 \rangle \right) \left(\phi-\phi_i \right)$.
Solving (\ref{equ:tra})  leads to the power spectrum  of curvature perturbations~\cite{Green:2009ds}
\beq
\Delta_\R^2 \, \approx\,  g^{8/3}\, \frac{H}{M} \left( \frac{M}{\Delta}
 \right)^{2/3} \ . \label{equ:DZtrap}
 \eeq
Using (\ref{equ:Mtra}), this can be written as
\beq
\Delta_\R^2 \, \approx\,   g^{9/4} \, \left( \frac{H}{\Delta}\right)^{1/2} \left( \frac{H^2}{\dot{\phi}}\right)^{1/4}  \ ,
\eeq
and the spectral tilt is
\beq
n_s - 1 = \frac{\dot H}{H^2} - \frac{1}{4} \frac{\ddot{ \phi}}{H \dot{\phi}} \ .
 \eeq
For the specific example studied in~\cite{Green:2009ds}, the tilt was found to be $n_s = 0.99$, but in general the tilt depends on the details of the model,  such as the shape of the potential, the density of particle production events,  and the properties of  the particles that are produced.

If the power spectrum of tensors is dominated by vacuum fluctuations, cf.~eq.~(\ref{equ:Dh}), then the tensor-to-scalar ratio is
\beq
r = g^{-8/2} \, \frac{HM }{\Mp^2} \left( \frac{\Delta}{M}\right)^{2/3} \ .
\eeq
For the regime of parameters that is consistent with constraints on the scalar fluctuations, ref.~\cite{Green:2009ds} finds $r \ll 10^{-4}$.

\item[$\triangleright$]  {\it Equilateral non-Gaussianity.}---The nonlinear couplings between the inflaton $\phi$ and the extra fields $\psi_i$ lead to non-Gaussianity in the primordial curvature perturbations.
The bispectrum peaks in the equilateral
configuration and has amplitude
\beq
\fnl^{\rm equil} \simeq \frac{M^2}{H^2}\ .
\eeq
The bispectrum for trapped inflation still satisfies the single-field consistency condition~\cite{Maldacena:2002vr, Creminelli:2004yq}, as proved in~\cite{LopezNacir:2012rm}.

\item[$\triangleright$]  {\it Secondary gravitational waves.}---The produced $\psi$ particles can also be a {\it{classical}} source of gravitational waves. Refs.~\cite{Senatore:2011sp, Cook:2011hg, Barnaby:2012xt} studied EFT variations of trapped inflation in which this source of tensor fluctuations can sometimes be competitive with the quantum-mechanical result~(\ref{equ:Dh}).  Their examples go beyond the simplest models of trapped inflation and have not yet been realized in string theory. Moreover, it remains to be checked whether the regime that produces large tensors is consistent with existing constraints on non-Gaussianity.
\end{itemize}

\vskip 4pt
\noindent
{\it Unwinding inflation.}---The  phenomenology of unwinding inflation is just beginning to be explored, and more detailed realizations in string theory will be required to solidify the predictions of the model.  The  signatures depend strongly on the D-brane dimension $p$: for $p=3$, fluctuations in open string  production provide the dominant source of perturbations, but the compactification must be highly anisotropic, while for $p=4$ ($p=5$) the scalar perturbations receive a 10\% (1\%) contribution from  open strings.\footnote{In the case of $p=6$  the Lorentz factor exceeds the limit of eq.~(\ref{equ:GCon}),  while $p=7$ and $p=8$  are clearly incompatible with  metastable compactification \cite{DAmico:2012sz}.}
Possible  signatures
include a modest level of tensor perturbations ($r\lesssim 10^{-2}$),  equilateral non-Gaussianity,  and (for $p=4$)  oscillations in the spectrum from modulations of the open string pair production rate.

\vskip 4pt
\noindent
{\it Chromo-natural inflation.}---String theory realizations of chromo-natural inflation
are not yet developed enough to make robust predictions for observables.  Taken at face value, the original model (\ref{equ:Schromo2}) is in conflict with the CMB data~\cite{Adshead:2013nka}:
it either predicts a spectral tilt that is too red, overproduces gravitational waves, or both.
Nevertheless, it remains interesting to explore whether large Chern-Simons couplings can arise in consistent string compactifications and if models with viable phenomenology can be constructed.

\chapter{Conclusions and Outlook}
\label{sec:Outlook}

\begin{quote}
{\footnotesize Our mistake is not that we take our theories too seriously, but that we do not take them seriously enough. It is always hard to realize that these numbers and equations we play with at our desks have something to do with the real world. Even worse, there often seems to be a general agreement that certain phenomena are just not fit subjects for respectable theoretical and observational effort.} \vskip 0.1pt \hfill
{\footnotesize Steven Weinberg, {\it on the Big Bang model}~\cite{Weinberg:1977ji}.}
\end{quote}

\vspace{0.5cm}

Consistent theories of quantum gravity do not grow on trees.  After a search spanning nearly a century, string theory is the only known example of such a theory.  Of course, it does not follow that string theory describes our universe: mathematical consistency is a necessary requirement, but it is far from sufficient.  To connect string theory to particle physics and cosmology,  we must seek guidance from terrestrial experiments and from observations of the cosmos.  One should not be surprised that experimental evidence is elusive, for quantum gravity is naturally relevant at scales many orders of magnitude beyond those accessed on Earth.  Running the theory to low energies and extracting predictions that are sensitive to its high-scale origin has proved challenging.  However, the early universe provides an arena where ideas about quantum gravity can be tested, and the initial singularity of the Big Bang model is a prime example where a theory of quantum gravity is compulsory.
Quantum fluctuations of the metric during inflation, imprinted in primordial B-mode perturbations of the CMB, are the most vivid evidence conceivable for the  reality of quantum gravity, and for the  significance of quantum gravity in the early history of our universe.

Inflation defers the singularity problem, allowing us to make predictions for the initial conditions that emerge from the aftermath of the Big Bang.  However, as we  have shown, the inflationary mechanism retains
a subtle sensitivity to Planck-scale interactions.   This is both a challenge for microscopic theories of inflation, as well as an opportunity for using the early universe as a window on Planck-scale physics.  To fulfill this promise, inflationary scenarios in string theory must be developed to an unprecedented level of completeness and sophistication.

\vskip 4pt
The last decade of research on inflation in string theory has witnessed a number of significant advances.
The development of methods of moduli stabilization has led to vastly improved technical capabilities, and in turn to a sharply improved understanding of metastable string compactifications and of the associated inflationary models.  In special cases it has been possible to characterize the Planck-suppressed corrections to the inflaton action,  leading to the first existence proofs of inflation in string theory.  Furthermore, the symmetry structures required for large-field inflation are now better understood.  In addition, techniques for studying the dynamics of theories with many moduli have recently emerged.
Moreover, string inflation has expanded and refined our ideas for inflationary mechanisms in effective field theory. Consistency conditions in string theory have suggested that certain classes of models conceived as effective theories may not admit ultraviolet completions, and at the same time, confronting these restrictions has led to novel ideas for consistent low-energy field theories. Thus---as in many problems outside cosmology---string theory has frequently yielded solutions with unanticipated properties, and has served as a generating function
for ideas that were hard to perceive from a purely low-energy point of view.

\vskip 4pt
On the other hand, many
critical challenges still remain.
Our understanding of reheating and of the connection between the inflationary sector and the Standard Model degrees of freedom is tenuous.
Non-supersymmetric solutions of string theory, particularly de Sitter solutions, remain much less controlled than their supersymmetric counterparts.  This continues to be a zeroth-order challenge for deriving inflation from string theory, and  has stymied many attempts to develop inflationary scenarios outside of type IIB  string theory.
Moreover, time-dependent solutions in string compactifications  are barely understood beyond the adiabatic approximation.
Furthermore, in most cases the Planck-suppressed corrections to the inflationary action are only partially characterized.
Finally, and most importantly,  there is not a single observation  that gives direct evidence for a string-theoretic origin
for inflation: although an  unambiguous detection of gravitational waves  produced by quantum fluctuations of the metric during inflation would  directly prove the quantization of the gravitational field,  discerning  the character of the quantum gravity theory requires more refined observations.
At present, we are led to inflation in string theory  by a web of inference  involving the success of inflation in  effective field theory, the naturalness principle in particle  physics, and
the unique status of string theory as an ultraviolet completion of gravity.

\vskip 4pt
A striking feature of present observations is the extraordinary simplicity of
the primordial curvature
fluctuations, which
are approximately Gaussian, adiabatic, and nearly scale-invariant.
In contrast, the ultraviolet completions presented in this book are complex, involving many interacting fields and a landscape of quantized parameters.
Should the  `simple' observations  be read as  evidence against `complicated'  models of inflation in string theory?
We do not believe so:
although the simplicity of the data motivates considering simple {{\it effective}} theories of inflation, it does not constrain the  ultraviolet completions in the same way.  As an analogy, the Fermi theory of beta decay is far simpler --- in terms of a counting of parameters --- than the Standard Model, but is merely a  low-energy effective description. Indeed, the whole point in using effective theories is that they are simpler to use than their ultraviolet completions.
Even so, it remains important to understand
whether the simplicity of the data can
emerge from the apparent complexity of the ultraviolet completion: one should determine which details of the short-distance physics decouple and which leave subtle traces in the data.

\vskip 4pt
\index{eternal inflation} \index{measure problem}
We have largely avoided discussing deep issues involving the initial conditions for inflation, including the global view of eternal inflation~\cite{Vilenkin:1983xq, Linde:1986fd, Linde:1993xx, Linde:1993nz}, the associated {\it measure problem}~\cite{Vilenkin:1994ua, Garriga:2005av, Garriga:2008ks, Bousso:2006ev, Bousso:2006ge, Bousso:2008hz, DeSimone:2008if, Gibbons:2006pa}, and the geodesic incompleteness of inflation in the past~\cite{Borde:2001nh}.  String theory has inspired
several compelling
approaches to these questions, but
no complete
solutions have been advanced.
Many authors have noted that these unresolved problems threaten the predictivity of the inflationary paradigm.
Of course, once an inflationary phase begins in a particular region of field space, clear and specific predictions do emerge.
On the other hand, it is an important open problem to determine the relative probabilities of different inflationary models in a broader setting.
More generally, deriving specific predictions from the string landscape as a whole, rather than from individual models, is a distant goal that could
require a new approach to the measure problem.

\vskip 4pt
In closing, we would like to emphasize that the study of inflation in string theory has advanced to a stage where a properly-constructed model can be falsified: indeed, many models have already been falsified by recent observations, while others are under observational pressure.
Proper construction of models of string inflation, however, is a subtle art.  We have railed against incorrect predictions rooted in oversimplified effective theories, and have catalogued the pitfalls in attempts to compute observational signatures.  Our hope is that the reader will use the ideas and techniques presented here to derive predictions that illuminate the history of the universe and shed light on the nature of quantum gravity. 



\bibliographystyle{utphys}
\bibliography{References}

\end{document}